# The Computer System Trail

# The Computer System Trail

---

## Selected Works in System Design

Sushant Kumar Gupta

*Author*
Sushant Kumar Gupta

# About the Author

Sushant Kumar Gupta is a software engineer specializing in the design and development of large-scale, distributed data infrastructure. He has a strong foundation in distributed systems and system performance optimization, complemented by extensive practical experience in networking, blockchain technologies, and cloud computing. He is also an active researcher and published author in the field of computer systems.

# Preface

Welcome to **The Computer System Trail**, your guide to understanding computer systems through the lens of influential research papers!

The book delves into the core of modern computing by exploring key areas like datastores, big data, operating systems, and distributed systems. Unlike AI summaries, the chapters here are crafted from my extensive notes taken while reading hundreds of research papers and engaging in numerous discussions within academic and industry circles. The presented papers represent a blend of academic and industry perspectives, offering insights into both theoretical foundations and practical implementations. The academic papers would emphasize theory and scientific rigor, while industry papers would focus on practical considerations and implementation details.

To maximize your learning, I suggest following this order for each chapter:

1. Read the research paper *at least* twice from start to end (including any appendices).
2. Then, read the corresponding book chapter.
3. Finally, read the research paper one more time.

The key point is that thoroughly reading the papers is essential; the notes merely offer supplementary explanations. Without engaging with the papers themselves, these clarifications will have limited impact on deepening your understanding. The iterative approach will help you progressively build your understanding, moving from the original source to a guided explanation and back again. Though not strictly required, having a basic foundation in computer systems concepts, perhaps from your college studies, will be beneficial. A quick refresher might be a good starting point.

Keep in mind that the ideas presented in these papers are interconnected. You'll often find that concepts from different papers relate to and build upon each other. Don't hesitate to revisit previous chapters as new connections emerge. It's perfectly normal if some details aren't immediately clear. By working through all the chapters, you'll gradually develop a robust and interconnected understanding of these critical concepts. Expect to reread chapters as your knowledge deepens. Ultimately, you'll gain a strong, holistic grasp of the subject matter.

The book is not intended to replace system design courses. Instead, it aims to offer something more profound. If you're just beginning your journey into system design, I would highly recommend prioritizing the reading of these foundational papers over enrolling in online courses. Here's why:

1. **Detail-Oriented:** Reading the original papers provides a complete picture, explaining not just the "what" and "how" of a system, but also the crucial "why" behind its design choices. Research papers offer a deep dive into system design, evaluation, and related work, backed by the rigorous review of leading researchers. While their detailed nature can initially make them challenging to navigate, the effort pays off. As you become more comfortable with reading them, your understanding of the system will align with the insights of its very creators.
2. **Beyond Mediocrity:** Engaging with these fundamental ideas will push you beyond surface-level understanding and foster a deeper, more nuanced perspective.
3. **Critical Thinking Enhancement:** Analyzing research papers actively cultivates your critical thinking skills, enabling you to evaluate and synthesize complex information.
4. **Reading Habits and Patience:** This process will significantly improve your reading comprehension and develop patience, essential skills for any technical discipline.

If you're new to reading research papers, anticipate spending around one week on each paper, assuming a commitment of 1-2 hours of reading per day. Keep in mind that some chapters will involve reading multiple papers. Factoring in off-weeks and busier periods, completing all the material will likely take approximately one year (or 800 hours). While this time commitment may seem significant, the long-term benefits to your understanding and career will undoubtedly be worth the investment.

To find answers to questions about these papers, I would recommend starting with a Google search, as many queries can be resolved by reading relevant web chapters.

In fact, exploring other people's explanations of the same paper from forums like StackOverflow would provide valuable alternative perspectives.

If you have any other questions, feel free to reach out to me directly, and I will do my best to answer them. While I've read and discussed these papers extensively, I don't claim to be an expert. My familiarity comes from repeated readings and group discussions. Therefore, please understand that I may not be able to answer every detailed question about specific paragraphs within the papers.

Lastly, **the book is free**, reflecting my belief that high-quality education should be freely available on the internet. I often find that commercially driven education caters quick success at the cost of depth.

I hope you are ready to start! I wish you the best on the journey.

<div align="right">SUSHANT KUMAR GUPTA<br>*Author*</div>

# 1. Moving Beyond End-to-End Path Information to Optimize CDN Performance

(1) Krishnan, R.; Madhyastha, H. V.; Srinivasan, S.; Jain, S.; Krishnamurthy, A.; Anderson, T.; Gao, J. Moving beyond End-to-End Path Information to Optimize CDN Performance. In Proceedings of the 9th ACM SIGCOMM conference on Internet measurement; 2009; pp 190–201.

This influential paper from Google was authored by Rupa Krishnan et. al. The paper came out in 2009, the time when computer networks were constantly being improved for the next generation of the Internet. The paper explores Content Delivery Networks (CDNs), which became hot in the subsequent decade. Several internet-giants like Google, Facebook, and Netflix rely on CDNs to deliver content to the end users.

In this insight, we will start with OASIS - the inspiration behind CDN architectures. Then we will delve into some networking concepts relevant to the paper and finally look into the mitigation techniques applied by the authors to reduce end-to-end latency of Google's CDN. As a bonus, we will also go through the architecture of Coral CDN - which introduced a multi-layered architecture to CDNs.

### 1.1. OASIS

OASIS [39] was developed by M. Freedman et. al. It addresses the critical challenge for the Internet: locating the service replica with the lowest latency for a given client.

Prior to OASIS, clients naively pinged every service replica to determine the fastest one based on round-trip time (RTT). This was highly accurate, however, this approach suffered from excessive probing and computationally expensive comparisons.

#### *1.1.1. Architecture*

OASIS introduced a two-tier architecture as shown in Figure 1. OASIS maintains replicas that redirect a client to the nearest replica of any service.

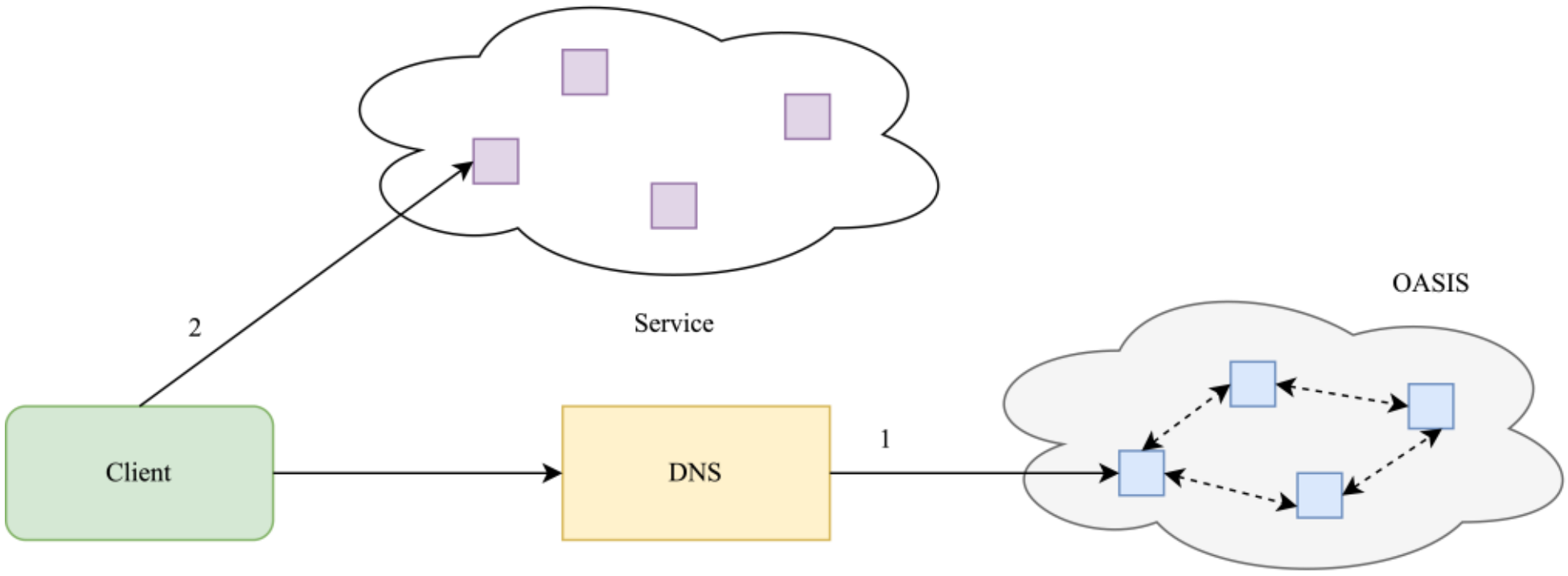

**Figure 1**: The OASIS architecture.

The OASIS replicas:

1. Possess global membership information about all service replicas.
2. Employ **epidemic gossiping** to detect and handle failures.
3. Utilize **consistent hashing** for efficient replica placement.

Indeed, the 2000s witnessed a surge in decentralized protocols, such as those employing consistent hashing. OASIS is an example of such an approach.

OASIS was seamlessly integrated into existing protocols like DNS and HTTP, demonstrating its practicality and potential for widespread adoption.

*1.1.2. Algorithm*

**Step 1**

The client's IP address is converted to an **IP prefix**. An IP prefix, or network prefix, is a group of IP addresses that identifies a network. IP prefixes help organize IP addresses and the devices connected to the Internet. For instance:

**18.26.4.9 → 18.0.0.0/8**

**18.0.0.0/8** is a CIDR notation representing a network address in routing. The **/n** suffix in CIDR indicates the number of bits in the subnet mask that are set to **1**. In this specific case, **/8** signifies a subnet mask of **255.0.0.0**. To derive the network prefix from an IP address, we need to perform a bitwise AND operation between the IP address and the subnet mask.

**Step 2**

This IP prefix is mapped to a unique geographic proximity. There is a strong correlation between geographic proximity and RTT, the OASIS replica redirects the client to the geographically closest service replica.

If multiple replicas of service reside in the same geographic region, the client probes only those replicas to select the one with the lowest RTT, significantly reducing the probe space.

## 1.2. Google CDN Architecture

Google's CDN architecture shares similarities with OASIS, but with a key distinction: it utilizes the IP prefix of the DNS server initiating the request for redirection, rather than the client's IP address.

### *1.2.1 Goal*

The authors contend that solely relying on RTT (a.k.a. end-to-end path information) for redirection, may not consistently deliver the best possible **Quality of Service (QoS)** to the client.

They delve into various factors that can contribute to suboptimal QoS, beyond simple RTT. The paper then explores methods for identifying and mitigating these contributing factors.

Before delving deeper into their proposed techniques, it's crucial to establish a common understanding of the key network terminologies employed throughout the paper.

## 1.3. Networking Architecture

### *1.3.1. Routing*

Computers within a Local Area Network (LAN), such as your school network or a cluster of machines in a data center, communicate directly. These networks typically employ network switches to efficiently forward packets to their intended destinations based on their unique **Media Access Control (MAC)** addresses, which operate at the Ethernet layer.

However, when a packet needs to traverse beyond the confines of a local network, routing becomes essential. By examining the destination IP address of each packet, routers consult their routing tables to determine the optimal next hop in the network, sending the packet closer to its ultimate destination. These routing tables are dynamically calculated and maintained through sophisticated distributed algorithms.

### *1.3.2. Point of Presence*

A Point of Presence (PoP) is a designated entry point where end users can connect to an **Internet Service Provider (ISP)**. Each ISP maintains a network of PoPs.

Within a single PoP, multiple routers may be present, allowing for various paths for data packets to travel from the end users to PoP.

### *1.3.3. Autonomous System*

An Autonomous System (AS) is a collection of interconnected networks under the control of a single administrative entity. Each AS operates with its own distinct routing policies, determining how traffic flows within its boundaries. Common examples of entities operating ASes include ISPs, large corporations (like Google), universities, and government agencies.

Google's internal network, connecting its vast infrastructure (like the Borg cluster), functions as an AS with its own internal routing policies.

**Border Gateway Routers (BGP Routers)** are specialized routers residing at the edges of an AS, responsible for exchanging routing information with other ASes.

### *1.3.4. Traceroute*

The performance of Internet traffic can be impacted by the stability and condition of routers along the path, some of which may be older or less reliable. Traceroute [40] is a widely used network diagnostic tool that reveals the sequence of routers traversed by packets traveling from a source to a destination. The router hops displayed by traceroute signifies the number of routing tables consulted along the path, providing insights into network topology.

Traceroute operates by sending a series of ICMP [41] (Internet Control Message Protocol) packets with increasing Time-to-Live (TTL) values as shown in Figure 2.

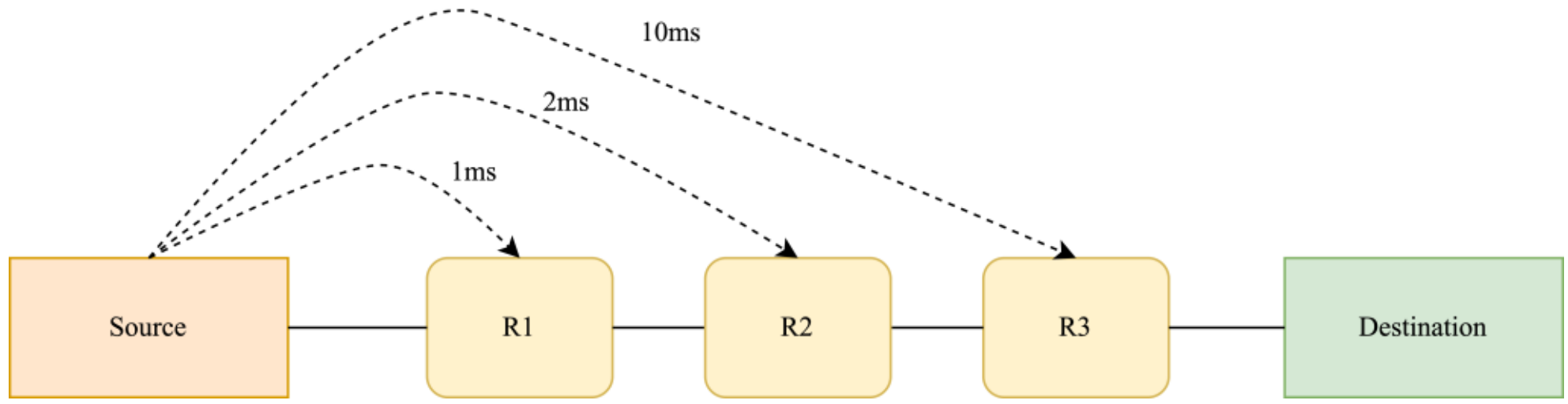

**Figure 2**: Example working of Traceroute.

A fundamental characteristic of the Internet is that the path taken by packets from source **A** to destination **B** may differ significantly from the return path from **B** to **A**. For example:

- **A → X → Y → B**
- **B → Y → Z → X → A**

Traceroute can only measure the RTT for each hop, not the one-way travel time. Thus, while traceroute effectively reveals the forward path, determining the reverse path requires more complex analysis.

### 1.4. iPlane

iPlane is a system designed to predict path properties between any two points on the Internet. iPlane uses traceroutes to all Internet prefixes from numerous vantage points (e.g., PlanetLab [42] servers).

By analyzing the collected traceroute information, iPlane clusters routers into PoPs. Routers are grouped together if they exhibit the following characteristics:

- Respond to traceroutes with the same source IP address.
- Display similar RTT values across different vantage points.

This clustering technique enables iPlane to effectively model and predict Internet path characteristics, providing valuable insights into network topology and performance.

### 1.5. Latency Cause Detection Techniques

This paper explores several techniques for detecting latency issues in Google's CDN:

**1. Redirection**

**Issue:** Clients may experience increased latency when they are not redirected to the geographically closest CDN node. This can occur, for example, when the closest node is overloaded.

**Detection:** CDN nodes themselves can detect mismatched redirection by flagging IP prefixes that they are not responsible for serving.

**2. Prefix Latency Inflation**

**Issue:** Even when multiple IP prefixes are served by the same CDN node, some prefixes may exhibit significantly higher latency than others.

**Detection:** This is identified by performing traceroute analysis. For instance, if the route path is:

**CDN → R1 (1ms) → R2 (2ms) → R3 (100ms) → Client**

A significant delay between **R2** and **R3** (100ms) is considered anomalous, especially when the delay between CDN and **R2** is only 2ms and the expected inter-hop delay is typically less than 40ms anywhere on earth. This suggests a circuitous path back from **R3**.

**3. Queueing Delays**

**Issue:** Even within a single prefix, a significant difference between the median and minimum RTT observed by clients can indicate either different paths or queueing delays.

**Observation:** RTT inflation within a prefix is unaffected by:

- Changes in routes throughout the day.
- Changes in the PoP paths (determined using iPlane data).
- Changes in the AS path.

The authors conclude that persistent RTT inflation is likely caused by queueing delays within the network.

Only (1) and (2) are addressed by the authors. In summary,

(1) identifies anomalous prefixes facing redirection, and
(2) identifies prefixes within a CDN node facing anomalously high RTT.

### 1.6. Mitigation

The paper explores several approaches to address the observed inflated RTT issue:

- **Direct Peering:** Establishing direct peering connections between Google and the relevant ASes for the affected prefixes. This solution is exemplified in Case 1, where a new peering link was added between Google and PhilISP1.
- **Increased Peering Capacity:** Enhancing the capacity of existing peering links between Google and ASes. This is demonstrated in Case 3, where a peering connection between Google and PhilISP2 existed but suffered from insufficient capacity.
- **Routing Configuration Optimization:** Correcting routing configurations on border routers within Google's or the AS's network. Case 4 illustrates this approach, where Google's network administrators adjusted routing configurations to enable shorter reverse paths between Google and a JapanISP.
- **Traffic Engineering Techniques:** Implementing various traffic engineering approaches to optimize network traffic flow.

### 1.7. Bonus: Coral CDN

Among the related works discussed in the paper, Coral CDN [43] stands out. Coral CDN was developed by the authors of OASIS. It was a free service, designed to mirror web content.

Coral CDN's primary use case was to mitigate **slashdotting**, a phenomenon where a website, particularly smaller ones, experiences a sudden and overwhelming surge in traffic after being linked by a popular website.

The most distinctive feature of Coral CDN is its utilization of a Distributed Sloppy Hash Table [44] (DSHT). Unlike traditional **Distributed Hash Tables (DHTs)** that rely on a single hash ring, DSHT employs multiple concentric Chord hash rings arranged hierarchically. DSHT can be adapted to work with other hashing protocols like Kademlia.

*1.7.1. Interface*

Coral provides a simple interface for higher-level applications:

- **put (key, val, ttl)**: Inserts a key-value mapping with a specified TTL for the entry.
- **get (key)**: Retrieves a subset of values associated with the given key.

*1.7.2. Implementation*

The Coral CDN implementation utilizes a three-tiered hierarchical structure, as shown in Figure 3:

- **Regional Layer:** Connects nodes within the same geographical region.
- **Continental Layer:** Connects nodes within a continent.
- **Global Layer:** Connects nodes globally.

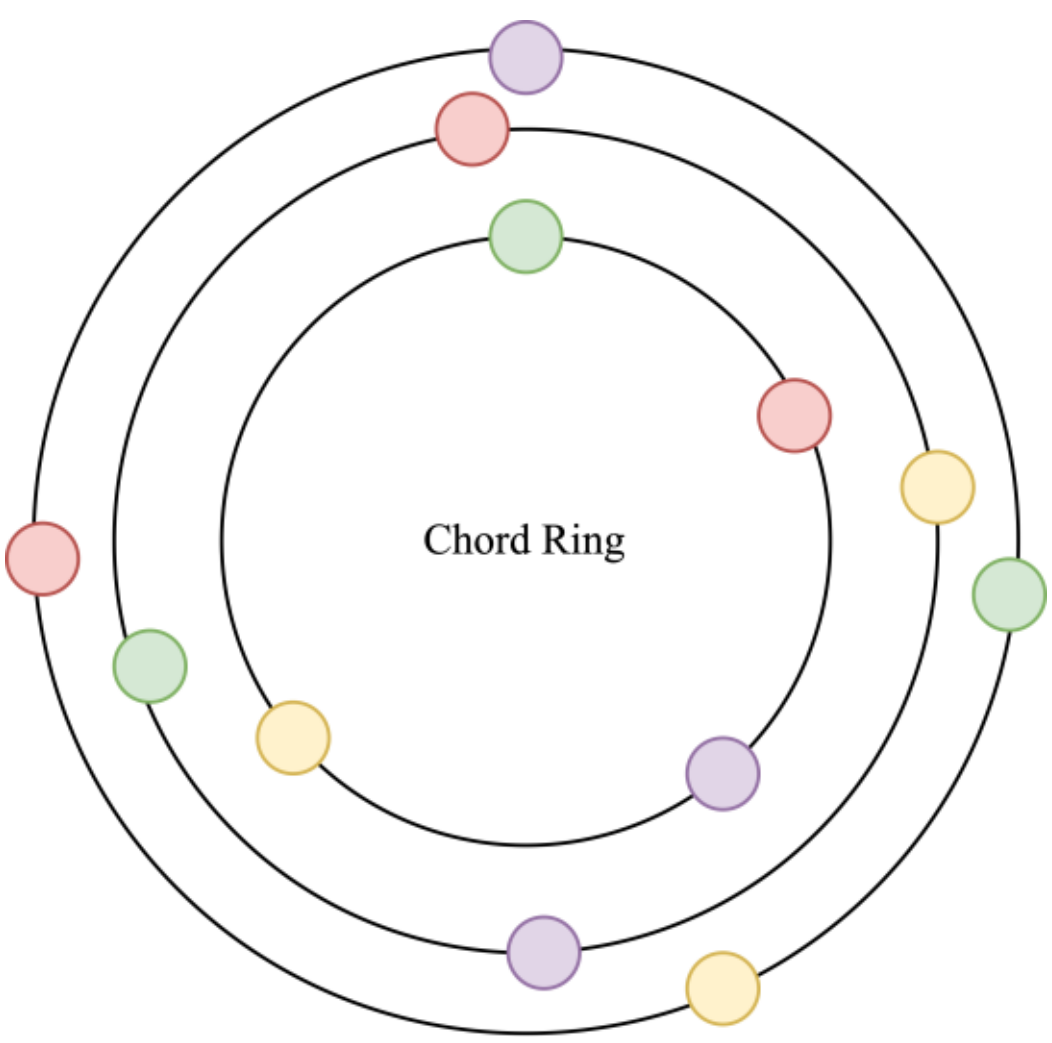

**Figure 3**: Coral CDN hierarchical structure.

Nodes have the same Id across layers, but their assigned ranges vary due to variation in the hashing modulo. During the **put** operation, the value is replicated across all three layers, ensuring redundancy. Conversely, during the **get** operation, the system prioritizes retrieving the value from the local regional layer.

### 1.8. Paper Remarks

The paper does not propose novel networking architectures; instead, it offers a robust methodology for network analysis. Furthermore, it provides significant insights into Google's extensive global infrastructure. The analysis of queuing delays and the characterization of circuitous reverse paths are especially compelling aspects of the work.

# 2. A Comparison of Software and Hardware Techniques for x86 Virtualization

(2) Adams, K.; Agesen, O. A Comparison of Software and Hardware Techniques for X86 Virtualization. ACM Sigplan Notices 2006, 41 (11), 2–13.

Following our discussion on CDNs, we will move on to an advanced computer system concept: virtualization. Learning how virtualization works provides great insight into how operating systems function and how they facilitate hardware interactions. While most standalone workstations use a single native operating system, almost all operating systems in data centers today run within a virtualized environment. In our journey to understand virtualization, we will also learn about the microprocessor and memory - the two most important components of a computer system; a computer can function effectively with just these two, even if all other components are removed.

This paper in particular was authored by the esteemed Keith Adams and Ole Agesen at VMWare. Written in 2006, the paper discusses software virtualization and compares it with hardware virtualization.

In this insight, we will begin with a deep dive into microprocessors, exploring instruction sets and architecture. We will also briefly touch upon memory, though the next chapter will contain a more rigorous discussion on that topic. Following this, we will introduce the operating system - the central software that layers the interaction between hardware and user applications. Finally, we will explore the different implementations of virtualization and compare them. This insight assumes the reader has a basic familiarity with microprocessors and operating system concepts.

### 2.1. Microprocessor

At the heart of every computer lies the microprocessor, the engine responsible for executing **instructions** and performing computations. In essence, it's the "brain" of the system. In this chapter, we will use the term **microprocessor** and **CPU** interchangeably.

Each instruction consists of two fundamental parts: the **opcode**, which specifies the operation to be performed (e.g., addition, subtraction), and one or more **operands**, which represent the data to be manipulated. While instructions are stored and processed in binary form, **assembly language**, a low-level symbolic representation, provides a more human-readable format.

### *2.1.1. Instruction Set Architecture*

The microprocessor's "manual", known as the **Instruction Set Architecture (ISA)**, defines the complete set of instructions it can execute. Think of it as the control panel of a machine, outlining every possible action. For example, the x86 ISA is a widely adopted standard, supported by a diverse range of processors, including the 8086 (the original x86 microprocessor), Pentium, Celeron, Xeon, Atom, and the Core series.

This chapter focuses on SISD microprocessors, which execute a single instruction on a single data stream. This contrasts with GPUs, which utilize SIMD architecture to apply a single instruction to multiple data streams.

### *2.1.2. Hardware Architecture*

The term "microprocessor" often refers to a single processing core. This core is the fundamental unit of execution. Microprocessors are typically integrated into **sockets**, which can accommodate one or more microprocessors, as shown in Figure 2.1. While most personal computers utilize a single socket, server-grade processors like Intel Xeon frequently reside in multi-socket systems.

A microprocessor's internal architecture comprises several key components, as shown in Figure 4:

- **Execution Unit:** This is where the actual computations occur, primarily through the **Arithmetic Logic Unit (ALU)**, which performs arithmetic and logical operations.
- **Control Unit:** The control unit acts as the orchestrator for the flow of instructions and data within the microprocessor. It fetches instructions from memory, decodes them, and generates the necessary control signals to coordinate the other components.

- **Registers:** Registers are small, high-speed storage locations within the microprocessor that hold data and instructions temporarily during processing. They are crucial for rapid access and manipulation of frequently used data.

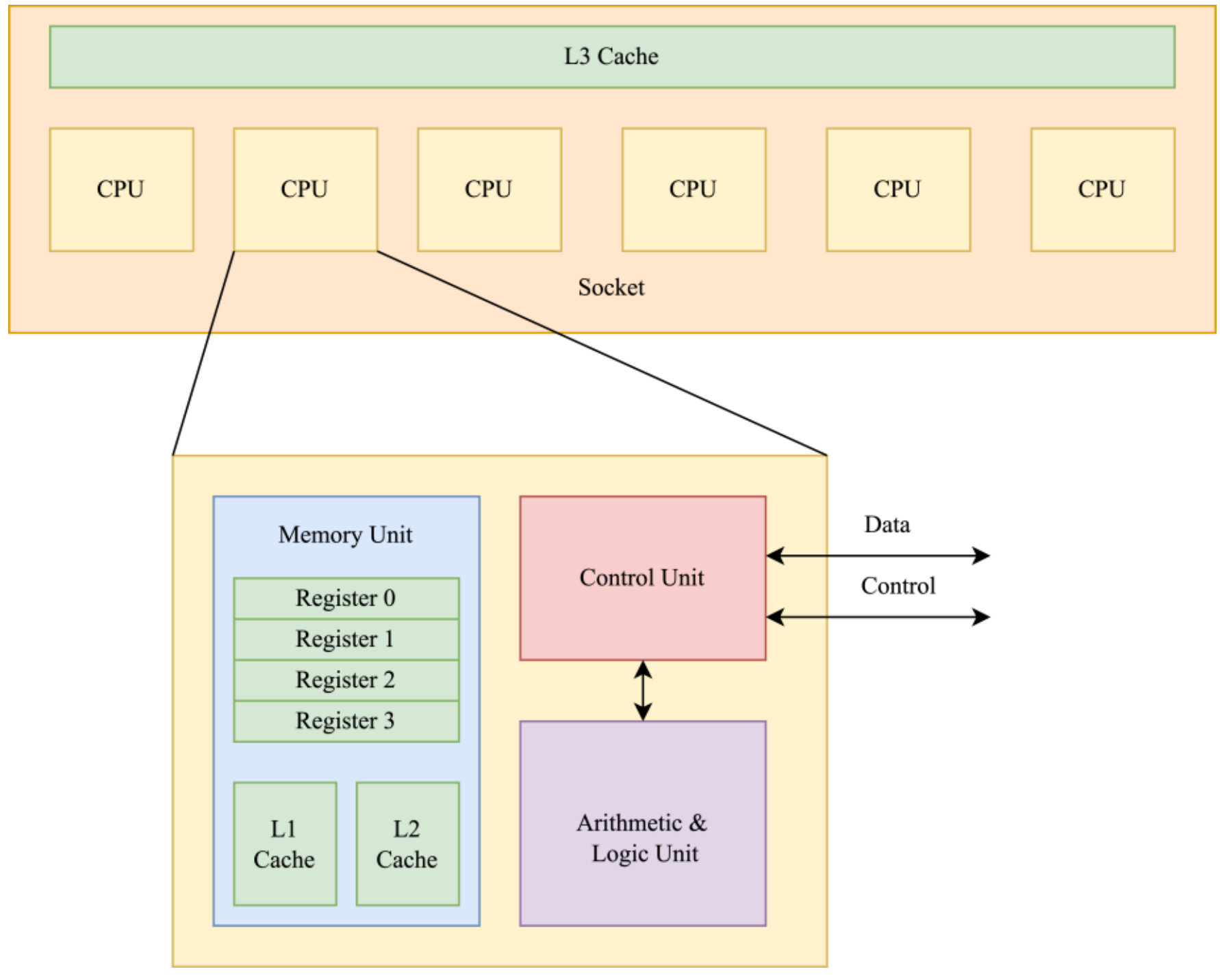

**Figure 4**: Hardware architecture of a CPU.

*2.1.3. Fetch-Decode-Execute Cycle*

The execution of instructions on a single microprocessor follows the fetch-decode-execute cycle:

1. **Fetch:** The control unit retrieves an instruction from memory.
2. **Decode:** The control unit deciphers the instruction to determine the operation to be performed.
3. **Execute:** The execution unit performs the specified operation.

To enhance performance, modern microprocessors employ **pipelining**, a technique that allows multiple instructions to be processed concurrently in different stages of the cycle. This overlap significantly increases the throughput of the processor.

#### 2.1.4. CPU Cache

Caches are small, high-speed memory units that store frequently accessed data and instructions, enabling faster retrieval. There are typically multiple levels of cache:

- **L1 Cache:** The fastest and smallest level of cache, dedicated to each individual core within the microprocessor. It is often split into **L1 instruction cache** (L1i) and **L1 data cache** (L1d). L1 cache sizes vary depending upon the CPU, but are often in the tens of KBs.
- **L2 Cache:** A larger and slightly slower cache than L1, often shared between cores on the same chip. L2 cache sizes also vary greatly, but are commonly in the hundreds of KBs, or multiple MBs.
- **L3 Cache:** The largest and slowest level of cache, typically shared by all cores on the socket. L3 cache sizes can range from several MBs to tens of MBs.

By utilizing cache memory, microprocessors can significantly reduce the time required to access data and instructions, thereby improving overall performance.

To prevent data inconsistencies when a single process runs on multiple CPU cores, each with its own local cache, cache coherence is vital. This complex problem is resolved by hardware-level mechanisms.

#### 2.1.5. CPU Registers

CPU registers are essential components that hold data and instructions that the CPU is actively working on. There are a limited set of registers that are available. Let's go through some of the registers that x86 supports.

##### 2.1.5.1. General-Purpose Registers

These are versatile registers used for various operations, including arithmetic, logical, and data movement. They are often labeled as **R0**, **R1**, **R2**, etc., or have specific names like **%a**, **%b**, **%c**, **%d** in the x86 architecture.

The naming conventions of x86 registers are deeply rooted in the architecture's evolution. Register **%a** for instance, originated as the **accumulator register**, a central component optimized for arithmetic operations. However, the specific roles of these registers can differ considerably based on the compiler being used. Nowadays, **%a** is commonly designated for storing function return values, this isn't a strict, universally applied rule.

In addition to the more historically **named registers**, the x86 architecture also incorporates **numbered registers** from **%r8** to **%r15**, which offer increased flexibility for modern programming tasks.

In x86 architecture, registers are often identified by prefixes indicating their size:

- The prefix **r** and suffix **x** for named ones denotes a 64-bit register (e.g., **%rax**, **%rbx**, **%rcx**, **%rdx**).
- The prefix **e** and suffix **x** for named ones and a suffix **d** for numbered ones signifies the 32-bit version (e.g., **%eax**, **%ebx**, **%ecx**, **%edx, %r8d, %r9d, %r10d, %r11d, %r12d, %r13d, %r14d, %r15d**). It is important to understand that the 32-bit registers are simply the lower 32 bits of their corresponding 64-bit registers.
- The suffix **x** for named ones and the suffix **w** for numbered ones denotes a 16-bit version (e.g., **%ax**, %**bx**, %**cx**, %**dx, %r8w, %r9w, %r10w, %r11w, %r12w, %r13w, %r14w, %r15w**).
- Finally, the suffix **l** for named ones and the suffix **b** for numbered ones denotes is used for 8-bit version (e.g., **%al**, **%bl**, **%cl**, **%dl**, %**r8b**, **%r9b**, **%r10b**, **%r11b**, **%r12b**, **%r13b**, **%r14b**, **%r15b**).

The complete list of general-purpose registers  is shown in Table 1.

**Table 1**: List of x86 general-purpose registers.

| 64-bit | Lower 32 bits | Lower 16 bits | Lower 8 bits |
|---|---|---|---|
| %rax | %eax | %ax | %al |
| %rbx | %ebx | %bx | %bl |
| %rcx | %ecx | %cx | %cl |
| %rdx | %edx | %dx | %dl |
| %r8 | %r8d | %r8w | %r8b |
| %r9 | %r9d | %r9w | %r9b |
| %r10 | %r10d | %r10w | %r10b |
| %r11 | %r11d | %r11w | %r11b |
| %r12 | %r12d | %r12w | %r12b |
| %r13 | %r13d | %r13w | %r13b |
| %r14 | %r14d | %r14w | %r14b |
| %r15 | %r15d | %r15w | %r15b |

**Note**: There is also a suffix h that corresponds to the upper 8 bits of the lower 16 bits.

#### 2.1.5.2. Index Registers

Index registers are frequently used in conjunction with loop counters to iterate through data structures. Common examples include the **%si** (Source Index), **%di** (Destination Index) registers. These are used in conjunction with the **%bp** (Base Pointer) register. In the 32-bit x86 architecture, their extended counterparts are **%esi**, **%edi**, and **%ebp**, respectively. In the 64-bit architecture, they are **%rsp**, **%rdi**, and **%rbp**.

#### 2.1.5.3. Special-Purpose Registers

- **Program Counter (PC):** Holds the **relative** memory address of the next instruction to be executed. Instructions are fetched from this address. In the x86 architecture, the PC is known as the **Instruction Pointer (IP)** (**%ip, %eip, %rip**).
- **Stack Pointer (%sp, %esp, %rsp):** Points to the top of the stack in memory, used for function calls and temporary data storage.
- **Status Register (or Flags Register):** Contains flags that indicate the status of the CPU and the results of operations. Common flags include:
  - **Carry Flag (C)**: Indicates a carry or borrow in an arithmetic operation.
  - **Zero Flag (Z)**: Indicates that the result of an operation is zero.
  - **Overflow Flag (O)**: Indicates an overflow in an arithmetic operation.
  - **Sign Flag (S)**: Indicates the sign of the result.

#### 2.1.5.4. Segment Registers

A program's memory space is organized into distinct segments, each serving a specific purpose. Essential memory segments include:

- **Code segment:** This segment holds the program's executable instructions.
- **Data segment:** This segment stores initialized global and static variables. These variables have a fixed memory location throughout the program's execution.
- **BSS (Block Started by Symbol) segment:** This segment holds uninitialized global and static variables. These variables are typically initialized to zero or null before the program begins execution.

- **Stack segment:** This segment manages the dynamic allocation of memory for local variables and function call information. It operates on a Last-In, First-Out (LIFO) principle.
- **Heap segment:** This segment is used for dynamic memory allocation during program execution, allowing programs to request and release memory as needed. This is where dynamically allocated memory from functions like **malloc()** or **new** are stored.

Corresponding to each of these segments, we have registers storing their addresses:

- **%cs:** Points to the **code segment**. It is used in conjunction with the IP to determine the memory address of the next instruction to be executed. The CS register also encodes the current privilege level (described later) of the CPU.
- **%ds**: Points to the **data segment**.
- **%es**: Points to **extra data segment**. Many compilers use extra data segment for strings.
- **%fs and %gs:** General-purpose segment register that can point to any data segment. However, its most common use is to point to thread-local storage [45].
- **%ss**: Points to the stack segment.

There are no dedicated registers for BSS and heap.

#### 2.1.5.5. Descriptor Table Registers

Segment information is stored in descriptor tables:

- **Global Descriptor Table (GDT):** This table, managed by the operating system, contains memory segment information for the entire system.
- **Local Descriptor Table (LDT):** Each process maintains its own LDT, which stores memory segment information specific to that process.

The base addresses of these tables are stored in the following registers:

- The **GDTR** (**Global Descriptor Table Register**) holds the base address of the GDT.

- The **LDTR** (**Local Descriptor Table Register**) holds a selector that points to an entry in the GDT, which in turn points to the base address of the current process's LDT.

Additionally, the **Interrupt Descriptor Table (IDT)** stores information about interrupt handlers. The base address of the IDT is held in the **IDTR** (**Interrupt Descriptor Table Register**).

#### 2.1.5.6. Control Registers

Control registers are used to control the behavior of the microprocessor. With different settings, the microprocessor may behave differently. There are control registers **%cr0** to **%cr15**. However, not all registers are usable. Some of the commonly used control registers are:

- **%cr0**: Controls various processor modes and flags, including protected mode, paging, and enabling/disabling features.
- **%cr2**: Stores the address that caused a page fault.
- **%cr3**: Holds the address of the page table (a.k.a. **Page Table Register**), essential for memory paging.
- **%cr4**: Enables various processor extensions and features, such as PAE and virtualization extensions.
- **%cr8**: **Task Priority Register**, used for prioritizing external interrupts.

### *2.1.6. Addressing Modes*

Addressing modes are crucial aspects of CPU architecture that determine how the operands of an instruction are accessed. They define how the effective address of an operand is calculated. There are several different types of addressing modes.

#### 2.1.6.1. Immediate Addressing

The operand itself is directly embedded within the instruction. For example, the following instruction sets the value of **10** to **%rax**.

```
mov %rax, $10
```

#### 2.1.6.2. Register Addressing

The operand is located in a CPU register. For example, the following instruction sets the value of **%rax** to **%rbx**.

```
mov %rax, %rbx
```

2.1.6.3. Direct Addressing (Absolute Addressing)

The instruction contains the actual memory address of the operand. For example, the following instruction moves the value from memory address **1000** into the **%rax**.

```
mov %rax, [$1000]
```

2.1.6.4. Indirect Addressing

The instruction specifies a register that holds the address of the operand. For example, the following instruction moves the value from the memory address pointed to by the **%rbx** into the **%rax**.

```
mov %rax, [%rbx]
```

2.1.6.5. Indexed Addressing

The effective address is calculated by adding an index register's value to a base address (which can be a constant or another register). For example, the following moves the value from the memory address calculated by adding the **%rbx** (base) and **%rsi** (index) into the **%rax**.

```
mov %rax, [%rbx + %rsi]
```

2.1.6.6. Base-Indexed with Displacement Addressing

The effective address is calculated by adding the values of a base register, an index register, and a displacement (constant offset). For example, the following moves the value from the memory address calculated by adding the **%rbx**, **%rsi** registers, and **10** into the **%rax**.

```
mov %rax, [%rbx + %rsi + $10]
```

2.1.6.7. PC-Relative Addressing

The effective address is calculated relative to the current PC. For example, the following instruction makes the program jump **10** steps from the current instruction address.

```
jmp +$10
```

*2.1.7. Example Program: isPrime*

Let's go through an example program to understand the concepts re-using the same **isPrime** program that shows up in the paper as well.

```
1.  int isPrime(int a) {
2.    for (int i = 2; i < a; i++) {
3.      if (a % i == 0) return 0;
4.    }
5.    return 1;
6.  }

====================================

1.  isPrime:   mov %ecx, %edi
2.             mov %esi, $2
3.             cmp %esi, %ecx
4.             jge prime
5.  nexti:     mov %eax, %ecx
6.             cdq
7.             idiv %esi
8.             test %edx, %edx
9.             jz notPrime
10.            inc %esi
11.            cmp %esi, %ecx
12.            jl nexti
13. prime:     mov %eax, $1
14.            ret
15. notPrime:  xor %eax, %eax
16.            ret
```

The **isPrime** function determines if a given number is prime by iterating from 2 up to one less than the number, checking for divisibility at each step. If any number divides the input, it's not prime; otherwise, it's prime.

Here is a brief refresher on how function call works: before a function is called, the caller's register values are saved to the stack. Function arguments are passed either via the stack or, for a limited number of arguments, in registers. The function's return value is placed in a specific register, and after the function finishes, the stack is popped to restore the caller's register state. The compiler is responsible for doing all these linkages among functions.

Coming back to the example, the function's logic is divided into four labeled blocks:

- **isPrime**: This initial block effectively handles the special case where the input number is 2.
    - o **%ecx** is set to **%edi**, where **%edi** holds the function's argument "a".
    - o **%esi** (the loop counter) is initialized to 2.
    - o **%esi** is compared to **%ecx**.
    - o If **%esi** is greater than or equal to **%ecx** (meaning the input is 2), the program jumps to the **prime** label; otherwise, it proceeds.
- **nexti**:
    - o **%eax** is set to **%ecx** (the input number).
    - o **cdq** sign-extends **%eax** into **%edx**, preparing for signed division.
    - o **idiv** performs the division of **%edx:%eax** by **%esi** (the loop counter), with the remainder in **%edx**. The remainder will be negative for negative input.
    - o The remainder is checked for zero.
    - o If the remainder is zero (meaning **%ecx** is divisible by **%esi**), the program jumps to the *notPrime* label.
    - o **%esi** is incremented (loop counter).
    - o **%esi** is compared to **%ecx**.
    - o If **%esi** is less than **%ecx**, the program jumps back to the **nexti** label (loop continuation).
- **prime**:
    - o **%eax** is set to 1 (indicating prime).
    - o The stack is popped (return).
- **notPrime**:
    - o **%eax** is set to 0 (indicating not prime).
    - o The stack is popped (return).

Note that on line 15, “**xor** %eax, %eax” is used instead of “**mov** %eax, $0”. Historically, and sometimes still, **xor** can be faster than **mov** due to how it's handled by the CPU.

### 2.2. Operating Systems

An operating system can be viewed as a layer of software that possesses privileged access to the CPU's hardware. The operating system acts as an intermediary, orchestrating the execution of user programs. When a user program requires system resources or performs privileged operations, it makes a request to the operating system. The operating system, in turn, grants or denies these requests, ensuring system stability and security. Once the operating system completes its task, it yields control back to the program.

### *2.2.1. Protection Rings*

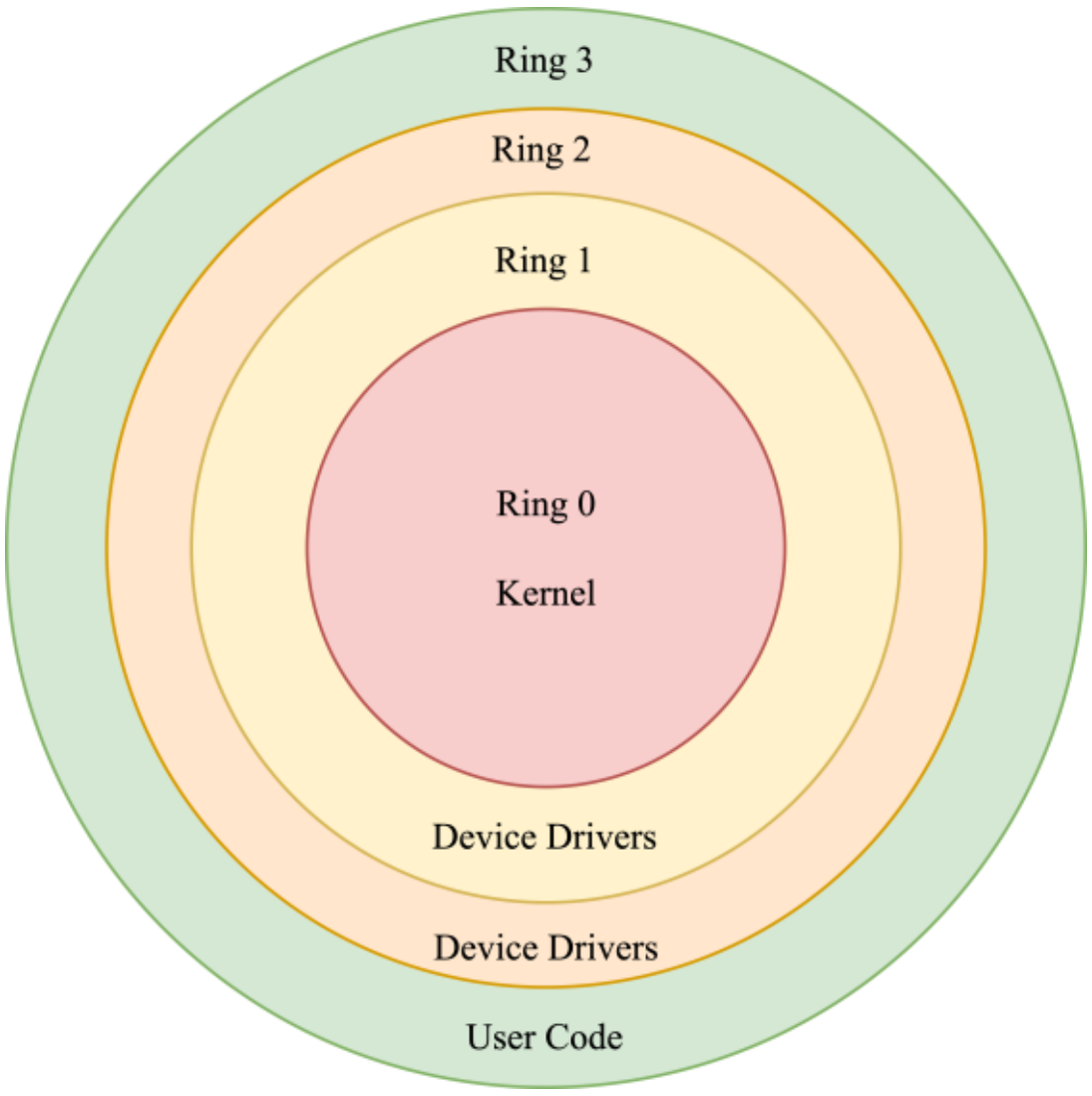

**Figure 5**: Protection rings in a CPU.

To enforce security and manage access to hardware resources, modern CPUs implement a system of privilege levels, commonly known as **protection rings**, as shown in Figure 5. These rings define the level of access that software has to the CPU and its peripherals. The most common arrangement is:

- **Ring 0 (Kernel Mode):** This is the highest privilege level, reserved for the kernel. The kernel, being the core of the operating system, has unrestricted access to all CPU instructions and hardware resources.
- **Ring 1 & 2 (Device Driver Mode):** These rings are typically used for device drivers, which are software components that enable the operating system to communicate with hardware devices. Using different rings for drivers allows for further fine grained control of driver access. Ring 2 is rarely used in modern operating systems.

- **Ring 3 (User Mode):** This is the lowest privilege level, where user applications run. Applications in Ring 3 have limited access to hardware resources and are restricted from executing privileged instructions. Any attempt to execute a privileged instruction from Ring 3 results in a trap, which transfers control to the operating system kernel.

The kernel, residing in Ring 0, holds the ultimate authority over the system. It can execute any CPU instruction, whereas applications in Ring 3 are restricted from accessing certain privileged instructions.

The protection ring level in which the CPU is currently operating is also called **Current Privilege Level (CPL)** and is encoded in the **%cs** register as described before.

### *2.2.2. Trap*

When an application in CPL 3 attempts to execute a privileged instruction, a **trap** occurs. This trap is a **hardware-generated interrupt** that forces the CPU to switch to kernel mode (CPL 0) and execute a kernel routine, allowing the operating system to handle the interrupt and failing the program if required.

### *2.2.3. System Calls*

A **system call**, like a trap, also causes the CPU to switch to kernel mode. Internally, it is implemented as a trap to the kernel. The kernel then executes the requested operation on behalf of the user application.

User applications are often built using libraries, which provide pre-written code for common tasks. These libraries can be linked to the application in two ways:

- **Static Linking:** The library code is incorporated directly into the application's executable during compilation. This results in a larger executable but eliminates the need for an external library at runtime.
- **Dynamic Linking:** The library code is stored in separate files (e.g., .dll files in Windows, .so files in Linux) and loaded into memory at runtime. This allows for smaller executable files and enables multiple applications to share the same library code.

Most core operating system functionalities, such as input/output and network access, are provided through dynamically linked kernel libraries. These libraries encapsulate system calls.

### 2.3. Virtual Memory

Virtual memory provides programs with an abstraction of physical memory, creating the illusion of a vast, contiguous address space, as shown in Figure 6.

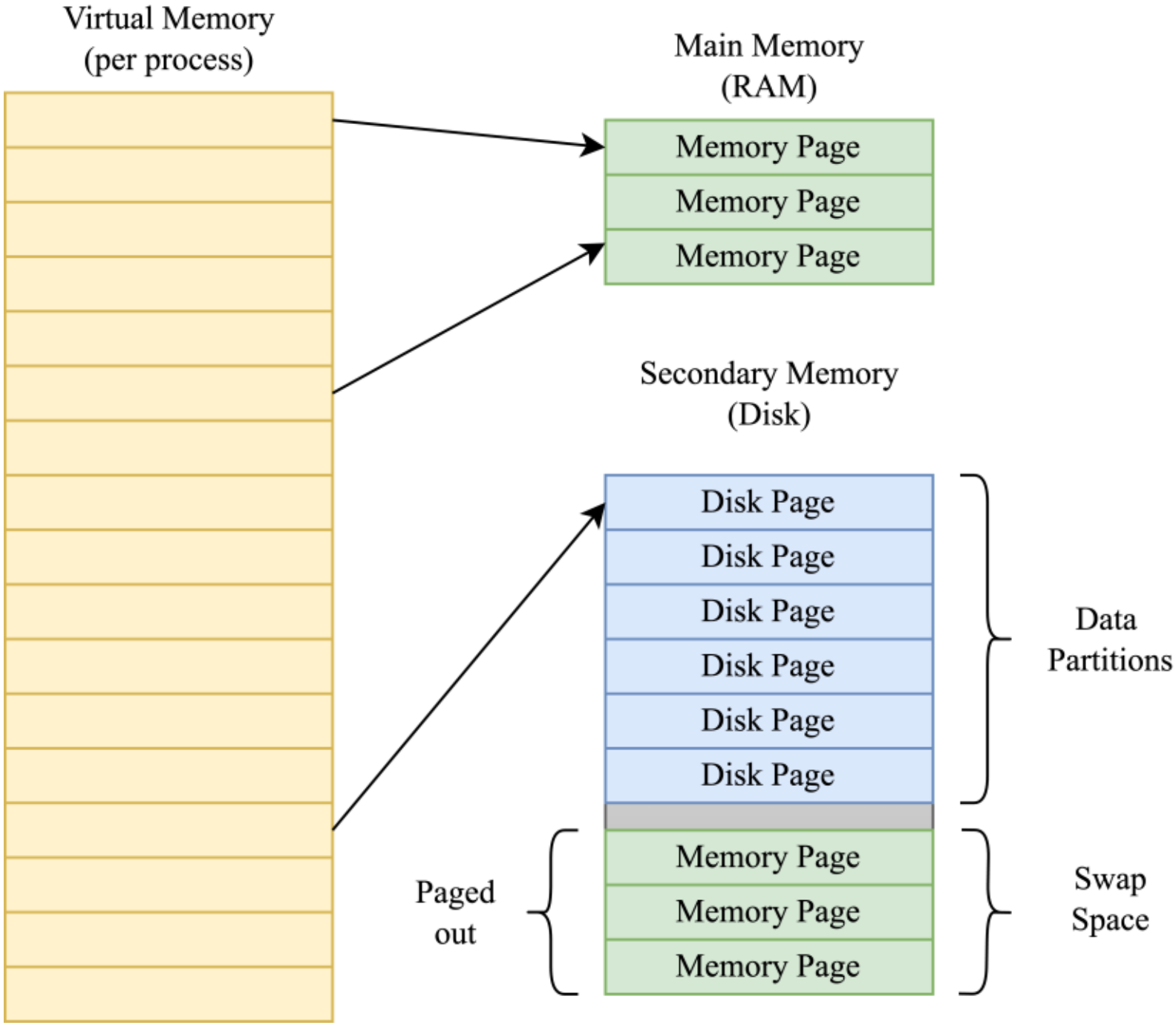

**Figure 6**: Virtual memory managed by an operating system.

Operating systems manage this virtual address space, translating virtual addresses to physical addresses. A virtual address can map to any location: physical memory, secondary storage (like a hard drive), or even to a non-existent location, as the virtual address space may exceed the available physical resources.

#### *2.3.1. Paging*

Direct byte-level mapping of virtual to physical addresses would require an impractically large translation table. To address this, memory is divided into fixed-size units called **pages**, typically 4 KB.

A **page table** is a data structure used for translating virtual page numbers (VPNs) to physical page numbers (PPNs) in main memory:

$$\textbf{Page Table}: \textbf{VPN} \rightarrow \textbf{PPN}$$

Historically, operating systems used only fixed-size pages, which still resulted in large page tables. To reduce the overhead, the concept of superpages [46] (or huge pages) was introduced. Superpages group multiple contiguous pages into a single, larger page, reducing the number of entries in the page table.

Each entry in the page table (shown in Figure 7) also contains some bits (apart from the physical address of the page):

- **Present Bit:** Indicates whether the corresponding page is currently present in physical memory. If this bit is not set, a page fault occurs, and the operating system must load the page from secondary storage.
- **Protection Bit:** Controls whether the page can be read from or written to. This allows the operating system to enforce memory protection.
- **Mode Bit:** Determines whether the page can be accessed by user-level programs or only by the operating system kernel.
- **Dirty Bit:** Indicates whether the page has been modified since it was loaded into memory. This is used by the operating system to determine whether a page needs to be written back to disk.

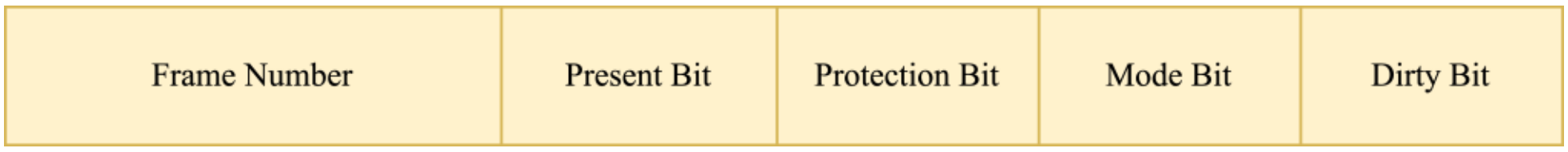

**Figure 7**: Page table entry format.

#### 2.3.1.1. Storing Page Table

The page table can itself be stored by the operating system in a specific memory location. The address of the page table is stored in **Page Table Register** (**%cr3**).

### *2.3.2. Page Swapping and Page Caching*

Accessing a page within main memory is relatively fast. The CPU has a direct memory bus that facilitates rapid data transfer. When data is read from main memory, it's fetched and often cached in CPU cache lines.

Conversely, accessing a page located in secondary storage (like a hard drive or SSD) is significantly slower. The CPU does not have a direct connection to secondary storage. Therefore, a page must first be transferred from secondary storage to main memory before it can be accessed by the CPU.

When main memory becomes full (i.e., there are no free pages), the operating system employs swapping. **Swapping** involves moving inactive or less frequently used pages from main memory to secondary storage, freeing up space for active pages. This process is essentially the transfer of pages that were residing in main memory to available space within secondary storage. When swapped out pages are accessed, it results in a **page fault** which brings back the page into the memory.

Opposite to swapping is the concept of **caching**, where disk pages residing in secondary storage (where they should be residing) are brought into main memory. This process can help speed up reads and writes to files. Note that, caching is different from **demand paging**, which brings in those pages from the secondary storage which are supposed to reside in main memory but were previously swapped out.

It's crucial to understand that the page table only contains entries for pages that are currently mapped to physical memory. Pages that exist only in secondary storage or have not yet been allocated will not have corresponding entries in the page table.

### *2.3.3. Memory Management Unit*

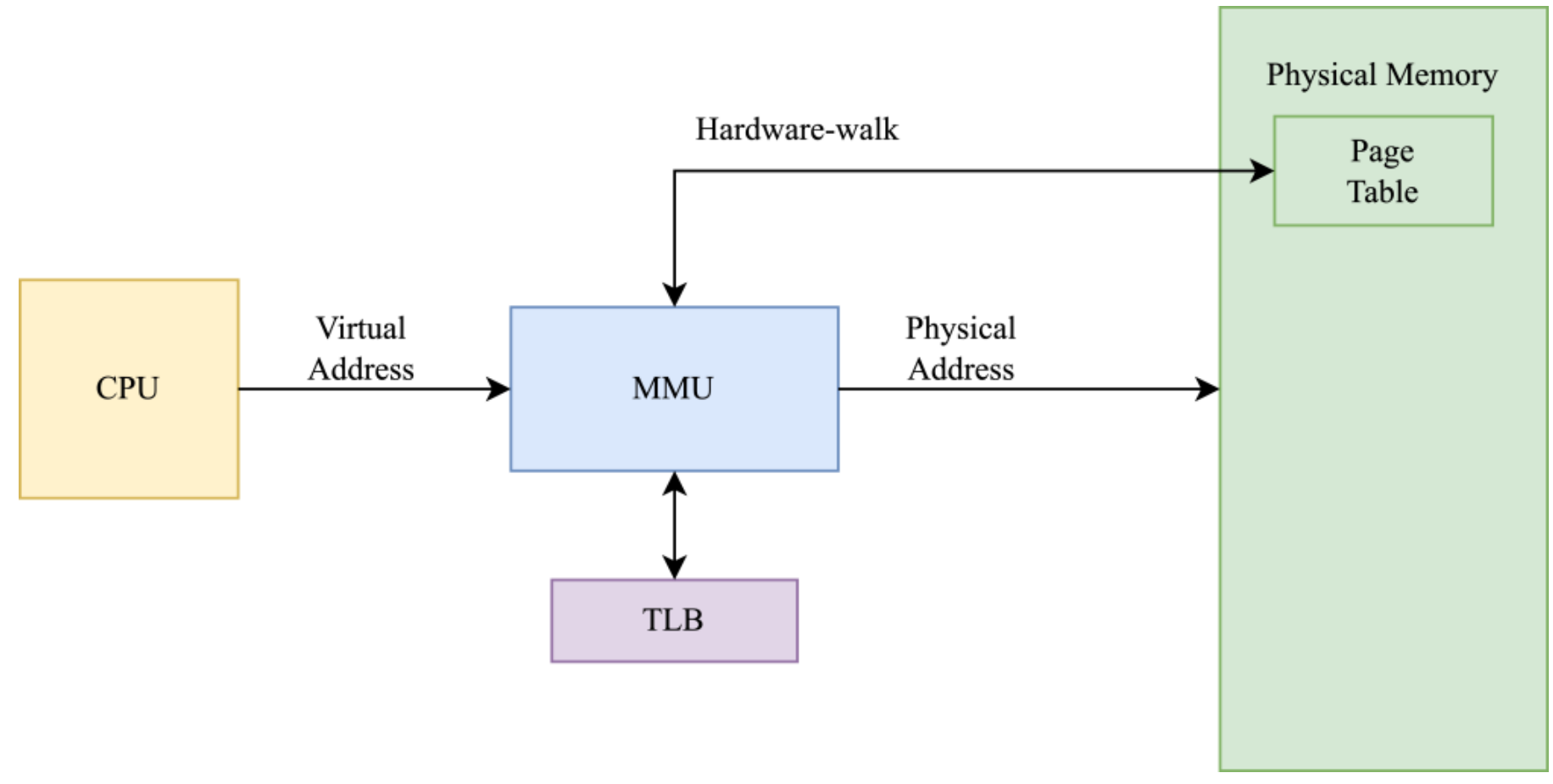

**Figure 8**: Virtual address translation using MMU.

The Memory Management Unit (MMU) is a dedicated hardware component integrated into the CPU, playing a pivotal role in virtual memory management, as shown in Figure 8.

#### 2.3.3.1. Translation Lookaside Buffer

It might seem that every virtual address translation would require the operating system to intervene, as it alone manages page table entries. This would entail a context switch on every memory access, causing significant performance overhead due to the expensive process of saving and restoring CPU state, and flushing CPU caches. To mitigate this, CPUs implement a hardware cache called the **Translation Lookaside Buffer** (TLB).

The TLB acts as a high-speed translation cache, storing recently used virtual-to-physical page mappings. When the CPU accesses a virtual address, the MMU first checks the TLB. If a matching translation is found (a TLB hit), the physical address is retrieved directly, bypassing the need for a page table lookup.

However, the TLB has a limited capacity, typically holding 4,000 entries. If the required translation is not found (a TLB miss), a page fault is generated. This fault triggers a trap, transferring control to the operating system. The operating system then performs a page table walk to locate the physical address, updates the TLB with the new mapping, and resumes the process.

##### 2.3.3.1.1. Side Note: TLB, Caches & Context Switches

Upon a process's departure from the CPU, all TLB entries associated with that process are invalidated (flushed). This is crucial to prevent subsequent processes from accessing the previous process's memory pages through stale TLB entries.

The **invlpg** instruction is typically used for this purpose.

Because TLB entries are cleared, they must be repopulated as the new process executes. This TLB flushing and subsequent repopulation contribute significantly to the overhead of context switches between processes. On top of that, CPU caches also become ineffective after a context switch, further impacting performance.

Consequently, many operating systems employ **CPU affinity**, assigning processes to dedicated CPU cores, to minimize context switching and improve efficiency.

#### 2.3.3.2. Hardware Walking Page Tables

In addition to operating system-driven page table walks during page faults, modern MMUs often feature hardware walking capabilities. This means the MMU itself can automatically traverse the page table structure to locate missing translations.

The physical address of the current process's page table is stored in a special CPU register, the **Page Table Register (PTR)**. On x86, this is register **%cr3**. This register allows the MMU to locate the active page table during address translation.

When a TLB miss occurs, the MMU's hardware logic takes over, fetching the necessary page table entries from memory. This process eliminates the overhead of a context switch to the kernel, significantly speeding up address translation. The MMU is designed to understand the page table structure, allowing it to efficiently navigate through even hierarchical page tables.

## 2.4. Virtualization

Virtualization essentially involves creating an illusion for a guest operating system, making it believe it's running directly on physical hardware. The goal is complete transparency, where the guest operating system cannot distinguish its virtualized environment from a native one. This is achieved by emulating hardware behavior and maintaining consistency through **shadow data structures**.

### *2.4.1. Strawman*

A basic virtualization approach utilizes a **trap-and-emulate** model. In this setup, the guest operating system itself runs as a standard application within the host operating system, executing in CPL 3. Consequently, any privileged instruction executed by the guest operating system, such as I/O operations, triggers a trap, transferring control to the host operating system  The host operating system then emulates the requested operation on behalf of the guest.

However, I/O is just one aspect of virtualization. Memory management presents another challenge. The guest operating system is allocated a limited memory space, which, being a process within the host, also has its own virtual address space. This necessitates a two-tiered address translation:

- The guest operating system's virtual addresses are translated to the actual physical addresses.

- And then the applications within the guest must have their own page tables to translate their virtual addresses into the guest operating system's physical addresses.

This double translation introduces significant performance overhead.

While this strawman model provides a functional virtualization framework, it's inherently inefficient.

*2.4.2. Full Virtualization v/s OS-level Virtualization*

Virtualization can itself be categorized into Full virtualization and OS-level virtualization.

2.4.2.1. Full Virtualization

Full virtualization is complete virtualization of the actual hardware to allow software environments, including a guest operating system and its apps, to run unmodified. The guest operating system (a.k.a VMs) remains unaware of its execution within a virtual environment. VMs operate on top of **hypervisors**, which can be categorized into two types, as shown in Figure 9:

- **Type 1 hypervisors:** Run directly on bare-metal hardware, such as ESX Server.
- **Type 2 hypervisors:** Run on top of another operating system, such as VirtualBox [47].

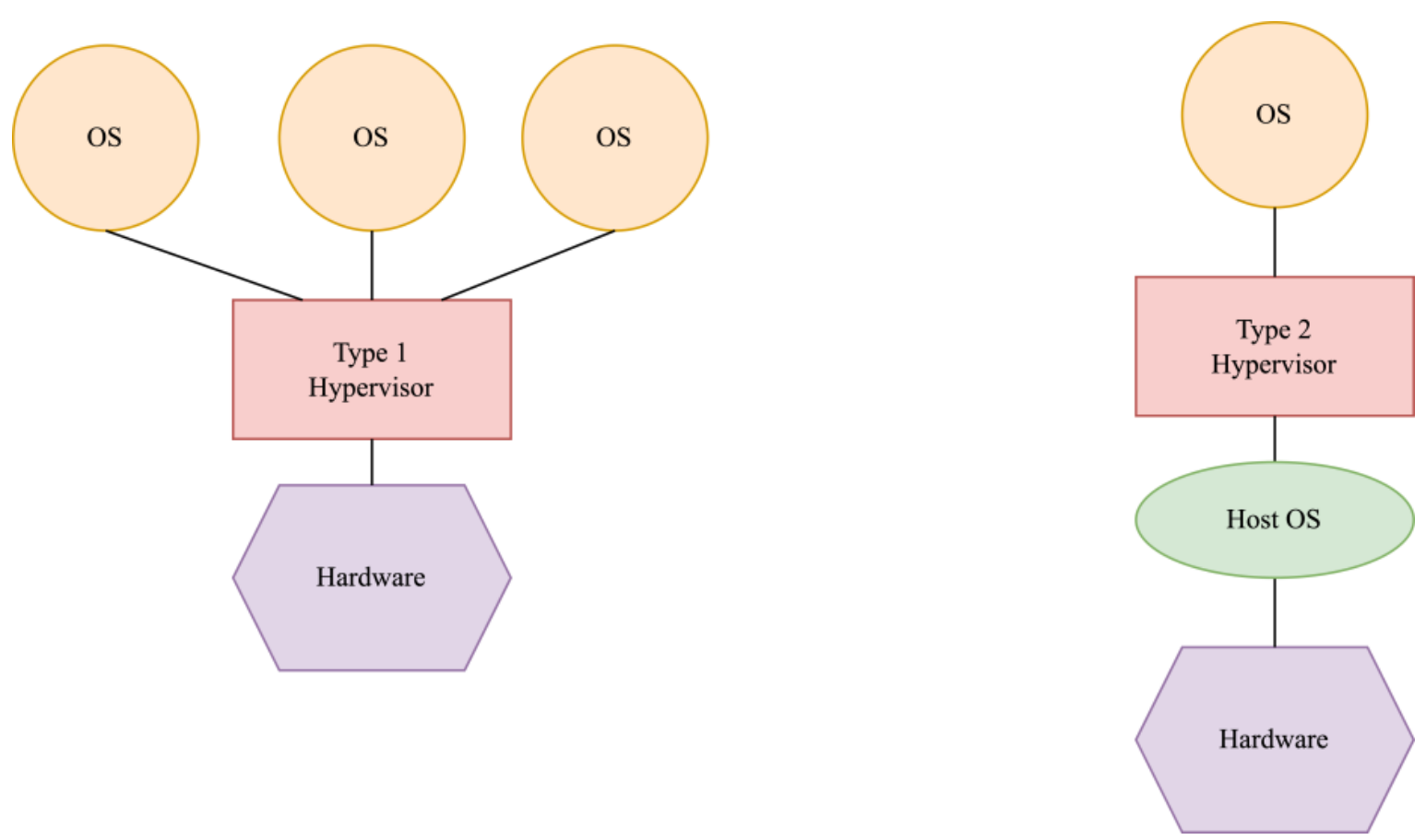

**Figure 9**: Type 1 v/s Type 2 hypervisors.

#### 2.4.2.2. OS-level Virtualization

OS-level virtualization isolates a pool of resources within the operating system. This encompasses technologies such as containers (like Docker [48]), jails (like FreeBSD jail [49]), chroot [50], and virtual environments. The most popular among them today are containers.

For the purposes of this paper, the term "virtualization" is used to denote full virtualization only.

## 2.5. Classical Virtualization

To ensure clarity in our discussion of virtualization techniques, let's establish key definitions:

- **Host** will consistently refer to the VMM, which can operate either as an application on a host operating system (Type 2) or directly on bare metal (Type 1).
- **Guest operating system** refers to the operating system running on top of the VMM.
- **Guest process** refers to any process executing within the guest operating system environment.

### *2.5.1. VMM Characteristics*

Popek and Goldberg defined three essential characteristics of a VMM for classical virtualization:

- **Fidelity:** The guest operating system software must execute identically to its execution on the native hardware, requiring no modifications.
- **Performance:** A significant majority of guest operating system instructions should execute directly on the CPU, minimizing traps.
- **Safety:** All hardware accesses must be mediated by the VMM.

### *2.5.2. Refining Strawman*

Let's refine our basic virtualization model (4.1), still using the trap-and-emulate approach. Guest instructions execute in a deprivileged mode, and privileged instructions trigger traps, transferring control to the VMM for emulation.

#### 2.5.2.1. Shadow Page Table

One key optimization involves the page table translation process. Instead of double translation, we introduce a **shadow page table** that directly maps the virtual address space of a guest process to machine (host) physical page numbers.

- **Guest Page Table:** VPN → PPN
- **Shadow Page Table:** VPN → MPN

Here, PPN refers to the guest's known physical page number and MPN refers to the machine's physical page number, i.e., the physical page number known to the host.

The shadow page table is used for actual address translation during guest process execution, eliminating the overhead of double translations. The guest operating system remains aware only of its own guest page table.

#### 2.5.2.2. Hidden Fault Handling

Whenever the guest operating system modifies its page table entries (adding or removing mappings), the VMM may not immediately reflect these changes in the shadow page table. Accessing a page not present in the shadow page table results in a **hidden fault**. Upon a hidden fault, the VMM updates the shadow page table.

#### 2.5.2.3. Tracing

Hidden faults effectively populate the shadow page table with missing entries. However, we must also address how to remove entries from the shadow page table when they are removed from the guest page table. Without this, a guest process might access invalid pages that have been reallocated, leading to potential security breaches.

The solution is to utilize **access bits** on the guest page table. The guest page table is itself stored on some physical page known to the host. That physical page is write-protected. Updates to it will trap to the host. This is how the VMM detects modifications to the guest page table and can then update the shadow page table accordingly. This page-protection method is known as **tracing**.

### *2.5.3. Performance*

Despite its functionality, classical virtualization exhibits performance bottlenecks primarily arising from the trap-and-emulate mechanism and the overhead associated with tracing memory access operations.

### *2.5.4. Trap-and-Emulate is Non-Classical*

The trap-and-emulate virtualization approach also deviates from classical virtualization, as defined by Popek and Goldberg, in two key ways:

- The guest operating system can detect it's not running in Ring 0 by reading the code segment register.
- Instructions like **popf** bypass traps, allowing direct modification of status flags without VMM intervention. These deviations violate the principles of fidelity and safety, undermining the transparency and controlled hardware access that classical virtualization aims to provide.

## 2.6. Software Virtualization

Software virtualization fundamentally relies on **binary translation**. Instead of direct CPU execution, guest instructions are intercepted and translated by the host into safe equivalents. This allows for features like virtualized flags (e.g., from **popf**) and spoofed CS registers.

Binary translation is not a novel concept; tools like Valgrind [51] utilize it for dynamic bug detection.

However, this translation overhead can significantly impact performance if each guest instruction is translated individually (like an interpreter). To mitigate this, just-in-time (JIT) compilation is employed, similar to the JVM's JIT [52]. This dynamic translation occurs at runtime, on demand, enabling optimizations.

**Side Note: JIT Benefits**

JIT compilation is a common compiler technique and offers performance enhancements. Notably, it can rearrange code blocks (as shown in Figure 10) to improve instruction cache locality, reducing cache misses caused by frequent jumps in the original code.

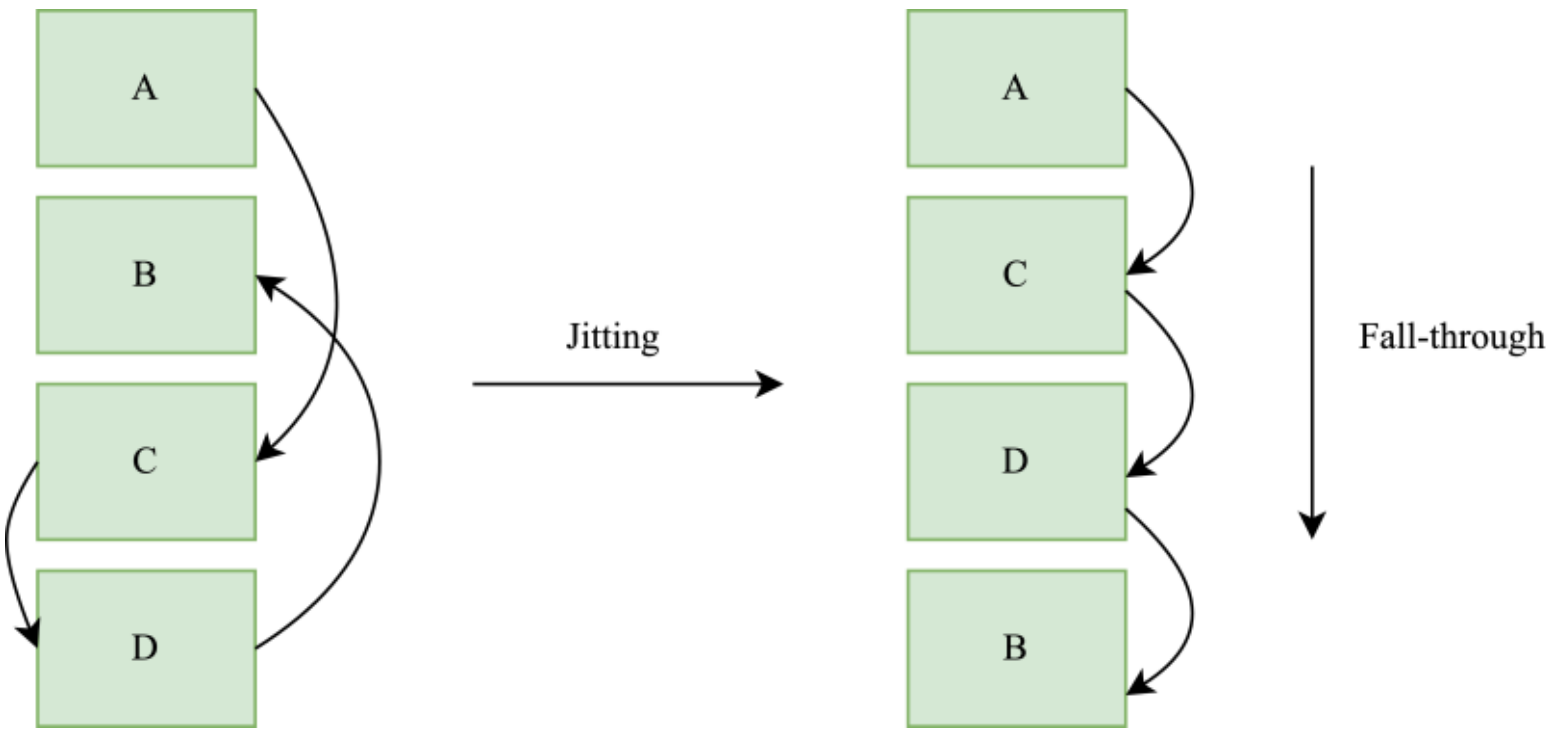

**Figure 10**: Code block rearrangement during JIT compilation.

### *2.6.1. Binary Translation*

The software virtualization approach executes guest code at a less privileged level (e.g., CPL 1). This allows most guest instructions to execute natively without full translation, minimizing overhead.

The translation process operates on **translation units** (TUs), which are basic blocks of guest code. Each TU is translated into a **compiled code fragment** (CCF). TUs are typically limited in size (e.g., 12 instructions) and terminate with a jump or return instruction.

The authors demonstrate that translated instructions offer a performance advantage over the conventional trap-and-emulate approach. In a specific instance, a read timestamp counter (**rdtsc**) operation required 2030 cycles using trap-and-emulate, while the same operation completed in only 216 cycles with TC emulation.

#### 2.6.1.1. IDENT vs. NON-IDENT Instructions

- **IDENT instructions:** These instructions require no translation and can be executed directly. The majority of guest instructions fall into this category.
- **NON-IDENT instructions:** These instructions necessitate translation. Examples include:
  - **PC-relative addressing** - This is because the translated code may not be in the same place as the original code.
  - **Control flow** - Jump (e.g., **jge**, **jmp**) and return instructions.
  - **Privileged instructions** - Such as **popf**.
  - **Non-privileged instructions** - Such as **load**, **store**, etc (adaptive).

#### 2.6.1.2. Lazy Translation

As a consequence of just-in-time compilation, only executed code paths are translated. Unexecuted code remains untranslated.

#### 2.6.1.3. Chaining Optimization

The translation process may alter instruction addresses, requiring adjustments to target addresses. If a jump target is determined to be the next sequential instruction, the jump is replaced with a no-op instruction, a technique known as **chaining optimization**.

#### 2.6.1.4. Example

Let's run through the translated *isPrime* example again on input 49 which is not a prime.

```
1.  isPrime’:  mov %ecx, %edi
2.             mov %esi, $2
3.             cmp %esi, %ecx
4.             jge [takenAddr]
5.
6.  nexti’:    mov %eax, %ecx
7.             cdq idiv %esi
8.             test %edx, %edx
9.             jz notPrime’
10.
11.            inc %esi
12.            cmp %esi, %ecx
13.            jl nexti’
14.            jmp [fallthrAddr3]
15. notPrime’: xor %eax, %eax
16.            pop %r11
17.            mov %gs:0xff39eb8(%rip), %rcx
18.            movzx %ecx, %r11b
19.            jmp %gs:0xfc7dde0(8*%rcx)
```

Let's break down each point:

- **Line 4:** The jump instruction's target address has been altered. While it was intended to point to a **prime** block, that block was never executed for the given input (49), hence its translation was skipped.

- **Lines 5 & 10:** The jump instruction **jmp [fallthrAddr]** was removed because the "fall-through" address was the very next instruction. This is chaining optimization.
- **Line 14:** The jump **jmp [fallthrAddr3]** was retained. However, since the **prime** block was never translated, the jump's target remains unresolved at this stage.
- **Line 16 - 19:** The **ret** instruction is translated:
- **Line 16:** The **pop** instruction retrieves the return address from the stack into **%r11**.
- **Line 17:** The **%rcx** register is saved to a memory location. The memory address is calculated by adding a fixed displacement of **0xff39eb8** to **%rip** and this will be the relative address within the segment pointed by **%gs** register.
- **Line 18:** The lower 8 bits of the **%r11** register (**%r11b**) are extracted and placed into the **%ecx** register, with the remaining bits of **%ecx** being set to zero.
- **Line 19:** Probably, relative address **0xfc7dde0** within **%gs** segment is the data structure (such as an array) which stores the return addresses to the callee. The index into the array is coming from the **%rcx** register.

Note that this translation is performed by the compiler. It may seem complex but that is how compilers really work. For instance spilling **%rcx** on line 17 was required, because its value was used to store another variable.

#### 2.6.1.5. Running the Translator

The translator runs as a part of VMM. The translator needs to store its data structures used for address translation. This data structure is stored in a memory segment mapped by **%gs** register. This memory segment is mapped to the guest operating system (to make it accountable).

### *2.6.2. Adaptive Binary Translation*

Apart from privileged instructions, there are also non-privileged instructions that can result in a trap. For example, **load** and **store** instructions that access memory can potentially cause a trap if there is a page fault.

In such cases, the authors utilize **adaptive translation**, where they monitor these CCFs that cause frequent traps and then replace them with cheaper alternatives.

Figure 1 in the paper illustrates an example of this. On the left side, jumping to CCF1 results in a page fault. The translator monitors the faults and then replaces CCF1 with CCF5. All jumps to CCF1 are now adapted to CCF5 which doesn't cause page faults.

### 2.7. Hardware Virtualization

Hardware-assisted virtualization significantly alters virtualization assumptions. Intel's VT extensions provide hardware support for this technique.

#### *2.7.1. Guest Mode*

In hardware-assisted virtualization, CPUs introduce a **protected mode** (also called **guest mode**) alongside the **real mode**. The VMM configures guest mode using the **Virtual Machine Control Block (VMCB)**.

Guest code executes directly on the CPU at the same CPL as VMM (CPL 0). However, based on the VMCB settings, specific instructions trigger traps, transferring control to the VMM.

The **vmrun** instruction initiates guest mode, and traps result in exits back to the VMM.

#### *2.7.2. Virtual Machine Control Block*

The VMCB is a data structure used by the host and CPU to facilitate virtualization. It contains:

- **Control Bits:** Used to intercept reads and writes to control registers (**%cr0 - %cr15**), generating traps.
- **Debug Bits:** Similar 32-bit fields for debug registers (**%dr0 - %dr15**).
- **Intercept Bits:** 8 bits to intercept reads and writes to IDTR, GDTR, and LDTR.
- **Instruction Intercepts:** Bits to intercept **pushf**, **popf**, **vmrun**, **hlt**, **invlpg**, **int**, **iret**, and **in/out** operations to specified ports.

- **Page Table Intercepts:** Because the page table base is stored in a control register (**%cr3**), the VMM can program the VMCB to generate exits on guest page faults and TLB flushes, enabling shadow page table management.

There are numerous additional bits for various virtualization controls.

When the guest operating system relinquishes control to the VMM, the guest's state (register values) is saved.

*2.7.3. Example: fork(2) with Hardware Virtualization*

The paper illustrates the **fork(2)** system call's behavior in hardware virtualization. Let's examine this process and compare it to software virtualization.

It's crucial to understand that both parent and child processes are in an identical state when **fork(2)** is called, and their states only begin to diverge afterward.

The **fork(2)** system call initiates a child process that, as an optimization for speed, initially shares the parent's page table. This approach avoids the time-consuming process of immediately copying the entire parent's memory to a new location before the child can begin execution. Instead, the operating system marks all of the parent's memory pages as read-only, effectively write-protecting them. When either the child or the parent process attempts to modify a write-protected page, the operating system intervenes and creates a separate copy of that page for the modifying process. The write-protected, shared pages represent the portion of their shared state that has not yet diverged between parent and child process.

When a guest process calls **fork(2)**:

- The system call transitions the CPL from 3 to 0, as the guest operating system handles it. Since the guest runs at CPL 0, it can process the call without direct VMM intervention. In software virtualization, this would trigger a VMM trap.
- The guest operating system write-protects the parent's page table. Accessing the page table register (**%cr3**) triggers a VM exit, transferring control to the VMM. The VMM then write-protects the corresponding shadow page table entry.

- The guest operating system resumes and schedules the child process. Scheduling the child requires updating the page table register. This again causes a VM exit. The VMM updates the shadow page table register.
- From then, hidden page faults are also vectored to the VMM, allowing it to maintain the shadow page table.

### 2.8. Comparison

Turning to the paper's central objective: a comparative analysis of hardware and software virtualization techniques. At first glance, hardware virtualization might appear superior, particularly when VM exits are infrequent. However, the authors' experiments challenge this assumption, revealing several scenarios where binary translation in software virtualization outperforms hardware virtualization.

It's important to acknowledge a potential bias: as VMware developed binary translation, there's a natural inclination towards showcasing the strengths of software virtualization.

#### *2.8.1. Experiment Results*

The experimental design focused on scenarios triggering privileged operations, as compute-intensive benchmarks demonstrated near-native performance (Figure 2 in the paper). Notably, the **mcf** application exhibited performance exceeding native execution, attributed by the authors to virtualization-induced improvements in TLB and cache access patterns.

Figure 3 in the paper illustrates that hardware VMMs occasionally outperform software VMMs. However, given the complexity of operating system interactions, attributing definitive conclusions is challenging. The authors' explanations for these variations appeared somewhat inconclusive.

In the fork/wait experiment, software virtualization completed 40,000 fork/wait operations in 36.9 seconds, while hardware virtualization required 106.4 seconds, representing a 4.4x latency increase.

Microbenchmark results further highlighted performance differences:

- **syscall:** Software virtualization incurred a significant performance penalty, adding approximately 2,000 cycles compared to hardware or native execution.
- **in:** Software virtualization excelled by translating the I/O port read instruction, resulting in rapid execution.
- **cr8wr:** Software virtualization again demonstrated superior performance by translating the privileged register write into a more efficient instruction sequence.
- **call/ret:** Binary translation introduced overhead, slowing down control flow due to the translation process.
- **pgfault:** Software virtualization outperformed hardware, as software-handled page faults proved less costly than VM run/exit cycles.
- **divzero:** Hardware virtualization exhibited superior performance, as the CPU-generated trap was handled more efficiently.
- **ptrmod:** Software virtualization excelled in manipulating page table entries, as it avoided traps and translated the instruction into a more efficient variant.

### 2.9. Paper Remarks

This paper offers an exceptionally dense exploration of compilers and operating systems, particularly regarding the complex interactions between kernels and the VMM. Ultimately, despite its density, the paper serves as a rigorous resource for those seeking to deepen their technical command of systems architecture.

# 3. Memory Resource Management in VMware ESX Server

(3) Waldspurger, C. A. Memory Resource Management in VMware ESX Server. ACM SIGOPS Operating Systems Review 2002, 36 (SI), 181–194.

The previous chapter focused extensively on microprocessors, providing only a brief overview of memory. In this chapter, we will explore memory in significantly greater depth. This paper, released by VMware in 2002 and presented at ACM Special Interest Group in Operating Systems (SIGOPS), details the memory management techniques employed by VMware ESX Server, a Type 1 hypervisor enabling full virtualization. There are lots of beautiful techniques discussed in the paper.

The insights will first examine how a Linux process organizes its memory and how usage is accounted for. We will then explore auxiliary concepts such as virtual memory (not to be confused with "virtualized memory") and the interaction between I/O devices and memory. Finally, we will take an in-depth look at how memory is managed within a virtualized environment.

**Recommended Read**: **2. A Comparison of Software and Hardware Techniques for x86 Virtualization.**

### 3.1. Process Memory

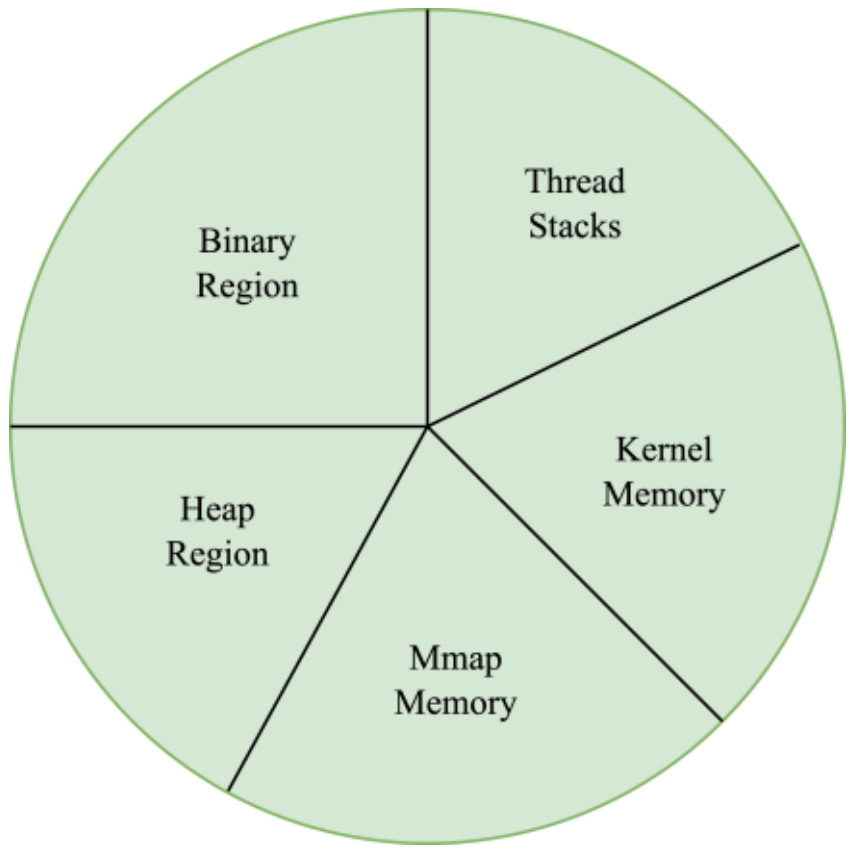

**Figure 11**: Process memory regions.

A Linux process operates in a memory space composed of several distinct regions as shown in Figure 11. Let's examine each component individually.

### *3.1.1. Binary Region*

A Linux process is built from various files, including the executable, object code, and shared libraries. These modules adhere to a common format known as the **ELF (Executable and Linkable Format)**. On Windows, the equivalent format is **Portable Executable**.

The ELF format structures a module into key sections:

- **Header**: Contains fundamental information about the file, such as its type, architecture, and entry point.
- **Program Header**: Describes the segments of the program that should be loaded into memory during execution, including their permissions and locations.
- **Section Header**: Provides a table detailing the various sections within the file, including their names, sizes, and offsets.
- **Sections**: These contain the actual data and instructions of the program:
    - **.text**: The code segment, holding the executable instructions of the program.
    - **.data**: The data segment, storing the program's initialized global and static variables.
    - **.rodata**: The read-only data segment, typically used for storing string literals and constant values.
    - **.bss**: The Block Started by Symbol (BSS) segment, reserved for uninitialized global and static variables, which are set to zero by the kernel at program start.

### *3.1.2. Thread Stacks*

Beyond the program's binary image, each thread within a process has its own stack, typically a few megabytes in size (e.g., 2MB default). The stack is allocated upon thread creation. It serves as storage for local variables and maintains the execution state during function calls.

### *3.1.3. Heap Region*

The heap is a large memory region dedicated to dynamic memory allocation. Objects created at runtime are allocated on the heap. Various memory allocators manage this space, with popular choices including the default PTMalloc, Google's TCMalloc [53], and Facebook's JEMalloc [54].

Crucially, all allocators keep track of the available heap regions using various metadata.

#### 3.1.3.1. PTMalloc

PTMalloc, the default memory allocator in C libraries like GNU, manages memory by dividing the heap into chunks of varying sizes. These chunks are organized into bins based on their size, allowing for efficient lookup.

When a program requests new memory for an object, PTMalloc searches these bins for a free chunk of the appropriate size. However, in a multi-threaded environment, multiple threads might try to allocate memory simultaneously. This can lead to contention, where threads compete for access to the bins, potentially slowing down memory allocation.

#### 3.1.3.2. TCMalloc

TCMalloc (as shown in Figure 12) also organizes the heap into slabs of fixed sizes. Each thread maintains a local lock-free free list of available slabs in a cache for efficient object allocation. A central, locked free list serves as a fallback for threads that exhaust their list in thread-local cache. Additionally, a central heap handles allocations of very large objects.

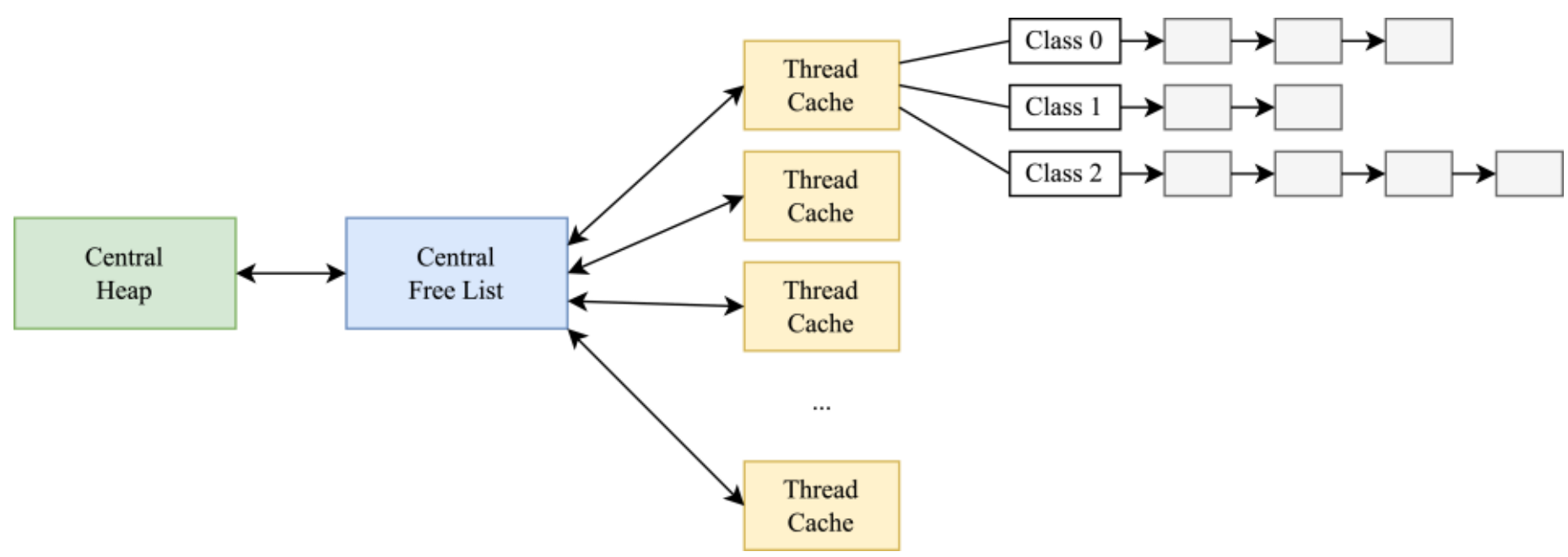

**Figure 12**: TCMalloc architecture.

#### 3.1.3.3. JEMalloc

JEMalloc (as shown in Figure 13) categorizes memory requests into three main sizes:

- **Small Objects**: 8, 16, 32, 128 bytes, up to 512 bytes. Directly handled by the thread-local cache, allowing for fast allocation.
- **Large Objects**: Ranging from 4 KB up to 4 MB. Leverages thread-local cache or **arenas**. Each arena is an independent memory pool, typically 4 MB in size, that manages its own set of memory chunks.
- **Huge Objects**: Any allocation 4 MB or larger (e.g., 4 MB, 8 MB, 12 MB). These are allocated directly from the underlying operating system's free memory. JEMalloc uses a global red-black tree to track these huge allocations.

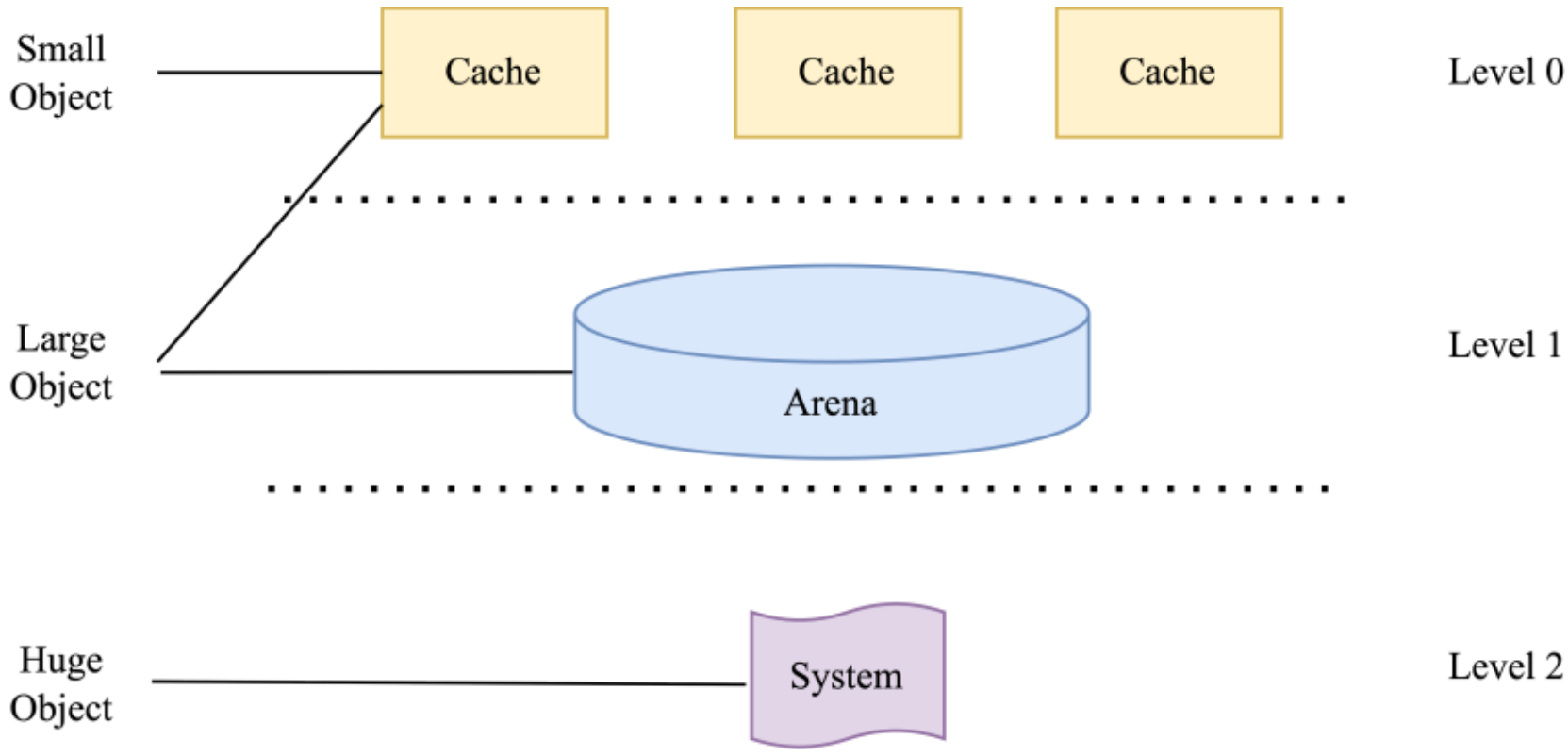

**Figure 13**: JEMalloc architecture.

### *3.1.4. MMap*

*Not to be confused with Memory-Mapped I/O.*

Mmap [55] is a mechanism for mapping files from disk directly into a process's address space. This creates a memory region linked to the file. MMap regions can be:

- **Clean Region**: Represents files that have only been read. These regions can be readily discarded from memory as their content is identical to the on-disk version and can be reloaded if needed.

- **Dirty Region**: Contains data that has been written to within the mapped memory region. These changes are not immediately reflected on disk and require a sync operation to persist the modifications.

#### *3.1.5. Kernel Memory*

Finally, each process is associated with a portion of kernel memory. This memory is managed by the operating system kernel on behalf of the process and often includes I/O buffers, such as network TCP/IP buffers, used for communication.

### 3.2. Paging

The system's memory, both primary (RAM) and secondary (disk), is organized into fixed-size blocks called **pages**. We can broadly categorize these pages based on their primary location and purpose, as shown in Figure 14:

- **Memory Pages**: These are pages intended to reside in the main memory (RAM) for active use.
- **Disk Pages**: These are pages primarily stored on secondary storage (disk), often corresponding to files or other persistent data.

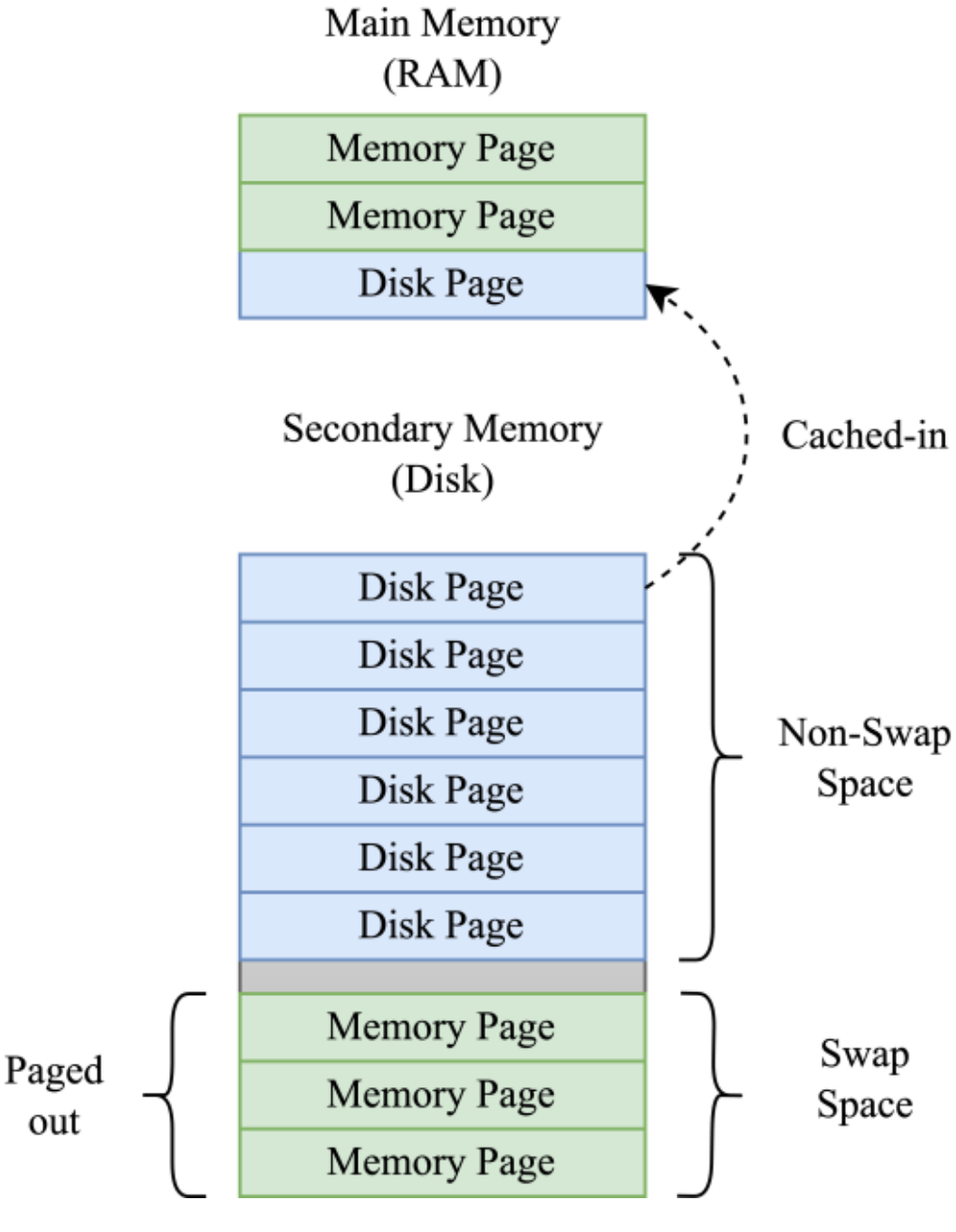

**Figure 14**: Memory paging.

### *3.2.1. Swapping*

When the system's primary memory becomes low, a process called **swapping** occurs. During swapping, memory pages that are currently in RAM are moved to a dedicated area on the secondary storage called the **swap space**. This frees up physical RAM.

When the data on a swapped-out page is needed again, it is read back from the swap space into the main memory. This on-demand retrieval is known as **demand paging**.

### *3.2.2. Page Caching*

**Page caching** is a mechanism where disk pages (pages originating from files on disk) are read into the main memory. This is often done through mmap() discussed above. The goal of page caching is to improve performance by keeping frequently accessed file data readily available in RAM, avoiding the slower process of repeatedly reading from the disk.

### *3.2.3. Non-Anonymous v/s Anonymous Pages*

Pages in memory can be further classified into two types:

- **Non-Anonymous Pages (File-Backed)**: These are the pages that constitute the page cache – they correspond to files on disk. Because their content is ultimately backed by a file, clean (unmodified) non-anonymous pages can be discarded from memory if space is needed, as their original state is preserved on disk. Dirty (modified) non-anonymous pages must be written back to their corresponding disk files before they can be evicted from RAM.
- **Anonymous Pages**: All other pages in main memory that are not file-backed are considered anonymous. These pages typically hold data that doesn't have a direct persistent storage on disk, such as the heap, stack, and binary segments of processes. When the system runs out of memory, anonymous pages are the ones written to the swap space to free up RAM.

## 3.3. Private v/s Shared Memory

Processes have a considerable amount of memory that is exclusively their own, allowing them unrestricted read and write access. However, a notable portion of memory is also shared across different processes. This shared memory primarily consists of:

- Commonly used ELF files, such as **shared libraries**.
- Clean pages residing in the page cache.

## 3.4. Memory Accounting

Linux employs several metrics to track memory usage by processes.

### *3.4.1. Resident Set Size*

Resident Set Size (RSS) represents the portion of a process's memory that is currently residing in physical RAM.

**Includes:**

- All non-swapped binary, stack, and heap pages.
- Disk pages currently held in the page cache.
- Shared memory segments. **Note**: this can lead to double counting when multiple processes share the same memory.

**Excludes:** Kernel memory used to support the process. This memory is managed from a central kernel memory pool and is not attributed to individual processes.

### *3.4.2. Proportional Set Size*

Proportional Set Size (PSS) is similar to RSS but accounts for shared memory more accurately. Instead of counting the entire shared memory segment for each process, PSS divides the size of each shared memory segment equally among all the processes sharing it. This provides a more realistic view of a process's "fair share" of memory usage.

### *3.4.3. Virtual Memory Size*

Virtual Memory Size (VMS) encompasses the total virtual address space used by a process.

**Includes:** All memory the process has access to, regardless of whether it's currently in RAM or on disk (swapped out). This includes the resident pages (counted in RSS) and the non-resident pages.

**Excludes:** Kernel memory used to support the process.

### 3.5. Virtual Memory

Virtual memory provides programs with an abstraction of physical memory, creating the illusion of a vast, contiguous address space, as shown in Figure 15. A virtual address can map to any location: physical memory, secondary storage (like a hard drive), or even to a non-existent location, as the virtual address space may exceed the available physical resources. Virtual address space is also divided into pages of the same size as the physical pages.

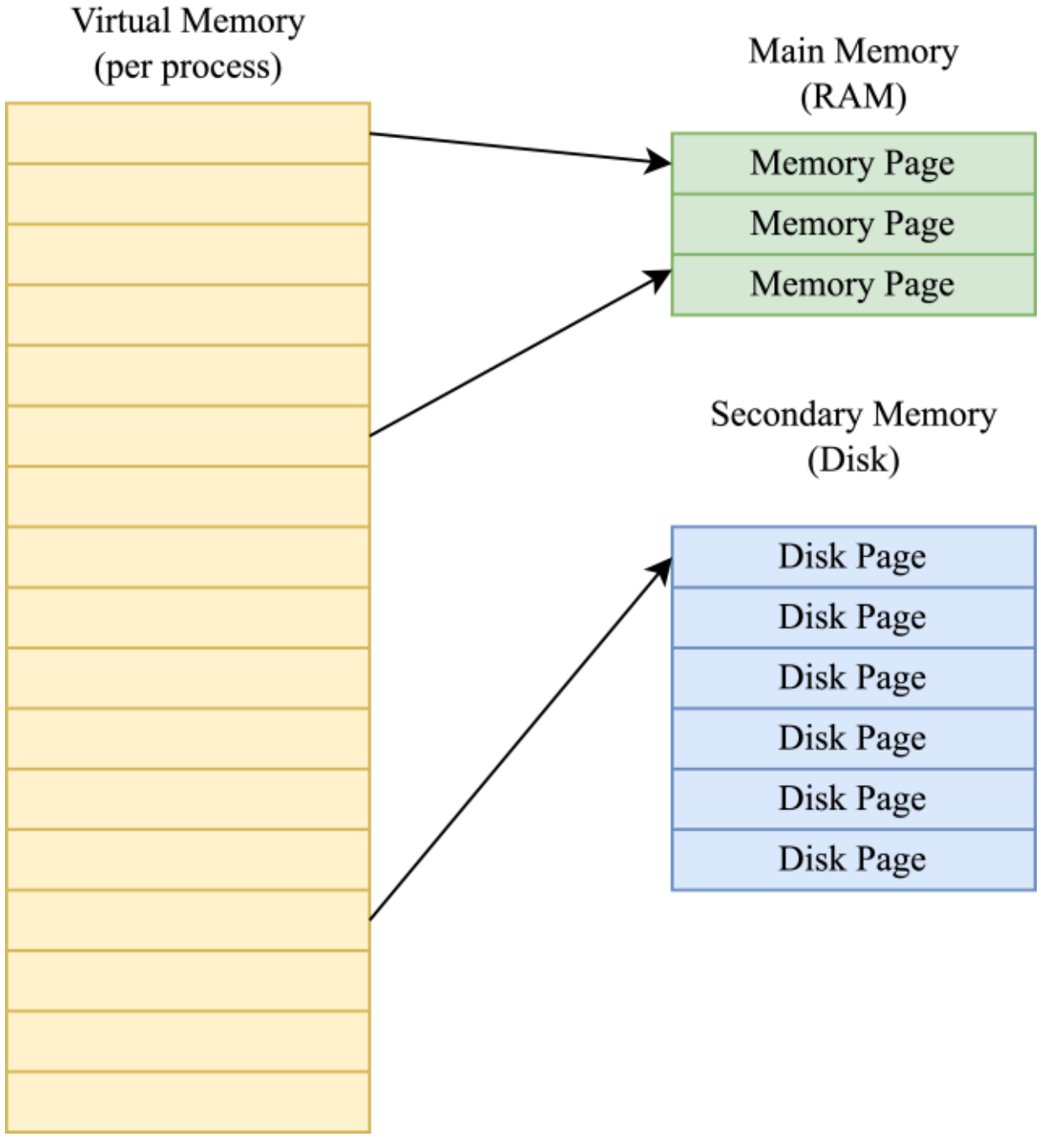

**Figure 15**: Virtual memory of a process.

A **page table** is a data structure used for translating **virtual page numbers (VPNs)** to **physical page numbers (PPNs)** in main memory:

$$\textbf{Page Table}: \textbf{VPN} \rightarrow \textbf{PPN}$$

The **Memory Management Unit (MMU)** is a dedicated hardware component integrated into the CPU, playing a pivotal role in virtual memory management, as shown in Figure 16. It helps in translating virtual addresses from CPU to physical address. The translation makes use of a page table along with support from **translation look-aside buffer (TLB)** to speed up.

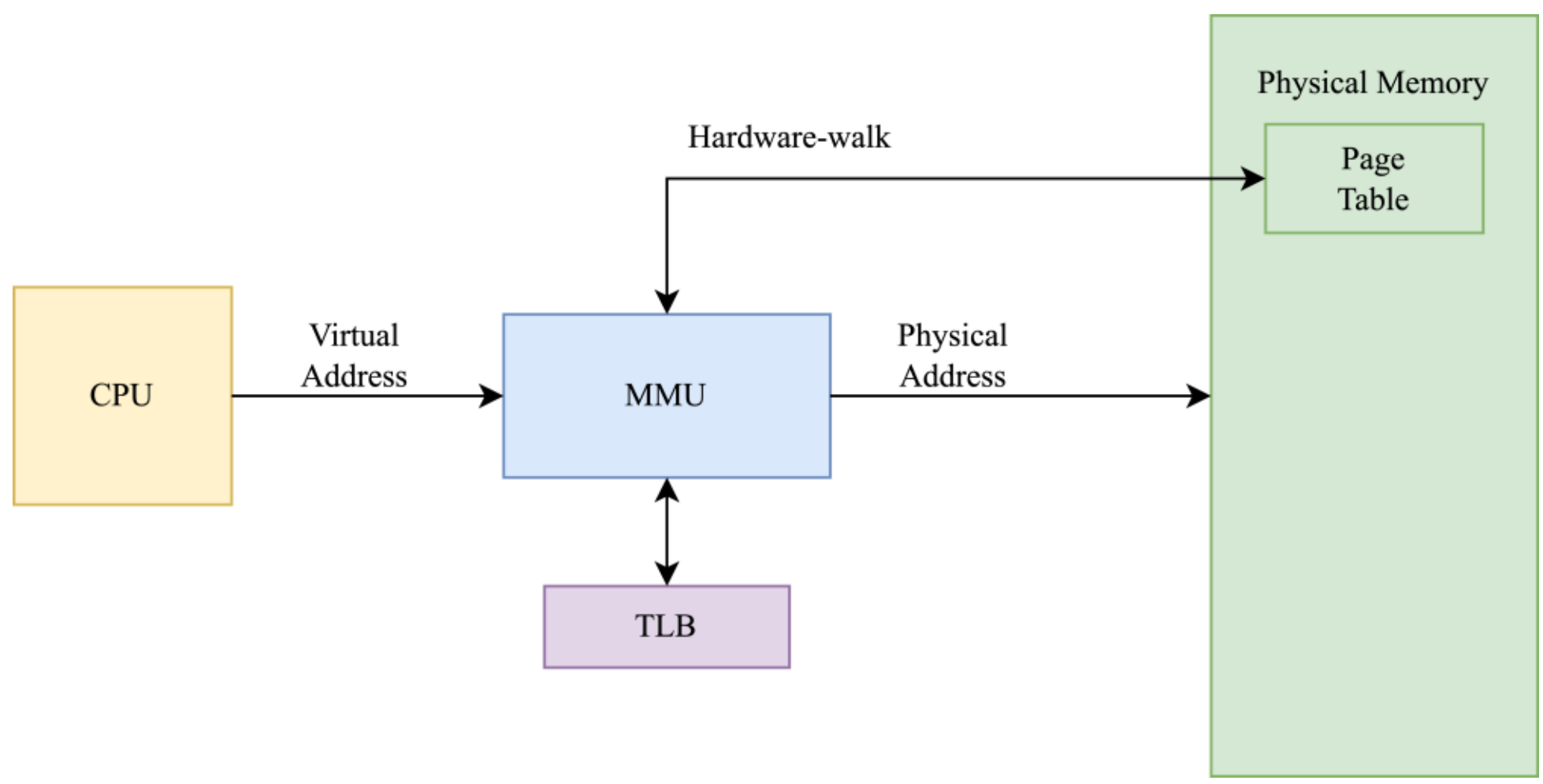

**Figure 16**: Virtual address translation using MMU.

### 3.6. Memory Mapped I/O

*Not to be confused with mmap.*

**Port-Mapped I/O** was a common method for CPU communication with peripheral devices in earlier computing systems. This technique involved the CPU using specialized instructions to interact with dedicated I/O ports of the devices. For instance, reading data from a device might involve the following x86 assembly instructions:

```
mov %dx, 0x3F8
in %al, %dx
```

The instructions above loads the I/O port address into **%dx** and then reads a byte from the port in **%dx** into **%al** register. Similarly, writing data to a device could be accomplished with instructions like:

```
mov %dx, 0x3FC
mov %al, 0x01
out %dx, %al
```

Port-Mapped I/O is less efficient. For example, transferring data from memory to a disk would require the CPU to read data from memory into its registers and then write that data out to the disk's I/O port. The CPU is directly involved in every step of the data transfer. The in and out instructions specifically target the I/O ports of the connected devices.

In contrast, **Memory-Mapped I/O** simplifies device interaction by mapping the physical registers and memory of peripheral devices into the kernel's virtual address space, as shown in Figure 17.

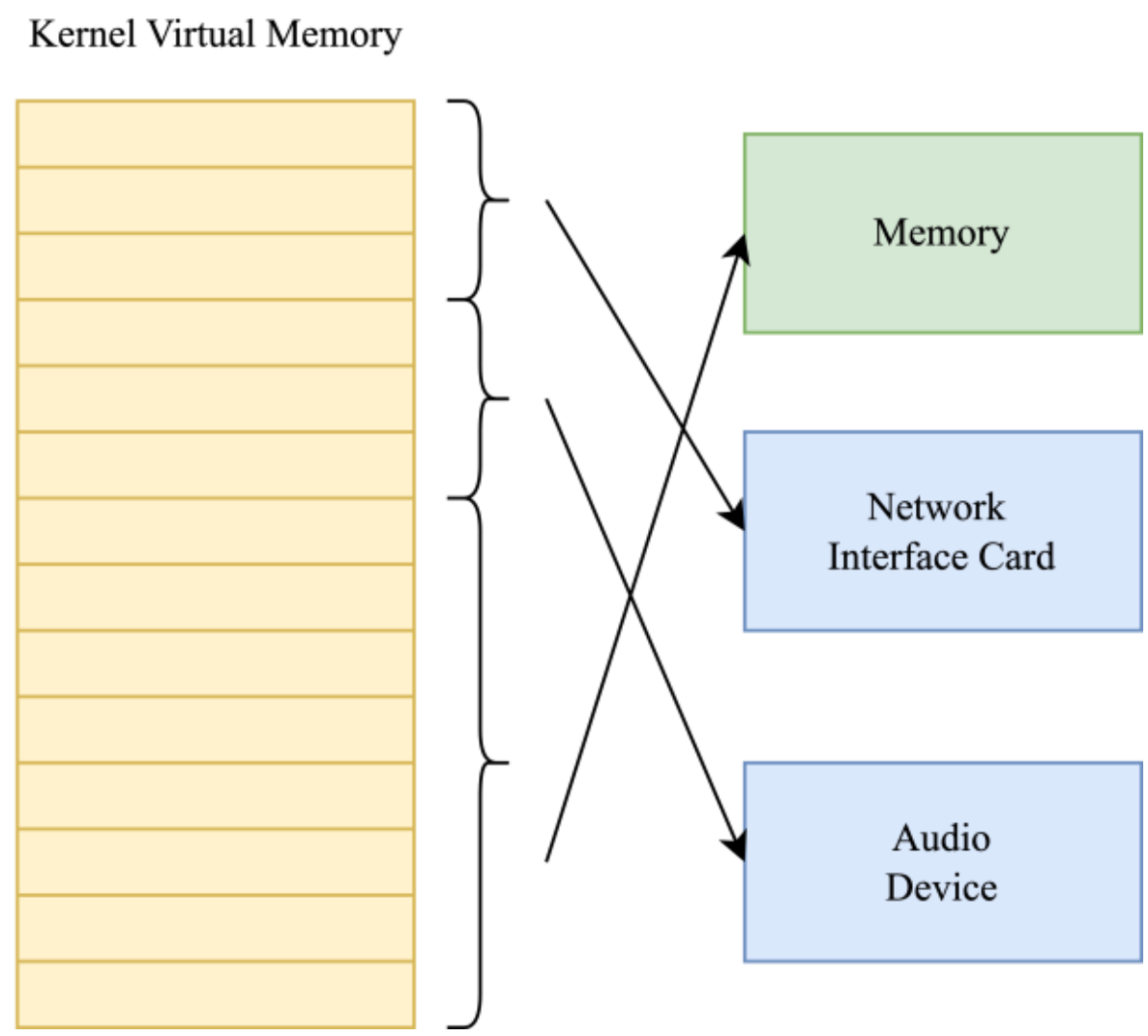

**Figure 17**: Memory-mapped I/O.

Once this mapping is established, communication with these devices occurs through standard memory read and write operations. For example:

```
mov %eax, 0x01
mov 0xf0, %eax
```

**Advantages**:

- A single address bus serves both memory and all I/O devices, eliminating the need for separate I/O ports and buses.
- Devices can benefit from CPU and bus optimizations designed for memory access.

The kernel's virtual memory space allocated to each memory-mapped device can be logically divided into input regions (for data coming into the device), output regions (for data going out from the device), and control, as shown in Figure 18.

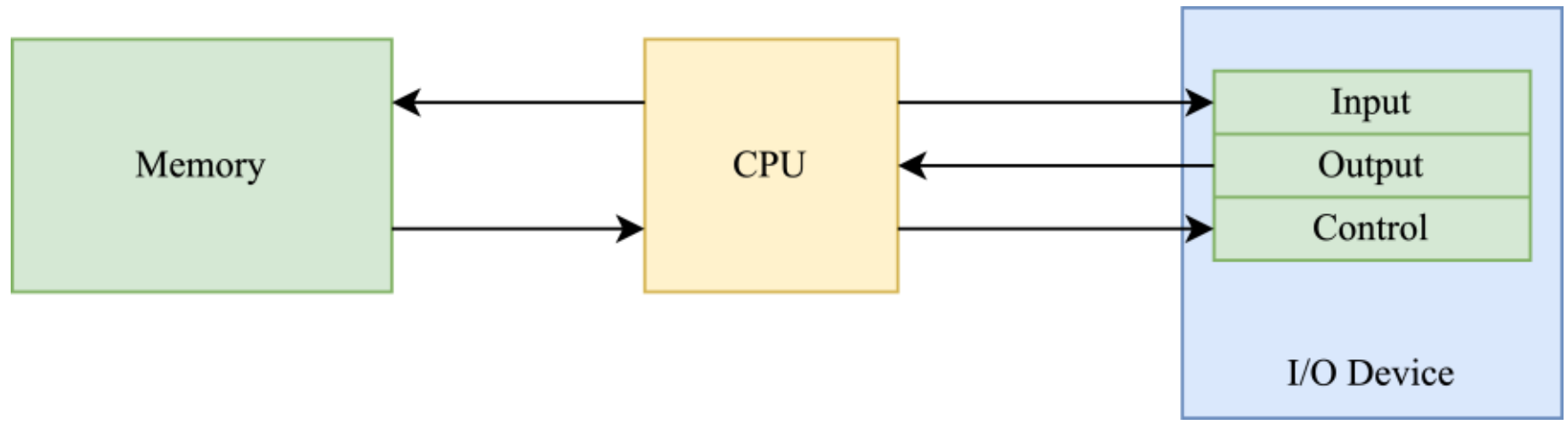

**Figure 18**: Device memory segmentation.

### 3.7. Direct Memory Access

Following the widespread adoption of Memory-Mapped I/O, **Direct Memory Access (DMA)** controllers became prevalent. DMA enables peripherals to directly access system RAM without constant CPU intervention, as shown in Figure 19. The CPU initiates a DMA transfer by providing the DMA controller with the physical address in RAM, the I/O device's input/output region, and the number of bytes to transfer. The DMA controller then handles the data transfer autonomously. Note that traditional DMA controllers operate using physical addresses and do not inherently understand virtual addresses.

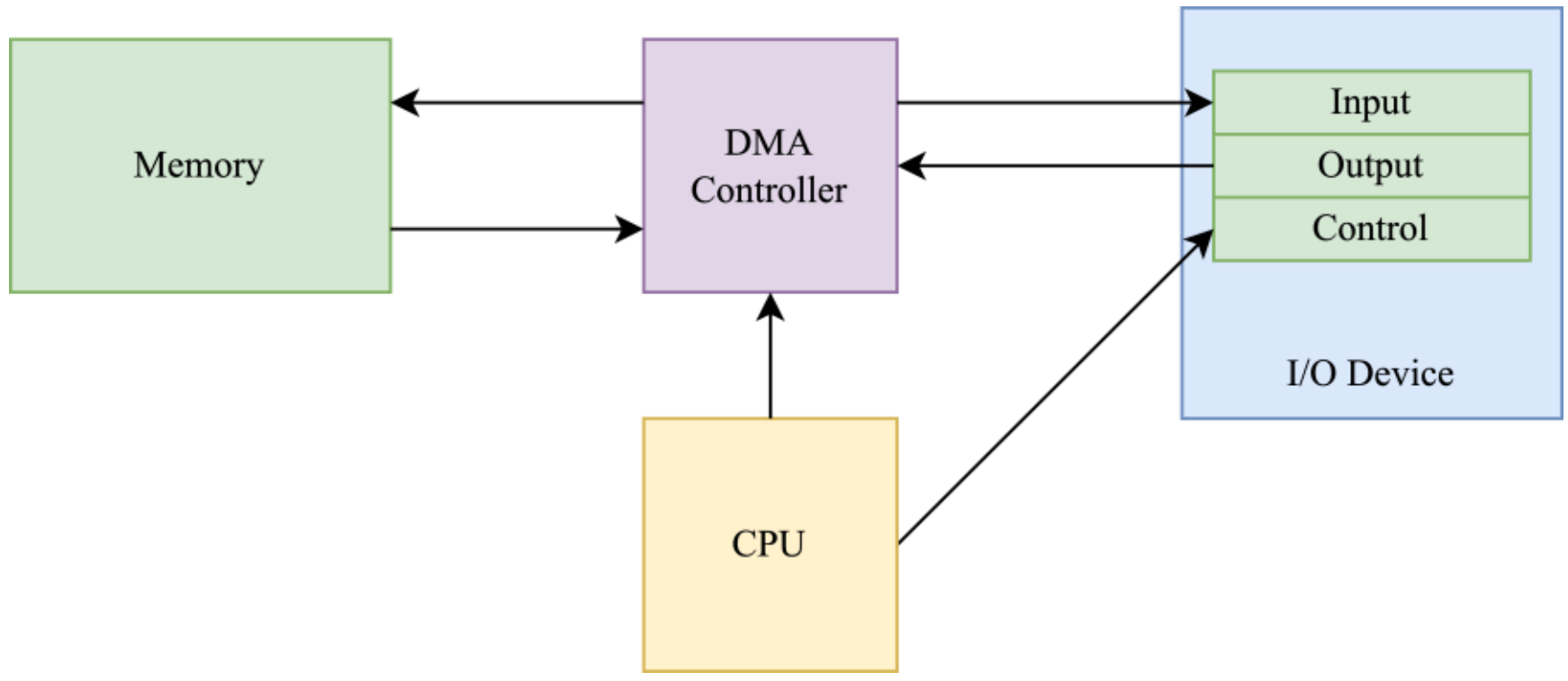

**Figure 19**: I/O via DMA controller.

Some older DMA controllers had limitations, such as only being able to address the first 4 GB of physical memory due to 32-bit addressing. In such cases, the CPU might need to re-copy data between this lower physical memory region and the actual memory locations used by user applications.

To overcome this limitation and enhance flexibility, **Input/Output Memory Management Units (IOMMUs)** were integrated, as shown in Figure 20. An IOMMU functions similarly to the CPU's MMU but performs address translation for I/O devices, allowing DMA controllers to work with virtual addresses.

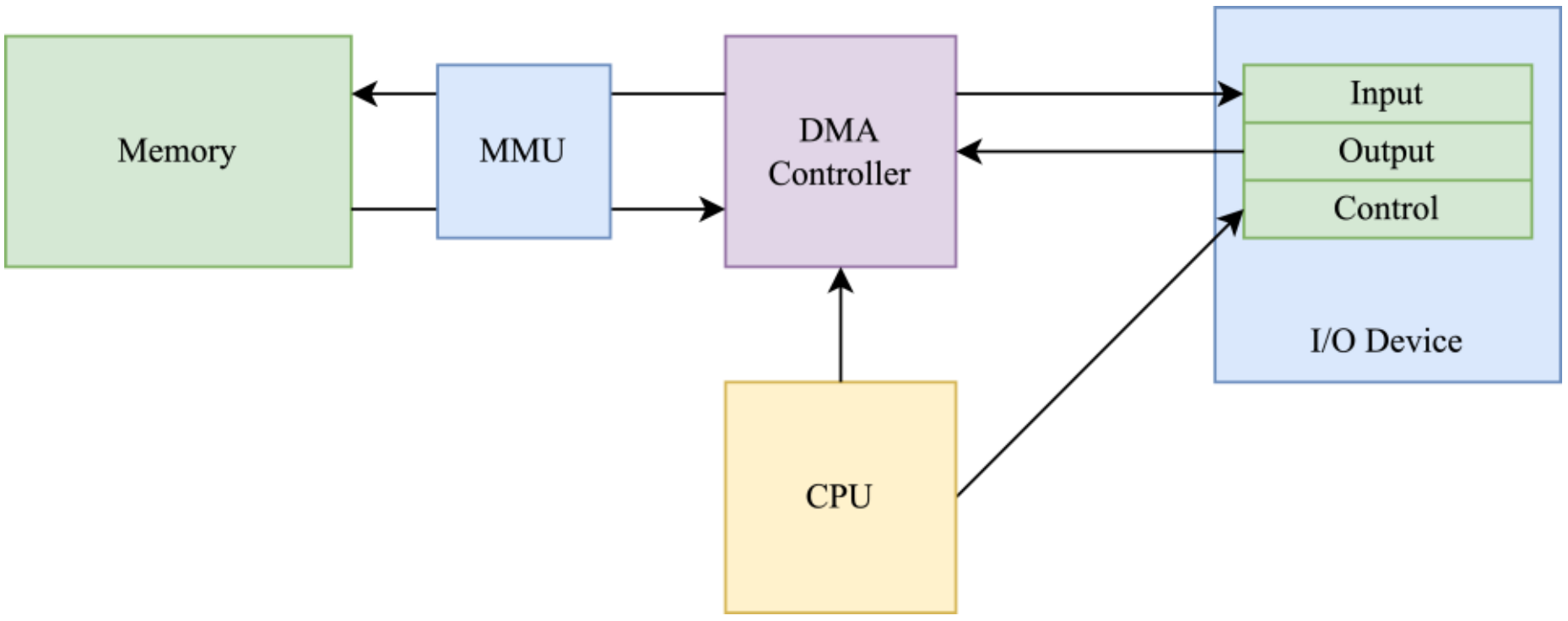

**Figure 20**: IOMMU integrated DMA controller.

### 3.8. Device Drivers

Device drivers are essentially software modules containing instructions that enable the operating system kernel to interact with specific I/O devices.

In the era of Port-Mapped I/O, device drivers often contained extensive low-level code to manage the intricacies of communicating through dedicated I/O ports. However, with the advent of Memory-Mapped I/O and DMA, device drivers have become significantly simpler. Their primary tasks involve:

- Reading control and status bits from device registers (via memory-mapped I/O).
- Instructing the DMA controller to initiate data transfers between specified physical RAM addresses and the I/O device's input/output regions (address and number of bytes).

Device drivers are responsible for translating the virtual addresses used by applications into the physical addresses required by DMA controllers.

Note that in systems without a DMA controller, device drivers will directly perform data transfers to and from memory-mapped I/O regions using standard memory access instructions.

Also note that, user-space applications are generally restricted from directly accessing I/O devices. Instead, they interact with devices by reading and writing to buffers within the kernel's address space. The kernel then initiates the actual I/O operations to the device based on these buffer accesses.

## 3.9. Memory Management in ESX Server

ESX Server operates as a type-1 hypervisor (a.k.a. Virtual Machine Monitor) meaning it runs directly on the host hardware, controlling all hardware resources and mediating access for VMs. A critical function of this type-1 hypervisor is memory virtualization, which involves creating an abstract memory space for each VM.

### *3.9.1. Recap: Memory Virtualization*

Every process on a VM running on a type-1 hypervisor possesses its own distinct virtual address space. Within this space, a **guest page table** is maintained by the guest operating system. This guest page table is responsible for translating VPNs into PPNs. From the perspective of the guest operating system, these PPNs represent the actual physical addresses of memory.

$$\textbf{Guest Page Table}: \textbf{VPN} \rightarrow \textbf{PPN}$$

However, since the VM is running on top of a hypervisor, the PPNs understood by the guest operating system are not the true physical addresses. They are, in fact, virtualized physical addresses. To access the actual hardware memory, these PPNs must be further translated into machine page numbers (MPNs), which correspond to the host's physical memory addresses.

To avoid the performance overhead of a double translation (VPN → PPN → MPN), there is a **shadow page table**. The shadow page table directly maps the virtual address space of a guest process to machine (host) physical page numbers.

$$\textbf{Shadow Page Table}: \textbf{VPN} \rightarrow \textbf{MPN}$$

The guest operating system remains oblivious to the existence and operation of the shadow page table; it continues to manage only its own guest page table.

In addition to the guest and shadow page tables, hypervisors maintain a third data structure: the **pmap**. The pmap maps guest PPNs to MPNs.

**pmap**: **PPN → MPN**

3.9.1.1. Hidden Page Fault

A crucial aspect of this system is how changes are handled. When the guest operating system modifies its guest page table entries (e.g., adding or removing page mappings), the hypervisor may not immediately update the corresponding shadow page table entries. If a guest process attempts to access a page that is mapped in the guest page table but not yet reflected in the shadow page table, it results in a **hidden fault**. Upon detecting such a hidden fault, the hypervisor updates the shadow page table to include the necessary mapping.

With that background, let's focus our attention on various strategies used by ESX Server for optimization.

*3.9.2. Strategy 1: Memory Overcommitment and Reclamation Strategies*

**Overcommitment** is a widely adopted practice in cloud and cluster computing. It involves allocating more virtual resources to VMs than the physical resources available on the host machine. For example, a machine with 100 GB of RAM might allocate 50 GB to each of three VMs, resulting in a total commitment of 150 GB. This strategy is predicated on the assumption that not all VMs will simultaneously demand their full allocated memory capacity.

Furthermore, the memory limits assigned to each VM are typically soft limits. This means that VMs are not rigidly capped at their allocated memory unless the system experiences memory pressure and the VMs are exceeding their limits.

When the host machine's physical memory resources become constrained, pages from a VM's memory need to be reclaimed. This usually involves swapping these pages out to disk. In a bare-metal operating system, the operating system itself has knowledge of its memory usage and can intelligently select the "best" pages to swap out, often employing policies like LRU (Least Recently Used).

However, a significant challenge arises because ESX Server is a type-1 hypervisor. The VMs are opaque to the hypervisor; it cannot "see" inside a VM to determine which pages are truly useful or recently accessed by the guest operating system and its applications. This lack of insight makes it difficult for the hypervisor to make informed decisions about which pages to reclaim.

#### 3.9.2.1. Ballooning

Ballooning is a clever technique specifically designed to address the opacity issue. A balloon driver is installed in each guest operating system. This balloon driver communicates with the hypervisor through an appropriate channel (e.g., inter-process communication or hypercalls).

When the hypervisor determines that memory needs to be reclaimed, it instructs the balloon driver within a guest VM to "pin" certain pages. In response, the guest operating system is compelled to start paging out its non-pinned pages to its own swap space. Crucially, the pages "pinned" by the balloon driver are then effectively returned to the hypervisor, making that physical memory available for allocation to other VMs. This setup is shown in Figure 1 in the paper.

This technique introduces an element of **paravirtualization**. Paravirtualization occurs when the guest operating system is aware of the hypervisor's presence. The balloon driver essentially paravirtualizes the guest setup.

Simply having the balloon driver pin pages is not enough. To ensure that the guest operating system truly cannot access these reclaimed pages, the hypervisor takes an additional step: it removes the corresponding entry in the shadow page table. If the guest operating system then attempts to access an MPN that has been reclaimed (and whose entry has been removed from the shadow page table), it will result in a fault. This fault can then be handled by the hypervisor, which might redirect the access to a new page (e.g., a zeroed-out page or a page with random values) that is no longer part of the VM's active memory.

#### 3.9.2.2. Random Demand Paging

Even after ballooning has been employed, there might be scenarios where pages still need to be forced out of memory. For example, the ballooning mechanism might be disabled by the guest operating system (the host doesn't have much control over what the guest can do; the guest can choose against paravirtualization for security reasons).

In such situations, ESX Server resorts to demand paging itself, directly paging out memory to the swap space. However, the ESX Server does not use an LRU policy for this.

The reason for avoiding LRU is to prevent a "LRU + LRU" scenario. If both the hypervisor and the guest operating system were to use LRU, a page deemed LRU by the hypervisor and swapped out could simultaneously be deemed LRU by the guest operating system. When the guest operating system then attempts to access this same page, it would cause a page fault, forcing the hypervisor to page it back in (and mark it as recently used). As a result, ESX Server opts for random paging when it directly performs demand paging to avoid these conflicts.

#### 3.9.2.3. Evaluation

The paper's presentation of its evaluation is somewhat unconventional. Instead of a dedicated, consolidated section for evaluating all techniques, it discusses the evaluation of each technique immediately after its technical explanation.

Figure 2 in the paper provides a comparative analysis. On the x-axis, it represents different VM sizes. The gray bars illustrate performance when ballooning is actively reducing the VM's memory footprint to the indicated size. The black bars, in contrast, depict performance when a hard memory limit is set to that specific size.

The evaluation indicates that performance with ballooning remains notably close to raw (native) performance. A slight overhead observed is attributed primarily to the CPU usage involved in the ballooning process itself. Additionally, the paper mentions that there are some memory overheads associated with the larger data structures that ESX Server maintains for memory management.

### *3.9.3. Strategy 2: Content-Based Page Sharing*

The second technique: sharing pages across VMs when they have byte-identical content. This happens when pages load common library code that can be reused across different VMs.

The system regularly scans pages across different VMs, looking for matches. It hashes each page for an initial comparison, then performs a full comparison to prevent hash collisions from causing issues.

If a page isn't found to be shared, it gets a **hint frame** with the following fields:

- **hash**: The hash of the page content.
- **MPN**: The machine page number.

- **VM**: The VM it belongs to.
- **PPN**: The physical page number known to the VM.

A page with only a hint frame can be modified freely by the VM. Once the scan finds another matching page (using a full match), the hint frame is converted to a **shared frame** with a reference count. Additionally, the page is marked as protected in the shadow page table of all VMs sharing it. As a result, if the page is written to, it will be copied to a fresh machine page number, and the reference count will be decremented. The shadow page table is then updated to map the VPN to the new common MPN. These interactions are shown in Figure 3 in the paper.

#### 3.9.3.1. Evaluation

When running identical VMs, Figure 4 in the paper shows that 60% of the memory is shared. There's a gap between the *shared* and *reclaimed* lines because at least one page needs to exist for all its copies. The absolute gap between shared and reclaimed is constant. The percentage gap is large with two machines but closes as the number of VMs sharing the page increases, demonstrating the effectiveness of sharing.

Even with one VM, there's a 5 MB memory saving from shared zero pages (pages with entirely zero content).

In a real-world scenario, page sharing reclaimed one-third of all RAM usage. This is very useful because cloud providers can offer and sell more space while incurring fewer charges, at the cost of a slight CPU overhead.

#### 3.9.3.2. Considerations

A question that comes up is how this impacts cache performance and interacts with huge pages. Contiguous virtual page numbers should ideally be assigned contiguous machine page numbers. However, with shared pages, the machine page numbers might be random. This could affect L2 and L3 cache performance.

### *3.9.4. Strategy 3: Proportional Memory Distribution*

This technique addresses how the ESX Server efficiently distributes memory among various VMs.

#### 3.9.4.1. The Problem of Memory Distribution

A straightforward approach to memory management might involve simply allowing oversubscription and allocating memory proportionally to what each VM requests. This is typically achieved through **share-based allocation**, where each task is assigned a number of shares, and memory is distributed accordingly.

The paper introduces a refined metric: the **share-per-page ratio**. This ratio essentially indicates how many shares a VM is willing to "burn" for each page of memory it's allocated. The core idea is to incentivize active memory utilization. If a page is actively used, fewer shares are paid for it. Conversely, if a page is not actively used (idle), it costs more shares. This mechanism encourages VMs to actively utilize the memory they've been allocated.

Once the share-per-page ratio ($\boldsymbol{\rho}$) is determined for a VM, the pages are allocated proportional to it.

#### 3.9.4.2. Calculating Share-Per-Page ($\boldsymbol{\rho}$)

The share-per-page ratio ($\boldsymbol{\rho}$) for a VM is calculated using the following formula:

$$\boldsymbol{\rho} = \frac{\mathbf{S}}{\mathbf{P} * \left(\mathbf{f} + \mathbf{k} * (\mathbf{1} - \mathbf{f})\right)}$$

**S**: Total number of shares assigned to the VM.
**P**: Total number of pages allocated to the VM without idle page adjustment.
**f**: The fraction of pages that are actively used.

Here, **k** represents a price constant for idle pages. You can think of **k** as a multiplier that increases the cost in shares for keeping idle pages. It's expressed as:

$$\mathbf{k} = \frac{\mathbf{1}}{\mathbf{1} - \boldsymbol{\tau}}$$

where $\boldsymbol{\tau}$ is the tax rate.

- If $\tau = 0$, then $k = 1$. In this scenario, idle pages and used pages have the same price in shares.
- If $\tau = 0.99$, then $k = 100$. This means idle pages are 100x more expensive in terms of shares.

With a high tax rate, like $\tau = 0.99$, a VM with a significant amount of idle memory will have a very high share-per-page ratio, leading to a much lower RAM allocation. This can trigger ballooning much faster, reclaiming those idle pages.

#### 3.9.4.3. Determining the Fraction of Idle Pages (**f**)

The last crucial step is to determine **f**, the fraction of idle pages for a VM. This is done through **statistical sampling**: the system randomly picks **N** pages, removes their entries from the shadow page table, and then monitors for faults upon access.

The system maintains three estimations for **f**:

- A slow exponentially weighted moving average of the ratio of faults to samples (**t/n**) over many samples.
- A faster weighted average that adapts more quickly to changes.
- Another version of the faster average that incorporates samples from the current period.

The system then takes the maximum of these three estimations. Taking the maximum incorporates hysteresis during downscaling: upscaling (increasing memory allocation) is rapid, while downscaling (decreasing memory allocation) is slower and more cautious.

#### 3.9.4.4. Evaluation

Figure 7 in the paper illustrates the impact of this technique:

- **No Taxes ($\tau = 0$)**: When no taxes are applied, each VM is allocated a similar amount of memory, even if one VM is idle and not processing anything.
- **Tax ($\tau = 0.75$, so $k = 4$)**: With a tax rate of 0.75, idle pages become four times more expensive than used pages. In this scenario, if VM2 has nearly 100% utilization, its memory limit is increased. Conversely, if VM1 has idle pages, its memory limit is decreased because it's paying a higher price per page. If both VM1 and VM2 are given the same number of shares, VM1 will exhaust its shares on fewer pages.

The evaluation shows that with a tax rate of 0.75, the ratio of memory allocation between a busy VM and an idle VM can become 2.5:1, instead of the 1:1 ratio observed without taxes. This highlights the effectiveness of proportional memory allocation in penalizing idle memory and reallocating it to actively used VMs.

### *3.9.5. Strategy 4: Adhering to the User Limits*

Beyond proportional memory distribution using shares, it's crucial to respect the **min** and **max** RAM limits set by the user for each VM.

Even if the calculated page allocation (based on shares and share-per-page ratio) is very small - for example, due to all pages being idle - the minimum RAM guarantee will still be honored. Conversely, the allocated memory will not exceed the maximum RAM limit.

The specified minimum for a VM actually includes a 32 MB overhead for various data structures like the pmap, shadow page table, and frame buffer. Since the ESX server only guarantees minimum memory availability, any memory between the maximum and minimum limits can be paged out to disk. This means a corresponding amount of swap space needs to be reserved for the difference (**max** - **min**).

#### 3.9.5.1. Min Memory and Reclamation

It's worth shedding some light on how min memory works with reclamation. The min setting specifies the minimum amount of memory a VM gets. No pages below that will be reclaimed. Pages above the min limit may be reclaimed as follows:

- First, ballooning will kick in to reclaim pages. The guest operating system will be forced into believing that it is using too much memory (almost close to its max), and it will itself trigger demand paging.
- If ballooning doesn't work, then the hypervisor will itself page out the pages.

#### 3.9.5.2. Page Sharing and Memory Accounting

It seems like the ESX server follows the definition of RSS. Shared pages are a transparent optimization; the memory accounted to each VM includes the shared pages as well.

### 3.9.5.3. Memory Pressure States

The system operates under different memory pressure states, which dictate its response to memory availability:

- **High (6% free)**: Plenty of free memory; no constraints are applied.
- **Soft (4% free)**: Ballooning is initiated to reclaim pages and deflate VMs that are operating above their minimum memory limits. Again, note that only the minimum limit is strictly honored.
- **Hard (2% free)**: The system begins paging out memory to disk.
- **Low (1% free)**: Execution of VMs is blocked, preferably those that are above their maximum limits.

### 3.9.5.4. Evaluation

An evaluation was conducted using five VMs, each running Windows with an application. The maximum memory for these VMs was set to 256 MB×3 and 320 MB×2. The minimum was set to half of the maximum. The total maximum memory, including the 32 MB×5 overheads, amounted to 1462 MB. Memory overcommitment was at 60%, and shares were distributed proportionally to the maximum memory.

All techniques - ballooning, proportional share allocation, demand paging, and page sharing - were enabled. Figure 8 in the paper illustrates the complex interactions when these techniques are combined. Here's a high-level overview of the observations:

- **Startup**: Upon boot, Windows accesses all pages to zero them out. This causes all VMs to initially access their maximum memory limits. Since many pages are zeroed, this also creates opportunities for page sharing, which helps keep overall memory usage in check.
- **Post-Startup**: After the initial startup phase, ballooning activates to reclaim pages allocated during start-up. This is necessary because otherwise, VMs would continue to use up to their maximum allocated limits. Page sharing becomes less effective after startup as more pages contain non-zero data.
- **SQL Server Behavior**: When Microsoft SQL Server is idle, it is still allocated its minimum memory limit. However, when it processes a large query, many of its pages become active. At this point, proportional share allocation ensures it receives more pages.

#### *3.9.6. Strategy 5: I/O Page Remapping*

As discussed earlier, memory-mapped I/O operations occur in the lower memory regions of a machine, while an application's allocated machine pages might reside in higher memory ranges. This often necessitates the use of *bounce buffers* to copy I/O bytes from the lower, I/O-mapped regions to the application's higher memory ranges.

The ESX server employs a clever technique to avoid this extra copy. It tracks virtual addresses within VMs that are frequently copied to these low I/O-mapped memory regions. Once identified, the ESX server directly maps those specific virtual addresses to the low memory regions. This direct mapping eliminates the need for bounce buffers and the associated copying overhead.

Naturally, not all VMs can be granted access to these lower memory regions.

### 3.10. Paper Remarks

The paper is a compelling study that presents several innovative frameworks for memory resource management within shared and virtualized environments. By detailing the practical implementation of these concepts within the ESX Server, it serves as an exemplary experience paper. It is an essential read for those seeking a deeper understanding of modern operating systems.

# 4. Firecracker: Lightweight Virtualization for Serverless Applications

(4) Agache, A.; Brooker, M.; Iordache, A.; Liguori, A.; Neugebauer, R.; Piwonka, P.; Popa, D.-M. Firecracker: Lightweight Virtualization for Serverless Applications. In 17th USENIX symposium on networked systems design and implementation (NSDI 20); 2020; pp 419–434.

The virtualization techniques discussed in previous chapters were highly popular and widely adopted across data centers throughout the 2000s and 2010s. However, traditional operating systems are often bulky; running multiple instances on the same hardware proved costly, particularly for cloud providers. Consequently, more lightweight virtualization alternatives were developed to improve efficiency.

This paper, co-authored by Alexandru Agache and distinguished AWS scientist Marc Brooker, along with other researchers, was presented at the esteemed Networked Systems Design and Implementation (NSDI) '20 conference. It introduces Firecracker, a lightweight virtualization framework designed to host AWS Lambda, Amazon's serverless computing service.

In this insight, we will begin by establishing fundamental cloud computing concepts and introducing the specific tools utilized in the paper. We will then conduct a technical deep dive into Firecracker.

**Recommended Read**: **2. A Comparison of Software and Hardware Techniques for x86 Virtualization** where several virtualization concepts were introduced.

### 4.1. Serverless Computing

Serverless computing is a cloud computing paradigm where the cloud provider dynamically allocates machine resources, especially compute, as needed. The cloud provider manages the underlying servers.

AWS offers a comprehensive serverless stack with a diverse range of products, including:

- API Gateway [56]: For creating and managing APIs.
- Lambda [57]: For executing serverless compute functions.
- Simple Storage Service (S3) [58]: For serverless object storage.

This paper primarily focuses on Lambda, a serverless compute platform. However, a foundational understanding of cloud object stores like S3 is also crucial.

**Recommended Read**: **13. Delta Lake: High Performance ACID Table Storage over Cloud Object Stores** for a deeper dive into cloud object stores.

Before delving deeper into serverless computing, let's briefly review some relevant technologies.

### 4.2. Virtualization

Virtualization is a technology that enables the creation of virtual resources from underlying machine resources. Broadly, there are two types of virtualization:

#### *4.2.1. Full Virtualization*

**Recommended Read**: **2. A Comparison of Software and Hardware Techniques for x86 Virtualization** where full virtualization is discussed in detail.

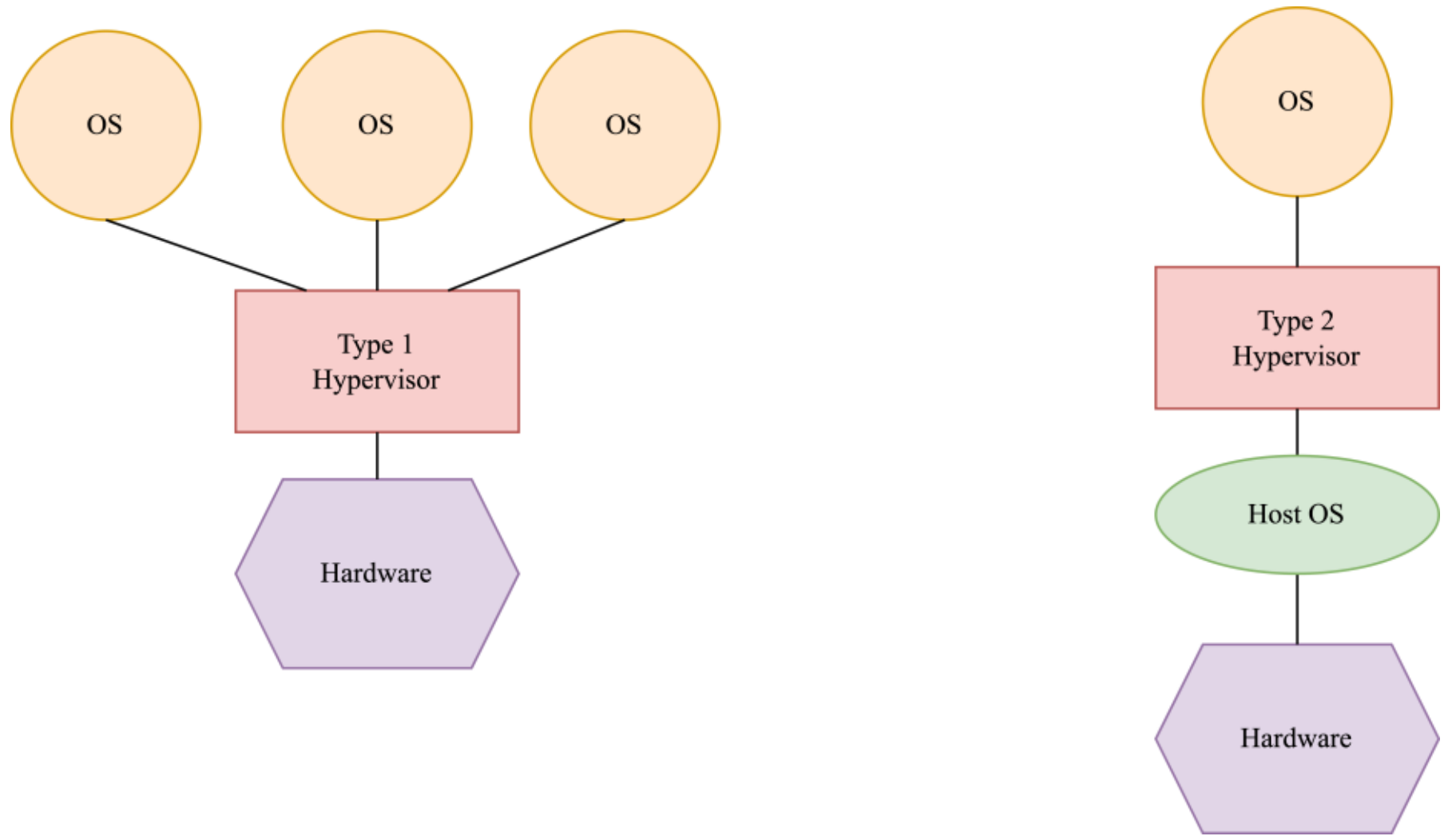

**Figure 21**: Type 1 v/s Type 2 hypervisors.

Full virtualization is complete virtualization of the actual hardware to allow software environments, including a guest operating system and its apps, to run unmodified. The guest operating system (a.k.a VMs) remains unaware of its execution within a virtual environment. VMs operate on top of **hypervisors**, which can be categorized into two types, as shown in Figure 21:

- **Type 1 hypervisors:** Run directly on bare-metal hardware, such as ESX Server.
- **Type 2 hypervisors:** Run on top of another operating system, such as VirtualBox [47].

**Note:** A hypervisor is also known as a **Virtual Machine Monitor (VMM)**.

Two common techniques for full virtualization include:

- **Binary translation (as in classical virtualization):** Translates privileged guest instructions into safer alternatives on-the-fly. Some early x86 virtualization solutions relied heavily on binary translation. The hypervisor intercepted privileged instructions (like those accessing hardware directly) and dynamically translated them into safe instructions that could be executed within the virtual machine's context.
- **Hardware-assisted virtualization:** The CPU (e.g. Intel VT-x and AMD-V) itself provides virtualization support. It utilizes specialized CPU instructions to accelerate virtualization processes.

Both Type 1 and Type 2 hypervisors can support both techniques of full virtualization. Specifically, in binary translation, the core functionalities provided by the hypervisor are:

- **Instruction translation**: Guest operating system instructions are executed in non-privileged mode on the CPU. However, when a privileged instruction (e.g., accessing hardware directly) is encountered, it triggers a trap, transferring control to the hypervisor for translation.
- **Page table translation**: Guest operating system's page table entries are translated into the corresponding entries within the host operating system's page table.

These translations are not required in hardware-assisted virtualization.

**Note:** These core functionalities have numerous variations and optimizations across different hypervisor implementations.

#### *4.2.2. OS-level Virtualization*

OS-level virtualization isolates a pool of resources within the operating system. This encompasses technologies such as containers (like Docker [48]), jails (like FreeBSD jail [49]), chroot [50], and virtual environments. The most popular among them today are containers.

The operating system has the complete control over the following:

- **Hardware capabilities:** CPU and network interface card (NIC) characteristics.
- **Data:** Access to files and folders.
- **Peripherals:** Connected devices like webcams, printers, and scanners.

Containers, however, restrict these system capabilities.

### 4.3. Linux Containers

Linux containers utilize OS-level virtualization techniques to provide application isolation. Key mechanisms enabling this isolation include:

- cgroups [59]: Isolate and limit resource consumption (CPU, memory, I/O) for individual containers.
- namespaces [60]: Create isolated environments for processes, network, and user ids.
- seccomp-bpf [61]: Restrict the system calls available to a container, enhancing security.
- **chroot:** Isolate the container's file system view, limiting its access to specific parts of the host system's file system.

Do not confuse "Containers" with "Linux Containers". "Containers" is a broader term encompassing the concept of isolating applications within lightweight, portable units. "Linux Containers" specifically refers to the mechanisms used to create and run isolated applications within the Linux operating system. In essence, containers like Docker are built upon the foundation of Linux containers.

### 4.4. Serverless Computing v/s Containers

Serverless computing focuses on event-driven functions, where code executes in response to specific triggers (e.g., API calls, data changes).

Containers, on the other hand, package applications and their dependencies into self-contained units, providing consistent execution environments across different systems.

### 4.5. KVM and QEMU

KVM [62] (Kernel-based Virtual Machine) is a Type 1 hypervisor that operates directly on the host hardware. It leverages hardware-assisted virtualization extensions like Intel VT-x and AMD-V. KVM is tightly integrated within the Linux kernel.

QEMU [63] (Quick Emulator) is a versatile open-source software that emulates various hardware components, including CPUs, disks, and network devices.

When combined, KVM and QEMU create a powerful virtualization solution. QEMU provides the necessary I/O emulation layer, while KVM harnesses the host hardware's virtualization capabilities, resulting in high-performance virtual machines.

With that background in mind, let's jump into the details of the paper.

### 4.6. Introduction to Firecracker

Firecracker is a lightweight VMM that runs on top of KVM. It is designed for serverless computing.

**Q. Why Firecracker doesn't use Linux containers (or other OS-level virtualization)?**

Firecracker prioritizes strong security isolation over the flexibility offered by Linux containers. Virtualization, with its inherent hardware-level isolation, provides robust defense against a wider range of attacks, including sophisticated threats like **side-channel attacks**. Containers offer some isolation, but can still be vulnerable to these types of attacks.

**Q. Why doesn't Firecracker use QEMU?**

QEMU is bulky. Firecracker was designed to be lightweight and efficient, enabling higher server density and minimizing overhead for serverless functions.

## 4.7. Firecracker VMM

Firecracker is a Type 1 hypervisor built upon KVM.

### *4.7.1. The KVM Layer*

Firecracker leverages KVM for hardware-assisted virtualization, enhancing performance and security:

- All CPU instructions are directly executed by the KVM layer (with hardware-assistance).
- KVM also handles thread and process scheduling within the guest environment.
- KVM provides robust CPU isolation at the hardware level, preventing unauthorized access to host resources.

### *4.7.2. The VMM Layer*

On top of KVM, instead of using the full-fledged QEMU, Firecracker employs a lightweight VMM implementation (re-using crosvm [64]) that supports a limited set of essential devices:

- **Serial ports** for basic input/output.
- **Network interfaces** for communication.
- **Block devices** for storage.

Serial ports have a lightweight implementation, requiring less than 250 lines of code. All network and block I/O operations within Firecracker are trapped into the virtio [65]. Virtio provides access to network and block devices via a serial API, which has a concise implementation of under 1400 lines of code.

### *4.7.3. Rate Limiters*

CPU and memory resources are constrained by modifying the **cpuid** instruction within KVM, limiting the available cores and memory. This approach ensures a homogeneous compute fleet. However, it is significantly less sophisticated than Linux container's cgroup mechanism, which offers advanced features such as CPU credit scheduling, core affinity, scheduler control, traffic prioritization, performance events, and accounting.

Virtual network and block devices also incorporate in-memory token bucket-based rate limiters [66]. These limiters control the receive/transmit rates for network devices and I/O operations per second (IOPS) for block devices.

### 4.8. Firecracker Internals

Firecracker's core architecture is shown in Figure 3 in the paper.

#### *4.8.1. Firecracker Architecture*

- **MicroManager:** The central component of Firecracker, responsible for spawning and managing VMs (or, as they call it, **MicroVMs**).
- Each VM is spawned as a process running on the KVM layer.
- This process itself has subcomponents:
    - **Slot:** Contains the guest operating system kernel. Within the guest kernel, the lambda function runs as a user-space process. The lambda binary includes a **λ-shim** that facilitates communication with the MicroManager over TCP/IP.
    - **Firecracker VMM:** Handles all virtual I/O operations.

#### *4.8.2. Lambda Function*

- **Slot Affinity:** Each lambda function is tightly bound to a specific slot (VM), ensuring consistent execution within the same environment.
- **Event-Driven:** Lambdas are designed to be event-driven, activating only when events (such as incoming requests or messages) arrive. This minimizes resource consumption during idle periods.
- **Slot States:**
    - ***Idle*:** The VM is descheduled, consuming minimal CPU resources but retaining its memory state.
    - ***Busy*:** The VM is actively processing events, utilizing both CPU and memory.
    - ***Dead*:** The VM has been terminated and removed from the system.

This paper explores specific scheduling constraints for lambda functions within the Firecracker environment to enhance performance and resource utilization. These constraints are detailed in paper's section 4.1.1.

### 4.9. Multi-tenancy

A significant portion of lambda functions reside in an idle state, consuming approximately 40% of the available RAM. This characteristic can help in machine oversubscription.

The economic advantage of lambda functions stems from their ability to support multi-tenancy.

For example, consider a machine with 10GB of RAM capacity. If 10 lambda functions are scheduled on this machine, each allocated 5GB of capacity, and the compliance offering is 80% (guaranteeing at least 4GB of RAM to each lambda), the machine is oversubscribed fourfold.

Assuming all lambdas can acquire their required resources upon request, the efficiency of the system can be calculated as:

$$\textbf{Efficiency} = \frac{\textbf{Total Allocated Capacity}}{\textbf{Machine Capacity}} = \frac{\textbf{40GB}}{\textbf{10GB}} = \textbf{4x}$$

This high level of efficiency contributes significantly to revenue generation.

### 4.10. Fast Boot time

The boot time for a Firecracker microVM is approximately 125ms.

**Little's Law**: Keep at least **creation rate** $*$ **creation latency** slots ready.

For example, if the desired creation rate is 100 VMs per second and the boot time (creation latency) is 125ms (0.125s), the minimum number of ready VMs in the pool should be:

**100 VMs/second * 0.125s = 12.5 VMs**

Therefore, we need to maintain at least 13 ready VMs in the pool to accommodate the expected workload.

## 4.11. Evaluation

### *4.11.1. Memory Overhead*

This analysis focuses solely on the memory consumption of the VMM process within each MicroVM, as this constitutes the primary overhead of the MicroVM. Given that all VMMs are initiated as processes, the code is loaded into shared memory segments and consequently shared across all VMMs. This static overhead is excluded from the evaluation. Therefore, the analysis only considers the sum of non-shared memory segments, as identified by the pmap [67] command.

Firecracker demonstrates exceptional lightweight characteristics, not only in terms of binary size but also due to its minimal per-VM RAM overhead, which is remarkably low at 3 MB.

### *4.11.2. I/O Performance*

Like QEMU, Firecracker VMM acts as an I/O emulator, handling all I/O operations within the virtual machine. This allows for software-level rate limiting of I/O traffic.

#### 4.11.2.1. Block Device I/O Performance

Block devices (such as SSDs and HDDs) operate at block-level. In Firecracker VMM, the reads and writes to blocks are serialized. This serial access significantly impacts performance compared to QEMU or bare-metal systems (increased latency and reduced throughput).

Large-block writes are efficient but that is because of Firecracker's lack of flush-on-write. Without flush-on-write, all writes must be explicitly flushed to the underlying storage, introducing additional latency.

#### 4.11.2.2. Network Device Performance

Emulated network devices within Firecracker also exhibit lower throughput compared to dedicated hardware.

## 4.12. Security

The paper discusses several types of attacks that can potentially affect Lambda functions:

- **Side-Channel Attacks:** These attacks exploit unintended information leakage through channels other than the primary data flow.
    - ***Cache Side-Channel Attacks*:** Multiple VMs might share the same physical cache. By analyzing cache access patterns, an attacker could potentially infer sensitive information, such as cryptographic keys.
    - ***Timing Side-Channel Attacks*:** These attacks involve measuring the execution time of another process to glean information.
    - ***Power Consumption Side-Channel Attacks*:** These attacks attempt to extract information by monitoring the power consumption of a target process.
- **Meltdown Attacks:** These attacks exploit a race condition in modern CPUs that allows a process to bypass privilege checks and potentially read the memory of other processes. The race condition happens between memory access and privilege checking during instruction processing.
- **Spectre Attacks:** These attacks use timing of speculative execution and branch prediction to infer sensitive data like cryptographic keys and passwords.
- **Zombieload Attacks:** By carefully timing the execution of specific instructions, attackers can observe the side effects of "zombie loads" (a type of memory access) and potentially extract sensitive information.
- **Rowhammer Attacks:** These attacks exploit a physical limitation of DRAM where repeated access to a specific memory row can cause unintended bit flips in neighboring rows, leading to data corruption.

The paper also provides the following remedies:

- **Disabling Simultaneous Multithreading (SMT a.k.a Hyperthreading):** SMT allows multiple instruction streams to execute concurrently on a single physical core. Disabling SMT can help reduce the potential for one process to observe the activity of another, thereby mitigating certain types of side-channel attacks.
- **Kernel Page-Table Isolation (KPTI):** KPTI isolates the kernel's page table from user-space page tables. This separation helps to prevent meltdown attacks, which exploit vulnerabilities in memory access control mechanisms.

- **Indirect Branch Prediction Barriers (IBPB):** IBPB prevents previously executed code from influencing the prediction for future indirect branches. This helps to mitigate speculative execution attacks like Spectre.
- **Indirect Branch Restricted Speculation (IBRS):** IBRS is a hardware-based mitigation technique that aims to restrict the impact of speculative execution on sensitive data.
- **Cache-Based Attacks:** Techniques like Flush + Reload and Prime + Probe exploit cache sharing between processes. Mitigating these attacks often involves avoiding shared files and implementing appropriate memory access controls.
- **Disabling Swap:** Disabling swap can help to reduce the attack surface by minimizing the amount of sensitive data that resides in volatile memory.

### 4.13. Paper Remarks

The paper is exceptionally dense, distilling numerous complex concepts into a concise format. It utilizes jargons common in cloud computing and cybersecurity, which may initially overwhelm readers unfamiliar with these domains. However, it serves as an excellent introduction to modern lightweight operating system implementations, demonstrating both application performance and industry-grade security.

# 5. The Design and Implementation of a Log-Structured File System

(5) Rosenblum, M.; Ousterhout, J. K. The Design and Implementation of a Log-Structured File System. ACM Transactions on Computer Systems (TOCS) 1992, 10 (1), 26–52.

Moving beyond virtualization, we will now focus our attention on file systems - an essential subcomponent of the operating system that governs the logical organization and management of data on storage media. Over the next four chapters, we will examine various types of file systems and explore the underlying mechanics of modern storage hardware.

This paper was authored by M. Rosenblum (co-founder of VMware) and J. Ousterhout in 1992. It explores Log-Structured File Systems (LFS). While LFS was previously considered obsolete, the rise of Solid State Drives (SSDs) has rekindled interest in its core principles, particularly the concept of immutability. Furthermore, modern file systems, such as RAID 5, incorporate principles from LFS. HP's commercial AutoRAID [68] product, for example, is based on LFS.

In this insight, we will begin with the fundamentals - defining what a file is, what a file system is, and how a file system integrates into the broader operating system. We will then discuss some basic file systems before conducting a deep dive into LFS.

### 5.1. File

A file is an ordered collection of bytes. Files can reside in various locations, such as on disk, in memory, or across a network. This chapter focuses on disk-based files.

The Von Neumann architecture can work with just processors and memory. However, the need for files arose from the desire for persistence. Files provide a mechanism to save the results of a program so they can be retrieved and used later, essentially preserving data across sessions.

Essentially **File** is also an interface exposing crucial APIs:

```
class File {
    static void Open(char* path, char* mode);
    void Read(void* buffer, int nbytes);
    void PRead(int pos, void* buffer, int nbytes);
    void Write(void* buffer, int nbytes);
    void PWrite(int pos, void* buffer, int nbytes);
    void Flush();
    void Seek(int pos);
    int Tell();
    void Close();
};
```

When a program interacts with a file, it uses a File object, which maintains an internal cursor to track the current position within the file. The File interface typically provides the following methods:

- **Open(path, mode):** Opens the file at the given path in the specified mode. Common modes include "r" (read), "w" (write), and "a" (append), with variants like "w+" (create if it doesn't exist).
- **Read(buffer, nbytes)**: Reads *nbytes* from the current position into buffer.
- **PRead(pos, buffer, nbytes)**: Reads *nbytes* from the specified position into buffer.
- **Write(buffer, nbytes)**: Writes *nbytes* from buffer at the current position.
- **PWrite(pos, buffer, nbytes)**: Writes *nbytes* from buffer at the specified pos.
- **Flush()**: Forces all pending writes to disk, committing them.
- **Seek(pos)**: Moves the cursor to the specified pos. Subsequent Read() and Write() operations will start from this new position.
- **Tell()**: Returns the current cursor position.
- **Close()**: Closes the file, releasing associated resources (including the File object itself).

The File interface is one of the most misused interfaces in Linux with several unconventional implementations. However, as long as the interface's contract is adhered to, such manipulations are sometimes acceptable for achieving performance gains.

### 5.2. File Inode

Files stored on disks are divided into equally-sized **blocks**. The **inode** data structure stores the locations of these blocks for each file. Inode data structures vary across file system implementations.

The most common approach is a table format, where entries are typically ordered by the file's block numbers. To prevent inode data structure overflow when files use an excessive number of blocks, inodes are often implemented as linked structures, as shown in Figure 22.

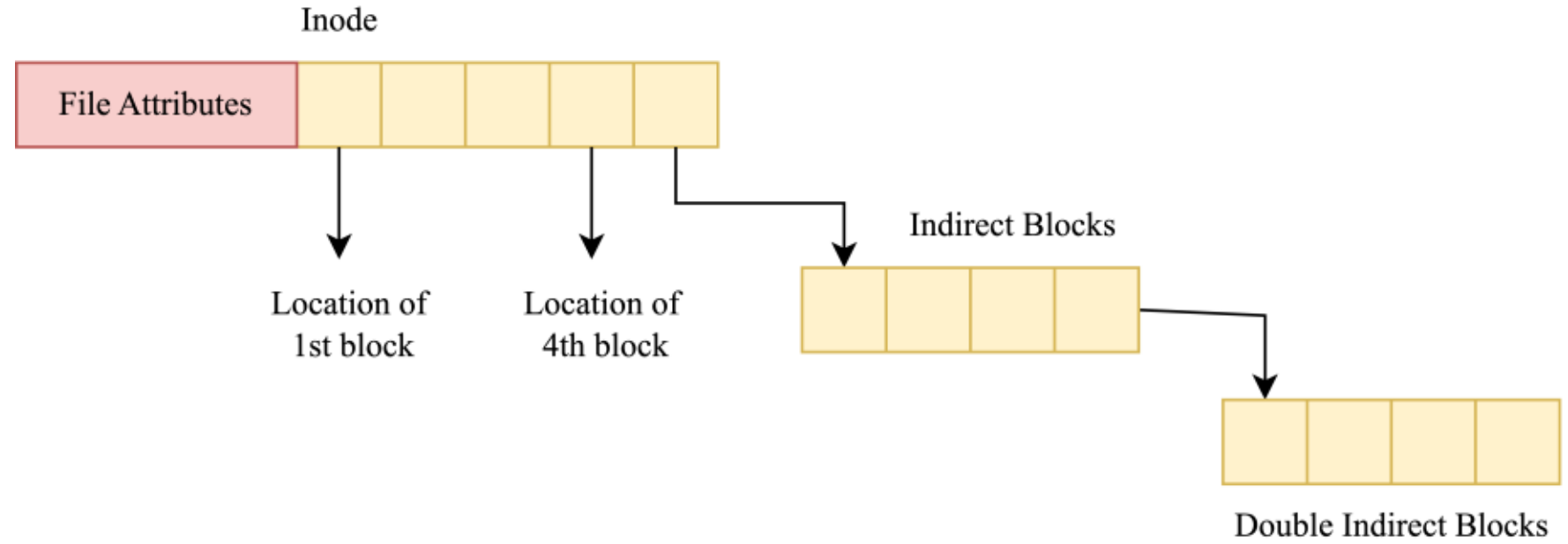

**Figure 22**: Linked inode structure.

Apart from the location of blocks, inodes also store **file attributes/metadata** such as owner, group, mtime (last modified time), etc.

**Note**: Inodes themselves are also stored on disk.

Each inode includes a **generation number**, a unique identifier for each version of a file. If a file is deleted and recreated, the new file receives a distinct generation number, differentiating it from the previous instance.

### 5.3. Directories

Files are organized into directories. In BSD-based Linux systems, a directory is itself a special type of file. Crucially, a directory file contains entries **<filename, inode>** that represent the files and subdirectories within it.

Directory entries, or **dentries**, can be stored as a straightforward table, but this approach leads to inefficient lookups. To address this, file systems use optimized data structures.

*5.3.1. Multi-Level Hash Tables*

These structures utilize hash functions for fast lookups. The hash tables are organized into multiple levels, with each level containing a growing number of dentry blocks, as shown in Figure 23. A lookup starts at level 0, and if the entry is not found, the search continues to the next level. The bits of the hash value guide the search to the appropriate bucket within each level. Unbalanced hash tables can result in O(N) worst-case lookup complexity.

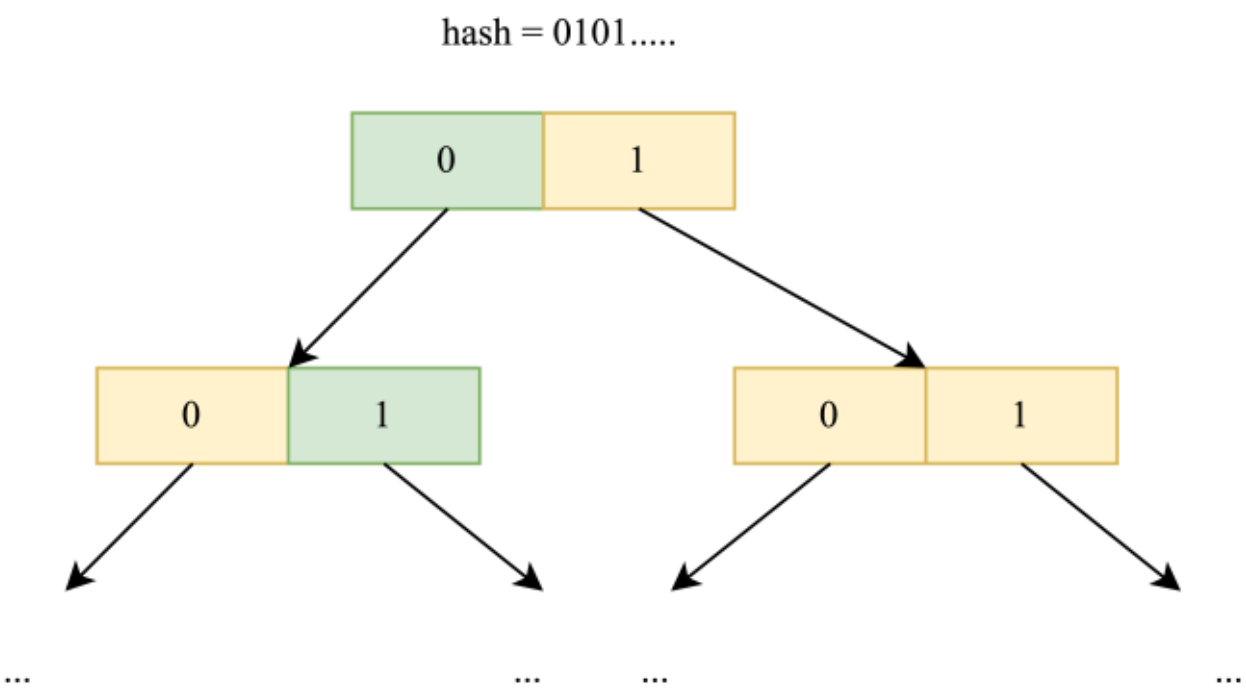

**Figure 23**: Multi-level hash table for dentries.

*5.3.2. B-trees*

Alternatively, B-trees or similar balanced tree structures are used, as shown in Figure 24. B-trees are self-balancing, maintaining sorted data and providing logarithmic time complexity for searches, sequential access, insertions, and deletions. This makes them highly efficient for large directories.

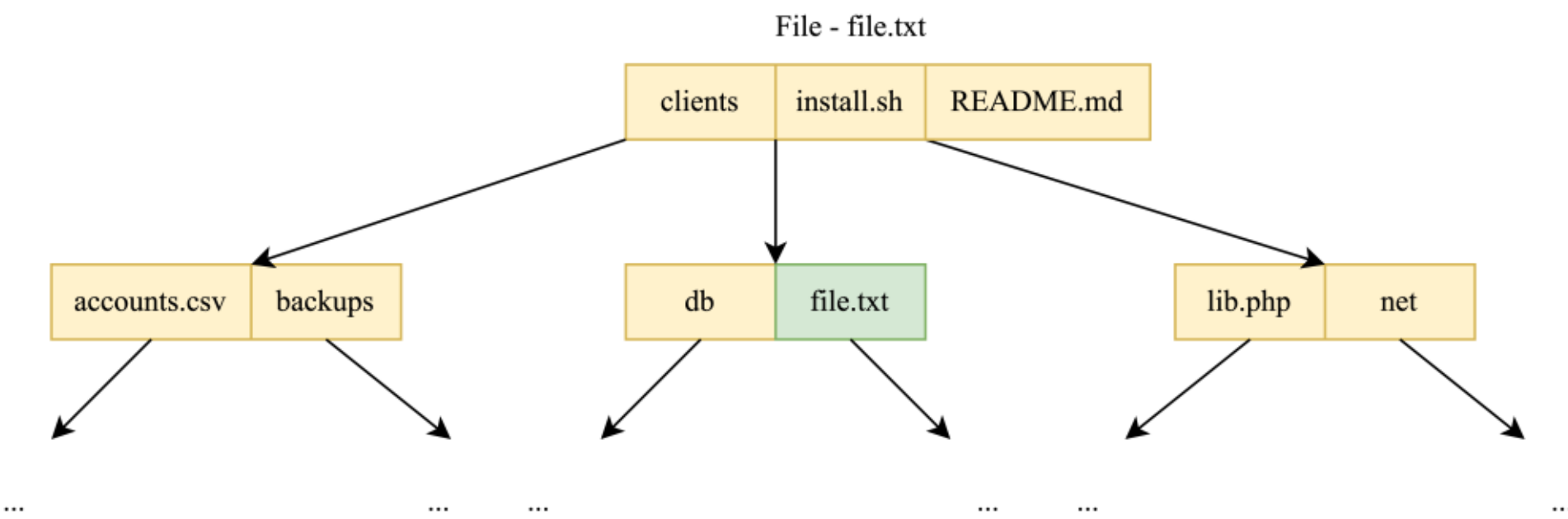

**Figure 24**: B-tree for dentries.

### 5.4. Reading a File

The root directory has a known, fixed inode location, let's say **1**. To access data within the file **/dir/file.txt**, the following steps are taken, as shown in Figure 25:

1. The directory data of the root directory is read to find the inode location of the **dir** directory.
2. The **dir** directory's inode is retrieved to determine the disk block locations of the directory's contents.
3. The directory entries are searched to find the inode location of **file**.
4. The **file** inode is retrieved to determine the disk block locations of the file's data.
5. The data blocks for **file** are fetched.

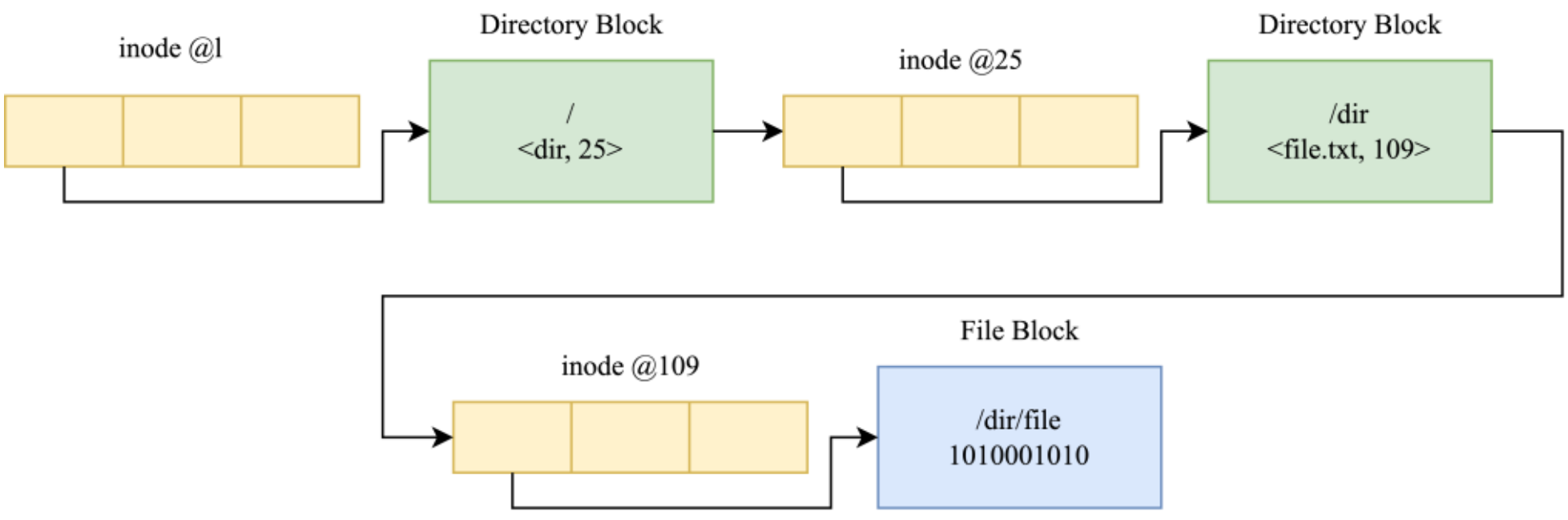

**Figure 25**: Steps in reading a file.

### 5.5. Writing to a File

When a file is written, several updates must be persisted to disk:

1. The file's data is written to allocated file blocks (possibly new ones).
2. The file's inode is updated to reflect any new data block locations and other file attributes (e.g., size, mtime, etc).
3. If the file is new:
    a. A directory entry for the new file is created within its parent directory block.
    b. The parent directory's inode is updated. Creating a new file also updates the parent directory's modification time (mtime).

## 5.6. Storage Media

Hard disk drives (HDDs) are a prevalent storage solution, offering substantial data capacity at a lower cost than alternatives such as solid-state drives (SSDs). Their operation relies on a rotating magnetic disk and a fixed read/write head that positions itself to access specific locations on the disk.

This chapter will concentrate on the characteristics and operation of HDDs.

**Recommended Read**: **6. F2FS: A New File System for Flash Storage** to learn more about how SSDs and their file systems work.

## 5.7. File System

A file system defines how files are organized and managed on storage media, specifically how file blocks are arranged on the disk.

Each file system has a **superblock** at a fixed location on disk that contains static configuration information about the file system including the type. A basic file system would store the root inode directly within the superblock. From there, the file system could navigate the entire directory tree to locate files.

Storage devices commonly support **partitioning**, a process that creates multiple logical volumes on a single physical disk. Each partition can host a distinct file system and utilize a defined subset of the disk's capacity. Partition details are stored within the superblock.

Several types of file systems exist, including:

- FAT (File Allocation Table)
- NTFS [69] (New Technology File System) - a journaling file system.
- ext4 [70] (fourth extended file system) - another journaling file system.

## 5.8. FAT File System

FAT is a relatively simple file system.

### *5.8.1. Disk Organization*

A FAT-formatted disk is organized into three primary regions, as shown in Figure 26:

1. **FAT Region:** This region stores several File Allocation Tables.
2. **Root Directory Region:** This region holds the entries for the root directory.
3. **Data Region:** This region contains the data blocks (clusters) for files and directories.

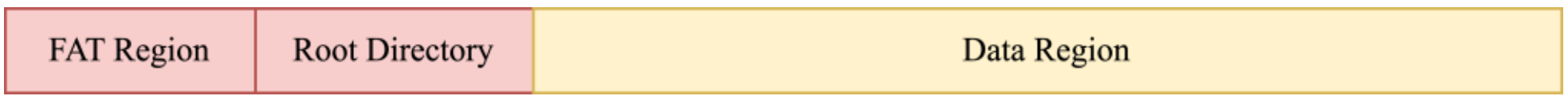

**Figure 26**: Disk organization in FAT.

*5.8.2. File Allocation Table*

Files and directories are divided into fixed-size units called **clusters**. Clusters belonging to a file are linked together in a chain. The FAT itself is a table that contains entries for each cluster. Each entry indicates one of the following:

- The cluster number of the next cluster in the chain.
- A special End-of-Cluster (EOC) marker, indicating the end of the chain.
- A marker indicating a bad cluster.
- A zero, signifying an unused cluster.

The FAT functions similarly to an inode, residing in a fixed, dedicated region on the disk.

*5.8.3. File Writes*

Note that clusters for a file or directory are not necessarily contiguous. The clusters for a single file can be scattered across the disk. As a result, write operations in the FAT file system are generally random. This is because file data blocks, directory blocks, and FAT entries can be located at disparate positions on the disk.

*5.8.4. FAT Variants*

FAT has evolved through several variants:

- **FAT12:** Uses 12-bit cluster entries, supporting volumes of limited size (e.g., floppy disks).
- **FAT16:** Uses 16-bit cluster entries, allowing for larger volumes.
- **FAT32:** Uses 32-bit cluster entries, supporting significantly larger volumes and files.

*5.8.5. Usage*

The FAT file system, particularly FAT32, is widely used on:

- Removable storage devices (USB flash drives, SD cards) due to its broad compatibility.
- Embedded systems and devices with limited resources.

### 5.9. Journaling File System

**Important note**: Journaling is not a type of file system but a feature of a file system.

A journaling file system employs a journal (or log) in addition to the main file system. This journal records pending updates, which greatly aids in recovery after a system crash or power outage.

Common journaling file systems include ext4, ext3, JFS, BRTFS, NTFS, and HFS+.

**Recommended Read**: **7. Rethink the Sync** to learn more about journaling file systems.

### 5.10. Role of Operating Systems

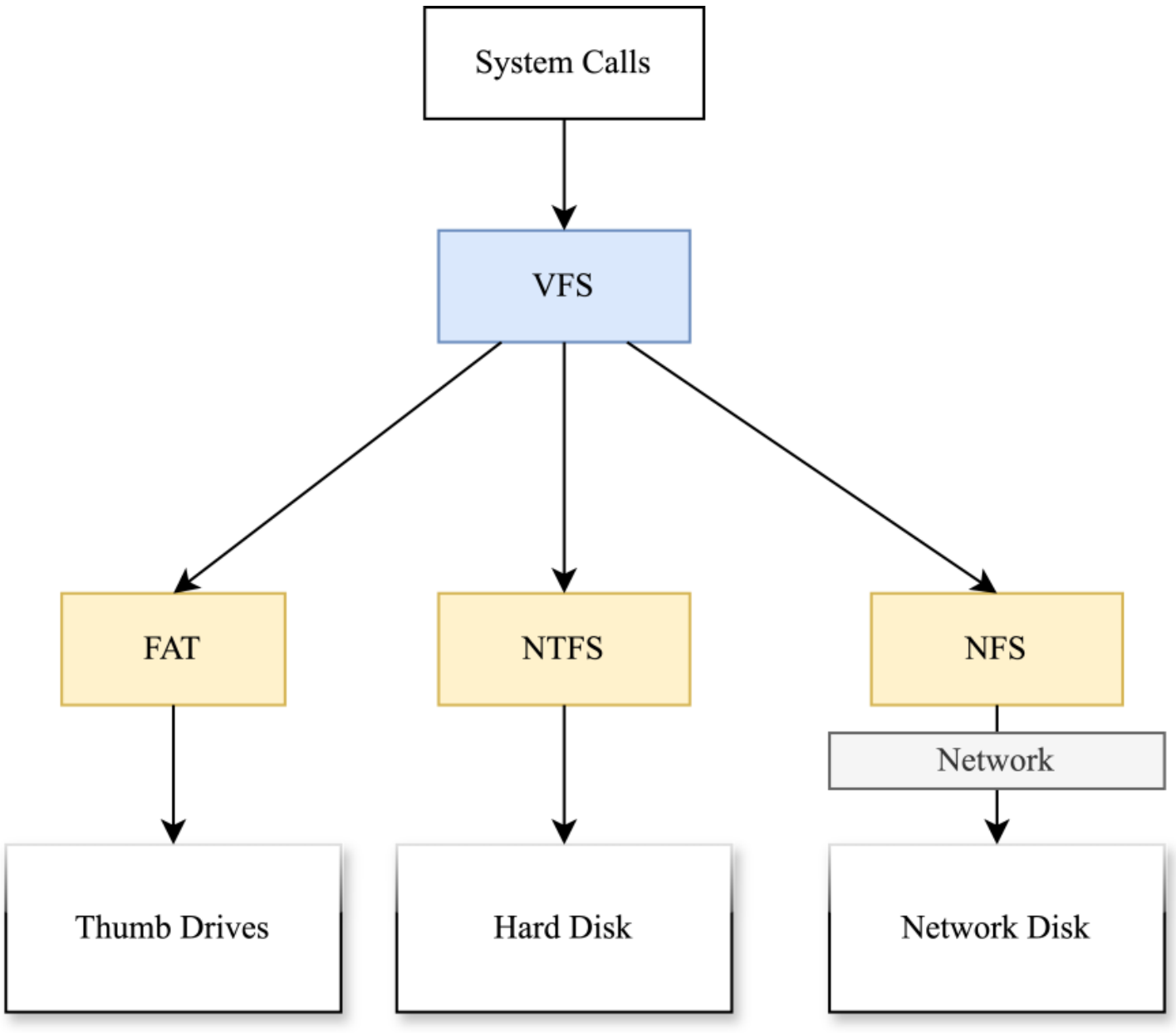

**Figure 27**: Virtual file system interface.

The operating system's primary role in file management is to interact with the file system. This interaction is facilitated by a virtual file system [71] interface, which acts as an abstraction layer, as shown in Figure 27. The file system itself provides the concrete implementation of this interface, and all client calls are routed through it.

### *5.10.1. Everything is a File*

The concept that **everything is a file** holds a significant place. This principle extends to terminals and hardware devices as well.

Terminals and hardware devices are represented as special files known as **device files**. These device files reside in the **/dev** directory. This means that the operating system interacts with terminals using the same file I/O operations (like read and write) that it uses for regular files.

Terminals are specifically classified as **character devices** because they handle data character by character. This is in contrast to **block devices** (like hard drives) that handle data in blocks.

When one interacts with a terminal, the following three files used:

- **Standard input (stdin)**: Where the system receives input from the user.
- **Standard output (stdout)**: Where the system displays output to the user.
- **Standard error (stderr)**: Where the system displays error messages.

### *5.10.2. File Descriptor*

When a process opens a file, it receives a **file handle**. This file handle contains a file descriptor.

The **file descriptor** is an integer that serves as an index into the process's file descriptor table. Each process maintains its own independent file descriptor table.

Unix-like systems define three standard file descriptors for each process: 0 - stdin, 1 - stdout, and 2 - stderr.

Other file descriptors within the table may reference file inodes located on disk.

#### *5.10.3. File & File System Commands*

Beyond standard file operations like **Open**, **Read**, **Write**, **Flush**, and **Close**, and file system operations such as **MkDir** and **Create**, operating systems provide utilities for maintenance and data integrity.

##### 5.10.3.1. fsck

**fsck** is a utility used in Unix-like operating systems (e.g., Linux) to examine and repair file system inconsistencies. It scans for errors, including corrupted data structures, bad blocks, and incorrect metadata. Its purpose is to ensure file system integrity and stability.

##### 5.10.3.2. fsync

Simply flushing a file (e.g., using a **Flush**() operation) may not guarantee immediate persistence to disk. Operating systems often optimize performance by initially writing data to in-memory buffers. The **fsync** command tells the operating system to write all these in-memory buffers to the physical storage device.

##### 5.10.3.3. sync

**sync** is similar to **fsync** but operates across all files. Additionally, the operating system will also instruct the disk buffers (described next) to be flushed. This action ensures that all pending write operations are truly committed to disk.

### 5.11. Log-Structured File System

LFS treats the entire file system as a journal. LFS was developed in the 1990s, a time when:

- Disk technology was improving in cost and capacity, but not proportionally in transfer bandwidth and access time (shown in Figure 28). Disk seek time, in particular, was a major performance bottleneck. Hard drives rely on a physical read/write head, and the disk must rotate to position the head correctly. Numerous small writes can result in excessive disk seeks, significantly impacting performance. Because disks rotate in one direction, a missed seek requires a full rotation to return to the correct position.

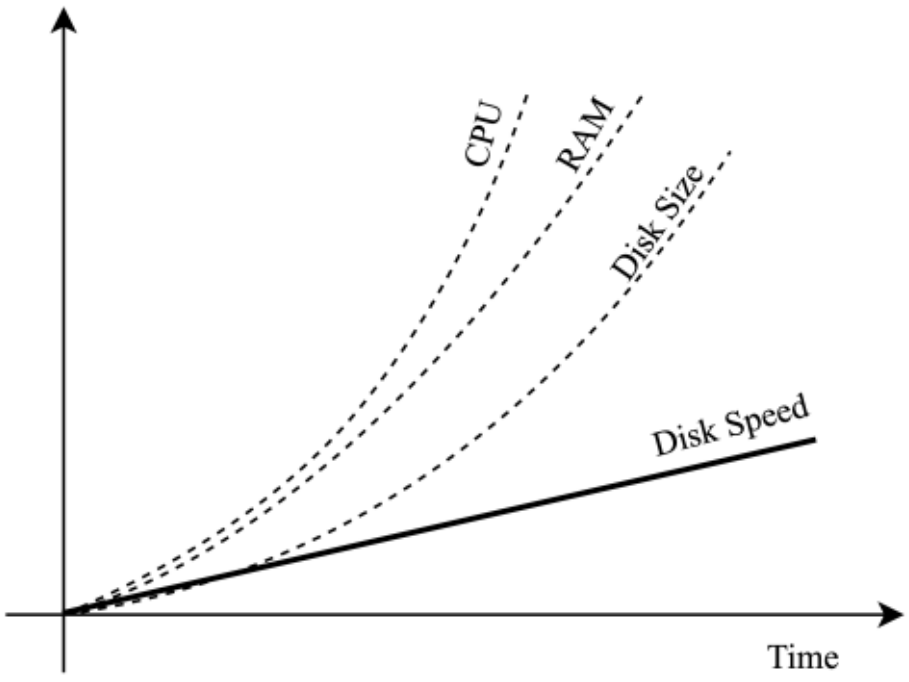

**Figure 28**: Improvement of CPU, RAM, disk size and disk speed over time.

- To mitigate the cost of disk seeks, disks often included large caches to buffer file writes, enabling more efficient block writes. However, buffered data is vulnerable to loss in the event of a system crash.

### *5.11.1. A Hint on External Consistency*

**Disk buffers** are just one example of buffering at various levels:

- **Program Buffer**: The program itself uses a buffer in its memory. **Write**() operations initially store data in this buffer. **Flush**() typically transfers data from the program buffer to the operating system buffer.
- **Operating System Buffer**: The operating system also maintains a buffer to further optimize disk writes. Files often provide a **Fsync**() API to flush data from the operating system buffer to the disk.
- **Disk Buffer**: As mentioned earlier, the disk itself has a buffer. The **sync** command triggers a disk-level synchronization command to ensure all buffered writes are committed to the physical disk.

External consistency (or external sync) is making sure that the written bytes are on the disk before an externally visible event occurs. Think about **Flush()**. One might expect data to hit the disk right away, but the operating system is often playing a smarter game for optimization. It might hold off till there is an externally visible event. Say, then, an email is sent. The operating system will then flush that file data before sending your message.

**Recommended Read**: **7. Rethink the Sync** to learn more about how external consistency is ensured by operating systems.

### *5.11.2. Random Writes to Sequential Writes*

Traditional file writing methods often require multiple disk seeks. For example, the paper asserts that a typical file system (FFS) uses five seeks to write a single file. This is because writes are often random, necessitating multiple physical movements of the disk head.

A key optimization strategy is to convert these random writes into sequential ones. The core idea is to buffer a series of write operations and then commit them to disk in a single, contiguous write. This sequential write requires only one disk seek, a significant improvement. This is precisely the approach taken by LFS.

### *5.11.3. Segmentation*

The disk is organized into segments, and each segment is written sequentially from beginning to end. Segment size is around 512KB to 1 MB. A segment effectively acts as a write-ahead log, supporting sequential writes.

A segment is composed of blocks, which are the smallest units of data read from and written to the disk. Block sizes typically range from 512 B to 4 KB. The file blocks, directory blocks, inodes, and other new data structures are logged as blocks on the segments.

We will explore how segments are created in more detail shortly.

### *5.11.4. Writing to a File - A Straw Man*

When a file is written:

1. The file's data is written to allocated file blocks (possibly new ones), which are then buffered.
2. The file's inode is updated to reflect the updated block locations within the log. This updated inode is also buffered.
3. If the file is new:
    a. The directory block is updated to include an entry for the new file and buffered.
    b. The directory's inode is updated to reflect the new directory block location and buffered.

All these buffered updates are then written sequentially in one shot. This write operation translates to a small number of sequential disk writes, depending on how many segments are involved.

However, this initial approach presents a key challenge. Because file and directory inodes are now written to new locations within the log, it becomes necessary to efficiently locate them when needed.

### *5.11.5. Inode Map*

Log-structured file systems require new data structures to manage the log-structured writes. A crucial component is the **inode map**, which tracks the location of inodes within the log.

Whenever an inode is rewritten to a new location, its corresponding entry in the inode map is updated. The inode map is designed to be compact enough such that at least its active portions can reside in memory, avoiding disk lookups. The inode map uses inode numbers (unique file identifiers) as keys and the on-disk locations of inodes as values.

With the inode map in place, directory blocks are also slightly different. They now contain entries of **<filename, inode number>**. The root directory's inode is assigned a fixed inode number.

Note that the inode map itself is logged to the segments as blocks. The fact that the inode map is itself written to the log creates a new problem: how do we find the inode map's location? We will see later how checkpoint regions help.

### *5.11.6. Reading a File*

Reading a file now involves the following steps:

1. The current directory's inode number is looked up in the inode map to get its on-disk address.
2. The directory's inode is retrieved to find the location of its directory blocks.
3. The directory blocks are read to find the inode number of the target file or subdirectory.
4. Steps 1-3 needs to be repeated until the desired file's inode number is found.

5. The file's inode number is looked up in the inode map to get its on-disk address.
6. The file's inode is retrieved to determine the location of its data blocks.
7. The file's data blocks are read.

Note that the inode map can be stored in-memory to save some costs of lookups.

#### *5.11.7. Writing to a File*

Writing a file is also streamlined, as shown in Figure 29:

1. The file's data is written to allocated file blocks (possibly new ones), and then buffered.
2. The file's inode is updated to reflect the updated block locations within the log. This updated inode is also buffered.
3. If the file is new:
    a. The directory block is updated to include an entry for the new file and buffered.
    b. The directory's inode is updated to reflect the new directory block location and buffered.
4. The inode map is updated to reflect the new inode locations and buffer it.
5. All buffered updates are then written sequentially to segments.

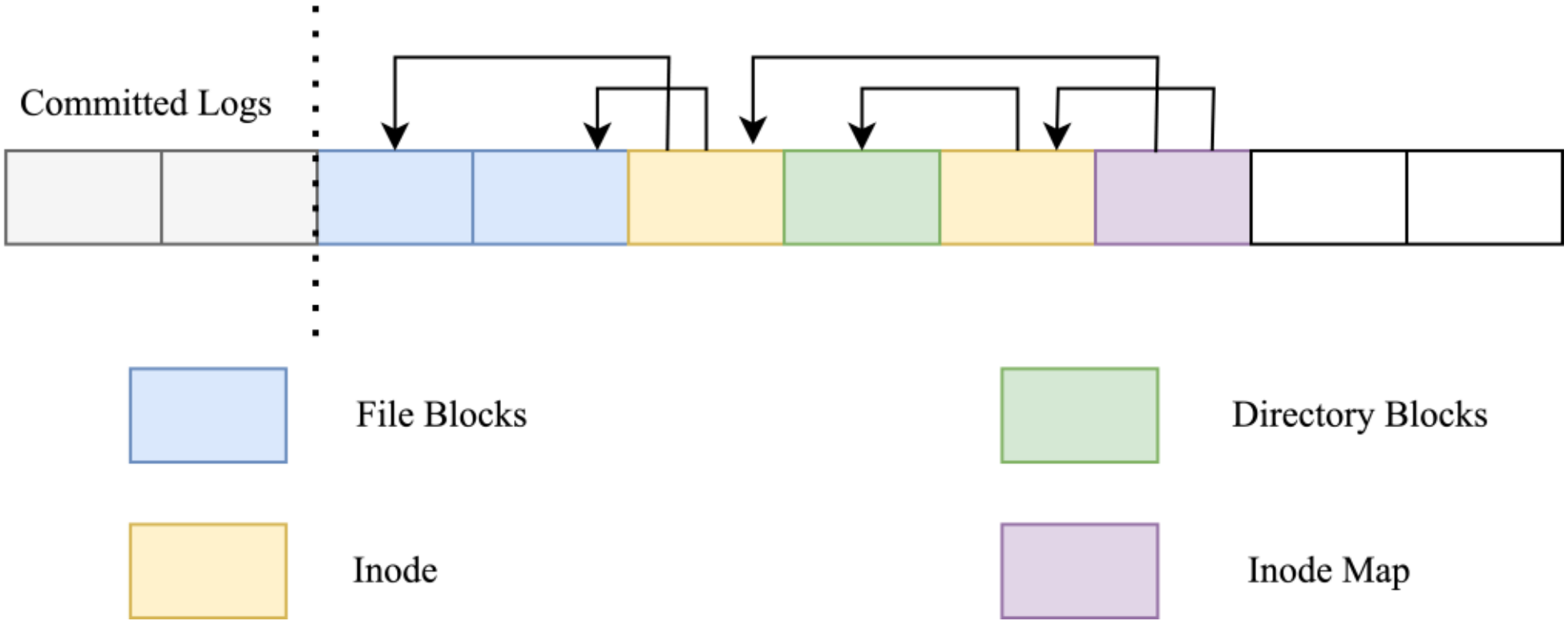

**Figure 29:** Writing to a file in LFS.

### *5.11.8. Immutability at Play*

The core principle behind LFS is leveraging the immutable nature of committed log entries. Once data is written to the log, it is not overwritten (though obsolete data can be reclaimed through cleaning, as we'll see). Immutability is a powerful concept with broad applications in system design. See Pat Helland's "Immutability Changes Everything" [72] for further exploration of this topic.

### *5.11.9. A Note on Locality*

File system writes can exhibit two types of locality:

- **Logical Locality:** This refers to the arrangement of file blocks contiguously on the disk. This allows the entire file content to be read sequentially in a single operation, regardless of the order in which the blocks were originally written. For example, a client might write to block 1 of a file, then write data to other files, and later write to block 2 of the same file. With logical locality, these blocks, though written out of order, are stored adjacent to each other on the disk, enabling sequential reads.
- **Temporal Locality:** This refers to the arrangement of file blocks according to the order in which they were written. In the example above, blocks of the same file would be scattered across the disk, reflecting the order of writes. Reading the entire file would then require multiple, non-sequential (random) disk accesses.

LFS prioritizes temporal locality of writes, meaning blocks are written in the order they are received.

## 5.12. Segment Cleaning

Segments are not necessarily stored contiguously on disk; they can be threaded together, as shown in Figure 30.

As file blocks, updated inodes, and updated inode map blocks are written to new locations, the older versions become candidates for garbage collection.

Segment cleaning involves identifying a segment containing obsolete data and copying all the *live* (still in use) blocks within that segment to a new, active segment. This process frees up the original segment, making it available for new writes.

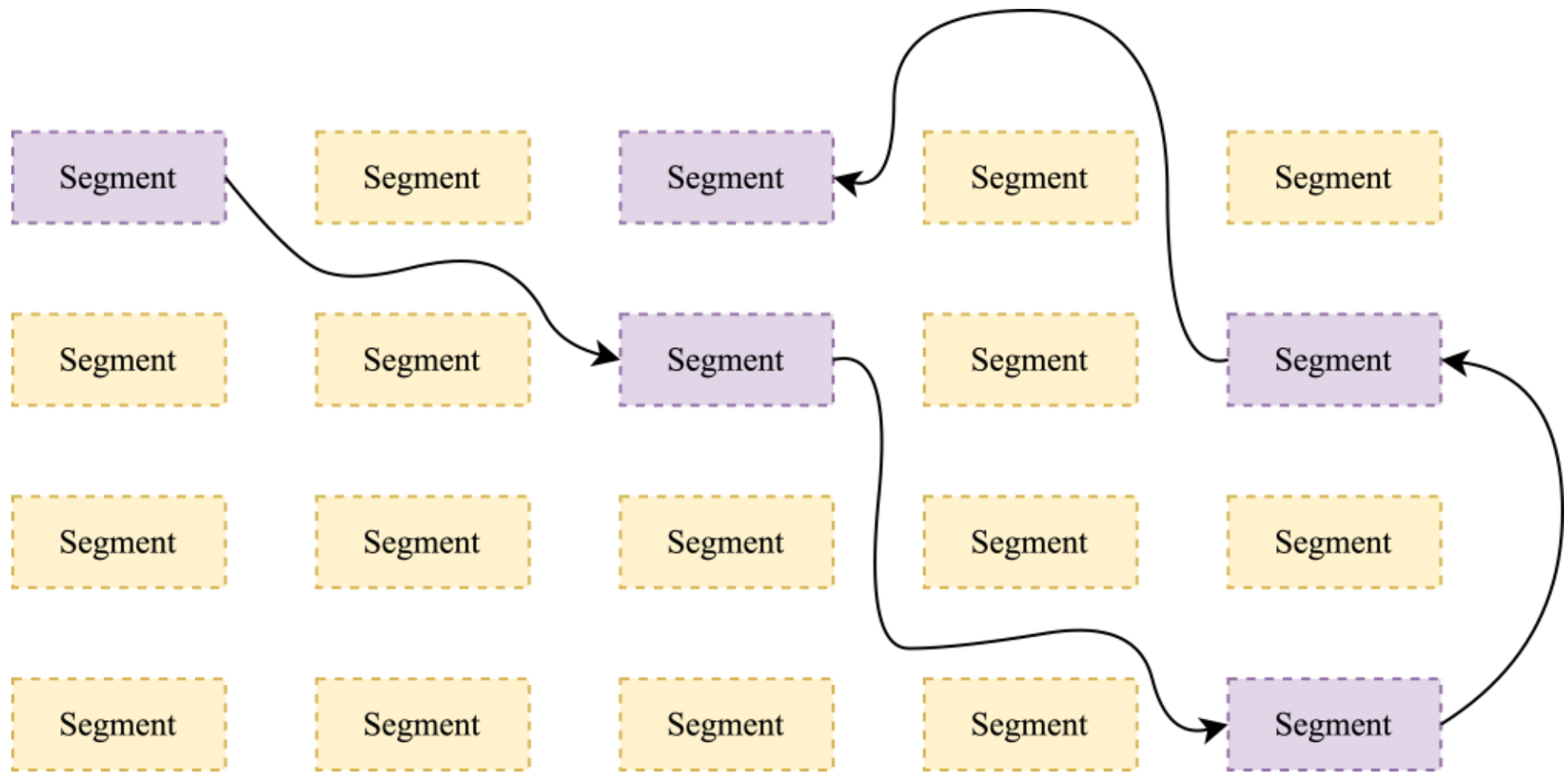

**Figure 30**: Segment organization in LFS.

Because writes to a segment must be sequential from beginning to end, clean segments are essential for continued operation.

The cleaning process selects **M** segments for cleaning and results in **N** clear segments, $\mathbf{M} > \mathbf{N}$.

*5.12.1. Write Cost*

Segment cleaning introduces overhead to the write process. The total write cost isn't just the expense of writing new log data; it also includes the cost of relocating live blocks during segment cleaning.

The authors provide a conservative estimate of this cost based on segment utilization (**u**) of live blocks. Consider **N** segments with an average utilization **u**:

- Reading the **N** segments costs **N**.
- Writing the live data out to new segments costs $\mathbf{N} * \mathbf{u}$.
- The cleaning process yields $\mathbf{N} * (\mathbf{1} - \mathbf{u})$ free space for new data.

The **effective write cost** is defined as the ratio of total disk reads and writes to the writes of *new* data only. This represents the write amplification factor.

$$\mathbf{Write\ cost} = \frac{\left(\mathbf{N} + \mathbf{N} * \mathbf{u} + \mathbf{N} * (\mathbf{1} - \mathbf{u})\right)}{\left(\mathbf{N} * (\mathbf{1} - \mathbf{u})\right)} = \mathbf{2} * (\mathbf{1} - \mathbf{u})$$

As segment utilization increases, so does the write cost. This formulation omits rotational latency and disk seek time, as these are considered negligible for the sequential reads and writes characteristic of log-structured file systems.

The formula shows that the write cost increases as utilization increases. For example, as shown in Figure 3 in the paper, the write cost reaches 14 when segment utilization reaches 80%.

5.12.1.1. Optimizing Write Cost

Files are categorized into two groups based on write frequency:

- **Hot files:** The 10% of files that account for 90% of write operations.
- **Cold files:** The remaining 90% of files, which account for only 10% of writes.

During segment cleaning, blocks are sorted by age and re-written. Stable blocks (older, less frequently modified) are written to new segments first, followed by newer blocks. This sorting leads to a rough division of segments into two categories:

- **Hot segments:** These segments contain recently written data and are more likely to have blocks invalidated frequently.
- **Cold segments:** These segments contain stable, valid data blocks.

Hot segments are more likely to contain hot files, while cold segments are more likely to contain cold files.

Now, we will see how segments are selected for cleaning.

**Approach 1: Greedy**

One approach to segment cleaning is to use a fixed cleaning threshold. A segment is cleaned if its utilization falls below this predefined threshold.

Figure 5 in the paper illustrates this: below the cleaning threshold, there are no segments because those segments would have already been cleaned and freed. The curve slopes downward after the cleaning threshold, indicating that most hot segments tend to have lower utilization (just above the cleaning threshold).

The primary disadvantage of a fixed cleaning threshold is that the utilization of hot segments drops rapidly as data becomes obsolete (note that segment utilization only accounts for live blocks). These segments quickly reach the cleaning threshold and are cleaned frequently. This high cleaning frequency is inefficient.

Conversely, cold segments experience less rapid utilization decline. Once cleaned, they tend to remain stable and require less frequent cleaning.

When a cold segment is cleaned, its live blocks are likely moved to another cold segment, which itself will then be less likely to require immediate cleaning. However, when a hot segment is cleaned, its blocks are moved to another hot segment, which will likely also need cleaning soon.

**Approach 2: Cost Benefit**

We can model the benefit of cleaning a segment as a function of two factors:

1. The amount of free space in the segment ($\mathbf{1-u}$). More free space translates to a greater benefit from cleaning.
2. The "age" of the segment, defined as the time elapsed since the most recent modification of any block within that segment. Younger segments are less beneficial to clean.

The cost of cleaning a segment is $\mathbf{1+u}$ (1 represents the cost of reading the segment, and $\mathbf{u}$ represents the cost of rewriting the live blocks).

Therefore, the cost-benefit of cleaning a segment can be expressed as:

$$\mathbf{Cost - benefit} = (\mathbf{1-u}) * \frac{\mathbf{age}}{\mathbf{1+u}}$$

The cost-benefit approach results in a bimodal distribution of segment utilization (shown in Figure 6 in the paper): hot segments tend to have lower utilization, while cold segments have higher utilization.

The fixed cleaning threshold approach focused solely on maximizing free space (cleaning as much as possible). The cost-benefit approach, however, balances the cost of cleaning with the benefit of the resulting free space.

If cold segments have limited free space, the cost-benefit policy favors cleaning hot segments, even though they have a higher cleaning cost, to maximize the amount of free space gained. Conversely, if cold segments have sufficient free space, they are prioritized for cleaning due to their lower cleaning cost.

This policy achieves a good balance between cleaning cost and benefit. The write cost is typically between 2 and 4 when the overall utilization is around 80%.

#### *5.12.2. Segment Usage Table*

All these optimizations require a new data structure called **segment usage table** to track the usage of each segment (**u**) and their last write time. The blocks of the segment usage table is itself written to logs. It is essential to track these blocks.

### 5.13. Checkpointing

The checkpoint region stores the locations of the inode map and the segment usage table. It also records the last checkpoint written to the log, which is crucial for recovery.

Unlike data blocks, inodes, and the inode map, the checkpoint region resides at a fixed location on the disk.

A checkpoint represents a point in the log where all file system structures are consistent and complete. Checkpoints are written to two distinct locations on disk for redundancy. This replication is essential because checkpoint loss would lead to file system corruption.

Replication also enables atomic checkpoint updates. To write a checkpoint:

1. All checkpoint blocks are written.
2. A timestamp is written atomically to a specific block.

Until the timestamp is successfully written, any partial checkpoint data is considered invalid. Upon system restart, the operating system selects the checkpoint with the highest (most recent) timestamp, ensuring that only complete and consistent checkpoints are used. This atomic timestamp update guarantees checkpoint integrity.

### 5.14. Roll Forward Recovery

Checkpoints are written every 30 seconds. Consequently, a crash might occur before the latest checkpoint is written, leaving up to 30 seconds of log data untracked. It's sometimes possible to recover some of this lost information through a process called **roll-forward recovery**.

*5.14.1. Segment Summary*

The **segment summary** block describes the contents of its associated segment, including file numbers (inode numbers) and offsets for each block within the segment. Critically, the segment summary block itself is written to the log, not stored at a fixed disk location.

The segment summary is essential for identifying all live (valid) blocks within a segment.

*5.14.2. Directory Operation Log*

A **directory operation log**, a special record within the log, captures each directory modification. This record contains an operation code, the location of the directory entry, the entry's content (name and inode number), and the updated reference count for the named inode. Like other log entries, the directory operation log is stored within the log itself.

Crucially, the directory operation log is written *before* the corresponding file inode or directory block. This ordering facilitates reference fixing during recovery.

*5.14.3. Recovery Steps*

The recovery process for an LFS involves the following steps:

1. The recovery program scans all segment summary blocks (by reading segments). When a summary block reveals a new inode, that inode's information is added to the inode map. The segment usage table is also updated based on the information in the segment summary blocks, reflecting the current state of each segment. If an inode for a file was never written, its associated data blocks are ignored during recovery.
2. Each inode maintains a reference count, indicating the number of directories referencing it. A key challenge is ensuring consistency between directory entries (filename - inode number mappings) and the newly recovered inodes. If a crash occurred after writing an inode but before updating its corresponding directory entry, the directory operation log comes into play.

These logs are used to correct the contents of directory blocks, ensuring they accurately reflect the current state of files.

The recovery program then appends the modified directory blocks, their inodes, the file inodes, the updated inode map, and the updated segment usage table to the log. Finally, a new checkpoint region is written to include these changes.

Similar recovery processes exist in other file systems. This explains, in part, why the operating system boot process can take a noticeable amount of time.

### 5.15. Summary of Disk Organization in LFS

Apart from the superblock, checkpoint regions are located at the beginning and end of the disk, as shown in Figure 31.

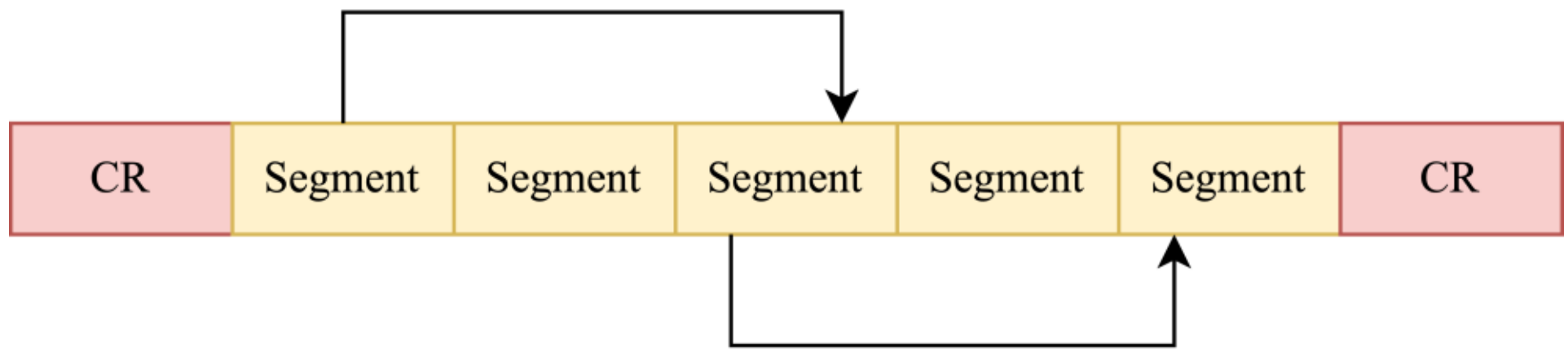

**Figure 31**: Disk organization in LFS.

The remaining disk space is divided into equal-sized segments, linked together in a list. Each segment contains blocks that can be classified as live (in use), obsolete (no longer needed), or free.

Segment cleaning is the process of identifying and consolidating live blocks to reclaim space. This concept is analogous to garbage collection in programming: Valid file/directory blocks are those reachable from the root, just as valid objects in a program are those reachable from root objects, stack globals, or the stack itself.

### 5.16. Summary of New Data Structures in LFS

Log-structured file systems introduce several key data structures:

- **Inode map:** Maps inode numbers to the inode's on-disk locations.
- **Segment summary:** Describes the contents of a segment, including file numbers and block offsets.

- **Segment usage table:** Tracks the amount of live data within each segment and the last write time for data in those segments.
- **Checkpoint region*:** Stores the locations of the inode map and segment usage table.
- **Directory change log:** Records directory operations to maintain consistency of inode reference counts.

* - Located at a fixed location.

### 5.17. Evaluation

The evaluation of LFS performance reveals the following:

- **Small File Performance:** LFS demonstrates a significant performance advantage for small file operations, exhibiting up to a 10x speedup for new file writes. Reads are also accelerated due to the sequential arrangement of file blocks for multiple small files.
- **Large File Performance:**
    - Sequential writes to large files are faster in LFS. Sequential file writes in traditional file systems can lead to random disk writes due to inode and directory updates. Instead, LFS converts all writes to sequential operations.
    - Random writes to large files are, as expected, considerably faster with LFS.
    - Sequential and random reads of large files experience a performance degradation because LFS does not preserve logical locality after writes.
- **Cleaning Overhead:** The observed write cost is significantly lower than predicted by simulations. Measured write costs ranged from 1.2 to 1.6, compared to the simulated range of 2.5 to 3. The bimodal distribution of segment utilization is highly skewed, with peaks at the extreme ends. This indicates the presence of many extremely cold segments that remain unmodified.
- **Recovery Time:** Recovering 50,000 1 KB files required a maximum of 132 seconds. The recovery time increases by approximately one hour for every 70-second increase in the checkpoint interval.

## 5.18. Paper Remarks

This paper capitalized on the prevailing trend of the time, addressing the bottleneck of slow disk access speeds. This is a well-written paper with clever ideas, fundamentally based on the concept of immutability. It is highly recommended as an excellent introduction to file system concepts.

# 6. F2FS: A New File System for Flash Storage

(6) Lee, C.; Sim, D.; Hwang, J.; Cho, S. F2FS: A New File System for Flash Storage. In 13th USENIX Conference on File and Storage Technologies (FAST 15); 2015; pp 273–286.

In the previous chapter, we explored the inner workings of a Log-Structured File System (LFS). In this chapter, we will apply those concepts to explore a file system designed specifically for Solid State Drives (SSDs) - the most prevalent storage media in modern data centers.

This 2015 paper, authored by researchers at Samsung, Korea, was presented at the highly regarded USENIX File and Storage Technologies (FAST) conference and presents the design of a file system for SSDs. This paper is of particular interest for several reasons: it provides valuable insights into contemporary SSD-optimized file systems and it highlights the increasing importance of SSDs in an era of escalating DDR5 costs and exponential data growth.

The insight begins with a deep dive into how SSDs function - exploring their strengths, their limitations, and how those limitations are addressed by hardware. We will then examine how the LFS-based F2FS file system operates effectively with SSDs, addressing the unique challenges that these hardware presents.

**Must Read**: **5. The Design and Implementation of a Log-Structured File System** where several basic file system concepts were introduced.

### 6.1. Computer Memory

Computer memory can be broadly divided into two main types: **volatile** and **non-volatile**. Volatile memory, exemplified by Random Access Memory (RAM), requires continuous power to maintain stored data; data is lost upon power interruption. Conversely, non-volatile memory retains data persistently, even when power is removed.

#### *6.1.1. Non-Volatile Memory*

Non-volatile memory can be further categorized into distinct types:

- **EEPROM (Electrically Erasable Programmable Read-Only Memory):** Data can be electrically written and erased, providing flexibility.
- **EPROM (Erasable Programmable Read-Only Memory):** Data is electrically written but requires ultraviolet light exposure for erasure.
- **PROM (Programmable Read-Only Memory):** Data can be written only once, typically by the end-user.
- **ROM (Read-Only Memory):** Data is permanently written during manufacturing and cannot be erased.

## 6.2. Flash Memory

Flash memory is a type of EEPROM that has become ubiquitous in modern technology, finding applications in solid-state drives (SSDs), mobile phones, memory cards, and USB flash drives. EEPROM, in turn, is built upon MOSFET technology.

### *6.2.1. Memory Format*

The memory format of an SSD is shown in Figure 32.

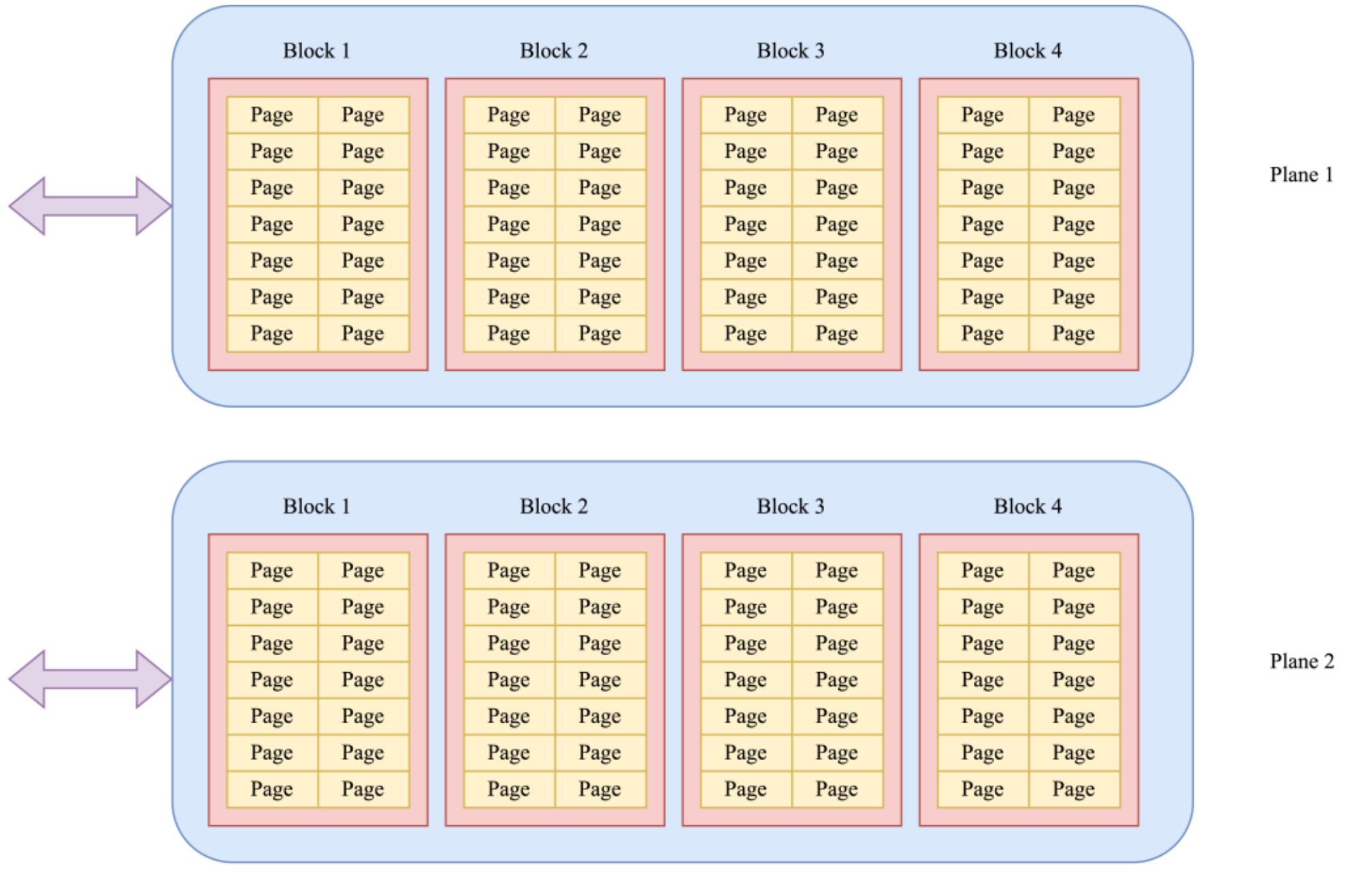

**Figure 32**: Flash memory format.

#### 6.2.1.1. Flash Cells

A flash cell is the fundamental unit for storing a bit of data. Flash cells can store varying numbers of bits:

- SLC (Single-Level Cell): 1 bit per cell
- MLC (Multi-Level Cell): 2 bits per cell
- TLC (Triple-Level Cell): 3 bits per cell
- QLC (Quad-Level Cell): 4 bits per cell
- PLC (Penta-Level Cell): 5 bits per cell

Higher bit densities per cell increase storage capacity but generally result in slower read speeds.

#### 6.2.1.2. Pages

A page is a collection of flash cells, typically ranging from 2 to 4 KB in size. The page is the smallest unit that can be read and programmed (written) in flash memory.

#### 6.2.1.3. Blocks

A block is a group of pages, typically containing 32 to 128 pages. The block is the smallest unit that can be erased in flash memory. Erase is required before re-writes.

#### 6.2.1.4. Planes

Blocks are organized into planes. Reads and writes can be performed concurrently across different planes, increasing parallelism.

#### 6.2.1.5. Dies

A die consists of one or more planes. Multiple dies are placed on a circuit board. A controller circuit is integrated to manage the storage area and communicate with the device driver.

#### 6.2.1.6. Packages

A package combines multiple dies into a single physical unit.

### *6.2.2. NOR vs. NAND Flash Memory*

Flash memory is broadly categorized into two primary types: **NOR** and **NAND**. These types differ in their underlying logic gate arrangements, resulting in variations in operational performance.

While a detailed discussion of the architectural differences is beyond the scope of, the following key distinctions can be made:

- NOR flash typically offers lower storage density compared to NAND flash, meaning NAND can store more data in the same physical space.
- NOR flash is generally more expensive than NAND flash.
- NOR flash excels at random read operations, allowing fast access to individual memory locations.
- NOR flash exhibits slower data erasure speeds compared to NAND flash.
- NOR flash is commonly used for program storage, where random read access is crucial. NAND flash is predominantly used for data storage, where high density and sequential write/read performance are prioritized.

## 6.3. NAND Flash

NAND flash memory is a cost-effective and widely adopted form of flash storage. However, it exhibits two significant limitations:

- **Block-Level Erasure:** Erase operations can only be performed at the block level. Data is erased at the block level, followed by write operations at the page level.
- **Limited Endurance:** SSDs have a finite write endurance, typically expressed as a capacity multiplier (e.g., 600x). Overwrite operations contribute to wear and tear, eventually leading to device failure.

Repeated writes to the same block can accelerate wear. To mitigate this, SSD controllers employ **wear-leveling** techniques, distributing write operations across different physical blocks. Consequently, the **logical block addresses** presented to the operating system differ from the actual **physical block addresses**.

This concept bears similarity to the log-structured file system (LFS) approach.

To manage the dynamic mapping between logical and physical block addresses, a translation mechanism is required. This mechanism, often implemented within the SSD controller's firmware, maintains a mapping table to track the current physical location of each logical block.

### 6.4. Flash Translation Layer

The **Flash Translation Layer** (FTL) is firmware responsible for dynamically translating logical block addresses (LBAs), as seen by the operating system, into physical block addresses (PBAs). This translation is performed to:

- Avoid wear-leveling, by distributing write operations evenly across the flash memory.
- Ensure that write operations are directed to free physical blocks, which have been previously erased and are ready for data.

Similar to memory, the FTL could have been a mapping of logical page numbers (LPNs) to physical page numbers (PPNs). However, due to the memory overhead associated with maintaining page-level mappings (e.g., 4 bytes per entry), FTLs often implement block-level mappings:

$$\textbf{FTL}: \textbf{Logical Block Number } (\textbf{LBN}) \rightarrow \textbf{Physical Block Number } (\textbf{PBN})$$

#### *6.4.1. Log Flash Blocks*

While the FTL effectively manages wear-leveling, it introduces write amplification. To rewrite any page within a logical block, the FTL must create a new physical block, involving the relocation of valid data from the original block. This process is performed by the SSD's hardware.

To address the performance impact of block-level erasure during page overwrites, SSD controllers utilize log flash blocks. When a page within a block is modified, the updated page is written to a log flash block. Periodically, the contents of the log flash block are committed to their corresponding physical blocks.

The mapping between log flash blocks and physical flash blocks employs different associativity schemes:

- **Block Associative:** A log flash block is dedicated to a specific physical flash block.
- **Fully Associative:** A log flash block can be used for any physical flash block.
- **Set Associative:** A log flash block is dedicated to a set of physical flash blocks, as shown in Figure 33.

Fully associative or set associative schemes are typically employed to optimize performance under random write workloads.

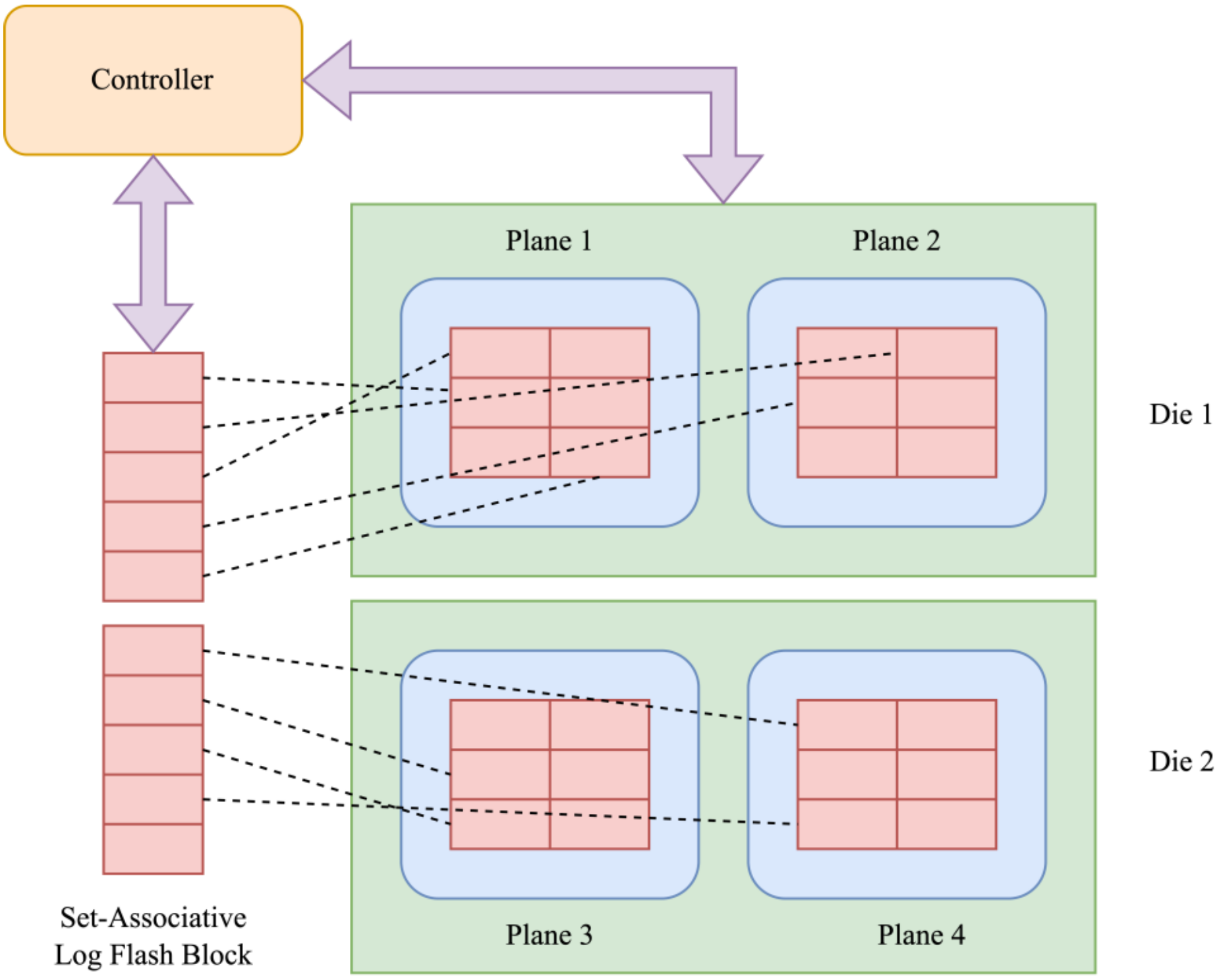

**Figure 33**: Set associative log flash blocks.

### 6.5. SATA v/s PCIe

SSDs utilize different hardware interfaces for data communication, notably **SATA** and **PCIe**.

SATA (Serial Advanced Technology Attachment) is a legacy interface, originally developed for hard disk drives, that was subsequently adopted for SSDs to maintain backward compatibility with existing operating systems.

PCIe (Peripheral Component Interconnect Express) represents a more modern interface, providing increased data transfer rates through the use of four parallel data lanes, unlike the single lane of SATA.

## 6.6. Virtual Memory

Virtual memory provides an abstraction of physical memory resources, allowing programs to operate as if they have access to a much larger, contiguous memory space, as shown in Figure 34.

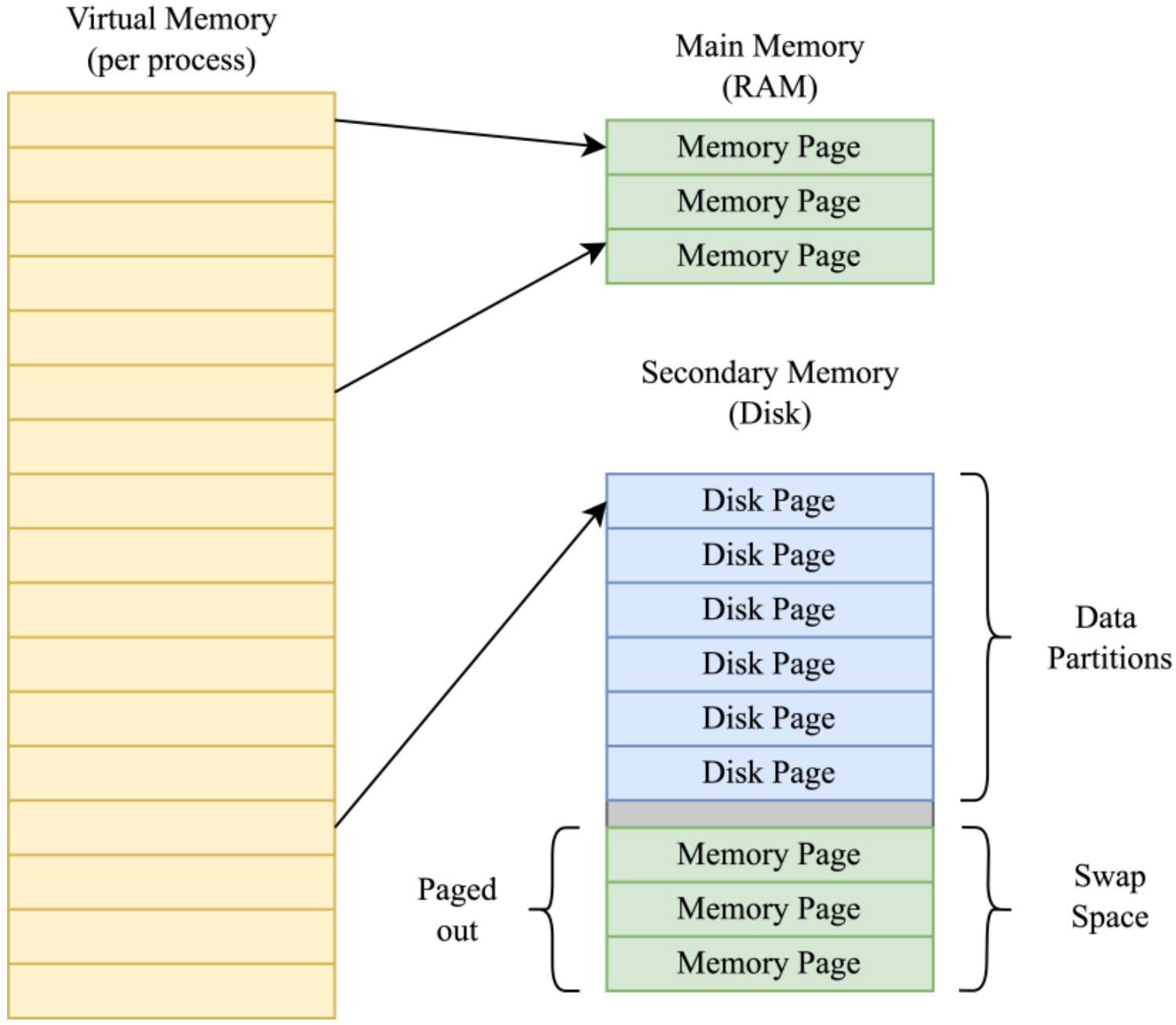

**Figure 34**: Virtual memory managed by an operating system.

The operating system, in conjunction with hardware like the **Memory Management Unit**, handles the translation between virtual addresses (used by programs) and physical addresses (actual memory locations). This system allows for the sharing of memory between processes, improves security by isolating memory spaces, and enables programs to utilize more memory than is physically installed.

### *6.6.1. Memory v/s Disk Pages*

Virtual memory uses "pages" to divide both physical memory (RAM) and disk storage. A virtual memory pages are of two types, as shown in Figure 35.

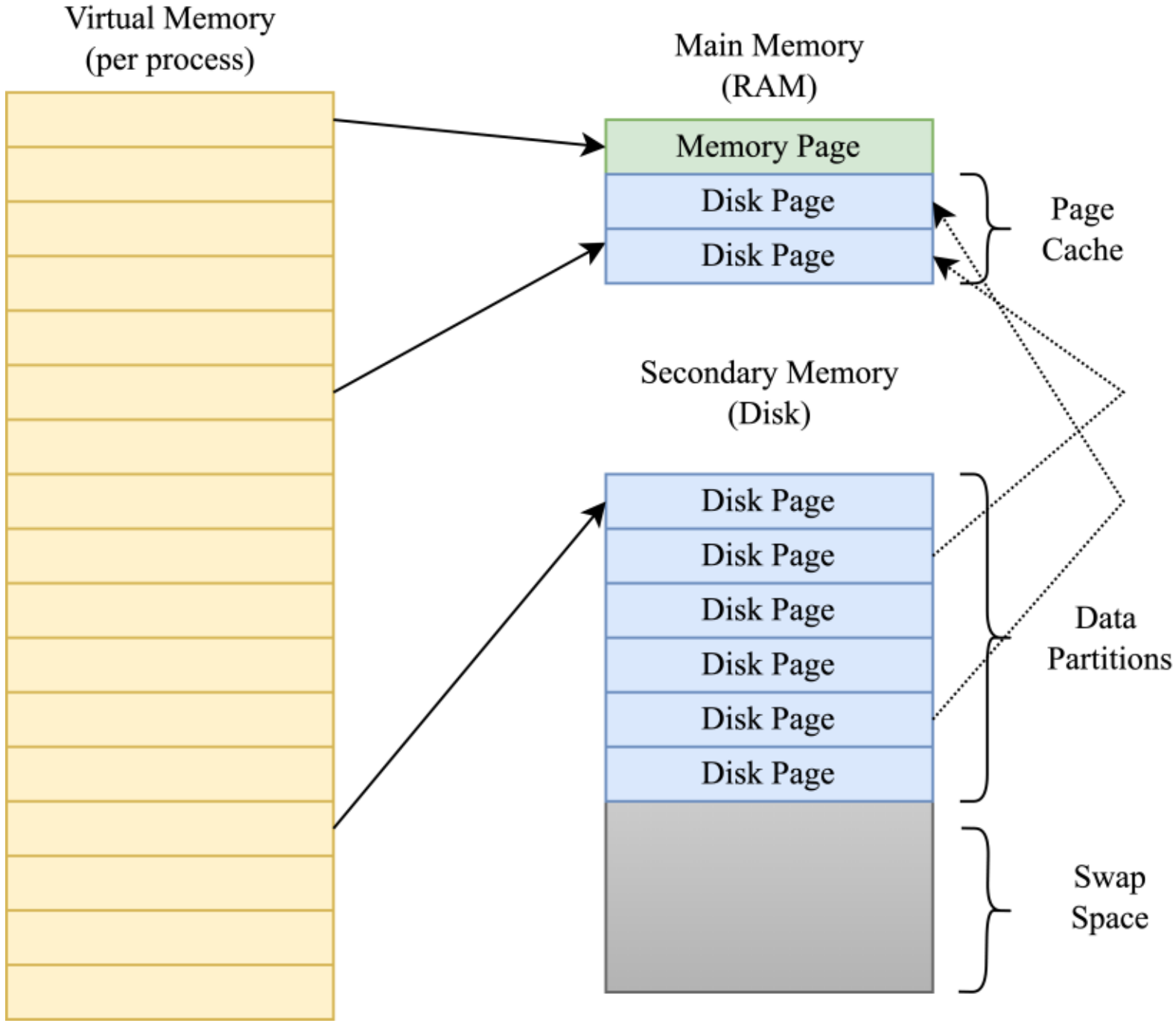

**Figure 35**: Memory pages v/s disk pages.

- **Memory Pages:**
  - These pages hold actively used data and program instructions. They provide fast access for the CPU.
  - **The primary location is RAM**. However, when RAM is full, less frequently used memory pages may be **paged out** (written) to disk.
  - If a program tries to access a page that has been paged out, a **page fault** occurs. The operating system then retrieves the page from disk and loads it back into RAM.
- **Disk Pages:**
  - These pages store data that has been committed to the disk.
  - **The primary location is the disk**. However, disk pages can be loaded into RAM if and only if free RAM space is available.

RAM is the primary location for actively used, uncommitted data, while disk is the primary location for committed data. Memory pages are only written to disk when RAM is full, and this is done to free ram, not to commit data. Disk pages are loaded into RAM when free ram is available, increasing performance.

### *6.6.2. Page Cache*

The portion of RAM that stores the disk pages is called a **page cache** or a **disk cache**. It significantly speeds up access to frequently used data on disk.

## 6.7. SSD Performance

**Random Reads**: Maximum random read throughput is achieved with 64 KB read sizes. Throughput decreases for read sizes smaller than 64 KB.

**Sequential Writes**: Optimal linear write performance is observed with write sizes of 64 KB or greater. Write performance degrades significantly for write sizes of 8 KB or less, as these writes fall below the page size, necessitating read-modify-write operations.

**Random Writes**: Random write performance exhibits a sharp decline for write sizes below 4 MB. This degradation is primarily attributed to write amplification, which arises from the need to locate and manage free blocks for smaller, scattered writes. Typically, 4MB is a block size, therefore writes below that size cause many block operations.

Sequential writes are more efficient because they write data sequentially to a block. Once a block is erased, it can be filled completely with linear write data, minimizing overwrite operations. Random writes, conversely, can result in numerous small writes distributed across multiple blocks. This leads to substantial write amplification, as even a small write to a block may trigger extensive block-level operations.

## 6.8. LFS at Rescue!

LFS improves write performance by treating the disk as a sequential log. This conversion of random writes to sequential writes is particularly advantageous for SSDs. Although this paper diverges from a strict LFS implementation, it adapts certain LFS concepts to better suit SSD workloads.

## 6.9. Flash-Friendly File System

F2FS (Flash-Friendly File System) is based on the principles of LFS, with specific adaptations for flash memory.

### *6.9.1. Disk Organization*

It partitions the disk into two distinct areas - **sequential** and **random**, as shown in Figure 1 in the paper.

#### 6.9.1.1. Sequential Write Area (Main Area)

This area is dedicated to log-structured writes of data, including file blocks and inodes. All writes are converted into sequential log entries. Segments within this area are categorized into:

- **Node Segments:** Store inodes (aka "node blocks").
- **Data Segments:** Store file/directory data blocks.

#### 6.9.1.2. Random Write Area (Metadata Area)

This area stores critical file system metadata at fixed logical addresses:

- **Superblock:** Contains fundamental partitioning information.
- **Checkpoint:** Similar to LFS checkpoints, replicated twice, and stores the addresses of the Node Address Table (NAT) and Segment Information Table (SIT).
- **Node Address Table (NAT):** Similar to an inode map in LFS or the File Allocation Table in FAT, it locates all inodes (node blocks) in the main area.
- **Segment Information Table (SIT):** Analogous to the segment usage table in LFS, it tracks the number of valid blocks and a bitmap of valid blocks within the main area, facilitating segment cleaning.
- **Segment Summary Area (SSA):** It maps data blocks to inodes.

The metadata residing in the random write area has fixed logical addresses. This can lead to potential wear-leveling issues on the physical blocks where metadata is stored. However, the FTL effectively mitigates this by translating these fixed logical writes into writes to new physical blocks, ensuring wear-leveling.

To further reduce repetitive writes, the NAT and SIT incorporate **journaling mechanisms.**

Also note that the NAT, SIT, and Checkpoint data are written to two distinct locations on the disk, providing redundancy. During checkpoint creation, the updated version is committed to disk. The SSA, written to a single location, can be reconstructed from other available information.

#### 6.9.2. Reading a File

Reading a file is similar to that in LFS, except that the inode number is looked up in the NAT instead of inode map:

1. The current directory's inode number is looked up in the NAT to obtain its on-disk address.
2. The directory's inode is retrieved to find the location of its directory blocks.
3. The directory blocks are read to locate the inode number of the target file or subdirectory.
4. Steps 1-3 are repeated until the desired file's inode number is found.
5. The file's inode number is looked up in the NAT to get its on-disk address.
6. The file's inode is retrieved to determine the location of its data blocks.
7. The file's data blocks are read.

#### 6.9.3. Writing to a File

Writing to a file is also similar to that in LFS, with the notable exception that NAT and SIT updates are journaled and buffered until a checkpoint is performed:

1. The file's data to is written allocated file blocks (possibly new ones) is buffered.
2. The file's inode is updated to reflect the updated block locations within the log. This updated inode is also buffered.
3. If the file is new:
    a. The directory block is updated to include an entry for the new file and buffered.
    b. The directory's inode is updated to reflect the new directory block location and buffered.
4. A journal entry is added to update the NAT and SIT and buffered.
5. All buffered updates in step 1-3 are then written to segments. Data blocks go to data segments while node blocks go to node segments.

**fsync**

F2FS utilizes the page cache, where file data is initially written to in-memory disk pages upon a flush operation. However, these in-memory changes are not persistently committed to disk until **fsync** is called on the file.

Even after **fsync**, only the file's data and inode are written to their respective data and node segments. Crucially, the NAT and SIT updates, which are maintained as journal entries, remain buffered. These vital metadata updates are only committed to disk during checkpoint operations.

### *6.9.4. Inodes*

An inode contains (shown in Figure 2 in the paper):

- File attributes and metadata, including the file name, inode number, size, and modification time (mtime).
- Direct pointers to the file's data blocks or, for small files, inline data.
- Indirect pointers, which point to blocks containing further pointers to the file's data blocks.

Given a disk page size of 4 KB, small inodes are inefficient. To maximize space utilization, data for files smaller than 3692 bytes is inlined within the inode.

**Problem: Wandering Tree Problem in LFS**

The inode map in LFS only tracks the location of the inode itself, not its associated indirect blocks. The addresses of these indirect blocks are solely maintained within the inode.

This design introduces a critical challenge: when a new entry is added to an indirect block, the block must be written to the log, resulting in a change of its on-disk address. Consequently, the inode, which stores the indirect block's address, must also be updated and written to the log. Again, writing the inode to the log causes its own address to change, necessitating an update to the inode map. This, in turn, requires the inode map to be written to the log, creating a cascading update problem.

**Solution: Inode Numbers for Indirect Blocks**

In F2FS, indirect pointer blocks are also assigned inode numbers, requiring translation to physical addresses via the NAT.

The use of inode numbers for indirect blocks, registered within the NAT, addresses the wandering tree problem associated with LFS. When an entry is added to an indirect block, only the NAT needs to be updated; the inode itself remains unchanged.

*6.9.5. Directory Structure*

A 4 KB directory block (also known as a **dentry** block) is organized as follows:

- **Bitmaps (27 bytes/216 bits):** These bitmaps indicate the validity of each slot within the directory block.
- **Dentries (216 slots x 11 bytes per slot):** Each slot contains (shown in Figure 36):
    - Hash value
    - Inode number
    - Name length and name
    - Type (file or directory)

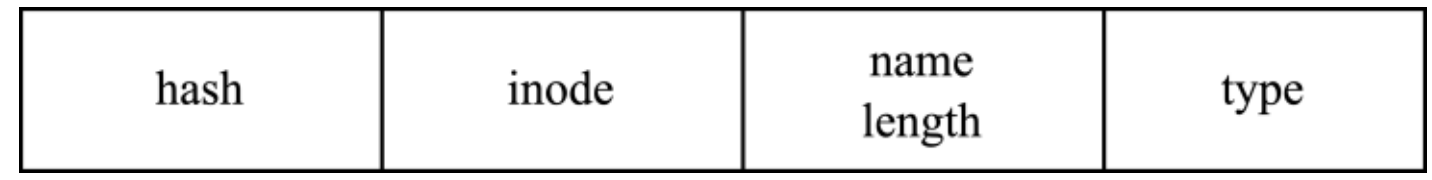

**Figure 36**: Dentry structure.

- **Overflow File Names (216 slots x 8 bytes per slot):** This section accommodates file names that exceed the space allocated within the dentry itself.

These directory blocks are structured as a multi-level hash table, which exhibits O(N) worst-case complexity. In contrast, file systems like XFS [73] employ B-tree structures, offering a more efficient O(log N) worst-case complexity.

*6.9.6. Segment v/s Sections*

Data is organized into **segments**, each a 2 MB contiguous block group, serving as the primary management unit.

A **section** is composed of segments and is a unit for garbage collection and cleaning. It functions similarly to segments in LFS.

### 6.10. Logging

#### *6.10.1. Multi-head Logging*

SSDs, unlike HDDs, eliminate the mechanical head, enabling parallel writes. Consequently, multiple segments can be written concurrently, with sequential writes occurring within each segment.

F2FS enhances data management by classifying segments into three categories: **hot**, **warm**, and **cold**, expanding upon LFS's two-category approach.

This classification impacts node and data blocks as follows:

- **Node Blocks:**
    - Direct node blocks for directories are classified as hot.
    - Direct node blocks for files are classified as warm.
    - Indirect blocks are classified as cold.
- **Data Blocks:**
    - Directory entry blocks are classified as hot.
    - Newly created data blocks are classified as warm, analogous to the hot category in LFS.
    - Data blocks moved during cleaning are classified as cold, analogous to the cold category in LFS.

The log, along with its corresponding segments, is partitioned into six distinct sections, each managing the log for one of the above categories. To maximize parallelism, F2FS maintains six open segments concurrently, one from each category (**multi-head logging**).

F2FS maps each of these six partitions to separate **zones**. While the paper lacks specific details, these zones likely correspond to distinct planes within the SSD. This zoning allows the log flash block to maintain set associativity, ensuring that its blocks map to a single zone.

### *6.10.2. Adaptive Logging*

When the file system exceeds 95% capacity, F2FS initiates **threaded logging**, writing directly to invalid blocks before cleaning can free sections. Since all blocks are uniform in size, external fragmentation is avoided.

Threaded logging introduces random writes which can overwrite logical blocks. FTL saves the logical block overwrites by mapping them to new physical blocks. FTL mitigates the performance penalties of cleaning near full capacity by trading-off the write performance. This trade-off was deemed beneficial by the authors.

## 6.11. Cleaning

Section cleaning in F2FS is divided into foreground and background processes, each following these steps:

- **Victim Selection:**
  - Foreground cleaning employs a greedy strategy, selecting the section with the fewest valid blocks. The number of valid blocks is obtained from the SIT.
  - Background cleaning uses a cost-benefit curve, choosing sections based on the number of valid blocks and their age. This information also comes from the SIT.
- **Valid Block Identification and Movement:**
  - The SIT is used to locate all valid blocks.
  - The SSA is then utilized to identify valid node blocks associated with the data blocks. The SSA ensures that when a data block is cleaned, the corresponding node blocks are updated to reflect the data block's new location.
  - The updated locations of the node blocks are **buffered into the NAT and SIT journals**.
  - **Lazy Migration:** Flushing valid blocks doesn't immediately write to new sections. Instead, they are buffered into the page cache, organized into the disk pages of the new section, and marked as dirty. The operating system then flushes the entire new section to disk.
- **Mark Victim as Pre-free:**
  - The victim section is not immediately freed. It is marked as pre-free and remains unavailable until a checkpoint is written.

- The checkpoint ensures that all updates to the NAT and SIT are committed (in order to honor **fsync**). Without this, the NAT and SIT might still reference outdated locations. Therefore, the section is only freed after the checkpoint completes.

### 6.12. Checkpointing and Recovery

It's crucial to reiterate that **fsync** only guarantees the persistence of data blocks and node blocks. Updates to the NAT and SIT are recorded as journal entries within a buffer, not written directly to disk.

A checkpoint operation then performs the following:

- All dirty node and dentry blocks residing in the page cache are flushed to disk. All dirty file data blocks are also flushed at this stage.
- Metadata modifications are temporarily suspended to ensure consistency.
- The NAT and SIT are written to dedicated, non-active regions on the disk. SSA is also written.
- The Checkpoint Pack (CP) region is written, containing:
    - NAT and SIT bitmaps, along with any buffered journal entries.
    - The location of the SSA blocks.
    - **Orphaned inodes**: These represent inodes for files that were opened but not properly closed, requiring later garbage collection. This can occur, for example, when a file is opened by multiple processes and deleted by one before the others close it.

The validity of a checkpoint is determined by examining its header and footer regions. These regions contain information that indicates if the checkpoint was completed successfully.

#### *6.12.1. Roll-back Recovery*

Roll-back recovery proceeds as follows:

- The most recent valid checkpoint region is located.
- The active NAT and SIT are retrieved from this checkpoint.
- Any journal entries are applied to the retrieved NAT and SIT.
- Orphaned inodes and their associated data blocks are deleted.
- A new checkpoint is created to finalize the recovery process.

#### *6.12.2. Roll-forward Recovery*

**Rollback recovery alone is insufficient for complete data integrity.** While it establishes a consistent checkpoint state, it fails to account for writes committed via fsync that occurred after that checkpoint. This is because the journaled and buffered updates to the NAT and SIT are only committed during checkpoints, leaving fsync writes potentially unrecorded in the restored metadata.

Therefore, a roll-forward recovery process is required to recover these post-checkpoint fsync writes. The objective is to identify all node blocks written after the established checkpoint, as the corresponding data blocks can be recovered from them.

To facilitate this identification, each node block contains a special marker, typically the checkpoint number. By scanning the log for node blocks with a marker value greater than the checkpoint number, the system can identify and recover the necessary post-checkpoint writes.

### 6.13. Evaluation

This study compares the performance of four file systems: F2FS, ext4 [70] (journaling), BTRFS [74] (journaling), and NILFS2 (LFS implementation), across mobile and server workloads.

Btrfs, a modern Linux file system, utilizes a Copy-on-Write (COW) mechanism, enhancing data integrity through snapshot capabilities.

TRIM [75], a command for SSD optimization, plays a role in the evaluation.

**Mobile Systems:**

- **Inzone:**
  - F2FS outperformed EXT4 by 3.1x due to its ability to convert random writes to sequential writes.
  - F2FS surpassed NILFS2 by avoiding synchronous NAT and SIT writes.
  - Btrfs also outperformed EXT4 and NILFS2, leveraging sequential writes during COW, but it did not match F2FS's performance.
  - Read and sequential write performance was consistent across all file systems.

- **SQLite:**
  - F2FS excelled due to its log-structured nature, ideal for SQLite's sequential WAL log writes.
  - Btrfs exhibited the poorest performance due to significant copy overhead.

**Server Systems (SATA and PCIe):**

- **SATA:**
  - Sequential read/write performance was consistent (e.g., videoserver).
  - EXT4's numerous small TRIM commands hindered performance, while F2FS's section-based cleanup was more efficient (fileserver).
  - Log-structured file systems (F2FS, NILFS2) benefited varmail's frequent sync operations.
- **PCIe:**
  - Performance mirrored SATA results, except in fileserver, where concurrent buffered writes normalized performance across file systems.

**TRIM Behavior (paper's Figure 5 Analysis):**

- Figure 5 in the paper is hard to read due to the small plots, and magnified sections. It shows writes as blue crosses, reads as red circles, and discards as gray squares.
- F2FS and NILFS2 demonstrated increasing logical block addresses with each write.
- EXT4 and Btrfs exhibited higher TRIM command frequencies.

**Multi-Head Logging (paper's Figure 6 Analysis):**

- With 6 heads, F2FS achieved a bimodal distribution in cleaning efficiency (note that Figure 6 in the paper is a cumulative distribution).
- F2FS requires at least 2 heads (data and nodes) and so 1-node analysis was not conducted.

**Adaptive Logging:**

- F2FS demonstrated superior random write performance at 94% device fill.

- F2FS's adaptive logging helped maintain performance parity with EXT4 at 100% device fill.

### 6.14. Paper Remarks

This paper highlights the power dynamic between hardware and software: hardware leads the innovation cycle, and software must adapt. The hardware trends will continue to dictate the direction of systems research, with quantum computing being the next major milestone for software architects.

Samsung's vertical integration is a key takeaway; their deep understanding of SSD "nitty-gritty" allowed them to build a storage-aware file system that third-party developers might struggle to replicate. The paper is an easy, informative read and a great primer on SSDs.

# 7. Rethink the Sync

(7) Nightingale, E. B.; Veeraraghavan, K.; Chen, P. M.; Flinn, J. Rethink the Sync. ACM Transactions on Computer Systems (TOCS) 2008, 26 (3), 1–26.

Having understood how file systems work, it is now time to focus our attention on a critical aspect: modifying their state. File system state modification encapsulates several types of changes, the most important being how updates to a file are committed. One approach is to synchronously commit every file change to the storage media. However, this can lead to an enormous performance hit - for example, when many small changes are made in rapid succession. Periodic bulk commits can improve performance, however, they come at the cost of potential consistency loss. Systems researchers have spent years developing methods to commit changes that maximize performance without sacrificing consistency.

This paper, presented at the highly regarded Operating System Design and Implementation (OSDI) '06 conference, represents a significant contribution to this research. Authored by Nightingale et al. (University of Michigan), it distinguishes itself by providing empirical evidence of a magnitude of performance improvement in file systems that had previously eluded researchers. Beyond this quantitative achievement, the paper also introduces a paradigm shift by altering fundamental operating system behavior based on external observation.

In this insight, we will begin by introducing how processes communicate with each other in Linux. We will then take a deeper look into Speculator, one of the tools the authors use extensively. Then, we will dive into what a "file" in Linux actually represents followed by what a journaling file system is. Finally, we will examine the various layers of the operating system that intermediate state modification - which can lead to consistency issues - and walk through the design of xsyncfs to see how it addresses these challenges while maintaining performance.

**Must Read**: **5. The Design and Implementation of a Log-Structured File System** where several basic file system concepts were introduced.

### 7.1. Inter Process Communication in Linux

Linux provides several mechanisms for processes to communicate with each other:

### *7.1.1. Pipes*

Pipes enable one-way data flow between processes, typically used to chain commands.

**Example:** `ls -l /etc | grep conf`

This command pipes the output of `ls -l /etc` (listing directory contents) to `grep conf` (filtering for lines containing "conf").

### *7.1.2. Unix Domain Sockets*

Unix Domain Sockets (UDS) facilitate local communication between processes on the same system similar to what network sockets offer across systems. They offer various communication modes:

- **SOCK_STREAM** (reliable, connection-oriented, similar to TCP [76])
- **SOCK_DGRAM** (connectionless, datagram-based, similar to UDP [77])
- **SOCK_SEQPACKETS** (connection oriented, sequenced packets, similar to SCTP [78])

### *7.1.3. Signals*

Signals are used to notify processes of events. Common signals:

- **SIGINT** (Ctrl+C): Interrupt signal.
- **SIGTERM**: Graceful termination request.
- **SIGKILL**: Forceful termination.
- **SIGSEGV**: Segmentation fault (invalid memory access).
- **SIGHUP**: Hangup signal (often used for configuration reloading).

### *7.1.4. Shared Memory*

Shared memory allows multiple processes to access a common memory region, enabling fast data exchange. Synchronization mechanisms (e.g., semaphores, mutexes) are essential to prevent data corruption.

## 7.2. Speculator

The speculative mechanism [79] originated in distributed file systems.

Suppose a read is performed by a client process:

- Upon a **Read**() call, a child process is forked. It retrieves data from the local cache, and continues execution from the point where the parent left off.
- The parent process waits for the server to complete the read operation.
- If the server's response differs from the cached data, the child process is terminated, and the parent continues.
- If the responses match, the parent process is terminated.

This optimization leverages the assumption that cached data is typically accurate, minimizing latency. In cases of discrepancies, the child process is terminated to prevent propagation of incorrect values.

Newer versions of the speculator have expanded its capabilities to include process checkpointing and rollback, analogous to branch prediction [80] in CPUs. This allows for speculative execution based on predicted outcomes. If the prediction is incorrect, the process can be rolled back to a previous checkpoint, facilitating recovery.

### *7.2.1. Speculation and Dependency Tracking*

Speculator maintains a record of processes and kernel objects that depend on the success or failure of a speculated execution. As the processes interact with kernel objects (e.g., UDS), the speculator tracks causal dependencies.

**Example: Inter-Process Communication (shown in Figure 37)**

Extending our previous client example in distributed file system:

- If the child process, after a speculative read, sends a message to another process via IPC, the recipient process is also tracked as speculated execution.
- If the speculative read fails, both processes are terminated or rolled back to a previous state.

The speculator ensures that speculated states are never visible to external observers (e.g., users), maintaining system consistency.

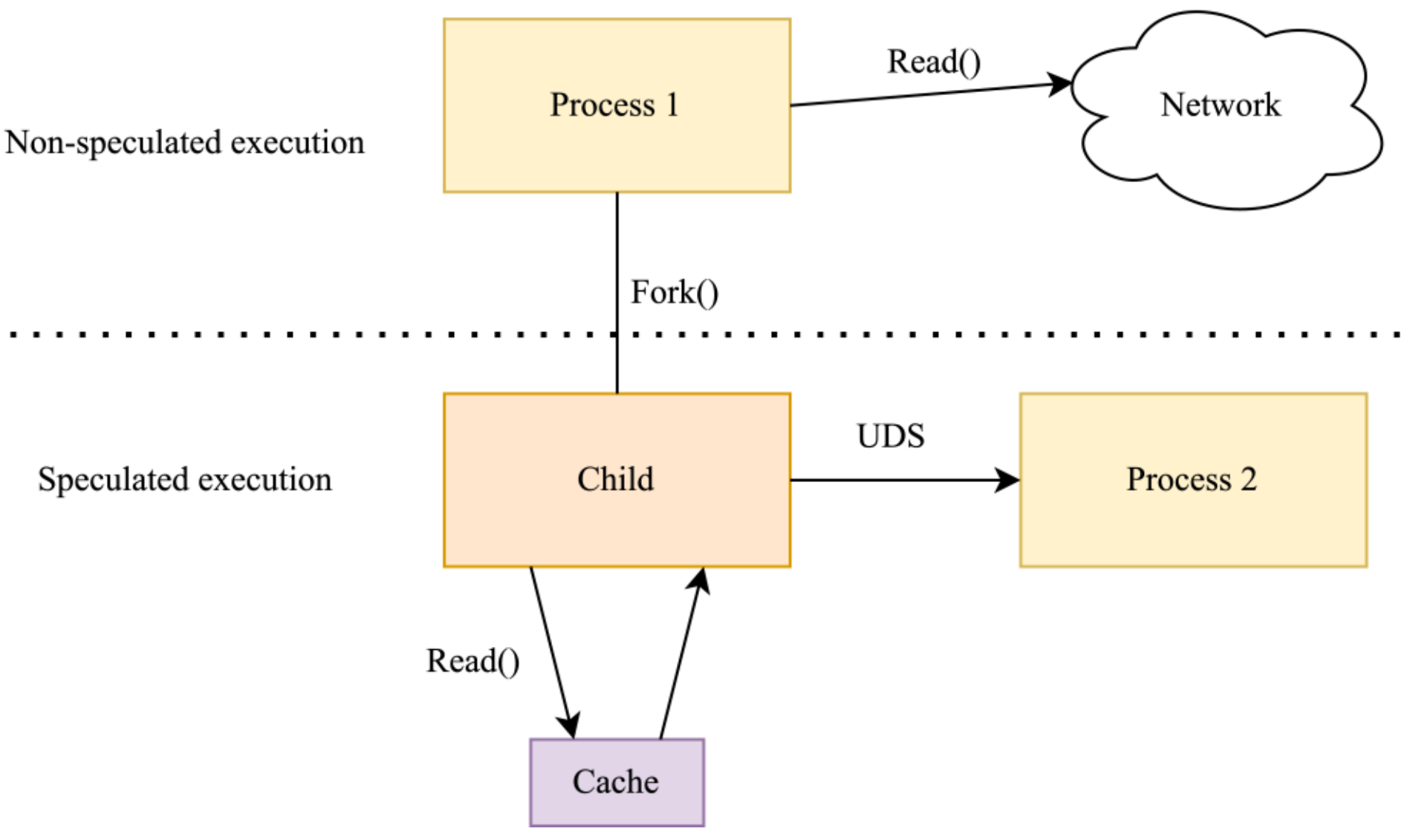

**Figure 37:** Example of dependency tracking using speculator.

### 7.3. Everything is a File

**Everything is a file** is a core concept in Unix-like operating systems, including Linux. It captures a fundamental design philosophy. A key aspect is the use of file descriptors. These are integers that represent open files or resources. Because many system resources are treated like files, file descriptors provide a unified way to access them.

All of the following are treated as a file in Linux and assigned a descriptor when opened:

#### *7.3.1. Regular Files*

- **Text files**: These contain plain text data. For example - **/etc/passwd**, **/home/user/alice/file.txt**
- **Binary files**: These contain executable code or other non-text data. For example - **/usr/bin/python3**, **/usr/share/image.png**

#### *7.3.2. Directories*

Directories themselves are files that contain metadata about the files and subdirectories they hold. You can use commands like **ls** to list their contents.

For example - **/home**, **/etc**, **/var**, **/tmp**

*7.3.3. Device Files*

- Hard drives or storage devices. For example - **/dev/sda**, **/dev/sdb**
- Terminals and pseudo-terminals. For example - **/dev/tty**, **/dev/pts/***
- Random number generators. For example - **/dev/random**, **/dev/urandom**
- **/dev/null** - a special file that discards all data written to it, and provides no data when read from.

*7.3.4. Unix Domain Sockets*

Represented as file system entries. These allow local inter-process communication. For example - **/var/run/my_daemon.sock**

*7.3.5. Procfs and Sysfs*

The **/proc** file system provides information about running processes and the kernel. The **/sys** file system provides information about hardware devices and kernel subsystems.

### 7.4. Journaling File System

The hierarchical organization of a file system can be represented as a tree data structure, where file and empty directory nodes constitute the leaf nodes.

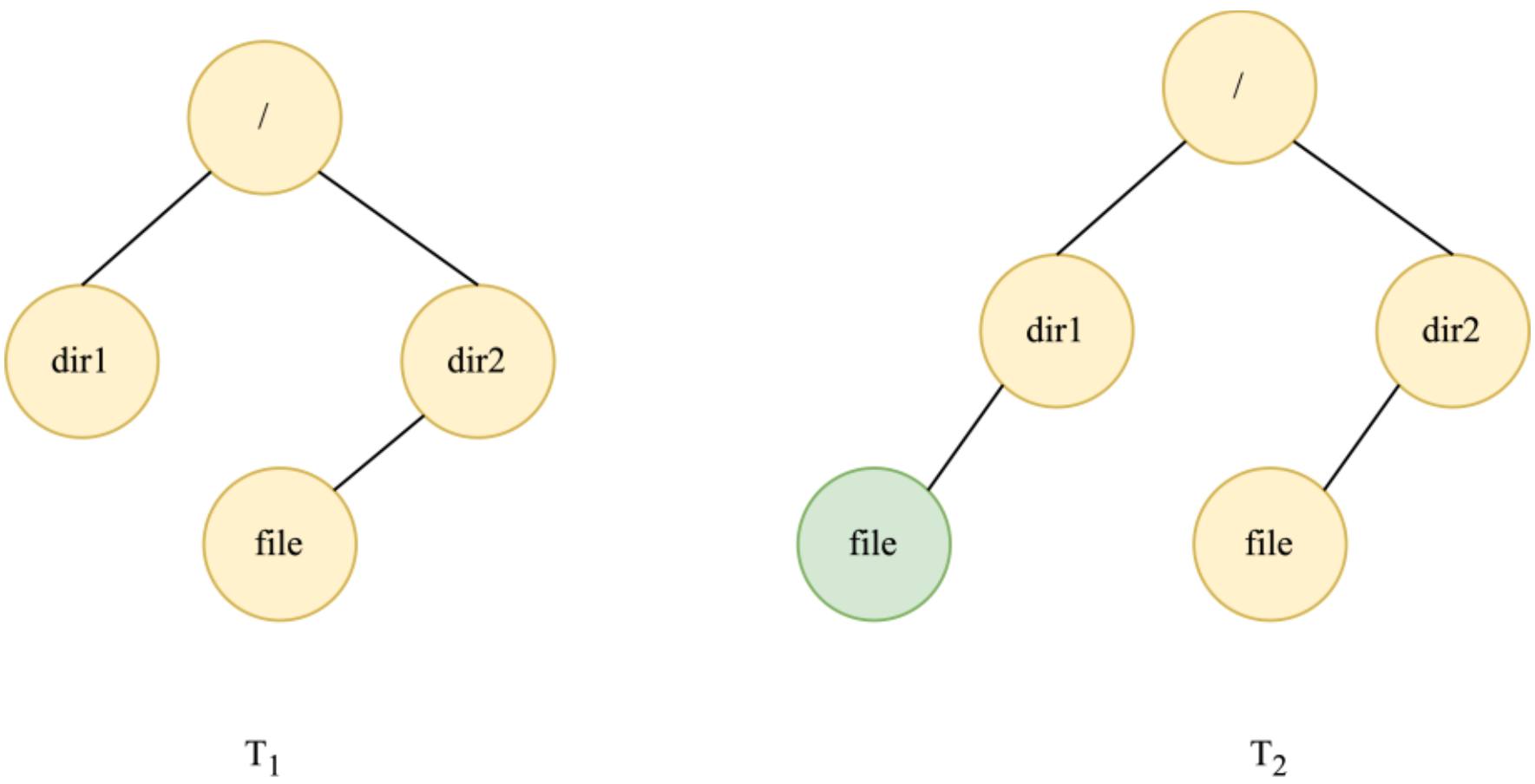

**Figure 38**: Temporal file system states in a journaling file system.

Given two temporal states, $\mathbf{T_1}$ and $\mathbf{T_2}$, of this tree (shown in Figure 38), the delta between them can be serialized into a journal, a log of structural modifications to the file system's data representation.

A journaling file system is built upon a base file system. Changes to the file system are initially logged in memory as a **journal**. When a checkpoint occurs or a sync operation is initiated, these logged changes are permanently written to the underlying disk. After successful disk commitment, the journal entries are removed.

Popular examples of journaling file systems are NTFS [69], ext4 [70], and XFS [73].

#### *7.4.1. Idempotency at Work*

Journaling file systems rely on the idempotency of file system operations. This allows recorded journal entries to be replayed repeatedly until successful, enabling recovery from crashes at any point. The journal, also known as a write-ahead log (WAL), functions similarly to a transaction log by recording changes before they are committed.

#### *7.4.2. Operation Modes*

Journaling file systems typically offer multiple modes:

1. **Ordered mode:** Only metadata (inode blocks) is journaled. Data blocks are written directly to disk.
2. **Journaled mode:** Both data and metadata are journaled.

Metadata integrity is crucial because it provides the pointers to data blocks. Data blocks must be written to disk before their corresponding metadata.

### 7.5. Commit Buffering

When we call **Flush**(), the bytes are not written immediately. There are several layers of buffering that prevent the commits.

#### *7.5.1. Commit Buffer Layers*

There are 2 primary buffer layers - the operating system layer (**page cache**) and the disk layer (**drive cache**), as shown in Figure 39.

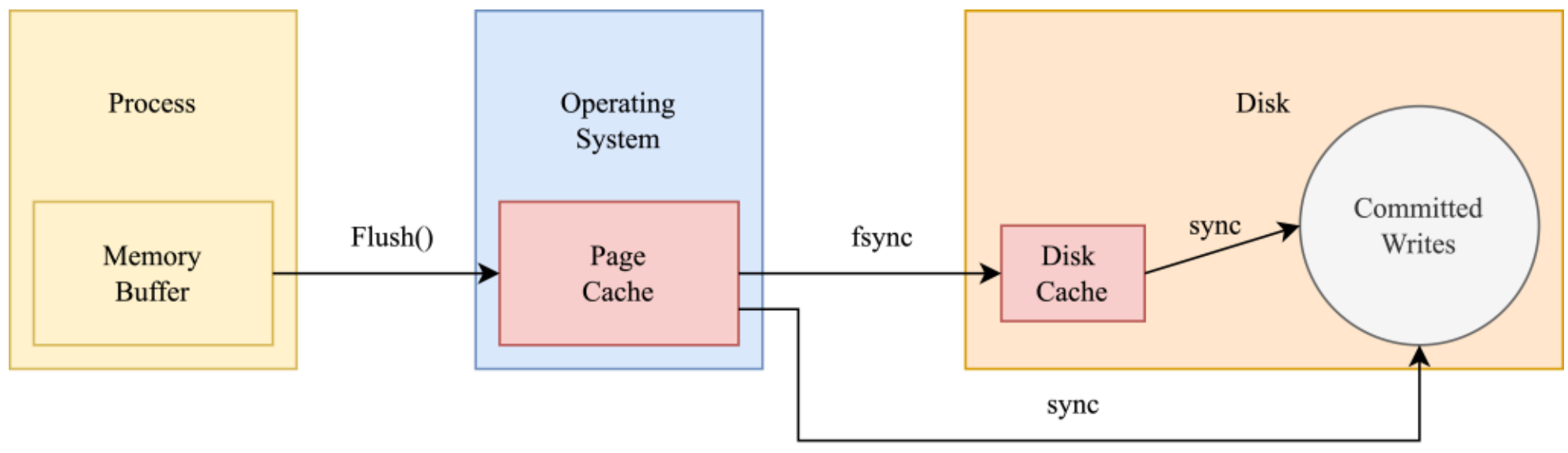

**Figure 39**: Commit buffer layers.

7.5.1.1. Page Cache

Operating systems implement virtual memory, which maps virtual addresses used by processes to physical addresses in RAM. When RAM is fully utilized, less frequently used pages are swapped out to secondary storage (e.g., disk). Conversely, when RAM has available space, pages from disk can be brought into memory. This mechanism is known as **page caching**, where disk pages are stored in RAM to accelerate read operations.

7.5.1.2. Drive Cache

Hard drives incorporate an internal cache to buffer write operations. This drive cache reduces rotational latency for both writes and reads. This cache forms an additional buffering layer.

Once writes are buffered in the drive cache, the actual commit to the disk may happen out-of-order. For example, if there were two sequential writes - **A** followed by **B**, then, it is possible that **B** is written first followed by **A**.

**Immediate Write Reporting**: If a drive uses immediate write reporting, it tells the operating system that the write is done as soon as the data is buffered in its cache. This can significantly improve perceived write performance. However, writes may be lost if there is a power failure. In fact, there are several hard drives which don't allow disabling immediate write reporting.

**Battery-Backed Drives**: To avoid write loss in the event of power failures, some hard drives have built-in battery packs to ensure that all writes from cache are committed. However, this battery is itself prone to failures.

### *7.5.2. Flushing Data - Flush()*

The **Flush**() system call, or the implicit flush upon file **Close**(), transfers data from a process's buffers to the operating system's page cache (in-memory buffers). These pages are marked as "dirty", indicating they are not yet committed to disk. While **Flush**() protects against process crashes, it does not guarantee data persistence in the event of an operating system or power failure.

### *7.5.3. Synchronizing File Data - fsync*

The **fsync** system call is invoked with a file descriptor. It instructs the operating system to commit the contents of in-memory buffers to the physical disk. The operating system, in turn, interacts with the file system, which may employ optimizations. Fortunately, most file systems reliably honor **fsync**, ensuring data is written to disk. **fsync** provides resilience against operating system failures but is still vulnerable to power outages.

### *7.5.4. Forcing Disk Commit - sync*

Even after fsync, the disk itself maintains a buffer. A power failure can result in data loss from this disk buffer. The sync system call forces the disk to flush its buffer, guaranteeing that all data is physically committed to the storage medium. This provides the highest level of data persistence, mitigating the risk of power failures.

## 7.6. Consistency in File System

Consider a disk state, **D**. Three sequential changes, **A**, **B**, and **C**, are made to the disk. In the event of a crash during these writes, the recovered disk state must be a consistent prefix of these changes: **D** + {**A**}, **D** + {**A**, **B**}, or **D** + {**A**, **B**, **C**}. An inconsistent state, such as **D** + {**B**, **C**}, is unacceptable. This prefix guarantee is the minimum consistency requirement for an operating system.

### *7.6.1. Failure Scenarios*

Two primary failure scenarios impact disk consistency:

- **Operating System Crash:** An operating system crash should not lead to file system inconsistency. While data loss is inevitable, the file system's structure must remain valid. This is because the operating system can maintain the sequential order of changes.

- **Power Failure:** Power failures pose a significant risk due to drive-level caching and delayed writes. Drives may report writes as complete before they are physically committed. The buffered writes are physically written out-of-order.

File systems such as Ext3, ReiserFS, and JFS can encounter consistency issues due to these drive-level complexities.

## 7.7. The External Sync

The concept of **external consistency** is important for file systems. Fundamentally, any computer system aims to maintain external consistency. This principle drives numerous operating system optimizations.

### *7.7.1. Optimizations Based on Observed Behavior*

Let's begin with some fun, hypothetical examples:

- Consider the scenario of writing a file. If the operating system can reliably determine that the file will never be accessed again, the commit operation could be omitted! However, if the user shares the file's name with someone else (e.g., via email) who may later want to read the file, the operating system would commit the write to storage.
- Let's imagine another scenario where the computer has an AI webcam which can tell if the user is looking at the screen or not. The operating system could delay displaying the output of **cat** command until the user's attention is directed towards the screen, optimizing resource usage without affecting the observed result!

The above are all hypothetical examples. But here are some common examples where writes are buffered by real-world operating systems:

- **External Drive Ejection (Windows):** Serial I/O through USB is often expensive and so the operating system buffers the writes to external drives in memory. Before removing external drives, they need to be **ejected**. The eject function forces sync operation, ensuring all buffered writes to the external drive are committed to disk before its removal. This prevents data loss.

- **System Shutdown:** Post shutdown, there would be no way for the operating system to recover all buffered writes. The shutdown process involves flushing and synchronizing uncommitted disk pages, maintaining data integrity. That's why shutdown takes time!

These optimizations are grounded in the principle of preserving externally observable consistency. The operating system does not guarantee that bytes will be committed when an external drive is pulled out suddenly or when the system loses power. All commitments are delayed as much as possible. Only when there is an externally observable expectation of an action, such as ejecting a drive or performing a clean shutdown, are the bytes committed.

We will see more examples of such optimizations.

#### *7.7.2. Paper Idea*

This paper aims to identify all instances where disk commitment is strictly required, enabling the operating system to maximize the delay of physical writes. This approach seeks to optimize performance while ensuring external consistency.

### 7.8. Design Overview

External synchronous I/O (**xsync**) refers to the observable behavior of an I/O operation. An I/O operation is considered xsync if its external output is indistinguishable from that of a strictly synchronous I/O operation.

To ensure correctness, xsync I/O must produce the same output as synchronous I/O, maintaining the same causal order of events. The concept of **causal order** is critical for preserving consistency. For example, if the user wrote **A** and then after successful flush wrote **B**. If a crash happens, after restart, if **B** is committed, then **A** must also be committed (**A** → **B** is the causal order).

#### *7.8.1. User-Centric vs. Application-Centric Consistency*

The paper distinguishes between **user-centric** and **application-centric** approaches to consistency. In the application-centric view, a process treats all other processes as external entities. In contrast, the user-centric view considers all process states as internal to the system, with the operating system responsible for maintaining external consistency.

Re-iterating our first hypothetical example, if a written file is never accessed again, the operating system may defer flushing in-memory buffers. From the application's perspective, the write appears committed. If the application itself reads the file again, it will receive the bytes from the in-memory buffers. However, from the user's perspective, the write is not truly committed until the operating system deems it necessary. The determination of "necessity" is a key aspect of xsync.

#### *7.8.2. Batching Commitments*

Xsync does not eliminate disk writes; it optimizes them by delaying commits to maximize performance. Modifications are grouped and journaled into a buffer, then atomically applied to the file system. Atomicity can be achieved through various mechanisms, such as checkpointing in LFS.

Batching offers significant advantages:

- **Amortized Costs:** It reduces the overhead of individual commits.
- **Elimination of Transient Writes:** Operations on temporary files that are created and deleted before commitment are avoided, minimizing unnecessary disk I/O.

This approach balances performance optimization with the requirement for externally observable consistency.

### 7.9. Commit Dependencies

A **commit dependency** represents a many-to-many causal relationship between a process's outputs and uncommitted file system modifications. A process with one or more commit dependencies is considered "uncommitted". Uncommitted output is not visible to external observers.

#### *7.9.1. Outputs and Output-Triggered Commits*

An **output** is defined as any event that triggers commits when released. And a commit triggered by a released output is termed an **output-triggered commit**. There are several outputs that can trigger commits.

**1. Screen Output**

```
echo hello > file.txt
cat file.txt
```

Executing **echo** followed by **cat** necessitates the file's contents being displayed on the screen. The operating system must commit the output to satisfy the user's expectation of consistent data visibility (causal relationship). This is a user-centric approach, where the user expects the same data to be available after a system restart, even after a catastrophic failure.

**2. Network Packets**

Network packets are not allowed to be sent until all associated dependencies are committed. This is because the receiving process cannot be determined beforehand (which can be within the same host).

**3. ioctl**

*90% of the operating system are device drivers.*

The **ioctl** system call provides a generic interface for controlling device-specific operations. Device drivers commonly implement **ioctl**, with each implementation tailored to the hardware. Consequently, **ioctl** behavior varies significantly; for instance, an **ioctl** on a network card might trigger network packet transmission. Due to the device-specific nature of **ioctl** and the inability to predict accessed data, all dependencies must be committed before ioctl operations.

**4. /proc Access**

The **/proc** file system, which provides process information through virtual files, requires all dependencies to be committed before its data is accessed.

**5. Reboot**

All dependencies need to be committed before shut down.

**6. sync**

All dependencies are committed upon **sync**.

**Note**: Apart from these output events, there are periodic commits from the operating system.

### *7.9.2. Challenges*

Since commits happen only before their corresponding output is released, it can lead to challenges:

- **Recovery Complexity:** Managing recovery from catastrophic failures becomes complex. For example, if an application commits data and attempts to send a network packet (an output), but the device fails to commit before the packet is sent, conveying the failure to the process after initial acknowledgment is problematic.
- **Cross-Filesystem Atomicity:** Achieving atomic commits across multiple file systems is inherently difficult.

### *7.9.3. Tracking Dependencies*

The authors use a Speculator to track commit dependencies:

- When an output needs to be released, the speculator first commits all corresponding buffered output.
- Multiple outputs and commit dependencies can originate from a single process.
- A process can be marked "committed" after all its commit dependencies are resolved.

### *7.9.4. Inheriting Dependencies Across Processes*

Different processes can inherit each other's commit dependencies as follows:

- **Shared Memory**: When two processes or threads access shared memory, a bidirectional dependency is established. If one process has write permission and another has read permission, the reader inherits the writer's dependencies (unidirectional). While more granular dependency tracking was possible, the authors prioritized simplicity.
- **Pipes:** When process P1 pipes its output to process P2, P2 inherits P1's dependencies (**unidirectional**).

- **Unix Sockets:** The process that receives the message inherits dependencies from the sender.
- **Signals:** Dependency inheritance via signals is **unidirectional**, similar to shared memory.
- **Fork:** The child process inherits all dependencies from the parent process, as they share the same virtual memory map.

*7.9.5. Examples*

**Example 1**

Consider two processes sharing memory. If one process writes to and closes a file, immediate disk commitment is not strictly necessary. As long as the other process accesses the file through shared memory or the operating system's in-memory buffers, external consistency is maintained.

However, if a process commits a file and then transmits a network packet, immediate disk commitment becomes imperative. The network packet could initiate file access by a remote process (which could be within the system itself), requiring commit.

**Example 2 (shown in Figure 40)**

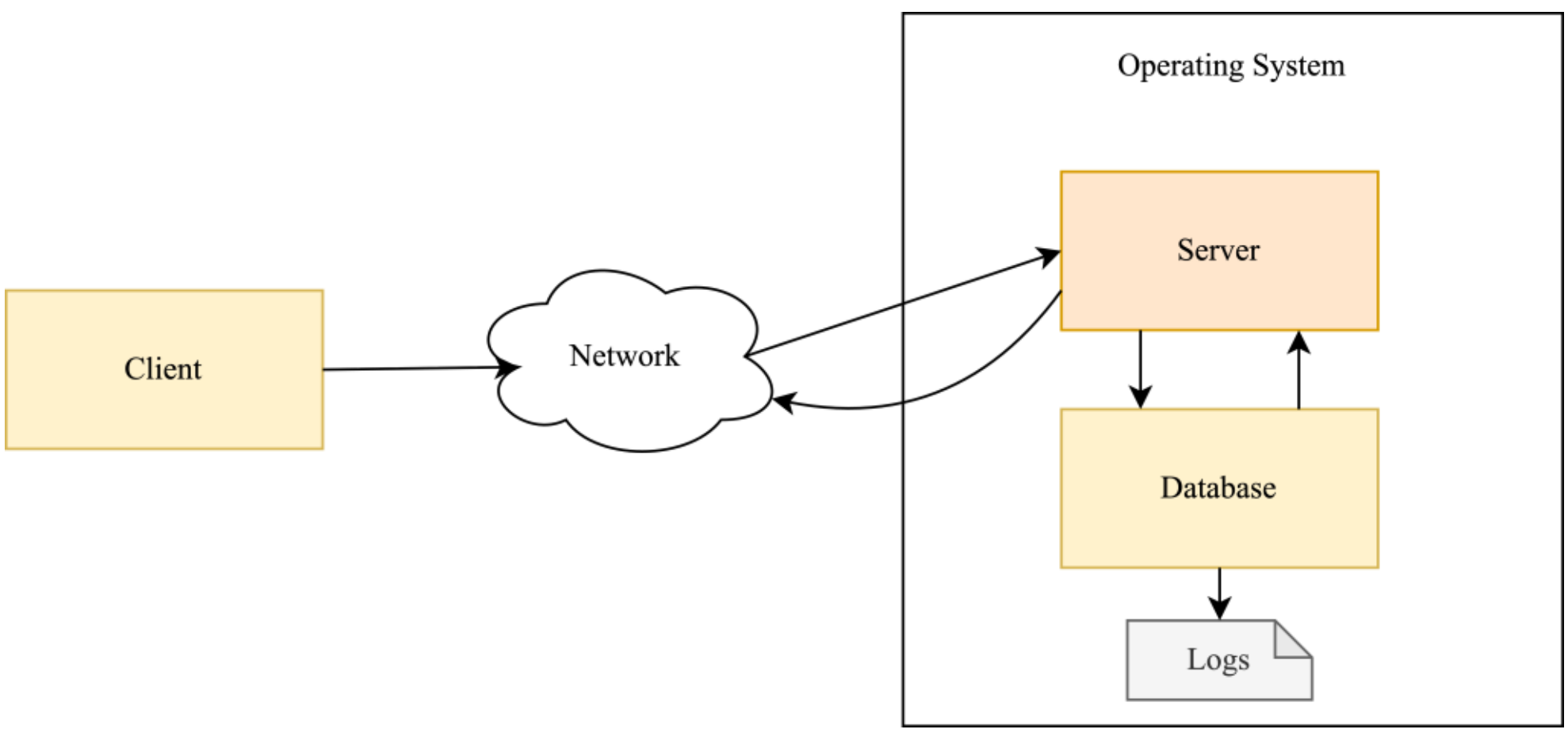

**Figure 40**: Multi-transaction commit.

Suppose a client sends a request to a web server running on a machine with its own database (running locally). As a result of the request, the server is supposed to commit multiple transactions. All these transactions are committed by the database but not immediately flushed to storage by the operating system. They are flushed to storage only when the server sends back the reply packets to the client (the output). As a result, multiple transaction logs can be batched together for commitment.

## 7.10. Xsync in Journaling File System

As previously stated, journaling file systems maintain a journal that records modifications.

Xsync also maintains an undo log for each running process, as shown in Figure 2 in the paper. Each entry in the undo log corresponds to an output. The outputs themselves have many-to-many commit dependency relationships with uncommitted file system modifications.

For instance, when a file is created, the modification is recorded in the journal, and a corresponding undo log entry is added to delete that file.

When a process generates output (e.g., screen output or a network packet), all commit dependencies associated with that output are flushed. Subsequently, corresponding entries from the process's undo log are removed.

Journal entries are grouped into two types: **committing** and **active**. Committing entries form a prefix of the journal, indicating that they have been or will be durably written, while active entries represent pending modifications.

### *7.10.1. Durability Guarantees*

Consider a scenario where **D** + {**A**, **B**, **C**} represent journal entries in that order. In journaled mode, if entry **C** is committed, then entries **A** and **B** are guaranteed to be committed as well. However, this guarantee does not hold true in ordered mode. In ordered mode, data blocks are written directly to the disk (instead of journal), which can result in out-of-order commitments to the disks.

## 7.11. Rethinking Sync

The **sync** command serves as a crucial commit point, ensuring all buffered writes are flushed, encompassing both operating system and drive caches. Consequently, **sync** is a resource-intensive operation. In xsyncfs, extensive dependency inheritance

- resulting from numerous buffered writes - can significantly prolong sync completion.

The architecture of xsyncfs involves the following layers, as shown in Figure 41:

- **Virtual File System:** The operating system's virtual file system layer initiates calls to the speculator.
- **Speculator:** The speculator acts as an intermediary, managing commit dependencies. Upon receiving a **sync** command, the speculator commits all pending dependencies. Otherwise, buffered writes remain in their current state.
- **xsyncfs:** xsyncfs operates transparently to application developers, functioning as a layer between the operating system and the underlying file system. This design allows xsyncfs to be compatible with various file systems.

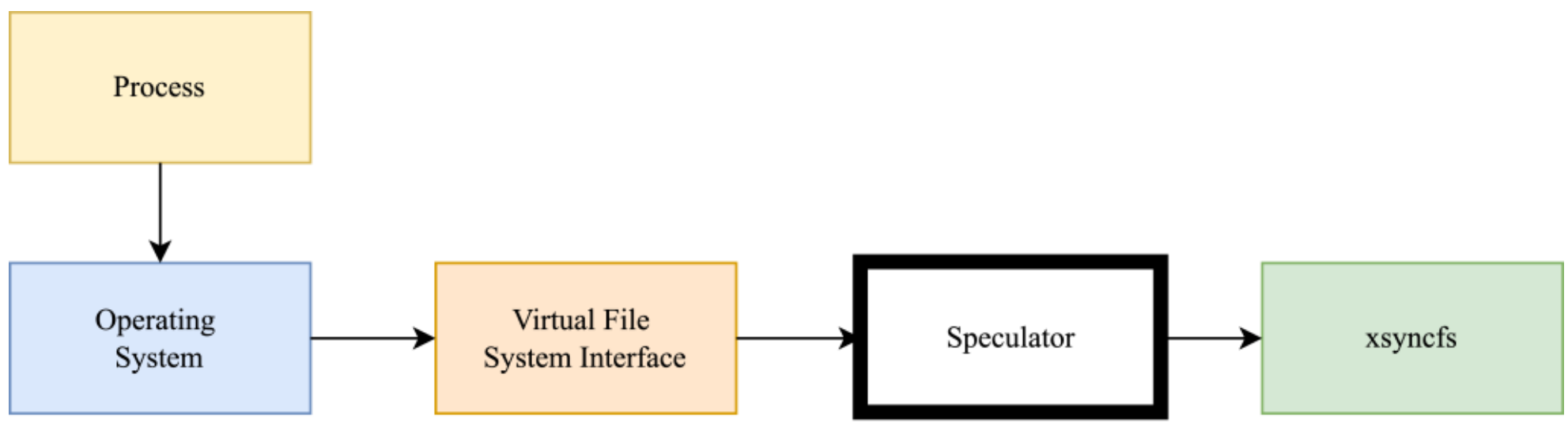

**Figure 41**: Xsyncfs architecture layers.

## 7.12. Evaluation

### *7.12.1. Systems Under Comparisons*

Figure 3 in the paper illustrates different levels of data durability:

- **Asynchronous:** Offers no durability*. Data may be lost even after an **fsync** call due to drive caching.
- **Synchronous:** Also lacks durability*, same as asynchronous behavior.
- **Synchronous with Write Barriers:** Provides durability*. Write barriers enforce a strong synchronization, ensuring data is committed.
- **External Synchrony:** Achieves full durability*.

* - The key to understanding these distinctions lies in the definition of durability: if a remote computer logs a write operation, but the test computer lacks the written data, durability is considered failed. Therefore, "durable" here signifies that data is committed to storage before the remote confirmation packet is sent, which is indeed the case in external synchrony.

#### *7.12.2. Postmark Benchmark (I/O-Bound)*

The Postmark benchmark, an I/O-intensive workload, involves creating 10,000 files, executing 10,000 transactions (reads, writes, creates, deletes), and subsequently removing all files.

- As depicted in paper's Figure 4 (logarithmic scale), xsyncfs exhibits a 100x performance improvement over sync with barriers, attributed to the substantial overhead of write barriers.
- xsyncfs also outperforms sync by 10x, as drive cache limitations (e.g., 2MB) can lead to rapid blocking.
- While xsyncfs is slightly slower than asynchronous I/O, it provides the critical guarantee that all externally observable outputs are committed.

#### *7.12.3. Apache Benchmark (CPU-Bound)*

- Figure 5 in the paper illustrates that xsyncfs performance approaches that of asynchronous I/O.
- Furthermore, its performance is comparable to RAMFS (in-memory filesystem), indicating minimal I/O blocking for CPU-bound processes.

#### *7.12.4. MySQL Benchmark*

MySQL employs its own group commit strategy, necessitating the disabling of the drive cache for data durability.

- xsyncfs demonstrates superior performance compared to ext3 with barriers.
- The performance gap between xsyncfs and other approaches narrows as transaction parallelism increases.

#### *7.12.5. Specweb99 Benchmark (Network-Intensive)*

Specweb99, a network-intensive benchmark, generates numerous synchronization operations.

- xsyncfs maintains performance with only an 8% overhead compared to asynchronous I/O.
- xsyncfs outperforms the barrier based sync.

### 7.13. Paper Remarks

This paper is a standout. A concept that initially seems impossible is successfully implemented, forcing a total reconsideration of how system state is handled. The central theme - that operations lacking observable consequences should be avoided or at least delayed - introduces a vital user-centric lens to systems engineering. It is a challenging read that requires focus to fully grasp, but the insights into I/O and consistency are well worth the effort. It is highly recommended for anyone focused on high-performance system design.

# 8. The Google File System

(8) Ghemawat, S.; Gobioff, H.; Leung, S.-T. The Google File System. In Proceedings of the nineteenth ACM symposium on Operating systems principles; 2003; pp 29–43.

Having explored a range of file systems in previous chapters, we now turn our attention to their distributed counterparts. Specifically, we will examine the Google File System (GFS).

This paper was written by Sanjay Ghemawat et. al. Sanjay Ghemawat is a Google Fellow. His contributions extend far beyond GFS, encompassing core libraries and systems such as MapReduce. It is difficult to envision the current state of Google's infrastructure without his significant contributions.

The paper, presented in Symposium on Operating Systems Principles (SOSP) '03, marked the introduction of a system at Google that has since evolved into a critical infrastructure component known as Colossus [81]. This system supports a vast array of Google's operations and is of immense importance to the company's overall functionality. The architecture of Colossus has undergone significant transformations since its initial conception, developing into a highly sophisticated and robust platform.

Despite its current complexity, the design of its predecessor, GFS, prioritized **simplicity** in its large-scale distributed architecture. This focus on simplification was deliberate, aimed at achieving performance and efficient operations. By making specific simplifying assumptions, the system was able to accelerate its processes and handle the demands of Google's growing data needs.

In this insight, we will start by understanding what a network file system is and how a distributed file system differs from it. Following that, we will take a detailed look at GFS.

**Extended Read**: Colossus under the hood: a peek into Google's scalable storage system [81]

**Recommended Read**: **5. The Design and Implementation of a Log-Structured File System** where several basic file systems concepts were introduced.

### 8.1. Network File System

Sun Microsystems pioneered the concept of network file systems with their influential paper - “Design and Implementation of Sun Network Filesystem” [82].

At its core, Network File System (NFS) enables access to files across a network. A server maintains the physical storage, and clients interact with it via Remote Procedure Calls (RPC).

From the client's perspective, network file system operations are designed to be seamless. In Linux environments, for instance, NFS implements the virtual file system [71] interface (shown in Figure 42), allowing them to be mounted without the user necessarily being aware of their network-based nature.

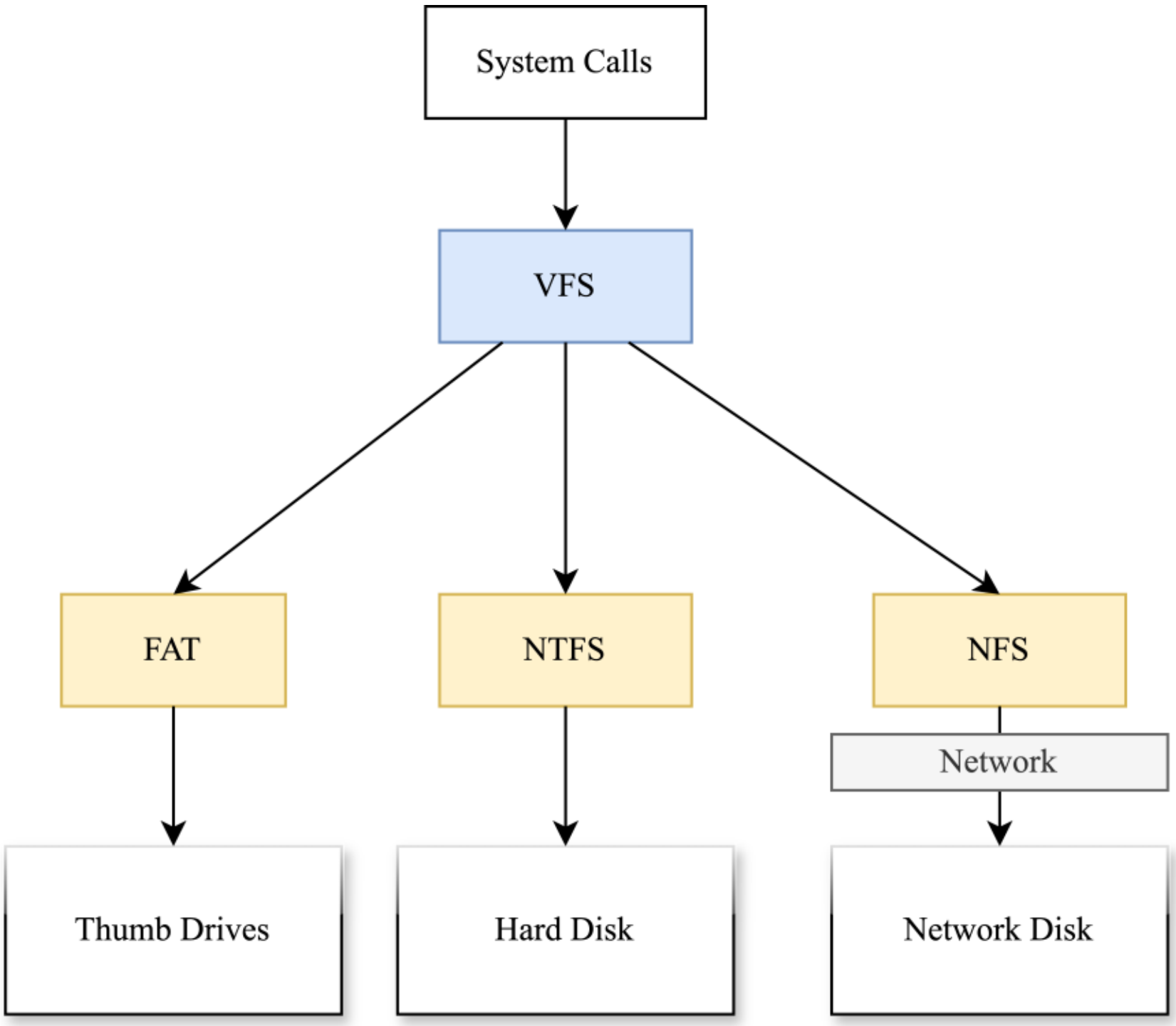

**Figure 42**: Virtual file system interface.

NFS was widely adopted for file sharing in the past. However, with the emergence of distributed file systems like GFS and Hadoop Distributed File System [83], their prevalence has diminished for large scale workloads.

### *8.1.1. Network File System v/s Distributed File System*

NFS centralizes file storage on a single server. Client requests are routed over the network to this server. This creates a shared environment. However, it is not a distributed one offering fault tolerance. Consequently, if the server fails, the entire file system becomes inaccessible.

In contrast, a distributed file system distributes the file system itself across multiple servers. Data is replicated and spread across these servers, providing inherent fault tolerance. If one server fails, the system continues to operate using the replicated data on other servers.

## 8.2. The Google File System

**Disclaimer**: In this chapter, the discussion centers on the original GFS and not its successor, Colossus. The original GFS paper refers to the metadata server as the **master**. In Colossus, this component has been redesigned and is now known as the **curator**, leveraging Bigtable for persistent storage.

GFS is built on several key principles: **immutability** and **simplicity**.

### *8.2.1. Immutability*

Echoing the same concept found in log-structured file systems, GFS operates on an **append-only model** - data, once written, whether successfully or not, cannot be modified. Files accept only **append-writes**. Additionally, a file can only accept append-writes if it is not **frozen** (immutable).

**Important Note**: While the paper mentions file overwrites, GFS has evolved to strictly enforce appending to the files only. In this chapter, we will also focus only on appends and so all mutations correspond to append-writes only.

### *8.2.2. Simplicity*

GFS stands out as a remarkably simple large-scale distributed system, making it an ideal introductory topic in any distributed system course.

This simplicity stems from deliberate design choices, such as:

- **Consistency over Availability -** GFS enforces a single master instance, sacrificing availability for consistency. Even then, GFS only offers **eventual consistency** which greatly simplifies design.
- **No Atomicity -** GFS employs a straightforward replication strategy during append-writes - data is synchronously replicated across *all* replicas. If any (but not all) replica fails to commit the append, it results in an "inconsistent" (not fully committed) state of the corresponding byte range. This inconsistent state is permanent and needs to be handled by applications.
- **No Isolation** - GFS readers may be able to read uncommitted data while some append-write is still in progress. GFS does not provide isolation guarantees, and therefore cannot be externally consistent.

These choices contrast with systems like ZooKeeper and Spanner, which do provide high availability, atomicity, and external consistency. However, the choices allowed for a much simpler implementation.

**Note**: Colossus has evolved significantly. Unlike the original GFS, modern Colossus achieves high availability and supports several replication schemes.

### 8.3. File Structure

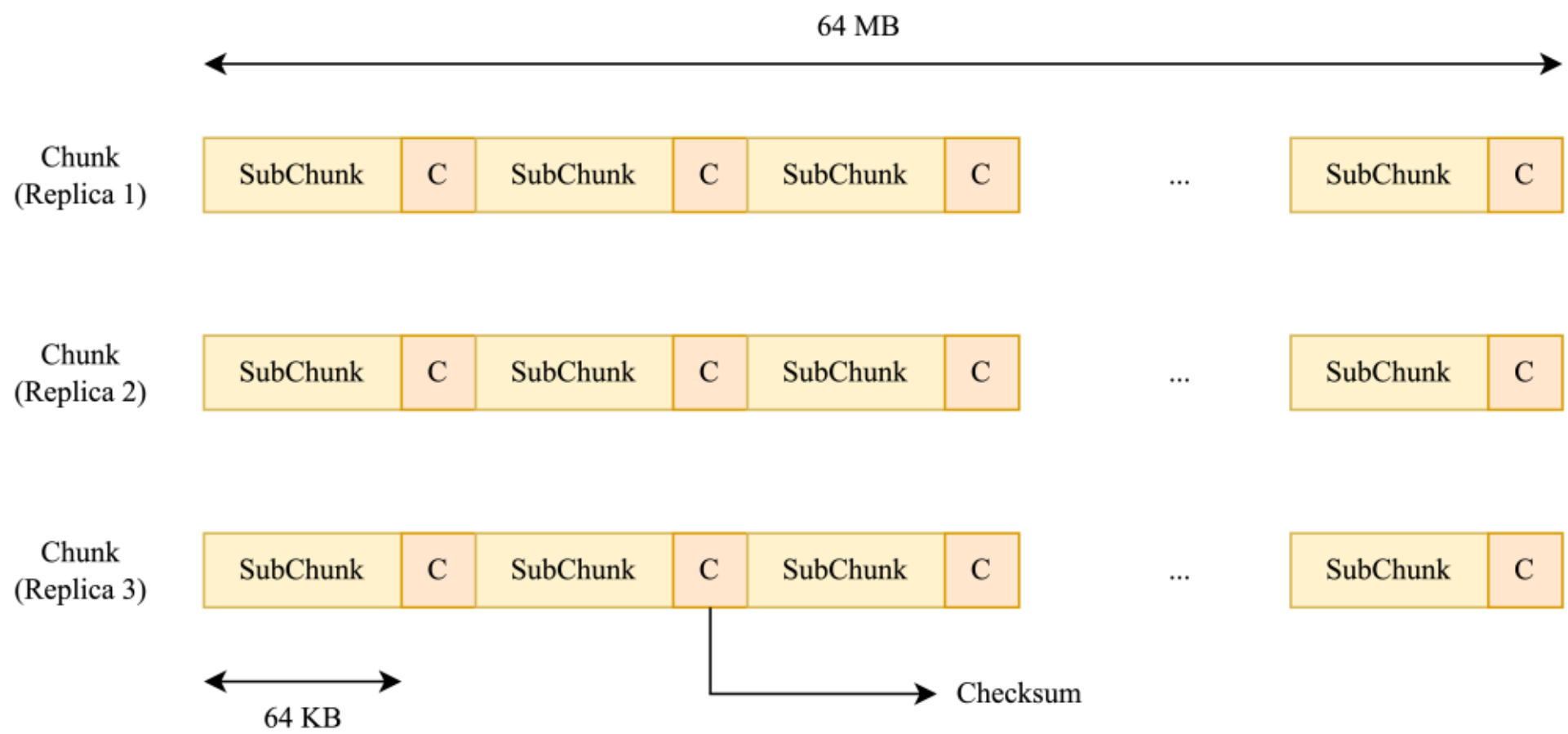

**Figure 43**: File structure in GFS.

GFS organizes files into chunks and sub-chunks, as shown in Figure 43:

- **Chunks:** Files are segmented into 64 MB chunks. Each chunk is replicated across multiple chunk servers (described later) for fault tolerance. Chunk servers store these chunks within their local Linux file systems, potentially as individual files on disk.
- **Sub-Chunks:** Each 64 MB chunk is further divided into 64 KB sub-chunks. Checksums are calculated and stored at the sub-chunk level. These checksums are crucial for detecting data corruption.

#### *8.3.1. Choice of Chunk Size*

Chunk size, a configurable parameter, is set to 64 MB based on workload analysis. This decision reflects a trade-off:

- **Metadata Efficiency:** Larger chunks reduce metadata operations during reads and mutations, and minimize master metadata storage requirements.
- **Hotspot Potential:** Conversely, larger chunks concentrate data on fewer servers, increasing the risk of hotspots, particularly during tail reads.

#### *8.3.2. Choice of Replication Factor*

The replication factor significantly impacts performance. A low replication factor (e.g., 3) may limit read scalability under heavy workloads. Conversely, a high replication factor (5 or more) can slow down mutations and significantly impact tail latency.

### 8.4. Architecture

GFS employs a multi-tiered architecture.

#### *8.4.1. Global Setup*

GFS deployments consist of independent clusters distributed globally. These clusters operate autonomously, with no inter-cluster communication. Therefore, the core distributed aspects of GFS are primarily confined to the cluster level. The remainder of this chapter will focus on the internal workings of a single GFS cluster.

#### *8.4.2. Cluster Setup*

Like any file system, GFS manages both file data (chunks) and metadata. In traditional file systems, metadata is often stored in inodes. GFS adopts a similar concept but distributes it across two key components:

1. **Master:** This server manages the file metadata, including file attributes and chunk location information.
2. **Chunk Servers:** These servers store the raw file data, divided into fixed-size chunks.

#### *8.4.3. The Master*

The master manages the file metadata, ensuring the integrity and availability of the file system.

##### 8.4.3.1. Namespace Management

The master stores the entire directory structure. Internally, GFS represents the namespace as a **flat key-value store** and not as a traditional tree. For example, if there is a file **/a/b/file.txt**, then the keys would be **/a**, **/a/b**, and **/a/b/file.txt**. Values associated with these keys include:

- File/directory metadata (attributes).
- Read/write locks. Acquiring locks on a file requires read locking all parent directories.

For files, the value also contains information about its chunks. For each chunk, it has:

- The chunk sequence number.
- The latest version number (described later) for the chunk.
- Address of the chunk servers holding the latest version number of the chunk.

**Note**: In Colossus, Bigtable is used as the backing store for the key-value store (which itself uses Colossus).

##### 8.4.3.2. Replication and Logging

Master state is replicated for reliability using a distributed log, periodically flushed to disk. Only one **active master** handles mutations and all background tasks, and hence is also responsible for logging all the metadata modifications to the log. **Shadow masters** serve read requests based on the logs that the active master generates.

Distributed logs provide a mechanism for global serialization as well, say, when there are multiple modifications on the same file

**Example**: Say there are 2 requests, one to create a new file **/a/b/file.txt** and another to rename the file to **/a/b/file2.txt**, the distributed log would look something like:

```
create /a/b/file.txt
rename /a/b/file.txt → /a/b/file2.txt
```

While executing the first instruction, it will hold read locks on /a and /a/b and write lock on /a/b/file.txt. After the operation has been executed, it will release the locks. Locks help in concurrent execution of the next instruction while the previous one is in progress. In this example, the locks for the second mutation will be held only after the first mutation completes.

**Note**: Chunk information is not materialized to the logs. They are dynamically rebuilt upon master startup by polling the chunk servers and using heart beat messages (described later).

#### 8.4.3.3. Checkpointing

In addition to the distributed log, the master's state is periodically snapshotted (serialized B-tree data structure) and stored on disk. During recovery, snapshots are loaded, and subsequent log entries are applied.

**Recommended Read**: **21. ZooKeeper: Wait-free Coordination for Internet-Scale Systems** where we discuss how applying updates on snapshots work.

#### 8.4.3.4. Failure Recovery

Upon primary master failure, an external monitoring system initiates a new primary master. The new primary recovers by reading the logs, catching up to the latest state, and then polling chunk servers to rebuild its chunk location information.

GFS guarantees that there would only be one and only one master at a time, sacrificing availability during change.

**Note**: Colossus replaces the GFS master with Curator, a sharded system offering higher availability.

#### *8.4.4. The Chunk Servers*

Each chunk has a designated **primary** chunk server. All other replicas for the chunk are **secondaries**. The primary orders all mutations to that chunk, ensuring consistency across secondary replicas.

The master maintains a dynamic view of chunk server availability through periodic **heartbeat messages**. Additionally, upon startup, the master server reconstructs chunk location information by polling all chunk servers.

##### 8.4.4.1. Chunk Placement Policy

Data centers often organize machines into racks, each with a shared network switch (Top-of-the-Rack switch). Failure of a rack switch can render all chunk servers within that rack unavailable.

To mitigate rack-level failures, GFS strategically places replicas of the same chunk on chunk servers residing in different racks. This ensures higher fault tolerance against correlated failures, such as network switch failures.

##### 8.4.4.2. Chunk Version Number

For each chunk, the chunk servers track the **chunk version number**. Although not directly mentioned in the paper, it is implicit that the version number of a chunk on the chunk server is incremented whenever the chunk is modified.

The heartbeat messages also contain the version number of all the chunks on a chunk server. This version number is used by the master to distinguish between **stale** and **up-to-date** replicas for each chunk.

### 8.5. Consistency Model

This and the next section are the core of the GFS design discussion and so it is very important to understand them. The rest of the system's design builds upon these foundations.

#### *8.5.1. Metadata Consistency*

Metadata operations are atomic. This atomicity is ensured by the single master server, which handles all metadata operations sequentially.

### *8.5.2. Chunk Consistency*

A range of bytes in a chunk can be:

- **Consistent:** All replicas of the chunk contain the same data for the byte range.
- **Inconsistent:** Different replicas of the chunk contain different data for the byte range.

If a mutation to a chunk succeeds, the byte range achieves a consistent state across all replicas.

However, a failed mutation, for instance, if the data was appended to a subset of replicas (at least one), can result in an inconsistent byte range within the chunk. Such mutations are not fully committed from the client's perspective. Note that once an append-write to a byte range fails on any (but not all) replica, that byte range is permanently inconsistent. The client's next retry will append to the subsequent byte range.

Readers may perform relaxed reads from a single replica, where uncommitted data might be visible. GFS does not guarantee external consistency, acknowledging the possibility of such inconsistencies. Clients are expected to handle such inconsistencies.

GFS provides a specific, albeit limited, guarantee regarding mutations: **at-least-once atomic-append**. This guarantee means that if a client repeatedly attempts to append data to a file, at least one of those append operations will eventually succeed and be applied atomically.

## 8.6. Leases and Mutation Order

As previously discussed, each chunk replica has a designated primary server. The primary is granted a **lease** by the GFS master. This primary server is the central coordinator for all mutations to the chunk, responsible for establishing a total order of these operations.

The lease has a default timeout of 60 seconds. To maintain its lease, the primary can request extensions, which can be piggybacked into the regular heartbeat messages exchanged with the master.

To enforce a total order, the primary assigns a serial number to each mutation. This serial number is a unique sequence number generated by the primary. This serial number provides a globally consistent ordering mechanism across all mutations applied to the chunk, guaranteeing that all replicas converge on the same final state.

### 8.7. Fault Tolerance for Chunk Servers

The availability of chunk servers presents a different set of challenges. Chunk servers can experience various failures, including -

- Crashes, leading to redundancy loss for all the chunks.
- Data corruption for specific chunks, for example, when disks fail.

These incidents render affected chunks temporarily unavailable for mutations, causing operations to fail continuously until recovery. The master server responds to these scenarios with specific actions for each affected chunk.

#### *8.7.1. Primary Chunk Server Failure*

When a primary chunk server fails, the master must select a new primary. A new primary is elected as follows:

- Upon primary failure, the master queries remaining replicas for their chunk version numbers.
- Among the replicas with the latest version (up-to-date replicas), the master selects a new primary.
- The chunk version number is then incremented, and this updated version number is passed to all up-to-date replicas.

The chunk becomes available for append-writes only after all the steps above are complete.

#### *8.7.2. Secondary Chunk Server Failure*

When a secondary chunk server fails, the master assigns the affected chunks to a new chunk server. This triggers a re-replication process (described next), where the new server fetches the current chunk state from other replicas.

After the chunk state is fetched by the new chunk server, the chunk becomes available again for mutations.

### *8.7.3. Re-replication and Rebalancing*

Re-replication is the process of restoring lost replicas and is crucial for maintaining data redundancy. When a new chunk server is assigned responsibility for previously unowned chunks, it retrieves the latest chunk data from existing replicas.

Re-replication is subject to resource constraints:

- **Bandwidth limitations**: There's a limit on the bandwidth a chunk server can dedicate to re-replication.
- **Prioritization**: Re-replication prioritizes chunks based on importance:
    - Chunks actively requested by clients take precedence to restore mutation availability quickly.
    - Chunks with higher redundancy loss are prioritized over others.

Similar to replica restoration, chunks are moved across chunk servers to balance load and optimize disk space.

### *8.7.4. Stale Replica Removal*

Stale replicas arise when a chunk server fails to receive mutations, for instance, when having been offline for an extended period. This scenario occurs when the master reassigns the chunk to another chunk server, and subsequent mutations occur while the original server is still offline.

Consequently, the returning server's replica will lack the current chunk version number. The master leverages this version number to identify and remove these outdated replicas.

## 8.8. Reading a File

Reading a file in GFS is straightforward:

- The client first obtains chunk information, including the latest version number, from the master and caches it.
- Then, using the replica locations, the client selects the closest chunk server for data retrieval. Closeness is determined by IP address proximity, leveraging cluster's network topology to minimize latency.

*8.8.1. Read Failure Scenarios*

During read operations, two potential issues can arise:

- **Stale Replicas**: A replica might be stale, meaning it lacks the most recent mutations.
- **Inconsistent Byte Range**: Alternatively, it might contain a mutation that is either partially committed or failed completely resulting in an inconsistent byte range.

Stale replicas are relatively easy to handle. When the client refreshes metadata from the master, it will obtain the locations of up-to-date replicas and retrieve the data.

However, handling uncommitted mutations is more complex. Readers must employ application-level mechanisms like checksums to validate byte ranges and detect duplicates (which can occur if a writer retries a failed append-write). Data record formats, such as Record I/O, can automate these checks and handle these situations without requiring manual intervention from the user.

### 8.9. Writing a File

Writing to a file in GFS is a multi-stage operation, as shown in Figure 2 in the paper. An append-write can target a single chunk or span multiple chunks, with the process repeating for each affected chunk.

*8.9.1. Chunk Replica Identification (Steps 1 - 2)*

The client first requests the primary and secondary replica locations for the target chunk from the master server. The master, upon receiving this request, checks for an existing primary. If none exists (for instance, when it is a new chunk), it grants a lease to one of the replicas before responding to the client.

*8.9.2. Data Upload (Step 3)*

The client then uploads the data to all chunk server replicas of the chunk. This step ensures all replicas have the data before the commit phase begins.

Notably, the client uploads the data only once. The data is replicated in a linear chain across the replicas to optimize bandwidth utilization. Each chunk server in the chain forwards the data to the next closest server based on network topology, likely determined by IP address prefixes, allowing for hop count inference.

#### 8.9.2.1. Time to Upload

Let's analyze the time required to upload data in GFS, considering factors like bandwidth, latency, and replication.

Assume **B** bytes are uploaded, **T** is the available bandwidth in the cluster fabric, **L** is the latency between any two chunk servers, and **R** is the replication factor.

Time to establish connections along the replication chain = $\mathbf{R} * \mathbf{L}$.
Time to transfer the **B** bytes is $\mathbf{B/T}$.
Therefore, the total time is approximated by $\mathbf{B/T + R * L}$.

It's important to note that this calculation is a simplification. It doesn't account for network congestion, which can significantly impact actual bandwidth utilization. In real-world scenarios, the effective bandwidth may be lower than the theoretical maximum.

For example, with a 100 Mbps bandwidth and a 1ms latency, uploading 1 MB of data would ideally take 80ms. However, network congestion and other factors could increase this time.

#### 8.9.2.2. LRU Buffer Cache

The received data is temporarily stored in an LRU buffer cache on each replica, not yet committed to the chunk. These cached entries are eventually evicted after a timeout if the mutation is not finalized.

### *8.9.3. Commit (Step 4 - 7)*

The client sends a commit request to the primary replica. The primary assigns a serial number to the mutation and commits the bytes. It is important for primary to commit the bytes first because the primary must maintain the latest chunk version.

Then the primary forwards the request along with the serial number to the secondary replicas. Each secondary replica independently applies the mutation to its copy of the chunk and sends a confirmation back to the primary. Once the primary receives acknowledgments from all secondaries, it sends a final acknowledgment back to the client.

### *8.9.4. Write Failure Scenarios*

A failed append-write may result in an inconsistent state for the byte range to which the append-write was supposed to be committed.

**Q. What if some secondary replica fails?**

If a secondary replica fails during the data upload, the entire mutation can be retried by the client. All good, no inconsistency till this point.

If it fails after receiving a mutation request from the primary, then it results in an inconsistent state for the byte range. Any readers of the byte range may be able to see the appended bytes (if they were in contact with another secondary which may have appended the bytes successfully) or see garbage. Eventually the failure would be detected by the master and the chunk will be assigned to a new chunk server. Re-replication will be performed. Only then will the mutations start succeeding again.

**Q. What if the primary replica fails?**

If the primary fails at any point before forwarding the request to secondaries, the entire process can be simply retried by the client. All good, no inconsistency till this point.

If the primary fails after sending the request to secondaries, there will be an inconsistent byte range. The next successful retry (after primary re-assignment) will append bytes to a new range.

**Q. What if the client is interacting with an old primary while a new primary has been assigned?**

A clock drift between the primary and the master can create a scenario where the primary believes its lease is still valid while it has actually expired.

If a mutation is received by an old primary, it will not be able to commit it on any of the secondary replicas as the chunk version number would have already been incremented. Eventually, it will be detected as stale and removed.

## 8.10. Miscellaneous Details

### *8.10.1. Client Library*

While Linux local file systems adhere to the virtual file system interface, GFS does not. Instead, GFS provides a client library for applications to interact with the file system.

### *8.10.2. File Snapshots*

GFS file snapshots employ a **copy-on-write** mechanism similar to Linux's **fork(2)** system call. In **fork(2)**, a child process initially has exactly the same state as the parent. Instead of fully copying all memory pages for children, the parent's pages are marked write-protected. When either process modifies a shared page, a copy is created, and the modification occurs on the copy, effectively diverging the process states.

Similarly, a GFS file snapshot creates a new file that initially shares the existing file's underlying chunks. When the chunks are modified for any of the files, a copy-on-write approach is used.

Here's the detailed process:

1. **Lease Revocation:** Upon receiving a snapshot request, the master revokes all leases on the target file. This prevents any further modifications, akin to write-protecting memory pages.
2. **Metadata Duplication:** The master creates a new file entry, replicating all metadata and chunk information from the original file. For each shared chunk, a **reference counter** is incremented.
3. **Copy-on-Write:** Any modification to a chunk with a reference count greater than one triggers a copy. The new chunk copy is then leased, its metadata is updated, and its location is provided to the client. To speed up the copy process, the new chunk is typically created on the same chunk server.

### *8.10.3. Garbage Collection*

GFS employs a **soft-delete** mechanism for files. When a file is deleted, the master server doesn't immediately remove its data. Instead, it renames the file to a hidden name with an old timestamp, marking it as deleted. This allows for potential recovery of accidentally deleted files.

Subsequently, a background scan by the master server identifies and removes the metadata associated with these deleted files. Chunk servers, during their regular heartbeat communications, receive a list of unknown chunks from the master, prompting them to delete those chunks.

While this approach might seem slow, it offers several advantages:

- **Eventual Consistency:** This method ensures eventual consistency in a straightforward manner. Periodic scanning is a common technique in distributed systems for synchronizing state across different components, ensuring that all states eventually converge. This method avoids the complexity of tracking real time diffs, which can be unreliable when the two components are external systems which are directly modified.
- **Accidental Deletion Recovery:** The inherent delay in garbage collection provides a window for rolling back accidental file deletions.

*8.10.4. Checksum Verification*

During append-writes, if the operation encounters a partially filled sub-chunk at the end of the file, the new data is appended after the existing content. The checksum for this partial sub-chunk is then incrementally updated to reflect the added data, as shown in Figure 44.

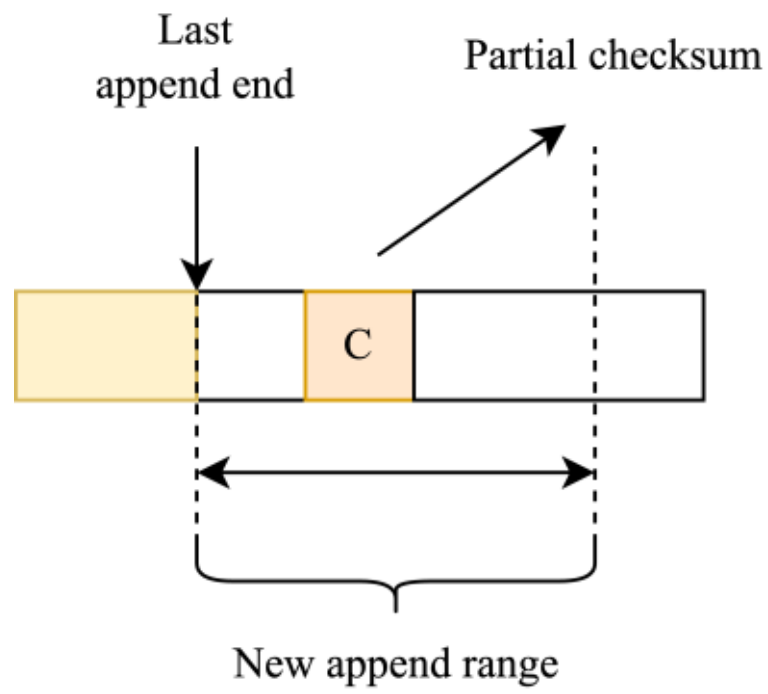

**Figure 44**: Incremental checksum after append in GFS.

This mechanism also serves as a corruption detection tool; if the partial sub-chunk was previously corrupted, the updated checksum will not match.

### 8.11. Paper Remarks

The paper is a comprehensive exploration of several key concepts in distributed systems. Even with its detailed nature, the paper is easy to follow and gives a great look at how distributed file systems are designed. Its massive impact is clear, as seen in its more than 10,000 citations. As a staple of technical literature, it is frequently discussed in reading groups.

# 9. CacheSack: Admission Optimization for Google Datacenter Flash Caches

(9) Yang, T.-W.; Pollen, S.; Uysal, M.; Merchant, A.; Wolfmeister, H. CacheSack: Admission Optimization for Google Datacenter Flash Caches. In 2022 USENIX Annual Technical Conference (USENIX ATC 22); 2022; pp 1021–1036.

In the previous chapter, we explored a distributed file system known as the Google File System (GFS), also referred to as Colossus. Data centers primarily rely on hard disks due to their high capacity and low cost. However, reading data directly from a disk can be expensive because the disk head supports only limited I/O operations. Consequently, frequently accessed data must be cached on faster storage media like RAM or flash to reduce read latency and costs. On the other hand, caching data in RAM or flash storage is expensive due to their limited capacities. Solving this problem requires a careful balance to achieve an optimal solution.

This paper was presented by Google at the USENIX Annual Technical Conference (ATC) '22, which is recognized as one of the premier conferences in the field of computer systems. The paper identifies and solves a compelling challenge regarding caching in file systems. While machine learning models are commonly employed for statistical problem-solving, this paper solves it using traditional linear programming methods.

In this insight, we will begin with the fundamental building blocks used in the paper: the fractional knapsack problem, the convex hull, and the algorithms used to construct a convex hull. Next, we will go through an overview of flash memory and examine the various caching layers of Colossus. Finally, we will delve into CacheSack, which formulates the caching challenge as a knapsack problem and solves it using linear programming technique.

**Recommended Reads**:

- **6. F2FS: A New File System for Flash Storage** where flash storage was introduced.
- **8. The Google File System** where a distributed file system (GFS) was introduced.

### 9.1. Fractional Knapsack Problem

The problem involves a knapsack with a capacity of $\mathbf{N}$, and a set of items, each with a cost $\mathbf{C}_i$ and a size $\mathbf{S}_i$.

**Goal**: Minimize the total cost of filling the knapsack completely, with the key feature that items can be taken in fractions.

**Solution**:

1. Calculate the cost-to-size ratio ($\mathbf{C}_i/\mathbf{S}_i$) for each item.
2. Sort the items in ascending order based on their cost-to-size ratio.
3. Fill the knapsack by taking items in this sorted order, taking fractions of items as needed to completely fill the knapsack.

### 9.2. Convex Hull

For a given set of points, the **convex hull** is the smallest convex polygon that encloses all the points, as shown in Figure 45.

The **lower convex hull** is the portion of the convex hull that, when projected onto the x-axis, casts the same shadow as the entire convex hull. Essentially, it's the "bottom" part of the hull, as shown in Figure 46.

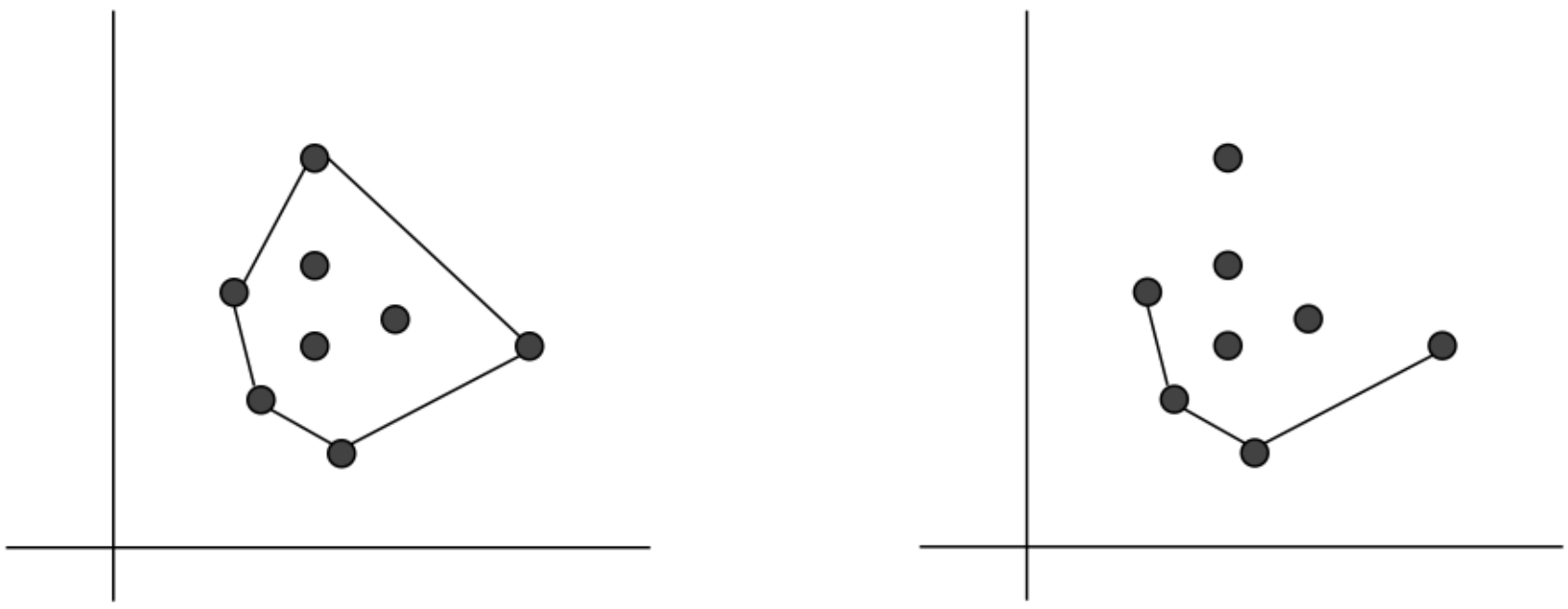

**Figure 45**: A convex hull. **Figure 46**: A lower convex hull.

#### *9.2.1. Graham's Scan Algorithm*

Graham's scan is a classic algorithm to compute the convex hull of a set of points, as shown in Figure 47:

1. **Initialization:**
    a. Find the point with the lowest y-coordinate (and leftmost if there are ties). This point is guaranteed to be part of the convex hull.
    b. Push this lowest point onto a stack.
2. **Sorting:**
    a. Sort the remaining points in counterclockwise order around the initial lowest point. This sorting is based on the angle formed by the x-axis and the line connecting the initial point to each other point.
3. **Hull Construction:**
    a. Iterate through the sorted points:
        i. For each point, push it onto the stack.
        ii. While the stack contains at least three points and the last three points on the stack form a non-left turn (i.e., a right turn or straight line), pop the second-to-last point from the stack.
    b. Once the iteration is complete, the points remaining on the stack form the convex hull.

The non-left turn check ensures that the hull remains convex. A left turn indicates that the points are forming a convex angle.

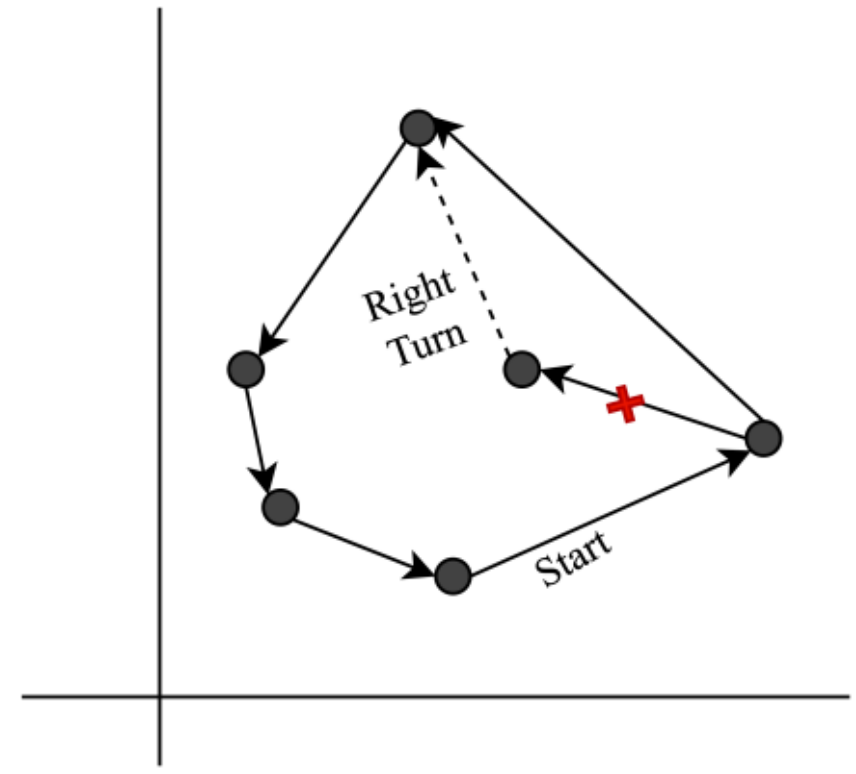

**Figure 47**: Graham's scan for convex hull.

### 9.3. Flash Memory

As discussed in a previous chapter, flash is the technology underpinning SSDs. We will go over Flash technology briefly in this chapter.

### 9.3.1. SSD Organization

SSDs organize data into pages, typically 2-4 KB (analogous to blocks in HDDs), which are grouped into blocks of 32-128 pages, as shown in Figure 48.

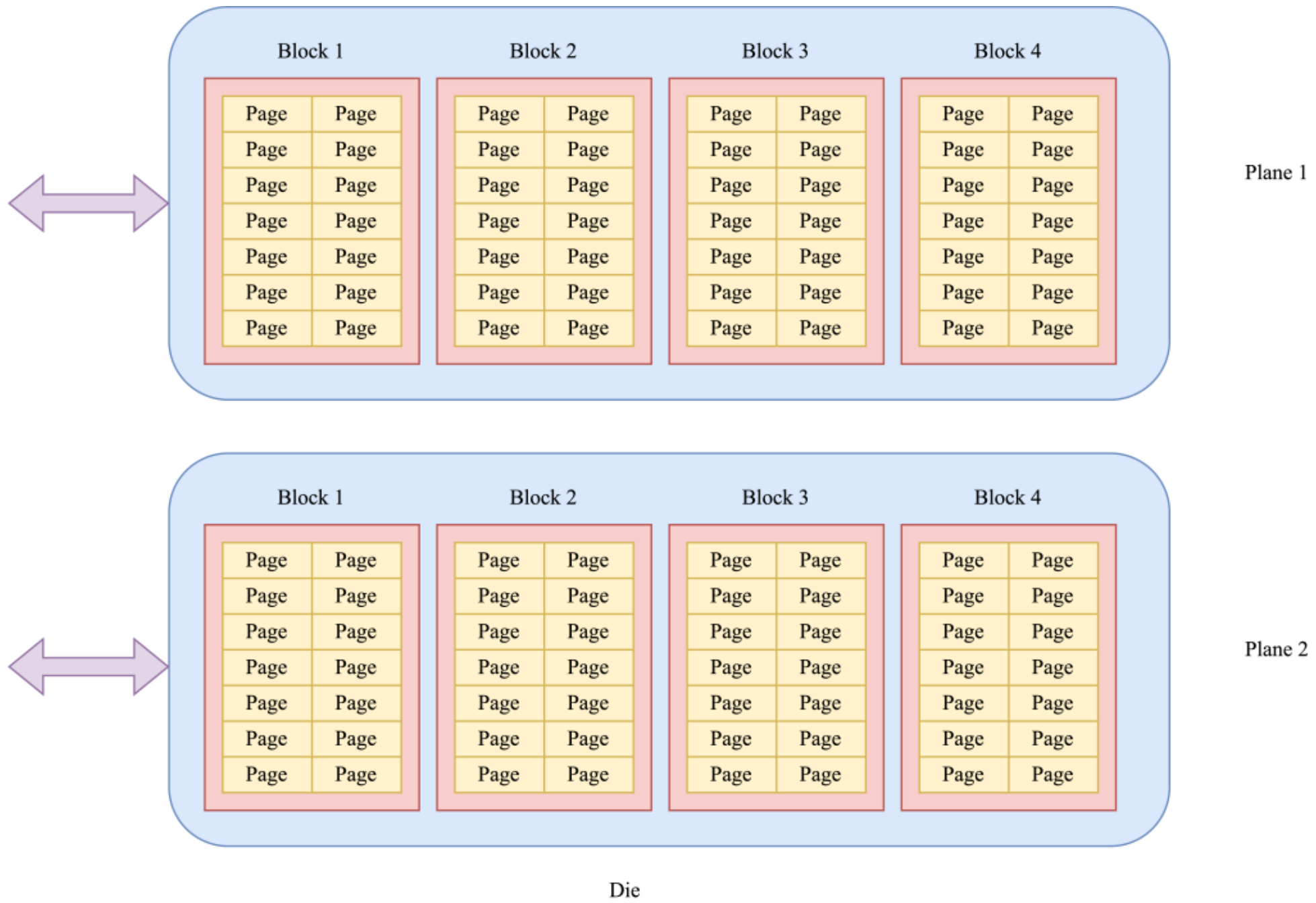

**Figure 48**: SSD internal organization.

A key characteristic of flash memory (which utilizes EEPROM) is that a flash cell must be erased explicitly before it can be re-written. With SSDs, there is a disparity between write and erase operations: writes occur at the page level, while erasures can be performed at the block level only.

### 9.3.2. Write Performance

SSDs employ a **Flash Translation Layer (FTL)** to map **Logical Block Addresses** from the operating system to **Physical Block Addresses** (PBAs). New blocks are mapped to free blocks. When free blocks are depleted, the SSD reclaims space by finding blocks where data has been deleted (trimmed), allowing for block erasure.

FTL effectively manages **sequential writes.** However, **random writes** present a challenge. Modifying even a single byte in an already-written block triggers a block copy with the updated byte to a new block. This is known as **write amplification**.

*9.3.3. Read Performance*

SSDs perform reads in parallel and at significantly higher speeds than HDDs due to the absence of a physical read head.

*9.3.4. Life Span*

Flash memory cells have a finite limit on the write/erase cycles before wear out, as a result SSD's lifespan is limited. SSDs employ FTL to distribute write operations evenly across all blocks, extending the drive's lifespan. However, it can still wear out after continuous writes.

### 9.4. Colossus Caching

Note that the Google File System has been rebranded as Colossus [81].

Problems with SSDs:

- Have less capacity and are more costly as compared to HDDs.
- Additionally, they wear-out after a certain number of writes.

Problem with HDDs:

- Throughout is limited due to the use of a spindle head.

In the vast infrastructure of data centers like Google's, the sheer scale of storage demands makes an all-SSD approach economically impractical. Therefore, Colossus chunk servers (D servers) primarily rely on HDDs for their storage capacity.

To mitigate HDD's read bottleneck, caching becomes essential. Caching frequently accessed data blocks allows for improved read speeds without the need to deploy a prohibitively large number of HDDs just to get good read performance. This strategy exemplifies how software effectively addresses hardware limitations in response to ever-changing industry needs.

The caching system for Colossus has multiple layers: the D Buffer Cache and the Flash Cache, as shown in Figure 1 in the paper.

### *9.4.1. D Buffer Cache*

The D Buffer Cache stores blocks directly in DRAM, offering fast access. However, due to DRAM's high cost, its capacity is inherently limited.

### *9.4.2. Flash Cache*

The Flash Cache, on the other hand, uses SSDs to store recently accessed blocks.

Optimizing this layer presents significant challenges, particularly in selecting an appropriate eviction algorithm. System research has extensively explored the nuances of SSD caching. For instance:

- The FIFO (First-In, First-Out) algorithm tends to produce sequential write patterns on SSDs, which are efficient.
- Conversely, algorithms like LRU (Least Recently Used) can lead to random write patterns, resulting in substantial write amplification and reduced SSD lifespan.

## 9.5. Colossus Flash Cache

The addition of a flash cache layer significantly enhances Colossus's performance. However, its limited capacity necessitates strategic caching decisions.

### *9.5.1. Total Cost of Ownership*

The primary optimization goal for this layer is minimizing the Total Cost of Ownership (TCO). In this context, TCO represents the overall expense incurred by Google for hosting the SSDs used for caching. A key factor influencing TCO is the lifespan of the SSDs, which is inversely proportional to the number of write operations. Therefore, the caching algorithm must intelligently select which data to cache, aiming to reduce writes and extend SSD longevity.

### *9.5.2. Client Interaction*

All reads from clients are sent to cache service before Colossus. This design allows the cache service to accurately track cache misses and efficiently retrieve the required data blocks from the underlying HDD storage into the SSD cache.

By carefully managing the data stored within the limited SSD capacity, the Colossus Flash Cache strives to balance performance gains with cost efficiency, ultimately minimizing the overall TCO.

### *9.5.3. Categorization & Admission Policies*

The flash cache system allows users to define categories for data, enabling fine-grained control over caching behavior. For example, databases like BigTable and Spanner have categories based on table name and locality group combinations.

Each category can have four admission policies for blocks, determining when a block is added to the cache:

- **AdmitOnWrite:** Blocks are inserted into the cache immediately upon being written or when a read request results in a cache miss.
- **AdmitOnMiss:** Blocks are cached only when a read request results in a cache miss.
- **AdmitOnSecondMiss:** Blocks are cached after experiencing two consecutive cache misses.
- **NeverAdmit:** Blocks are never cached, effectively bypassing the flash cache layer.

## 9.6. CacheSack

The goal of CacheSack is to find an optimal assignment of **<category, policy> → size**, i.e. the space that each **<category, policy>** would be assigned in the cache.

### *9.6.1. The Fractional Knapsack*

CacheSack is essentially a fractional knapsack problem:

- **Knapsack**: Represents the flash cache with a total space of **N**.
- **Items:** Represent the different user-defined categories. Each category is associated with the size for each of its four possible admission policies (**AdmitOnWrite**, **AdmitOnMiss**, **AdmitOnSecondMiss**, **NeverAdmit**).

This problem involves assigning cache space to categories and policies. Unlike the traditional knapsack problem:

- The cost is not known beforehand, necessitating an online learning approach.
- Cost can be computed at the category level only. However, the space allocated to a category must be distributed among its associated policies.

*9.6.2. Online Learning*

The solution begins with a random initial allocation of cache space across all **<category, policy>** pairs. Subsequently, the system monitors events for a 5-minute interval, using these events to calculate the cost. Based on this calculated cost, a knapsack optimization is performed to reallocate cache space to categories and policies.

For example, if a category experiences frequent reads of recently written blocks, and its **AdmitOnWrite** policy has a small initial allocation, the optimization process will likely increase the size allocated to that category's **AdmitOnWrite** policy.

*9.6.3. Category Cost Computation*

The cost function incorporates two key factors:

- **HDD Reads:** The number of disk reads, representing HDD spindle resources time burn.
- **SSD Writes:** The volume of data written to SSD, contributing to SSD wear.

The cost metric must balance these competing concerns.

9.6.3.1. LRU Approximations

CacheSack approximates flash cache behavior as an LRU cache. It employs a threshold **D**. This threshold determines whether a block access is considered a cache hit or miss:

- If the time elapsed since the last access, **d,** is less than or equal to **D** ($\mathbf{d} \leq \mathbf{D}$), the access is assumed to be a cache hit.
- If the time elapsed since the last access, **d**, is greater than **D** ($\mathbf{d} > \mathbf{D}$), the access is considered a cache miss.

The paper asserts that this LRU approximation is crucial and effective.

To maintain the last access time and, crucially, the last two access times for the **AdmitOnSecondMiss** policy, the system utilizes a **ghost cache**. It's important to note that this ghost cache is an approximation, as it cannot feasibly store every single entry. Instead, it maintains access time data for the most recent four hours. Any block that remains unaccessed for a period exceeding four hours has its access information lost. This means that if a block experiences a miss, and then, after a four-hour interval, experiences another miss, the system requires one additional miss to occur before that block becomes eligible for admission into the cache under the **AdmitOnSecondMiss** policy.

**D Buffer Cache Simulator**

Note that the LRU approximation above is only applicable to the flash cache. The authors make use of a blackbox tool called **D buffer cache simulator** to determine if a flash cache miss also caused a D buffer cache miss. Only then is it added to HDD read cost.

9.6.3.2. Cost Computation

Within each category, individual blocks are assigned to specific policies. This assignment can be probabilistic, proportional to the amount of cache space allocated to that policy.

Once a block is assigned a policy, the cost associated with accessing that block is calculated according to the chosen policy's behavior, as follows:

**Disk Read Cost**

- **AdmitOnWrite:** 0 if the block was written or accessed within the last **D** minutes (LRU approximation); 0 if the block was found in D buffer cache (simulation); 1 otherwise.
- **AdmitOnMiss:** 0 if the block was accessed within the last **D** minutes (LRU approximation); 0 if the block was found in D buffer cache (simulation); 1 otherwise.
- **AdmitOnSecondMiss:** 0 if the block was accessed twice within the last **D** minutes (LRU approximation); 0 if the block was found in D buffer cache (simulation); 1 otherwise.

- **NeverAdmit:** 0 if the block was found in D buffer cache (simulation); the total number of read requests from the client otherwise.

**Bytes Written Cost**

- **AdmitOnWrite:** The total number of write operations from the client.
- **AdmitOnMiss/AdmitOnSecondMiss:** 1 if a write operation occurred within the monitoring interval (5 minutes) as a result of miss/second miss.
- **NeverAdmit:** 0 (blocks are never written to the cache).

*9.6.4. Solving the Knapsack*

As noted earlier, one difference between regular knapsack and Cachesack is the fact that there are 4 fractions that need to be computed for each item (i.e., category). This makes the algorithm a little bit more involved to optimize.

The mathematical derivation is provided in Appendix A. It may not seem very trivial and so the following sections help walk through some of the complex bits.

9.6.4.1. Formal Cost Representation

Essentially, we have the following cost representation:

$$\mathbf{V_C^P}(\mathrm{D}) = \text{Cost of } \mathbf{S_C^P}(\mathbf{D}) + \textbf{Cost} \text{ of } \mathbf{W_C^P}(\mathbf{D})$$

Here **D** is the modeled threshold described previously.

The total cost $(\mathbf{V_C^P})$ for category **C** and policy **P** is calculated as the sum of the disk read cost $(\mathbf{S_C^P})$ and the bytes write cost $(\mathbf{W_C^P})$.

Let $\boldsymbol{\alpha}_\mathbf{C}^\mathbf{P}$ is the proportion of blocks admitted under a category **C** and policy **P**. The total cost is then the weighted sum of cost for each category and policy. This total cost needs to be minimized through the linear programming:

$$\min_{\alpha_C^{AOM},\alpha_C^{AOSM},\alpha_C^{AOW},\alpha_C^{NA}} \textstyle\sum_C \left(\boldsymbol{\alpha}_\mathbf{C}^\mathbf{AOM}\mathbf{V}_\mathbf{C}^\mathbf{AOM}(\mathbf{D}) + \boldsymbol{\alpha}_\mathbf{C}^\mathbf{AOSM}\mathbf{V}_\mathbf{C}^\mathbf{AOSM}(\mathbf{D})\right.$$

$$\left. + \boldsymbol{\alpha}_\mathbf{C}^\mathbf{AOW}\mathbf{V}_\mathbf{C}^\mathbf{AOW}(\mathbf{D}) + \boldsymbol{\alpha}_\mathbf{C}^\mathbf{NA}\mathbf{V}_\mathbf{C}^\mathbf{NA}(\mathbf{D})\right)$$

This cost minimization needs to happen under the following constraints which are fairly trivial to understand. The second constraint ensures the fractions sum to one, while the third limits the total cache size.

$$0 \leq \boldsymbol{\alpha}_{\mathbf{C}}^{\mathbf{AOM}}, \boldsymbol{\alpha}_{\mathbf{C}}^{\mathbf{AOSM}}, \boldsymbol{\alpha}_{\mathbf{C}}^{\mathbf{AOW}}, \boldsymbol{\alpha}_{\mathbf{C}}^{\mathbf{NA}} \leq \mathbf{1},$$

$$\boldsymbol{\alpha}_{\mathbf{C}}^{\mathbf{AOM}} + \boldsymbol{\alpha}_{\mathbf{C}}^{\mathbf{AOSM}} + \boldsymbol{\alpha}_{\mathbf{C}}^{\mathbf{AOW}} + \boldsymbol{\alpha}_{\mathbf{C}}^{\mathbf{NA}} = \mathbf{1},$$

$$\sum_{C}(\boldsymbol{\alpha}_{\mathbf{C}}^{\mathbf{AOM}}\mathbf{U}_{\mathbf{C}}^{\mathbf{AOM}}(\mathbf{D}) + \boldsymbol{\alpha}_{\mathbf{C}}^{\mathbf{AOSM}}\mathbf{U}_{\mathbf{C}}^{\mathbf{AOSM}}(\mathbf{D}) + \boldsymbol{\alpha}_{\mathbf{C}}^{\mathbf{AOW}}\mathbf{U}_{\mathbf{C}}^{\mathbf{AOW}}(\mathbf{D}) + \boldsymbol{\alpha}_{\mathbf{C}}^{\mathbf{NA}}\mathbf{U}_{\mathbf{C}}^{\mathbf{NA}}(\mathbf{D})) \leq \mathrm{U}_{\mathrm{total}}.$$

Now, we need to define **S** and **W** for each category and policy. They are trivial for **AdmitOnWrite** and **NeverAdmit**. Let's walkthrough the case of **AdmitOnMiss** and **AdmitOnSecondMiss**

Under the **AdmitOnMiss** policy:

- A disk read (**S**) occurs when the last access time exceeds the threshold **D**, and the block is missing in the buffer cache (**B**).
- The cache-byte time usage ( **U** ) is calculated as the block's byte size multiplied by its **residency time**. The residency time is the sum of all the times during which the block was in the cache hence it is a summation over, for all accesses **i** the minimum of $\mathbf{d}_i$ (calculated as the time difference between consecutive access times, $\mathbf{t}_i$ and $\mathbf{t}_{i-1}$) and **D**.
- Lastly, the write-cost (**W**) is calculated as block's byte size multiplied by number of actual reads. A read will happen whenever $\mathbf{d}_i$ (time since last access) exceeds **D**.

$$\mathbf{B}_{\mathbf{b}}^{\mathbf{AOM}}(\mathbf{D}, \mathbf{i}) = \begin{cases} \mathbf{1}, & \textbf{Buffer Cache Hit at } \mathbf{t}_i, \textbf{ using AOM.} \\ \mathbf{0}, & \textbf{Buffer Cache Miss at } \mathbf{t}_i, \textbf{ using AOM.} \end{cases}$$

$$\mathbf{S}_{\mathbf{b}}^{\mathbf{AOM}}(\mathbf{D}) = |\{\mathbf{i}: \mathbf{d}_i > \mathbf{D}, \mathbf{B}_{\mathbf{b}}^{\mathbf{AOM}}(\mathbf{D}, \mathbf{i}) = \mathbf{0}\}|,$$

$$\mathbf{U}_{\mathbf{b}}^{\mathbf{AOM}}(\mathbf{D}) = \mathbf{Size}(\mathbf{b}) * \sum_{i} \mathbf{min}(\mathbf{d}_i, \mathbf{D}),$$

$$\mathbf{W}_{\mathbf{b}}^{\mathbf{AOM}}(\mathbf{D}) = \mathbf{Size}(\mathbf{b}) * |\{\mathbf{i}: \mathbf{d}_i > \mathbf{D}\}|.$$

These are then summed over all the blocks to get the corresponding value for a policy.

Under the **AdmitOnSecondMiss** policy:

- A block will be in the cache if two consecutive accesses are within the threshold **D** (i.e., $\mathbf{d}_{i-1} \leq \mathbf{D}$ and $\mathbf{d}_i \leq \mathbf{D}$). Otherwise, when $\mathbf{max}(\mathbf{d}_i, \mathbf{d}_{i-1}) > \mathbf{D}$) and if the block is not in the buffer cache (**B**), a read is required.

$$\mathbf{B_b^{AOSM}(D, i)} = \begin{cases} \mathbf{1}, & \textbf{Buffer Cache Hit at } \mathbf{t}_i\textbf{, using AOSM.} \\ \mathbf{0}, & \textbf{Buffer Cache Miss at } \mathbf{t}_i\textbf{, using AOSM.} \end{cases}$$

$$\mathbf{S_b^{AOSM}(D)} = \left|\left\{\mathbf{i}: \mathbf{max}(\mathbf{d}_{i-1}, \mathbf{d}_i) > \mathbf{D}, \mathbf{B_b^{AOSM}(D, i)} = \mathbf{0}\right\}\right|.$$

- And for $\mathbf{U}$, the residency time is $\mathbf{d}_i$ if it was a cache hit, or it is $\mathbf{D}$ (the max) if the block was inserted at $\mathbf{t}_{i-1}$.
- The reasoning about $\mathbf{W}$ is left as an exercise to the readers.

9.6.4.2. Solving the Linear Program using Convex Hull

The linear program is solved using a convex hull approach, which can be broken down into the following steps:

- For each category, we calculate $\mathbf{U^p}$ (usage) and $\mathbf{V^p}$ (cost) for each policy $\mathbf{P}$.
- These values are then plotted as points ($\mathbf{U^p}$, $\mathbf{V^p}$) on a graph, with each category generating its own plot. Since there are four policies, each category's plot will contain four points.
- For each category's plot, we construct the lower convex hull. This hull represents the set of efficient policy combinations for that category, minimizing cost for a given usage. We identify the lines (segments) that make up the lower convex hulls of all categories.
- These lines are then sorted in decreasing order of the negative slope ($-\mathbf{V^p}/\mathbf{U^p}$). This slope represents the cost-to-usage ratio, or the cost reduction per unit usage.
- We greedily assign fractions to the points on these lines, starting with the line having the steepest negative slope, until the total usage constraint ($\sum \mathbf{U_C^P}$) is met. Essentially, at each step, we are choosing the category and policy combination that offers the greatest cost reduction for each unit of usage.

For example, as shown in Figure 49, the lines are selected in order of decreasing slopes across convex hulls of different categories.

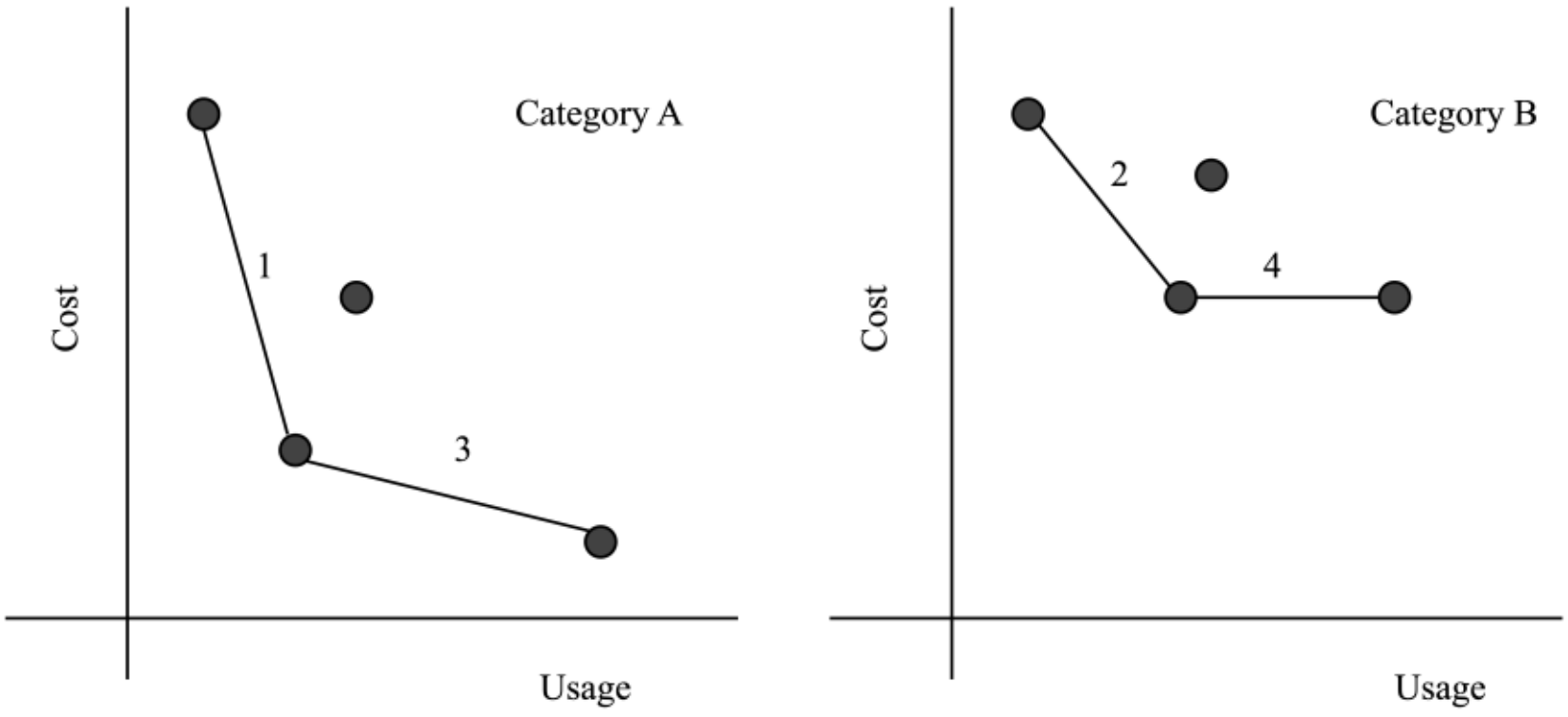

**Figure 49**: Linear program solution using convex hull.

*9.6.5. Online Learning*

The above knapsack formulation is solved for all possible values of **D**. The one that obtains the minimum cost is chosen and the assignment corresponding to it is the final assignment.

### 9.7. Evaluation

Note that users are not given the ability to define their own space allocation for categories. This control is entirely managed by the flash cache, meaning that the allocation of cache space is determined solely by the system's automation. This lack of user-configurable tuning actually serves to incentivize users to adopt the cache more readily, as they are relieved of the burden of manual optimization.

The implementation of CacheSack has demonstrated a reduction in the overall TCO. In production environments, CacheSack has consistently outperformed the cost associated with statically assigned policies, indicating its effectiveness in resource management.

Simulation results provide further insight into CacheSack's adaptive behavior. Specifically:

- For smaller cache sizes, the simulations suggest that **AdmitOnSecondMiss** should receive a larger portion of the available space. This strategy is good for minimizing the impact of transient read blocks within the cache.
- Conversely, for larger cache sizes, the simulations indicate that **AdmitOnMiss** and **AdmitOnWrite** should be prioritized.

CacheSack's learning algorithm effectively navigates these varying scenarios by dynamically adjusting allocations based on the cost of each operation, thereby maintaining a high hit ratio across different cache configurations.

### 9.8. Paper Remarks

The paper presents a distinctive approach to the caching problem by utilizing a mathematical optimization framework based on the fractional knapsack problem and linear programming. This methodology stands in contrast to many contemporary research efforts in this domain, which often rely on sophisticated machine learning models. The paper provides a rigorous introduction to the complexities of caching within distributed systems.

# 10. MapReduce: Simplified Data Processing on Large Clusters

(10) Dean, J.; Ghemawat, S. MapReduce: Simplified Data Processing on Large Clusters. Communications of the ACM 2008, 51 (1), 107–113.

We now take a digression from operating systems to focus on data processing.Data processing is a fundamental requirement for modern industrial applications. When datasets grow too large to fit within the memory or storage of a single machine, distributed computing becomes essential. Before the 2000s, data processing was mostly ad hoc; however, post 2000s, it has evolved into a set of standard frameworks. Utilizing these has become common practice in system design.

Authored by Google's Jeff Dean and Sanjay Ghemawat, this paper was presented at Operating System Design and Implementation (OSDI) in 2004. Since then, the MapReduce model has been adopted into several data processing frameworks across the industry.

In this insight, we will begin by understanding different data processing models. We will then examine the MapReduce model and see how it can be executed in a distributed environment. Finally, we will look at how the model is implemented in Google’s version of MapReduce. As a bonus, we will explore various implementations of the shuffle phase, which is the most widely studied phase of the MapReduce lifecycle.

### 10.1. Data Processing

Data processing involves performing operations on collections of data, often represented as **records**. These records can reside in various locations, such as database tables or files. Two primary approaches to data processing are:

#### *10.1.1. Batch Processing*

Processes data in discrete batches or groups. Batch processing is suitable for tasks that can be processed periodically, such as batch updates, nightly reports, and data warehousing. The focus of this chapter will be on batch processing.

### *10.1.2. Stream Processing*

Processes data continuously as it arrives, one record at a time. Stream processing is ideal for real-time applications like fraud detection, stock market analysis, and IoT data processing.

To learn about stream processing, refer:

- **11. Apache Flink™: Stream and Batch Processing in a Single Engine**
- **12. The Dataflow Model: A Practical Approach to Balancing Correctness, Latency, and Cost in Massive-Scale, Unbounded, Out-of-Order Data Processing**

## 10.2. The MapReduce Programming Model

MapReduce is a programming model designed for processing massive datasets across distributed computing environments. It operates through three core phases:

- **Map:** Transforms each input record into a key-value pair. This operation is stateless, meaning it relies solely on the input record for its output.
- **Shuffle:** Redistributes and groups the key-value pairs based on their keys. This step prepares the data for the reduce phase, ensuring that all values associated with the same key are processed together.
- **Reduce:** Aggregates the values associated with each key, producing a final output.

### *10.2.1. Shuffle*

The shuffle phase is a critical phase of MapReduce, responsible for redistributing data to ensure that all values associated with the same key are processed by the same reducer.

**Partitioning Function -** A partitioning function determines which reducer will process each key-value pair. It takes an input record and outputs a partition number. Crucially, it must ensure that all records with the same key are assigned to the same partition.

### *10.2.2. Reducer Properties*

A key requirement for reducers is that the operation they perform must be have the following properties -

- **Associativity** - Ensures that the grouping of elements doesn't change the outcome.

$$\mathbf{reduce}(\mathbf{reduce}(\mathbf{a},\mathbf{b}),\mathbf{c}) \;=\; \mathbf{reduce}(\mathbf{a},\mathbf{reduce}(\mathbf{b},\mathbf{c}))$$

- **Commutativity** - Ensures that the order of elements doesn't change the outcome.

$$\mathbf{reduce}(\mathbf{a},\mathbf{b}) \;=\; \mathbf{reduce}(\mathbf{b},\mathbf{a})$$

In essence, these properties guarantee that the order in which elements are aggregated is irrelevant.

*10.2.3. Illustrative Example: Word Frequency Counting*

Consider the task of counting the frequency of each word in the sentence: "the quick brown fox jumps over the lazy dog".

**1. Map Phase**

The map function processes each word, emitting a key-value pair where the word is the key and 1 is the value: (the, 1), (quick, 1), (brown, 1), (fox, 1), (jumps, 1), (over, 1), (the, 1), (lazy, 1), and (dog, 1).

**2. Shuffle Phase**

The shuffle phase groups the key-value pairs by key, creating a list of values for each unique key: (brown, [1]), (dog, [1]), (fox, [1]), (jumps, [1]), (lazy, [1]), (over, [1]), (quick, [1]), and (the, [1, 1]).

**3. Reduce Phase**

The reduce function aggregates the values for each key, summing them to produce the final word counts: (brown, 1), (dog, 1), (fox, 1), (jumps, 1), (lazy, 1), (over, 1), (quick, 1), and (the, 2).

**Scalability and Applications**

This simple example demonstrates the core concept. MapReduce's power lies in its ability to scale to handle massive datasets. When dealing with billions or trillions of words, the input is distributed across multiple machines. The map, shuffle, and reduce phases are executed in parallel, enabling efficient processing.

#### *10.2.4. MapReduce and SQL*

MapReduce is fundamental to many large-scale data processing systems, including those used for executing SQL queries. Data analytics often involves processing millions of rows, which aligns perfectly with MapReduce's map-shuffle-reduce paradigm. Platforms like BigQuery [84], Dremel [85], and F1 utilize MapReduce pipelines to execute complex SQL queries efficiently.

### 10.3. Distributed Execution of MapReduce

While a MapReduce pipeline can run on a single machine, its true power lies in distributed execution across multiple machines, known as **workers**. This parallel processing significantly accelerates computation.

#### *10.3.1. Components*

- **Workers:** These machines execute the map and reduce functions.
- **Driver Program:** This program orchestrates the entire MapReduce job, scheduling tasks and managing workers.

#### *10.3.2. Execution Flow*

All distributed execution of MapReduce has roughly the following stages:

- **Input Partitioning:** The driver program initially divides the input dataset into smaller, manageable partitions. These partitions are distributed among the map workers. Ideally, each input partition should fit within a worker's memory.
- **Map Phase:** Each map worker processes its assigned input partition, generating key-value pairs. The number of output records from the map phase directly corresponds to the number of input records processed.
- **Shuffle and Partitioning:** The key-value pairs are shuffled and partitioned. This step groups values with the same key and distributes them to the appropriate reduce workers. Typically, partitioning is integrated into the map and the reduce phase itself.

- **Reduce Phase:** Reduce workers scan the map output partitions, retrieving all key-value pairs relevant to their assigned key space. They then perform the reduce operation, aggregating values for each key.

#### *10.3.3. Statelessness*

The key to distributed MapReduce execution is ensuring map and reduce functions are stateless. Stateless functions always produce the same output for the same input and are therefore deterministic.

This is important because it lets the system safely re-run tasks without changing the final outcome, which is essential for handling failures and keeping the system consistent.

#### *10.3.4. Underlying File System*

A crucial aspect of distributed MapReduce is the underlying file system. It facilitates data storage, retrieval, and distribution across the worker nodes. We will explore how Google's MapReduce leverages its file system to achieve this.

### 10.4. Google's MapReduce

Google's MapReduce system is designed for large-scale data processing across a cluster of machines. Here's a breakdown of its key components and execution flow (shown in Figure 1 in the paper):

**1. Input Splitting**

Input files are divided into **M** partitions, typically ranging from 16 MB to 64 MB.

**2. Master (Driver)**

A central master coordinates all MapReduce tasks, acting as the orchestrator and scheduler for workers in a MapReduce cluster.

**3. Map Phase**

- Map workers are preferably run on machines where the input already resides.
- Map workers read their assigned input splits.
- Map output is buffered in memory.

- Periodically, the buffered output is written to the local disk, partitioned based on the keys (**partial shuffles**).
- The locations of these partitioned files are communicated to the master.

**4. Shuffle Phase**

- Reduce workers are assigned specific key space.
- Reduce workers make RPCs to read the required data from the map workers' local disks.
- Because the partitioning done by mappers was only local to their memory buffer, the reduce phase performs a global scan (**total shuffle**) to fetch all keys that it has been assigned.

**5. Reduce Phase**

- Reduce workers perform the reduction operation.
- The final output is written to one or more output files.

*10.4.1. Example (Figure 50)*

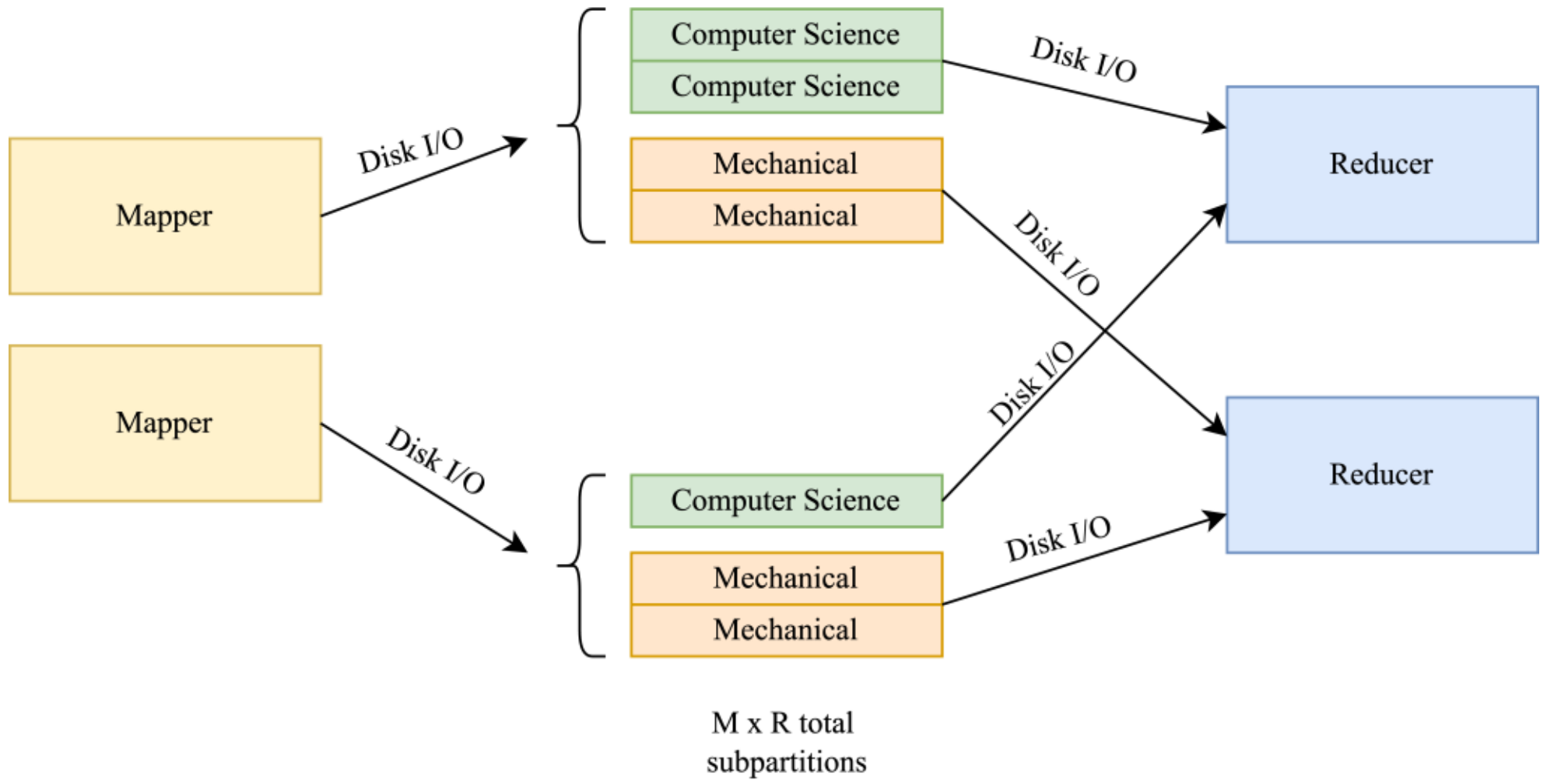

**Figure 50**: MapReduce example.

Consider the SQL query:

```
SELECT COUNT(*) FROM PROFESSORS GROUP BY DEPARTMENT;
```

The **GROUP BY** clause necessitates shuffling to group all professor records by their respective departments before counting (reduction).

The input table is divided into input partitions, each assigned to a mapper. Each mapper processes its assigned partition, extracting rows and grouping them by department, which becomes the key. These grouped records, representing partial shuffles, are then written to the mapper's local disk.

Each reducer, responsible for specific departments (keys), retrieves the corresponding sub-partitions from the local disks of all mappers and aggregates the count.

### *10.4.2. Complexity Analysis*

Consider **M** map tasks, each producing **R** partitions, one for each reducible key. These partitions represent partial shuffles.

Consequently, each mapper performs one I/O operation to read its input and writes **R** sub-partitions to the local disk in another I/O operation. Each reducer requires **M** I/O operations to read its complete partition and one I/O to write its output, resulting in a total I/O cost of $\mathbf{M} + \mathbf{M} + (\mathbf{M} * \mathbf{R}) + \mathbf{R} = (\mathbf{2} + \mathbf{R}) * \mathbf{M} + \mathbf{R}$.

The master node maintains metadata for **M** map tasks and **R** reduce tasks, leading to a memory complexity of $O(M + R)$.

### *10.4.3. Key Features*

#### 10.4.3.1. Master as Central Coordinator

The master program maintains the state of all map and reduce tasks and facilitates file location sharing. The state of the master is periodically snapshotted to save the state and facilitate recovery.

#### 10.4.3.2. Output Storage

Map output (immediate output) is stored on local disks, while reduce output (actual output) is stored on Google File System.

10.4.3.3. Partitioning Function

The default partitioning function is $\mathbf{hash}(\mathbf{key}) \% \mathbf{R}$, where **R** is the number of reducers. Users can also define custom partitioning functions.

10.4.3.4. Combiner Function (Partial Shuffling)

After the map stage, the combiner function divides the output into partitions. Because each mapper task processes only a subset of the input data, these partitions represent **partial shuffles**. The reducer tasks then aggregate these partial shuffles to create the final, totally partitioned shuffle.

10.4.3.5. Locality Optimization

Google's MapReduce attempts to schedule map tasks on machines that hold the input data locally, minimizing network traffic.

10.4.3.6. Stragglers

The system addresses "stragglers" (slow workers) by running backup tasks. Examples of Stragglers are:

- Machines with bad disks.
- Machines with software bugs such as operating system errors.

The backup task increases the overall computational resources for the operation by no more than a few percent.

10.4.3.7. Fault Tolerance

Since map output is on local disk, machine failures can lead to data loss. The master program re-executes failed map tasks. Map and reduce tasks are meant to be stateless, enabling safe retries without affecting correctness.

When a map task finishes, it sends a completion notification to the master. The master ensures atomicity by marking the task as completed and ignoring any subsequent completion reports for that task, including those from backup tasks. For reduce tasks, completion involves atomically renaming the output file on GFS.

#### 10.4.3.8. Side-Effects

While we've primarily discussed stateless and deterministic map/reduce tasks, they can also produce side-effects, such as modifying external system states. To ensure system reliability, especially in the face of failures that trigger task retries, application developers must guarantee that these side-effects are idempotent.

## 10.5. Bonus: Shuffle Implementations

There are different implementations of shuffle architecture for MapReduce.

### *10.5.1. Pull-Based Shuffle*

Mappers write their output to the local disk. Reducers pull their required partitions from the mappers' local disks. Mappers perform $\mathbf{M}$ I/O operations, and reducers perform $\mathbf{M} * \mathbf{R}$ I/O operations.

This is exactly what is used in Google's MapReduce.

This is inefficient due to numerous small I/O operations, impacting disk throughput. Reliability and scalability are also concerns.

### *10.5.2. Pre-Shuffle Merge*

Mappers on the same machine merge their outputs before writing to disk. This reduces the number of I/O operations. If $\mathbf{F}$ mappers are bundled, mappers perform $\mathbf{M/F}$ disk I/O, and Reducers perform $\mathbf{M} * \mathbf{R/F}$ disk I/O. This decreases the total number of I/O operations. This is used in Riffle [86] by Facebook.

### *10.5.3. Push-Based Shuffle*

Introduces a dedicated shuffle service between mappers and reducers, as shown in Figure 51. Mappers push their sub-partitions to the shuffle service, in addition to writing to local disk. Reducers retrieve their partitions from the shuffle service. This decouples mappers and reducers, improving fault tolerance and scalability. This is used in Magnet [87] by LinkedIn.

The shuffle service stores the shuffled partitions on disk to prevent memory overload. Reducers can perform a single large I/O operation to read their assigned partition from the shuffle service.

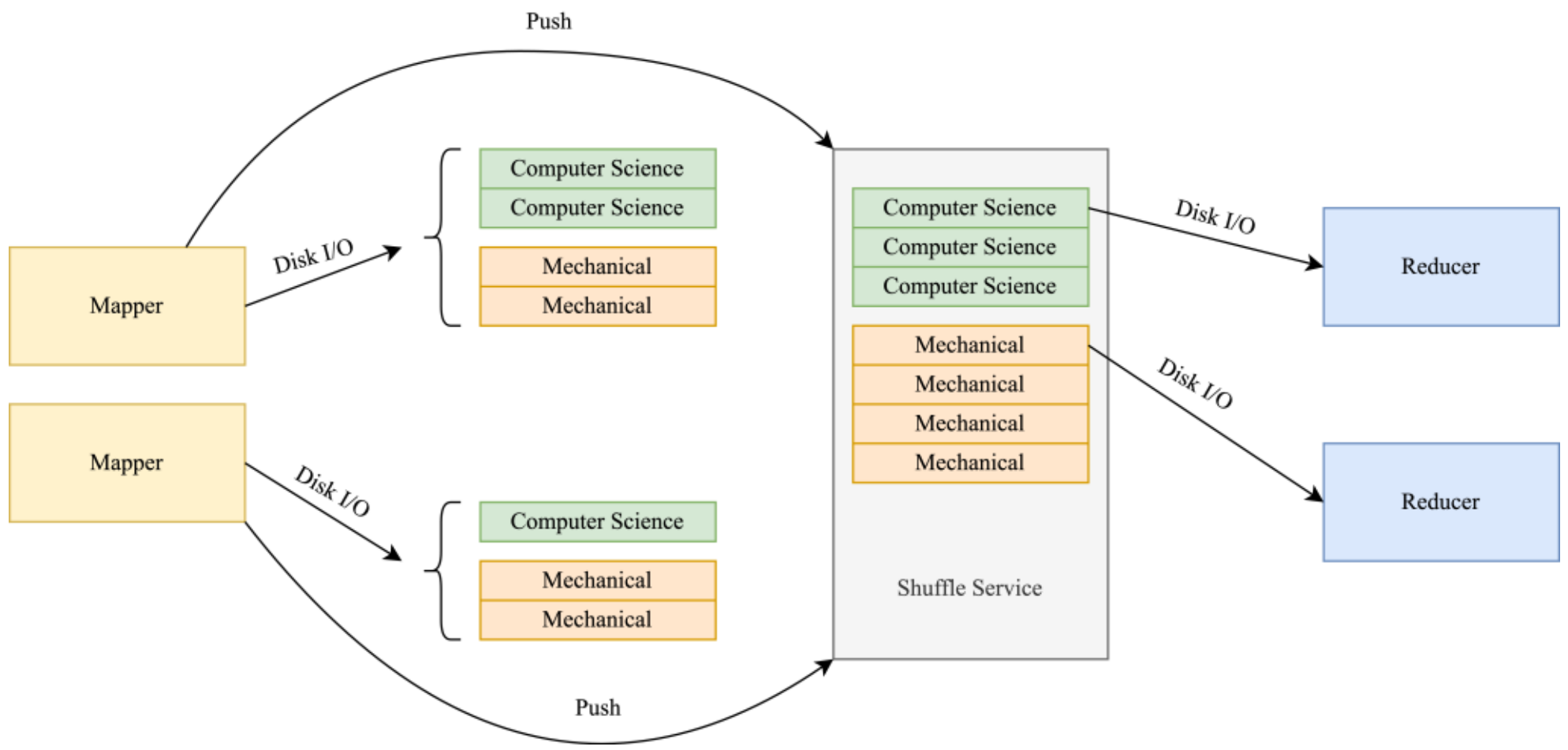

**Figure 51**: Push-based shuffle in MapReduce.

Note that, mappers push to the shuffle service is best effort. If the push fails, reducers can fallback to pulling from the mappers local disk.

### 10.6. Paper Remarks

This paper is one of the easiest papers in distributed systems. All of the concepts and the system itself can be easily grasped. The work has also been highly influential, as it served as the backbone for FlumeJava [88] and various batch processing frameworks, such as Apache Hadoop and Apache Spark.

# 11. Apache Flink™: Stream and Batch Processing in a Single Engine

(11) Carbone, P.; Katsifodimos, A.; Ewen, S.; Markl, V.; Haridi, S.; Tzoumas, K. Apache Flink: Stream and Batch Processing in a Single Engine. The Bulletin of the Technical Committee on Data Engineering 2015, 38 (4).

In the previous chapter, we explored batch processing. In this chapter, we will look at stream processing in greater detail. Data processing frameworks evolved toward stream processing primarily to meet the demand for real-time information. The 2010s saw a growing interest in stream processing, leading to the creation of several specialized platforms.

This paper was presented at the IEEE International Conference on Cloud Engineering (IC2E) Workshop in 2015. Workshops provide a forum for focused discussions on practical research; this specific paper introduced Flink, an engine designed for both stream and batch processing

In this insight, we will first revisit the differences between batch and stream processing. Next, we will briefly explore the history of how various data processing frameworks came into existence. Following that, we will dive deeper into stream processing by discussing data flow graphs and operators. Finally, we will examine the specific features of Apache Flink and learn how it reliably implements stream processing.

### 11.1. Batch and Stream Processing

Data processing involves performing operations on collections of data, often represented as **records**. These records can reside in various locations, such as database tables or files. Two primary approaches to data processing are:

#### *11.1.1. Batch Processing*

Processes data in discrete batches or groups. There are certain primitive operations in batch processing:

- **Map:** Applies a function to each individual record within a batch, producing an output for each input.

- **<u>Reduce</u>:** Aggregates the results of the map operation, often by combining values associated with the same key.
- **<u>Shuffle</u>:** (Optional) Redistributes data between processing stages, typically between the map and reduce phases, to improve data locality and efficiency.

**Example:** Counting word frequencies across multiple documents:

- **Map:** Count occurrences of each word within each document.
- **Shuffle:** Group all occurrences of the same word from different documents.
- **Reduce:** Sum the counts for each word across all documents.

Batch processing is suitable for tasks that can be processed periodically, such as batch updates, nightly reports, and data warehousing.

Refer **10. MapReduce: Simplified Data Processing on Large Clusters** to learn more on batch processing.

#### *11.1.2. Stream Processing*

Processes data continuously as it arrives, one record at a time. The data records processed by stream processing systems are also called **events**. Prime characteristics are:

- **Low latency**: Processes events with minimal delay.
- **Real-time analysis**: Enables immediate responses to events.

Stream processing is ideal for real-time applications like fraud detection, stock market analysis, and IoT data processing.

### 11.2. A Bit of History

Early systems, such as FlumeJava [88] (2010) built upon Google's MapReduce framework, primarily focused on batch processing. Spark [89], another prominent early player, also initially centered around batch processing capabilities.

Around 2013, the landscape shifted towards stream processing. Spark Streaming [90] emerged that addressed stream processing by dividing the incoming data stream into short intervals and processing each interval as a mini-batch. This was effective but this approach retains some characteristics of batch processing.

On the other hand, Google's Millwheel [91] pioneered true stream processing by enabling continuous, real-time processing of data with features like stateful operations (e.g., incrementing a counter for each incoming record).

The terms "batch-native" and "stream-native" are often used to categorize systems. Batch-native systems, like early versions of Spark, were initially designed for batch processing and later extended to support stream processing. Conversely, stream-native systems, such as Apache Flink, were inherently designed for stream processing and subsequently adapted to handle batch processing as a special case.

Apache Flink draws significant inspiration from Millwheel and exemplifies a stream-native approach. Its core architecture is fundamentally designed for stream processing, with batch processing effectively treated as a specialized form of stream processing.

## 11.3. Building Blocks of Stream Processing

All stream processing systems support a common construct: a **data flow graph** consisting of **operators** through which events flow. Some operators require event collection (such as windowing), which in turn requires the pipeline to understand the notion of time. This is achieved using **watermarks**.

### *11.3.1. Data Flow Graph*

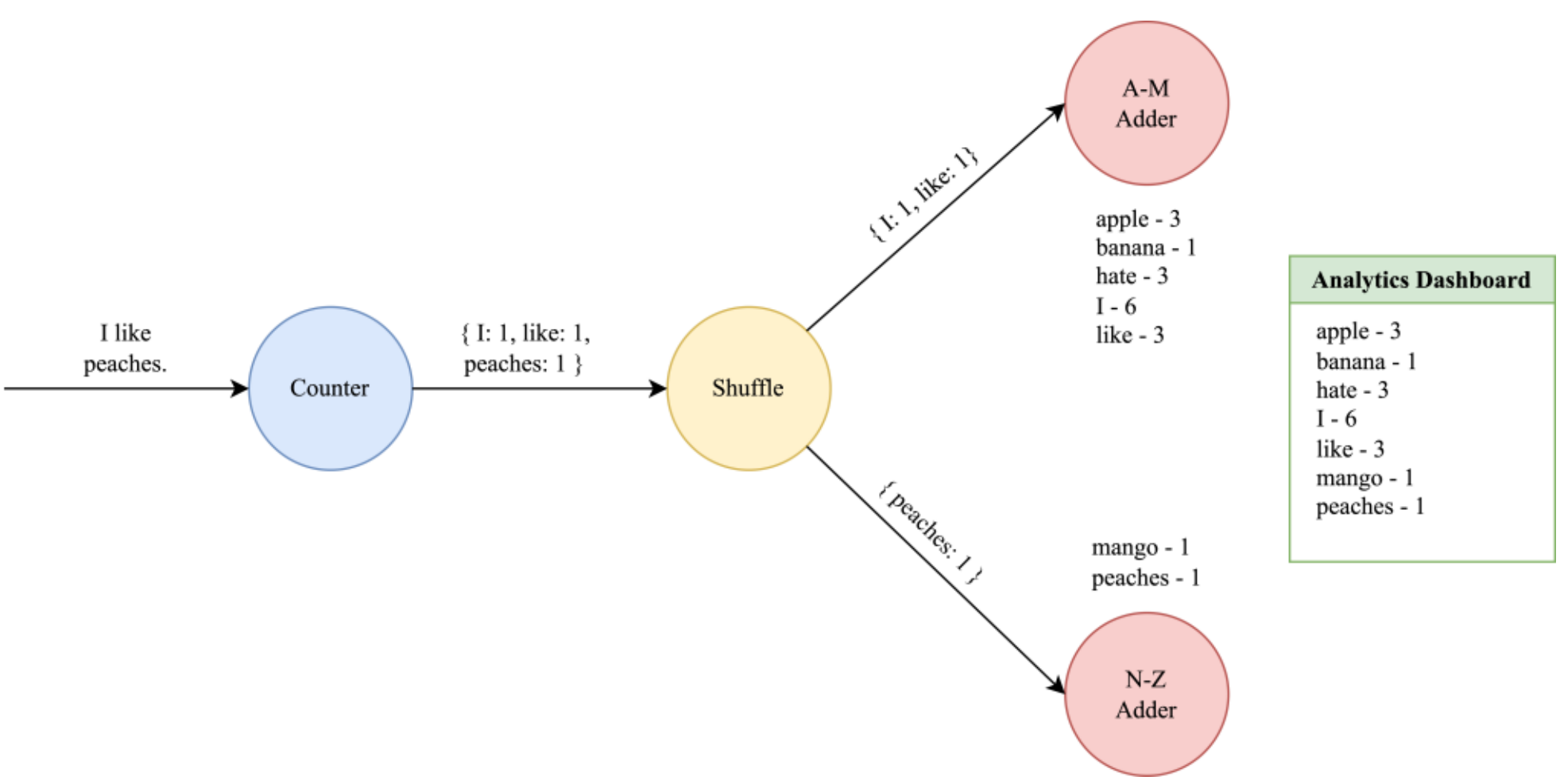

**Figure 52**: Example of data flow graph for stream processing.

A data flow graph is a directed acyclic graph (DAG) consisting of stateful operators, as exemplified in Figure 52.

This example illustrates data flow in a stream processing system by extending the word frequency count scenario to real-time processing:

- **Counter:** This is a stateless operator. It receives a document as input and produces a map of words to their frequencies within that document.
- **Shuffle:** Another stateless operator, it distributes individual words from the word-frequency maps to corresponding "Adder" operators based on the word's alphabetical range. This ensures efficient distribution of data.
- **Adder:** A stateful operator responsible for maintaining the current count of each word across all processed documents.
    - It receives word-frequency maps from the "Shuffle" operator and then updates its internal state by adding the frequencies of incoming words to the existing counts.
    - It can be sharded (e.g., one Adder for words A-M, another for N-Z) to improve performance and scalability.
- The current state of the Adders (word counts) can be visualized on an analytics dashboard to provide real-time insights.

This example demonstrates a basic data flow in a stream processing system. However, real-world systems incorporate a much wider range of operators, enabling complex data transformations and analyses.

### *11.3.2. Pipelined Operators*

In stream-native processing systems, operators must be designed to support continuous data flow, enabling efficient pipelining of events. Key examples of pipelined operators include:

- **Window**:
    - Collects a stream of events within a defined time window (e.g., sliding window).
    - Pipelined because the collection of subsequent windows can proceed concurrently, without waiting for the processing of previous windows to complete.

- In systems like Flink, batch processing can be emulated by applying a window operator to collect all events within a specific batch interval.
- **Map**:
    - Applies a function to each incoming event independently.
    - Highly pipelined as each event can be processed concurrently without relying on other events.
- **Group By (Shuffle)**:
    - Redistributes events based on a key.
    - Pipelined because events are continuously forwarded to the appropriate downstream operator (e.g., reducer) responsible for the corresponding key.
- **Reduce**:
    - Aggregates incoming events.
    - Supports "rolling reduce" for continuous aggregation, requiring an **associative** function (e.g., sum, count).
    - The output represents the ongoing reduction of all events seen thus far.
- **Interval Join**:
    - Joins events within a specific time interval.
    - The only type of join that can be effectively pipelined.
    - Essentially equivalent to applying a window operator followed by a join on the accumulated events within that window.

#### *11.3.3. Non-Pipelined Operators*

Certain operators are inherently non-pipelined, meaning they cannot process data continuously and require access to the entire dataset before producing results.

**Example:** A sort-merge join that requires sorting the entire dataset before performing the join operation is an example of a non-pipelined operation.

#### *11.3.4. Watermarks*

Stream processing involves two notions of time:

- **Processing Time:** The time at which an event is processed by the system.
- **Event Time:** The time at which the event occurred in the real world.

Windowing operations rely heavily on these time concepts.

- **Processing Time Windows:** These windows are defined by the system's processing time. Events arriving within a specific processing time window are grouped together. This approach can be problematic when event arrival times are unpredictable, as processing time may not accurately reflect the true event times.
- **Event Time Windows:** These windows are defined by the event times themselves. Events with event times falling within a specific window are grouped together. This approach provides a more accurate representation of real-world event occurrences.

For example, consider a stream of application logs from mobile devices. Due to network latency, these logs may experience delays before reaching the processing system. This discrepancy between the event's actual occurrence time (event time) and its processing time can lead to inaccurate results when using processing time windows.

Watermarks address this challenge. A watermark signals that no further events with an event time earlier than the watermark's timestamp are expected to arrive. This allows the system to proceed with window processing even if some late events might still be in transit.

In other words, processing time advances monotonically with the processing system's clock. In contrast, event time requires explicit mechanisms like watermarks to progress, ensuring timely window closure and subsequent processing.

#### 11.3.4.1. Limitations of Watermarks

While watermarks effectively advance event time processing, they can be less effective in handling significantly late events (e.g., events arriving long after the watermark has passed).

In our example, what happens if a user disconnects from the internet? Logs generated during this period might be delayed significantly. The watermark may have already progressed? That is where Apache Beam [92] is useful and has the capability of handling late events.

**Recommended Read**: **12. The Dataflow Model: A Practical Approach to Balancing Correctness, Latency, and Cost in Massive-Scale, Unbounded, Out-of-Order Data Processing**

11.3.4.2. Examples

- **Spark Streaming:** Only works with processing time windows.
- **Flink:** Supports both processing time and event time windows, providing greater flexibility.
- **Apache Beam:** Offers robust support for handling late events, providing more sophisticated mechanisms for dealing with data arriving out of order.

### 11.4. Apache Flink

Apache Flink implements the distributed dataflow graphs described previously, leveraging watermarks as its core mechanism for handling the notion of time. In the following section, we will explore the specific architectural features of Flink.

#### *11.4.1. Data Exchange*

Operators in Flink communicate by exchanging data.

- **Pipelined Data Exchange:** Characterize continuous, uninterrupted data flow from non-window operators (e.g., Counter, Shuffle), ideal for stream processing.
- **Blocking Data Exchange:** Requires the entire window to be collected before processing, typically used in batch processing.

Operators exchange data through serialization and buffering (accumulating data before transmission). Buffers are periodically flushed to the consumer, improving throughput but potentially introducing latency.

#### *11.4.2. Control Events*

In addition to data records, control messages can flow through the data flow graph, enabling the triggering of specific conditions or actions within the processing pipeline.

#### *11.4.3. Checkpoint Barriers*

Flink guarantees **exactly-once processing** semantics, ensuring that each event in the stream affects the system state exactly once. To achieve this, Flink leverages **Asynchronous Barrier Snapshotting (ABS)**.

##### 11.4.3.1. ABS Mechanism

Checkpoint barriers are propagated through the operator pipeline. Upon encountering a barrier, each operator records its current state (e.g., the current count in a counter operator) to a persistent storage. This creates a consistent snapshot of the system's state across all operators.

##### 11.4.3.2. Restart and Recovery

In case of a failure, operators can recover from the last successful checkpoint. They resume processing from the point where the checkpoint was taken, replaying events and applying updates to their state accordingly.

##### 11.4.3.3. Limitations of ABS

ABS ensures exactly-once processing within the Flink system itself. However, it cannot prevent duplicate external triggers by the processing logic. For instance, if an operator initiates an external transaction (e.g., booking a flight ticket) as part of its processing, recovering from a checkpoint will result in the transaction being executed again, leading to duplicates.

##### 11.4.3.4. Extending Exactly-Once Semantics Beyond Flink

Since 2017, Flink has extended its exactly-once guarantees to external systems. This is achieved through a two-phase commit protocol:

1. **Prepare Phase:** When the last operator in the pipeline receives a checkpoint barrier, it initiates a two-phase commit with all downstream sinks. Before committing any data, the operator buffers all output records in memory.
2. **Commit Phase:** Once all operators have successfully recorded their checkpoints, the last operator commits all buffered output records to the sinks.

#### 11.4.3.4.1. Example: Kafka Sink

For a Kafka sink, the last operator starts a Kafka transaction when a checkpoint barrier arrives. Subsequent barriers trigger the completion of the previous transaction, ensuring that each event is written to Kafka exactly once.

### *11.4.4. Windowing Algorithms*

Flink offers several windowing algorithms for grouping and processing data streams:

#### 11.4.4.1. Tumbling Windows

- Divides the data stream into fixed-size, non-overlapping intervals.
- Suitable for applications requiring regular, discrete time intervals for analysis, such as hourly or daily aggregations.

#### 11.4.4.2. Sliding Windows

- Divides the data stream into fixed-size windows that overlap.
- Enables continuous analysis with overlapping intervals, useful for rolling averages or trend detection.

#### 11.4.4.3. Session Windows

- Dynamically sized windows defined by periods of inactivity in the data stream.
- Ideal for analyzing user sessions or other scenarios where activity patterns are irregular.

#### 11.4.4.4. Global Windows

- Unbounded windows that encompass the entire data stream.
- Typically used for global aggregations or analyses requiring consideration of all available data.
- Require an external trigger (e.g., a manual command or a scheduled event) to close the window and trigger processing.

### 11.5. Bulk Synchronous Parallel v/s Stale Synchronous Parallel

Bulk Synchronous Parallel (BSP) (shown in Figure 53) requires all workers to synchronize at the end of each iteration, leading to potential bottlenecks due to barrier synchronization.

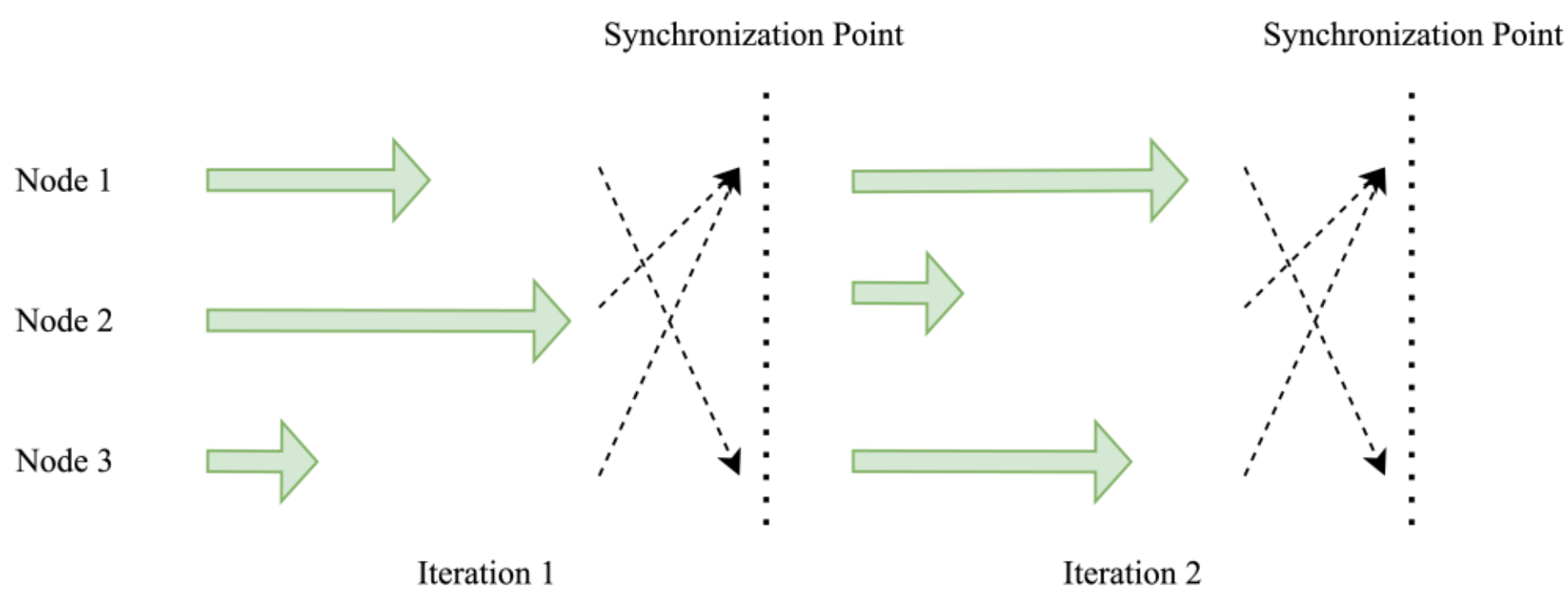

**Figure 53**: Bulk Synchronous Parallel.

Stale Synchronous Parallel (SSP) (shown in Figure 54) allows workers to proceed independently, improving speed but potentially compromising consistency.

SSP is suitable for algorithms like stochastic gradient descent where some degree of staleness is acceptable.

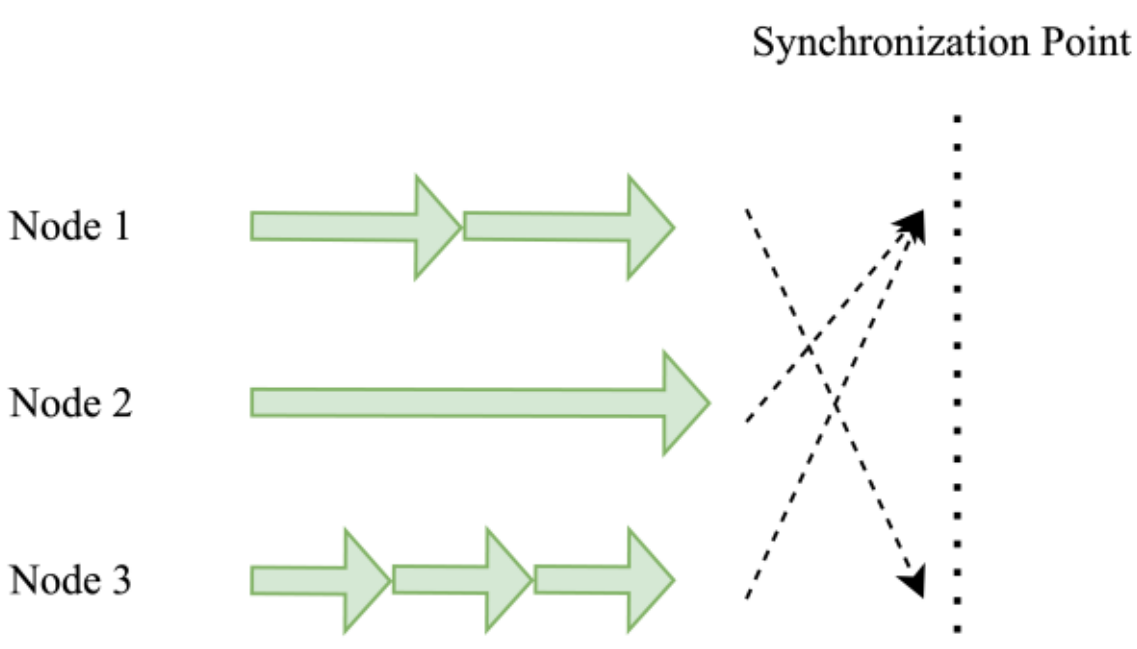

**Figure 54**: Stale Synchronous Parallel.

Flink's execution model allows for any type of structured iteration logic to be implemented on top, by using iteration-control events.

### 11.6. Bonus: Difference between Apache Flink, Apache Spark, Apache Storm?

**Apache Flink:** Designed primarily for stream processing with strong support for stateful computations and event-time semantics. Ensures data is processed exactly once, even in case of failures.

**Apache Spark** [93]**:** Excels at batch processing tasks, with stream processing capabilities added later. Known for its fast and general-purpose nature, supporting a wide range of data processing workloads.

**Apache Storm** [94]**:** Focused solely on real-time stream processing with high throughput and low latency. Guarantees that each message will be processed **at least once**, but may be processed multiple times in case of failures.

### 11.7. Paper Remarks

This paper offers a clear introduction to stream processing systems, making it easy to follow. The system it describes has been widely implemented and is used in the industry, and hence the paper is an interesting read for anyone enthusiastic about big data.

# 12. The Dataflow Model: A Practical Approach to Balancing Correctness, Latency, and Cost in Massive-Scale, Unbounded, Out-of-Order Data Processing

(12) Akidau, T.; Bradshaw, R.; Chambers, C.; Chernyak, S.; Fern\'andez-Moctezuma, R. J.; Lax, R.; McVeety, S.; Mills, D.; Perry, F.; Schmidt, E.; others. The Dataflow Model: A Practical Approach to Balancing Correctness, Latency, and Cost in Massive-Scale, Unbounded, out-of-Order Data Processing. Proceedings of the VLDB Endowment 2015, 8 (12), 1792–1803.

Having learned about stream processing systems, we will now focus on one crucial aspect: out-of-order processing. Processing streaming events out of order presents unique challenges that must be addressed to ensure correctness while simultaneously ensuring that latency remains unaffected.

This influential paper from Google, presented at VLDB 2015, is a landmark in Data Engineering. Authored by Tyler Akidau, it explores groundbreaking concepts. Akidau's work on Millwheel [91] significantly influenced Apache Flink, while his Dataflow Model laid the groundwork for Apache Beam [92]. Notably, Google Cloud Dataflow [95] implements the Apache Beam framework.

In this insight, we will begin by understanding the problem of unbounded, out-of-order processing in greater detail through a practical example. Then, we will dive deep into the Dataflow model itself. To wrap up our exploration of data processing systems, we will take a look at some industry-standard data processing systems.

Note that this paper does not describe a specific system; rather, it is a conceptual framework.

**Recommended Read: 11. Apache Flink™: Stream and Batch Processing in a Single Engine**, which explains the key differences between batch and stream processing.

### 12.1. Motivating Example

In a streaming system, events arrive continuously and are processed in an ongoing manner. The core concept of this paper is to process these events based on their "event time" - the actual time when the event occurred.

#### *12.1.1. Challenges of Event Time Processing*

Processing events based on their event time often leads to out-of-order processing. This is because events may not arrive at the processing system in the same order as they were generated.

##### 12.1.1.1. Example:

Consider a stream of logs generated by end-user devices. To process these logs in batches, we can define a window (e.g., 2 minutes). All events arriving within this window are collected and processed together. However, we have multiple notions of time:

- **Processing Time:** The time when a log reaches the processing system is referred to as "processing time".
- **Event Time:** The actual time when a log was generated is its "event time".

For example, if a log was generated at 9:58 AM and reached the system at 9:59 AM, its event time is 9:58 AM and its processing time is 9:59 AM.

Processing events based on the processing time window is relatively simpler. Systems like Spark Streaming [90] operate based on processing time semantics, collecting events into batches based on their arrival time.

Processing events based on the event time window is complex. With billions of devices generating logs, it's difficult to determine when all events within a specific event time window will arrive at the processing system. For instance, if we want to process all events generated before 10:00 AM, we need to know when all events from that timeframe have been received.

#### *12.1.2. Watermarks*

To address this challenge, the concept of "watermarks" is introduced. A watermark is a control message sent to the system to indicate that all events generated before a certain time are likely to have arrived. For example, a watermark at 10:01 AM may suggest that all events generated before 10:00 AM have probably been received, allowing the system to proceed with processing the batch.

#### 12.1.2.1. Limitations of Watermarks

Even with watermarks, there's still a possibility of "late" events - events that arrive after their expected event-time based processing window has closed. For instance, a log generated at 9:59 AM might arrive at the system at 10:05 AM. Since the watermark for the 10:00 AM window has already passed, the system will be forced to ignore this late event.

Apache Flink makes use of watermarks and has a similar limitation.

### *12.1.3. The Dataflow Model*

To address the issue of late events, the Dataflow model provides mechanisms to handle them appropriately. This ensures that the impact of late events is accurately reflected in the system's processing, even though they arrive after their designated batch has been processed.

## 12.2. Dataflow

The Dataflow model represents a novel approach to stream processing, offering a unique set of advantages:

- **Event Time Processing:** Processes events based on their actual occurrence time, not just their arrival time.
- **Fault Tolerance:** Resilient to system failures, ensuring data integrity and continuous processing.
- **Eventual Correctness:** A ***novel*** idea that allows for controlled trade-offs between processing latency and accuracy.
- **Scalability:** Designed to handle massive volumes of data with high throughput.
- **Low Latency:** Achieves minimal processing delays for time-sensitive applications.

The core of the Dataflow model lies in its ability to flexibly adjust the balance between latency and correctness.

### 12.3. Event time v/s Processing time

**Event Time:** The time at which the event actually occurred.

**Processing Time:** The time at which the event is processed by the system.

Ideally, processing time should perfectly align with event time, meaning events are processed immediately upon occurrence. However, in reality, events may arrive at the processing system with a delay. Figure 55 shows the difference between ideal and actual relation between event and processing time.

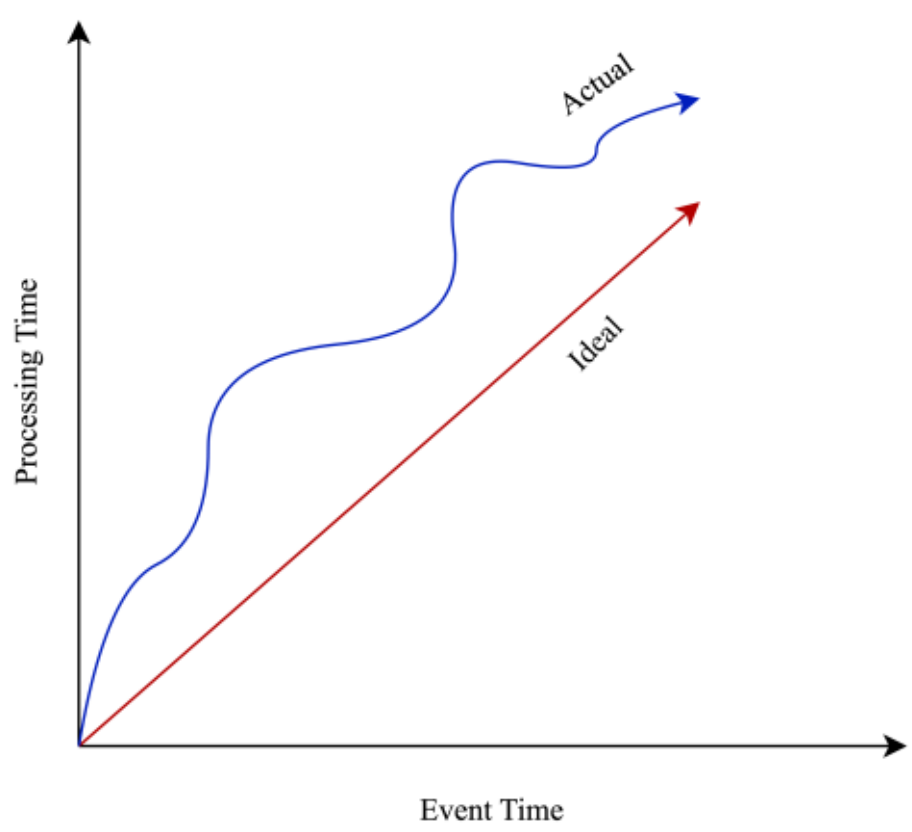

**Figure 55**: Event time v/s processing time.

#### *12.3.1. Challenges of Out-of-Order and Late Events*

Delays can cause processing discrepancies. Late events, those arriving significantly after their expected time, complicate the situation. Maintaining the correct processing sequence becomes challenging.

For example, in Figure 56, an event appearing late can result in out-of-order processing. Both Spark and Flink ignore this event as "late" leading to incorrectness.

For most real-world applications, it's common to encounter out-of-order, late events. Effectively handling these situations is critical for ensuring the accuracy and reliability of the stream processing system.

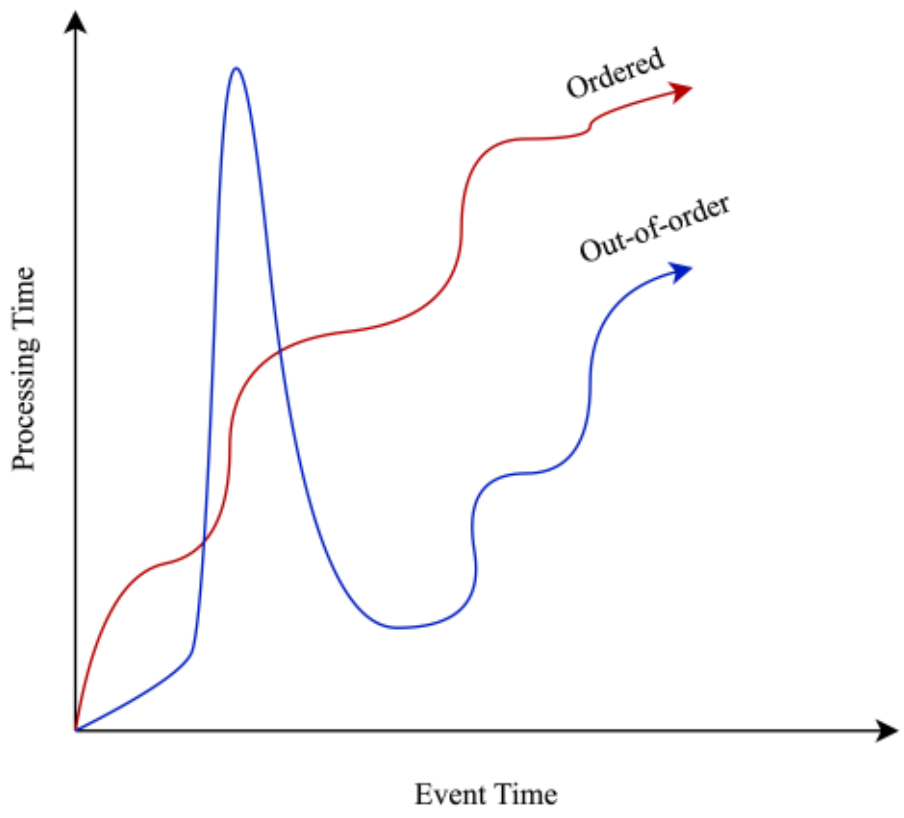

**Figure 56**: Event time v/s processing time for late events.

### 12.4. Watermarks

Watermarks are mechanisms (either internal or external to the system) that signal the system's knowledge of event arrival. Watermarks always correspond to event time. They represent specific points in the event time space, indicating that all events occurring before that point are believed to have arrived at the system, as shown in Figure 57.

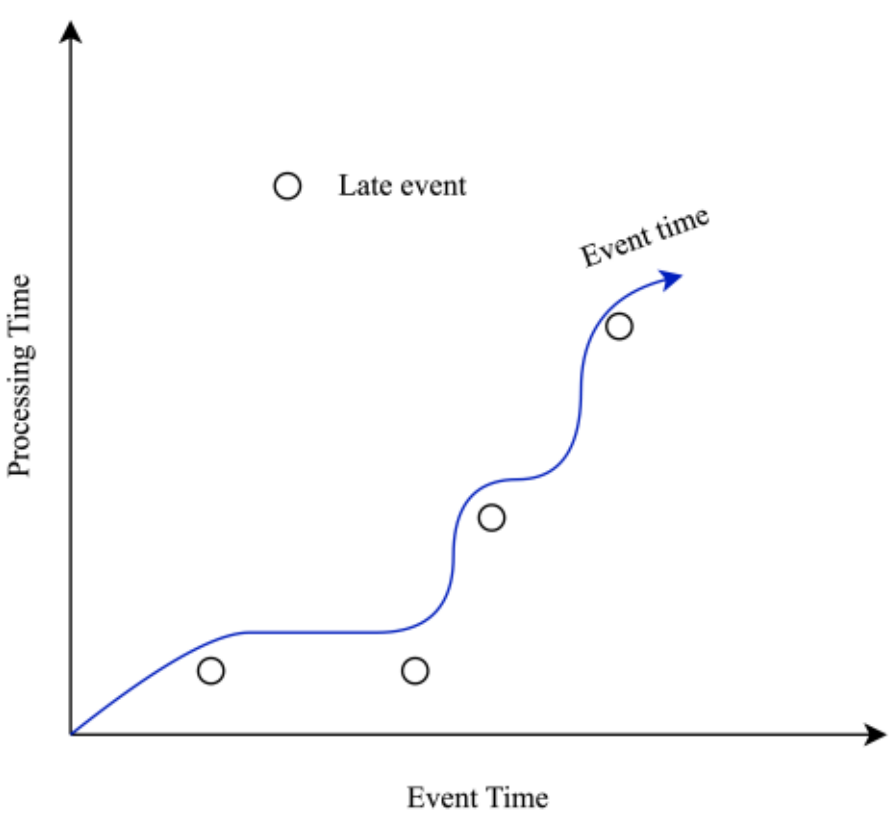

**Figure 57**: Watermarks marking event time progress.

Note that watermarks may not always be accurate reflections of event arrival. Events can arrive late leading to out-of-order arrivals. Despite this, the watermark itself must always monotonically increase. This ensures that the system's understanding of event arrival progresses forward in event time.

## 12.5. Windows and Windows Pane

Windows are defined as segments or regions within the event time space. They can be:

- **Overlapping:** Windows can partially overlap with each other.
- **Non-overlapping:** Windows have no overlap.
- **Aligned:** All windows cover the same event time bounds across all event keys.
- **Unaligned:** Window event time bounds may vary across different event keys.

### *12.5.1. Types of Windows*

- **Fixed Windows (Tumbling Windows):** Non-overlapping windows with a fixed duration.
- **Sliding Windows:** Overlapping windows with a fixed duration, sliding forward at regular intervals.
- **Session Windows:** Capture activity within a specific time period for a particular subset of data, unaligned.

The paper provides a detailed algorithm for creating windows from event streams in section 2.2. We will not delve into the specifics here, as the paper itself offers a clear explanation with illustrative examples.

### *12.5.2. Window Panes and Processing*

A window of events in event time space may be processed multiple times due to the arrival of late events. Each processing instance of a window is referred to as a "pane".

### *12.5.3. Triggering Window Pane Processing (Vertical Axis Bounds)*

To initiate the processing of a window pane, specific triggers are required, as :

- **Processing Time Intervals:** Trigger processing at regular intervals (e.g., every minute).
- **Num Events:** Trigger processing when a certain number of events have arrived.

*12.5.4. Triggering Window Processing (Horizontal Axis Bounds)*

To determine when a window is complete in event time, divide the event time space into equal intervals and then utilize watermarks to track the progress of event time and identify when a specific point in event time has likely been reached.

The processing machine has the knowledge of processing time using system clock, however, it lacks direct knowledge of event time progression. Watermarks are crucial for tracking the advancement of event time within the system.

*12.5.5. Triggering the Rectangle*

By combining triggers in both the horizontal (processing time) and vertical (event time) dimensions, the system can effectively determine when a rectangular window is ready for processing, as shown in Figure 58.

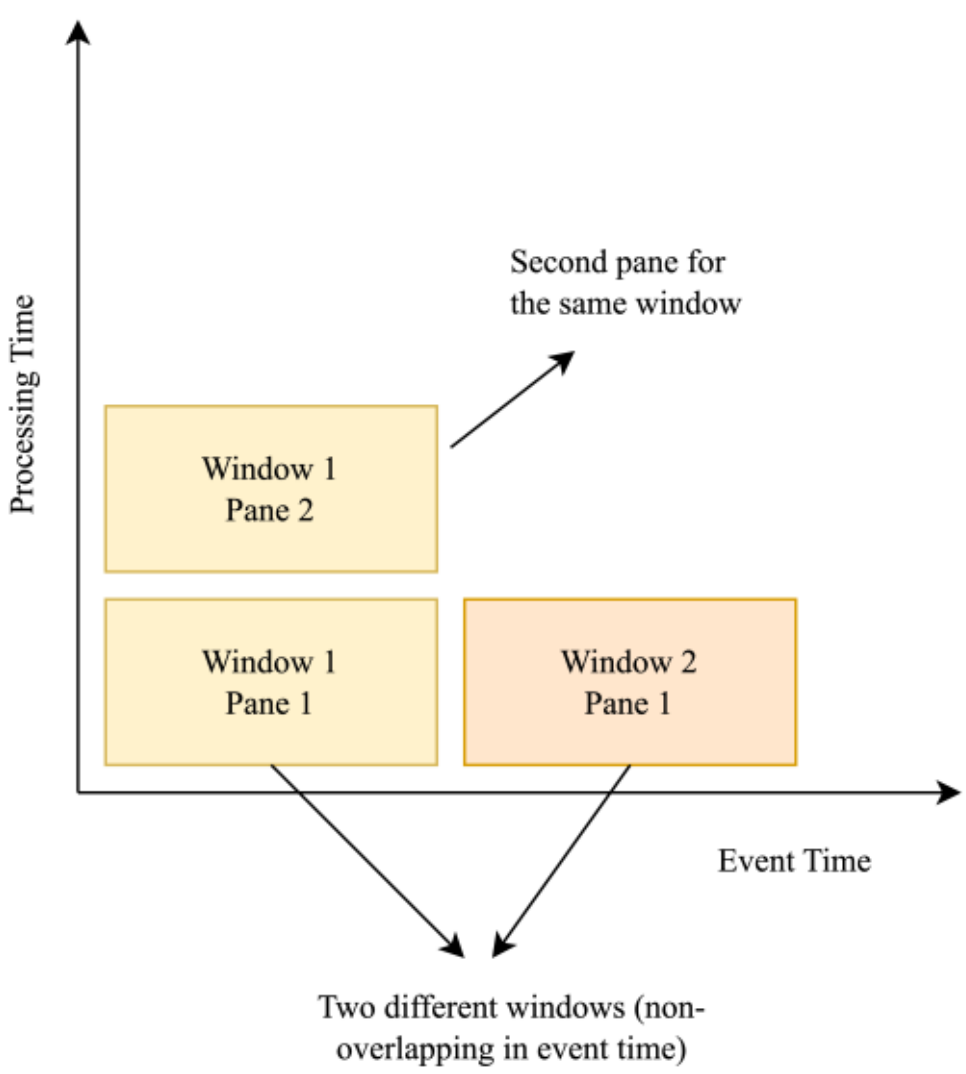

**Figure 58**: Triggering window and window pane.

The paper provides great examples in section 2.4, along with API usage to explain how windowing mechanism works for different window types.

## 12.6. Incremental Processing

Since the same window may have multiple panes due to late events, the system needs an efficient mechanism to handle these incremental updates. Three common approaches are:

- **Discarding:** Discard all previously processed panes for the window and process only the events in the new pane. This is the most efficient method.
- **Accumulating:** Accumulate all events across all panes of the window and process them together. This approach is suitable only if the processing logic overwrites previous results with the latest known value.
- **Accumulating and Retraction:** This two-step process involves first retracting (or reversing the effect) of the previously processed panes and then processing the new pane, which now includes all accumulated events for the window. This method is powerful but requires that the consumers of the processed results can effectively reverse the impact of previous processing. This is not feasible if the processing results have already been committed as part of a transaction.

## 12.7. Bonus: Related Systems

### *12.7.1. Google's Big Data Processing Systems*

The concepts presented in this paper have been materialized in Google Cloud Dataflow, a powerful stream and batch processing service built upon the foundations of FlumeJava and MillWheel.

- **FlumeJava [88] (2010):** An early batch processing system built on the MapReduce framework. It provides a set of APIs, specifically immutable parallel collections, for writing MapReduce programs. FlumeJava acts as a compiler framework, optimizing user-defined MapReduce applications.
- **MillWheel (2013):** A low-latency stream processing engine where user-defined logic is represented as a directed graph. Each node in the graph represents a processing step. MillWheel guarantees exactly-once processing semantics.
- **Dataflow (2013):** A unified framework that seamlessly integrates batch processing (leveraging MapReduce) and stream processing (utilizing MillWheel). It offers a simplified programming model and high-level abstractions. Dataflow is open-sourced as Apache Beam and is also available as a managed service on Google Cloud as Cloud Dataflow.

Google also developed Photon, a stream processing engine that evolved from Ubiq [96]. A key distinction of Photon is its multi-homed architecture, where jobs are replicated globally while maintaining exactly-once semantics through PaxosDB-based transactions.

Google has a suite of powerful query engines that makes use of data processing system under the hood:

- **Dremel** [85]**:** For interactive query processing. It is the brain behind Google's BigQuery [84]. Underneath it makes use of MapReduce.
- **F1 Query:** Google's most advanced query engine, also leveraging MapReduce for efficient query execution.

*12.7.2. Misc Big Data Processing Systems*

- **Hadoop** [97]**:** A foundational framework for distributed storage and processing of large datasets.
- **Pig** [98]**:** A high-level language for expressing dataflow programs that execute on Hadoop. (Relationship: Pig to Hadoop :: FlumeJava to MapReduce)
- **Hive** [99]**:** Built on top of Hadoop, enabling SQL-like data querying and analysis, similar to Dremel.
- **Spark** [89]**:** A fast and general-purpose cluster computing system that extends the MapReduce model.
    - ***Spark Streaming*** [90]**:** An extension of Spark for stream processing, primarily focusing on micro-batching.
- **Apache Storm** [94]**:** A distributed stream processing platform that provides at-least-once processing guarantees.

**12.8. Paper Remarks**

The paper does not propose a fundamentally new architecture, rather, it creatively leverages existing streaming systems to introduce a new paradigm for balancing correctness and latency. The core problem statement may be challenging to grasp at first, however, it provides the necessary context for the rest of the well-structured argument. This is a must-read for anyone looking to deepen their understanding of Big Data processing.

# 13. Delta Lake: High Performance ACID Table Storage over Cloud Object Stores

(13) Armbrust, M.; Das, T.; Sun, L.; Yavuz, B.; Zhu, S.; Murthy, M.; Torres, J.; Hovell, V.; Ionescu, A.; uszczak, A.; others. Delta Lake: High-Performance ACID Table Storage over Cloud Object Stores. Proceedings of the VLDB Endowment 2020, 13 (12), 3411–3424.

From here, we begin our journey into database systems - systems built on top of file systems to provide a more structured and efficient way to manage data. The history of databases dates back to the 1970s, and they have evolved significantly over the decades. While there are several hundred types of databases in existence, this series will explore a few of the major ones in widespread use today. Most of the databases we will cover are distributed as these are more prevalent than standalone ones because they offer essential data redundancy.

This paper, in particular, was presented at the 2020 Very Large Data Bases (VLDB) conference. It is written by Michael Armbrust of Databricks, with co-authors including CEO Ali Ghodsi and CTO Matei Zaharia. The paper introduces a database architecture built on top of Cloud Object Stores, which can be conceptualized as a serverless database.

In this insight, we will begin by defining the differences between a database and a data warehouse. We will then break down a database into its constituent layers and explore each one. Next, we will examine the various file formats databases use to store data. Following that, we will define ETL pipelines and look at how they are utilized within the industry. Finally, we will explore Cloud Object Stores in detail and see how Delta Lake was built on top of them. As a bonus, we will also walk through several related systems described in the paper.

### 13.1. Databases, Data Warehouses, and Data Lakes

Let's start with a discussion about the differences between these frequently used jargons in the computer science industry.

#### *13.1.1. Databases*

While the term *database* is broad, it commonly refers to **transactional** **SQL databases** **(OLTP)**. These SQL databases prioritize transactional integrity and offer varying levels of isolation. A fundamental principle of SQL databases is that data must always maintain consistency.

For example, in a bank's table, the total balance across all accounts must remain constant regardless of the number of transfers. If two transactions - one at time $\mathbf{T_1}$ and another at time $\mathbf{T_2}$ each add $1 to every account, the expected behavior is:

Before $\mathbf{T_1}$: No accounts have $1 added.
After $\mathbf{T_1}$: All accounts have $1 added.
After $\mathbf{T_2}$: All accounts have $2 added.

**Note**: There are also NoSQL databases which often sacrifice some isolation guarantees for improved performance. In this chapter, we will focus on SQL databases only.

#### *13.1.2. Data Warehouses*

Designed for **Online Analytical Processing (OLAP)**, data warehouses prioritize historical data for analysis. The data in a data warehouse must also be consistent, however, it does not necessarily reflect real-time changes.

See **32. Napa: Powering Scalable Data Warehousing with Robust Query Performance at Google** to learn more about how data warehouses work.

#### *13.1.3. Data Lakes*

Data lakes are repositories for storing large volumes of raw data in various formats (structured, semi-structured, unstructured). This raw data can then be processed through **ETL** pipelines (described below) to create data warehouses.

### 13.2. Database Layers

Most databases are built as layers on top of other databases, as shown in Figure 59, though there are many exceptions.

At the foundation are **core transaction engines**, often provided as libraries, that manage transactions on the underlying data files. Examples include InnoDB [100] and MyISAM [101] for SQL databases, and WiredTiger [102], BerkeleyDB [103], and RocksDB [104] for NoSQL databases. Building upon these engines are standalone databases like MySQL [105] and MongoDB [106].

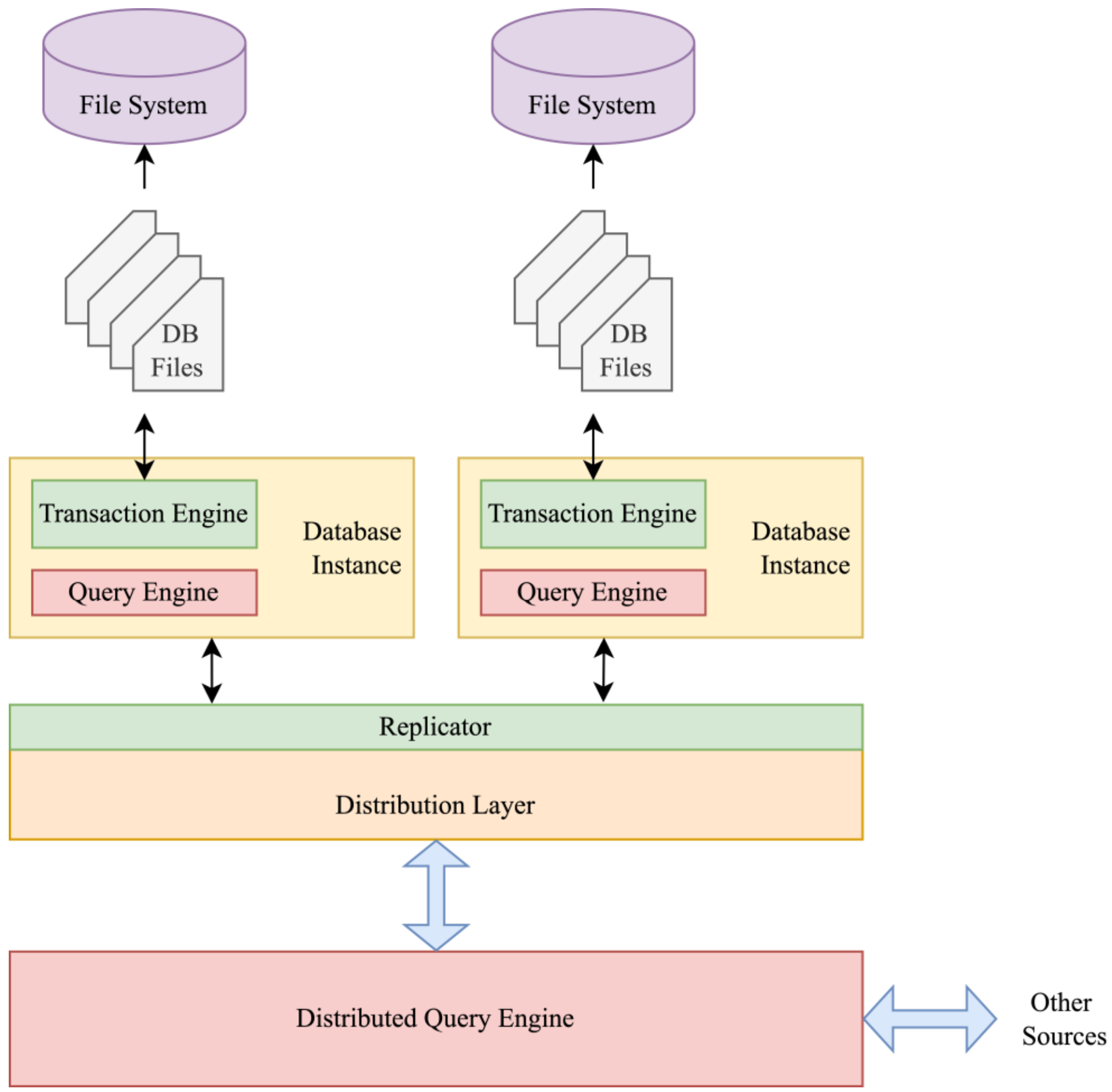

**Figure 59**: Database layers.

Above the core engines and standalone databases, a **distribution layer** can be implemented to distribute data across multiple instances. Amazon Aurora, for instance, distributes data across MySQL instances. Finally, distributed **query engines** handle query processing and can operate across various types of databases. Note that standalone databases can also incorporate query engines to process queries within their own data.

Examples of exceptions are:

- PostgreSQL [107] - Unlike MySQL, PostgreSQL has its own query engine.
- Spanner - Unlike Amazon Aurora, Spanner distributes the core engine itself instead of building on top of standalone engines.

## 13.3. Database Storage Formats

All databases (except in-memory ones) persist data records in files on disk. The way these records are arranged within the files varies, leading to different storage formats.

### *13.3.1. Record I/O (Row-Oriented)*

Record I/O organizes data by storing all column values for a single record (row) contiguously on disk, as shown in Figure 60. The record data might be spread across multiple files. An index structure, commonly a B-tree, facilitates efficient record location. The B-tree structure generally ensures fast index updates during writes.

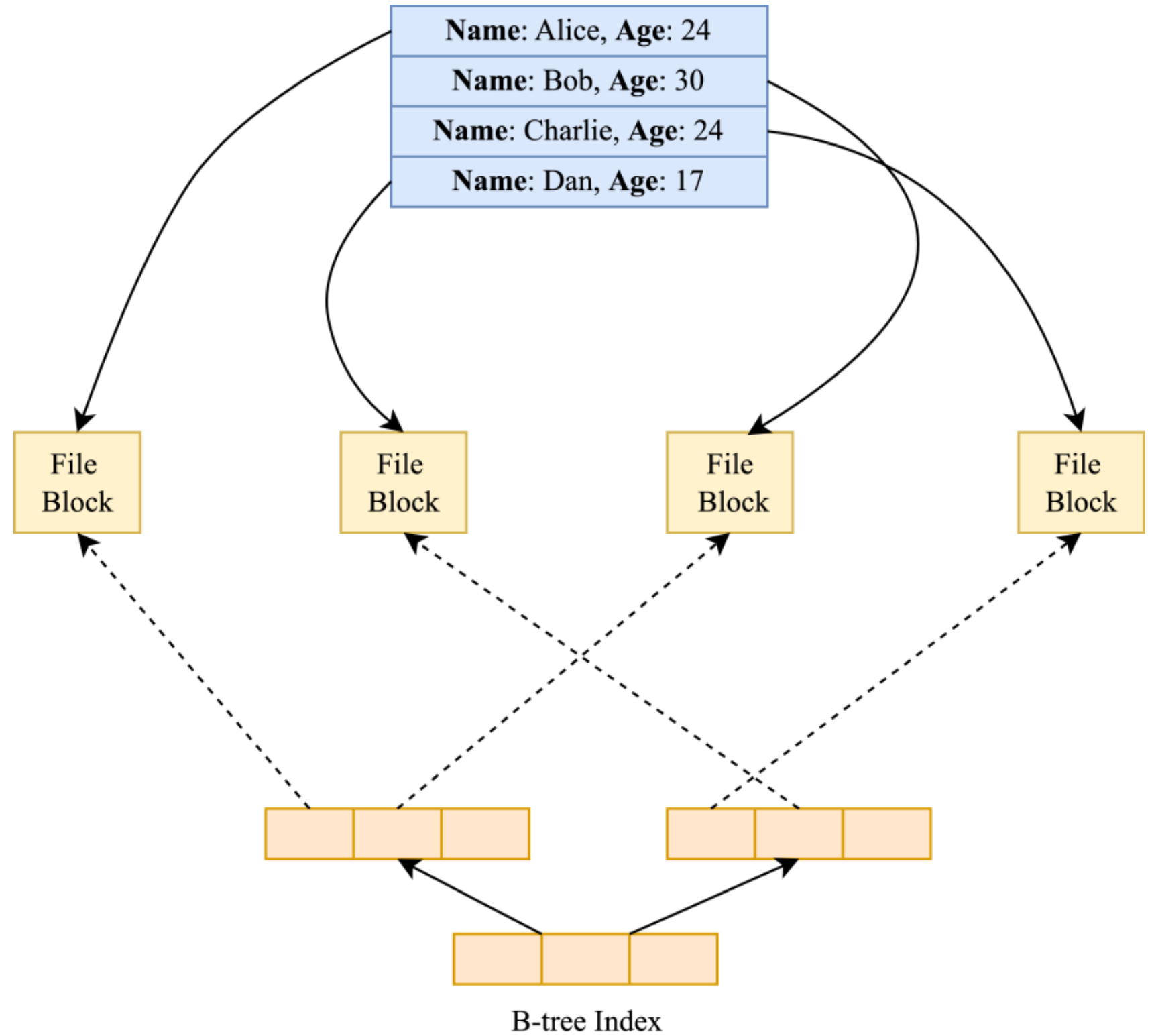

**Figure 60**: Record I/O format.

Record I/O is used in write-heavy databases where write optimization is crucial.

**Example Implementations** - CSV and Apache Avro [108].

**Advantages**

- Efficient retrieval of specific records by directly following the index and reading the record.

**Disadvantages**

- Inefficient for projections, requiring multiple small I/Os to read specific columns.
- Limited compression due to the storage of diverse data types together within a row.

*13.3.2 Columnar (Column-Oriented)*

Introduced in 2005 by Stonebraker et al. in their influential VLDB paper on C-Store [109], columnar storage arranges data by storing all values for a single column contiguously on disk, as shown in Figure 61.

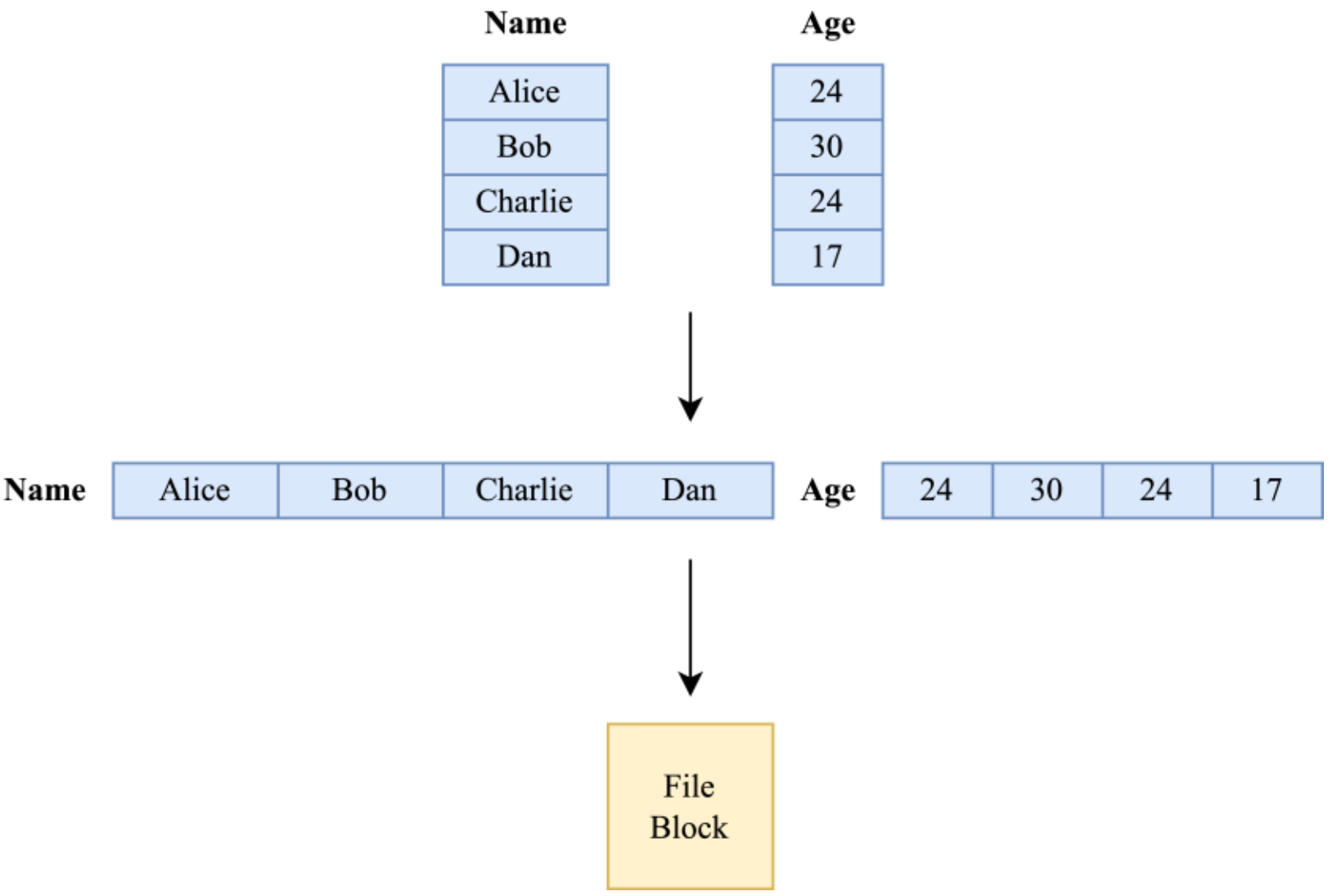

**Figure 61**: Columnar format.

Columnar storage is well-suited for read-heavy databases. It compresses each column individually and relies on header information to define column boundaries and store summary statistics (like key ranges for index columns).

Example Implementations - Apache Parquet [110].

**Advantages**

- High compression ratios are achievable due to the homogeneous data types within each column (e.g., booleans as bit vectors), reducing storage and improving I/O.
- Efficient reading of specific columns.

**Disadvantages**

- Difficult record reads, requiring multiple small I/Os to reconstruct a single row.
- Slower transaction processing.

*13.3.3. Capacitor*

Announced by Google's BigQuery in 2016, the Capacitor [111] format enhances the columnar approach. Its key innovation is splitting records into partitions and then storing these partitions in a columnar fashion, as shown in Figure 62. It also excels at handling structured data efficiently.

BigQuery leverages Capacitor's ability to dynamically reorganize and re-encode data based on usage, optimizing storage and query performance through techniques like run-length and dictionary encoding.

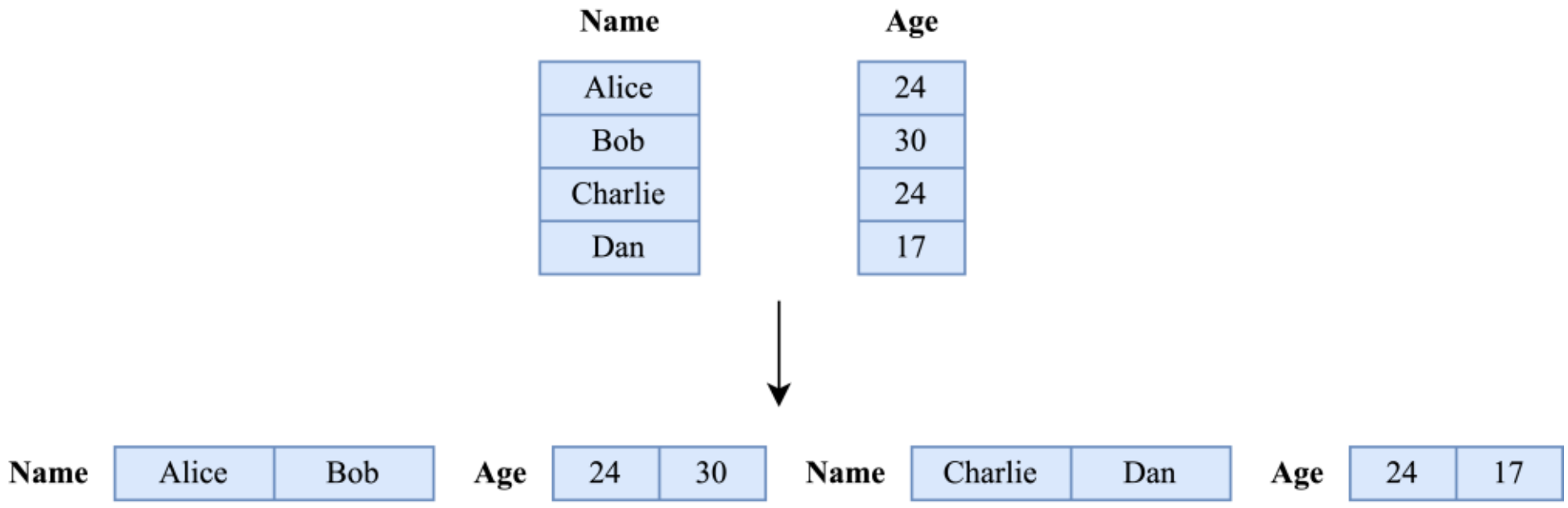

**Figure 62**: Capacitor format.

Another notable format is the **SSTable**, where keys are stored in sorted order. See **14. Bigtable: A Distributed Storage System for Structured Data** to learn about SSTable and the corresponding LSM-tree structure.

### 13.4. ETL Pipelines

ETL pipelines are processes used to **extract, transform, and load** data into data warehouses. Data sources for ETL pipelines typically include other databases, data lakes, or other data warehouses.

Let's take a couple of examples to understand how ETL is used.

**Example 1 (Figure 63)**: Consider an e-commerce platform that processes numerous online orders. To gain insights into sales trends, the business needs an analytical data warehouse that summarizes total revenues on an hourly basis. The raw order data might reside in various sources: transactional SQL databases, periodic batch files from external partners, and even large datasets in diverse formats within a data lake. An ETL pipeline would then be crucial for connecting to these disparate systems, extracting the relevant order information, transforming it (e.g., calculating hourly sums), and loading it into the data warehouse. The pipeline would run every 24 hours and process data for all orders in the last 24 hours.

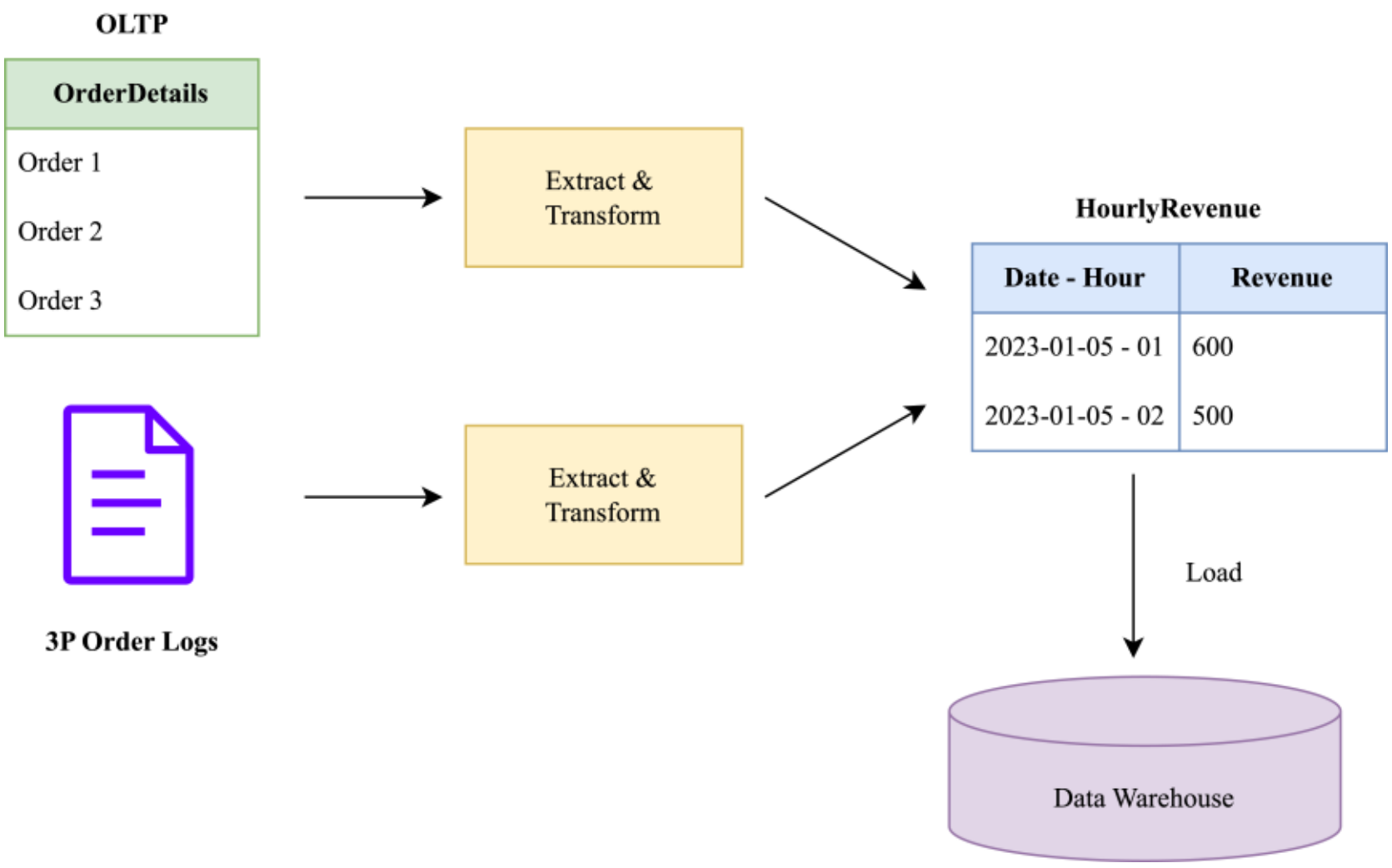

**Figure 63**: ETL for OLTP and logs.

**Example 2 (Figure 64)**: Consider a scenario where a system maintains a detailed analytical table with hourly revenue metrics. To facilitate higher-level analysis and reporting, a daily summary table is also required. An ETL pipeline could be designed to run each day, reading the hourly data from the analytical table, aggregating it to a daily level, and then populating the daily summary table.

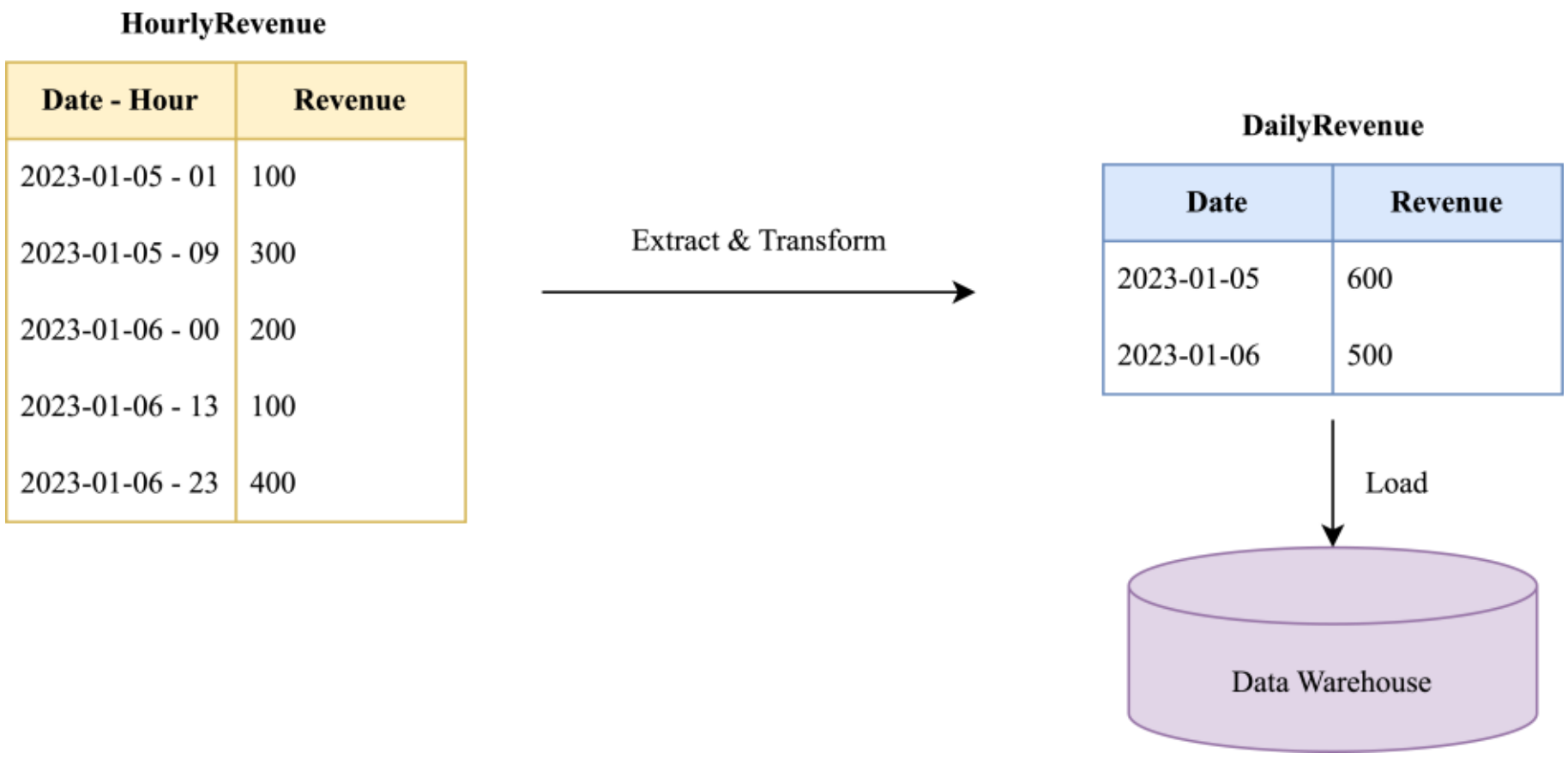

**Figure 64**: ETL for daily aggregation.

**Example 3**: Let's consider a company that gathers customer feedback from various channels, such as online surveys, social media posts, and call center transcripts. To understand overall customer sentiment, an ETL process could extract this textual data, transform it by performing sentiment analysis, and then load the aggregated sentiment scores into a reporting database. This allows the company to track customer satisfaction trends over time.

### 13.5. Cloud Data Store

The paper effectively describes the design of a cloud data store. Due to its key-value nature and simple API, it has seen wider adoption than a fully-fledged distributed file system. Popular examples of cloud data stores include Google Cloud Storage [112], Amazon S3 [58], and Azure Blob Storage [113].

### *13.5.1. Design Points*

- **Key-Value Store with Eventual Consistency**: Cloud data store functions as a key-value store with eventual consistency. Keys resemble file paths (strings) while values can be bytes ranging from a few kilobytes to terabytes.
- **Data Immutability**: In most cloud stores, data is immutable. Appends are possible but generally not optimal.
- **Data Partitioning (Buckets)**: Keys are partitioned into buckets, analogous to directories in a distributed file system.

### *13.5.2. Limited Functionality*

The key-value store is designed for simplicity. Transaction support, especially across multiple keys, is notably missing.

For example, consider a scenario where key **A** has the value 10 in the store. Transaction support would be necessary to atomically read the value of **A** and then write its double (20) back to the store. Without transaction support, another write operation could concurrently modify the value of A to 30 before the first operation completes, leading to an incorrect result (20 instead of 60).

### *13.5.3. Performance Characteristics*

Typical latencies range from 5 to 10ms. Achievable throughput typically falls within the 50-100 MB/s range.

### *13.5.4. Cloud Data Store v/s Distributed File System*

Most cloud applications are composed of microservices, each with distinct *storage* and *compute* components:

- **Storage Layer**: Comprises the underlying storage medium (e.g., disks) where data is persisted and accessed for computation.
- **Compute Layer**: Handles user requests, performs computations. Typically involves a replicated service running continuously.

This architecture is well-suited for low-latency, real-time services, allowing independent scaling of storage and compute resources.

Distributed File Systems have gained widespread popularity since the advent of Google File System (GFS, also known as Colossus [81]). A distributed file system offers a more comprehensive file system interface ideal for database systems. Colossus and Hadoop Distributed File System [83] are well-known examples of distributed file systems.

However, Distributed File Systems have inherent compute overhead. Distributed file systems incorporate a compute layer consisting of *chunk servers* (handling read/write requests) and *metadata servers* (managing metadata updates). These services run constantly, consuming compute resources even when idle.

On the other hand, Cloud Data Stores leverage serverless architectures like Lambda. The compute layer is invoked on-demand, only when data is read or modified. This pay-per-use model offers cost-effectiveness compared to the continuous compute consumption of distributed file systems, databases, or data warehouses.

### 13.6. Delta Lake

In this paper, the authors aim to construct an **ACID table** on top of a cloud data store. This approach can be extended to any eventually consistent key-value store given that the key-value store is not transient.

It is crucial to remember that this creates an ACID table, not an ACID database. Therefore, transactions spanning multiple tables may still violate ACID properties.

#### *13.6.1. Data Storage*

Delta lake stores records in Parquet files which are themselves stored on a cloud data store.

```
table/date1/f1.data
            f2.data
            ...
     /date2/f1.data
            f2.data
            ...
```

#### *13.6.2. Transactions*

To handle data modifications, a log-based approach is employed. When a record is updated, the existing Parquet file is not directly altered. Instead:

- A new Parquet file is created to accommodate the changes.
- The original file is *soft-deleted*, meaning it's marked as inactive but not immediately removed.

Log records maintain a history of additions and deletions. Each log entry corresponds to a specific Parquet file. To enhance system stability and performance, these log records are periodically snapshotted into checkpoints.

To further optimize performance, data files are partitioned. This partitioning strategy improves the efficiency of the aforementioned operations.

#### *13.6.3. Read-Only Transactions*

By default, all reads are performed as snapshot reads. Serializable reads are not natively supported. To achieve serializable read behavior, a workaround is necessary: a dummy write operation must be executed.

#### *13.6.4. Performance*

Due to its current implementation, the system can only process one transaction at a time, significantly limiting its throughput. This limitation may be acceptable if the database is intended for use in environments with low transaction volumes.

#### *13.6.5. Query Processing*

As noted earlier, Parquet files store data in a columnar format. Each Parquet file includes summary statistics for each column, enabling efficient query processing by allowing the system to quickly identify and skip files that do not contain the relevant data. This approach is particularly effective when data is sorted by the queried column.

However, if data records are not sorted based on the query column, performing range queries on that column can become significantly more expensive.

**Use a Secondary Index?**

Building and maintaining a secondary index requires cross-table transactions, which are not supported by Delta Lake.

**Solution: Z-Ordering**

Z-ordering is a technique used to optimize the physical layout of data records within Parquet files.

Instead of sorting data along a single dimension, Z-ordering considers multiple attributes simultaneously. It aims to group together data points that are spatially close to each other in the multi-dimensional space.

Consider a 2D space with points represented by (X, Y) coordinates (shown in Figure 65).

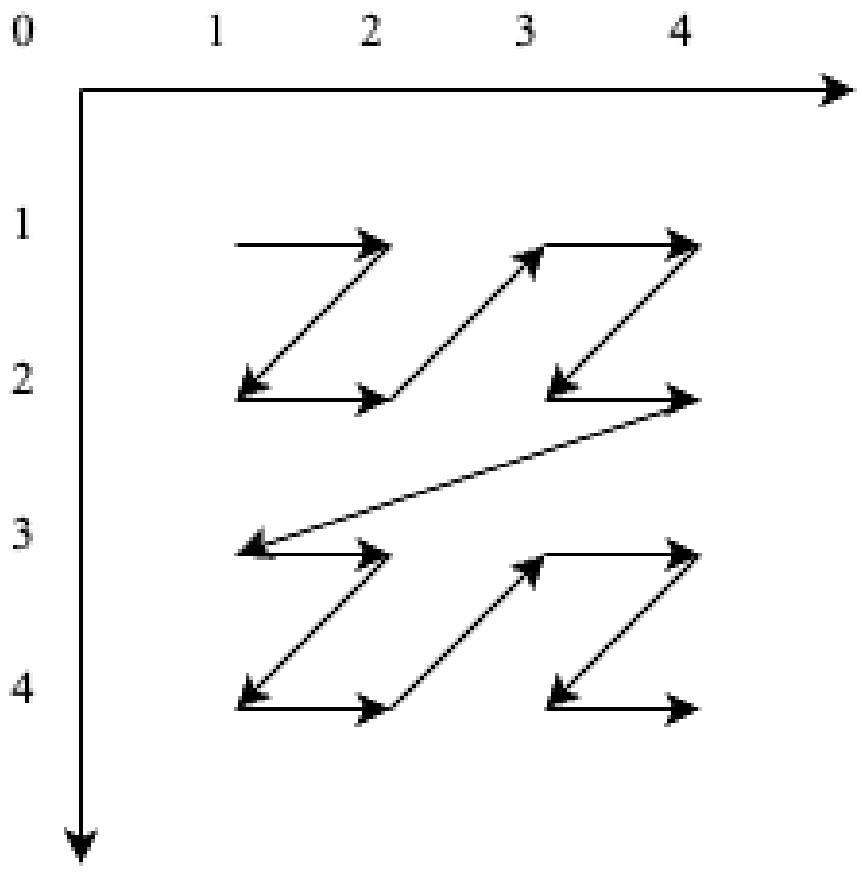

**Figure 65**: Z-order of 2-D coordinates.

A simple Z-order might traverse the space in the following order:

(1, 1), (1, 2), (2, 1), (2, 2), (1, 3), (1, 4), (2, 3), (2, 4),
(3, 1), (3, 2), (4, 1), (4, 2), (3, 3), (3, 4), (4, 3), (4, 4)

Z-ordering significantly reduces the search space when filtering on multiple attributes, leading to faster query execution. By grouping related data together, Z-ordering improves data locality, which can lead to better data compression and faster read/write operations.

Z-ordering can be extended to higher dimensions, making it suitable for complex datasets with multiple attributes.

### 13.7. Alternative Considerations

Alternative approaches to building ACID stores on top of cloud data stores often involve employing a separate transactional store for object metadata, analogous to metadata servers in distributed file systems. However, Delta Lake addresses this challenge by seamlessly integrating metadata management within the object store itself.

### 13.8. Bonus: Related Systems

The paper mentions several key cloud systems and technologies:

#### *13.8.1. Data Warehousing*

- **Google Cloud BigQuery** [84]**:** A data warehouse service built on top of Dremel [85], leveraging Colossus for distributed storage. It offers columnar storage and separates storage from processing for scalability.
- **Amazon Redshift** [114]**:** A columnar-oriented data warehouse similar to BigQuery.
- **Snowflake** [115]**:** A cloud-agnostic data warehouse with a multi-cluster shared architecture.
- **Apache Hive** [99]**:** An open-source data warehouse built on Hadoop, providing a SQL-like query interface.

#### *13.8.2. Columnar Data Formats*

- **Apache Parquet:** A widely used columnar storage format offering efficient compression and encoding schemes.
- **Apache Iceberg** [116]**:** A data format for analytic tables, enabling SQL-based queries across various engines like Spark, Flink, Hive, and Impala.

#### *13.8.3. NoSQL Databases*

- **Apache Kudu** [117]**:** A columnar-oriented data store within the Hadoop ecosystem, optimized for OLAP workloads and utilizing a relational schema.
- **Apache HBase** [118]**:** A non-relational, distributed database supporting wide-column storage, inspired by Google Bigtable.

#### 13.8.4. Query Service

- **Amazon Redshift Spectrum** [119]**:** Enables querying data directly in Amazon S3 without prior loading into Redshift, decoupling storage from processing and optimizing costs.
- **AWS Athena** [120]**:** An interactive query service that allows users to analyze data directly in Amazon S3 using standard SQL.

#### 13.8.5. Competitor

- **Apache Hudi** [121]**:** An open-source framework for managing data lakes, offering features like transactions, upserts, deletes, indexing, clustering, and compaction, positioning it as a direct competitor to Delta Lake.

### 13.9. Paper Remarks

The paper devotes a considerable portion (two pages) to discussing business use cases for Delta Lake, specifically emphasizing the need for millions of partitions. This is a usual characteristic of an industry-led paper that focuses on solving real-world business use cases. While the core concept of the paper seems straightforward, the actual implementation likely presented significant engineering challenges. Overall, the paper provides an excellent introduction to cloud data stores and other relevant big data technologies, making it a recommended read.

# 14. Bigtable: A Distributed Storage System for Structured Data

(14) Chang, F.; Dean, J.; Ghemawat, S.; Hsieh, W. C.; Wallach, D. A. and B., Mike; Chandra, T.; Fikes, A.; Gruber, R. E. Bigtable: A Distributed Storage System for Structured Data. ACM Transactions on Computer Systems (TOCS) 2008, 26 (2), 1–26.

The world of databases has evolved over several decades. We will now explore one of the early distributed databases, which was built by Google for internal use. At the time of its development, this system served as the backend for numerous Google services. While it is still in use today, its design principles have been exceptionally influential for future database architectures.

This paper was presented at the Operating Systems Design and Implementation (OSDI) 2006 conference. Co-authored by luminaries such as Jeffery Dean and Sanjay Ghemawat, with significant contributions from Mike Burrows (the architect of Bigtable and creator of Chubby [122]), this paper has become a landmark in the field. Burrows's work on Chubby was also presented at OSDI 2006, making the conference a pivotal event for distributed systems. Adding to the event's impact, the paper **Rethink the Sync** was also presented and shared the Best Paper award with Bigtable.

In this insight, we will begin with the background concepts that the paper builds upon. We will start with Bloom filters. We will then explore the various data structures employed by databases for efficient reads and writes - specifically, we will dive into Log-Structured Merge trees. Afterward, we will conduct a deep dive into the internals of Bigtable.

### 14.1. Bloom Filters

Bloom filters are probabilistic data structures used in big data pipelines for efficient **set membership** testing. They provide:

- A definitive **NO** for non-membership
- A probable **YES** for membership.

A deterministic full hash set would consume excessive memory. Conversely, bloom filters are memory-efficient by sacrificing the determinism for true positives.

### *14.1.1. Implementation*

A basic implementation uses a bit array. Elements are mapped to array indices via hash functions, setting the corresponding bits to 1 for potential membership.

To minimize false positives, multiple hash functions are employed, as shown in Figure 66. If all corresponding bits are 1 after hashing, the element *might* be a member. If any bit is 0, it's definitely not. The number of hash functions and the size of the bit array affect the false positive rate.

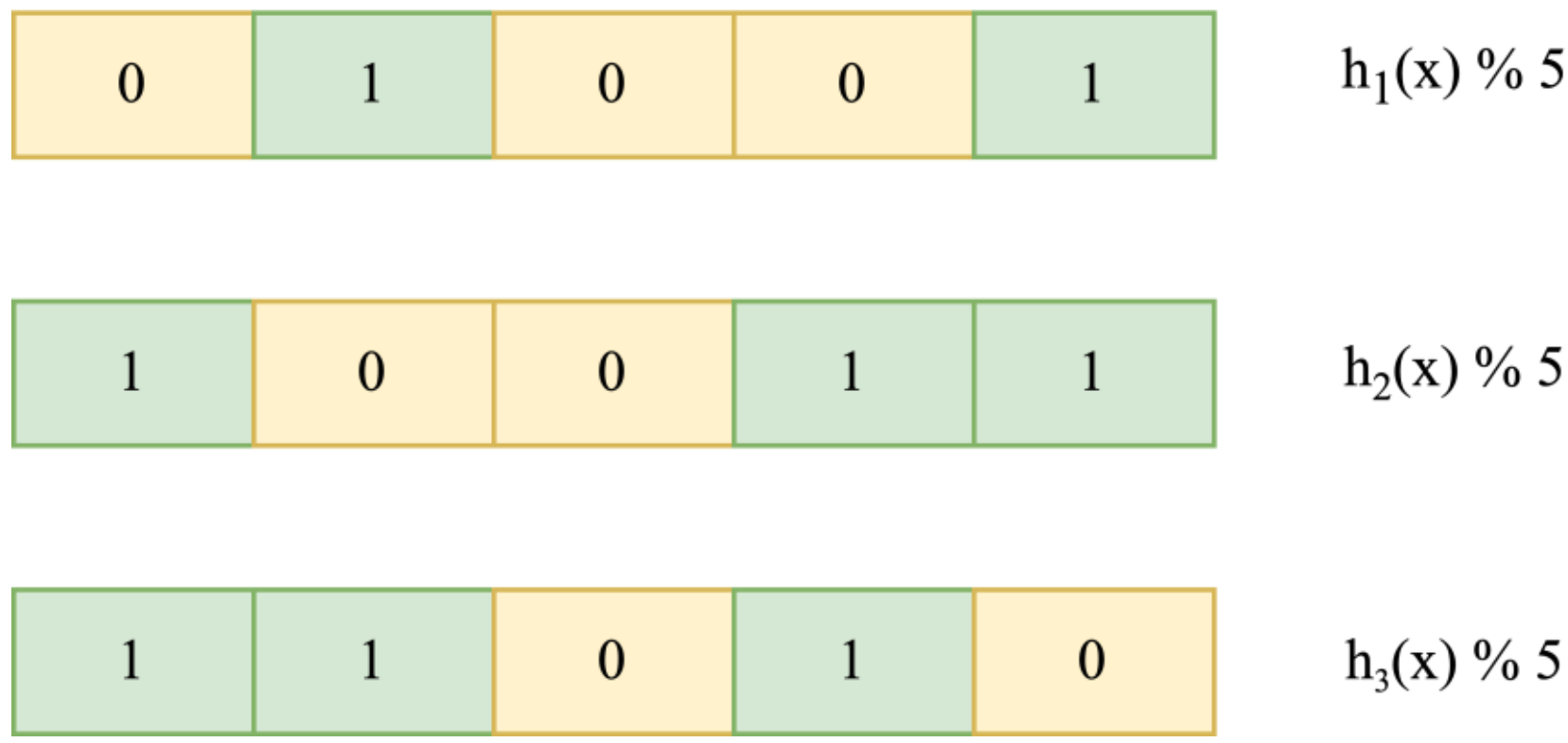

**Figure 66**: Bloom filter.

## 14.2. SQL Databases

SQL databases are a category of databases which are inherently consistent. This implies that data integrity is always upheld. For instance, in a banking database, the cumulative balance across all accounts must remain unchanged at any time regardless of the number of same bank transfer transactions.

To ensure this data consistency (the C in ACID), SQL databases necessitate support for atomic operations (the A in ACID), transaction isolation (the I in ACID), and durable storage (the D in ACID).

### *14.2.1. Classification*

SQL databases can be categorized into two types:

- **Real-time, Strongly Consistent:** Examples include Spanner, AuroraDB, and non-distributed versions of MySQL. These databases guarantee immediate visibility of transaction results, ensuring strong consistency in real-time.

- **Non-real-time, Consistent:** Data warehouses like Mesa/Napa, and Snowflake [123] fall under this category. They maintain consistency, however, they may exhibit some data staleness as all transactions might not have been immediately applied.

**Recommended Reads**:

- **24. Spanner: Google's Globally-Distributed Database**
- **23. Amazon Aurora: Design Considerations for High Throughput Cloud-Native Relational Databases**
- **32. Napa: Powering Scalable Data Warehousing with Robust Query Performance at Google**

#### *14.2.2. Applications*

Both Online Transaction Processing (OLTP) and Online Analytical Processing (OLAP) applications critically rely on SQL databases. NoSQL databases shouldn't be used for transactional or analytical workloads.

### 14.3. NoSQL Databases

Databases that do not adhere to the SQL semantics discussed above are generally categorized as NoSQL databases. These databases are sometimes referred to as BASE databases, contrasting them with ACID databases. They prioritize faster request processing by relaxing some of the stringent ACID guarantees.

For example, Bigtable, a prominent NoSQL database, supports atomicity only for modifications within a single row.

#### *14.3.1. Classification*

NoSQL databases exhibit diverse architectures and can be broadly classified into the following categories:

- **Key-Value Stores:** These are the generalized NoSQL databases, offering a simple key-value data model. Example - Dynamo.
- **Document Databases:** Specialized key-value databases in which the values are structured documents, often in formats like JSON. Examples include MongoDB [106] and CouchBase [124].

- **Wide-Column Databases:** These databases allow for a dynamic number of columns in a table structure. Prominent examples include Cassandra, HBase [118], and Bigtable, all of which are heavily influenced by the data structures and design principles of Google's Bigtable.
- **Graph Databases:** Like Facebook's TAO and Neo4j [125] excel at representing and analyzing relationships within complex networks. They are designed to efficiently store and query highly interconnected data.

**Recommended Reads**:

- **15. Dynamo: Amazon's Highly Available Key-value Store**
- **16. Cassandra - A Decentralized Structured Storage System**
- **28. TAO: Facebook's Distributed Data Store for the Social Graph**

#### *14.3.2. Underlying Model*

The **key-value model** serves as the underlying model for virtually all NoSQL databases. All other NoSQL models are, in fact, built upon this foundational key-value structure.

To illustrate, document databases utilize document IDs as keys and the documents themselves as values. Furthermore, even wide-column databases, such as Bigtable, are internally structured as key-value stores.

#### *14.3.3. Transactions*

Traditional database systems define a transaction as the atomic application of multiple actions. This ensures that a series of operations on multiple data items are treated as a single, indivisible unit, guaranteeing data consistency. However, key-value stores adopt a different approach, limiting transaction support to individual data items only. Consequently, multi-item transactions are not supported.

Document databases, like MongoDB, commonly implement single-row transactions. This is practical, as many real-world application transactions primarily involve operations on single records. For instance, a user updating their profile details within a **Users** table typically modifies only one row.

A **single-data item transaction** can be fundamentally understood as a **read-write operation** on a specific data item. This significantly simplifies the implementation compared to multi-data items transaction systems. This simplicity is a core advantage and selling point of NoSQL databases, which often prioritize ease of use and performance by eschewing complex concurrency control, locking, and rollback mechanisms.

Conversely, **multi-data items transactions**, which are essential for SQL databases, necessitate sophisticated database components. These transactions require robust concurrency control to manage concurrent access, locking mechanisms to prevent data conflicts, and rollback support to maintain data integrity in the event of failures.

Note that certain NoSQL databases, such as Dynamo, Cassandra, and TAO, completely forgo transaction support. These systems rely on **blind writes**, where modifications are applied without any transactional guarantees. That is, a client cannot decide to write a value based on the value it read (as in read-write transactions). This design choice prioritizes high availability and partition tolerance (ALPS principles) at the expense of strong consistency.

## 14.4. Data Structures Behind Databases

All persistent database systems, whether SQL or NoSQL, require durable storage mechanisms. To optimize read and write performance, various data structures are employed, such as **B-Trees** and **Log-Structured Merge Trees**.

### *14.4.1. B-Trees*

B-trees are a prevalent data structure, particularly in SQL databases. In essence, B-trees are self-balancing tree data structures optimized for disk-based storage, crucial for databases and file systems. They allow for efficient search, insertion, and deletion of data by maintaining sorted keys within nodes and minimizing disk I/O through their balanced structure, as shown in Figure 67.

However, updating objects within a B-tree involves significant random I/O, as traversal and modification require accessing non-sequential disk locations.

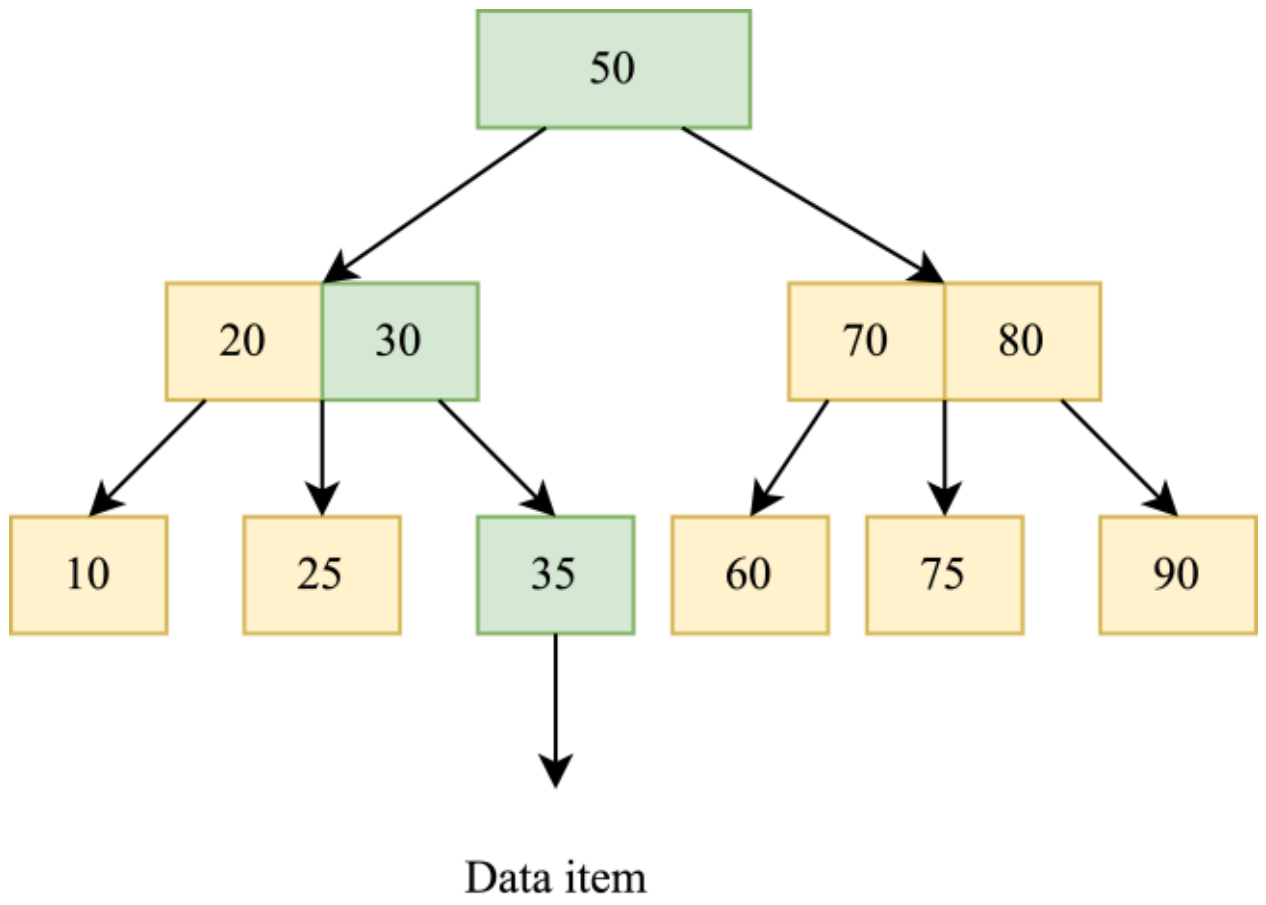

**Figure 67**: B-Trees in databases.

To minimize random disk access, **Write-Ahead Logs (WAL)** are used in SQL databases to buffer transaction writes until commit. Example of log entries:

```
start 100013
write A, 50
write B, 100
commit 100013
```

B-trees remain the preferred data structure for read-heavy databases like MySQL, PostgreSQL [107], and Spanner, as they excel at **range queries**, a common SQL operation. Even popular NoSQL databases like Dynamo, MongoDB (which makes use of the WiredTiger [102] engine) and CouchDB [126] also make use of B-trees.

The embedded database SQLite [127] too makes use of B-trees.

*14.4.2. Log-Structured Merge Trees*

Log-Structured Merge (LSM) trees are widely used in NoSQL databases. Their fundamental components are:

- **In-memory Memtables**
- **On-disk SSTables**

as shown in Figure 68.

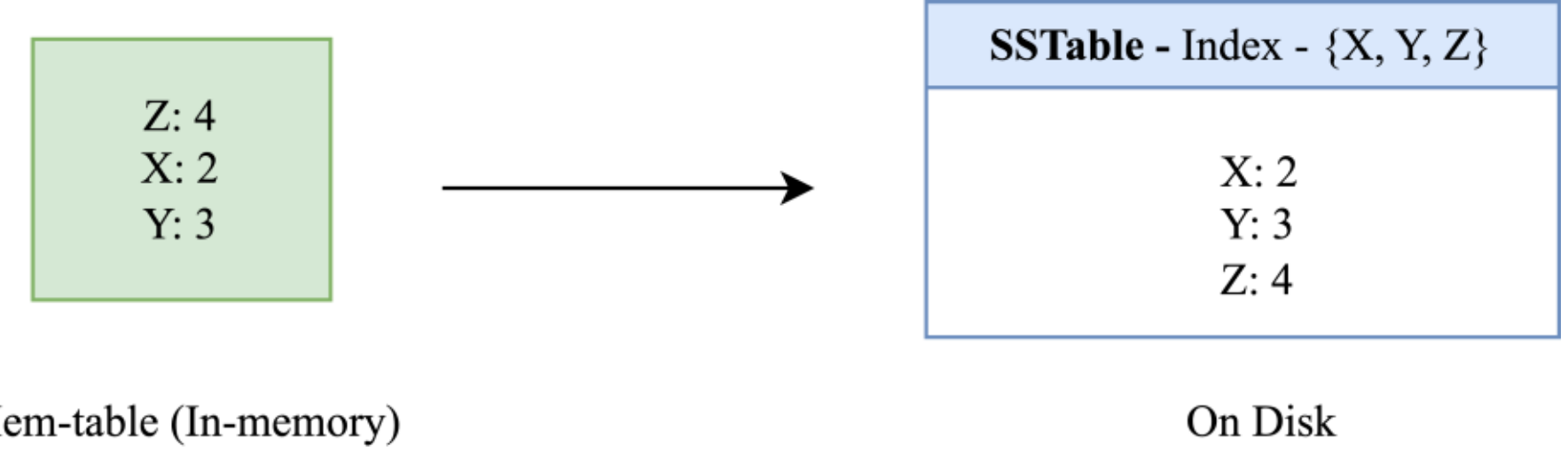

**Figure 68**: LSM tree components.

A **memtable** is often implemented as a hash map. Incoming writes are buffered in memory within the memtable. A write-ahead log ensures memtable durability in case of system crashes. When the memtable reaches a size threshold, it is serialized into an SSTable, and the memtable is cleared.

**SSTables** are file structures that store key-value pairs in sorted order. This sorted organization facilitates efficient indexing and searching. Each SSTable also includes an index to locate keys within the file.

Once materialized, SSTables are immutable. There may be several SSTables. Newer SSTables may contain newer versions of existing keys, effectively overriding older values. Therefore, the effective dataset comprises the current memtable and all SSTables, with newer SSTables taking precedence. Deletes are represented as tombstones.

SSTables are merged through a process called **compaction**, which produces a new, sorted SSTable. The algorithm to merge two SSTable $\mathbf{S_1}$ and $\mathbf{S_2}$ is similar to merging two hashmaps:

- Iterate through $\mathbf{S_1}$ and $\mathbf{S_2}$, adding the key-value pair with the smaller key to $\mathbf{S_{merged}}$.
- If keys are equal, keep the newest value.
- Append remaining elements from either SSTable.

Compaction can create a leveled structure, with larger SSTables at deeper levels. This tree structure is what is known as an **LSM tree**, as shown in Figure 69.

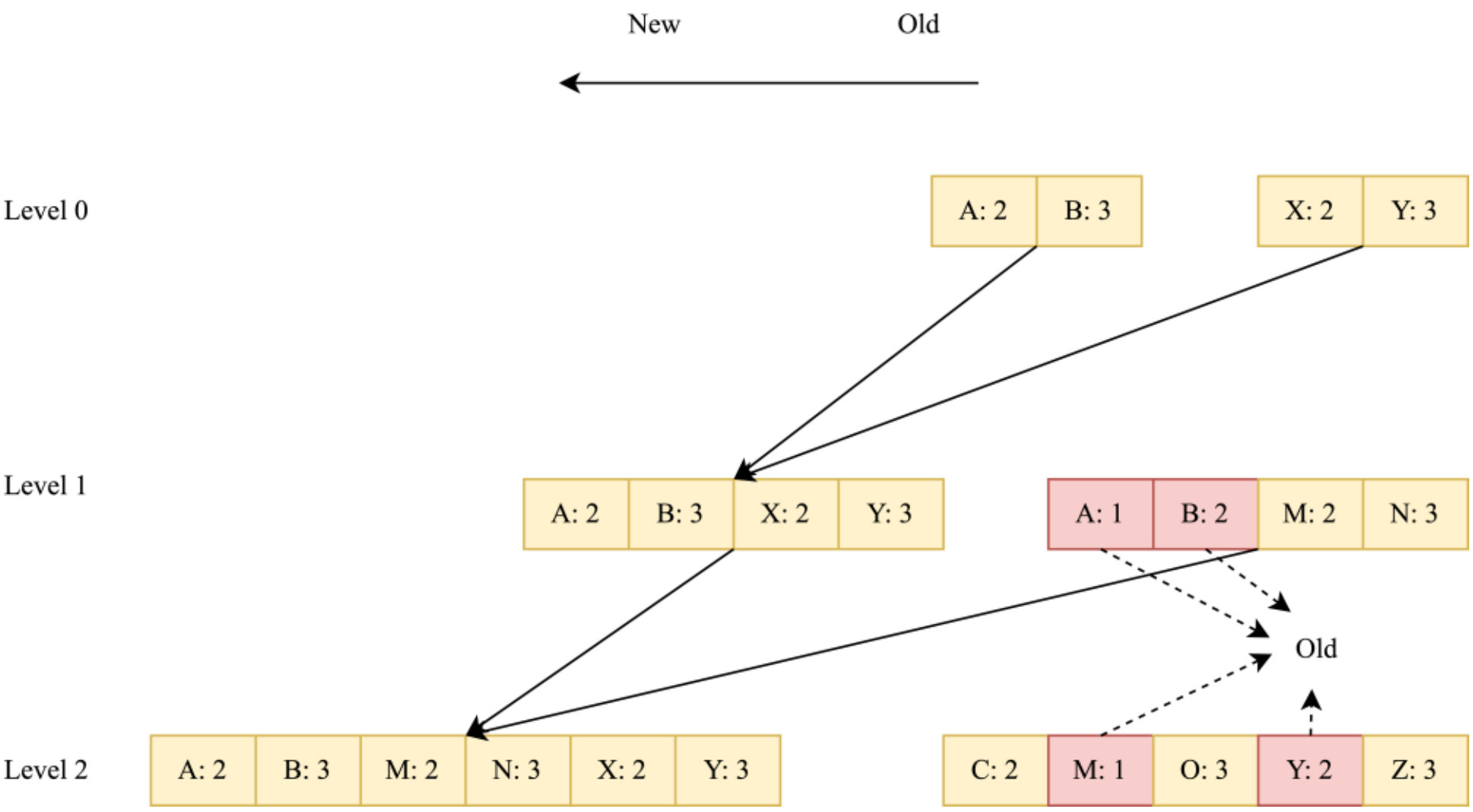

**Figure 69**: An LSM tree.

Unlike B-tree updates, LSM tree writes are sequential as the materialization to SSTable happens in a single shot sequential write. These LSM trees are optimal for write-heavy data storage.

Read performance can be less efficient, as reads may require searching multiple SSTables. Additionally, there is increased disk space consumption due to the storage of older key versions.

To improve read performance in LSM trees, several strategies are employed:

- **Key-Range Caching:** Caching the key ranges of SSTables to eliminate unnecessary searches.
- **Bloom Filters:** To determine whether a key might exist in an SSTable.

LSM trees are well-suited for write-intensive database workloads, as demonstrated by their use in systems like Bigtable, HBase, and Cassandra. Additionally, embedded databases like LevelDB [128] (by Google) and RocksDB [104] (a successor of LevelDB) also make use of LSM trees. Interestingly, all such databases are inspired by Bigtable's design.

### 14.5. Bigtable

Bigtable is a wide-column storage system.

**Q. But what exactly is wide-column storage?**

Unlike traditional relational databases that organize data into rows with predefined columns, wide-column stores organize data into columns where the columns aren't necessarily fixed or predefined. Instead, they can be dynamically added, allowing for flexible schemas and sparse data representation.

In addition to its wide-column nature, Bigtable is also a distributed system, meaning data is spread across multiple servers for scalability and fault tolerance.

Bigtable is used by Google Analytics, Google Search, and Google Earth.

## 14.6. Data Model

As previously discussed, NoSQL databases, at their core, can be viewed as key-value stores. This fundamental principle extends to Bigtable, whose data model can be represented as:

$$< \mathbf{row}: \mathbf{string}, \mathbf{column}: \mathbf{string}, \mathbf{time}: \mathbf{int64} > \rightarrow \mathbf{string}$$

The key component of this model is multifaceted and requires further explanation. Essentially, Bigtable leverages a key-value store to *emulate* a tabular data structure. The **row** and **column** components serve to represent the table's dimensions. The **time** field, however, introduces the capability to store time-versioned data, enabling the retrieval of historical values.

Note that the **time** dimension is flexible and not uniformly applied across all rows and columns. This flexibility distinguishes Bigtable from data warehouses (such as Mesa/Napa) or time-series databases (such as Monarch [129]). A given row and column combination may have a single timestamp, while another may possess millions, resulting in a sparse data structure.

### *14.6.1. Rows*

Row keys are arbitrary strings, up to 64 KB in size. These row keys form the basis for data partitioning, as we will explore later.

*14.6.2. Columns*

Unlike row keys, column keys adhere to a structured format: **family:qualifier**. The **qualifier** component can be an arbitrary string, while the **family** component is predefined and limited. Column keys are grouped into **column families**.

Column families serve multiple purposes:

- They represent the smallest unit of access control.
- They function as a unit of data compression, with data from the same family being compressed together.
- They define the scope for garbage collection settings, allowing for the specification of retention policies (e.g., retaining the last **N** versions or versions within up to a specific time in the past).

*14.6.3. Data Organization*

Bigtable organizes data in a table according to the following principles:

- Data is lexicographically ordered by row keys.
- Data is partitioned into **tablets**, with each tablet containing a range of row keys. These tablets form the basis for data distribution and load balancing.
- Within each tablet, columns are organized by column families, facilitating efficient data compression and retrieval.
- Within a specific **<row, column>** combination, data is ordered by descending timestamp, ensuring that the most recent versions are accessed first.

All these organizational structures are abstractions built upon the underlying key-value store. **The core mental model remains that of a key-value store.**

*14.6.4. Transactions*

Bigtable supports transactions at the row level. This capability is made possible by the fact that Bigtable partitions data by row key, and all keys for a single row reside within the same tablet. A tablet is owned by only one server (described below).

**14.7. Architecture**

Having established an understanding of the Bigtable APIs, let's now examine its architectural design.

Bigtable's architecture features a single **master** server and multiple **tablet servers**. This design paradigm bears a strong resemblance to the Google File System (GFS), reflecting a prevalent architectural preference within Google at the time. Around 2007, the broader industry was exploring decentralized architectures, as exemplified by systems like Dynamo and IPFS.

The master server is responsible for crucial coordination tasks, including assigning tablets to tablet servers, detecting load imbalances, and initiating tablet reassignment. Each tablet server manages a collection of tablets, typically ranging from tens to thousands per server. Client applications communicate directly with tablet servers for read and write operations.

Similar to GFS, Bigtable adopts a per-cluster architecture. Each cluster operates independently, with no inter-cluster communication.

Within each cluster, there are multiple tables. Each table is further subdivided into **tablets**, which represent contiguous ranges of row keys.

#### *14.7.1. Tablet Exclusive Assignment*

A fundamental principle of Bigtable's design is that each tablet must be exclusively owned by a single tablet server. This constraint is essential for ensuring row-level transaction support.

Internally, Bigtable relies on Chubby, a distributed coordination service, to manage and track tablet servers. Chubby's functionality is analogous to ZooKeeper, providing a mechanism to guarantee that, at any given time, only one node can hold ownership of a specific key. While the FLP impossibility theorem highlights the inherent challenges of achieving perfect consensus in distributed systems, Chubby proves effective in practical applications.

Bigtable leverages Chubby to monitor tablet server availability, as shown in Figure 70. Upon startup, a tablet server acquires a lock in Chubby. The Bigtable master server periodically scans these locks to discover active tablet servers. Conversely, when a tablet server fails, its corresponding Chubby lock is released. The master then reassigns the tablet to another available tablet server.

See Chubby's white paper [122] for a deeper understanding of its operation.

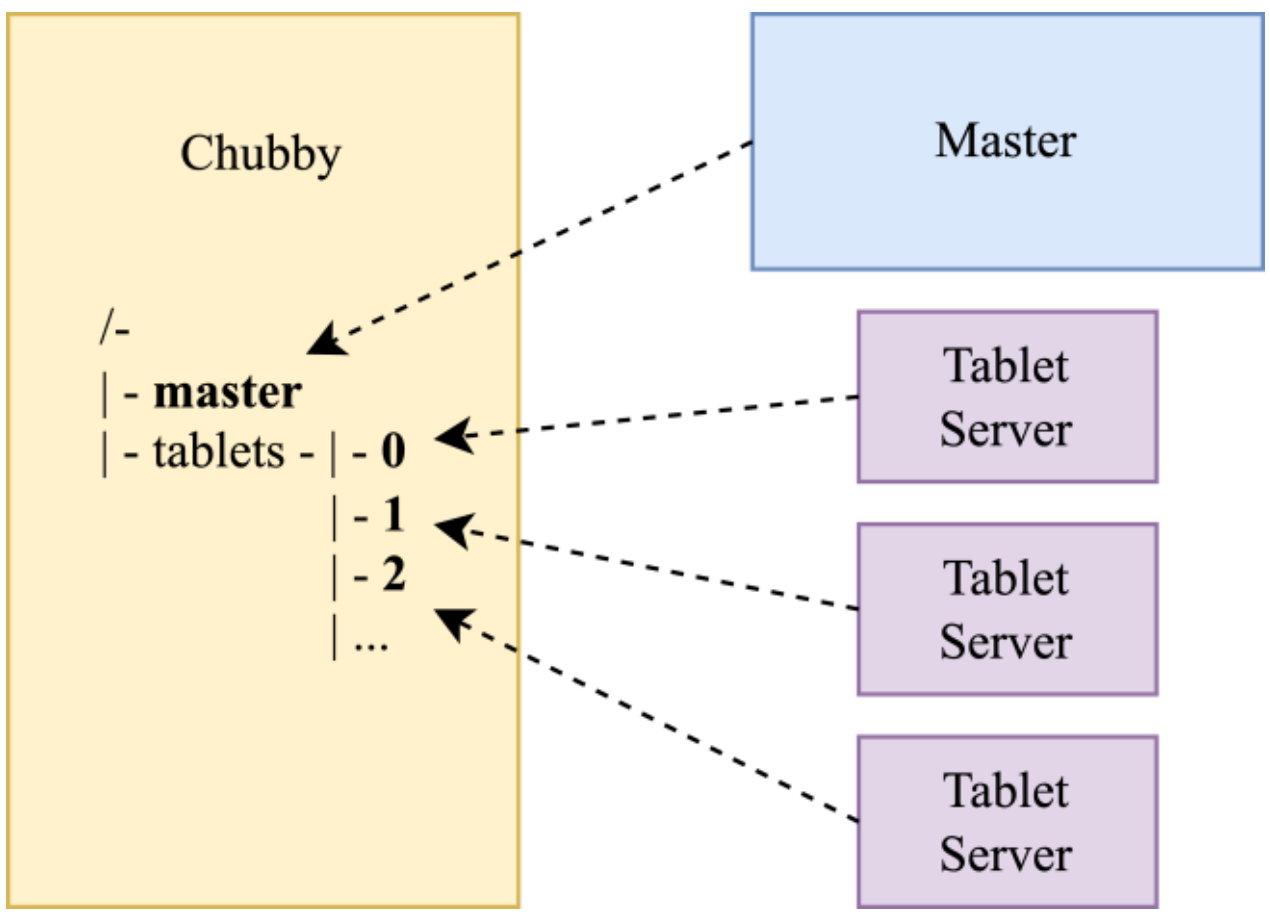

**Figure 70**: Table assignment in Bigtable using Chubby.

The Bigtable master server itself acquires a unique lock within Chubby, ensuring that only one master instance is active at any time. This, combined with the master's ability to accurately track active tablet servers through their Chubby locks, guarantees that each tablet is exclusively owned by a single tablet server.

*14.7.2. Tablet Location Hierarchy*

Bigtable employs a three-level hierarchical structure to locate all tablets within the system, as shown in Figure 4 in the paper:

1. **Root Tablet:** This tablet serves as the entry point for the tablet location system. It stores the location of all METADATA tablets within a special METADATA table.
2. **METADATA Tablets:** These tablets contain the locations of the actual data tablets. Each entry within a METADATA tablet points to the location of a specific row key range. Given that each metadata entry is approximately 1 KB in size, a 128 MB METADATA tablet can store the locations of a substantial number of user tablets.
3. **User Tablets:** These are the actual data tablets that hold the user's table rows.

The location of the root tablet is itself stored within a file in Chubby, providing a reliable and consistent starting point for the tablet location process.

## 14.8. Tablet Internals

In Bigtable, tablets are stored on the GFS, providing a robust foundation for data reliability. This separation of storage and computation allows tablet servers to function primarily as a computing layer. Consequently, if a tablet server fails, it can be readily replaced, and the new server can seamlessly access the data from the point where the previous server left off. This architectural design is known as a **shared-disk architecture**.

### *14.8.1. Splitting & Merging*

Bigtable employs dynamic tablet splitting and merging to manage data distribution. When a tablet's size exceeds a predefined threshold, it is split into two or more smaller tablets. This involves partitioning the row key range into smaller, contiguous segments. Conversely, when tablets become underutilized and their size diminishes, they may be merged to optimize resource usage.

The process of splitting or merging tablets is designed to maintain data availability. Tablets *remain available* during these operations.

Let's examine the tablet splitting process in detail:

1. The tablet server responsible for the large tablet initiates the splitting operation.
2. The tablet server creates new metadata entries for the resulting smaller tablets.
3. The tablet server updates the METADATA table to reflect the new tablet locations.
4. The master server is notified of the split, and updates its information.
5. The new tablets are served to clients, while the original tablet is deleted.

### *14.8.2. Structure*

Each tablet is implemented as an LSM tree, aligning with Bigtable's optimization for write-heavy workloads. Bigtable's read operations are often performed by batch jobs (MapReduce), which can tolerate higher query latencies.

Bigtable employs multiple compaction cycles for tablet's SSTable to manage data and reclaim storage space:

- **<u>Minor Compaction</u>:** Minor compaction involves merging smaller, recently written SSTables into larger SSTables. This process reduces the number of SSTables that need to be read during queries, improving read performance. It also helps to prevent an excessive accumulation of small SSTables, which could lead to increased disk I/O.
- **<u>Major Compaction</u>:** Major compaction involves merging all SSTables within a tablet into a single, consolidated SSTable. This process eliminates deleted data (tombstones) and older versions of data, reclaiming disk space and improving read efficiency.

Bigtable utilizes two levels of caching to enhance read performance:

- **Scan Cache:** Caches key-value pairs retrieved from SSTables.
- **Block Cache**: Caches raw byte blocks of SSTables from GFS.

Bigtable also leverages Bloom filters to efficiently determine the potential presence of keys within SSTables, reducing unnecessary disk reads.

### 14.9. Locality & Compression

Each locality group (a set of related column families) is stored in a separate SSTable. This design offers several advantages:

- Independent reading of locality groups, enhancing throughput.
- Fine-grained tuning of locality groups, enabling features like in-memory storage for frequently accessed data.

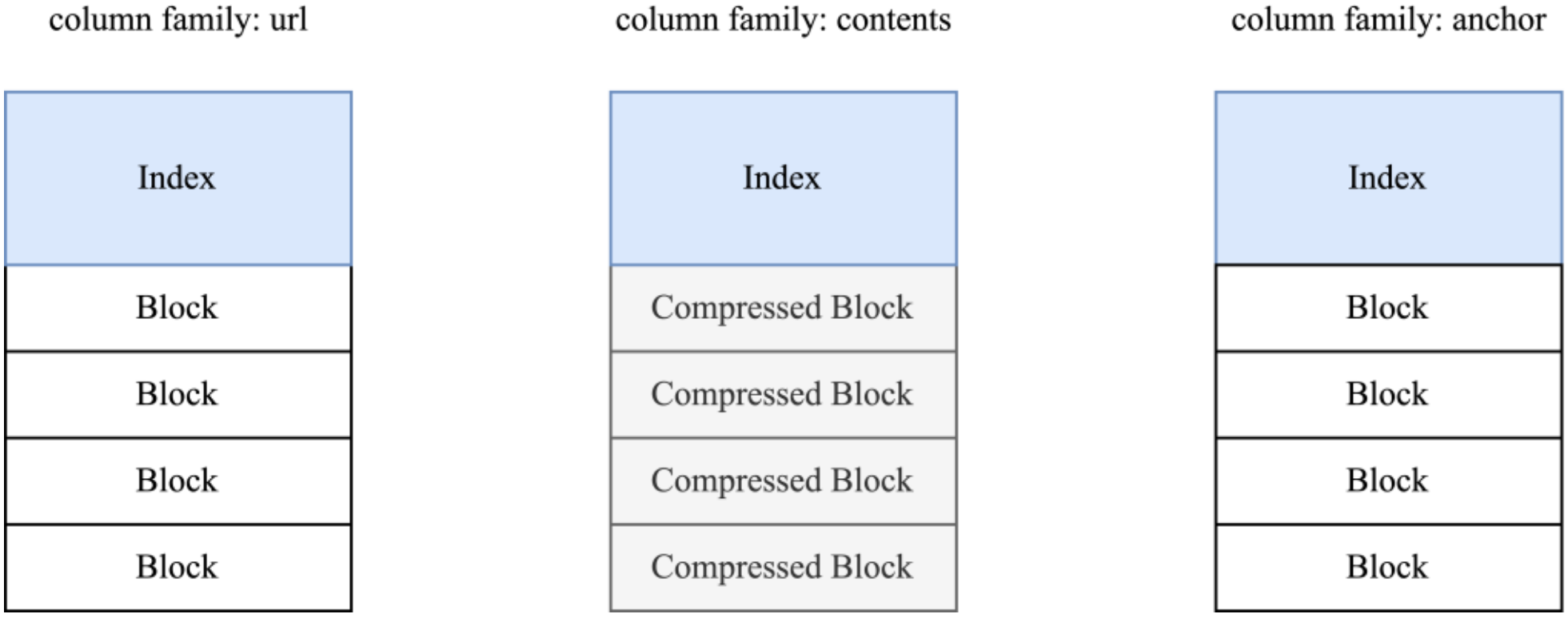

**Figure 71**: Locality groups block organization in Bigtable.

Within each SSTable, keys are divided into blocks, and each block is compressed using a user-defined compression format, as shown in Figure 71. This per-block compression minimizes decompression overhead during single key-value pair reads, as only the relevant block needs to be decompressed.

### 14.10. Commit Log

A single commit log is used for all tablets residing on a tablet server. This approach amortizes the cost of group commits, as multiple transactions across tablets can be written to the log in a single append operation before acknowledgment.

However, the interleaving of tablet logs can pose challenges during tablet server failures. When tablets are reassigned to different servers, each server must read through the previous server's logs to retrieve the entries corresponding to its newly assigned tablets.

To mitigate this, the commit log is externally sorted [130] before being read by the new servers.

### 14.11. Immutability at Play

As Pet Helland emphasized in "Immutability Changes Everything" [72], Bigtable's high write throughput relies on the immutability of SSTable files. This design ensures that once an SSTable is written, only new SSTables are created to reflect subsequent writes, effectively employing a **copy-on-write** approach.

### 14.12. Evaluation

A sign of a good paper is when the performance evaluation calls out both the good and the bad of the design.

**Single Tablet Server Performance:**

- **Random Reads:** Random read performance is notably slow, achieving approximately 1,200 reads per second. The authors argued that this performance was sufficient, as it saturated the NIC bandwidth of the tablet server at the time. However, with modern NICs offering significantly higher speeds (at least 100x faster), it's crucial to re-evaluate the validity of this claim in current environments.

- **Writes (Random and Sequential):** Both random and sequential writes exhibit strong performance, attributed to Bigtable's LSM tree-based design, which prioritizes write throughput.
- **Sequential Reads:** Sequential reads outperform random reads, as expected.

**Scalability:**

- Increasing the number of tablet servers enhances the system's overall throughput.
- However, the scalability is not perfectly linear, due to:
    - Load imbalances across tablet servers.
    - Resource contention with other processes for CPU and network resources.
- Random reads demonstrate poorer scaling, with only a 100x improvement observed for a 500x scale-up.
    - This diminished scaling is attributed to the amplification effect of retrieving 64 KB blocks for each random read, which places significant strain on shared network links.

### 14.13. Paper Remarks

This paper is considered an essential starting point for the study of NoSQL databases. Foundational concepts introduced in this work have significantly influenced the design of later systems, including Cassandra and HBase. Furthermore, the development of embedded databases like LevelDB and RocksDB can be traced back to Bigtable’s use of LSM trees. The paper is easy to follow; in particular, Section 9 is highly recommended for its unique insights into system design lessons.

# 15. Dynamo: Amazon's Highly Available Key-value Store

(15) DeCandia, G.; Hastorun, D.; Jampani, M.; Kakulapati, G.; Lakshman, A.; Pilchin, A.; Sivasubramanian, S.; Vosshall, P.; Vogels, W. Dynamo: Amazon's Highly Available Key-Value Store. ACM SIGOPS operating systems review 2007, 41 (6), 205–220.

In the last chapter, we explored Bigtable. In this chapter, we will examine another NoSQL database developed around the same time: Dynamo. Developed by Amazon, it differs significantly from Bigtable.

This paper, presented at Symposium on Operating Systems Principles (SOSP) 2007, has become a cornerstone in the field of computer systems, profoundly influencing subsequent research and development. The ideas in this paper served as a blueprint for numerous NoSQL databases, including prominent examples like MongoDB [106], Cassandra, and Azure Cosmos DB [131].

In this insight, we will first walk through the basic building blocks of Dynamo, which include Distributed Hash Tables and Consistent Hashing. Next, we will learn how Quorum systems work (Dynamo is a quorum-like system), followed by a discussion on Vector Clocks and their application in resolving data conflicts in Quorum systems. Next, we will see how Merkle Trees can be used to ensure consistency between replicas in a Quorum system. We will then briefly touch on failure detectors, although a more rigorous analysis will be provided in the next chapter. Finally, we will bring all these pieces together to understand the architecture of Dynamo.

### 15.1. Distributed Hash Table

A Distributed Hash Table (DHT) is a decentralized system that provides a lookup service akin to a traditional hash table. Key characteristics of DHTs include:

- **Autonomy and Decentralization:** Nodes operate independently, forming the system without centralized control.
- **Fault Tolerance:** The system remains reliable even when nodes join, leave, or fail.

- **Scalability:** It efficiently handles systems with thousands or millions of nodes.
- **Anonymity:** Participant identities can be hidden.

### *15.1.1. Keyspace Partitioning*

DHTs distribute ownership of the keyspace among participating nodes. Common partitioning methods include:

- **Consistent Hashing:** This widely-used method partitions the keyspace based on the distance $\boldsymbol{\delta}(\mathbf{k_1}, \mathbf{k_2})$ between keys. Each node is assigned an identifier. The node with identifier $\mathbf{i_x}$ owns all keys $\mathbf{k}$ where $\boldsymbol{\delta}(\mathbf{i_x}, \mathbf{k})$ is minimal.
- **Rendezvous Hashing:** All clients employ the same hash function $\mathbf{h}(.)$ to map a key to one of the available servers.
- **Locality-Preserving Hashing:** This method assigns similar keys to similar nodes, enhancing data locality.

### *15.1.2. Overlay Network*

Each node maintains a routing table to direct requests to the node responsible for the specific key. This is called an **overlay network**.

A trade-off exists between routing table size and the number of hops required for a request to reach its destination. Larger routing tables reduce the number of hops.

## 15.2. Consistent Hashing

Consistent Hashing is used to partition a key-space, with node assignments determined by a distance function. This concept exhibits several variations.

### *15.2.1. Modulo Hash*

Assigns key $\mathbf{x}$ to node $\mathbf{H}(\mathbf{x})\ \%\ \mathbf{N}$, where $\mathbf{H}(\mathbf{x})$ is the hash of the key and $\mathbf{N}$ is the number of nodes.

**Problem:** Changes in **N** (e.g., node additions/removals) invalidate all cached values.

*15.2.2. Chord Hash*

Arranges hashed keys on a circular ring, as shown in Figure 72. Each node is responsible for a contiguous arc of the ring.

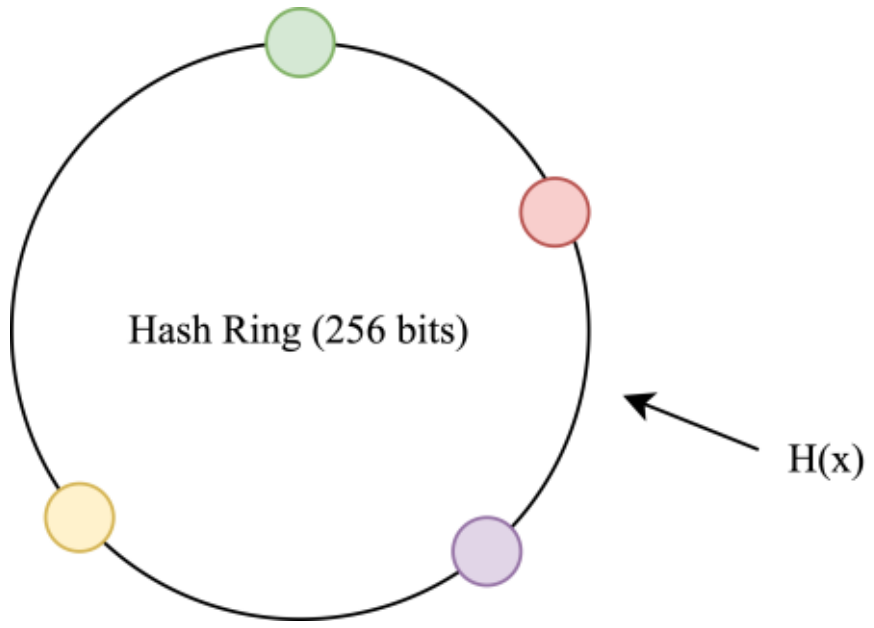

**Figure 72:** Chord hash.

**Advantage:** Node failures only impact data within their specific arc.

*15.2.3. Tree Hash*

Uses the bit values of **H**(**x**) to traverse a tree and locate the responsible node, as shown in Figure 73.

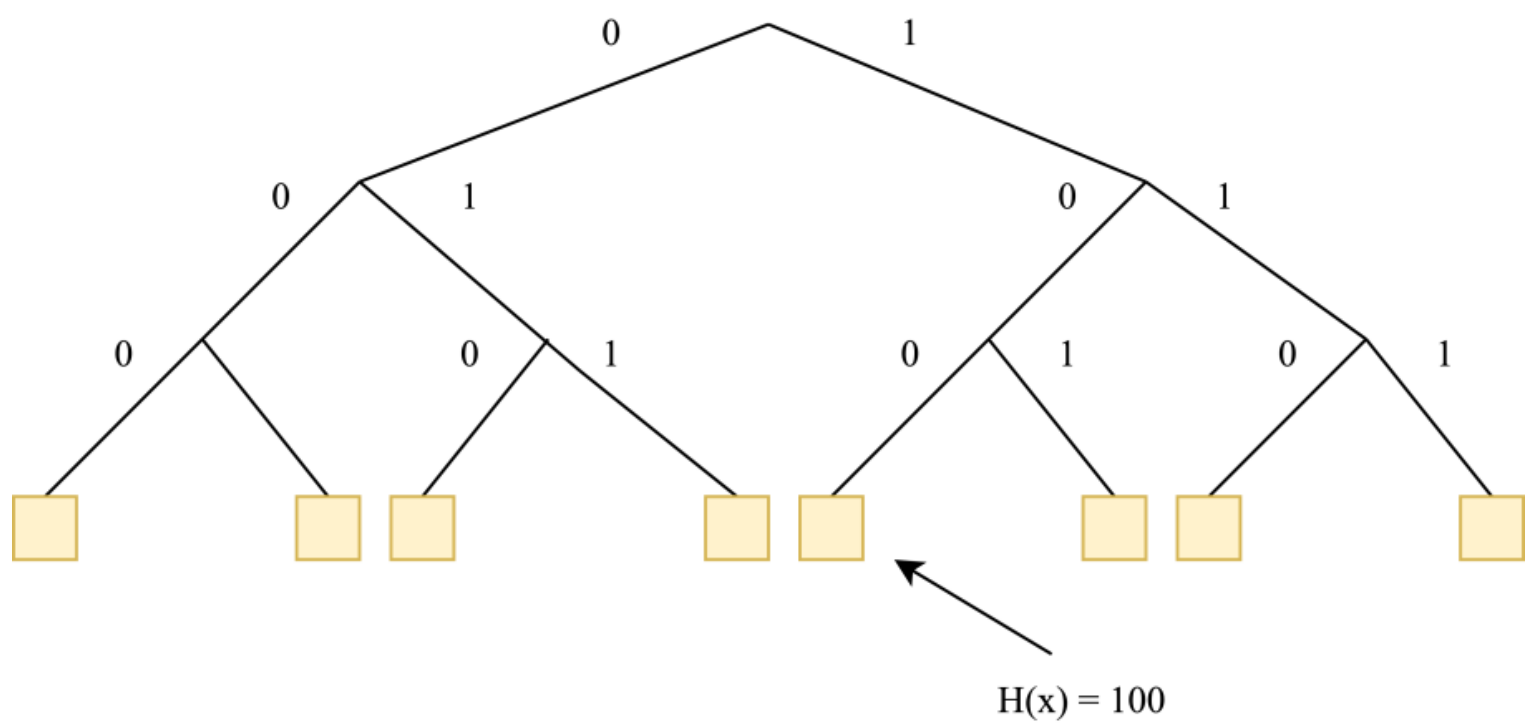

**Figure 73**: Tree hash.

**Challenge:** Node failures require efficient reassignment of the failed node's data.

*15.2.4. Kademlia Hash*

Stores key **x** on the node closest by XOR distance. XOR distance is a function which, for each bit, returns zero if the two bits are equal and one if the two bits are different.

**Flexibility:** Handles node additions/removals by dynamically redistributing data.

### 15.3. Load Distribution in Consistent Hashing

Chord hashing faces a significant challenge with node failures: the loss of a node necessitates moving all its data to the successor node, leading to load imbalance.

#### *15.3.1. Naive Solution: Hash Space Redistribution*

This approach involves adjusting the hash space to evenly distribute data among the remaining nodes.

**Drawback:** This can trigger substantial data movement across the network, impacting performance.

#### *15.3.2. Better Solution: Redundant Node Placement*

Both Chord and Kademlia employ a strategy of placing a node at multiple locations within their respective structures (multiple points in Chord, multiple leaf positions in Kademlia), as shown in Figure 74. This redundancy ensures that upon node failure, the keyspace is more evenly distributed among the remaining nodes, minimizing load imbalance.

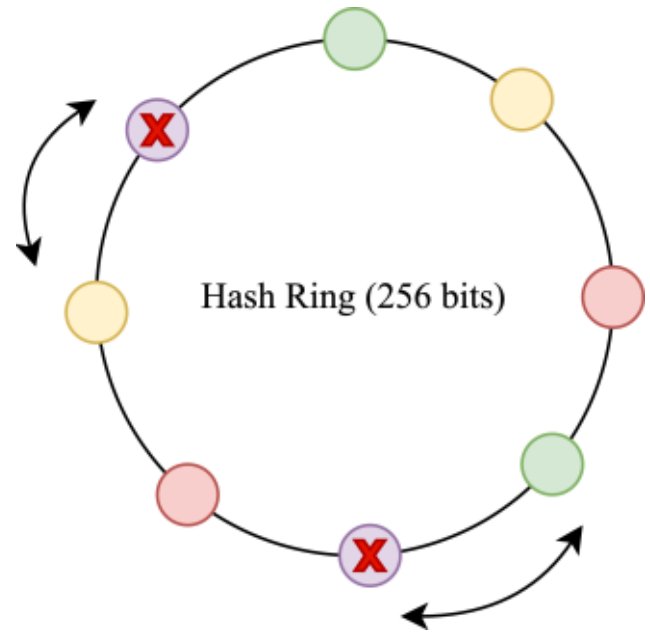

**Figure 74**: Redundant placement in Chord hash.

In Figure 74, the same node is placed at multiple positions on the hash ring. When the purple node goes down, the green and yellow nodes share the portion of the arc owned by the purple node.

**Another Advantage:** Not all machines will have the same capacity. Nodes with higher capacity can be placed at more positions, effectively increasing their share of the keyspace.

## 15.4. Cache Array Reading Protocol

Cache Array Reading Protocol (CARP) is another assignment algorithm that offers good load distribution.

Each machine is assigned a unique hash value: $\mathbf{h}_i$, where **i** ranges from 1 to **N**. For an input **x**, CARP calculates a hash value for each machine:

$$
\begin{aligned}
\mathbf{H_1} &= \mathbf{H(h_1, x)} \\
\mathbf{H_2} &= \mathbf{H(h_2, x)} \\
&\ldots
\end{aligned}
$$

Then, it maps input $x$ to the machine with the largest hash value $\mathbf{H}_i$.

**<u>Pros</u>:** When a machine fails, the load is automatically redistributed. Since the hash function is independent of machine failures, the next machine chosen for a given input will be determined randomly based on its hash value.

**<u>Cons</u>:** Mapping an input to a machine requires calculating **N** hash values, resulting in a linear time complexity of O(N). This can significantly impact performance, especially during high-traffic periods. In contrast, distributed hashing protocols like Chord and Kademlia offer logarithmic time complexity (O(log N)) for finding the responsible node, making them more efficient for large-scale systems.

## 15.5. Routing Algorithm in Consistent Hashing

In decentralized systems with millions of machines, several critical challenges arise:

- **Dynamic Membership:** Nodes frequently join and leave the network.
- **Decentralized Control:** Nodes are owned and operated by independent entities.

### *15.5.1. Challenges of Early Peer-to-Peer Systems*

- **<u>Napster</u>:** Relied on a centralized server to maintain a global index and route requests. This single point of failure made the system vulnerable to disruption.
- **<u>Gnutella</u>:** Employed a decentralized approach where each peer connected to a limited number of neighbors. However, locating resources often involved costly broadcasts to all connected peers, leading to significant overhead.

### 15.5.2. Consistent Hashing as a Solution

Consistent hashing provides an effective mechanism for routing requests in decentralized systems.

### 15.5.3. Routing Algorithm for Chord

Each node is assigned a unique identifier (e.g., a hash value). Each node maintains a "finger table" with O(log N) entries. The i-th entry in node **n**'s finger table points to the node that is $2^{i-1}$ identifiers away from **n** on the ring, as shown in Figure 75.

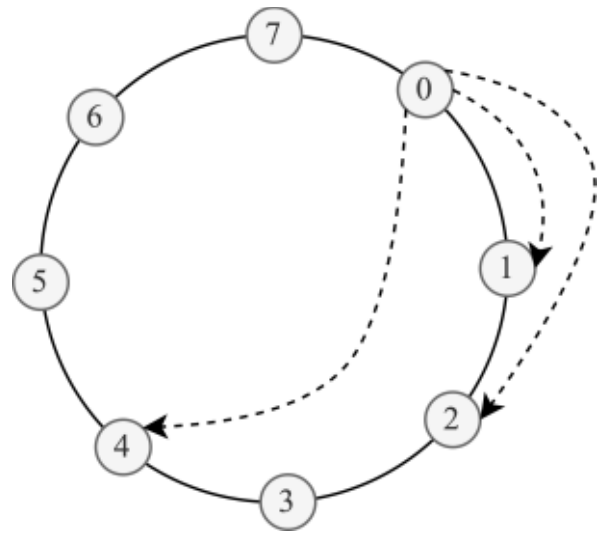

**Figure 75**: Routing table in Chord.

During lookup, the current node examines its finger table to find the closest node to the key's identifier.

### 15.5.4. Routing Algorithm for Kademlia

Each node is assigned a unique identifier (e.g., a hash value).

Each node maintains a routing table with entries based on its identifier's binary representation:

$\mathbf{b}'_1$: Entries for nodes whose first bit differs from the current node.
$\mathbf{b}_1\mathbf{b}'_2$: Entries for nodes whose second bit differs.
$\mathbf{b}_1\mathbf{b}_2\mathbf{b}'_3$: Entries for nodes whose third bit differs, and so on.

When a node receives a request for a specific hash, it checks if it owns the hash. If not, it forwards the request to a neighboring node that is "closer" in terms of their identifier's binary representation (i.e., differs in fewer bits).

The routing table has O(log N) entries. Finding the responsible node requires traversing the routing table, resulting in logarithmic lookup time (O(log N)).

In essence, this algorithm allows nodes to efficiently route requests towards the node responsible for the target hash by iteratively correcting one bit at a time.

*15.5.5. Problem*

In a system with a million nodes, Chord typically requires around 20 hops to locate a specific key. However, the failure of a single node along the routing path can cause the entire lookup to fail.

Kademlia addresses this issue through replication. Kademlia employs **r = 20**, meaning each key is stored on 20 closest nodes. To support replication, each node in Kademlia maintains 20 entries for each prefix of its own ID, each pointing to one of the $\mathbf{k^{th}}$ closest nodes for that prefix.

### 15.6. Koorde Hash

Koorde hash is an algorithm that can achieve constant size routing tables independent of **N**. In tree-based distributed systems, efficient routing can be achieved by strategically placing nodes within the tree structure, as shown in Figure 76.

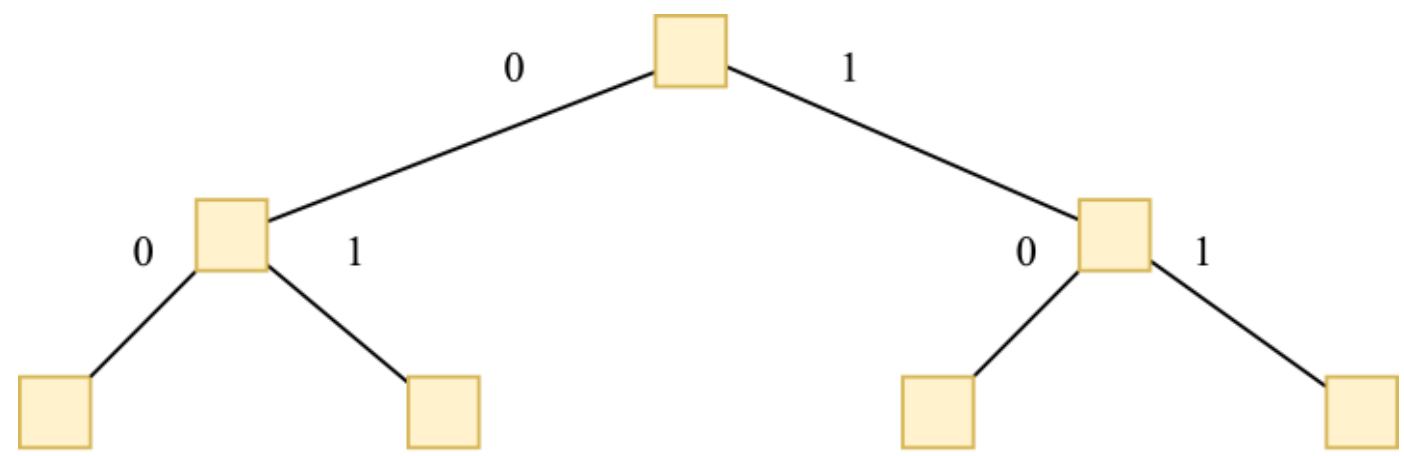

**Figure 76**: Koorde hash node placement.

By placing nodes at every non-leaf position of the tree, each node only needs to maintain pointers to its immediate children.

Each node requires a fixed number of pointers, independent of the total number of nodes in the system.

Finding any node within the tree can be accomplished in logarithmic time (O(log N)).

### 15.7. SQL vs NoSQL Databases

SQL databases are built upon the concept of relational data, SQL databases prioritize strong data integrity guarantees:

- **Atomicity:** Transactions execute as a single, indivisible unit, either succeeding or failing completely.
- **Consistency:** Ensures that all data modifications maintain the integrity of the database schema (e.g., data type constraints, referential integrity).
- **Isolation:** Transactions appear to execute serially, even when occurring concurrently, preventing unexpected interactions. This is one of the most important and distinctive features of SQL databases.
- **Durability:** Committed data persists permanently, even in the event of system failures.

**Examples:** MySQL, PostgreSQL, MS SQL Server.

However the SQL databases are limited by their scalability - it is difficult to distribute data across multiple nodes while maintaining ACID properties. Additionally, vertical scaling is limited by the capacity of individual servers.

NoSQL databases were developed to address the limitations of SQL databases, particularly in terms of scalability and flexibility.

There are different types of NoSQL databases supporting different data models.

- **Key-Value Stores:** (e.g., Redis [132], Memcached [133], DynamoDB [134]) Stores key-value data structures with fast read/write operations.
- **Document Stores:** (e.g., MongoDB, CouchDB [126]) Stores value as document-like structures.
- **Wide-Column Stores:** (e.g., Cassandra, HBase [118]) Stores datasets with sparse attributes.
- **Graph Databases:** (e.g., TAO, Neo4j [125]) Stores objects and relationships between them.

They have relaxed consistency models, often prioritizing **latency, performance, and availability (ALPS)** over strong consistency. Data will eventually be consistent across the system, but there may be temporary inconsistencies. Many NoSQL databases offer limited transaction support. Key-value stores typically support atomic operations on single keys. Some, like MongoDB, provide limited support for multi-document transactions.

NoSQL databases are inherently designed for distributed environments, enabling easier horizontal scaling. They utilize simpler consistency protocols (e.g., Dynamo's vector clocks) compared to more complex protocols like Paxos/2PC required by some SQL databases.

### 15.8. Quorum Systems

Quorum systems are used to ensure that multiple nodes in a system agree on a decision or state. Quorum systems ensure that operations on a single key appear to occur in a specific order. However, they are not suitable for linearizability, which guarantees that concurrent transactions appear to execute in real-time serial order.

#### *15.8.1. Transactions in Quorum Systems*

Transactions within a quorum system exhibit the following characteristics:

- **Atomicity:** Transactions either complete successfully or fail entirely.
- **Per-Key Sequential Consistency:** Operations on individual keys are executed in a well-defined order.

To enable transactions, a two-phase commit protocol is typically employed:

1. **Write Phase:**
    a. The value for the key is written to a specific number of replicas (C).
    b. Each replica provides a commit vote, essentially locking the key for other writes.
2. **Commit Phase:**
    a. Once enough votes are collected, the transaction is committed.

Transactions are useful in quorums but are not usually used. Maintaining consistent states across all replicas can be resource-intensive.

#### *15.8.2. Voting Protocol for Quorums*

The most common way of using quorum systems involves voting. In this scenario, atomicity of operations may not be guaranteed, but with proper voting configuration, the system can appear per-key sequentially consistent to the client, even if individual replicas are in inconsistent states.

Given **N** servers, **W** (write quorum), and **R** (read quorum), the system can achieve per-key sequential consistency if:

$$\mathbf{R + W > N}$$

Consider a system with three servers (A, B, C), where R = 2 and W = 2 (shown in Figure 77). Say, a write operation succeeds on server A but fails on the others.

- The overall write operation fails. However, because this action is non-atomic, node A is left in an inconsistent state.
- The client can read data from replicas A and B. Upon detecting conflicting values from A and B, it will read from C, receiving a version the same as A.
- Server A can then repair its state by fetching the correct value from another server.

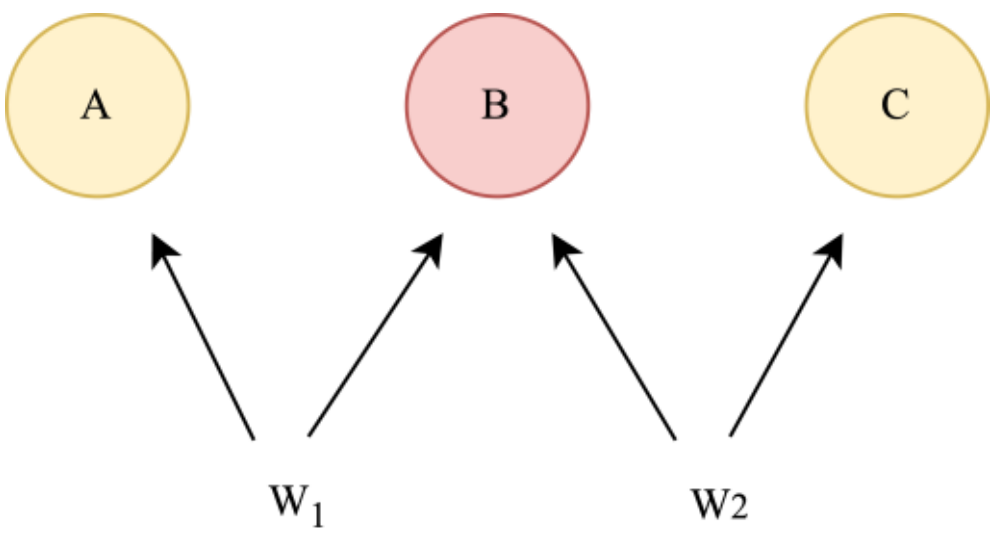

**Figure 77**: Quorum voting example.

Per-key sequential consistency cannot be guaranteed when $\mathbf{R + W \leq N}$. In such cases, conflicts can arise.

Say a write $\mathbf{W_1}$ succeeded at replica A and B and a write $\mathbf{W_2}$ to the same key succeeded at replica B and C. Since B had observed both writes, it can break ties. But if B is down, then there would be a conflict. In this case, W = 2 but R = 1 and hence R + W = 3 <= N.

## 15.9. Vector Clock

Vector Clocks are a powerful technique for establishing causal order among events in a distributed system. They enhance Lamport's clock [135] by maintaining a separate logical clock for each node within the system.

In an N-node system, a vector clock is an N-dimensional vector, where each element represents the logical clock of a corresponding node.

### *15.9.1. Event Handling*

- **Local Event**: Increment the local node's clock (i.e., the i-th element of the vector).

$$\mathbf{V[i] = V[i] + 1}$$

- **Send Event**: Increment the local node's clock. Transmit the current vector clock ($\mathbf{V[1...N]}$) along with the message.
- **Receive Event**:
  - Increment the local node's clock.

$$\mathbf{V[i] = V[i] + 1}$$

  - Update each element of the local vector by taking the maximum of the corresponding value in the received message and the current local value.

$$\mathbf{V[j] = max(V_{msg}[j], V[j]) \ \forall \ j \neq i}$$

### *15.9.2. Comparing Vector Clocks*

- **<u>Equality</u>:** Two vector clocks, $\mathbf{V_1}$ and $\mathbf{V_2}$, are equal if and only if all their corresponding elements are equal.

$$\mathbf{V_1 = V_2 \ iff \ V_1[i] = V_2[i] \ \forall \ i \in \{1, ..., N\}}$$

- **<u>Less Than or Equal To</u>:** Vector clock $\mathbf{V_1}$ is less than or equal to $\mathbf{V_2}$ if and only if every element of $\mathbf{V_1}$ is less than or equal to the corresponding element in $\mathbf{V_2}$.

$$\mathbf{V_1 \leq V_2 \ iff \ V_1[i] \leq V_2[i] \ \forall \ i \in \{1, ..., N\}}$$

## 15.10. Vector Clocks in Quorum System

While quorum systems with $\mathbf{R} + \mathbf{W} \leq \mathbf{N}$ cannot guarantee linearizability, vector clocks can be employed to effectively resolve conflicts and facilitate eventual reconciliation.

### *15.10.1. Approach*

- Each node maintains and updates its vector clock according to the rules described earlier.
- When a write operation occurs on a key, the current state of the vector clock is associated with the written value.

Example: **<key, value, <A: 1, B: 2>>** where **<A: 1, B: 2>** represents the vector clock at the time of the write.

### *15.10.2. Conflict Resolution*

- **Partial Ordering:** Vector clocks enable partial ordering of values.

$$\mathbf{V_1 \leq V_2 \text{ iff } V_1[i] \leq V_2[i] \ \forall \text{ nodes } i}$$

- **Conflict Resolution Strategies:**
    - Choose a value randomly.
    - Select the value associated with the most recent physical timestamp.

Neither of these conflict resolution strategies guarantees linearizability.

## 15.11. Anti-Entropy using Merkle Trees

Say, a set of replicas own a keyspace. Maintaining consistency among these replicas is crucial, a concept often referred to as anti-entropy.

Merkle trees offer a robust mechanism for detecting conflicts and efficiently updating outdated values within this keyspace. This is particularly valuable in quorum systems, where replica inconsistencies can frequently arise.

A merkle tree (shown in Figure 78) is a binary tree-based data structure where each leaf node holds the hash of some items. Internal nodes, in turn, hold the hash of their respective child nodes. In this case, the leaf nodes hold the hash of values of all the keys. The hierarchical structure enables efficient verification of data integrity and facilitates the identification of discrepancies between replicas.

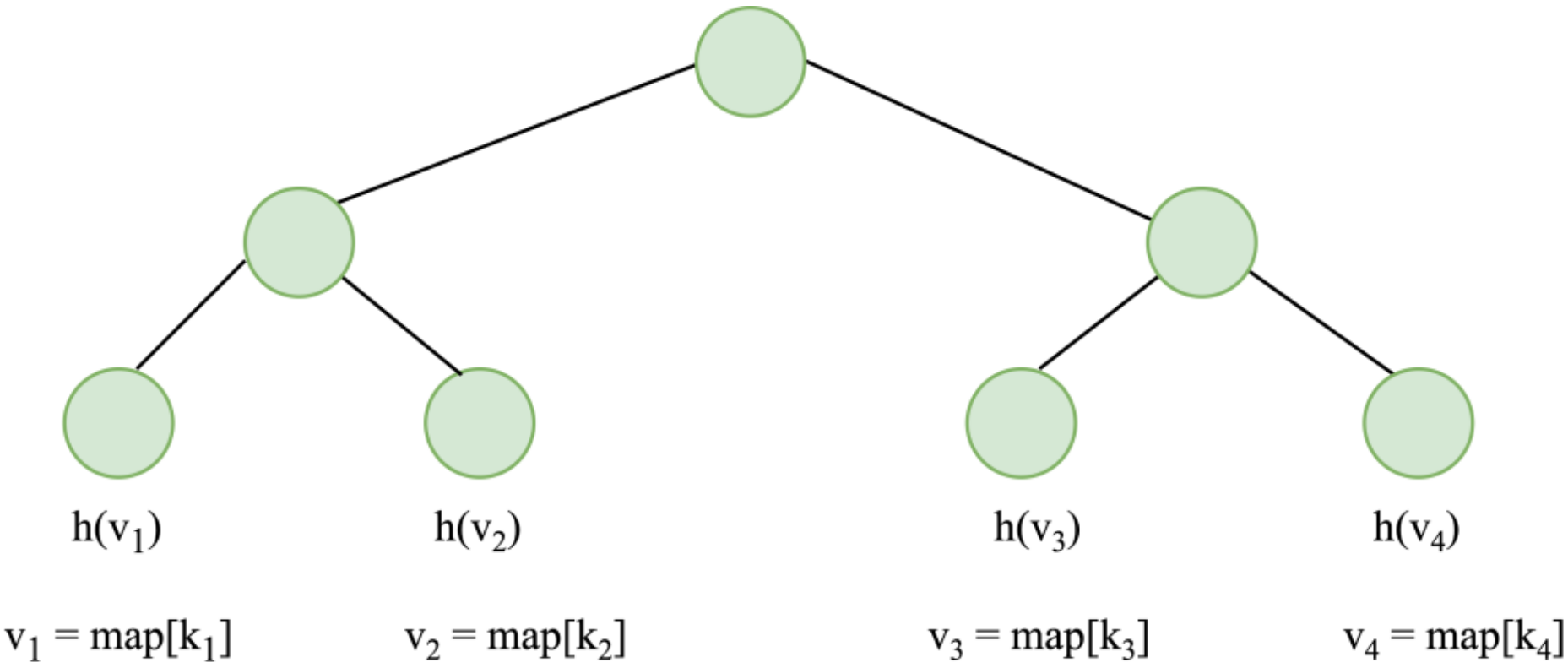

**Figure 78**: Merkle tree.

To compare the merkle trees of a keyspace across two replicas, we initiate a top-down traversal, starting at the root node, to find the keys which are not synced, as shown in Figure 79.

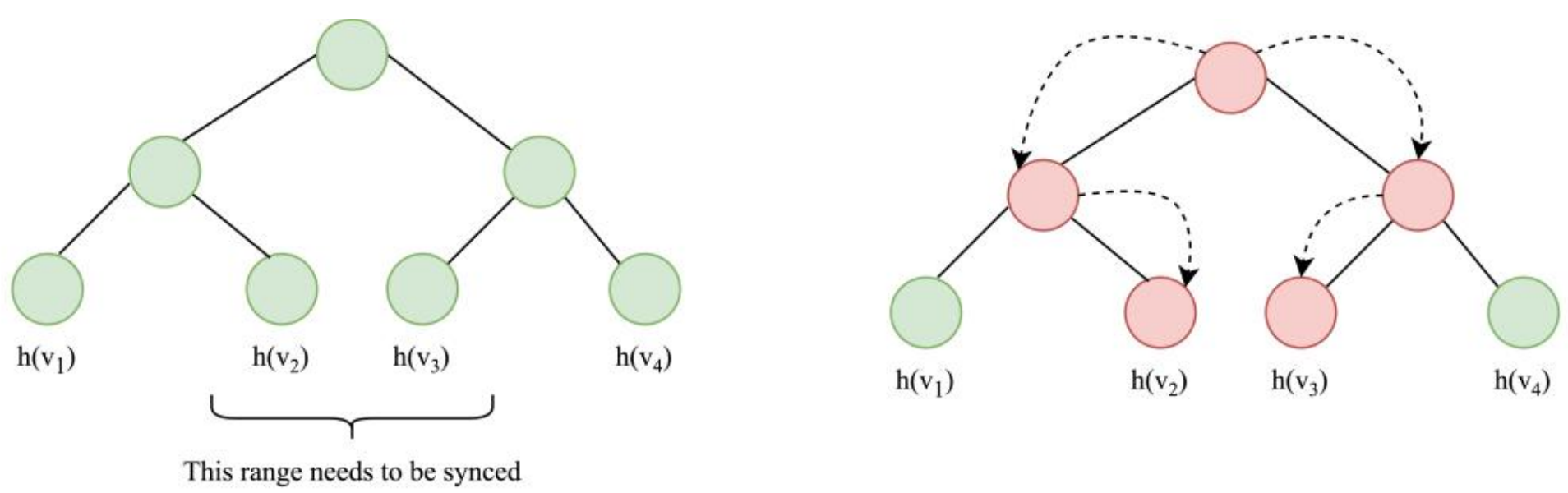

**Figure 79**: Keyspace sync using a Merkle tree.

### 15.12. Failure Detectors

Refer **16. Cassandra - A Decentralized Structured Storage System** for detailed description of failure detectors and gossip protocols.

Failure detectors are components that help in identification of faulty nodes in a distributed system. It is impossible to have a perfect failure detector. Consequently, practical implementations rely on imperfect approximations. Heartbeat-based detectors are a prominent example that leverages gossiped heartbeat information among nodes.

### 15.13. Dynamo

Dynamo, a pioneering key-value NoSQL database, has significantly influenced the landscape of modern NoSQL systems. A key innovation of Dynamo lies in its deliberate trade-off of strict consistency for achieving low-latency and high availability.

Dynamo is a critical dependency of numerous other Amazon services on Dynamo. Hence, the system prioritizes stringent tail latency requirements. Specifically, it focuses on achieving high performance for the 99.9th percentile of tail latency. This emphasis stems from the understanding that in many scenarios, an incorrect response is preferable to experiencing no response or delayed response. The additive nature of latency SLAs across interdependent systems underscores the potential for severe cascading failures if even a small fraction of requests experience prolonged latencies.

### 15.14. The API

Following the design principles of many successful distributed systems, Dynamo boasts a remarkably simple API:

- **Get(key) → (context, list of values)**: Retrieves the value(s) associated with the given key, along with contextual information.
- **Put(key, context, value) → void**: Stores the specified value under the given key, utilizing the provided context for conflict resolution.

The context parameter in these operations is an opaque byte string. This context information is crucial for resolving write conflicts using a vector clock mechanism.

### 15.15. Consistent Hashing in Dynamo

Dynamo was developed in 2007, during a period of intense interest in decentralized systems. Reflecting this trend, Dynamo employs a peer-to-peer architecture where all nodes are identical and utilize consistent hashing for efficient overlay network formation. This design adheres to the principles of incremental scalability, decentralization, symmetry, and heterogeneity.

Traditional consistent hashing presents challenges, such as key redistribution when a new node joins, potentially triggering expensive merkle tree recomputations across multiple nodes. That is why, Dynamo employs a variant of consistent hashing.

#### *15.15.1. Dynamo's Strategy*

- The keyspace is divided into $\mathbf{Q}$ equally sized partitions.
- If the system has $\mathbf{S}$ nodes, $\mathbf{Q/S}$ tokens are assigned to the nodes. The number of tokens are assigned based on the physical capacity of a node.

This approach offers several advantages:

- When a node leaves, only the new nodes responsible for its partitions need to recompute merkle trees, as the partitions themselves remain fixed.
- Partitioning (defined by $\mathbf{Q}$) and partition placement are decoupled.
- The routing table maintains a constant number of entries ($\mathbf{Q}$).

Additionally, we define a fairness metric as follows:

$$\textbf{Load balancing efficiency} = \frac{\textbf{mean loan}}{\textbf{max load}}$$

A node is considered out of balance if its load exceeds 1.15 times the mean load. Identifying and addressing unbalanced nodes is critical, as they significantly impact tail latencies.

Note, the same physical machines may end up at multiple positions on the ring, also known as **virtual nodes**.

### 15.16. Handling Client Operations

Each key **k** is assigned to a coordinator node. This coordinator node is responsible for replicating the keys that fall within its assigned range.

For redundancy, the coordinator replicates the key to $\mathbf{N - 1}$ successor nodes within the ring. This results in a system where each node is responsible for the region of the ring between itself and its **N**th predecessor.

The list of nodes responsible for storing a particular key is termed the preference list. Since each physical node may be responsible for multiple ranges on the ring, the preference list strategically skips positions on the ring to ensure that it only contains distinct physical nodes.

For example, in Figure 80, the key is placed on node 0 (the coordinator), 1, and 2.

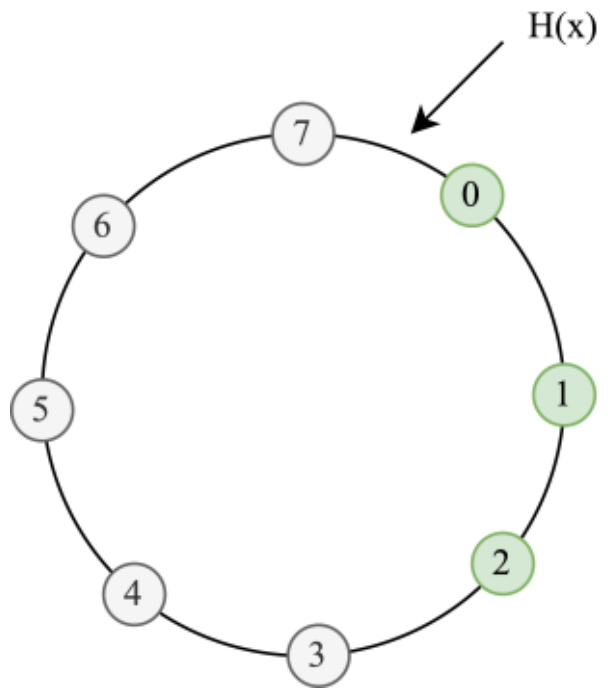

**Figure 80**: Key preference list in Dynamo.

Clients can interact with the system through two primary methods:

- **Directly using the assignment map:** This approach offers efficient and low-latency access.
- **Through a load balancer:** This may introduce additional latency.

If a client initially contacts a different node (not the key's coordinator), the request will be redirected to the correct coordinator.

Unlike traditional quorum systems that enforce linearizability by adhering to the $\mathbf{R} + \mathbf{W} > \mathbf{N}$ rule, Dynamo deviates from this approach. This deviation is intentional, as adhering to $\mathbf{R} + \mathbf{W} > \mathbf{N}$ can lead to significant latency increases, especially when $\mathbf{N}$ is large.

Instead of strict quorum rules, Dynamo utilizes vector clocks associated with each key's value to handle conflicting writes. To ensure durability, the coordinator synchronously replicates the write to at least one other node before acknowledging the client's request.

For applications demanding high-performance and consistent reads, a configuration of **R = 1** and **W = N** may be suitable.

Section 4.4 provides a detailed example of how vector clocks are employed to reconcile conflicting values for a key. While the specific details are well-explained in the paper, the core concept is that $\mathbf{D3}$ precedes $\mathbf{D4}$ because $\mathbf{Sx}(\mathbf{D3}) = \mathbf{Sx}(\mathbf{D2})$ and $\mathbf{Sy}(\mathbf{D3}) > \mathbf{Sy}(\mathbf{D2})$. $\mathbf{D3}$ and $\mathbf{D4}$ are considered conflicting versions because $\mathbf{Sy}(\mathbf{D3}) > \mathbf{Sy}(\mathbf{D4})$ and $\mathbf{Sz}(\mathbf{D4}) > \mathbf{Sz}(\mathbf{D3})$ . These conflicts are resolved through semantic reconciliation.

The paper also mentions that Dynamo truncates the vector clock to 10 elements by removing the oldest entry. This design decision raises concerns, as it may potentially lead to several unforeseen issues.

## 15.17. Handling Failures

### *15.17.1. Temporary Failure*

Dynamo introduces "hinted handoff", a technique that improves fault tolerance. When a node is unavailable, writes are redirected to its successor node on the ring. These "hinted writes" are stored in a separate file, facilitating efficient transfer when the target node becomes available again.

This approach inherently breaks the traditional quorum algorithm, leading Dynamo to adopt a "sloppy quorum" strategy.

### *15.17.2. Permanent Failure*

When a replica fails permanently, writes are still directed to at least $\mathbf{N} - \mathbf{1}$ other nodes, ensuring some level of data durability. This is where merkle trees prove invaluable.

Dynamo incorporates an anti-entropy replica synchronization protocol. In this protocol, the merkle tree serves as a data structure where leaf nodes represent the hashes of values of individual key-value pairs. During synchronization, a comparison of merkle trees between replicas reveals discrepancies, enabling the efficient transfer of only the out-of-sync values.

### 15.18. Ring Membership

Only administrators possess the authority to add or remove members from the ring. Upon any membership change, a gossip-based protocol propagates these updates throughout the system, ensuring eventual consistency in the membership view across all nodes. In this protocol, each node randomly selects a peer for communication every second.

Failure detection within the system also leverages a simple gossip-style mechanism. However, this detection is inherently local. Node A may deem node B as failed if it fails to receive responses from B, even though B might still be responsive to other nodes like C. To further reinforce failure detection, a periodic loop is implemented to actively check for the presence of node B.

### 15.19. Evaluation

- **Tail latencies:** The p999 hovers around 200ms, significantly higher than the average latency.
- **Load Balancing:** The new consistent partitioning strategy demonstrates load balancing efficiency exceeding 0.9.
- **Version Divergence:** Two primary scenarios contribute to version divergence - system failures and concurrent write operations. 99.94% of requests encounter a single version of the data. 0.00057%, 0.00047%, and 0.00009% encounter two, three, and four versions respectively.
- **Client-Driven Coordination:** Implementing client-driven coordination, where clients directly contact servers based on their knowledge of the key-to-preference list mapping, results in a 30ms reduction in both read and write latencies at p999.

### 15.20. Paper Remarks

The paper represents a seminal contribution to the field of distributed systems, introducing several groundbreaking paradigms that have profoundly shaped the NoSQL landscape. Given the technical density and breadth of the work, these insights serve only as a high-level overview. To truly appreciate the nuances of Dynamo's architecture, a rigorous and iterative study of the original text is highly recommended.

**Appendix: Towards Distributed SQL Databases**

Even with the rise of NoSQL, the need for strong consistency and data integrity remained paramount for critical applications like financial systems.

Google Spanner emerged as a groundbreaking achievement, becoming the first distributed SQL database to offer strong consistency guarantees. It pioneered the concept of "DRDBMS" – systems that are:

- **Scalable:** Capable of horizontal scaling to handle massive datasets and high throughput.
- **Consistent:** Adhering to ACID properties, ensuring data integrity and transactional isolation in a distributed environment.
- **Geographically Replicated:** Providing high availability and low latency by distributing data across multiple geographic locations.
- **SQL-compliant:** Supporting the familiar SQL language for data querying and manipulation.

**Recommended Read**: **24. Spanner: Google's Globally-Distributed Database**

Amazon Aurora, another significant player, attempted to enhance consistency by optimizing consensus protocols on top of existing standalone SQL databases like MySQL. However, this approach often faced limitations, with a single write node becoming a performance bottleneck.

**Recommended Read**: **23. Amazon Aurora: Design Considerations for High Throughput Cloud-Native Relational Databases**

CockroachDB [136], drawing inspiration from Spanner, offers a scalable and consistently replicated transactional store. However, it differentiates itself by:

- **Independence from Atomic Clocks:** Unlike Spanner, CockroachDB does not rely on tightly synchronized atomic clocks, making it more adaptable to diverse deployment environments.
- **Open Source and Platform Agnostic:** CockroachDB is open-source and can be deployed outside of Google Cloud Platform (GCP), providing greater flexibility and choice for organizations.

# 16. Cassandra - A Decentralized Structured Storage System

(16) Lakshman, A.; Malik, P. Cassandra: A Decentralized Structured Storage System. ACM SIGOPS operating systems review 2010, 44 (2), 35–40.

Following our exploration of Bigtable and Dynamo, we turn our attention to another NoSQL database developed during the same era: Cassandra. This research paper, authored by Avinash Lakshman (co-inventor of Amazon Dynamo) and Prashant Malik, originates from Facebook and dates back to 2008.

In terms of design, Cassandra acts as a synthesis of Dynamo (2007) and Bigtable (2006), drawing heavily upon the architectural concepts of both systems. Notably, this paper was published during the rapid rise of these influential distributed databases.

In this insight, we will begin with a detailed exploration of failure detectors and the various algorithms that enable them. Since both Dynamo and Cassandra rely on integrated failure detection components to maintain availability, understanding them is crucial. Afterward, we will dive deep into the specific architecture of Cassandra.

**Recommended Reads:**

- **14. Bigtable: A Distributed Storage System for Structured Data**
- **15. Dynamo: Amazon's Highly Available Key-value Store**

### 16.1. Failure Detectors

In the realm of computer science, the FLP impossibility result is a foundational concept. It states that in an asynchronous network where a single node can crash, it's impossible to simultaneously guarantee both safety and liveness (or termination). Crash faults are always a possibility in large systems and real-world networks are always asynchronous. This makes consensus, state machine replication and atomic broadcasts impossible problems to solve.

Algorithms like Paxos ensure safety but not termination. Achieving termination necessitates some degree of synchrony, whether it's partial or full synchronization within the network. Consequently, all practical consensus algorithms in real-world asynchronous networks rely on assumptions about network time bounds. These bounds are crucial for making progress and ensuring termination.

Failure detectors are essential components within these algorithms. They assist in approximately identifying and isolating faulty nodes, enabling the system to continue making progress despite the presence of failures.

Failure detectors can be classified based on their:

- **Degree of Completeness:**
  - **Strong Completeness:** Every faulty process is eventually suspected by every non-faulty process.
  - **Weak Completeness:** Every faulty process is eventually suspected by at least one non-faulty process.
- **Degree of Accuracy:**
  - **Strong Accuracy:** No non-faulty process is ever falsely suspected.
  - **Weak Accuracy:** Some non-faulty processes may be falsely suspected.
  - **Eventually Strong Accuracy:** No non-faulty process is suspected after a certain period.
  - **Eventually Weak Accuracy:** Some non-faulty processes may be falsely suspected initially, but eventually, all non-faulty processes are no longer suspected.

### *16.1.1. Algorithms*

#### 16.1.1.1. Scalable Weakly Consistent Infection Protocol

Scalable Weakly Consistent Infection Protocol [137] (SWIM) is a failure detection algorithm known for its strong completeness and weak accuracy. However, a detailed discussion of SWIM falls outside the scope of this chapter.

#### 16.1.1.2. Heartbeat-Based Failure Detectors

These represent the most fundamental class of failure detectors. They operate by monitoring the regular reception of heartbeat messages from other nodes. Heartbeat-based detectors exhibit strong completeness but only achieve eventual weak accuracy. This implies that while they will eventually detect a node failure, they may temporarily suspect healthy nodes as faulty.

#### 16.1.1.3. Ping-Ack Failure Detectors

Similar to heartbeat-based detectors, ping-ack mechanisms rely on the exchange of ping messages and acknowledgments (acks) to monitor node health. They share the same strong completeness and eventual weak accuracy characteristics as heartbeat-based detectors.

#### 16.1.1.4. Accrual Failure Detectors

Accrual Failure Detectors extend the concept of heartbeat-based mechanisms. They incorporate tunable parameters, allowing for adjustments to the balance between accuracy and completeness. This flexibility enables administrators to fine-tune the detector's behavior to meet specific system requirements.

**φ** represents a metric used to assess the likelihood of a machine failure which is inversely proportional to probability of receiving a heartbeat.

$$\boldsymbol{\varphi}(\mathbf{t}) = -\mathbf{log_{10}}(\mathbf{P}(\mathbf{t}))$$

$\boldsymbol{\varphi}(\mathbf{t})$ is the probability that a machine has failed at time **t**.

$\mathbf{P}(\mathbf{t})$ is the probability of receiving a heartbeat from the machine at time **t**.

A higher **φ** value indicates a higher probability of machine failure. For example:

- **φ** = 1 implies a 10% probability of receiving a heartbeat.
- **φ** = 2 implies a 1% probability of receiving a heartbeat.
- **φ** = 3 implies a 0.1% probability of receiving a heartbeat.

and so on.

How to define $\mathbf{P}(\mathbf{t})$? This is achieved by maintaining a sliding window of heartbeat intervals. The samples within this window collectively form a probability distribution, as shown in Figure 81. Then:

$$\mathbf{P}(\mathbf{t}) = \text{Integral of this distribution from } \mathbf{t} \text{ to } \infty.$$

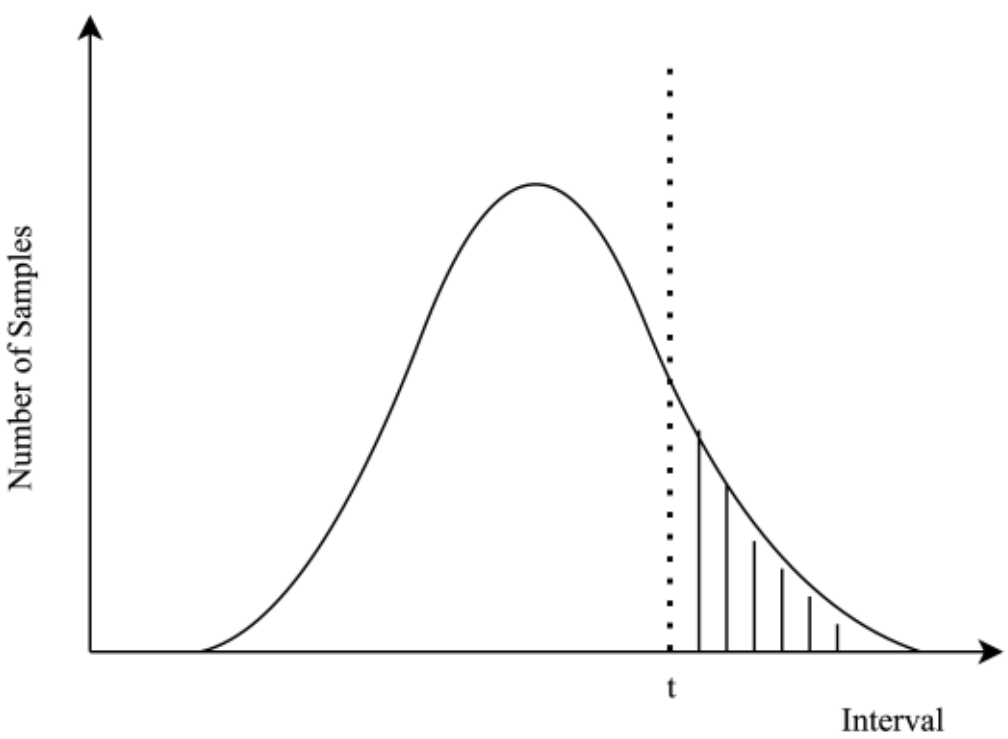

**Figure 81**: Heartbeat interval distribution.

According to the requirements, the application may utilize cutoffs for **φ** to determine whether to accept or reject as a failure.

## 16.2. Gossip Protocols

Gossip protocols constitute a class of broadcast protocols within peer-to-peer networks (other classes include point-to-point (one-to-one) and eager reliable (one-to-all) protocols).

The core principle of gossip protocols involves each node periodically transmitting a message to a randomly selected subset of other nodes. This decentralized approach ensures that, with high probability, a particular message will eventually propagate throughout the entire system. This characteristic finds valuable applications in various domains, including failure detection.

### *16.2.1. Types of Gossip Protocols*

Gossip protocols are typically characterized by two key tunables: the required time for message dissemination and network usage. Based on these parameters, several distinct types of gossip protocols have emerged:

- **Anti-Entropy:** In anti-entropy protocols, nodes transmit their entire message set to other nodes. This approach generally results in high bandwidth consumption. To mitigate this, diffing techniques can be employed to avoid redundant data transmission. The term "anti-entropy" originates from the protocol's objective: to minimize the information divergence between nodes by exchanging complete datasets.

- **Rumor-Mongering:** Rumor-mongering protocols, also known as dissemination protocols, focus on transmitting only updates or changes to the message set. This approach significantly reduces network usage, enabling more frequent data exchange.
- **Aggregation:** Aggregation protocols facilitate the exchange of system-wide aggregates through sampling techniques.

Fundamentally, all gossip protocols share a common approach: they disseminate information by having each node randomly transmit messages to a subset of its peers. This decentralized strategy relies on the probabilistic nature of these random transmissions to ensure that information eventually propagates throughout the entire network.

## 16.3. Cassandra

Apache Cassandra is a wide-column NoSQL database. These databases organize data into columns, but unlike traditional relational databases, the columns can vary from row to row.

### *16.3.1. Data Model*

Cassandra employs a **<row,column>** key format, closely resembling Bigtable (except for the time dimension). Notably, Bigtable, Cassandra, and HBase [118] are all fundamentally implemented as key-value stores.

Each row in a Cassandra table is uniquely identified by a string **row key**, typically 16 to 36 bytes long. Columns are organized into column families, and come in two forms: **simple** and **super**. Super column families can be conceptually understood as nested column families.

Regardless of the data model presented - be it tabular or graph-like - all NoSQL databases, including Cassandra, are fundamentally key-value stores.

### *16.3.2. API*

Cassandra's core API simplifies data interaction to three fundamental operations:

- **insert(table, rowkey, rowMutation)**
- **get(table, rowkey, columnName)**
- **delete(table, rowkey, columnName)**

This essentially boils down to key-value **put** and **get** operations.

### *16.3.3. Data Structures*

Exhibits strong similarities to Bigtable's tablet structures (implemented as an LSM tree):

- Utilizes an in-memory data structure to track updates concurrently with writing to a commit log.
- Periodically, this in-memory data is flushed to disk, accompanied by summary information and a Bloom filter to accelerate query performance. Although not explicitly mentioned in the paper, it's highly probable that the on-disk data files are stored in the **SSTable** format. An SSTable is a serialized representation of key-value pairs, where keys are arranged in sorted order alongside their corresponding values.
- Data files undergo periodic compaction to optimize storage and improve read performance.

### *16.3.4. System Architecture*

The architecture shares similarities with Dynamo, leveraging a chord-based consistent hashing scheme for data placement.

It's noteworthy that Bigtable leverages Google File System (GFS) for efficient data storage and retrieval. In contrast, Cassandra does not rely on a distributed file system. This design choice may be attributed to the lack of a GFS equivalent within Facebook's infrastructure at the time of Cassandra's development.

The presence of a distributed file system could have significantly simplified Cassandra's implementation. All replicas of a given partition could have seamlessly relied on a consistent view of the data stored within the distributed file system. However, without a distributed file system, the Cassandra team had to implement a more intricate solution: **consistent hashing** for data partitioning and replication, coupled with sophisticated membership control mechanisms involving **failure detectors** and **gossip protocols**.

### 16.4. Range Assignment

Cassandra makes use of consistent hashing to distribute keys. Each range is assigned to **N** nodes, one of which act as the coordinator for that range. This replication is required to ensure that there are no single points of failure. Note that the same machine can be a node for multiple ranges.

For example, in Figure 82, the range between red and purple dot is owned by the nodes.

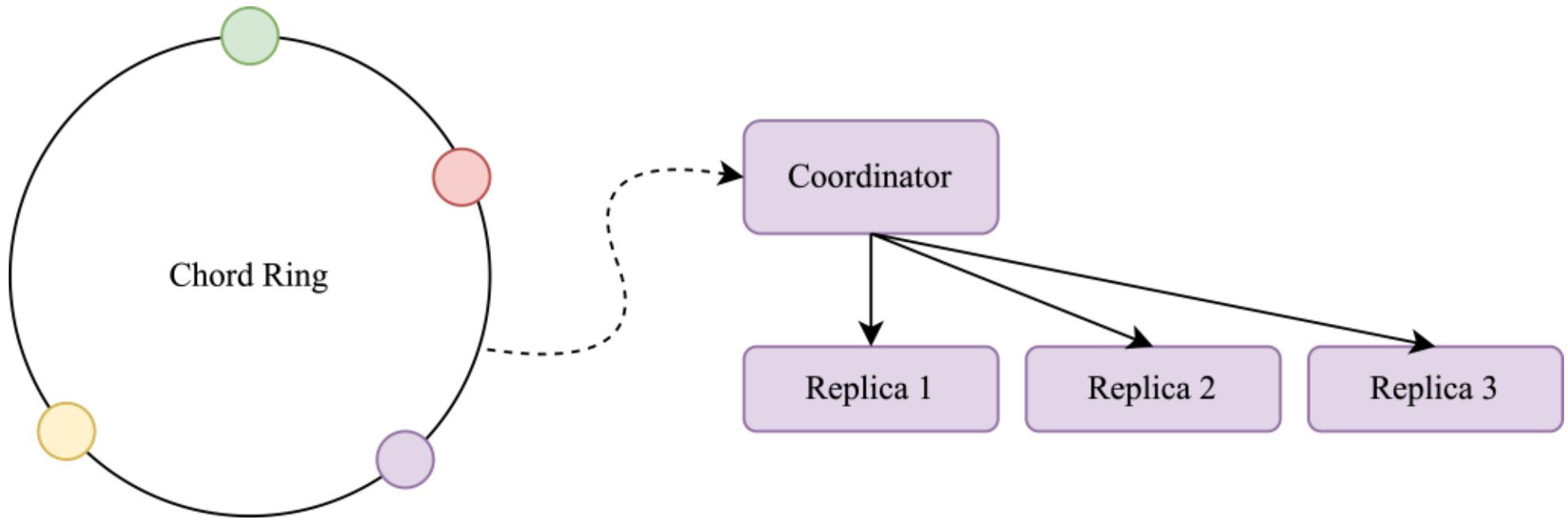

**Figure 82**: Range assignment in Cassandra.

It is interesting to note that during that time there was a trend of building truly decentralized systems where each node is essentially the same.

Cassandra employs the quorum mechanism. The replication factor is set to a threshold value, **N**. A write operation is considered successful only if the coordinator receives successful write acknowledgments from at least **N** replicas.

### 16.5. Cassandra and Failure Detection

Within a range, it is crucial for replicas to detect the failure of the coordinator, enabling the election of a new coordinator.

Cassandra utilizes an anti-entropy gossiping mechanism in conjunction with an accrual failure detector. In each exchange, a node transmits its entire collection of heartbeat information (received from other nodes) to its peers. This comprehensive exchange often results in shorter heartbeat intervals. Consequently, an exponential distribution typically provides a more accurate representation of heartbeat arrival times compared to a Gaussian distribution.

Post failure detection, Cassandra elects a leader amongst its nodes using Zookeeper, a distributed coordination system.

### 16.6. Paper Remarks

This paper is remarkably concise, covering a wide range of concepts within a relatively small number of pages. One familiar with Dynamo and Bigtable will find the paper easy to comprehend.

# 17. The Eternal Tussle: Exploring the Role of Centralization in IPFS

(17) Wei, Y.; Trautwein, D.; Psaras, Y.; Castro, I.; Scott, W.; Raman, A.; Tyson, G. The Eternal Tussle: Exploring the Role of Centralization in IPFS. In 21st USENIX Symposium on Networked Systems Design and Implementation (NSDI 24); 2024; pp 441–454.

We will now take a brief departure from database exploration to examine the world of decentralized systems. While "distributed" describes how components are physically spread out for efficiency, "decentralized" describes how power and control are shared among multiple independent parties. Note that while all decentralized systems are distributed, not all distributed systems are decentralized.

A primary example of this architecture is the InterPlanetary File System [138] (IPFS). The original white paper, “IPFS - Content Addressed, Versioned, P2P File System” [139], described a fully decentralized system designed for the next generation of the internet, often referred to as Web3. However, achieving full decentralization has faced challenges due to significant performance-related issues. As a result, IPFS is currently evolving toward a hybrid model that incorporates elements of centralization to improve efficiency.

The specific paper discussed here was presented at the Networked Systems Design and Implementation (NSDI) '24 conference, a prestigious venue for distributed systems research.

In this insight, we will begin with the necessary background on the XOR trie, a data structure essential to the components described in the paper. Next, we will explore Kademlia Hash in detail, which is used extensively in IPFS. Following that, we will explore the decentralized web and the inner workings of IPFS in its most decentralized form. Finally, we will examine how the transition toward a more centralized, hybrid model is detailed within the paper.

**Recommended Read**: **15. Dynamo: Amazon's Highly Available Key-value Store** where consistent hashing and its techniques were discussed.

### 17.1. XOR Trie

A binary trie is a trie data structure specifically designed to store binary strings. **Example**: A binary trie containing the strings "001", "101", and "111" would have a structure where each node represents a bit (0 or 1), and the paths to the leaf nodes represent the stored strings, as shown in Figure 83.

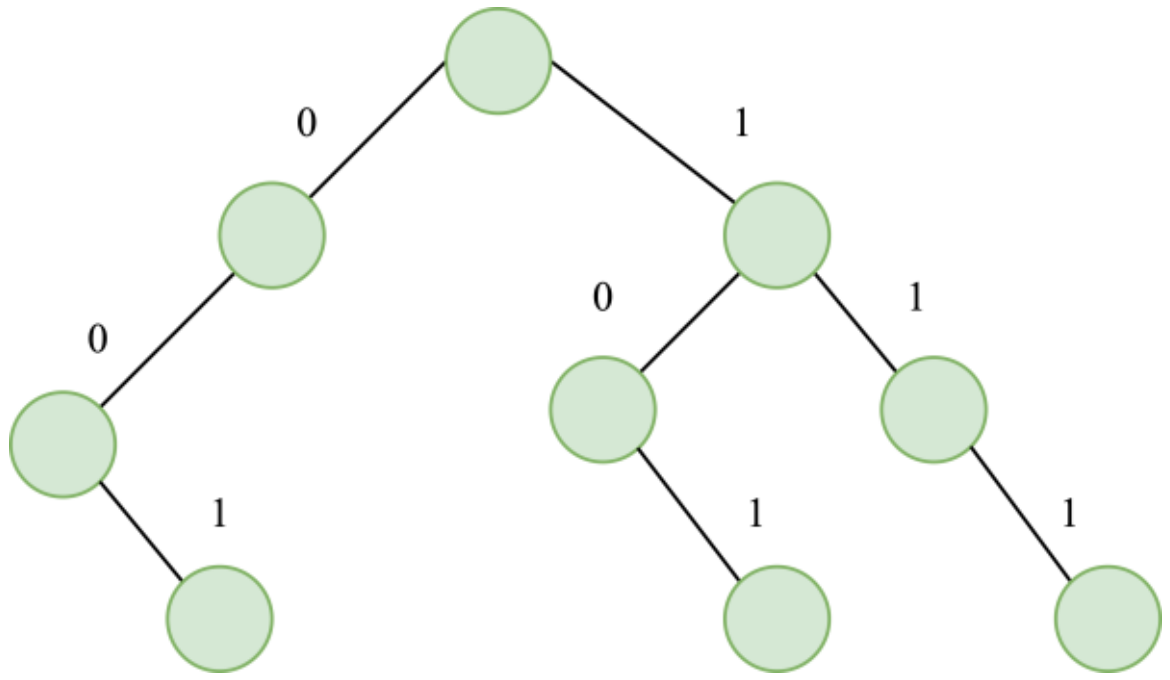

**Figure 83**: Binary trie.

The same tree can be represented in a more compact form, as shown in Figure 84. Then it becomes a XOR trie [140].

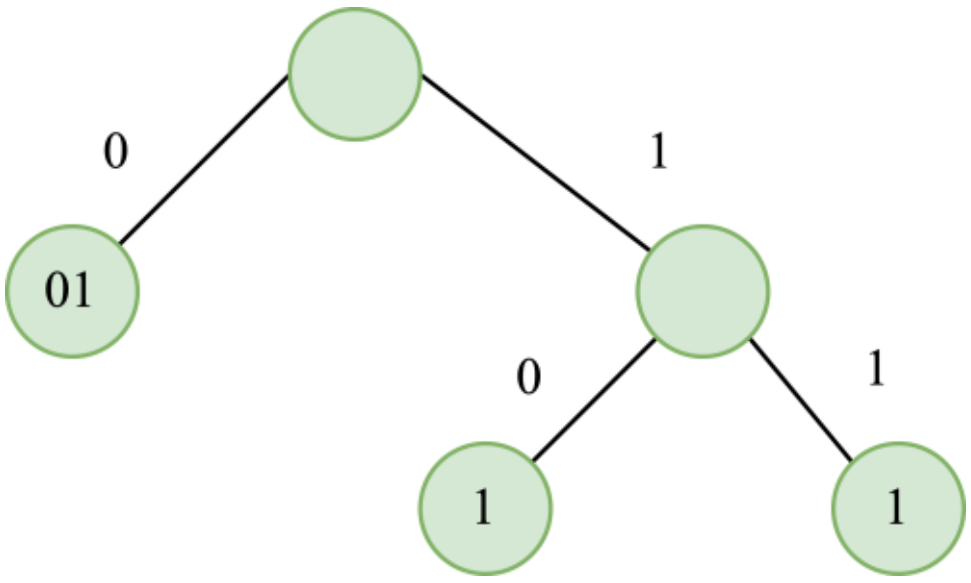

**Figure 84**: XOR trie.

XOR tries have the following key properties:

- All binary strings inserted into the trie must be of the same length.
- Both child branches of a node are either both empty (nil) or both non-empty.
- If both child branches are non-empty, the node itself has no key associated with it.
- If both child branches are leaf nodes, they both must have keys associated with them.

When inserting a new string into an XOR trie, the goal is often to minimize the increase in the tree's depth. In our example above, one strategy to achieve this is to insert a new node to the right of the leftmost node at depth 1, as shown in Figure 85.

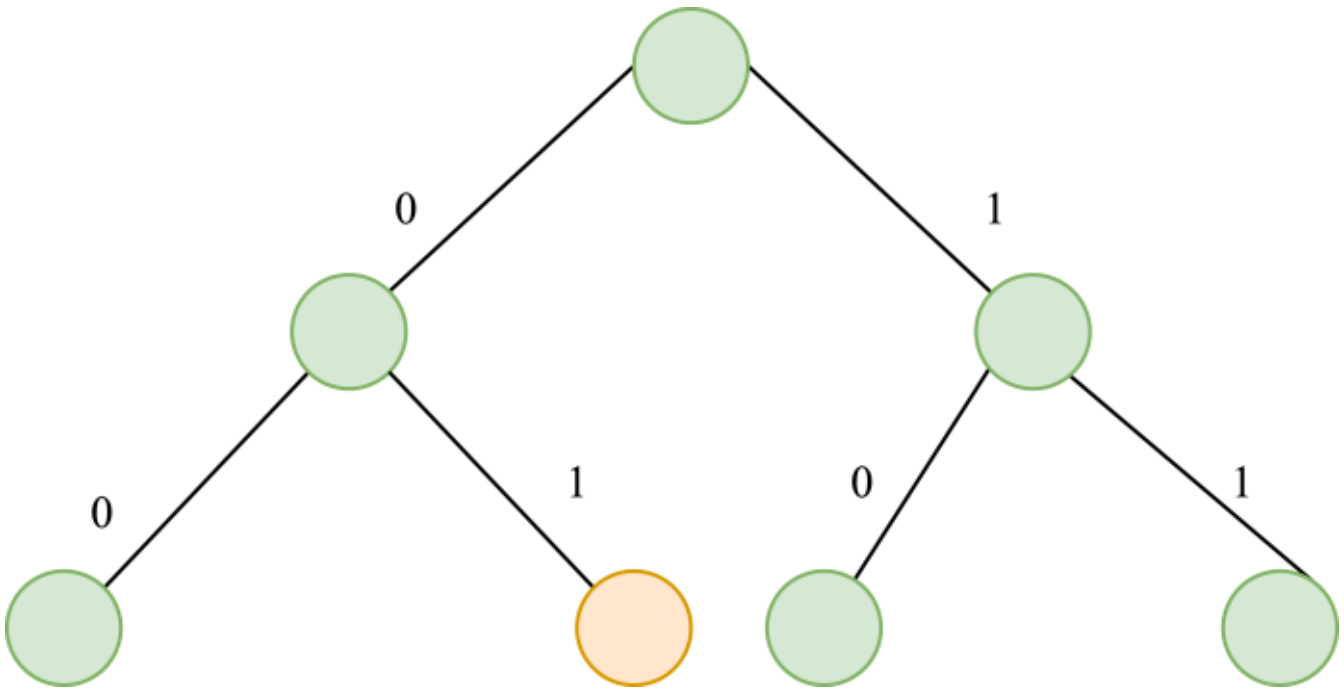

**Figure 85**: Key insertion in XOR trie.

### 17.2. Kademlia Hash

This section provides a high-level overview of Kademlia [141], a Distributed Hash Table designed for decentralized p2p networks. See **15. Dynamo: Amazon's Highly Available Key-value Store** for a deeper dive into consistent hashing and different hashing algorithms.

Kademlia, developed by P. Maymounkov and D. Mazières in 2002, has gained prominence through its use in trackerless BitTorrent and currently serves as the Distributed Hash Table (DHT) for IPFS.

Kademlia arranges the virtual nodes into leaves of a full binary tree, as shown in Figure 4. Note that the same physical machine may be placed at multiple leaves of the tree in the form of virtual nodes.

In Figure 86, all leaf positions are occupied by nodes. However, in reality, not all leaf positions may have a node present.

A key **x** is stored on the node that minimizes the XOR distance to **x**. XOR distance measures the number of differing bits between two identifiers. Kademlia employs a replication factor of 20, ensuring data redundancy by storing each key on the 20 closest nodes based on XOR distance.

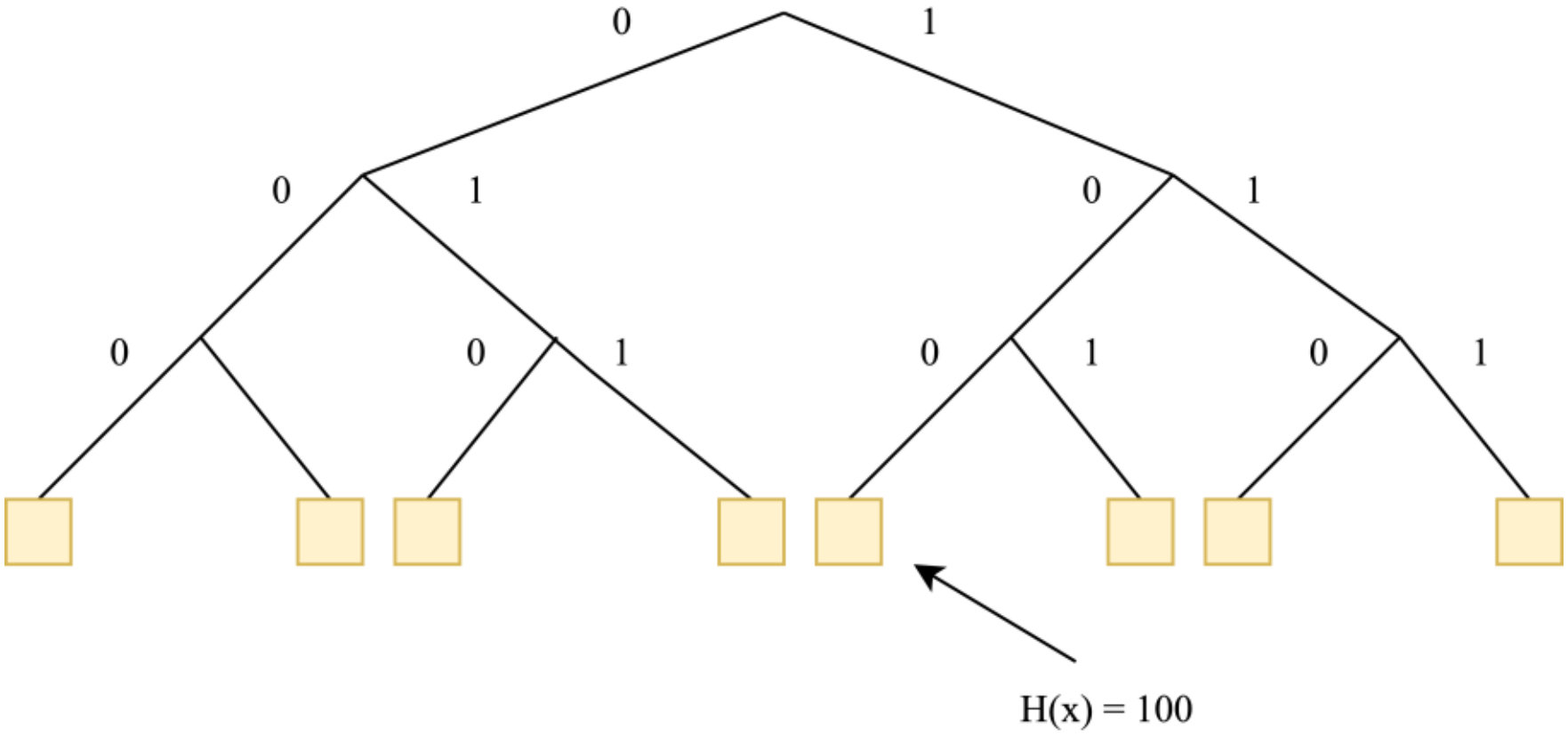

**Figure 86**: Node placement in Kademlia.

### *17.2.1. XOR Properties*

XOR was chosen as the distance metric between node IDs due to its desirable properties:

- The XOR distance between a node and itself is zero.
- **Symmetry:** The XOR distance between any two nodes (A and B) is the same regardless of the order (i.e., distance(A, B) = distance(B, A)).
- **Triangle Inequality:** For any three nodes (A, B, and C), the XOR distance between A and B is less than or equal to the sum of the XOR distances between A and C, and C and B.

### *17.2.2. Routing Algorithm*

Each node is assigned a unique identifier (e.g., a hash value).

Each node maintains a routing table with entries based on its identifier's binary representation:

$\mathbf{b}_1'$: Entries for nodes whose first bit differs from the current node.
$\mathbf{b}_1\mathbf{b}_2'$: Entries for nodes whose second bit differs.
$\mathbf{b}_1\mathbf{b}_2\mathbf{b}_3'$: Entries for nodes whose third bit differs, and so on.

When a node receives a request for a specific hash, it checks if it owns the hash. If not, it forwards the request to a neighboring node that is "closer" in terms of their identifier's binary representation (i.e., differs in fewer bits).

The routing table has $O(\log N)$ entries. Finding the responsible node requires traversing the routing table, resulting in logarithmic lookup time ($O(\log N)$).

### 17.3. Decentralized Web

Traditional websites are like toll free numbers - rely solely on the website owner to bear the costs of hosting and computation. In contrast, the decentralized web leverages a distributed network of nodes. This means that the computational burden, and therefore the associated costs, can be shared among numerous participants, regardless of the website's traffic load.

Decentralized web is popularly called Web3 [142].

### 17.4. Decentralized File System

A decentralized file system represents each file block by its cryptographic hash. This hash uniquely identifies the block's content. Subsequently, these hashes are mapped to locations on a decentralized network of servers.

### 17.5. InterPlanetary File System

The InterPlanetary File System (IPFS) is a decentralized file system that:

1. Divides files into blocks.
2. Generates a unique **Content Identifier (CID)** for each block using cryptographic hashing.
3. Stores the mapping of CIDs to their locations on a DHT.

Note: IPFS also supports directories. The directory contents (i.e., list of files) are hashed and stored with a unique CID on IPFS.

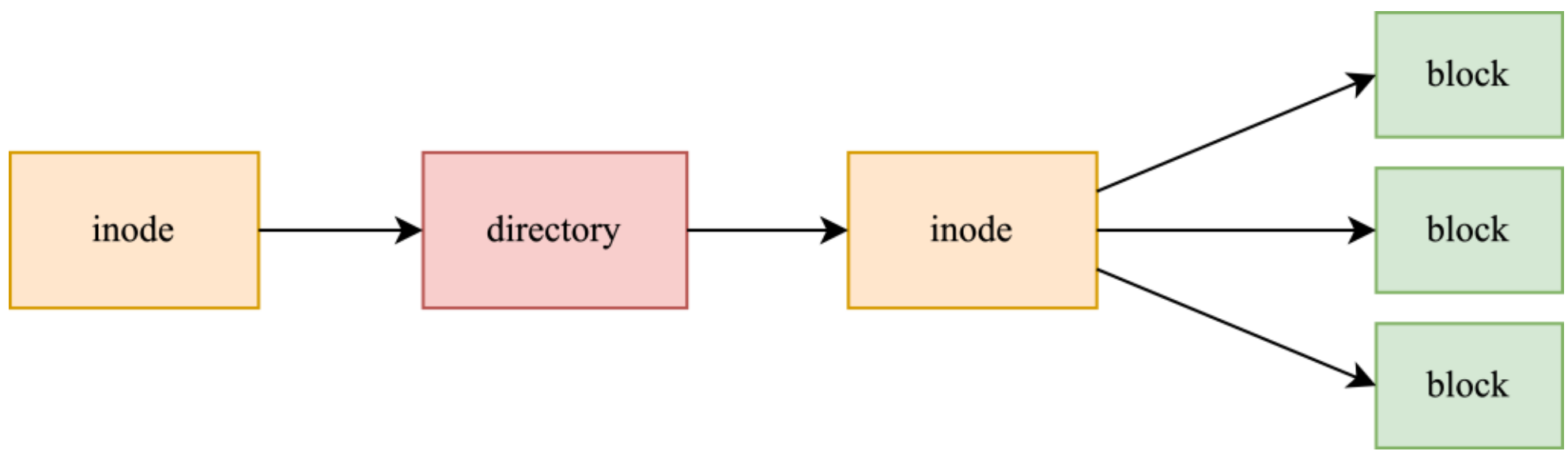

**Figure 87**: File blocks on IPFS.

As shown in Figure 87, clients initially retrieve directory content from IPFS. This content includes CIDs for the desired files. Subsequently, clients fetch the complete files by looking up these CIDs.

### *17.5.1. DHT Entries*

Two key entries within the IPFS DHT are:

- **Provider Records:** Map CIDs to the PeerIDs of nodes that possess the corresponding blocks.
- **Peer Records:** Map PeerIDs to their respective Multiaddresses. Multiaddresses enable flexible p2p communication across various protocols (e.g., IPv4, IPv6, HTTP).

### *17.5.2. Node Discovery*

While IPFS is a decentralized system, it relies on a set of special nodes known as bootstrap nodes [143]. Bootstrap nodes play a crucial role in network discovery:

1. They maintain a list of other nodes within the IPFS network.
2. When an IPFS client starts, it connects to one or more bootstrap nodes to obtain a list of potential peers.
3. The client then establishes connections with a subset of these peers, typically ranging from 500 to 1000 nodes.

This initial peer connectivity ensures that requests can be effectively routed and forwarded within the decentralized network.

### *17.5.3. Content Retrieval*

**Step 1: Local and Connected Peer Lookup:**

- Upon receiving requests for CIDs, the node first attempts to locate each block locally.
- If not found locally, it queries its connected peers using BitSwap (see 7.3) to see if any of them possess the desired blocks. This is called opportunistic lookup.

**Step 2: DHT Lookup (Kademlia):**

- If a block is not found among connected peers, the node initiates a DHT lookup using the Kademlia algorithm.
- This DHT lookup helps locate nodes within the network that are likely to have the requested block.

**Step 3: BitSwap:**

- **Want-List:** The requesting node sends a list of desired CIDs (the "want-list") to its connected peers (for opportunistic lookup) or to the peer owning the block (found after DHT lookup).
- **Block Sharing:** Each node tracks the blocks its peers have requested. When a node receives a new block, it checks if any of its peers are seeking that specific block.
- **Want-Have Responses:**
  - If a peer possesses a CID on the requesting node's want-list, it responds with a "want-have" message.
  - If a peer does not have the CID, it responds with a "dont-have" message.
- **Block Retrieval:** Upon receiving a "want-have" response, the requesting node sends a "want-block" message to the peer, triggering the transfer of the requested block.

*17.5.4. Content Caching*

In addition to retrieving content, IPFS nodes cache both file blocks and routing information locally. This caching mechanism significantly improves subsequent content retrieval performance

### 17.6. Challenges with Decentralized IPFS

The authors highlight several challenges in implementing a decentralized IPFS system:

**1. Slow Data Publishing:** The primary bottleneck in data publishing is the time-consuming process of replicating data records across 20 nodes in the DHT. The 95th percentile publishing time is a significant 66.73 seconds, primarily due to the overhead of DHT lookups to locate the 20 closest nodes for each CID. This slow publishing speed is unsuitable for real-time data streaming applications, although it may be acceptable for static data.

**2. Slow Data Retrieval:** Similar to publishing, data retrieval is also hindered by the time taken for DHT lookups. The average retrieval time is 4.42 seconds, which is considered too slow for many web3 applications that require rapid data access.

**3. Complexity of Setup:** Setting up and using IPFS currently requires a degree of technical expertise. Given that 58% of internet traffic originates from mobile devices and 92% of users access the web via smartphones, the complexity of IPFS setup presents a significant barrier to widespread adoption.

## 17.7. Towards Centralization

The authors propose a solution that introduces a degree of centralization to address the performance limitations of decentralized IPFS. It's crucial to emphasize that this centralization is limited in scope and focuses solely on performance improvements. The core decentralized nature of the IPFS network remains intact. Users can still bypass these centralized components if desired.

### *17.7.1. Centralized Components*

- **<u>InterPlanetary Network Indexers</u>:** These accelerate content publication and retrieval by optimizing data indexing and lookup processes.
- **<u>Hydra Boosters</u>:** These specialized nodes improve performance by reliably hosting a large number of provider records, enhancing the efficiency of locating data within the network.
- **<u>HTTP Gateways</u>:** These gateways provide an HTTP interface to IPFS, facilitating easier integration and adoption by simplifying access for developers and applications.

The authors also performed a massive surgery on the IPFS client code to integrate all these centralized components.

## 17.8. InterPlanetary Network Indexers

InterPlanetary Network Indexers are high-performance key-value stores that efficiently index provider records. The key is CID. The value is the physical addresses of peers that possess the content, along with the protocols to be used. By offloading the storage of provider records from the DHT, Indexers significantly reduce network overhead and improve query performance.

Pebble [144] (by Cockroach DB Labs), a high-performance key-value store, is used as the underlying data store.

### *17.8.1. Advertisement*

To be visible on the network, a peer must advertise its available content to the Indexers. This is achieved through a chain of immutable advertisements – a data structure that records the publication or deletion of content by a given provider.

Providers maintain these advertisement chains locally. Information about these advertisements is disseminated to Indexers via gossip-based announcement messages. Indexers enable efficient bulk publishing of content availability information.

### *17.8.2. Client Queries*

Indexer identities are publicly available. Clients can query Indexers with one or more CIDs. Indexers respond with a list of provider records that match the query.

## 17.9. Hydra Boosters

Hydra introduces shortcuts within the IPFS routing space to improve performance. It consists of:

- **Hydra Head Nodes:** These nodes are strategically placed within the DHT itself.
- **Shared Hydra Database:** This database, implemented using Amazon DynamoDB [134], stores mappings between CIDs and peer addresses. Hydra Head nodes have access to this database.

The Hydra database is distributed across multiple servers within Amazon's infrastructure, enhancing performance. However, it's important to note that this database is not fully decentralized.

### *17.9.1. Picking Node IDs for Hydra Heads*

Hydra Heads are strategically placed within the DHT to ensure efficient routing. To do this, the PeerIDs of Hydra Heads are tracked in a XOR trie. When adding a new node:

- Two random node IDs are generated (power of two choices [145]).
- The node ID that keeps the depth of the trie more regular is selected for the Hydra Head.

**Power of Two Choices**

Consider load balancing requests across **N** symmetric servers.

- **Option 1 (Random):** Distributing requests randomly across servers.
- **Option 2 (Two Choices):** Selecting two random servers and routing the request to the server with the shorter queue.
- **Option 3 (Three Choices):** Selecting three random servers and routing to the server with the lowest load.

Option 2 exhibits a significant performance improvement over Option 1. However, the improvement gained by moving from Option 2 to Option 3 is relatively minor.

## 17.10. HTTP Gateway

HTTP Gateways provide a bridge between the IPFS network and the standard HTTP protocol. These gateways leverage NGINX, a high-performance web server, to cache frequently accessed IPFS content. NGINX employs a Least Recently Used (LRU) cache replacement policy.

## 17.11. Evaluation

### *17.11.1. Indexer Performance*

- The indexer currently stores a massive 173 billion records, a 100x increase compared to the DHT.
- This data is primarily contributed by 604 major publishers utilizing NFTs or dedicated storage services.
- The indexer handles 2.5k lookup requests per second, while 5k requests still rely on DHT lookups.

- Requires a significant capital investment ($10k) and monthly operating costs ($1k) per indexer.
- The indexer cache has a 65.22% hit rate, significantly accelerating lookups.
- Even for the 34.78% of requests that miss the cache, latency remains lower than direct DHT lookups.

*17.11.2. Hydra Booster Performance*

- Hydra boosters contribute to improved lookup times, although the benefits are not uniform across all regions. Regions like eu-central, already exhibiting good performance, did not experience significant improvements.
- The placement strategy effectively positions Hydra Heads, ensuring that 96.6% of peers have at least one Hydra Head within their 20-proximity.
- Lookup speed is improved by -3.3% to 36.5% with Hydra Boosters. Performance at the tail of the distribution is still comparable to direct DHT lookups.
- This proximity significantly reduces DHT hops, often to a single hop.

*17.11.3. HTTP Gateway Performance*

- Cache hits, with 50.3% hit rate, achieve a 24ms p95 latency.
- An additional 27.6% of requests are resolved by contacting the first node directly.
- For the remaining requests, performance may degrade slightly due to the extra hop introduced by the gateway servers.

### 17.12. Problems

This centralized approach presents several critical issues:

- **Censorship Risk:** Indexers can potentially facilitate censorship of content within the IPFS network.
- **Hydra Booster Issues:**
    - The underlying Kademlia DHT is already vulnerable to malicious actors.
    - Hydra Boosters, while intended to improve performance, exacerbate this vulnerability by enabling malicious nodes to strategically position themselves close to several other nodes.

- There are additional performance concerns.
    - Indeed Hydra Boosters have now been turned down by IPFS.
- **Gateway Security Weaknesses:** HTTP gateways compromise the end-to-end cryptographic validation process, introducing security risks.
- False and slow advertisements can significantly degrade the user experience.

### 17.13. Paper Remarks

This return to a more centralized model, with components like Hydra Boosters and Indexers, resembles the Napster [146] era, where a central node exerted significant control over file sharing. This contradicts the core principle of decentralization. This inherent tension is aptly captured by the paper's title, "The Eternal Tussle", which highlights the constant oscillation between centralized efficiency and decentralized autonomy as the industry searches for an optimal balance. This paper serves as an excellent resource for understanding the complexities of the decentralized landscape and the evolving architecture of Web3. It provides a clear and rigorous look at the practical challenges of building peer-to-peer systems at scale.

# 18. Practical Uses of Synchronized Clocks in Distributed Systems

(18) Liskov, B. Practical Uses of Synchronized Clocks in Distributed Systems. In Proceedings of the tenth annual ACM symposium on Principles of distributed computing; 1991; pp 1–9.

We now focus on one of the most difficult questions in computer science: "What time is it?" Remarkably, no computer on Earth can answer that question with absolute accuracy. If two computers are queried in sequence, it is entirely possible for the second machine to return a timestamp less than the first. For computers, time does not always flow in a single, linear direction. It is primarily because we lack a single, perfect physical source of time; even atomic clocks are subject to "drift".

While perfectly synchronized clocks do not exist, distributed systems practitioners have successfully used near-synchronized clocks to build robust systems that this paper describes. This influential paper was authored by Barbara Liskov, a renowned computer scientist and a pioneer in the field of distributed systems.

The paper provides a valuable overview of several groundbreaking systems:

- **At-most-once delivery (SCMP):** This system ensures that a message is delivered at most once, preventing duplicate messages.
- **Authenticator Systems (Kerebos):** This system focuses on secure authentication within distributed environments.
- **Cache consistency (Echo):** This system addresses the challenges of maintaining data consistency across distributed caches.
- **Distributed Databases (Thor):** This system explores the design and implementation of distributed databases.
- **Replicated File System (Harp):** This system investigates the principles of replicating files across multiple servers for improved availability and performance.

While many of these concepts may seem outdated in the context of modern computing, studying them provides crucial insights into the foundational principles of distributed systems. Understanding how these early systems addressed challenges like message delivery, security, and data consistency helps us appreciate the advancements made in contemporary distributed computing technologies.

In this insight, we will begin by exploring synchronized clocks in detail and learning how the Network Time Protocol - the closest practical approximation for synchronized clocks - operates. Then, we will briefly touch upon external consistency before diving into a detailed exploration of each system described in the paper.

## 18.1. Synchronized Clocks

The existence of perfectly synchronized clocks across all computers would have dramatically simplified the challenges of distributed systems.

With perfect clock synchronization:

- All events occurring within the distributed system could be ordered based on their timestamps.
- All participating systems (assuming they are not exhibiting Byzantine failures) would agree on the same order of events.

This total ordering would have elegantly solved fundamental problems like atomic broadcast and consensus. Both safety (ensuring that certain undesirable states are never reached) and liveness (guaranteeing that the system makes progress) could have been achieved in a synchronous environment.

Unfortunately, perfect clock synchronization is unattainable in practice.

## 18.2. Network Time Protocol

The closest approximation we have is **Network Time Protocol** (NTP), which synchronizes clocks across geographically distributed systems over the network. While NTP achieves high precision, typically within a few milliseconds, perfect synchronization is still not possible.

### *18.2.1. Architecture*

NTP, designed by David L. Mills, employs a hierarchical network of systems to synchronize time across them, as shown in Figure 88. This hierarchical structure is organized into **strata**.

- **Stratum 0:** Consists of highly precise time sources such as atomic clocks and GPS receivers.
- **Stratum 1:** Comprises time servers maintained by governments and private organizations. These servers synchronize their time directly with Stratum 0 sources, with a few microseconds of difference.
- **Stratum 2 and above:** Include servers and workstations that synchronize their time with servers in lower strata (e.g., Stratum 2 servers synchronize with Stratum 1 servers).

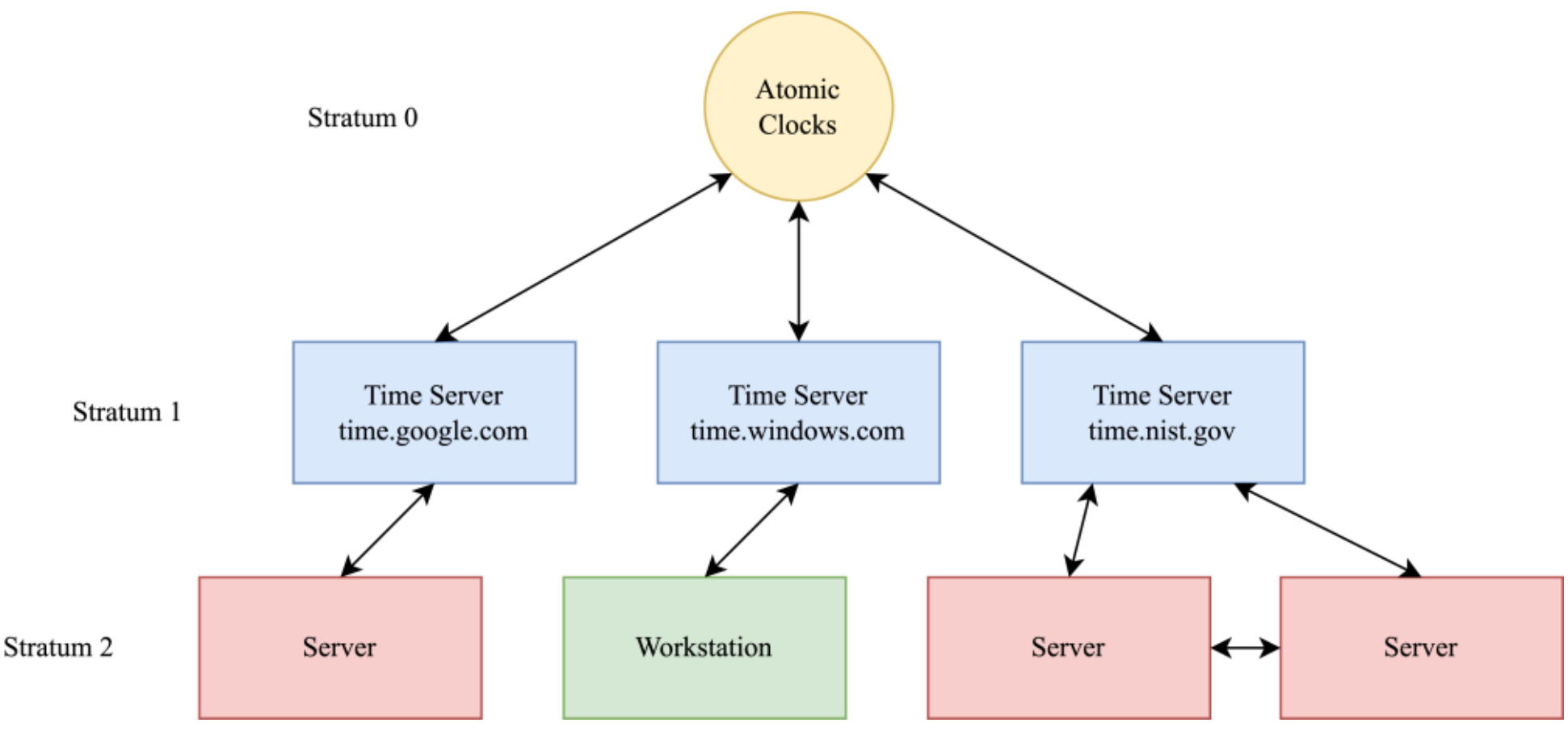

**Figure 88**: NTP's architecture.

### 18.2.2. *Algorithm*

NTP operates by exchanging timing information between clients and servers, as shown in Figure 89.

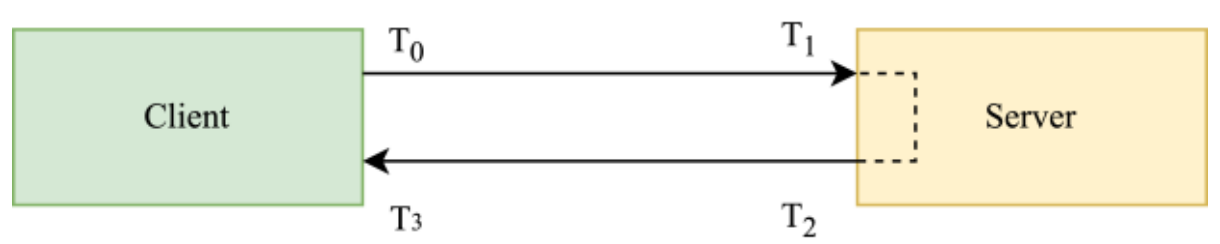

**Figure 89**: NTP message exchange.

- **Message Exchange:**
  - A client sends an ECHO message to a server at its local time $\mathbf{T_0}$.The server receives the message at its local time $\mathbf{T_1}$.
  - The server sends a reply message at its local time $\mathbf{T_2}$.
  - The client receives the reply at its local time $\mathbf{T_3}$.

- **Calculating Network Round Trip Time (RTT):** RTT is the difference between the time elapsed from receive to send as recorded by the client minus the time elapsed from send to receive as recorded by the server.

$$\mathbf{RTT} = (\mathbf{T_3} - \mathbf{T_0}) - (\mathbf{T_2} - \mathbf{T_1}) \quad \textbf{.. (i)}$$

- **Estimating Clock Offset ($\boldsymbol{\theta}$):** Assuming the client's clock is behind the server's clock by $\boldsymbol{\theta}$ ($\boldsymbol{\theta}$ will be negative if clock is forward). $\mathbf{T_0} + \boldsymbol{\theta}$ is the server's time when a message is sent by the client. When we add half of RTT to it, we should get $\mathbf{T_1}$. The following equation holds:

$$\mathbf{T_0} + \boldsymbol{\theta} + \frac{\mathbf{RTT}}{\mathbf{2}} = \mathbf{T_1} \quad \textbf{.. (ii)}$$

Substituting (i) in (ii), we get:

$$\boldsymbol{\theta} = \frac{\left((\mathbf{T_1} - \mathbf{T_2}) + (\mathbf{T_2} - \mathbf{T_3})\right)}{\mathbf{2}}$$

- **Clock Adjustment:**
  - The ECHO response includes timestamps $\mathbf{T_0}$, $\mathbf{T_1}$, $\mathbf{T_2}$, and $\mathbf{T_3}$.
  - The client uses these timestamps to calculate the clock offset ($\boldsymbol{\theta}$) and adjust its local clock accordingly.

*18.2.3. Assumptions and Limitations*

- **Constant Time Flow Rate:** The derivation assumes that the rate of time flow is identical for both the client and server clocks. This assumption may not always hold true in reality.
- **Bounded Message Delay:** The calculation relies on the assumption that message transmission delays are bounded. However, network conditions can cause unpredictable delays.

Due to these assumptions, NTP clock synchronization achieves high accuracy with a high probability but cannot guarantee perfect synchronization.

**Important Note:** While clock synchronization can improve performance in many distributed systems, it should not be considered a fundamental requirement for correctness. System design should prioritize mechanisms that function correctly even with imperfect clock synchronization.

### *18.2.4. Precision*

The algorithm synchronizes the clock with a skew of **ε**. At any given moment, a node's clock differs from the actual time by no more than **ε/2**.

## 18.3. External Consistency

External consistency, also known as **strict serializability** (for multi-object operations) or **linearizability** (for single-object operations), ensures that the effects of transactions are immediately visible to all clients in the order they were committed.

A simple analogy illustrates this concept:

- Suppose you add a new row to a table in a database and successfully commit the transaction.
- You then call up a friend and inform a friend about this change and ask them to read the database for the newly added row.
- If your friend's read fails to retrieve the row, the database violates external consistency.

Formally, if transaction $\mathbf{T_2}$ is submitted by a client only after transaction $\mathbf{T_1}$ commits, then the database must order $\mathbf{T_2}$ after $\mathbf{T_1}$.

A database that guarantees external consistency is considered **strictly serializable**.

## 18.4. Application 1: At Most Once Delivery

A messaging system facilitates the delivery of messages from a producer (sender) to a consumer (receiver). **At-most-once delivery** guarantees that a message will be delivered to the consumer at most one time. This differs from **at-least-once delivery**, which ensures that each message is delivered at least once, potentially leading to duplicates. At-most-once delivery prioritizes avoiding duplicate messages, even if it means that some messages might be lost during transmission or processing.

### *18.4.1. Naive Approaches*

- **<u>No Sender Retries</u>:** Eliminating sender retries prevents duplicate messages, but can result in significant message loss due to network delays.

- **Sender-based Deduplication:** Receivers maintain a table of active senders and track message IDs (which should be sequentially increasing). This allows for deduplication, but requires receivers to store sender information indefinitely, which is not infeasible without synchronized clocks and reliable sender retries.

*18.4.2. Slightly Better Approach*

**Handshake:** Before sending messages, a handshake is performed between sender and receiver to establish the receiver's current message reception position. This approach has high overhead for scenarios with only a few messages.

*18.4.3. The SCMP Protocol*

The SCMP protocol [147] utilizes synchronized clocks (though it can operate, albeit less efficiently, without them).

Receivers maintain records of recent communications. Senders include timestamps with each message. A message and its effects expire after a defined "lifetime interval" ($\mathbf{p}$).

Receiver's data structure ($\mathbf{G}$):

- $\mathbf{G.time}$: Receiver's time.
- $\mathbf{G.CT}$ **:** Maps connection IDs (unique identifiers for sender-receiver communication channels) to connection information (including the last known timestamp $\mathbf{ts}$ for that connection). Entries in $\mathbf{G.CT}$ are removed if their last known timestamp is older than $\mathbf{G.time - p - \varepsilon}$.
- $\mathbf{G.upper}$: The oldest timestamp that has been removed from $\mathbf{G.CT}$.
- $\mathbf{G.latest}$**:** A timestamp larger than the timestamp of all earlier accepted messages. It is usually set equal to $\mathbf{G.time + \beta}$ and committed to stable storage ($\boldsymbol{\beta}$ here is some increment).

A message on connection $\mathbf{C}$ is considered new if its timestamp is greater than $\mathbf{G.CT[C].ts}$ or $\mathbf{G.upper}$ if $\mathbf{C}$ is new.

Post-Crash Initialization**:** $\mathbf{G.upper}$ is initialized to $\mathbf{G.latest}$ after a crash.

#### 18.4.3.1. Limitations

If a connection **C** is forgotten by the receiver, **G. upper** might be used for recency checks instead of the last known timestamp for **C**. This can lead to valid messages being erroneously rejected as duplicates.

#### 18.4.3.2. Correctness

Clock synchronization is not strictly necessary for SCMP's correctness. A slow node's clock can cause its messages to be rejected due to their outdated timestamps.

## 18.5. Application 2: Authentication Tokens

### *18.5.1. Ticket-Based Authentication*

Ticket-Based Authentication is implemented by Kerebos [148]. Communication between a client (**C**) and a server (**S**) is controlled by a **ticket**. The client obtains the ticket from a **Ticket Granting Server (**TGS**)**. The TGS provides the client with:

- $\mathbf{T_{CS}}$**:** A ticket specific to the server (**S**).
- $\mathbf{K_{CS}}$**:** A secret session key shared between the client and the server.

This information is transmitted over an encrypted channel to ensure secure delivery only to the client.

In subsequent communications with **S**, the client presents the ticket (encrypted using **S** 's private key) and the session key.

Each ticket has an expiration time, after which communication using that ticket is prohibited. If **S** 's clock is significantly behind, it might erroneously accept expired tickets. This could necessitate **S** to verify each ticket with the TGS before processing requests, increasing overhead.

The server (**S**) must never accept and process an expired ticket. Synchronized clocks between **S** and **TGS** help ensure that **S** 's clock is within an acceptable range of the **TGS**'s clock, minimizing the risk of accepting expired tickets.

18.5.1.1. Correctness

There is a hard reliance on clock synchronization for absolute correctness. But the author calls out that the primary concern here is the theft of tickets from unattended workstations, which can be exploited by attackers. This concern is readily handled.

*18.5.2. Authenticators and Message Replay Prevention*

**Authenticators** are encrypted timestamps generated using the shared session key *(*$\mathbf{K_{CS}}$*)*, known only to the client and server. Each message includes an authenticator. Upon receiving a message, the server verifies the authenticity of the message and its timestamp to prevent replay attacks (where an old message is retransmitted).

Similar to at-most-once delivery systems, authenticators help ensure that each message is processed only once. Clock synchronization plays a role in minimizing the acceptance of duplicate messages.

**18.6. Application 3: Cache Consistency**

*There are only two hard things in Computer Science:*
*cache invalidation and naming things.*

*- Phil Karlton*

Clients maintain local copies of server objects as cached data. This caching mechanism enhances performance by eliminating the need to repeatedly fetch objects for access. However, it necessitates a mechanism to inform clients of server-side object modifications, triggering the invalidation of local copies and subsequent retrieval of updated values from the server. This is required by the Echo file system.

Caching strategies can be broadly categorized into two types:

- **Write-Through Cache:** Modifications are written to the cache and the source.
- **Write-Behind Cache:** Modifications are initially written to the cache before being committed to the source.

In the context of Echo, write-through caching involves the server invalidating object copies in all client caches prior to committing the modification. This approach necessitates a **lease mechanism** on the server to track clients holding references to the object. If a client becomes unreachable, the server must await lease expiration.

Leases are differentiated into read leases (held by multiple clients) and write leases (exclusively held by a single client). A new write lease is granted only after all existing leases have been either relinquished or expired.

Lease mechanisms can be implemented in two ways:

- **Explicit Client Acknowledgment:** Leases are assumed to have infinite expiration unless explicitly relinquished by the client. This approach relies on the unrealistic assumption of faultless processes and networks. This assumption is suitable for multi-processing environments, however, it doesn't work for distributed systems.
- **Clock Synchronization-Based Leases:** Leases are assigned timestamps based on synchronized clocks. However, clock desynchronization can render the system inoperable.

Neither of these lease mechanisms is entirely satisfactory. Clock synchronization is often adopted as an acceptable compromise to enable system functionality.

#### *18.6.1. Correctness*

Clock desynchronization compromises system correctness. This highlights a strong dependency on clock synchronization for proper system operation. However, for many caching applications, precise lease correctness is not a critical requirement. Applications utilizing such caching systems should be designed to function correctly even if lease correctness is not guaranteed, as demonstrated in the next application example.

### 18.7. Application 4: Atomicity

Transaction atomicity is a fundamental requirement in SQL databases, ensuring that a series of operations within a transaction either all succeed or all fail as a unit.

Traditionally, atomicity was achieved by using locks on the objects. However, the advent of **optimistic concurrency control (OCC)** introduced a new approach:

- OCC assigns version numbers to each object. The versioned objects are stored within a local client-side cache.
- Transactions can then be processed locally on the client.
- The client submits the completed transaction to the server. The server validates the transaction by checking if the version numbers of the objects involved still match the cached versions. If the versions match, the transaction is committed; otherwise, it is aborted and the client must retry.

OCC is supported by Thor. Clock synchronization can significantly enhance performance. Synchronized clocks contribute to better cache coherence across clients. More synchronized objects reduce the likelihood of version conflicts, leading to fewer transaction aborts and higher overall transaction commit rates.

*18.7.1. Correctness*

Clock synchronization is not necessary for the correctness of transaction commit as all commits ultimately occur on the server.

While the underlying cache may exhibit inconsistencies due to its reliance on an imperfect lease mechanism, the application layer, in the context of transaction atomicity, maintains correctness.

**18.8. Application 5: Commit Windows**

Harp [149] is a replicated file system that supports the virtual file system [71] interface and is designed to work with Network File System. Unlike file systems that support complex transactions (e.g. Transactional NTFS [150]), Harp treats each file operation as an independent atomic unit. This approach effectively transforms the file system into a key-value store where each file is treated as a key.

*18.8.1. Quorum-Based Replication*

To ensure data consistency, Harp employs a majority voting mechanism within a quorum of replicas for committing updates to any file.

*18.8.2. Membership Protocols and View Changes*

Harp operates within a "view" where all replicas are aware of each other. Views may change dynamically to accommodate server failures or additions, but always maintain a majority of the servers. This ensures that each new view includes at least one replica from the previous view, preserving the history of committed transactions.

Upon change, a new view elects a leader, which gathers the latest view information from all replicas before accepting new transactions.

### *18.8.3. Read Operations*

In such quorum-based key-value transactional stores, all reads are single key reads only. The reads can be further classified into two types:

- **Single-Key Strong Read (Read from Majority)**
  - Reads data from a majority of replicas.
  - Guarantees linearizability and external consistency, even during view changes.
    - If a view change occurs before, the transaction is simply retried.
- **Single-Key Relaxed Read (Read from Primary)**
  - Reads data only from the primary replica.
  - **Not linearizable** due to potential view changes that may leave the primary unaware of updates.
  - May return stale data.

### *18.8.4. External Consistency*

External consistency guarantees that the effects of all transactions up to a point **T** are visible at point **T**. Harp **does not guarantee external consistency**, with single-key relaxed reads.

### *18.8.5. Impact of Clock Synchronization*

Clock synchronization plays a crucial role in mitigating violations of external consistency. The primary replica can obtain short-lived leases from a sub-majority of backups. When a new view is established, the new primary must wait for all leases from the previous primary to expire before accepting new requests. This ensures that the primary holds valid leases from a sub-majority of backups during read operations.

18.8.5.1. Correctness

Even with clock synchronization, minor violations of external consistency can still occur.

*18.8.6. Bonus: SQL Multi-Key Strong Reads (Transactional Reads)*

For multi-key operations, SQL databases typically support **transactional reads** (also known as **multi-key strong reads**, **snapshot reads**, or **read-only transactions**). These reads guarantee that the values returned for all keys within the transaction reflect the state of the database at a consistent point in time, specifically the point at which the transaction commits.

### 18.9. Synchronized Rates

The paper also explores the concept of relying on synchronized rates of time flow, rather than absolute synchronized clocks, to address certain challenges.

Consider an event **e** occurring on node **O** (e.g., lease expiration). Another node **D** (e.g., the client holding the lease) is dependent on the occurrence of event e. For correctness, node **D** must perceive event **e** no later than node **O** in absolute time, i.e., $\mathbf{T_d} \leq \mathbf{T_e}$, where $\mathbf{T_d}$ is the absolute time of event **e** at node **D** and $\mathbf{T_e}$ is its absolute time of occurrence at node **O**. This principle is crucial for maintaining system consistency. For instance, the client node must relinquish the lease before the server does in absolute time.

Event **e** is preceded by events **g** on node **D** and **f** on node **O**, which serve as a common reference point for establishing rate synchronization. Once this reference point is established, a conservative estimate of $\mathbf{T_d}$ can be made to ensure $\mathbf{T_d} \leq \mathbf{T_e}$. The paper provides a straightforward derivation of this estimate.

Rate synchronization is only beneficial when a communication mechanism exists to establish a common reference point. In the absence of such a mechanism, traditional clock synchronization remains necessary.

### 18.10. Takeaways

This paper explores a wide range of topics and presents some intriguing observations.

- Correct system operation should not be contingent upon perfectly synchronized clocks. Systems must maintain correctness even when clock discrepancies occur or have a compensation for it. This principle applies universally across all systems examined in the paper. Notably, even in scenarios where external correctness is compromised, such as external consistency violations in Harp, internal correctness is always preserved.
- While transactions effectively ensure the internal consistency of databases, external consistency can only be guaranteed through the use of strong reads (both in single-key and multi-key scenarios). However, strong reads often incur significant performance overhead. Leases offer a potential solution for achieving strong reads, but their effectiveness is inherently tied to the accuracy of clock synchronization.

### 18.11. Paper Remarks

This paper, while seemingly straightforward, presents a nuanced and intricate exploration of several concepts. It delves into numerous systems developed by the MIT PDOS lab [151] during the 1990s, each of which could be a subject of independent research. This comprehensive work is an essential read for anyone passionate about distributed systems.

# 19. Kafka: A Distributed Messaging System for Log Processing

(19) Kreps, J.; Narkhede, N.; Rao, J.; others. Kafka: A Distributed Messaging System for Log Processing. In Proceedings of the NetDB; Athens, Greece, 2011; Vol. 11, pp 1–7.

A distributed systems course would be incomplete without addressing messaging systems, as messaging is the primary mechanism through which distributed components communicate.

This paper was authored by Jay Kreps, Neha Narkhede, and Jun Rao. This seminal paper, presented at the NetDB '11 workshop, laid the foundation for Apache Kafka [152], a highly influential open-source project in the realm of distributed systems. This particular work garnered significant attention and has had a profound impact on the field.

While the paper initially focused on a specific use case - log processing - Kafka has since evolved into a versatile and robust platform for general message delivery. Following the success of the project, both Kreps and Narkhede went on to co-found Confluent Inc. [153] to commercialize the technology.

In this insight, we will begin by defining what a messaging system is, exploring its various properties, models, and delivery guarantees. We will then dive deep into Kafka's core architecture. In the bonus section, we will touch upon other prominent messaging systems currently used in the industry.

### 19.1. Messaging Systems

Messaging systems facilitate the exchange of messages between different components within a distributed system. The sender of a message is known as the **producer** and the receiver is known as the **consumer**.

Two prominent types of messaging systems, as shown in Figure 90, are:

- **<u>Message Queues</u>:** These systems employ a point-to-point delivery model, where each message is consumed by only one consumer. This is analogous to a one-to-one communication channel.

- **<u>Publish-Subscribe (Pub/Sub)</u>:** In Pub/Sub systems, a message published by a producer can be received by multiple consumers. This enables one-to-many communication.

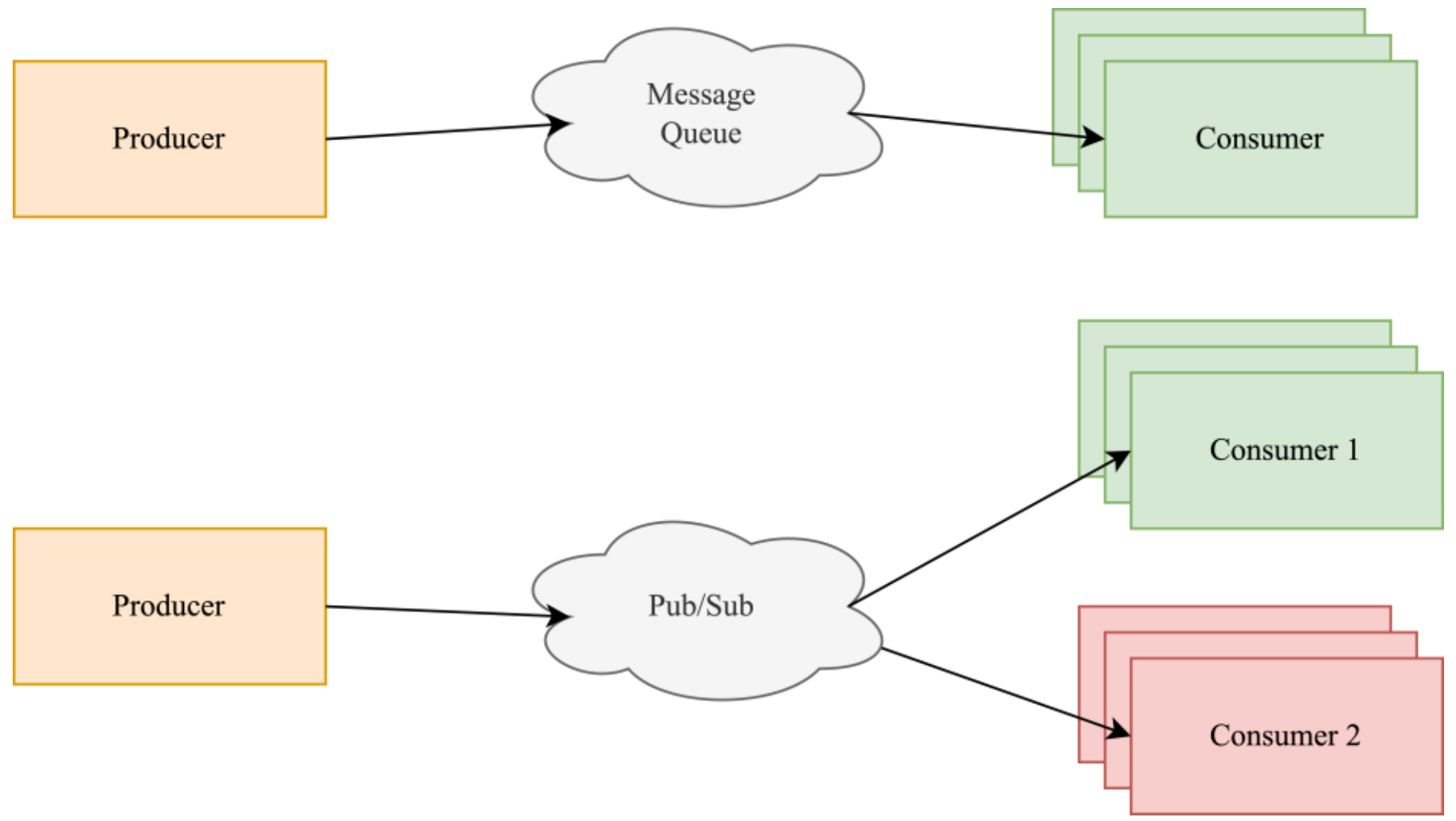

**Figure 90**: Message Queue v/s Pub/Sub.

Note that a "consumer" may also be a group, which may consist of multiple processes. These processes are typically replicas of the same executable and function identically. In case of message queues, a message will be delivered to exactly one process within this group, whereas, in case of pub/sub, a message will be delivered to exactly one process of each such group.

#### *19.1.1. Scenarios*

**Message Queue:** Consider an e-commerce scenario where customer orders are placed. These orders are enqueued for processing. A dedicated worker process retrieves each order from the queue and executes the necessary actions, such as deducting funds from the customer's account and forwarding the order to the supplier.

**Pub/Sub:** In an online shopping application, maintaining accurate inventory information is crucial. When an item's availability changes, a message is published by, say, a change detector. All micro-services that need inventory information (e.g., product pages, shopping cart) receive the message and update their states, ensuring consistent state across the system.

## 19.2. Delivery Guarantees

Messaging systems offer varying levels of delivery guarantees:

### *19.2.1. At-least-once Delivery (Recommended)*

This guarantee ensures that each message is delivered to consumers at least once, potentially multiple times. This simpler implementation allows for duplicate deliveries. To handle this, consumer actions must be idempotent - performing the same action multiple times has the same effect as performing it once. At-least-once delivery is widely supported (including by Kafka) and is the most common delivery guarantee.

For example, if a consumer maintains an in-memory key-value state and receives a new value for a key, applying that value updates the state; reapplying the same value has no further effect.

### *19.2.2. At-most-once Delivery*

This guarantee ensures that each message is delivered to consumers at most once. It might not be delivered at all. This is a weaker guarantee than at-least-once.

A simple approach would be "fire-and-forget" where the system attempts delivery once and then discards the message. This, however, may not handle failure scenarios or retry mechanisms effectively.

A more sophisticated approach to at-most-once delivery was proposed by Liskov in Practical Uses of Synchronized Clocks in Distributed Systems. This approach assumes synchronized clocks across systems with a maximum difference of **ε**. The receiver tracks the timestamp of the most recent message and discards older messages, accounting for the potential time difference.

### *19.2.3. Exactly-once Delivery (Transactional Delivery)*

This guarantee ensures that each message is delivered exactly once to the consumer. The consumer must complete the action associated with the message before sending an acknowledgment (ack).

The action itself must be transactional. If the ack fails, the effects of the action must be reversed, as the message will be redelivered. Conversely, the message cannot be acknowledged before the action is completed to prevent data loss.

Implementing exactly-once delivery is complex, requiring two-phase commit protocols and database support with locking mechanisms. This approach can be slow and may not handle high message volumes effectively. It's often recommended to use at-least-once delivery with idempotent operations or manual duplicate handling to achieve the desired effect without the complexity of true exactly-once delivery.

### 19.3. Ordering Guarantees

Messaging systems provide varying levels of message ordering guarantees:

- **Total Order:** Messages are delivered to consumers in the exact same order they were produced. This ensures strict sequential processing.
- **Partial Order:** Messages are delivered in a partially defined order. This order is typically determined for messages produced within a specific context, such as those originating from the same producer or within the same batch. Messages across different contexts may be delivered in any order relative to each other.
- **Out-of-Order:** Messages can be delivered to consumers in any order, regardless of their production sequence.

A key trade-off exists between ordering and performance. Out-of-order delivery can achieve lower latency because it avoids the need to wait for prior messages. In contrast, maintaining total or partial order can introduce head-of-line blocking, where subsequent messages are delayed while waiting for earlier messages to be processed.

### 19.4. Push vs Pull Model

Consumers generally have the option to either actively "pull" messages from a queue or passively receive them via a "push" mechanism, often implemented using a Remote Procedure Call (RPC) protocol. While push models offer lower latency, they can have lower throughput due to the overhead associated with processing each individual RPC, which consumes valuable computational resources. Conversely, pull-based models may introduce higher latency due to the need for periodic polling. However, they can improve throughput by enabling batch processing of messages, optimizing resource utilization.

Interestingly, many systems that internally support asynchronous push-based calls often implement their consumer libraries using a pull-based model under the hood. This design choice can significantly enhance overall throughput.

## 19.5. Pub/Sub Topics

In a pub/sub system, topics act as channels for routing messages to consumers. Producers (publishers) categorize each message by assigning it to a specific topic. Consumers (subscribers) then subscribe to the topics they're interested in and receive only the messages published to those topics.

To further enhance scalability and throughput, topics are often sharded into sub-topics (also called partitions). Producers write messages to one of these sub-topics, while consumers subscribing to the main topic receive messages from *all* of its sub-topics. This sharding allows different parts of the system (e.g., different servers or processes) to handle different sub-topics concurrently, thus distributing the workload and improving performance. Each topic shard (sub-topic) is typically managed by a distinct set of resources (e.g., jobs or processes) within the system.

## 19.6. Kafka Architecture

Kafka primarily functions as a pub/sub system.

### *19.6.1. Architecture*

Kafka utilizes a topic-based architecture. Each topic is further divided into multiple partitions. A dedicated server, known as a **broker**, is responsible for handling a specific partition. A broker may handle multiple such partitions.

### *19.6.2. Guarantees*

Kafka offers the following key guarantees:

- **At-least-once delivery:** Ensures that each message is delivered to the consumer at least once. In most scenarios, Kafka achieves exactly-once delivery. However, consumer restarts can potentially lead to duplicate message deliveries.
- **Partial ordering:** Guarantees that messages within the same partition are delivered to the consumer in the order they were produced.

### *19.6.3. Consumer Behavior*

Consumers can subscribe to specific topics. A consumer can create multiple "sub-streams" for a single topic. Messages destined for the topic are then multiplexed and delivered across these sub-streams.

Note: Kafka can also operate as a traditional message queue when there is only a single consumer for a given topic.

## 19.7. The Broker Layer

Brokers store messages in log files. Each log file is segmented into approximately 1 GB files. Messages are appended sequentially to the current segment file. When a segment file reaches its maximum size, it "rolls over" and a new segment file is created.

To optimize performance, message flushes to the segment file are performed periodically by a background process, amortizing the cost of disk I/O. A message is considered committed for delivery only after it has been successfully flushed to the segment file.

### *19.7.1. Message Addressing and Ordering*

Each message is uniquely identified by its "offset", a logical position within the log stream. This offset serves as the message ID. Consumers consume messages from a topic partition sequentially, ensuring total ordering within the partition and partial ordering within a topic. If a topic has only one partition, it guarantees total order for all messages within that topic.

### *19.7.2. Broker Statelessness*

When a consumer receives a message, it can independently calculate the offset of the next message. This client-side calculation eliminates the need for broker involvement in message tracking, allowing the broker to maintain a stateless architecture.

### *19.7.3. Message Deletion*

To manage storage space, brokers employ a simple time-based policy for message deletion. Typically, messages are retained for a period of 7 days before being automatically deleted.

## 19.8. Consumer Groups

Kafka employs the concept of "consumer groups", where a group consists of multiple consumers that collectively subscribe to a single topic. Each message within a topic is supposed to be delivered to exactly one consumer *within the group*. This does not guarantee exactly-once delivery across the entire system, as will be discussed later.

Kafka itself does not actively manage message delivery to specific consumers within a group. This responsibility lies with the consumers themselves, who must coordinate their actions. To facilitate this coordination, consumers leverage ZooKeeper, a distributed coordination service. ZooKeeper enables consumers to acquire and maintain exclusive ownership of specific partitions within a topic.

On a side note, ZooKeeper's ownership mechanism relies on timing assumptions and is not entirely foolproof.

### *19.8.1. Partition Assignment and Coordination*

Consumers coordinate to determine which partitions each member of the group will consume. This involves:

- **Maintaining Registries:** ZooKeeper hosts three key registries:
  - **Consumer Registry:** Tracks all consumers belonging to a particular group.
  - **Ownership Registry:** Records the "owner" (consumer) for each partition within the topic.
  - **Offset Registry:** Stores the offset of the last message consumed from each partition.
- **Partition Assignment:** When the consumer registry changes (e.g., a consumer joins or leaves the group), all consumers are notified by ZooKeeper. They then execute a deterministic algorithm to re-assign partitions among themselves. This algorithm ensures that all consumers arrive at the same partition assignment, regardless of the order in which they execute the algorithm. The algorithm divides the partitions into **N** groups (where **N** is the number of consumers) and assigns partitions to consumers in a round-robin fashion based on their IDs.

### *19.8.2. Offset Commits and Message Re-delivery*

Consumers are responsible for committing their offsets to the offset registry. If a consumer fails before committing an offset, the message associated with that offset may be re-delivered to a different consumer within the group.

For example, in Algorithm 1, the critical step involves assigning ownership of partition $\mathbf{p}$ to consumer $\mathbf{C}_i$ . This is the lock acquisition phase, mediated by ZooKeeper. ZooKeeper ensures that only the designated owner, $\mathbf{C}_i$ , can subsequently update the offset for partition $\mathbf{p}$ in the registry. However, existing owners might continue processing messages before they become aware of the ownership transfer (ZooKeeper cannot guarantee that). Therefore, it is crucial that message handling logic within each consumer is idempotent.

Achieving true exactly-once delivery is challenging and often impractical. The Kafka authors recommend relying on the at-least-once delivery guarantee.

## 19.9. Transfer Efficiency

Traditionally, sending data from local files to remote sockets involves a series of costly data copies:

- **File to Operating System Memory:** Data is read from the file system into an operating system memory page.
- **Operating System Memory to Application Memory:** The data is copied from the operating system memory page into the application's memory space.
- **Application Memory to Kernel Buffer:** The data is copied again, this time from the application's memory to a kernel buffer.
- **Kernel Buffer to Socket:** Finally, the kernel sends the data from the buffer to the network socket.

To circumvent these redundant copy operations, Kafka leverages the **sendFile** system call. **sendFile** enables the direct transfer of data from the file system to the network socket, bypassing the intermediate copies and significantly improving performance.

## 19.10. Evaluation

The paper's evaluation primarily focuses on a specific set of use cases. The authors acknowledge that their system achieved significant performance gains due in part to a reduced feature set compared to other systems (e.g., ActiveMQ, RabbitMQ).

- **Producer Throughput:** Exceeded 400k messages per second, reaching 800 Mbps, significantly higher than other systems.
- **Consumer Throughput:** Surpassed 20k messages per second.

Kafka lacks producer-side acknowledgments. Producers can send batches of messages without receiving individual acknowledgments from the broker. This omission can lead to data loss if the client crashes before committing the messages to Kafka.

Kafka utilizes the Avro [154] serialization protocol (similar to Thrift [155] and Protocol Buffers [156]). Avro offers significantly lower overhead (9 bytes) compared to other protocols (144 bytes).

## 19.11. Bonus: Related Systems

### *19.11.1. Message Queues*

- **RabbitMQ** [157]**:** A widely-used open-source message broker.
- **Apache ActiveMQ** [158]**:** Implements the Java Message Service [159] specification.

### *19.11.2. Publish-Subscribe Systems*

- **Google Cloud Pub/Sub** [160]**:** A cloud-based pub/sub service supporting both push and pull delivery models.
- **Apache Pulsar** [161]**:** A popular streaming platform optimized for geo-replication, enabling efficient data distribution across geographically dispersed services. Kafka can support the same through some extensions.

### *19.11.3. Legacy Systems*

The paper mentions other systems, such as Facebook Scribe [162] and Yahoo's Data Highway, which are no longer actively maintained.

### 19.12. Paper Remarks

This paper is easy to follow and serves as an excellent introduction to several fundamental concepts in distributed systems. It is a highly recommended resource for those seeking to understand the architectural trade-offs involved in high-throughput messaging.

# 20. Paxos Made Simple

(20) Lamport, L. Paxos Made Simple. ACM SIGACT News (Distributed Computing Column) 32, 4 (Whole Number 121, December 2001) 2001, 51–58.

We now turn our attention to a fundamental concept in computer science: consensus. Specifically, how can different computers in a distributed setup agree on a common value? In this context, there is no objective "right" or "wrong" - there is only agreement. To break the suspense: to date, there has not been a single "perfect" consensus algorithm. However, it is worth mentioning that solving this problem would revolutionize the world of distributed computing by making it significantly simpler. If we can make computers agree on a single value, we can make them agree on N values. Since each value can represent an instruction (as simple as a CPU instruction), consensus allows us to build perfectly replicated systems.

Because no consensus algorithm is perfect, each one incorporates specific assumptions about the system to simplify the problem.

In this paper, we will examine Paxos, a cornerstone consensus protocol within distributed systems. The title of the paper might imply a gentle introduction, however, Leslie Lamport's paper provides a comprehensive, dense treatment of Paxos.

In this insight, we will begin with some basic definitions related to distributed systems and explore consensus through a formal definition. We will then examine one of the most important results in computer science: the FLP impossibility. From there, we will dive deeper into Paxos to see how it achieves consensus under practical assumptions and how it can be used to build replicated state machines. In the bonus section, we will contrast Paxos with Raft, another consensus algorithm that has become widely popular and influential.

### 20.1. Basic Definitions

#### *20.1.1. Failure Types*

Two primary failure modes are commonly observed in distributed systems:

- **Crash Failures:** These occur when a node terminates unexpectedly, potentially due to hardware malfunctions, software errors, or power outages.
- **Byzantine Failures:** These involve nodes exhibiting arbitrary or malicious behavior, such as sending incorrect data or deviating from protocol specifications.

Handling Byzantine failures are crucial in internet-scale systems and decentralized protocols like blockchains. However, distributed systems within controlled environments, such as those within an organization's network, only need to think about crash-fault tolerance. This simplification assumes non-malicious node behavior, which is practical when nodes are under a single administrative domain.

#### *20.1.2. Network Types*

Furthermore, we distinguish between:

- **Asynchronous Network:** These networks have no upper bound on packet delivery times.
- **Synchronous Network:** These networks assume bounded packet delivery times.

Real-world networks are inherently asynchronous. Synchronous systems are primarily theoretical constructs. **Partially synchronous networks**, relevant for Byzantine fault tolerance, are outside the scope of this discussion.

### 20.2. Consensus

Consensus algorithms address the fundamental challenge of achieving agreement on a single value among distributed nodes. These algorithms must satisfy the following properties:

- **Termination (Liveness):** Every non-faulty node eventually decides on a value.
- **Integrity (Safety):** If all non-faulty nodes initially propose the same value, that value must be the decided value.
- **Agreement (Safety):** All non-faulty nodes agree on the same decided value.

Consensus algorithms are essential for distributed systems, serving as the foundation for coordination services like ZooKeeper. Their role is indispensable in building reliable distributed applications.

## 20.3. FLP Impossibility

Having defined consensus, we now examine the FLP (Fischer-Lynch-Paterson) impossibility result, as presented in "Impossibility of Distributed Consensus with One Faulty Process" [163]. This theorem establishes that -

**In an asynchronous network, no consensus protocol can simultaneously guarantee safety, liveness, and fault tolerance.**

The original paper provides a rigorous mathematical proof under the assumption of an asynchronous network where messages may be arbitrarily delayed and reordered but are eventually delivered exactly once. This is still the best-case assumption of asynchronous networks, which are often less reliable in practice. The FLP theorem demonstrates that **even with a single process crash fault, consensus cannot be achieved**.

Given the prevalence of crash faults and the inherent asynchronicity of real-world networks, the FLP result implies that no consensus algorithm really exists!

### *20.3.1. FLP Impossibility vs. CAP Theorem*

Although both FLP Impossibility and CAP Theorem address limitations in distributed systems, they differ significantly.

FLP states that in an **asynchronous network with eventual message delivery**, termination cannot be guaranteed if even one process may crash. CAP, conversely, addresses distributed data stores and posits that in a **network with potential message loss**, consistency and availability cannot be simultaneously guaranteed.

The key difference lies in their network assumptions. CAP makes stronger assumptions than FLP. Note that weaker assumptions lead to stronger impossibility results. FLP's weaker assumptions (eventual message delivery) yield a more powerful and widely recognized result.

**Recommended Read:** A Brief Tour of FLP Impossibility [164] for a deeper insight.

## 20.4. Practical Consensus Algorithms

Despite the theoretical limitations imposed by the FLP impossibility result, practical consensus algorithms have been developed that perform effectively in most real-world scenario, as listed in Table 2. These algorithms represent compromises that prioritize practicality.

While compromises are made, safety remains paramount. Liveness is often relaxed, relying on probabilistic termination, which typically occurs with sufficient randomness.

**Table 2**: List of a few consensus protocols.

| | Network | Safety | Liveness | Crash Faults | Byzantine Faults |
|---|---|---|---|---|---|
| **Paxos** | Asynchronous | Yes | No | Yes | No |
| **Raft** | Asynchronous | Yes | No | Yes | No |
| **Dolev-Strong** | Synchronous | Yes | Yes | Yes | Yes |
| **Honeybadger** | Asynchronous | Yes | No* | Yes | Yes |
| **PoW** | Asynchronous | No* | No* | Yes | Yes |

* - Actually the property does hold but is subject to probability. In PoW blockchain, for example, there's a tiny chance that two blocks might be created simultaneously, leading to a temporary fork in the blockchain. However, the protocol is designed so that the chain with the most accumulated work will eventually become the canonical chain, making disagreements highly improbable.

This chapter focuses on crash-fault tolerant algorithms, specifically **Paxos** and **Raft**. These algorithms are widely applicable to private network-based systems, which constitute a significant portion of industrial deployments.

## 20.5. Paxos

Following the introductory remarks, we will now delve into a comprehensive explanation of the Paxos consensus protocol. Paxos is fundamentally a consensus algorithm designed to function within an asynchronous network environment. This algorithm operates under the assumption that the system experiences **only crash faults** and does not exhibit Byzantine faults. The primary objective of Paxos is to enable agreement among nodes while guaranteeing safety, though not guaranteeing strict liveness.

*20.5.1. Node Roles*

Within the Paxos algorithm, each node is assigned one of three distinct roles:

- **Proposer:** This node initiates proposals for values that the system should agree upon.
- **Acceptor:** This node responds to proposals and votes on proposed values.
- **Learner:** This node learns the agreed-upon value.

*20.5.2. The Algorithm*

The overarching goal of this algorithm is to ensure that all participating nodes agree on a single value, denoted as **v**. The algorithm proceeds in two primary phases, as shown in Figure 91:

**PREPARE Phase**

- A proposer selects a unique proposal number, **N**, and sends a **PREPARE** request to all acceptors. Each acceptor, upon receiving a PREPARE request, may choose to accept the proposal number **N** if, and only if, **N** is greater than all proposal numbers it has previously accepted.
- If an acceptor accepts the proposal number **N**:
    - It promises not to accept any proposal number less than **N** in the future.
    - It sends a reply back to the proposer, which includes the highest proposal (along with any value) less than **N** that it has previously accepted.

**ACCEPT Phase**

- Once the proposer receives a majority of responses from acceptors, it associates the value **V** with the proposal number **N** and sends an **ACCEPT** request to all acceptors. The value **V** being associated with **N** has two possibilities:
    - The value **V** is the value that the proposer initially intended to propose. This is the ideal scenario and occurs when no other proposer has been able to complete the ACCEPT phase.
    - The value **V** is the value that has been previously accepted by a majority of acceptors.

- All acceptors that have accepted the proposal number **N** then associate the value **V** with **N**, given that they have not accepted another proposal number higher than **N**.

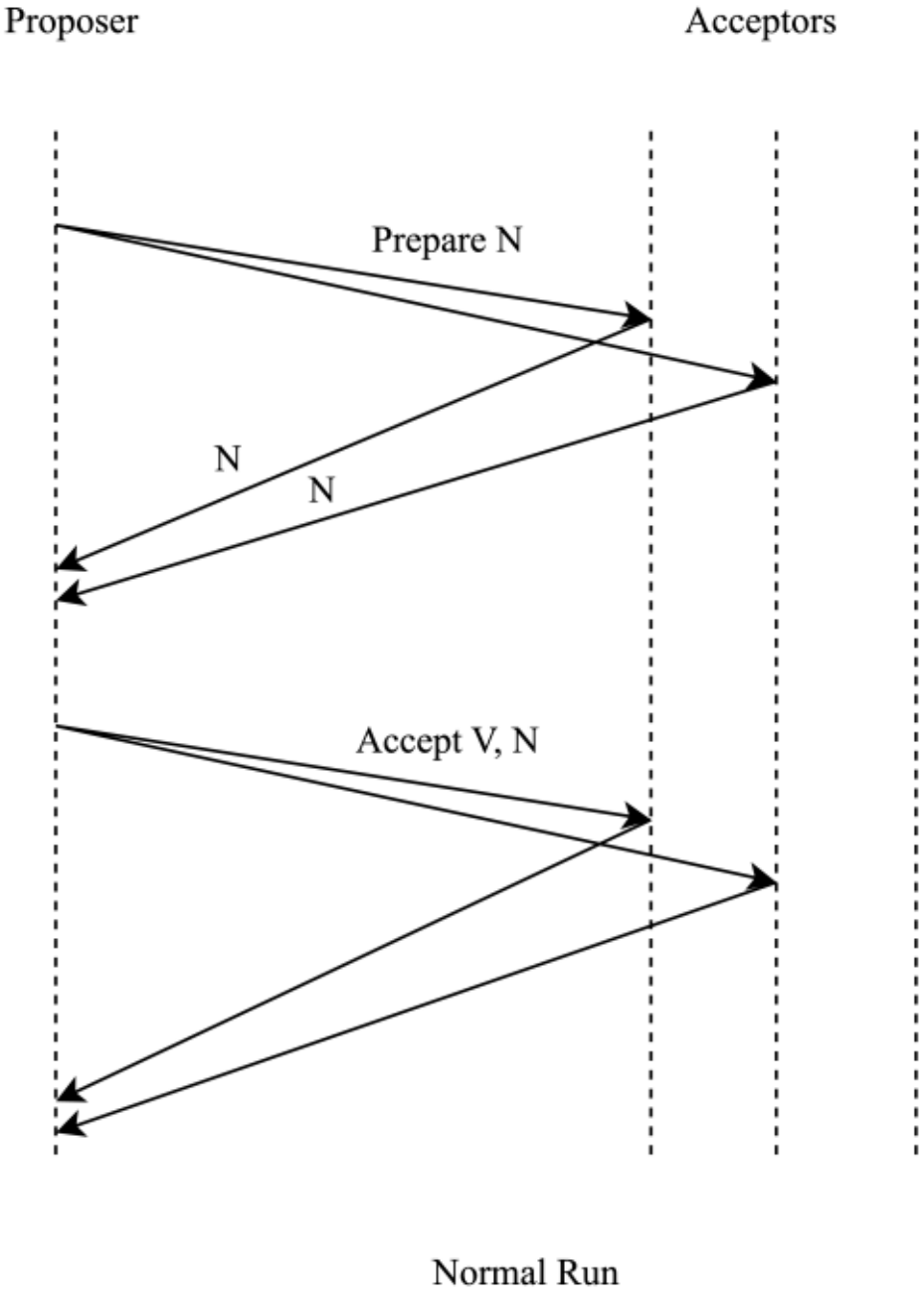

**Figure 91**: Paxos algorithm.

*20.5.3. Safety*

**The algorithm guarantees that the first accepted value (corresponding to any proposal number) will be the only accepted value (for any higher proposal number as well).**

Let's start with the requirement **P2(c)** from the paper and work backwards. **P2(c)** states that once a majority of acceptors have accepted a proposal number **N** and a corresponding value **V**, that value **V** is committed. This happens as follows (intuitive from the algorithm as well):

- The proposer, in the PREPARE phase, learns the highest proposal number that acceptors have already accepted. If the proposer's proposed number is less than the highest number known to the acceptors, it cannot proceed and has to try again.

- The acceptors promise (see **P1(a)**), not to accept any proposal number less than **N** in the future.
- As a result, once a majority of acceptors accept a proposal number **N**, it can be safely assigned a value **V**.

Using **P2(c)**, we can now derive **P2(b)** which states that once a value **V** is chosen for a number **N**, then all proposers with a proposal **M > N** will use the same value **V**:

- Say a second proposer with **M** begins just after **N** completes the ACCEPT phase. There would be at least one acceptor common between them due to majority. That acceptor will have the value **V** for number **N**.
- This forces the second proposer to use the same value as the first.

This scenario is shown in Figure 92.

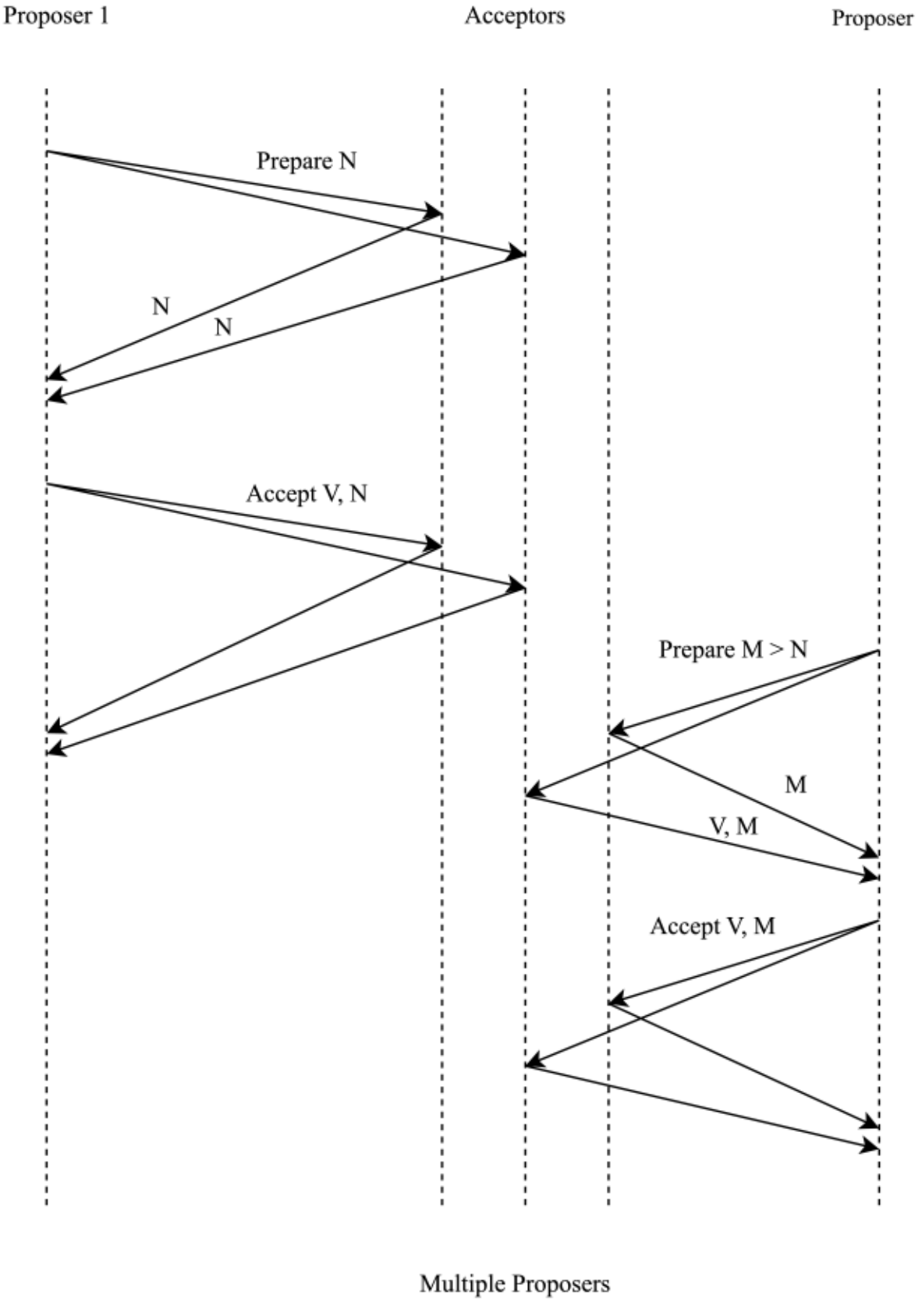

**Figure 92**: Multiple proposers in Paxos.

This is exactly what is proven inductively on page 3, last paragraph of the paper.

**P2(a)** is implied by **P2(b)**, and **P2** is implied by **P2(a)** which represents the total safety guarantee.

20.5.3.1. Optimization

To optimize the algorithm, acceptors can choose to remember only the highest proposal number they have accepted. **This is because all subsequent proposals after the first successful acceptance will result in the same value.** Additionally, once an acceptor has responded in the PREPARE phase, it has already sent back the old known value, thus it doesn't need to remember that value anymore. The responsibility of remembering this value shifts to the proposer.

It is important to note that the goal of consensus is to agree on a single value, not on a single proposal number. Different proposers may result in different proposal numbers being accepted by different acceptors (due to majority-only considerations). However, in the end, all acceptors will have the same value (i.e., the first accepted value) assigned to all their known highest proposal numbers.

*20.5.4. Liveness*

The liveness of the Paxos algorithm can be compromised in scenarios involving racing proposers.

For example, if an acceptor accepts a proposal number **N** but, before it receives a value for **N**, it receives and accepts a proposal number greater than **N**, then the ACCEPT phase will fail for the first proposer.

In such situations, the first proposer must retry with an even higher proposal number. However, if two or more proposers continuously race in this manner, each proposing higher numbers but never completing the ACCEPT phase, the system may never terminate.

To mitigate this issue, Lamport suggests using a single proposer in the system. This proposer can be elected through a consensus process (which is itself a form of value agreement). After the proposer is elected, it becomes responsible for all future proposals.

However:

- The initial election of the proposer (a.k.a. the leader election), can itself never terminate.
- Furthermore, if the single proposer fails, the system may become stuck until the Proposer recovers.

To address these scenarios and help the system make progress, timeouts and some randomness can be used. While it's still theoretically possible for the system to stall, this approach usually works in practice. **Even with the introduction of timeouts to fix liveness issues in practical scenarios, the safety of Paxos is never compromised**, which is a key feature of the algorithm.

### *20.5.5. Fault Tolerance*

The Paxos algorithm is designed to tolerate crash faults. Acceptors must persist the highest proposal number they have accepted and the associated value if the ACCEPT phase completes. Proposers must remember their highest proposal number and any values they receive during the PREPARE phase.

The algorithm's reliance on a majority of acceptors ensures that the failure of some nodes does not affect liveness. Additionally, because any node can become a proposer, and Paxos can handle multiple proposers safely, the failure of a proposer does not affect the algorithm.

In conclusion, Paxos is fault-tolerant to crash faults, provided that a majority of nodes remain operational. If a majority of nodes fail, no consensus algorithm can guarantee agreement anyway.

## 20.6. Replicated State Machines

Having established that Paxos can facilitate agreement on a single value among distributed nodes, we can extend its application to construct a **replicated state machine (RSM)**.

An RSM involves multiple machines executing an identical sequence of instructions. This architecture provides enhanced fault tolerance compared to a standalone state machine (a single computer). The instructions for an RSM are stored in a distributed log.

*20.6.1. Distributed Logs*

A distributed log is a sequence of ordered instructions that an RSM executes. Each log entry represents an instruction for the machines, ensuring that all nodes in the RSM execute these instructions in the same order.

For example, Figure 93 shows an RSM with a single variable **X** stored in it. The instructions for the state machines can be **blind-writes (X := Y)** or **read-writes (IF v(X) = Y THEN X := Z)**.

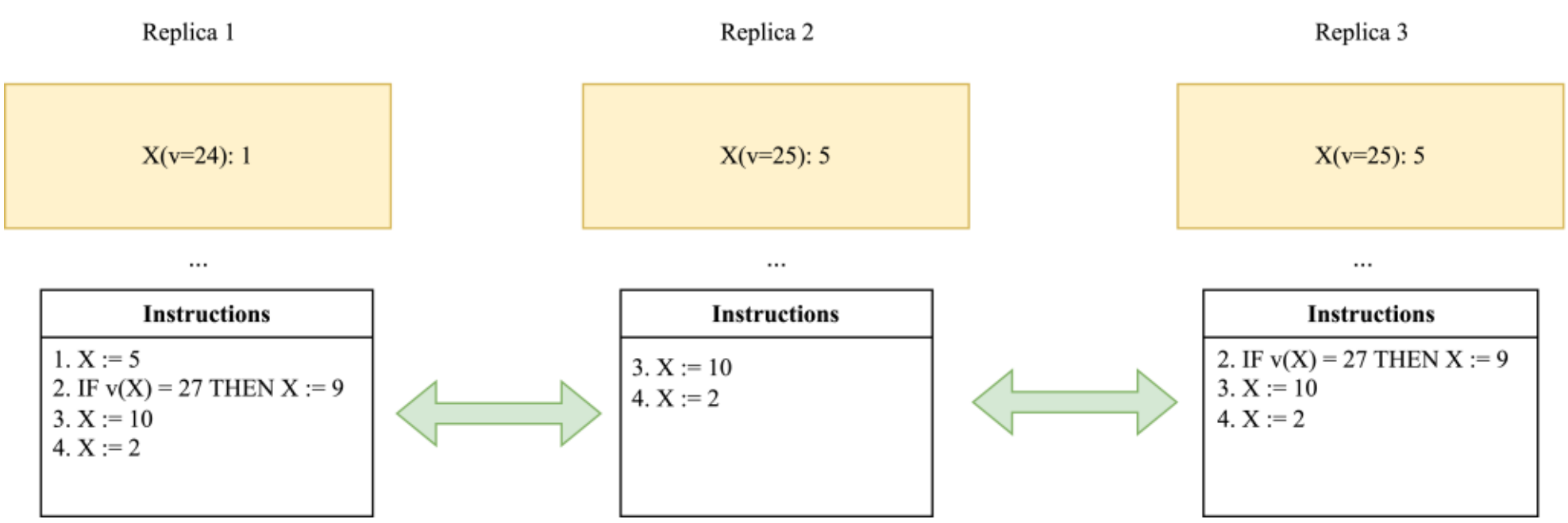

**Figure 93**: Distributed logs in RSM.

Note that different nodes of an RSM may be at different positions in the execution log but will finally converge to the same final state.

*20.6.2. Generating Distributed Logs Using Paxos*

Consensus algorithms, such as Paxos, can be employed to create distributed logs:

- The log is segmented into distinct entries.
- Each entry is treated as a value to be proposed and accepted via the consensus algorithm.
- Once an entry is proposed and accepted, it becomes the definitive value for that log position.

Paxos guarantees that once a value is accepted for a log entry, it remains unchanged, regardless of subsequent proposals for that entry. For a distributed log:

- When a proposer intends to append an entry to the log, it selects the next available position in the log and executes the Paxos algorithm to establish a value for that entry.

- If the proposer fails to insert its value into the chosen position, possibly due to a competing proposer successfully establishing its value first, the proposer retries with the subsequent available slot. This process continues until the proposer successfully adds an entry. If several attempts fail, the proposer may communicate this failure to the client.

This mechanism underscores the importance of Paxos's immutability property: once a value is accepted, it remains unchanged, regardless of future proposals for the same log position.

#### 20.6.2.1. Optimization

Executing both phases of the Paxos algorithm (PREPARE and ACCEPT) in real-time can introduce latency. To mitigate this, proposers can proactively execute the prepare phase to reserve log positions and subsequently propose values for them. This optimization is particularly effective when a single proposer reserves several slots in advance, significantly enhancing performance.

### 20.7. Bonus: Paxos v/s Raft

Raft is often perceived as more understandable than Paxos. I've consistently found Paxos to be significantly clearer, more logical, and concise in its presentation.

Raft [165] was authored by Ongaro and Ousterhout from Stanford University and presented at USENIX Annual Technical Conference '14. It is a very impactful paper as many systems like CockroachDB [166], MongoDB, and Kafka now use Raft.

The primary distinction between Paxos and Raft lies in their handling of proposers. In Paxos, concurrent proposers are permissible. Conversely, Raft designates a single leader responsible for log replication.

Here's a concise overview of Raft's operation:

**Leader Election**

- Raft elects a single designated leader to manage the system.
- In the event of leader failure, the remaining servers initiate an election to select a new leader.
- This election process employs timeouts and a voting mechanism to ensure the selection of a unique leader.

**Log Replication**

- The leader receives client requests and appends them as entries to its log.
- The leader then replicates these log entries to the follower servers.
- Upon receiving acknowledgments from a majority of followers for a log entry, the leader commits that entry.
- Committed entries are subsequently applied to the system's state.

**20.8. Paper Remarks**

This paper is notoriously challenging; its complex concepts require careful study and often several readings to fully comprehend. Despite its difficulty, it remains a cornerstone of computer science. It is a mandatory resource for distributed systems enthusiasts who wish to understand the fundamental logic governing the field.

# 21. ZooKeeper: Wait-Free Coordination for Internet-Scale Systems

(21) Hunt, P.; Konar, M.; Junqueira, F. P.; Reed, B. ZooKeeper: Wait-Free Coordination for Internet-Scale Systems. In 2010 USENIX Annual Technical Conference (USENIX ATC 10); 2010.

Having learned how replicated state machines can be created, it is time to put them into use. However, using distributed logs for every operation in a distributed system would make the architecture far too complex. Instead, we use distributed logs to develop a "coordination kernel". The goal of a coordination kernel is to handle the inherent quirks of distributed environments. All higher-level systems can then leverage this kernel to simplify their replication and redundancy strategies, significantly reducing their overall complexity.

This paper is one of the most significant contributions to the field of Distributed Systems. The work can be seen as an open-source counterpart to Google's Chubby [122]. Indeed, ZooKeeper's design is based on principles similar to Chubby, however, it also incorporates key differences as we shall we. This work was presented at USENIX Annual Technical Conference (ATC) '10. At the time of its publication, decentralized approaches, such as those using consistent hashing (e.g., Dynamo), were waning in popularity, and centralized systems were gaining traction. Both Google's Chubby and Apache ZooKeeper emerged during this shift.

In this insight, we will begin by revisiting distributed logs and examining how they are used to build single-item transactional systems that offer linearizability - the strongest consistency model. We will then explore another fundamental problem in distributed systems, atomic broadcast, and see how it is equivalent to the consensus problem. Finally, we will analyze the design of ZooKeeper in detail, exploring the architectural trade-offs it makes in light of the FLP impossibility result. We will also examine the various primitives that can be built using ZooKeeper; these primitives serve as the essential foundation for higher-level distributed systems, including distributed databases and messaging systems.

**Recommended Read**: **20. Paxos Made Simple** where Paxos, a consensus algorithm, was discussed in detail.

## 21.1. Distributed Logs

Distributed systems rely on distributed logs, which serve as the single source of truth for the operations performed. These logs are crucial for replicated state machines (RSM), where, to maintain consistency across nodes, it's essential that they execute the same set of operations in a consistent order. Typically, the operations recorded in a distributed log correspond to modifications of data stored on a node.

Figure 94 depicts three nodes, each holding a versioned data item, X. These replicas sequentially apply instructions from the distributed log. Instructions 1, 3, and 4 are blind-writes. Instruction 2 is read-write. While the state of the memory location holding X may differ across replicas at any given point in time, once all instructions from the log have been applied, all replicas will converge to the same final state.

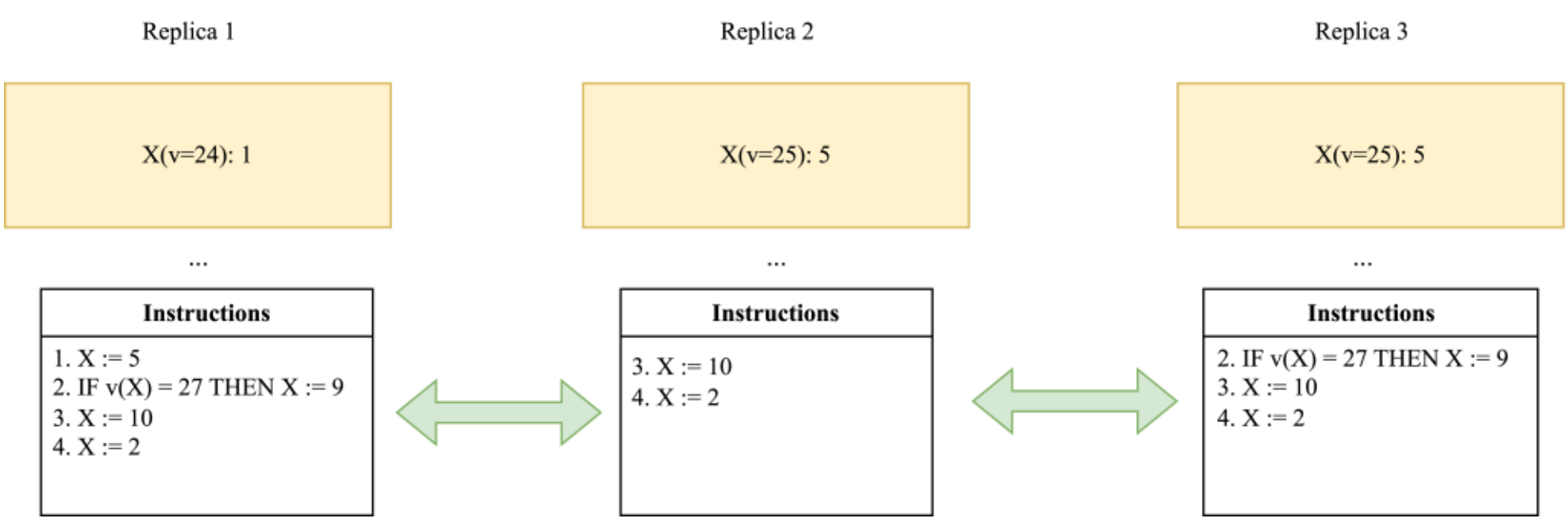

**Figure 94**: Distributed logs in RSM.

## 21.2. Single-Data Transactions

Transactions, in a distributed system, can take two forms:

- **Single-data transactions:** These transactions operate on a single object atomically.
- **Multi-data transactions:** These transactions affect multiple objects atomically.

This chapter focuses solely on single-data transactions. Multi-data transactions will be covered in future chapters on databases.

### *21.2.1. Implementation using Distributed Logs*

Single-data item transactions can itself take two forms:

- **Blind write (W):** A direct assignment of a value.
- **Read-then-write (R-W):** Reading a value and subsequently writing a new value. There may be some decision making based on the value read.

Distributed logs can implement both types of single-data item transactions. For a blind write (**W**), the log entry might be represented as

$$\textbf{key} := \textbf{value}$$

indicating that **value** should be assigned to the data item identified by the **key**. For a read-then-write (**R-W**), the log entry could be

$$\textbf{IF v(key)} = \textbf{version THEN key} := \textbf{value}$$

signifying a compare-and-set (CAS) operation: the value of the data item associated with **key** is updated to **value** only if the current version matches the **version**. The **version** signifies the value that was read.

Note that distributed logs are effective for single-data item transactions only, where the entire data item resides on the node(s) responsible for it. For multi-item transactions (or general database transactions), a combination of distributed logging and two-phase commit (2PC) is required. The discussion of 2PC is beyond the scope of this chapter.

#### *21.2.2. Consistency Models*

There are two ways to think about a consistency model in a single data item read-write system:

- Let's say there are clients observing **all** writes that are happening - what is the effective order in which the clients will see the writes?
- What would be the effective state of the objects in the data store once all the writes have been applied?

Several types of ordering are possible:

- **Total Order:** All transactions are ordered linearly and there is only a single valid order. All clients will see that same order and there will be only a single effective state.
- **Causal Order:** A partial order based on event causality. For example (shown in Figure 95), if client $\mathbf{C_1}$ performs write $\mathbf{W_1}$ and then sends a message to clients $\mathbf{C_2}$ and $\mathbf{C_3}$, and subsequently $\mathbf{C_2}$ performs write $\mathbf{W_2}$ and $\mathbf{C_3}$ performs write $\mathbf{W_3}$, then, according to Lamport's happened-before relationship, $\mathbf{W_1}$ happened before both $\mathbf{W_2}$ and $\mathbf{W_3}$. Therefore, $\mathbf{W_1}$ must appear before $\mathbf{W_2}$ and $\mathbf{W_3}$ in any valid causal order. Both $\{\mathbf{W_1}, \mathbf{W_2}, \mathbf{W_3}\}$ and $\{\mathbf{W_1}, \mathbf{W_3}, \mathbf{W_2}\}$ are valid causal orderings, but $\{\mathbf{W_2}, \mathbf{W_1}, \mathbf{W_3}\}$ is not valid.

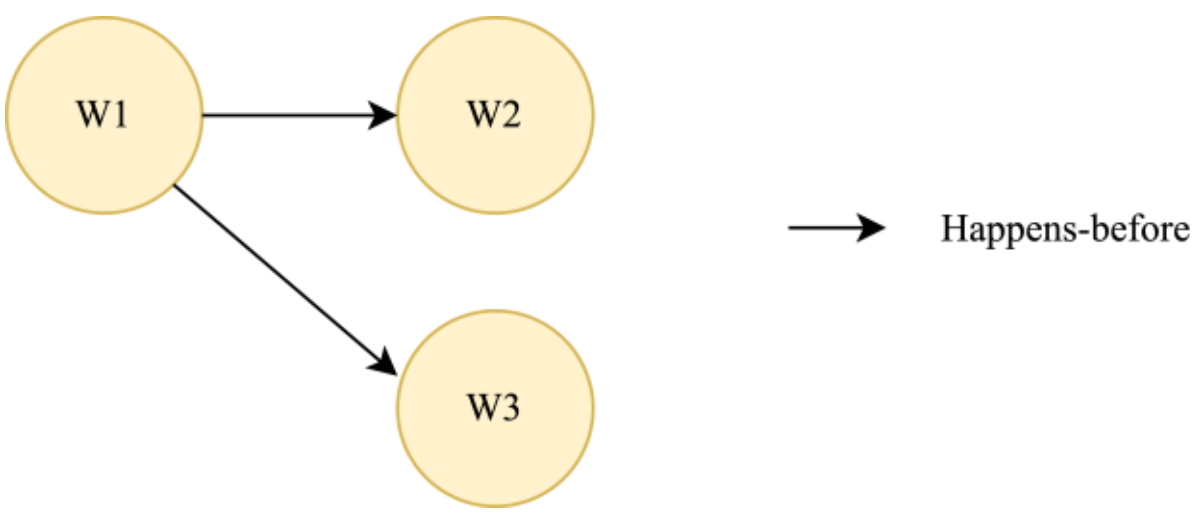

**Figure 95**: Causal order example.

- **FIFO Order:** A partial order ensuring that writes from the same client are ordered. For example (shown in Figure 96), if a client performs write $\mathbf{W_1}$ and then write $\mathbf{W_2}$, the final order must have $\mathbf{W_1}$ before $\mathbf{W_2}$. Writes from other clients (like $\mathbf{W_3}$ in the previous example) can occur at any point in the sequence relative to $\mathbf{W_1}$ and $\mathbf{W_2}$.

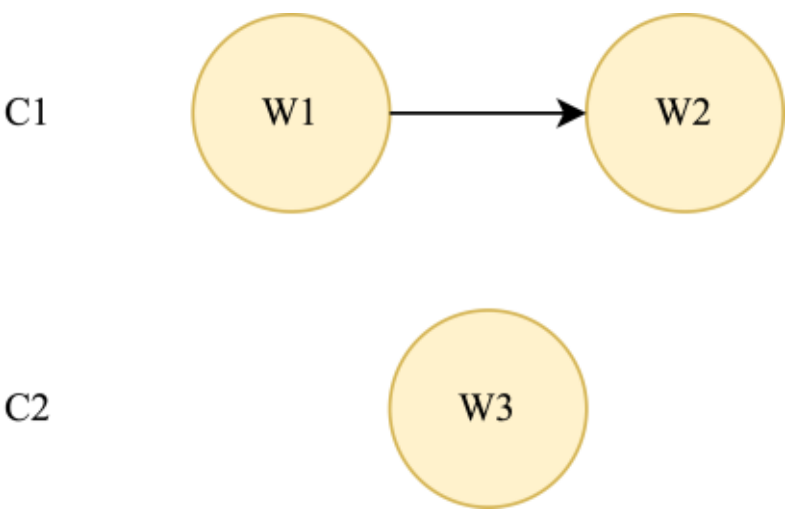

**Figure 96**: FIFO order example.

In this chapter, we will focus on total ordering only. For a rigorous insight into causal order, refer **27. Don't Settle for Eventual: Scalable Causal Consistency for Wide-Area Storage with COPS**.

#### 21.2.2.1. Linearizability (External Consistency) - Temporal Total Order

The strongest single-object consistency model, linearizability, guarantees immediate external visibility of transaction results (a.k.a. external consistency).

If a client executes transaction $\mathbf{T_1}$ and, only after its successful commit, executes transaction $\mathbf{T_2}$, then linearizability requires that the actual order of these transactions must reflect this client-observed real-time order: $\mathbf{T_1}$ must be ordered before $\mathbf{T_2}$.

Linearizability simplifies client reasoning about the system's behavior.

#### 21.2.2.2. Sequential Consistency - Total Order

Sequential consistency is a strong consistency model ensuring that operations appear to have occurred in some total order.

A helpful analogy illustrates the difference between linearizability and sequential consistency:

Suppose you successfully commit a transaction that adds a new item to a key-value store. You then call a friend, asking them to look up the new data item. If your friend's read fails, the store violates linearizability (external consistency).

Many distributed datastores, including ZooKeeper (with its relaxed reads, as we'll discuss), do not guarantee linearizability. However, they still maintain sequential consistency, ensuring that the final state reflects *some* total order of the executed transactions. For example, if your friend then attempts to add the same data items again (with the same key), the store must reject either yours or your friend's transaction depending on whose transaction the store decides to order first. This avoids duplicate keys and thereby makes the store at least sequentially consistent.

In other words, sequential consistency guarantees a total order of transactions without strict temporal (or externally-visible) semantics.

#### 21.2.2.3. Causal Consistency - Causal Order

At this level, the total order of transactions is broken, and only a partial order can be maintained based on causality. Causal consistency orders transactions according to their causal relationships. This weaker consistency model is **available during network partitions**.

See **27. Don't Settle for Eventual: Scalable Causal Consistency for Wide-Area Storage with COPS** to learn more about causal consistency.

21.2.2.4. PRAM Consistency (FIFO Consistency) - FIFO Order

PRAM consistency maintains partial ordering based on the principle that writes executed by the same process are totally ordered.

21.2.2.5. Eventual Consistency

Lastly, there are other consistency models, such as read-your-writes (RYW), and eventual. See **28. TAO: Facebook's Distributed Data Store for the Social Graph** to learn more about them.

21.2.2.6. Bonus: Strict Serializability

While this discussion focuses on single-item reads and writes, it's worth noting that extending linearizability to multi-item transactions yields strict serializability.

Figure 97 shows the different consistency models discussed so far.

### 21.3. Distributed Log Algorithms

Now that we understand the role of distributed logs in implementing replicated systems capable of single-data item transactions, let's explore the algorithms that enable their creation.

In a distributed system, nodes that function correctly (i.e., are non-crashing and non-Byzantine) are considered *non-faulty* nodes.

#### *21.3.1. Consensus*

A consensus algorithm addresses the problem of agreeing on a common value among the nodes. Such an algorithm must satisfy the following properties:

- **Termination (Liveness):** Every non-faulty node eventually decides on a value.
- **Integrity (Safety):** If all non-faulty nodes initially proposed the same value, then all correct nodes must decide on that value.
- **Agreement (Safety):** All non-faulty nodes must agree on the same value.

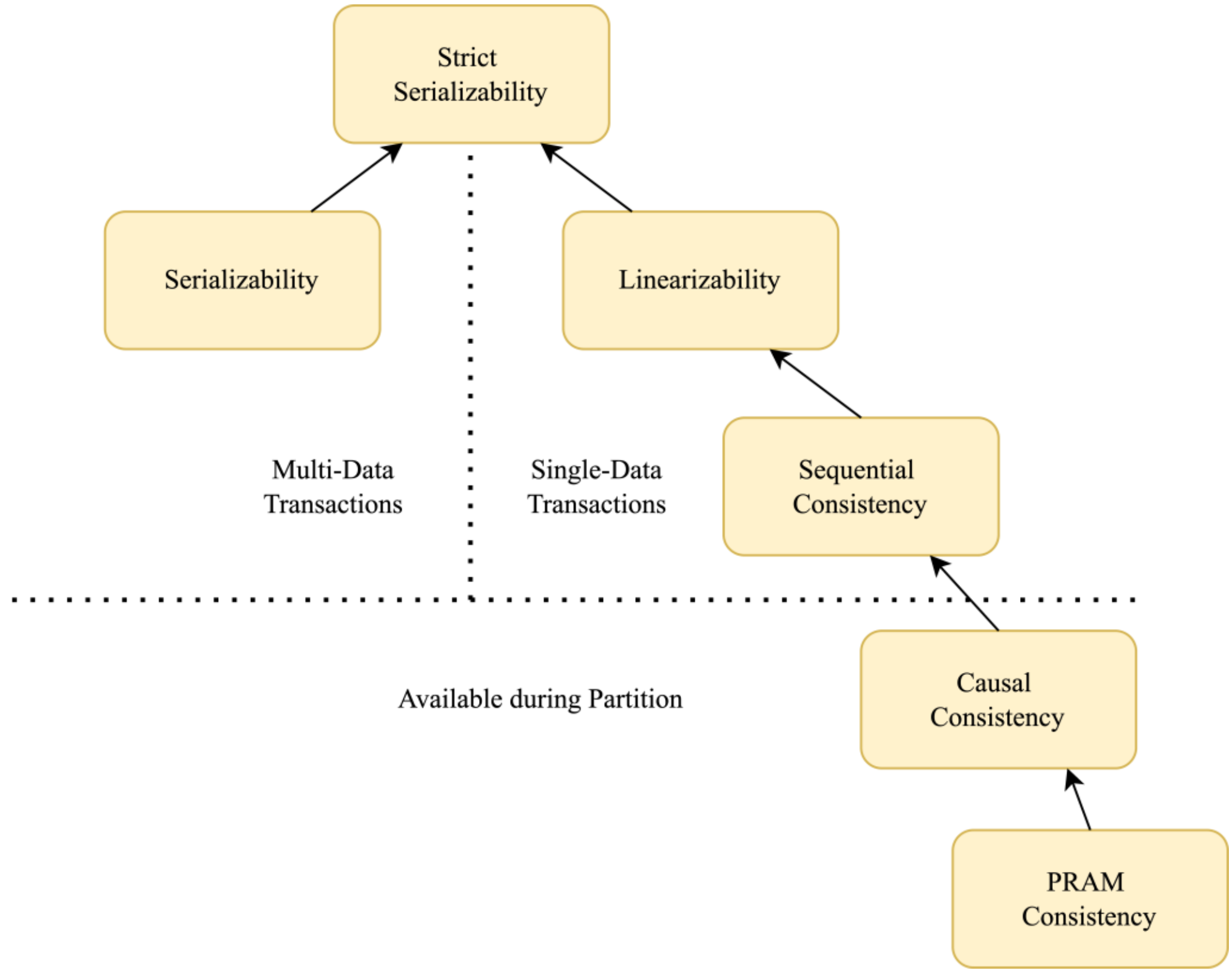

**Figure 97**: Consistency models hierarchy.

Consensus algorithms are fundamental to building distributed logs. For example, a distributed log can be divided into slots, where each slot holds a single log entry (or instruction for the distributed system). Consensus can then be used to ensure agreement on the value (instruction) to be placed in each slot.

*21.3.2. Atomic Broadcast*

Atomic broadcast is a message broadcast in distributed system which ensures that:

- **Validity (Liveness):** If a node broadcasts a message, all nodes eventually receive it.
- **Integrity (Safety):** Each node receives a given message at most once.
- **Agreement (Safety):** If one non-faulty node receives a message, all non-faulty nodes eventually receive it.
- **Total Order (Safety):** Messages are delivered to all non-faulty nodes in the same total order.

In essence, atomic broadcast guarantees that all correct nodes eventually receive all broadcast messages in a consistent total order. The use of "eventually" is crucial in the definition.

Beyond atomic broadcast, other broadcast protocols exist, each corresponding to a specific ordering guarantee:

- **FIFO Broadcast:** Messages originating from a single node are delivered to all nodes in the same order they were sent out. No ordering guarantees are provided for messages from different senders.
- **Causal Broadcast:** All messages are delivered in a causally consistent order.

#### *21.3.3. Equivalence of Consensus and Atomic Broadcast*

A distributed log, built using consensus, can represent the messages in an atomic broadcast, with each log entry corresponding to a broadcast message, thus ensuring total order. Conversely, an algorithm that solves atomic broadcast can also solve consensus. For instance, a distributed log can be implemented via atomic broadcast, where each instruction becomes a message. Therefore, consensus and atomic broadcast are equivalent problems. Algorithms capable of solving consensus can also solve atomic broadcast.

#### *21.3.4. No Perfect Algorithm Exists! - The FLP Impossibility*

Of the many principles governing distributed systems, the FLP (Fischer-Lynch-Paterson) impossibility result is perhaps the most significant. It states that in an asynchronous network where even a single node can crash, it's impossible to guarantee both **safety** (or correctness) and **liveness** (or termination) simultaneously.

Given that crash faults are always a possibility in large systems, and real-world networks are inherently asynchronous, this implies that no consensus algorithm can truly be considered universally applicable in practice. Even the most well-known algorithms, like Paxos and Raft [165], must make compromises and assumptions to function. Almost all practical consensus algorithms prioritize safety over liveness.

### 21.4. ZooKeeper

ZooKeeper is a system designed to help other distributed systems solve problems such as:

- **Distributed Locking:** Analogous to mutexes in parallel computing, distributed locks allow a node to acquire and release a lock on a specific object.
- **Leader Election:** A group of participants can elect a leader among themselves, with consistent knowledge of who that leader is.
- **Group Membership:** Participants can maintain a consistent view of the other members in their group.

All of these problems are equivalent to consensus. If leader election could be solved in a true FLP setting, so could consensus. Similarly, a perfect distributed locking algorithm would make consensus trivial, which is why consensus is relatively straightforward in parallel computing.

However, ZooKeeper relies on certain assumptions about the system and doesn't solve these problems perfectly. To achieve availability and performance, ZooKeeper compromises on both safety and liveness, as we'll explore later.

#### *21.4.1. A Coordination Kernel*

ZooKeeper itself doesn't directly solve end-user problems, nor is it a ready-to-use distributed application. Instead, it provides a **coordination kernel**, a framework upon which higher-level distributed systems can be built.

For example, Apache Kafka is a higher-level system that leverages ZooKeeper for its underlying coordination. Because ZooKeeper isn't perfect, Kafka inherits some of its limitations. However, it functions effectively in practice.

#### *21.4.2. Z-Nodes*

In ZooKeeper, data objects are called **z-nodes**. They are organized hierarchically, much like a file system, as shown in Figure 1 in the paper. Each z-node contains data and can optionally have child z-nodes. ZooKeeper thus acts as a data store that supports single-data-item (z-node) transactions.

Z-nodes come in two types:

- **Regular:** These z-nodes are explicitly created and deleted by clients using API calls.

- **Ephemeral:** These z-nodes are also created by clients but are deleted either explicitly or after a client-specified timeout. Ephemeral nodes are typically associated with client sessions; if the client disconnects, the ephemeral nodes it created are automatically removed.

ZooKeeper maintains the following information for each z-node:

- **Type:** Whether the z-node is regular or ephemeral.
- **Metadata:** Timestamps and version numbers.
- **Data:** The data stored in the z-node (default 1 MB, configurable).
- **Children:** The z-nodes that are children of this z-node.
- **Counter:** Used for creating sequential child z-nodes. Sequential child z-nodes share the same parent and are created in a specific sequence.

*21.4.3. Watcher*

ZooKeeper provides a "watch" mechanism, allowing clients to monitor a z-node for changes (including creation and deletion). The watch notifications only signal that an update has occurred; they do not include the actual modified data. The client must subsequently read the z-node to retrieve the updated information.

*21.4.4. Sessions*

ZooKeeper manages client connections using sessions, which are governed by timeouts. If a session times out, ZooKeeper assumes the client has closed and takes appropriate actions (e.g., deleting ephemeral nodes).

However, this assumption is flawed in real-world asynchronous environments, as network delays or transient failures can cause timeouts even if the client is still active. **This is a key area where ZooKeeper compromises safety.**

To mitigate session timeouts, the ZooKeeper client library sends heartbeats every **s/3** seconds and switches to a new server if it hasn't received a response within **2s/3** seconds, where **s** represents the session timeout length.

*21.4.5. Client API*

ZooKeeper provides the following z-node APIs:

- **create(path, data, flags)**: Creates a z-node at the specified path with the given data. Flags are used to specify the z-node type (e.g., ephemeral, sequential).
- **delete(path, version)**: Deletes the z-node at the specified path, but only if its current version matches the provided version. Version numbers act like generation numbers [167], preventing accidental deletions.
- **exists(path, watch)**: Returns true if a z-node exists at the specified path. The optional watch flag enables a watch on the z-node.
- **getData(path, watch)**: Returns the data and metadata of the z-node at the specified path. The optional watch flag enables a watch on the z-node.
- **setData(path, data, version)**: Sets the data of the z-node at the specified path to the given data, but only if its current version matches the provided version. This allows the operation to function as a CAS operation for read-write transactions.
- **getChildren(path, watch)**: Returns the list of child z-nodes for the z-node at the specified path. The optional watch flag enables a watch on the child list.
- **sync(path)**: Waits for all pending updates initiated before this operation to propagate to the server the client is currently connected to.

A key characteristic of these APIs is that they don't return a file handle or descriptor. There are no **Open**() or **Close**() operations on z-nodes. All operations are atomic transactions in themselves. No locks are held on z-nodes beyond the duration of the operation. All reads and writes are complete: full reads and full writes.

*21.4.6. Instruction Idempotency*

All client operations in ZooKeeper are designed to be idempotent. Idempotency is a highly desirable property in distributed systems. An idempotent operation can be executed multiple times without changing the result beyond the initial application. For example, the following operations are idempotent:

- $\mathbf{A := A * 1}$
- $\mathbf{A := A + 0}$
- $\mathbf{A := c}$ ($\mathbf{where\ c\ is\ a\ constant}$)

The following operation is not idempotent: $\mathbf{A := A + 1}$

ZooKeeper client instructions are idempotent at the server level. For example, the **setData**() operation includes a version number. After the operation is executed once, subsequent identical executions will have no effect because the version number will have changed.

It's important to note that while the server-side processing of ZooKeeper operations is idempotent, the client API calls themselves are not always idempotent. For instance, using the sequential flag to create sequential nodes results in non-idempotent behavior at the API level. However, the server internally translates these requests into idempotent operations.

*21.4.7. Guarantees*

ZooKeeper offers two guarantees:

1. **Linearizability:** All writes to z-nodes are linearizable. Because ZooKeeper stores single data items (z-nodes), linearizability is equivalent to strict serializability in this context. For reads, ZooKeeper provides two modes:
    a. **Synchronous Reads (or Strong Reads):** Offer linearizability but can be slower.
    b. **Relaxed Reads:** Do not guarantee linearizability but are generally faster.
2. **FIFO Client Order:** All requests originating from a single client are executed in the order they were sent by that client. This is relatively straightforward to implement on the client side.

**21.5. Implementation Details**

This section details ZooKeeper's internal workings.

For high availability, ZooKeeper is fully replicated across all its nodes.

At its core, ZooKeeper implements a distributed log, which contains a sequence of instructions to be applied to the z-node data structure. This entire z-node data structure is held in memory and periodically snapshotted to persistent storage. Each instruction in the distributed log corresponds to a modification of a specific z-node. A log entry is first committed to storage before it's applied to the in-memory data structure. A ZooKeeper replica only responds to a client request after all these steps are complete.

Implementing this distributed log requires a consensus algorithm. ZooKeeper uses the ZooKeeper Atomic Broadcast (ZAB) protocol for this purpose.

### *21.5.1. ZooKeeper Atomic Broadcast*

ZAB is the atomic broadcast algorithm used by ZooKeeper to construct its distributed logs. ZAB, like other atomic broadcast algorithms, relies on certain assumptions about the system.

ZAB operates in two phases:

1. **Leader Election Phase:** ZAB elects a leader among the ZooKeeper nodes. This process uses a Paxos-like approach. Each node proposes a value, and a consensus algorithm is run to select the node that proposed the highest value as the leader. **This initial consensus, like Paxos itself, is not perfect and can fail to terminate, thus potentially compromising the system's liveness.**
   The result of the initial leader election can be the first entry in the distributed log.
2. **Broadcast Phase:** The elected leader broadcasts messages containing log entries to all other ZooKeeper servers (followers). Because all modifications are routed through the leader, they are processed in a consistent order. A request is considered committed when a majority of followers respond successfully, i.e., when $\mathbf{f + 1}$ commits out of $\mathbf{2 * f + 1}$.

### *21.5.2. Handling Partition*

If the leader crashes or a majority of replicas become unreachable, a new leader election occurs. The distributed log will then contain an entry reflecting this new leadership. A node can only be elected leader if it has caught up with all log entries that were successfully committed **before** the view change (the transition to a new leader).

Prior to a view change, there are $\mathbf{2 * f + 1}$ nodes, of which at least $\mathbf{f + 1}$ are up-to-date.

After a view change (when $\mathbf{f + 1}$ nodes become unavailable), there will be at least one node that is fully caught up with the latest committed log entries.

Before a new leader is elected, the new set of replicas must have at least **f + 1** nodes that are caught up to this latest position. The new leader must also ensure that all lagging replicas receive all the log entries before the leader change entry itself can be committed. This is accomplished by having the most up-to-date replica broadcast the logs to everyone else.

This also means that duplicate delivery of messages is possible. However, because all ZooKeeper operations are idempotent, duplicate message delivery has no adverse effects.

*21.5.3. Fuzzy Snapshots*

Because ZooKeeper operations are idempotent, it's possible to create fuzzy snapshots. A fuzzy snapshot is taken without needing to acquire a full lock on the in-memory database.

For example, consider data items **A** and **B**, initially with values {**A: 1, B: 2**} -

- A fuzzy snapshot process begins, recording the current value of **A** (**A: 1**).
- A transaction updates **A := 4**. New state - {**A: 4, B: 2**}.
- Then, another transaction updates **B := 3**. New state - {**A: 4, B: 3**}.
- Finally, the fuzzy snapshot process records the current value of **B** (**B: 3**).

The resulting fuzzy snapshot appears as {**A: 1, B: 3**}. This state never actually existed at any point in time in the database.

However, due to the idempotent nature of the operations, the fuzzy snapshot can be brought up to the correct final state. By applying all log entries that occurred after the snapshot began, the fuzzy snapshot can be transformed to reflect the current state {**A: 4, B: 3**}.

A more detailed example specific to ZooKeeper is provided in Section 4.3 of the relevant paper.

*21.5.4. Relaxed v/s Linearizable Reads*

By default, ZooKeeper read operations are directed to a nearby replica, which might not have the latest updates from the distributed log. These **relaxed reads** can return stale data and are therefore not linearizable. The **zxid** (transaction id) on a replica indicates the last committed log entry. A client can detect a potentially stale read by comparing the zxid returned with the read to a previously known zxid.

To achieve linearizable reads, ZooKeeper provides the **sync** operation. This operation goes through the leader and is processed like any other instruction, although it's not actually committed to the log. sync effectively performs a read-only transaction, ensuring the client receives the latest value at that specific point in time. Having a single leader to coordinate read-only transactions ensures that such transactions fetch the latest value and reflect all writes up to that point. This single-leader approach also offers the advantage of consistent transaction ordering, aligning with client observations (i.e., external consistency).

The sync operation could be implemented straightforwardly by using the same distributed logging process as other instructions. However, ZooKeeper optimizes this process by assuming that the leader will not lose its leadership while the sync operation is in progress. This optimization relies on leader leases. **This assumption is, again, unsafe and can lead to non-linearizable reads, thus compromising correctness.**

While exchanging heartbeats, the client learns the latest zxid from the server. When switching to a new server, the client is guaranteed to connect only to servers with a zxid that is at least as recent as the last known zxid. This ensures the client can always find a server that is sufficiently up-to-date. These mechanisms are primarily for performance optimization and do not affect the fundamental correctness of the system.

### 21.6. "Wait-Free" v/s Chubby

The paper's title mentions "wait-free". Let's explore what that signifies.

As previously discussed, clients are notified of z-node modifications. These notifications are guaranteed for eventual delivery. "Eventually" means that all replicas will eventually catch up and notify their respective clients. This holds true even if a client switches to a new replica based on the zxid.

This eventual delivery ensures that write operations don't have to wait for every client to be notified. Write requests are committed to the transaction log and the response is immediately returned to the client. Watchers are notified asynchronously and eventually. This asynchronous notification mechanism is what constitutes "wait-free" behavior.

A contrasting approach would require write requests to block until all clients watching the modified z-node have been notified. Otherwise, the transactions would fail. This is precisely how Chubby operates. The authors argue that this synchronous notification is the reason for Chubby's slower performance and its lack of "wait-free" property. Chubby provides a very strong and consistent notification mechanism, but this comes at the cost of scalability, especially for fine-grained locking. Chubby only sends a response back to the client after notifying all watchers (though **it does incorporate timeouts, which, while improving responsiveness, introduce correctness issues**).

ZooKeeper, on the other hand, replies to the client before notifications are sent, achieving wait-free behavior. This is the key difference between Chubby and ZooKeeper.

### 21.7. ZooKeeper & FLP Impossibility

ZooKeeper violates the FLP impossibility result at several points:

- ZAB itself does not guarantee liveness during faults.
- ZooKeeper's lease mechanism, which relies on timeouts, also compromises safety. Timeouts depend on clocks, which are not guaranteed to be synchronized across processes in an asynchronous system. Therefore, there is no true consensus on clock values. Although this compromises safety, it's a widely accepted trade-off in practice.

Timeouts are one of two common ways to implement leases in asynchronous systems (the other being explicit lease relinquishment, often involving some variant of two-phase commit). A timeout can be conceptually viewed as a message received by the lease holder. In ZooKeeper, a timeout acts as a message to the leader to release locks on ephemeral nodes. This reliance on timeouts is where safety is compromised, but, again, this compromise is generally accepted in practice.

### 21.8. Primitives using ZooKeeper

ZooKeeper can be used to build a variety of distributed primitives. Among the most interesting are leader elections and distributed locks. The paper also explains other primitives, such as configuration management, group membership, and double barriers, which are relatively straightforward to understand.

### *21.8.1. Leader Election*

A simple leader election can be implemented by having clients attempt to create a specific z-node. The first client to successfully create the z-node is designated the leader.

A more sophisticated approach might require the leader to submit new configurations before assuming leadership. In this case, a special **ready** z-node can be used. Clients create z-nodes containing their portion of the new configuration. Once all configuration z-nodes are created, a client creates the ready z-node.

ZooKeeper's FIFO client order guarantee ensures that all configuration z-nodes will be created before the ready z-node. Consequently, any client that observes the ready z-node is also guaranteed to have seen all the updated configuration z-nodes.

### *21.8.2. Distributed Locks*

In its simplest form, a lock can be implemented using ephemeral z-nodes as **lock files**. A client releases the lock by either explicitly deleting the z-node or implicitly when its session terminates.

However, this simple approach suffers from the **herd effect**: many clients might be waiting to acquire the lock, all contending simultaneously when it becomes available. Also, this implementation provides only exclusive locks, not shared locks.

A more robust approach involves creating z-nodes with the sequential flag -

- Each client is assigned a position in a queue and acquires the lock in sequence.
- Each client watches the z-node of the client immediately preceding it in the queue. A crucial detail is the loop back to line 2 from line 6 in the **Lock**() function. This handles the scenario where the client ahead of the current client in the queue might have crashed (effectively leaving the queue).

To implement a shared lock (ReadLock), the algorithm ignores any read z-nodes ahead in the queue and only checks for a preceding write z-node.

### 21.9. Evaluation

ZooKeeper's performance has been significantly improved since its initial release, so the reported numbers may no longer be entirely accurate.

- **Read vs. Write Throughput (Figure 5 in the paper):** As expected, read throughput is higher than write throughput. This is because reads are typically relaxed (and thus faster), while writes must go through the atomic broadcast protocol and be logged.
- **Factors Affecting Write Throughput:** The atomic broadcast process is identified as the main limiting factor for write throughput. However, other factors, such as client communication overhead and ACL checks, also contribute.
- **Fault Tolerance:** The failure of a single follower does not significantly impact overall throughput. When the leader fails, the system recovers relatively quickly (within ~200ms).

### 21.10. Paper Remarks

This paper is a landmark in the distributed systems domain, providing a clear API specification and a comprehensive architectural walkthrough that bridges the gap between theoretical consensus and high-performance systems.

While software engineers rarely interact directly with consensus or atomic broadcast protocols, they frequently rely on the kernels and primitives that abstract these complexities. Understanding these underlying mechanisms remains invaluable. Ultimately, no coordination kernel - neither ZooKeeper nor Chubby - is perfect; they all operate under specific assumptions. However, despite their theoretical constraints, they perform remarkably well in practice and serve as the critical foundation for modern distributed systems.

# 22. A New Presumed Commit Optimization for Two Phase Commit

(22) Lampson, B.; Lomet, D. A New Presumed Commit Optimization for Two Phase Commit. In 19th International Conference on Very Large Data Bases (VLDB'93); 1993; pp 630–640.

While consensus ensures that replicated state machines work in unison, it cannot solve every problem. Distributed systems are often composed of heterogeneous components, and it is the two-phase commit (2PC) protocol that allows these different systems to agree on a single action. That is why 2PC is another critical concept in distributed systems.

This Lampson and Lomet's 1993 paper, from the now-defunct DEC Cambridge Research Lab, remains a classic. It was presented at the prestigious Very Large Databases (VLDB) '93. It explores several implementation strategies for 2PC, providing a deep dive into the concept.

In this insight, we will begin with an in-depth look at SQL transactions. First, we will understand what serializability entails. We will then examine the properties of atomicity and consistency, specifically how they are implemented in standalone SQL databases. Next, we will discuss their implementation strategy in distributed SQL databases and understand why 2PC is essential. We will also contrast 2PC with consensus to clarify their distinct roles in distributed environments. This will be followed by a detailed exploration of various techniques used to implement 2PC. Finally, we will revisit serializability and understand why 2PC alone is insufficient for ensuring transaction isolation in distributed SQL databases. We will briefly touch upon how synchronized clocks can assist with serializability - a topic we will analyze more thoroughly in our exploration of Spanner. As a bonus, we will also cover the three-phase commit protocol to understand its specific strengths and limitations.

### 22.1. Serializability

Transaction **serializability** guarantees that, while transactions may execute concurrently for performance reasons, the final outcome is effectively equivalent to some sequential execution of those same transactions. The "effectively" part means that the system ensures a consistent, serializable result even if the underlying execution is parallelized.

**Strict serializability** builds upon serializability by adding a temporal dimension. It mandates that once a transaction commits, its effects are immediately visible to all clients (a.k.a. **external consistency**). This differs from linearizability, which focuses on single-object operations, as strict serializability applies to multi-object transactions (e.g., updates across multiple rows of a table).

Serializability is a transaction isolation level, one of several including **read uncommitted**, **read committed**, and **repeatable read**. This discussion focuses exclusively on serializable transactions. Such transactions are implemented using either:

- **Locks** - Prevents concurrent execution of transactions that modify the same object. This only works for standalone non-distributed SQL databases.
- **Timestamp Ordering** - Implements a global ordering of transactions ordered by timestamps from a source of truth. This approach can offer serializability and indeed, strict serializability, however, such a source of truth is difficult to obtain for most systems.

There are other ways to improve the methods above. For example, versioned objects can be used for multi-version concurrency control [168] so that locks are not required. We will revisit serializability at the end of this chapter.

### 22.2. Single Data Transactions

Single-data transactions are the ones that atomically modify only a single data item (e.g., a single row). These are commonly offered by NoSQL databases, where multi-data transactions are not supported for performance. The item being modified in a transaction resides on a set of replicas. All replicas are expected to apply the same modification as a result of the transaction.

These single-item transactions are also commonly implemented by distributed lock services like ZooKeeper, using a distributed log built on consensus algorithms. These implementations also ensure linearizability.

In this chapter, we will focus on multi-data transactions. For more details on how to implement single-data transactions, refer **21. ZooKeeper: Wait-free Coordination for Internet-Scale Systems**.

## 22.3. Multi-Data Transactions

Multi-data transactions are the ones that modify multiple data items (e.g., multiple rows). The term "transaction" commonly refers to a multi-item transaction, which is a core feature of SQL databases.

Consider a banking scenario with accounts held by **A** and **B**. **A** transfers money to **B**, as shown in Figure 98.

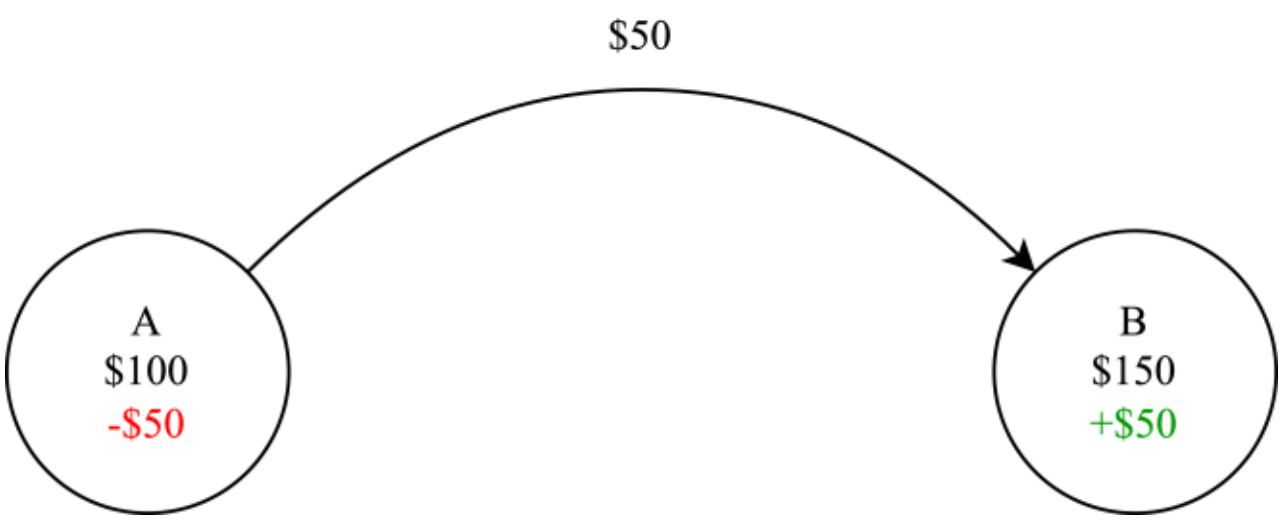

**Figure 98**: Intrabank transaction example.

The client code will be something like:

```
txn = Transaction.start();
balA = txn.Read(A);
balB = txn.Read(B);
txn.Write(A, balA - $50);
txn.Write(B, balB + $50);
txn.Commit();
```

The transaction needs to ensure:

- **Atomicity -** The transfer either completes fully or not at all.
- **Consistency -** The total balance across all accounts remains constant after any transfer.

### *22.3.1. Implementation in Non-Distributed SQL Databases*

In a non-distributed database, all data resides on a single server. B-trees, indexed by primary keys with leaf nodes pointing to row data, are a common data structure used in these systems.

Transactions are typically implemented by acquiring locks on the data items that are read/modified. The data items are then updated, and finally, the transaction commits. However, if a transaction fails before reaching the commit stage, a mechanism is required to revert any changes made. This is where **transaction logs** provide the necessary means to handle pre-commit failures.

#### 22.3.1.1. Transaction Logs

Transaction logs are essential for database recovery and ensure data consistency. Two primary types exist:

- **<u>UNDO Logs</u>:** These logs are written before data modifications. They capture the original state of the data, allowing for rollback operations in case of failures.
- **<u>REDO Logs</u>:** These logs are written after data modifications. They record the changes made to the data, enabling the system to reapply these changes during recovery.

UNDO logs contain the following types of records:

- **[start, TID]:** Indicates the beginning of a transaction with identifier **TID**.
- **[write, TID, X, old_value]:** Records the modification of data item **X** by transaction **TID**. **old_value** stores the original value of **X** before the modification.
- **[commit, TID]:** Signals the successful completion of transaction **TID**.
- **[abort, TID]:** Indicates that transaction **TID** has been aborted.

If a transaction fails (i.e., lacks a **commit** or **abort** entry in the UNDO log), the rollback system:

1. Iterates through the log entries for the failed transaction.
2. For each **write** entry, restore the original data value (*old_value*) to the corresponding data item.
3. Adds an **abort** entry to the UNDO log for the failed transaction, marking its termination.

This process ensures that the database remains in a consistent state even after unexpected failures. Such failures are retried by the client.

In our bank example above, assuming **A** had a balance of $100 and **B** had a balance of $150, the following would the entries in the UNDO logs for a committed transaction:

```
start 100024
write 100024 A $100
write 100024 B $150
commit 100024
```

It's crucial to understand that each entry in the UNDO log directly corresponds to a modification performed by the client within the scope of a single transaction.

*22.3.2. Implementation in Distributed SQL Databases*

Now, consider a system where data is distributed across multiple nodes. In our bank example, account balance **A** might reside on one server, while account balance **B** might reside on another.

The actions performed by the client within the scope of the transaction are sent to each node independently. The **start** and **commit** messages are sent to all the nodes. We can repeat the same process of having UNDO logs:

**Node 1**

```
start 100024
write 100024 A $100
commit 100024
```

**Node 2**

```
start 100024
write 100024 B $150
commit 100024
```

But, if the client crashes after committing a transaction to node 1 but before committing to node 2, we would be left in an inconsistent state.

**Node 1**

```
start 100024
write 100024 A $100
commit 100024
```

**Node 2**

```
start 100024
write 100024 B $150
commit 100024
```

Since node 1 has already committed the transaction, it will not rollback the changes made locally.

**This is exactly where we need <u>Two-Phase Commits (2PC)</u>.** We want all nodes to either commit or abort the transaction.

## 22.4. Consensus v/s 2PC

Consensus algorithms, such as Paxos, ensure that all replicas of a state machine execute the exact same sequence of actions or apply the same set of instructions. All replicas are equivalent. This guarantees a consistent state across all replicas.

In contrast, 2PC focuses on ensuring that all participants in a distributed transaction take the same action atomically. This means all participants either commit or abort the transaction together. Each participant is entirely a different entity.

Note that, in case of consensus, there is a concept of the majority of replicas being caught up to head. However, there's no concept of a majority decision in 2PC. All must commit or abort the transaction.

### *22.4.1. Example: Hotel Booking*

The example below is taken from Predrag's Blog [169]. See: How Paxos and Two-Phase Commit Differ [170].

Consider a hotel booking scenario. The booking process involves multiple steps:

1. Deducting money from the user's bank account.
2. Reserving the hotel room.
3. Sending a confirmation to the user.

These steps must be executed atomically. If any step fails, the entire transaction should be rolled back.

In this example:

- The participants include the booking website (**coordinator** and **cohort**), the bank (**cohort**), and the hotel (**cohort**).
- Each participant can be implemented as a highly available replicated state machine using algorithms like Paxos. This ensures that each participant itself can tolerate failures and remain operational.
- 2PC ensures that all participants either successfully complete all steps or none of them.

*22.4.2. Key Differences Summarized:*

- **Consensus:** Focuses on consistent state across replicas of a single state machine.
- **2PC:** Focuses on atomic actions across all participants in a distributed transaction.

## 22.5. The Two Phase Commit Protocol

The primary objective of 2PC is to ensure that all participants involved in a distributed transaction reach a unanimous consensus on whether to **commit** or **abort** the transaction.

Once all transaction instructions have been logged (written to transaction logs) and applied to the data structures on all nodes, the 2PC protocol determines whether the transaction should be committed or aborted.

*22.5.1. The Protocol*

2PC proceeds as follows:

**Phase 1: PREPARE**

A designated node (often the client) acts as the coordinator. It broadcasts **PREPARE** messages to all participating nodes (cohorts). At this stage, each cohort can still decide whether to commit or abort.

If locks are used in the transaction, a cohort will typically commit. However, in multi-version concurrency control, for example, a cohort might abort if read version numbers no longer match the latest version numbers after 2PC began.

If a cohort decides to commit, it sends a **COMMIT-VOTE** to the coordinator. Otherwise, it sends an **ABORT-VOTE**. Crucially, once a cohort sends a COMMIT-VOTE, it must be able to commit the transaction in the next phase. To guarantee this, cohorts must persistently record their commitment. Common strategies include:

- Writing the commitment to storage, such as adding a "pre-commit T" entry to the transaction log, indicating the cohort's commitment to transaction T.
- Using a Paxos group of replicas to act as a single, highly available cohort and logging the pre-commit message in the distributed log, ensuring all replicas in the group are aware of the commitment.

This paper focuses on individual nodes with local storage as cohort members, therefore we will only consider the first strategy (logging the commitment to local storage). The second strategy (using Paxos groups) is implemented by Spanner.

**Recommended Read**: **24. Spanner: Google's Globally-Distributed Database**

**Phase 2: COMMIT / ABORT**

If any cohort voted to abort, the coordinator broadcasts an **ABORT** message to all cohorts. If all cohorts voted to commit, the coordinator broadcasts a **COMMIT** message. Upon receiving a COMMIT message, each cohort must commit the transaction. After this, all locks can be released.

Cohorts then acknowledge the commit/abort message with an **ACK**.

The prepare phase can be viewed as a pre-commit phase, adding an extra step for all cohort members to agree on the transaction's final outcome.

*22.5.2. Under of Hood of 2PC*

Let's examine the details of 2PC, focusing on network communication and storage writes. Storage writes can be **forced** or **unforced**. Forced writes are expensive as they are not buffered by the operating system or by the disk write buffer cache. However, forced writes are guaranteed to be durable. On the other hand, unforced writes may still be lost.

#### 22.5.2.1. Cohort Responsibilities

Before replying to a PREPARE message: A cohort must force-log its COMMIT-VOTE. This ensures that even if the cohort crashes, it remembers its vote.

Before replying to a COMMIT or ABORT message: A cohort must force-log the fact that it has committed or aborted. This guarantees that the cohort knows the final outcome, even after a crash. These logs are crucial. They prevent cohorts from repeatedly querying the coordinator about the transaction's status. If the coordinator received the cohort's reply, it's guaranteed never to receive another query about that transaction from that cohort.

If a cohort crashes before logging the COMMIT or ABORT message, upon recovery, it must query the coordinator for the final outcome. Since the cohort previously voted to commit, and if the coordinator confirms a COMMIT, the cohort proceeds with the commit (it's the cohort's responsibility to honor its commitment). The coordinator, having not received a ACK reply, retains the transaction's final status until all cohorts have acknowledged it.

**The cohort doesn't need to keep anything in-memory. It just needs to force log COMMIT-VOTE and COMMIT/ABORT messages.**

#### 22.5.2.2. Coordinator Responsibilities

The coordinator, unlike a cohort, requires a combination of in-memory and persistent storage for its data.

##### 22.5.2.2.1. Protocol Database

The coordinator maintains an in-memory database (a.k.a. a protocol database) with the following information for each transaction:

- **TID**
- **stable**: A boolean indicating whether the transaction's existence has been persistently logged.
- **state**: The transaction's current state (initiated, preparing, aborted, or committed).
- **per-cohort information**: For each cohort, its ID and vote.

The per-cohort information can be deleted after the coordinator receives the ACK for the second phase message. The entire transaction record can be removed from memory once acknowledgements from all cohorts are received.

22.5.2.2.2. Storage Logs

The coordinator's log only needs to force record the final transaction state (COMMIT or ABORT). It doesn't need to retain other details. If the coordinator crashes and a cohort inquires about a transaction, the coordinator replies with the logged final state.

If the coordinator crashes before the end of the PREPARE phase (i.e., before the final state is logged), it can safely assume an ABORT and respond accordingly to any cohort queries.

Finally, the coordinator should have an unforced "end" record for each transaction. This is a performance optimization to prevent the coordinator from restarting already-completed transactions; it doesn't affect correctness.

22.5.2.3. Summary

**Coordinator**:

- Force-log the COMMIT or ABORT decision before sending the corresponding message.
- Maintain an unforced transaction end record.

**Cohort**:

- Force-log the COMMIT-VOTE before replying to the PREPARE message.
- Force-log the COMMIT or ABORT action before replying to the COMMIT or ABORT message.

*22.5.3. Applications of 2PC*

2PC is a critical protocol for ensuring atomicity in distributed systems, meaning that all participating systems either complete a transaction or none do. While commonly associated with databases, 2PC has applications in various scenarios where atomic actions across multiple systems are essential. Here are some key examples:

- **Exactly-once message delivery:** 2PC guarantees exactly-once delivery of messages, preventing duplicates or lost messages in distributed environments. The messages are delivered, processed, and acked in a single transaction.
- **Online booking systems:** In systems like flight or hotel reservations, 2PC ensures that all parts of a booking process (e.g., seat reservation, payment processing, confirmation) are completed together or not at all.
- **E-commerce transactions:** When you make a purchase online, 2PC ensures that the order, payment, and inventory updates are all synchronized.
- **Distributed caching:** 2PC can be used to maintain consistency across distributed caches, ensuring that all cache nodes have the same data.
- **Workflow management:** In complex workflows involving multiple systems, 2PC can coordinate the execution of tasks and ensure that all steps are completed successfully.

### 22.6. 2PC Variants

The 2PC variant we've been discussing is known as **Presumed Nothing (PrN)**. Other variants exist, including **Presumed Abort (PrA)**, **Presumed Commit (PrC)**, and **New Presumed Commit (NPrC)**, which is the focus of this paper. These variants are performance optimizations of the core 2PC protocol; they don't change the fundamental logic. However, because 2PC is used in so many distributed systems, even small performance gains can have a significant impact on overall system performance.

#### *22.6.1. Optimization Targets*

The goal of all these variants is to minimize the number of forced writes to disk and the number of messages exchanged. The primary targets for optimization are:

- Coordinator: The forced log of the ABORT record.
- Cohort: The forced log of the COMMIT or ABORT action and the reply sent after committing or aborting.

Note that the following operations cannot be optimized:

- The forced log of the COMMIT-VOTE is essential and cannot be avoided; cohorts must log their vote before sending it.

- Additionally, the reply to the PREPARE message is also necessary; cohorts must respond.
- Finally, forced log of the COMMIT message is essential; the coordinator must log committed transactions.

### *22.6.2. Presumed Abort*

In the PrA variant, the coordinator takes a more optimistic approach with regard to aborted transactions. Instead of force-logging an ABORT decision, the coordinator simply removes the transaction's entry from its in-memory database. The default assumption is that a transaction has been aborted unless proven otherwise. If the coordinator crashes and receives a query about a transaction that's no longer in its in-memory database and not in its COMMIT logs, it assumes the transaction was aborted and responds accordingly.

Furthermore, cohorts do not need to force-log ABORT action or acknowledge an ABORT message. Since the coordinator doesn't maintain any persistent state for aborted transactions, there's no action required upon receiving an ACK. Consequently, cohorts themselves don't even need to send an ABORT reply.

#### 22.6.2.1. Failure Recovery

Note that:

- The coordinator force-logs COMMIT decisions.
- The cohorts force-log COMMIT action and ACK to COMMIT messages.

**Coordinator Failure:**

- **Before COMMIT is logged**: The transaction is discarded. Any cohort querying the transaction's status is asked to abort.
- **Before all ACKs are received**: The transaction's status is COMMIT if and only if the COMMIT message was successfully logged before the crash; otherwise, the status is ABORT.

**Cohort Failure:**

- **Before receiving the decision message**: The cohort queries the coordinator for the decision.

- **After sending the ACK for ABORT (since ABORT logs are not forced)**: The cohort queries the coordinator for the decision and will idempotently receive ABORT again.

### *22.6.3. Presumed Commit*

In the PrC variant, the coordinator optimistically assumes that transactions will be committed.

However, PrC needs to log the PREPARE message. The coordinator does force-log both the PREPARE and COMMIT messages.

Furthermore, cohorts do not need to force-log COMMIT action or acknowledge an COMMIT message.

#### 22.6.3.1. Failure Recovery

Note that:

- The coordinator force-logs PREPARE and COMMIT messages.
- The cohorts force-log ABORT action and ACK to ABORT messages.

**Coordinator Failure**

- **Before PREPARE is logged**: The transaction is discarded.
- **Before COMMIT is logged**: The coordinator requests votes again from the cohorts.
- **Before the ABORT message is sent**: The coordinator requests votes again.
- **Before ACKs on ABORT**: The coordinator requests ACKs again.

**Cohort Failure**

- **Before receiving the decision message:** The cohort queries the coordinator.
- **After sending the ACK (since COMMIT logs are not forced)**: The cohort simply queries the coordinator again and will idempotently receive ABORT again.

Note: PrC is not always COMMIT only. It does ABORT as well based on the cohort's decisions.

### *22.6.4. ACKs and Garbage Collection*

Our focus here is on garbage collection (GC) of transaction data on the coordinator side, not on the cohorts.

#### 22.6.4.1. Presumed Abort

The final ACK for a COMMIT message triggers the removal of the COMMIT decision from persistent storage and the removal of the committed transaction from the protocol database.

ABORT decisions are never logged. Aborted transactions are removed from the protocol database as soon as any cohort sends an ABORT-VOTE.

#### 22.6.4.2. Presumed Commit

The final ACK for an ABORT message triggers the removal of the aborted transaction from the protocol database and the removal of the PREPARE message from persistent storage. The PREPARE message can also be removed after the COMMIT message is logged.

Upon receiving all ACKs for COMMIT, the COMMIT message can be removed, and the committed transaction can be removed from the protocol database.

A key difference is that in PrA, ACKs for ABORT are neither required for correctness nor for GC. In PrC, ACKs for COMMIT are not required for correctness, but they are necessary for GC.

#### 22.6.4.3. Post-Recovery Behavior and GC

**PrA**: If an ACK for COMMIT is not received, the coordinator will still have the COMMIT log and will query the cohorts for the transaction's outcome. ACKs for ABORT are never required. After recovery, aborted transactions are simply forgotten.

**PrC**: If an ACK for ABORT is not received, the coordinator will still have the PREPARE message. It will query the cohorts for the decision. Upon receiving confirmation of the abort, it will resend the ABORT message and collect ACKs again. If an ACK for COMMIT is not received, the coordinator will have the COMMIT message and will resend the COMMIT message and collect ACKs.

In both cases (PrA and PrC), eventual garbage collection of entries in both the protocol database and persistent storage on the coordinator is guaranteed.

### *22.6.5. Summary*

PrA and PrC optimize storage writes, as summarized in the Table 3 below (note again that cohorts always force-log their COMMIT-VOTE and reply, and coordinator always force-logs COMMIT message):

**Table 3**: Summary of various 2PC implementations.

| | **Action** | **PrN** | **PrA** | **PrC** | **NPrC** |
|---|---|---|---|---|---|
| **Coordinator** | PREPARE Log | No | No | Yes | No |
| | ABORT Log | Yes | No | No | No |
| **Cohort** | COMMIT Log + ACK | Yes | Yes | No | No |
| | ABORT Log + ACK | Yes | No | Yes | Yes |

At first glance, PrC might seem superior. If we assume that most transactions commit successfully, avoiding ACKs for commits appears advantageous. However, the assumption that most transactions will be committed is not always valid. Furthermore, PrC introduces an additional forced log write for the PREPARE message. The paper argues that PrA is often a better choice, and this summary highlights why.

## 22.7. New Presumed Commit

NPrC is a further optimization of PrC designed to reduce logging overhead, especially in the optimistic scenario where crashes are infrequent. It achieves this without affecting the correctness of the 2PC protocol.

### *22.7.1. Data Structures*

The following data structures are used in NPrC, as shown in Figure 99:

- $\mathbf{tid_l}$: The lowest transaction ID that is not recorded. Transactions with IDs lower than $\mathbf{tid_l}$ are presumed committed.
- $\mathbf{tid_h}$: The highest transaction ID that is not recorded. Transactions with IDs higher than $\mathbf{tid_h}$ are new.
- **REC**: The set of recent transactions.

$$\mathbf{REC} = \{\mathbf{tid} \mid \mathbf{tid_l} < \mathbf{tid} < \mathbf{tid_h}\}$$

- **COM**: The subset of **REC** containing transactions that are known to be committed. Committed transactions have forced COMMIT logs, as in PrC.
- **IN**: The set of transactions for which nothing has been written to storage. If a crash occurs, all transactions in **IN** are presumed aborted.

$$\mathbf{IN} = \mathbf{REC} - \mathbf{COM}$$

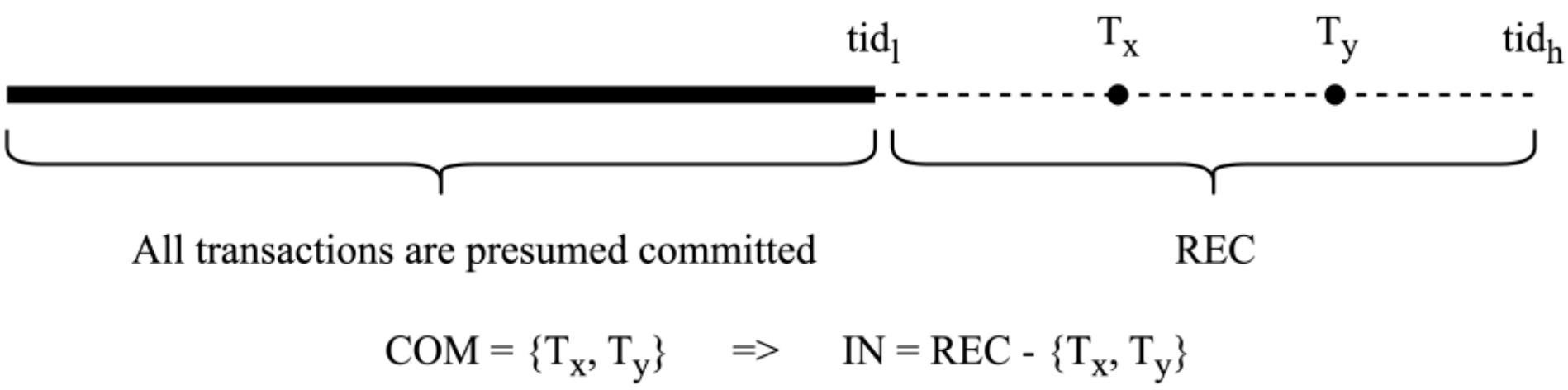

**Figure 99**: Data structures for NPrC.

*22.7.2. Logging $tid_h$*

Logged periodically (and can be forced, as it's infrequent) as new transactions with higher IDs arrive.

*22.7.3. Logging $tid_l$*

Logged when the transaction with the lowest ID in **REC** is committed. The actual logged value might be higher than the current $\mathbf{tid_l}$ because other transactions in the range might have already committed.

- The cost of logging $\mathbf{tid_l}$ can be amortized with the cost of logging the COMMIT for the transaction itself.
- If the transaction is ABORTed, $\mathbf{tid_l}$ can be logged unforced after all ACKs are received. Since all ACKs are received, it means no one will ever query about the transaction again, so $\mathbf{tid_l}$ can safely be advanced.

*22.7.4. Crash Recovery*

$\mathbf{tid_h}$ is set to a very high value to avoid overlap with previous transaction IDs. **IN** is rebuilt using the known values of $\mathbf{tid_l}$, $\mathbf{tid_h}$, and the **COM** set. Post this, the recovery is as follows based on the crash point of the coordinator:

- **Before $tid_l$ is advanced**: **COM** is used to determine if a record was actually committed.
- **After $tid_l$ is advanced**: The transaction is presumed committed.
- **Before $tid_l$ is advanced past the ID of an aborted transaction**: The transaction becomes part of **IN** upon recovery and is thus presumed as aborted.
- **After $tid_l$ is advanced past the ID of an aborted transaction**: All ACKs for the abort would have been received, and no one should query about the transaction again.

The paper discusses optimizations for persisting **IN** and quickly determining **IN** after recovery to minimize the chance of incorrectly presuming a transaction to be aborted. However, these optimizations are less relevant for modern systems with ample RAM. (See sections 4.2.3 and 4.3 of the paper for details).

*22.7.5. Garbage Collection*

Transactions between $tid_l$ and $tid_h$ are considered aborted unless they are within the **COM** set. Recall that:

- In PrA, abort decisions are not logged. If the coordinator fails before receiving votes, the transaction is considered aborted, and no garbage is generated.
- In PrC, if the coordinator fails before receiving votes, it has logged PREPARE messages for each transaction, allowing it to re-collect votes after recovery. No garbage is generated.

However, in NPrC, after recovery, only the **IN** range (transactions to be aborted) is available. Crucially, no PREPARE or COMMIT messages exist for them. This absence of information creates permanent garbage. The coordinator cannot determine the involved cohorts, so it must retain this **aborted transaction range** to respond with ABORT to future queries for transactions within that range.

*22.7.6. NPrC v/s PrC*

After a crash, an NPrC coordinator might have more work to do than a PrC coordinator because it has less persistent information. However, crashes are assumed to be rare, so this extra work during recovery is generally not a significant performance bottleneck.

## 22.8. Read-Only Transactions

Serializable queries require read-only transactions (also known as **snapshot transactions**). These transactions acquire locks, but their outcome (commit or abort) is irrelevant to the cohorts.

Cohorts respond to a PREPARE message with a READ-ONLY-VOTE. The coordinator, upon receiving only READ-ONLY-VOTEs from all cohorts, recognizes the transaction as read-only and omits sending COMMIT/ABORT messages.

In PrC, the coordinator doesn't need to log a COMMIT/ABORT message for a read-only transaction, but it must delete (unforced) the logged PREPARE message. NPrC offers further optimization: the PREPARE message is never logged, and the transaction is simply forgotten after all cohorts respond with READ-ONLY-VOTE. Crucially, no further communication about the transaction result occurs.

## 22.9. Serializability in 2PC

The basic 2PC protocol, in its various forms (PrN, PrA, PrC, and NPrC), can violate both serializability and strict serializability.

### *22.9.1. Violation in Read-Only Transactions*

The most straightforward violation occurs with read-only transactions. Participants in a read-only transaction release their locks after sending their READ-ONLY-VOTE. This can break serializability because subsequent transactions might modify the data before the read-only transaction's effective commit point.

Essentially, the read-only transaction might read a snapshot of the data that's no longer consistent with the actual position of the transaction in the order.

### *22.9.2. Violation in Read-Write Transactions*

Serializability requires some total order of all transactions. If a transaction $\mathbf{T_1}$ comes before another $\mathbf{T_2}$ in this order, $\mathbf{T_2}$ must see all of $\mathbf{T_1}$'s effects. Conversely, if $\mathbf{T_2}$ comes before $\mathbf{T_1}$, $\mathbf{T_2}$ **must not** see any of $\mathbf{T_1}$'s effects. 2PC can violate this.

Consider the bank database with accounts **A**, **B**, and **C** on nodes 1, 2, and 3. **A** has \$100, **B** has \$150, and **C** has \$0. Two transactions, $\mathbf{T_1}$ (transfer \$50 from **A** to **B**) and $\mathbf{T_2}$ (transfer \$200 from **B** to **C**), are submitted, with $\mathbf{T_2}$ following $\mathbf{T_1}$.

1. Transaction $\mathbf{T_1}$ is initiated.
2. Nodes 1 and 2 accept $\mathbf{T_1}$ and acknowledge the PREPARE message.
3. Node 2 commits $\mathbf{T_1}$ and releases its locks before Node 1 commits.
4. Transaction $\mathbf{T_2}$ is initiated.
5. Nodes 2 and 3 accept $\mathbf{T_2}$. Critically, Node 2 now reflects the updated balance of B (due to the partial commit of $\mathbf{T_1}$).
6. $\mathbf{T_2}$ commits on both Nodes 2 and 3.

The problem is that $\mathbf{T_1}$ is still uncommitted on Node 1 (and hence only partially committed) when $\mathbf{T_2}$ reads the updated balance of B on Node 2. This creates a **read uncommitted** isolation level, where $\mathbf{T_2}$ sees the effects of $\mathbf{T_1}$ before $\mathbf{T_1}$ is fully committed, violating serializability.

In essence, 2PC, in its basic form, doesn't prevent transactions from observing intermediate, inconsistent states of other transactions, thus failing to guarantee serializability and therefore strict serializability.

### *22.9.3. Solutions*

Two primary approaches address the serializability concern:

1. **Coordinator Timestamp:** The coordinator assigns commit timestamps. This approach requires that the coordinator cannot change, otherwise, timestamps may go out of order!
2. **Commit Timestamp:** Each COMMIT-VOTE is assigned a timestamp range. Locks are released when the upper bound of the COMMIT-VOTE timestamp range is reached. This eliminates the need for the coordinator to track anything. The cohort's COMMIT-VOTE specifies a permissible timestamp range, ensuring no conflicting transactions within that range. A new transaction would receive a timestamp range which will be higher than that for all existing transactions.

(2) is generally used as it allows any client to be the coordinator. NPrC is also compatible with this approach. It is interesting to note that this can also solve strict serializability as all transactions will receive globally consistent ordered timestamps. However, it requires near-perfect clock synchronization across all the cohorts. This is impossibly achieved by Spanner.

**Recommended Read**: **24. Spanner: Google's Globally-Distributed Database**

## 22.10. Bonus: Three-Phase Commits

2PC has significant vulnerabilities. It's not resilient to coordinator failures, cohort failures, or network partitions. Any of these events can halt 2PC progress.

A key concern arises after the coordinator has already logged a COMMIT or ABORT decision. While the coordinator must communicate the outcome to the client as soon as possible, it faces a blocking problem. Before it can respond, it needs acknowledgements from all cohorts:

- **PrN and PrA**: ACKs from all cohorts for the commit request.
- **PrC**: ACKs from all cohorts for the abort request.

### *22.10.1. Indefinite Blocking*

If a cohort crashes before the COMMIT phase, 2PC recovery is not possible until that cohort restarts. The coordinator can't finalize the transaction because it's waiting for confirmation from the unavailable cohort. The protocol is blocked indefinitely.

### *22.10.2. Three-Phase Commit*

The three phase commit (3PC) addresses this blocking issue by introducing a PRE-COMMIT phase between PREPARE and COMMIT.

- If the coordinator crashes before receiving PRE-COMMIT acks, the decision is always ABORT.
- If the coordinator has all PRE-COMMIT acks, the decision is always COMMIT.

This crucial change means that once the coordinator logs a COMMIT decision, it can safely commit regardless of cohort availability. Any cohort that misses the COMMIT message will eventually receive it.

The coordinator is no longer dependent on cohorts after logging the COMMIT decision, so it can immediately inform the client. Before logging the COMMIT decision, the decision is always ABORT. Therefore, cohort crashes no longer block the coordinator from responding to the client.

Of course, 3PC adds extra round trips. It's often preferable to use 2PC with highly resilient cohort members (like Paxos groups) where possible, mitigating the risk of cohort failures.

### 22.11. Paper Remarks

This paper can be quite challenging to read at times. The difficulty lies in grasping the nuances of every edge case, even though the foundational concepts appear straightforward. Despite its density, this paper is considered an invaluable first step for anyone aiming to master distributed SQL databases and is highly recommended.

# 23. Amazon Aurora: Design Considerations for High Throughput Cloud-Native Relational Databases

(23) Verbitski, A.; Gupta, A.; Saha, D.; Brahmadesam, M.; Gupta, K.; Mittal, R.; Krishnamurthy, S.; Maurice, S.; Kharatishvili, T.; Bao, X. Amazon Aurora: Design Considerations for High Throughput Cloud-Native Relational Databases. In Proceedings of the 2017 ACM International Conference on Management of Data; 2017; pp 1041–1052.

The previous chapter provided sufficient context for building a distributed SQL database using two-phase commit (2PC). However, before doing so, we will explore Amazon Aurora [171], a distributed SQL database that intentionally avoids 2PC.

This paper was published in 2017 and presented at the SIGMOD (Special Interest Group on Management of Data) conference by Alexandre Verbitsky et al. It presents a clever approach that enables distributed SQL functionality without the need for 2PC - a concept that may initially seem counterintuitive.

In this insight, we will begin by reviewing some foundational cloud computing concepts. We will then dig deeper into data replication and the various techniques used to achieve it. This will be followed by a revisit of the core concepts behind SQL databases before performing a comprehensive deep dive into Aurora's architecture.

**Recommended Read**: **22. A New Presumed Commit Optimization for Two Phase Commit** where 2PC was introduced.

### 23.1. Cloud-Native

The term "cloud-native" gained prominence in the late 2000s as cloud computing adoption accelerated.

**Cloud-native** refers to a methodology for building and running applications that fully leverages the distributed computing and scalability capabilities of cloud environments. It focuses on how applications are designed and deployed, not simply where they reside.

Aurora is a cloud-native database service, meaning it's constructed entirely from cloud components. Let's briefly cover the relevant AWS cloud services:

- **EC2 (Elastic Compute Cloud)** [172]**:** Virtual machines in the cloud.
- **S3 (Simple Storage Service)** [58]**:** Cloud object storage where data is stored as objects, each identified by a key. Object sizes can range from kilobytes to gigabytes. (For related reading, consider exploring **13. Delta Lake: High Performance ACID Table Storage over Cloud Object Stores**).
- **EBS (Elastic Block Store)** [173]**:** Provides block storage volumes that can be attached to EC2 instances.

### 23.2. Clusters & Availability Zone

A **cluster** is a group of interconnected computers, typically using high-bandwidth fabric networks. These clusters are often hosted within data centers.

**Availability Zones (AZs)** are distinct physical locations within a specific geographic region. Think of them as individual data centers, potentially composed of multiple machine clusters. AZs are themselves contained within a larger geographic region. As of February 2025, AWS Cloud spans 114 Availability Zones within 36 geographic regions. For example, the us-east-1 region includes AZs like use1-az1, use1-az2, and so on [174].

An AZ, or even an entire region, can become unavailable due to various factors such as natural disasters, network outages, and planned maintenance or repairs. Therefore, robust services must be replicated across multiple AZs and regions.

### 23.3. Correlated Failures

**Recommended Read**: **38. Availability in Globally Distributed Storage Systems**, an excellent paper for understanding data center failures.

Correlated failures are multi-point failures that share common root cause. Examples:

- **Operating System Bugs:** An operating system bug can impact numerous machines within a cluster or even an entire availability zone. This highlights the importance of staged rollouts for operating system upgrades: first a small set of machines, then a canary group, and finally, a broader deployment.

- **Hard Drive Failures:** A manufacturing defect in a batch of hard drives can lead to simultaneous failures across availability zones where those drives were deployed. Mitigation strategies include staggered hard drive replacements and using drives from different batches or manufacturers within a data center.
- **Network Switch Failures:** Servers are often organized into racks, with multiple machines connected within each rack. Communication between machines in different racks requires traversing a network switch. A failure of this switch can impact all machines within the affected rack. Therefore, when replicating data within a cluster, it's essential to ensure that each replica resides on a different rack. Ideally, no two replicas of the same data should be on machines within the same rack.
- **Natural Disasters (Fire, etc.):** These can incapacitate an entire availability zone or even a region.
- **Power Outages:** These can also take down an availability zone.
- **Undersea Cable Cuts:** In a worst-case scenario, an undersea cable cut can render an entire region unavailable.

### *23.3.2. Mean Time To Failure*

Mean Time To Failure (MTTF) measures the average time a product or system works before experiencing a failure.

$$\textbf{MTTF} = \frac{\textbf{Total operational time}}{\textbf{Number of failures}}$$

### *23.3.3. Mean Time To Repair*

Mean Time To Repair (MTTR) is a metric used to measure how long it takes to fix a broken system or device. It's a key indicator of a system's maintainability and repair efficiency.

$$\textbf{MTTR} = \frac{\textbf{Total repair time}}{\textbf{Number of incidents}}$$

### *23.3.4. Double Fault*

System design must account for the possibility of correlated failures. While even robust systems like Aurora are designed to withstand a single correlated failure (such as an AZ outage), managing multiple, simultaneous correlated failures (double faults) is significantly more challenging.

The key to minimizing the risk of double faults is ensuring that the **MTTR** << **MTTF**. This large disparity between MTTR and MTTF reduces the probability of a second failure occurring before the first is resolved.

### 23.4. Replication

Replication involves creating multiple copies of the same data, typically key-value pairs where the key identifies the value. The value itself is replicated across multiple machines. Maintaining consistency across these replicas is crucial, and several algorithms address this.

#### *23.4.1. Primary-Secondary Replication*

One replica is designated the primary, and all writes are directed to it. The primary then propagates these writes to the secondary replicas, often asynchronously, as shown in Figure 100. Relaxed reads at the secondaries can improve performance. This can be extended to multi-primary setups.

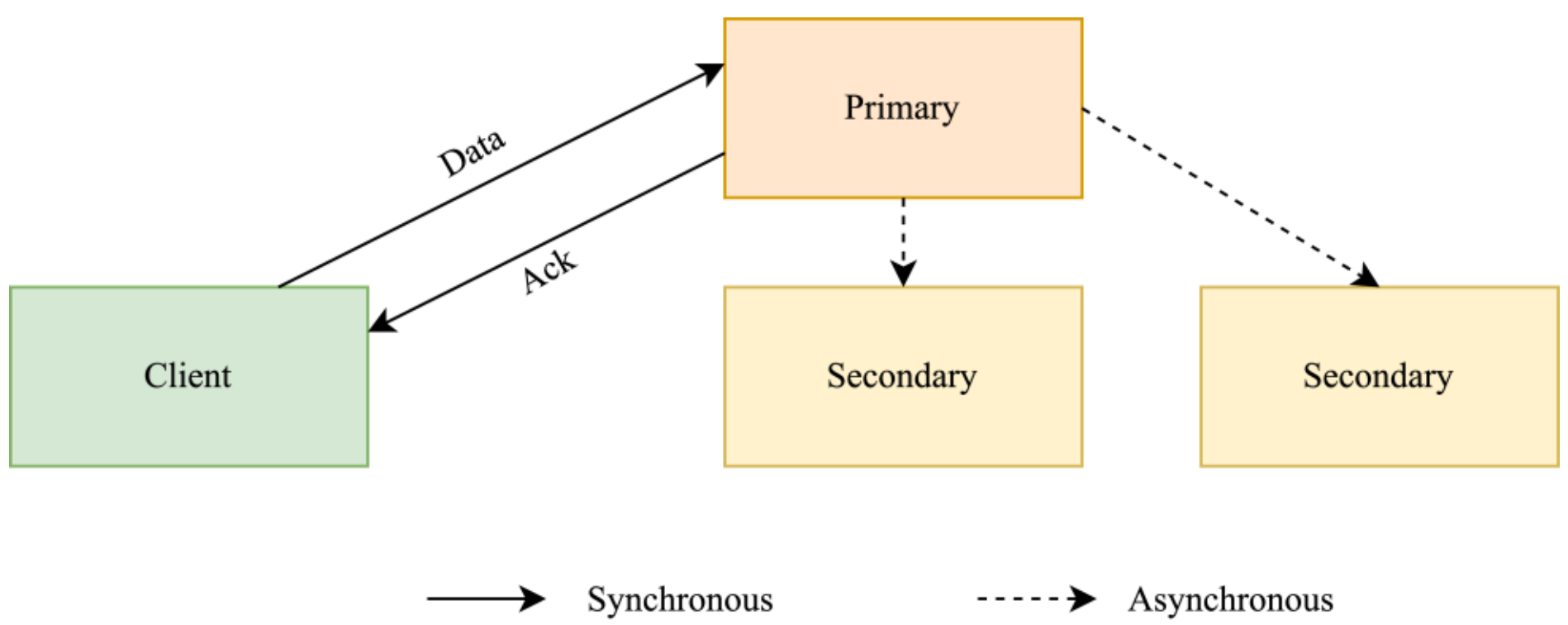

**Figure 100**: Primary-secondary replication.

#### *23.4.2. Leaderless Replication (Quorum)*

All replicas are equivalent. Each item has a version number that increments with each write. For a majority, at least $\mathbf{N/2+1}$ replicas must participate in both writes and reads at the same version number, as shown in Figure 101:

- **Writes:** The client sends writes to all **N** replicas. An acknowledgment is sent when $\mathbf{N/2+1}$ replicas have committed the write.
- **Reads:** The client queries $\mathbf{N/2+1}$ replicas. At least one of these will have the latest value, ensuring consistency.

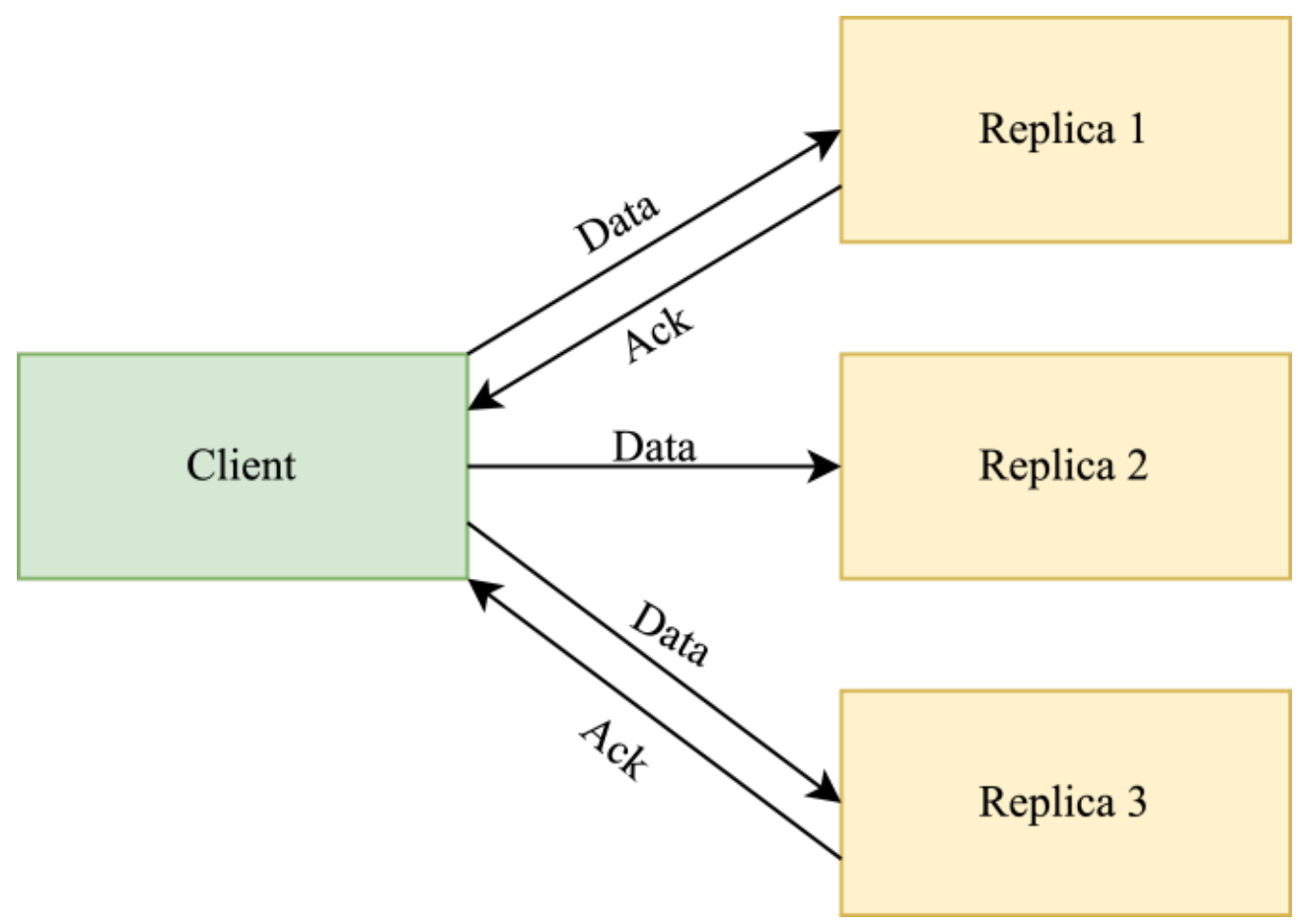

**Figure 101**: Leaderless replication.

Generally, if there are **N** replicas, **W** represents the write quorum, and **R** represents the read quorum, then $\mathbf{W+R>N}$ is required for consistency. Distributed file systems like Harp also use quorum replication, where the data item is a file or file sub-stripe. Quorum replication is linearizable, providing the latest value during consistent reads from **R** replicas.

**Recommended Reads:**

- **15. Dynamo: Amazon's Highly Available Key-value Store** for more details on how Dynamo achieves replication using quorum.
- **18. Practical Uses of Synchronized Clocks in Distributed Systems** for more details on how Harp replication works.

Note that quorum-based voting is used for replication only; it does not solve the consensus problem.

*23.4.3. Chain Replication*

In chain replication [175], replicas are arranged in a chain, as shown in Figure 102. All writes are sent to the **head**, and the **tail** sends the final acknowledgment.

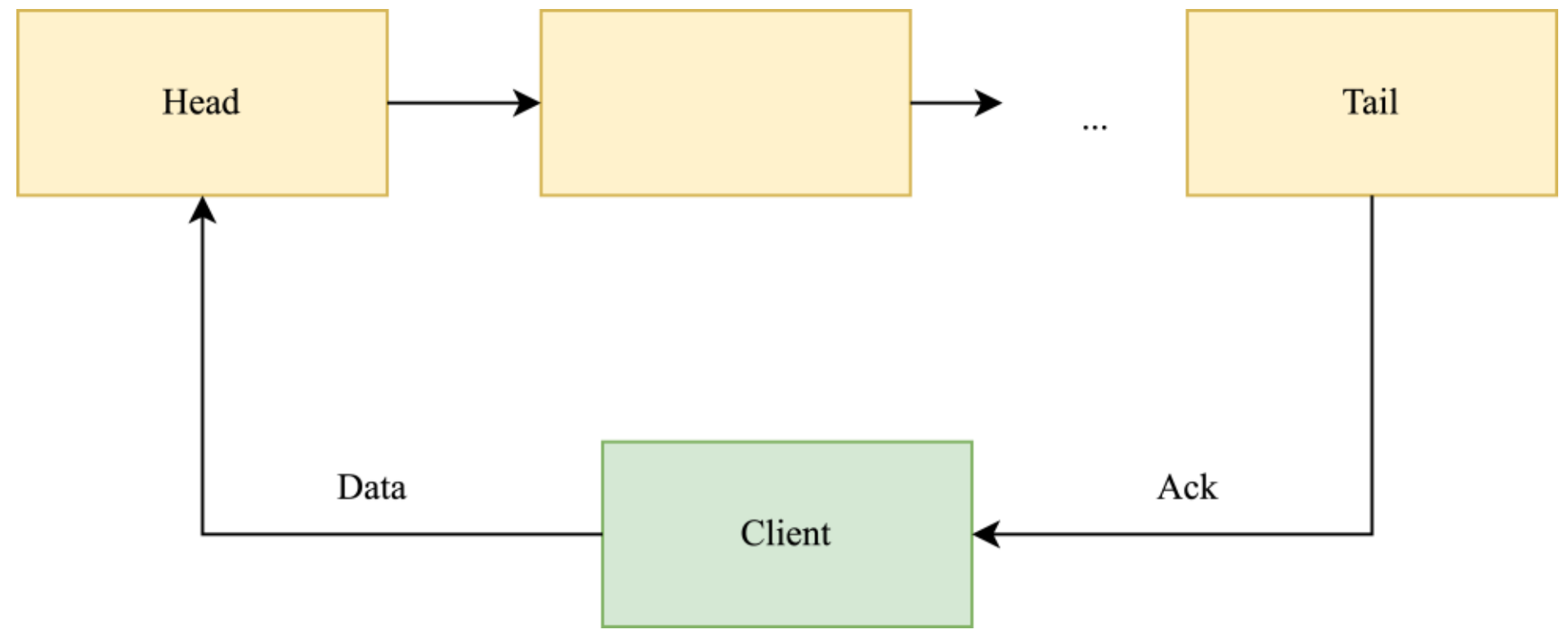

**Figure 102**: Chain replication.

Each node commits the write and forwards it to the next node in the chain. Each node has more information than its successor but less than its predecessor.

Each server maintains a list of updates it has sent but that haven't been committed at the tail. Upon reconnecting, the node forwards these uncommitted updates. This list is garbage collected when the tail periodically acknowledges all updates.

- If the head fails, the next node in the chain becomes the head. Pending writes are dropped because they were not yet committed.
- If a middle node fails, the two adjacent nodes connect with each other. Pending writes are propagated again from the point of failure.
- If the tail fails, its predecessor becomes the tail.

Consistent reads must go to the tail, as that's where the committed value resides.

**Advantages:** Lower network usage compared to quorum-based approaches, as clients don't need to send data to **W** members.

**Disadvantages:** Higher latency due to the sequential nature of writes.

Delta [176] is one example of an implementation of chain replication.

### 23.5. Relational (SQL) Databases

SQL databases guarantee Atomicity, Consistency, Isolation, and Durability (ACID) properties for transactions:

- **Atomicity**: A transaction either completes entirely or not at all. No partial updates are allowed.
- **Consistency**: Transactions maintain database integrity by adhering to all defined constraints. The database remains in a valid state before and after each transaction.
- **Isolation**: Transactions appear to execute independently, as if they were the only ones running. Different isolation levels exist, with serializable isolation ensuring transactions appear to occur in a specific serial order.
- **Durability**: Once a transaction is committed, the changes are permanent and survive even system failures.

Consider a bank transfer from account **A** to account **B**:

- Atomicity ensures both the debit from **A** and the credit to **B** occur, or neither does.
- Consistency ensures the total balance across all accounts remains constant.

A simple approach might involve locking the entire database during a transaction, however, this is highly inefficient. Modern databases support **concurrent transactions**, meaning multiple transactions can modify the same data simultaneously. This requires concurrency control mechanisms like:

- **Locks**: Data items (rows, cells) are locked in shared (read) or exclusive (write) mode to prevent conflicting modifications.
- **Version Numbers**: Data items are versioned, and modifications create new versions, allowing reads to occur on older versions while writes happen on newer ones.

To ensure durability despite potential server failures, databases use **transaction logs** (a.k.a. write-ahead logs). These logs record all changes before they are applied to the database itself. Two types exist: **UNDO logs** (covered in 2PC chapter) and **REDO logs** (used by Aurora, among others).

### *23.5.1. REDO Logs*

REDO logs contain records like:

- **[start, TID]**: Transaction **TID** has begun.
- **[write, TID, X, new_value]**: Transaction TID wrote **new_value** to item **X**.

- **[commit, TID]**: Transaction **TID** committed successfully.
- **[abort, TID]**: Transaction **TID** was aborted.

During a transaction, locks are held, and REDO log entries are made. For the bank transfer example, here are the REDO log entries, assuming an initial balance of $100 and $150 in **A**'s and **B**'s account respectively and a transfer of $50:

```
start 100346
write 100346 A $50
write 100346 B $200
commit 100346
```

Once the commit record is logged, the transaction is considered committed. The database state can be reconstructed solely from the REDO log by applying committed transactions in the order of their commit records. The log effectively becomes the source of truth.

**Interesting note**: Locks are only necessary before a transaction commits (before the **commit** entry is in the REDO log). After a transaction commits, its modifications can be replayed from the REDO log in commit order to rebuild the database state.

#### *23.5.2. Popular SQL databases*

- **MySQL** [105]**:** A widely used, open-source relational database known for its ease of use and reliability, making it a popular choice for web applications.
- **PostgreSQL** [107]**:** Another powerful open-source relational database, known for its extensibility and adherence to SQL standards. It's often preferred for applications requiring complex data management.
- **Microsoft SQL Server** [177]**:** A robust relational database developed by Microsoft, commonly used in enterprise environments for high-volume transaction processing and business intelligence.
- **SQLite** [127]**:** A lightweight, serverless relational database that's embedded within applications. It's often used for mobile apps and small-scale projects.

### 23.6. Distributed Relational (SQL) Databases

A distributed SQL database stores data across multiple servers, potentially geographically dispersed. This partitioning also applies to transaction logs. Building such a system requires sophisticated algorithms, such as 2PC.

Distributing a database is a notoriously difficult problem in computer science. Google's Spanner tackled this by distributing the entire database engine, relying heavily on 2PC. While 2PC can be slow and limit throughput, Spanner's use of TrueTime enabled it to achieve reasonable performance, making it a unique example of a true distributed database.

Aurora, however, took a different approach. It is a distributed RDBMS that achieves distribution without using 2PC. The next section will explore how Aurora accomplishes this.

### 23.7. Aurora

Aurora avoids distributing the database engine itself. Instead, it relies on replicating transaction logs. Since the transaction log effectively is the database (as it can be used to rebuild the database state), distributing the log achieves the desired distribution.

Aurora uses a three-server setup for each database: a **primary** instance that handles all transactions, and two **secondary** replicas that maintain copies of the database. The secondaries receive asynchronous updates from the primary after a transaction commits.

Each instance is a fully functional, independent SQL database server (e.g., MySQL).

Therefore, Aurora's approach differs from traditional distributed SQL databases. The transaction execution occurs on a single node (the primary). What makes Aurora "globally available" is the distributed replication of the transaction logs.

Let's delve deeper, but first, let's address some common questions.

#### *23.7.1. FAQ*

- **Doesn't distributed SQL require 2PC?** Yes, true distributed SQL databases do. However, Aurora's SQL engine itself is not distributed. Only the transaction logs are.
- **Is consensus required for transaction log replication?** No. Consensus is used to build distributed logs, where each entry represents a value agreed upon by multiple participants. In Aurora, the transaction log entries are determined by the single primary SQL instance. The underlying system's job is simply to replicate these pre-determined entries. Simple quorum replication is sufficient for this purpose.

### *23.7.2. Naive Approach to Transaction Log Replication*

A key goal in Aurora is replicating the transaction logs generated by the SQL engine. A naive approach, as described in the paper, involves the following steps:

1. Write logs to EBS.
2. EBS mirrors the logs (using software mirroring).
3. Forward the modifications to the replica SQL engine instance.
4. Write logs to the replica instance's EBS.
5. Replica instance's EBS mirrors the logs.

The paper also mentions additional logs, such as binary logs, stored in S3. Crucially, steps 1, 3, and 5 are sequential. A transaction cannot commit without the initial write to EBS (step 1). Replication cannot begin before the commit (and thus step 1). Similarly, the replica's EBS mirroring (step 5) cannot start before the logs are written to the replica's EBS (step 4, which is dependent on step 3).

These sequential dependencies introduce latency and excessive write operations.

## 23.8. Aurora's Architecture

Aurora employs a **service-oriented architecture (SOA)**, separating **compute** and **storage** layers, as shown in Figure 3 in the paper. This contrasts with a **shared-nothing architecture**. The storage layer handles transaction log storage, materialization, and replication, while the compute layer hosts the SQL engine instances.

### *23.8.1. Compute*

The compute layer consists of three instances across different availability zones. The primary SQL engine resides in one zone, and REDO logs are forwarded to the other two replicas using chain replication, although, in this case, the ack is sent by the primary instance itself and the replication is asynchronous.

### 23.8.2. Storage

The storage layer maintains six replicas of each log entry, with two replicas per availability zone. Replication is quorum-based, with a write quorum (**W**) of 4 and read quorum (**R**) of 3. The primary compute instance is responsible for replicating log entries, and an entry is considered committed once four out of the six replicas acknowledge receipt.

For performance reasons, all input/output operations are handled in batches. The storage layer's internal workings are more intricate, and further details can be found in section 3.3 of the paper.

## 23.9. Data Partitioning

The database volume is divided into 10GB **segments**. The SQL engines cache these segments, which hold the actual data. Each segment is replicated six ways in the storage layer. These replicated segments are called **Protection Groups (PGs)**.

Each segment stores multiple **pages** of the database. A page is the smallest unit of data that the database reads from or writes to storage at one time. It's a block of contiguous memory. The compute instances caches the database pages in the buffer cache. Upon miss, the database page is read from the storage.

For each page, the primary compute instance maintains the following metadata:

- The highest log entry number that modified the page. Since the primary commits log entries, it knows which entry last updated a given segment.
- The storage nodes that have a complete, materialized version of the page. As the primary receives acknowledgments from the storage replicas after writes, it tracks which storage nodes have all updates for each page.

## 23.10. Transaction Logs

The transaction log is the complete database.

### 23.10.1. Log Structure

Log records in Aurora are assigned a **Log Sequence Number (LSN)**. Because the SQL engine is a single instance, log entries are pre-determined, eliminating the need for 2PC or consensus during replication. Log entries are simply replicated using

Quorum. Additionally, those replicas that may miss an entry will receive it through gossip protocols.

Two important LSN values are tracked:

- **Volume Complete LSN (VCL):** The storage service guarantees log availability up to VCL. Log entries with LSNs higher than the VCL are not guaranteed to be available and can be discarded.
- **Volume Durable LSN (VDL):** The highest LSN for which the SQL engine guarantees database consistency. The VDL must be less than or equal to the VCL.

The VCL and VDL are independent because the compute and storage layers operate independently. The VCL is a storage layer concept related to log availability guarantees. The VDL is a compute layer concept, built upon the VCL, that guarantees database consistency up to that point.

When a new log entry is added, an LSN is generated using the current VDL and an upper bound called the **LSN Allocation Limit (LAL)**. The LAL limits how far LSNs can advance ahead of database consistency points, i.e., VDL.

While data is segmented, Aurora maintains a single, globally ordered transaction log. Each log entry includes a back pointer to the previous log entry relevant to a specific segment. These back pointers enable the system to efficiently identify all log entries associated with a given segment.

A key concept related to segments and logs is the **Segment Complete LSN (SCL)**. SCL indicates the point at which all log records for a particular replica of a segment are complete. This is used in gossip protocol to exchange log records.

*23.10.2. Log Materialization*

The storage layer not only stores transaction logs but also independently materializes them into database pages in the background. These materialized pages might not always be the most up-to-date version; new updates can be applied to them to reflect the latest state.

## 23.11. Transaction Processing

All transactions are executed by the primary instance. Since the primary cannot hold all database pages in memory, cache misses can occur. When a cache miss happens, the primary requests the complete page from a storage node that has it.

Critically, remember that there's only one primary SQL instance. All transactions flow through it. This means the primary always knows the highest sequence number that modified any given page.

The transaction flow is as follows:

1. The primary computes state updates (writes).
2. The primary writes the logs to the storage layer.
3. The primary waits for a write quorum from the storage nodes.
4. Storage nodes acknowledge as soon as the log is committed.
5. The primary sends REDO logs to the replica SQL instances.

The VDL advances as logs are written and acknowledged. Once the VDL advances past the commit log entry for a transaction, the primary sends an acknowledgment back to the client, confirming the transaction's completion.

## 23.12. Reads

Reads in Aurora are typically served from the buffer cache. A storage fetch is only necessary when the buffer cache lacks the requested page. Pages in the buffer cache are guaranteed to be the latest version, ensured by evicting a page only if its LSN is greater than or equal to the current VDL.

Aurora offers different read consistency levels:

- **Strong Reads (Serialized/Transactional Reads):** These reads are routed to the primary instance to guarantee the latest data, as the primary is the only instance fully up-to-date. Note that the SQL engine must also operate in serializable isolation mode to support these reads.
- **Relaxed Reads:** These reads can return slightly stale data. They can be served from any replica instance, which typically lag behind the primary by around 20ms. Replicas asynchronously consume log records with LSNs less than or equal to their respective VDLs.

- **Point-in-Time Reads:** These reads retrieve data as it existed at a specific LSN (the read point). This ensures the data is at least as current as the read point. The database can serve these reads from storage replicas whose page VDL meet or exceed the specified read point.

Finally, each segment has a **per-PG Minimum Read Point LSN (PGMRPL)**. This is the lowest LSN at which reads for that segment are in-process. The database maintains this value, which acts as a lower watermark. Log records older than the PGMRPL are no longer needed, allowing the storage layer to materialize pages up to the PGMRPL and garbage collect older log records.

### 23.13. Recovery

The storage layer handles log application, meaning the SQL engine doesn't need to replay logs after a crash. Upon recovery, the primary simply needs to recompute the VDL from the storage layer and resume logging new transactions from that point. The paper reports a recovery time of approximately 10 seconds.

### 23.14. Administration

Aurora has high fault tolerance:

- Writes to storage remain available even if an entire AZ goes down.
- Reads are possible with the loss of a single AZ. Furthermore, Aurora can withstand an additional failure affecting a subset of machines in another AZ, such as a switch failure.

Aurora's design simplifies administration in several ways:

- **Software Upgrades:** The fast recovery times enable zero-downtime patching. Because all state is persisted in the storage layer's logs, the compute layer can simply wait for active transactions to complete before restarting with the updated software (e.g., a new binary).
- **Simplified Storage Management:** There's no need to shut down SQL machines to update storage or adjust disk usage.

## 23.15. Evaluation

- Aurora's read and write throughput scales linearly with instance size, which is expected given the increased processing capacity of larger instances.
- Aurora demonstrates significantly higher throughput than MySQL, likely due to its sophisticated storage layer and batching mechanisms (though the specific reasons require further investigation). A comparison of latency under the same test conditions could have been valuable.
- Replica lag in Aurora is influenced by the write rate on the primary. In one test, a write rate of 10,000 writes/sec resulted in a replica lag of only 5.38ms. This is substantially lower than naive MySQL's replication lag, which reached 300s.

## 23.16. Paper Remarks

This paper is quite readable. The concepts are straightforward, especially once the design's niche is understood: the transaction log is the database. Global availability is simply achieved through transaction log replication. The paper also serves as an excellent resource for learning fundamental cloud computing concepts and operational details, providing practical insights into how cloud systems are managed.

# 24. Spanner: Google's Globally-Distributed Database

(24) Corbett, J. C.; Dean, J.; Epstein, M.; Fikes, A.; Frost, C.; Furman, J. J.; Ghemawat, S.; Gubarev, A.; Heiser, C.; Hochschild, P.; others. Spanner: Google's Globally Distributed Database. ACM Transactions on Computer Systems (TOCS) 2013, 31 (3), 1–22.

We have explored several databases up until now, most of which have been NoSQL. We also discussed one SQL database that is not truly distributed, as its transactions are executed on a single machine and the logs are subsequently replicated. Now, we will take a look at a database that is truly distributed, where transactions run in a distributed fashion.

This landmark paper, presented at Operating System Design and Implementation (OSDI) '12, has become one of Google's most significant contributions to distributed computing. It introduced TrueTime [178] that revolutionized system assumptions. Authored by J.C. Corbett, and with contributions from pioneers like Jeff Dean and Sanjay Ghemawat, it represents the leading edge of the field.

In this insight, we will first revisit the different types of databases. Then, we will review the CAP theorem, a well-known result that is widely understood in the industry. Next, we will dive deep into SQL databases, specifically discussing isolation properties. We will then cover serializability and understand how it can be achieved in distributed SQL databases. Following that, we will introduce Google's TrueTime, the mammoth system that serves as the backbone for solving serializability in distributed SQL databases. Then, we will examine Spanner and understand the intricacies of the system, running through examples to see how transactions work. Lastly, we will revisit the CAP theorem to understand how Spanner and TrueTime fit within its framework.

**Recommended Reads**:

- **18. Practical Uses of Synchronized Clocks in Distributed Systems** where we discussed why clock synchronization is necessary but not sufficient for external consistency.

- **22. A New Presumed Commit Optimization for Two Phase Commit** where two-phase commits (2PC) was introduced and how it is implemented in a distributed system.
- **23. Amazon Aurora: Design Considerations for High Throughput Cloud-Native Relational Databases** where Amazon Aurora was introduced that builds a (not quite) distributed SQL database without using 2PC.

### 24.1. Databases

Databases generally fall into two categories:

- **SQL Databases:** These adhere to ACID (Atomicity, Consistency, Isolation, Durability) properties to ensure reliable transactions.
- **NoSQL Databases:** These often follow BASE (Basically Available, Soft state, Eventually consistent) principles, prioritizing availability and scalability over strict consistency.

#### *24.1.1. SQL Databases*

SQL databases guarantee ACID properties, ensuring reliable data transactions:

- **Atomicity:** Transactions are treated as a single, indivisible unit of work. Either all changes within a transaction are applied, or none are.
- **Consistency:** A transaction maintains the integrity of the database by ensuring that all defined rules and constraints (like keys, values, and other constraints) are preserved.
- **Isolation:** Transactions are executed independently of each other. One transaction's operations won't interfere with another's.
- **Durability:** Once a transaction is committed, the changes are permanent and will survive even system failures.

Examples of popular SQL databases include MySQL [105], Microsoft SQL Server [177], Google Spanner (a distributed database), and Amazon Aurora (also a distributed database).

#### *24.1.2. NoSQL Databases*

NoSQL databases often prioritize availability and scalability over strict consistency, adhering to what's sometimes described as BASE principles:

- **Basically Available:** NoSQL databases often favor availability over strong consistency. This means the system remains operational even if some data is temporarily inconsistent. Conflicts arising from concurrent operations are typically resolved later.
- **Soft State:** The database's state isn't guaranteed to be consistent at any given moment. It's a "soft" state, meaning its accuracy is probabilistic.
- **Eventually Consistent:** The database strives to achieve consistency eventually. Given enough time, all updates will propagate throughout the system, and all nodes will see the same data.

**Recommended Reads**:

- **16. Cassandra - A Decentralized Structured Storage System**
- **15. Dynamo: Amazon's Highly Available Key-value Store**

In this chapter, our focus will be on SQL databases only.

### 24.2. The CAP Theorem

**Disclaimer**: FLP impossibility result is a more fundamental result than the commonly stated CAP theorem [179].

The CAP theorem states that:

**A distributed data store can only guarantee two out of the following three properties: Consistency, Availability, and Partition Tolerance.**

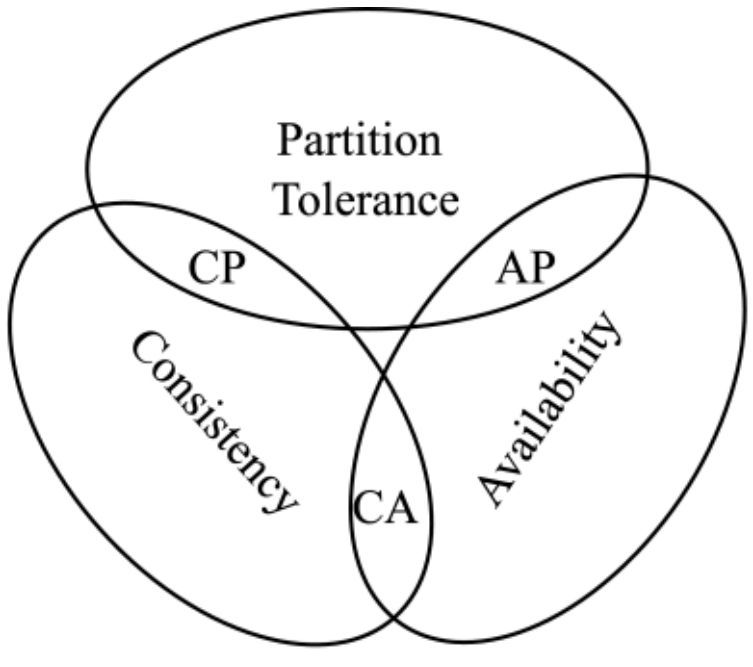

**Figure 103**: The CAP theorem.

This is shown in Figure 103. The CAP theorem leads to choices like:

- **Strong Consistency vs. Eventual Consistency:** How strictly do you need data to be consistent across all nodes?
- **Availability:** How important is it that the system remains operational and responsive, even during failures and partitions?
- **Partition Tolerance:** How well does the system handle network partitions?

ACID databases are categorized as **CP (Consistent and Partition-tolerant)**. This means that in the event of a network partition, some shards of the database might become unavailable to maintain consistency. BASE databases, on the other hand, are **AP (Available and Partition-tolerant)**. They prioritize availability during network partitions, even if it means temporarily sacrificing consistency.

Beyond CAP, other related theorems exist, such as:

- **PACELC** [180]**:** This theorem expands on CAP, stating that if there's a partition, you must choose between Availability (A) and Consistency (C), else (when there's no partition), you still have a trade-off between Latency (L) and Consistency (C).
- **Harvest-Yield** [181]**:** This is an older theorem that's less commonly discussed today.

### 24.3. Availability Numbers

System availability is often expressed using "nines". For instance:

- Three 9s: 99.9% availability
- Four 9s: 99.99% availability
- Five 9s: 99.999% availability

This percentage represents the portion of queries expected to be answered without a system failure (though the response might not be correct or consistent).

Typically, availability increases with the geographic scope of the system. Local replication might offer lower availability, while regional and global systems aim for higher availability. This is because larger, geographically distributed systems have more resources and are less susceptible to correlated failures that could affect a smaller, localized system. For example, a system might provide 99.9% availability within a single cluster, 99.99% within an availability zone, and 99.999% globally.

## 24.4. SQL Databases

As previously mentioned, SQL databases must guarantee ACID properties. Atomicity and Consistency are crucial, and these are often achieved through transaction isolation.

Consider a bank database and a transfer transaction that moves funds from account **A** to account **B**. This transaction must exhibit the following properties:

- **Atomicity:** The entire transfer must either complete successfully or fail entirely. The database (and any external systems) should never observe an intermediate state where the money has left account **A** but not yet arrived in account **B**.
- **Consistency:** The total balance across all accounts must remain constant after any transaction.

### *24.4.1. Transaction Isolation*

Transaction isolation determines how and when the changes made by one transaction become visible to other concurrent transactions. Essentially, it defines the level of segregation between transactions.

The simplest way to ensure isolation is to execute transactions serially (one after another). However, this approach is often too slow for practical applications, so databases allow concurrent transactions.

Concurrency introduces the possibility of one transaction seeing the partial, non-atomic effects of another. Returning to our bank transfer example, say **A** has an initial balance of $100 and **B** has an initial balance of $150, and $50 is transferred from account **A** to account **B**. The steps might look like this:

```
start 100014
write 100014 A $50 # Step 2
write 100014 B $200
commit 100014
```

These steps are also recorded in the transaction log. If another transaction runs concurrently, it could, after step 2, see that **A**'s balance has been reduced, but **B**'s balance hasn't yet been updated.

Several techniques are used to prevent this:

- **Locks**: The first transaction could acquire an exclusive lock on account **A** after the write operation. This prevents other transactions from reading **A**'s value until the first transaction commits. Locks are managed using a two-phase locking protocol [182] (not to be confused with the two-phase commit protocol).
- **Version Numbers**: Data items **A** and **B** could have version numbers. Any write operation creates a new version. These new versions are only committed when the transaction commits. The transaction can only commit successfully if the version numbers of **A** and **B** haven't changed by another transaction in the meantime.

24.4.1.1. Levels of Isolation

SQL databases offer different levels of transaction isolation, each with varying degrees of consistency and performance:

- **Serializable:** This is the highest isolation level, guaranteeing that the effect of concurrent transactions is equivalent to some serial execution order (also known as **multi-data sequential consistency**). This makes reasoning about transactions straightforward for users.
- **Repeatable Reads:** Similar to Serializable, but allows other transactions to insert new data items. This means that while a transaction might see a consistent snapshot of existing data, it might see newly inserted data items.
- **Read Committed:** Allows a transaction to read changes made by other committed transactions while it's in progress. While this might seem reasonable, it can lead to inconsistencies. Imagine two transactions: $\mathbf{T_1}$ transferring $50 from **A** to **B**, and another $\mathbf{T_2}$ reading the balances of **A** and **B**. A serializable read would return either {**A**: $100, **B**: $150} or {**A**: $50, **B**: $200}, depending on the order of execution. However, a read committed read could return {**A**: $100, **B**: $200}. This can occur if $\mathbf{T_2}$ reads **A**'s balance ($100), then $\mathbf{T_1}$ commits, and then $\mathbf{T_2}$ reads **B**'s balance ($200).
- **Read Uncommitted (Dirty Reads):** The lowest isolation level, allowing a transaction to read uncommitted values from other transactions. This can lead to reading data modified by transaction that is later rolled back, resulting in inconsistencies.

The mechanisms used by databases to achieve serializability at high concurrency are complex, involving concepts like serializable schedules, conflict serializability, view serializability, and recoverability. For the sake of brevity, those details are omitted here.

*24.4.2. Reads*

In the context of database transactions, a "data item" refers to the smallest unit of data that can be locked and is guaranteed to be read and written atomically. Typically, a data item can be a row in a table.

SQL databases support various types of reads:

- **Single Data Item Read:** Retrieves only one data item.
- **Multiple Data Item Read:** Retrieves multiple data items.

A database read is considered **consistent** if the retrieved values reflect the state of the data items at some point in time.

For instance, if accounts **A** and **B** initially have balances of $100 and $150, respectively, and a transaction transfers $50 from **A** to **B**, a consistent read should return either {**A**: $100, **B**: $150} (before the transfer) or {**A**: $50, **B**: $200} (after the transfer).

24.4.2.1. Consistently Reading a Single Data Item

Reading a single data item consistently is straightforward. Regardless of when the query executes, it will retrieve a consistent value for that single item. This is supported by all isolation levels excluding read uncommitted.

24.4.2.2. Consistently Reading Multiple Data Items

Consistent reads across multiple data items can be achieved using **read-only transactions**. These transactions often acquire locks on all involved data items before reading, and release the locks afterward. This is only guaranteed by the serializable isolation level.

### 24.5. Distributed SQL Databases

Distributed SQL databases are those where data is sharded across multiple nodes. Transaction execution within these systems is similar to how it works in standalone databases:

1. **Locking:** The necessary data items are locked on each node involved in the transaction.
2. **Commit:** The transaction is then committed. However, the commit process is more complex in a distributed setting, requiring a 2PC.

In our bank database example, if data items **A** and **B** are on node 1 and node 2, then the transfer transaction would be recorded as follows:

**Node 1**

```
start 100014
write 100014 A $50
commit 100014
```

**Node 2**

```
start 100014
write 100014 B $200
commit 100014
```

### 24.6. Serializability v/s Strict Serializability

Serializability ensures that the effects of all transactions are equivalent to some sequential execution order (also known as **sequential consistency**). Strict serializability (or **external consistency**) builds upon this by considering the temporal order of transactions.

A helpful analogy clarifies the distinction:

Imagine you successfully commit a transaction transferring money from your account to a friend's account. You then immediately call your friend to check their balance. If they don't see the transferred funds, the database violates strict serializability (external consistency).

However, the database could still maintain internal consistency (serializability). This means the final database state reflects some valid sequential order of transactions, even if it doesn't align with the real-time order of events. For example, if your friend later tries to withdraw $200, the success or failure of that withdrawal will depend on whether the database ordered their transaction after or before your transfer transaction, even if, in reality, your friend's request came after your transfer.

Another way to explain strict serializability is that the database's transaction ordering must respect the **temporal semantics** observed by external users. Formally, if transaction $\mathbf{T_2}$ is submitted by a client only after transaction $\mathbf{T_1}$ commits, then the database must order $\mathbf{T_2}$ after $\mathbf{T_1}$.

*24.6.1. Strict Serializability in Standalone Databases*

In serializable mode, SQL databases can guarantee serializability because all data resides within the database's boundaries. It's easy to implement locking mechanisms to ensure a transaction completes before the next one begins.

Even strict serializability is straightforward to achieve. Since all operations occur within the same process, the effect of a committed transaction is immediately visible to the next transaction.

*24.6.2. Strict Serializability in Distributed Key-Value Systems*

Data stores that operate on single data items (such as key-value stores, distributed file systems, and lock services like ZooKeeper) primarily rely on a distributed log to coordinate changes to the state of those items.

**Recommended Read**: **21. ZooKeeper: Wait-free Coordination for Internet-Scale Systems**

With single data items, strict serializability effectively becomes **linearizability**. The distributed log itself serves as the single source of truth for the ordering of transactions. Once a transaction is committed to the immutable distributed log, its effects become immediately visible to subsequent transactions.

All writes are linearizable because they must go through the leader. Even during leader changes, the new leader must catch up on all existing committed transactions in the distributed log before accepting new transactions. This ensures that the new leader's writes respect the prior ordering.

Linearizable reads are achieved by executing a read-only transaction at the leader. The leader guarantees that a transaction is acknowledged only after it's committed to the distributed log. Therefore, all read-only transactions through the leader see the latest state of the data store (because they are causally related to and ordered after previous transactions, as they go through the same leader).

It's important to note that relaxed reads (i.e., reads from a non-leader replica) might not be linearizable.

#### *24.6.3. Strict Serializability in Aurora*

Amazon Aurora achieves strict serializability because its transaction processing occurs within a single, non-distributed instance. This single instance handles all transactions and provides the latest committed value at any given point in time. This architecture is similar to a standalone database, making serializability, and even strict serializability, easy to implement.

However, as argued before, this architecture also means that Amazon Aurora isn't a truly distributed SQL database. All transactions must pass through a single primary instance, which can easily become a performance bottleneck.

#### *24.6.4. Strict Serializability in Distributed SQL Database with 2PC*

In a true distributed SQL database, data items are distributed across multiple participants, and these participants must coordinate commits using a protocol like 2PC. Each participant might itself be a Paxos group, composed of multiple nodes maintaining a distributed log for fault tolerance.

However, 2PC alone does not guarantee serializability. Consider the bank database with accounts **A**, **B**, and **C** on nodes 1, 2, and 3. **A** has \$100, **B** has \$150, and **C** has \$0. Two transactions, $\mathbf{T_1}$ (transfer \$50 from **A** to **B**) and $\mathbf{T_2}$ (transfer \$200 from **B** to **C**), are submitted, with $\mathbf{T_2}$ following $\mathbf{T_1}$.

1. Transaction $\mathbf{T_1}$ is initiated.
2. Nodes 1 and 2 accept $\mathbf{T_1}$ and acknowledge the PREPARE message.
3. Node 2 commits $\mathbf{T_1}$ and releases its locks before Node 1 commits.

4. Transaction $\mathbf{T_2}$ is initiated.
5. Nodes 2 and 3 accept $\mathbf{T_2}$. Critically, Node 2 now reflects the updated balance of B (due to the partial commit of $\mathbf{T_1}$).
6. $\mathbf{T_2}$ commits on both Nodes 2 and 3.

The problem is that $\mathbf{T_1}$ is still uncommitted on Node 1 (and hence only partially committed) when $\mathbf{T_2}$ reads the updated balance of **B** on Node 2. This creates a read uncommitted isolation level, where $\mathbf{T_2}$ sees the effects of $\mathbf{T_1}$ before $\mathbf{T_1}$ is fully committed, violating serializability.

The solution to achieving both serializability and strict serializability in a distributed SQL database involves establishing a globally consistent and externally consistent ordering of transactions. This is accomplished by timestamping every transaction upon entry into the system. When a transaction is received by any cohort, it's guaranteed to receive a timestamp greater than all previous transaction timestamps. Furthermore, all transactions will be ordered according to these timestamps.

However, generating such timestamps isn't trivial. It demands tight clock synchronization across all participating nodes. This is where Google's TrueTime comes into play.

### 24.7. TrueTime

Google's TrueTime [178] revolutionized the generation of globally ordered timestamps, representing a significant breakthrough in distributed systems and fundamentally altering common assumptions. This innovation is what made the Spanner paper a landmark contribution to the field.

As Lamport highlighted in another seminal paper [135], achieving perfect global clock synchronization across all machines is impossible. Clocks can drift (running fast or slow), become completely out of sync, or be nearly perfect but still insufficient for ensuring system correctness. Consequently, most distributed system algorithms avoid relying on clock synchronization for correctness. "Most" because some algorithms do use clock synchronization to simplify assumptions; without such simplifications, certain problems (like consensus) would be impossible to solve (see the FLP impossibility).

While clocks can drift, the error bound can be limited. When we use **System.Now()** system API to read the time, it might return a time **t** that is **ε** away from the true global time. The actual time would lie within the interval $[\mathbf{t} - \boldsymbol{\varepsilon}, \mathbf{t} + \boldsymbol{\varepsilon}]$. TrueTime leverages this concept, but on a scale that makes the error bound **ε** extremely small at a global level.

*24.7.1. API*

TrueTime offers a simple and intuitive API (a hallmark of well-designed systems):

- **TT.now()**: Returns a **TTInterval** object, representing a time interval: **[earliest, latest]**.
- **TT.after(t)**: Returns true if time **t** has definitely passed.
- **TT.before(t)**: Returns true if time **t** has definitely not yet arrived.

The absolute time assigned to an event **e**, denoted as $\mathbf{t_{abs}(e)}$, falls within the interval returned by **TT.now()** at the time of the event's invocation ($\mathbf{e_{now}}$):

$$\mathbf{tt.earliest} \leq \mathbf{t_{abs}(e_{now})} \leq \mathbf{tt.latest}$$

The interval itself is bounded by 2 * ε, as shown in Figure 104.

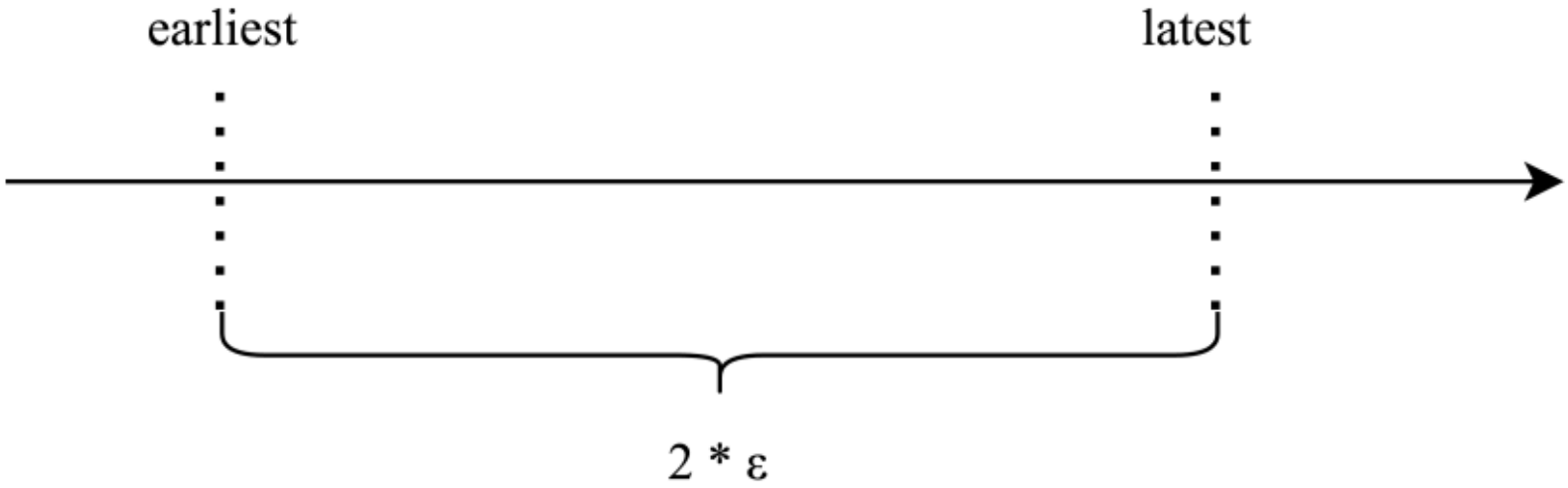

**Figure 104**: Error bound in TrueTime.

**TT.after(..)** and **TT.before(..)** are essentially wrappers around **TT.now()**.

*24.7.2. Architecture*

Google has invested heavily in achieving and maintaining the tight error bounds required by TrueTime.

**Recommended Read**: **18. Practical Uses of Synchronized Clocks in Distributed Systems** where internals of NTP were discussed.

TrueTime relies on GPS and atomic clocks as its time references. While GPS provides highly accurate time, it's susceptible to hardware failures and radio interference. Atomic clocks, while very stable, can also drift due to frequency errors. These potential sources of error contribute to the overall error bound.

Each data center has one or more time servers (or "time masters"). These masters fall into two categories:

- **GPS Time Masters:** Equipped with GPS receivers, these nodes receive time information directly from satellites.
- **Armageddon Masters:** Equipped with local atomic clocks, these nodes serve as a backup to GPS time masters in case satellite connections are unavailable.

Within each data center:

- There's at least one time master.
- Each machine runs a time slave daemon.

The masters synchronize their time references with each other. Armageddon masters advertise a slowly increasing time uncertainty, while GPS masters advertise a much smaller uncertainty (close to ±40ns).

The time slave daemons poll multiple masters to mitigate vulnerabilities. Each master provides a time interval, and these intervals must be merged to determine the largest common interval.

Marzullo's algorithm [183] is used for this interval intersection. It's a classic algorithm for finding the largest intersection of multiple intervals, which is exactly what's needed here. Any intervals that don't intersect with the largest common interval are discarded.

### *24.7.3. Calculating ε*

Each daemon advertises a slowly increasing clock uncertainty, derived from a conservative estimate of worst-case local clock drift.

The daemon polls time masters only every 30 seconds, assuming a maximum drift rate of 200μs per second. This polling interval contributes approximately 6ms to the uncertainty. An additional 1ms is added to account for communication delays to the time masters. Thus, ε is typically between 1 and 7ms.

With that understanding of TrueTime, let's now turn our attention to Google Spanner.

## 24.8. Spanner

Spanner was developed to replace Google's F1 [184] and Megastore [185] databases, which lacked both scalability and the characteristics of true distributed SQL databases. Spanner was designed to offer:

- Geographic replication for global availability.
- Massive scalability, handling tens of terabytes of data.
- External consistency.

When Spanner was created, no other distributed database offered external consistency (although the theoretical underpinnings were understood). Google assembled a team of scientists and engineers to make this groundbreaking system a reality.

### *24.8.1. The Architecture*

Spanner is a distributed SQL database that stores data in a distributed B+ tree structure. However, its architecture has some nuances.

A database resides within a **universe**. Within a database, tables exist. Spanner implements the following mapping within each table:

$$< \mathbf{key}: \mathbf{string}, \mathbf{timestamp}: \mathbf{int64} > \rightarrow \mathbf{string}$$

Each key has an associated timestamp, representing the transaction time when it was created or last modified. This timestamp is crucial for Spanner's **multi-version concurrency control** mechanism, which we'll discuss later.

This structure might resemble a key-value store, but Spanner is not simply a key-value store. It fully supports SQL semantics. Those familiar with Google Bigtable might find Spanner's organization somewhat similar. It's plausible that Spanner was built upon Bigtable technology with the long-term goal of eventually replacing it.

The keyspace is partitioned into **directories**, each with its own replication and placement policy. For instance, a geographic region could be incorporated into the key, and the keyspace could be divided into directories, each corresponding to the keys for a specific region. These directories could then be placed in the appropriate regions. Importantly, a directory contains a contiguous range of keys from the overall keyspace.

The keyspace is subdivided into tablets. A **tablet** is the fundamental unit of storage in Spanner. A tablet can contain one or more directories. Spanner co-locates frequently accessed directories within the same tablet for performance optimization.

#### 24.8.1.1. Interleaved Tables

Tables within a Spanner database can be interleaved, meaning one table can be designated as a child of another. The top-level table in this hierarchy is the **directory table**. Child tables share a key prefix with their parent table.

Here is an example of **Customers** directory table with **Orders** child table:

```
Table Customers {
  CustomerID INT64 PRIMARY KEY,
  Name STRING,
  ...
}

Table Orders {
  CustomerID INT64 REFERENCES Customers(CustomerID), # Shared
prefix
  OrderID INT64 PRIMARY KEY,
  OrderDate DATE,
  ...
}
```

Table interleaving is beneficial because it allows related rows across different tables to be stored within the same directory, improving data locality and query performance.

#### 24.8.1.2. Span Servers

Servers in Spanner are grouped into **zones**, which are similar to physically isolated data centers. Each zone contains hundreds or thousands of **spanservers**, and each spanserver can host between 100 and 1000 tablets.

A **zone master** is responsible for assigning tablets to spanservers. The paper has some other details on the **universe master** and the **placement driver**.

### *24.8.2. Paxos Groups and Transaction Manager*

A tablet isn't stored on a single machine. It's replicated according to the directory's replication policy. The set of machines holding a particular tablet forms a **Paxos state machine**, responsible for that tablet's data. The set of replicas within a Paxos state machine is called a **Paxos group**.

Directories within a tablet can be moved using the **movedir** background task. The Paxos group logs a transaction after the move completes, so **movedir** doesn't block operations. The entire directory can be replicated to another group, and then ownership updated via a transaction.

#### 24.8.2.1. Lock Table

The Paxos group's leader stores the **lock table**, which is used during transaction processing. The lock table maps key ranges to lock states. Implicitly, locks are held on a range of keys. The other Paxos group members participate in consensus and log transaction writes and commits.

#### 24.8.2.2. Transactions within Paxos Group

When a transaction stays within a single Paxos group (i.e., all read/modified data items reside within that group), 2PC is not required. The Paxos group's distributed log and the leader's lock table are sufficient to manage the transaction because all replicas within a Paxos group have the same data.

Locks are acquired at the transaction's start, modifications are logged to Paxos, and then a COMMIT log entry is added.

24.8.2.3. Transactions Across Paxos Groups

When a transaction spans multiple Paxos groups, 2PC is required. These Paxos state machines provide highly available cohorts for 2PC. (See consensus v/s 2PC). For 2PC, spanservers have a unit called **transaction manager**.

The leader of a Paxos group in 2PC is called the **participant leader**, while other replicas within the Paxos group are **participant slaves**. One of the Paxos group becomes the **coordinator**, and its leader becomes the **coordinator leader**. The leaders of the other Paxos groups involved in 2PC become **coordinator slaves**. The setup is shown in Figure 105.

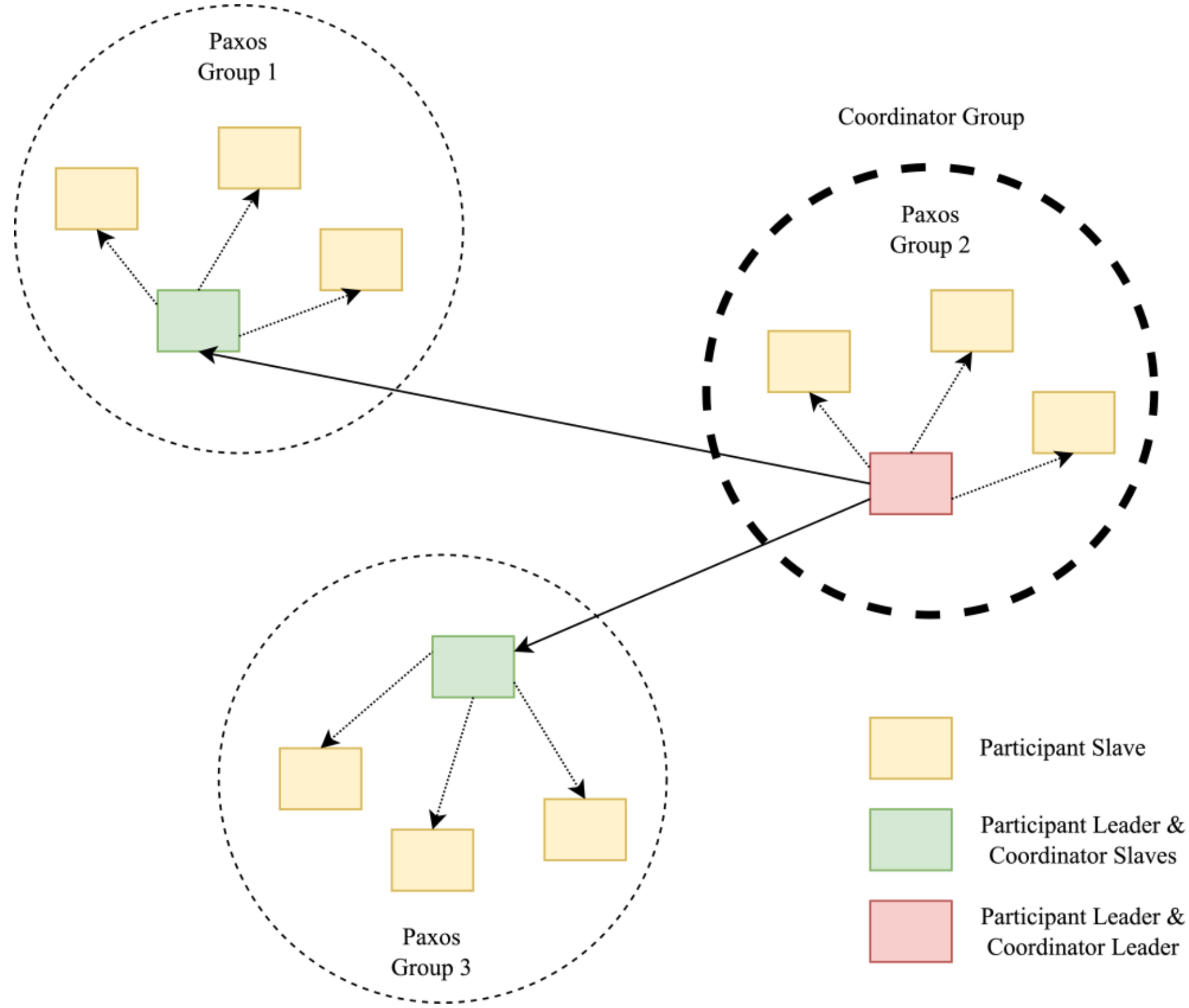

**Figure 105**: Paxos groups for 2PC.

24.8.2.4. Paxos Read-Only Transactions

Within a Paxos group, all read-only transactions must go through the leader for serializability. The leader ensures the transaction is ordered after all existing transactions. The leader also requires a quorum from a majority of replicas before a read is considered successful (e.g., in case of a view change). This overhead can be avoided by executing reads only at the Paxos leader, but this requires strong clock synchronization. Spanner uses TrueTime to make this possible.

To understand more on how clock synchronization helps, refer to the previous chapter, which explains how Harp [149] relies on reading from the primary. This process requires clock synchronization to maintain external consistency.

#### *24.8.3. Witness Nodes*

Witness replicas are a special type of node. They play a crucial role in ensuring high availability and fault tolerance. Unlike read-write and read-only replicas, witness replicas do not maintain a full copy of data, nor do they serve read requests.

Witness replicas do participate in voting for write commits. They contribute to the quorum required for successful writes, ensuring data consistency and durability. Witness replicas also participate in leader election, but they are not eligible to become the leader replica.

#### *24.8.4. Transaction Processing*

As mentioned earlier, implementing 2PC requires globally consistent, monotonically increasing timestamps for transactions. This is where TrueTime becomes essential. Let's examine how Spanner assigns timestamps to transactions.

Spanner supports two main transaction types:

- **Read-write (R-W) transactions:** These are the standard transactions that use locks (pessimistic concurrency control).
- **Read-only (R-O) transactions:** Transactions that don't perform any modifications. Because transactions have globally consistent timestamps and the data store is versioned, read-only transactions can run lock-free. We'll explore how this works shortly.

Spanner also offers another type of read called a **snapshot read** with bounded staleness.

24.8.4.1. Paxos Leader Leases

Spanner optimizes read-only transactions by having them typically execute only at the Paxos group leader. This is straightforward if the leader remains constant, but leader changes during a transaction may result in inconsistent values. Yet, single-node reads at the leader significantly improve performance and hence desirable.

Spanner achieves this optimization using leader leases (similar to Harp but with accuracy). The leader obtains a lease from the replicas, guaranteeing its leadership for the lease duration. These leases depend on accurate clock synchronization, which Spanner provides via TrueTime.

The lease mechanism is detailed in Appendix A of the paper. Here's a simplified explanation:

$\mathbf{v}_{i,\mathbf{r}}^{\mathbf{Leader}}$: The earliest time a vote request was sent by a potential leader **i** to a replica **r**.
$\mathbf{t}_{i,\mathbf{r}}^{\mathbf{end}}$: The time replica **r** grants a lease to **i** until, calculated as $\mathbf{TT.now().latest} + \mathbf{10s}$. Replicas won't grant another vote until $\mathbf{TT.after}(\mathbf{t}_{i,\mathbf{r}}^{\mathbf{end}})$ is true.

The new leader's lease (after receiving majority votes) is valid for the interval $[\mathbf{TT.now().latest}, \mathbf{min}_i(\mathbf{v}_{i,\mathbf{r}}^{\mathbf{Leader}}) + \mathbf{10s}]$. This ensures the leader's assumed lease expiration is always earlier than the actual expiration at any voter, preventing concurrent leader elections (**single-vote property**). With guaranteed single leadership (**leader-disjointness**), read-only transactions can safely execute at the leader only.

If a transaction with timestamp **T** is accepted, it must fall within the leader's lease bounds. Similarly, the leader relinquishes its lease only after $\mathbf{s}_{\mathbf{max}}$ has passed, where $\mathbf{s}_{\mathbf{max}}$ is the maximum timestamp of any transaction processed by that leader.

24.8.4.2. Assigning Timestamps

Consistent timestamp assignment during 2PC is crucial for Spanner's external consistency. Transactions arrive at a coordinator, which is also the Paxos leader. Because there's only one Paxos leader at any given time (leader disjointness), and that leader's timestamp assignment is constrained by its lease, the leader must assign timestamps greater than all previous transaction timestamps.

Spanner's external consistency guarantee is: **If transaction $T_2$ starts after transaction $T_1$ commits, then $T_2$'s commit timestamp must be greater than $T_1$'s**. This ensures that $T_2$ is ordered after $T_1$, maintaining external consistency. For example, if you execute $T_1$ and then tell a friend, they can subsequently execute $T_2$ and observe the effect of $T_1$.

Formally, let $e_i^{start}$ and $e_i^{commit}$ be the start and commit events of transaction $T_i$, and $s_i$ be the assigned timestamp. Then:

$$t_{abs}\left(e_1^{commit}\right) < t_{abs}\left(e_2^{start}\right) \Rightarrow s_1 < s_2$$

Spanner uses TrueTime to assign timestamps, ensuring that each timestamp is **TT.now().latest**. This makes it greater than any previously assigned timestamp. However, this alone isn't sufficient.

**Commit Wait**: After assigning a timestamp $s_i$, all 2PC participants wait until **TT.after($s_i$)** is true. This guarantees that any participant invoking **TT.now().latest** for a subsequent transaction will receive a timestamp greater than $s_i$. Therefore, all newer transactions will have higher timestamps.

This proof is detailed in Section 4.1.2 of the paper. It relies on the assumption that clients only send a second transaction after observing the commit of the first.

This timestamping mechanism applies to both read-only and read-write transactions.

24.8.4.3. Executing a R-W Transaction

Let's walk through an example of a read-write transaction in Spanner involving three data items, **A**, **B**, and **C**, residing in different Paxos groups. The transaction is as follows:

```
txn = new Transaction()
a = txn.Read(A)
b = txn.Read(B)
txn.Write(C, a + b)
txn.Commit()
```

The following sequence of actions takes place, as shown in Figure 106(a) and 106(b):

1. `txn.Read(A)`: The Paxos group leader for **A** returns the value of **A** and read-locks **A** to prevent concurrent modifications.
2. `txn.Read(B)`: The Paxos group leader for **B** read-locks **B** and returns its value.
3. `txn.Write(C, A + B)`: The client buffers the write to **C** locally. Spanner doesn't support **read-your-writes** (RYW) within a transaction, so writes can be buffered until commit.
4. `txn.Commit()`: The client initiates the commit. Any of the Paxos groups (**A**, **B**, or **C**) can act as the coordinator. Let's assume the Paxos group for **C** becomes the coordinator. The **Write(C)** and **Commit()** requests are sent to **C**'s Paxos group.
5. **Prepare Phase**: C's Paxos leader locks C, logs the new value, and sends a PREPARE message to the leaders of A and B's Paxos groups (these are the cohorts in the 2PC).
6. **Ack**: The leaders for **A** and **B** log the COMMIT message and acknowledge with their local TrueTime timestamps (**TT.now().latest**).
7. **Timestamp Assignment**: The leader at **C** determines the maximum of all received timestamps (including its own) and assigns this as the transaction's commit timestamp. This timestamp is guaranteed to be greater than any timestamp assigned to a previous transaction affecting the same data items.
8. **Commit Wait**: The leader at C waits until **TT.after(s)** is true, where **s** is the assigned commit timestamp.
9. **Commit Phase**: C sends the COMMIT message to A and B, releasing all locks. A and B also release their locks.

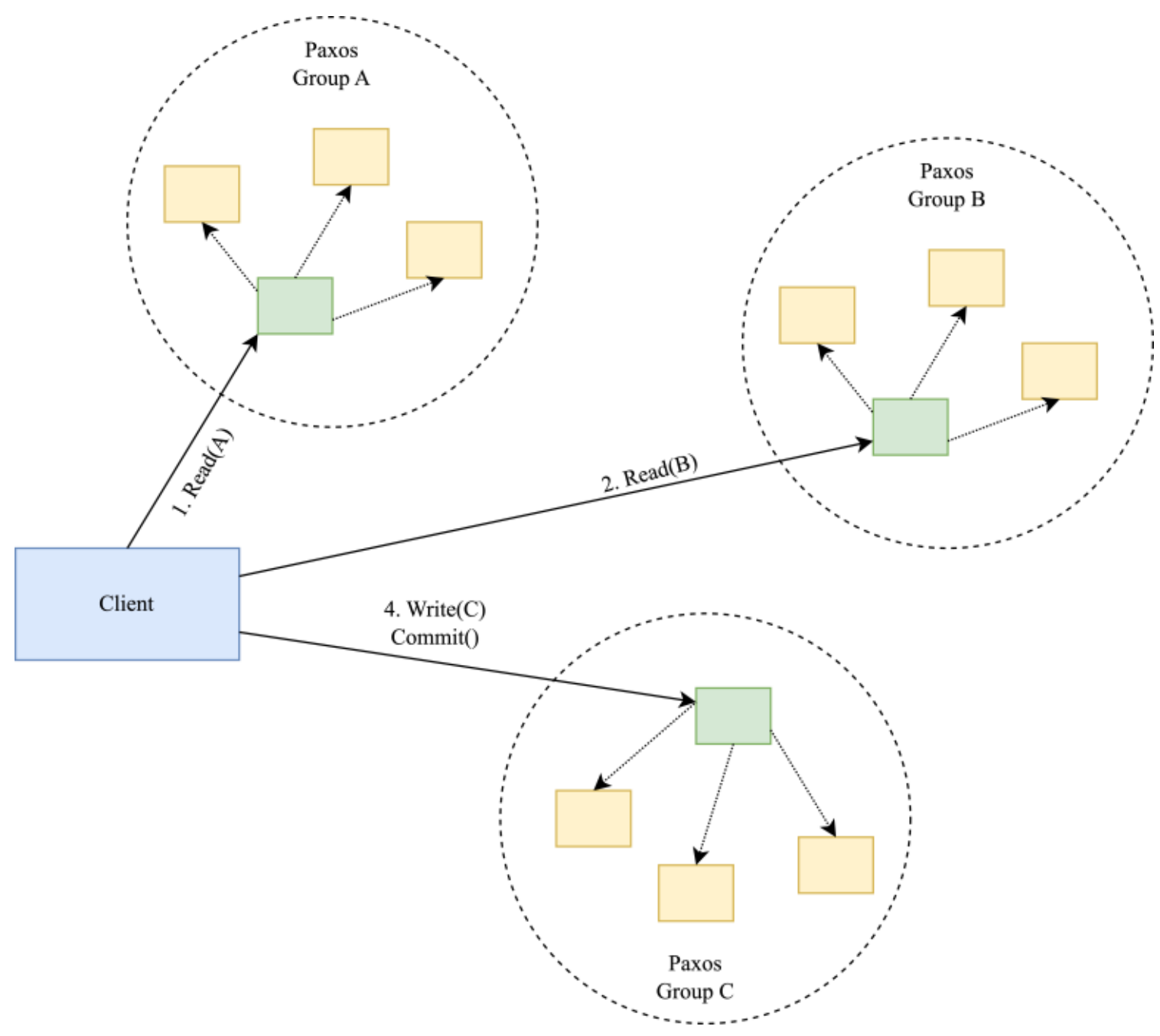

**Figure 106(a)**: Paxos groups in R-W transaction example.

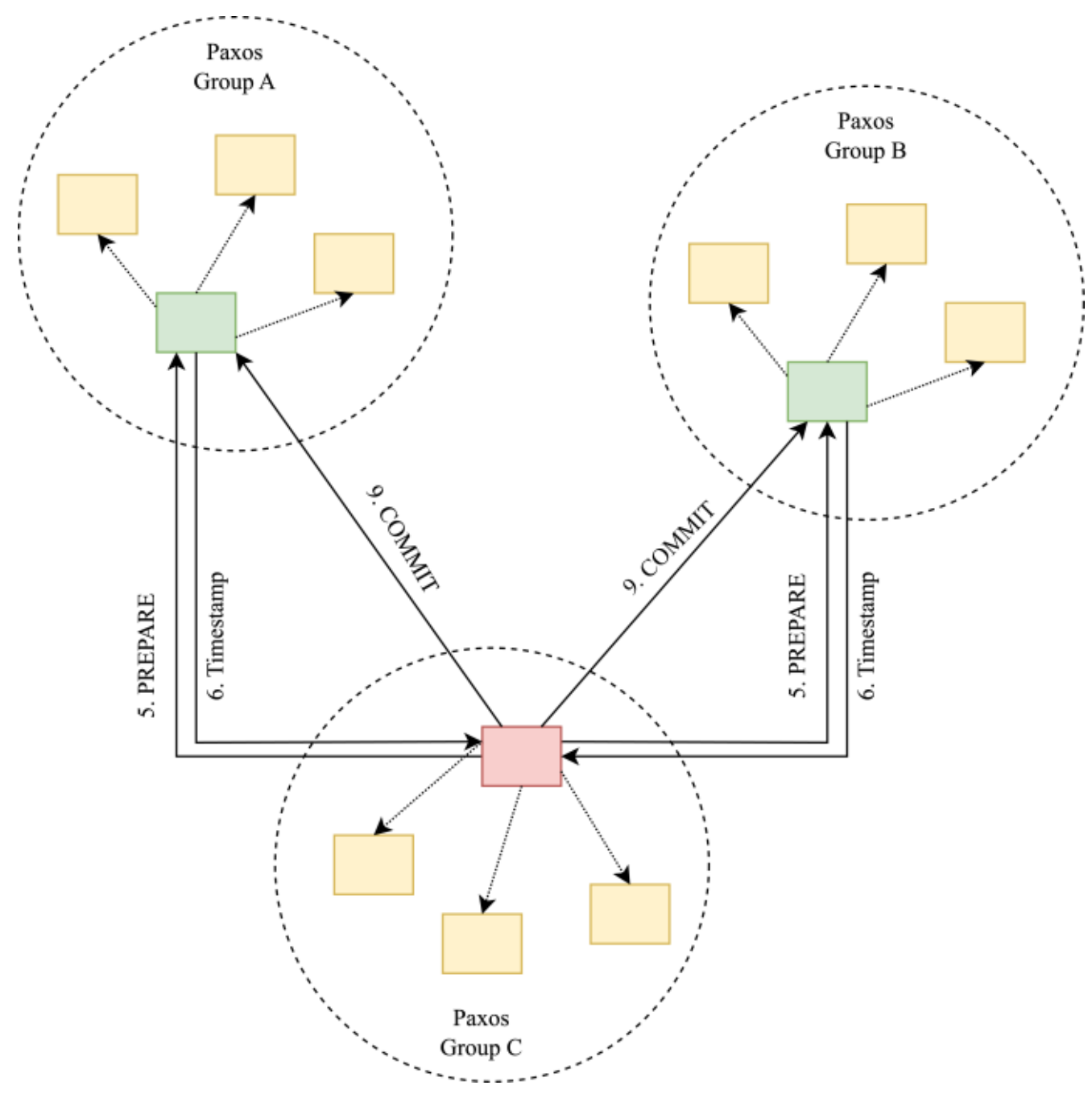

**Figure 106(b)**: 2PC in R-W transaction example.

24.8.4.3.1. FAQs:

**Q: What happens if two orthogonal transactions are submitted by two different clients at the same time?**

A: They might receive the same timestamp (even down to nanosecond precision). Ties may be broken arbitrarily. However, this doesn't affect external consistency, as neither transaction has committed before the other began.

**Q: What if the leader of any Paxos group goes down during 2PC after PREPARE reply?**

A: The COMMIT message would still be logged. In general, when a new leader comes up, it temporarily locks the entire database until all existing transactions have been cleared according to their timestamp order. This prevents LOCK messages for reads from being logged to Paxos. As noted in the paper, this approach relies on long-lived Paxos leaders for efficiency.

**Q: What if, between lock(A) succeeding and lock(B) succeeding, B is modified by another transaction?**

A: Nothing. The transaction that modified B will be ordered before the current transaction. Neither serializability nor strict serializability is impacted.

24.8.4.3.2. Deadlock Prevention

Because lock acquisition during individual transactions is unsynchronized, deadlocks are possible, as shown in Figure 107.

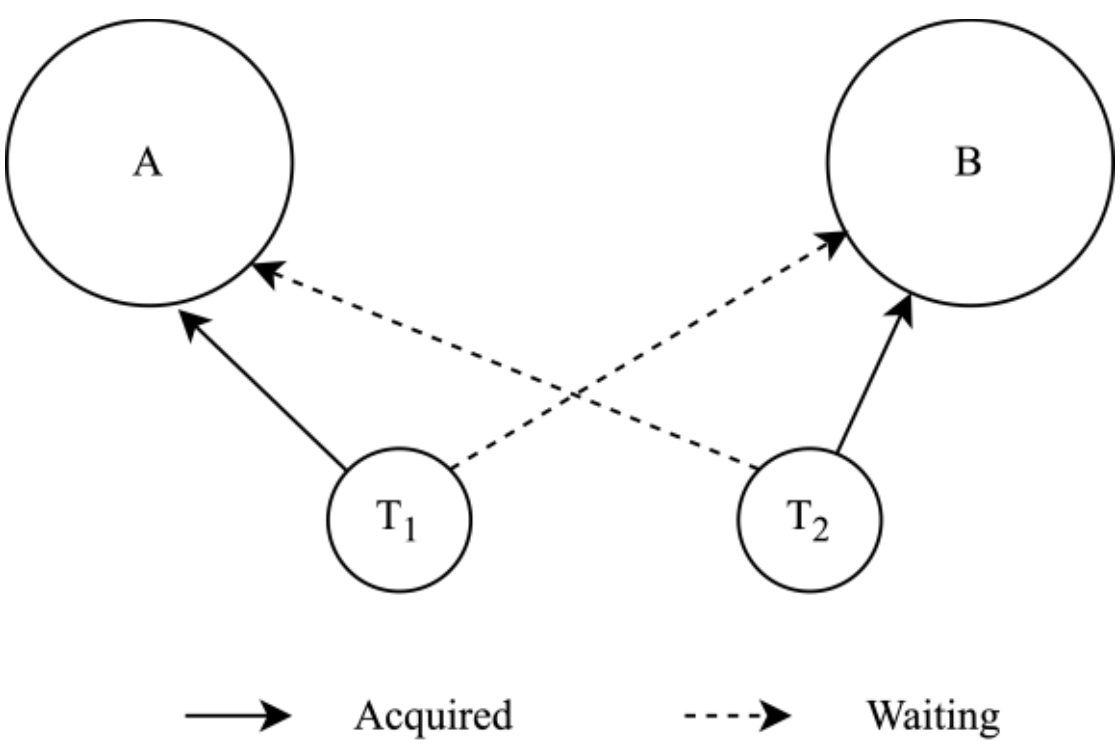

**Figure 107**: Transaction deadlock example.

Spanner implements transactions as sessions between clients and the leaders of the involved Paxos groups. Two deadlock prevention schemes are considered:

- **Wound-Wait:** An older transaction forces a younger one to abort and release a lock. A younger transaction waits for an older one to complete. Since timestamps haven't been assigned yet at this stage, ordering is determined solely by transaction start time (may be provided by client).
- **Wait-Die:** An older transaction waits for a younger one. A younger transaction requesting a lock from an older one is aborted.

Spanner makes use of the wound-wait algorithm to prevent deadlocks.

24.8.4.4. Executing a R-O transaction

Read-only transactions in Spanner are highly optimized. Because all transactions receive timestamps and Spanner stores versioned data, read-only transactions don't require locks. Once a timestamp is assigned to a read-only transaction, it can execute at any sufficiently up-to-date Paxos group replica. The read values will reflect a consistent snapshot at the assigned timestamp. This makes read-only transactions exceptionally fast and non-failing.

To determine which replicas can handle a read-only transaction with timestamp $\mathbf{t}$, each replica tracks a "safe time" ($\mathbf{t}_{\mathbf{safe}}$), representing the highest timestamp to which the replica's data is consistent. A replica can serve a read if $\mathbf{t} \leq \mathbf{t}_{\mathbf{safe}}$.

$\mathbf{t}_{\mathbf{safe}}$ is calculated as the minimum of two values:

- $\mathbf{t}_{\mathbf{safe}}^{\mathbf{Paxos}}$: The highest timestamp of a transaction committed within the Paxos group.
- $\mathbf{t}_{\mathbf{safe}}^{\mathbf{TM}}$: This accounts for pending transactions. If there are no pending transactions, $\mathbf{t}_{\mathbf{safe}}^{\mathbf{TM}}$ is infinite, meaning the replica is safe to read from as long as $\mathbf{t}_{\mathbf{safe}}^{\mathbf{Paxos}}$ is greater than $\mathbf{t}$. If there are pending transactions, $\mathbf{t}_{\mathbf{safe}}^{\mathbf{TM}}$ is calculated as:

$$\mathbf{t}_{\mathbf{safe}}^{\mathbf{TM}} = \mathbf{min}_{\boldsymbol{i}}(\mathbf{s}_{\boldsymbol{i}}^{\mathbf{prepare}}) - \mathbf{1}$$

Where $\mathbf{s}_{\boldsymbol{i}}^{\mathbf{prepare}}$ is the timestamp of the PREPARE message for pending transaction $\mathbf{i}$. Since pending transactions are assigned a timestamp that is the maximum of all cohort PREPARE timestamps, we find the minimum PREPARE timestamp across all pending transactions and set $\mathbf{t}_{\mathbf{safe}}^{\mathbf{TM}}$ to be less than that value.

For reads within a single Paxos group, the read-only transaction timestamp could be set to the timestamp of the last seen read-write transaction (**LastTS()**). For 2PC reads, the timestamp could be negotiated by querying all cohorts for their latest timestamps and taking the minimum. However, Spanner avoids these approaches by simply setting $\mathbf{s}_{\mathbf{read}}$ to **TT.now().latest**. While this simplifies the process, it might introduce some delay as the read operation may wait for all transactions up to $\mathbf{s}_{\mathbf{read}}$ to commit.

Section 4.2.4 of the paper details further refinements, focusing on efficiently determining the latest timestamp. The core idea is fine-grained tracking of PREPARE messages and the key ranges affected by transactions. This allows the system to exclude non-conflicting transactions when determining the timestamp for read-only transactions.

24.8.4.4.1. Executing Reads at a given timestamp

Spanner's versioned data and timestamped reads enable a powerful feature: snapshot reads. These allow reading the database as it existed at a specific point in time, even in the past. Critically, snapshot reads still provide a consistent view of all data items read at that point in time. This is equivalent to snapshot isolation. SQL queries often leverage snapshot reads (at a given timestamp) to retrieve consistent data without the overhead of locking.

### *24.8.5. Schema Changes*

Spanner's globally ordered timestamps simplify schema changes. A schema change is treated like any other operation and is assigned a timestamp. Each Spanner server then applies the schema change independently. Reads and writes against the new schema are only allowed after the schema change has been applied locally and **TT.after(s)** is true (where **s** is the schema change timestamp).

Importantly, schema changes are not implemented as a massive distributed transaction across all Spanner servers. Such a large transaction would be impractical to complete and could potentially lock the entire database.

### *24.8.6. Evaluation*

- In experiments with a single replica per Paxos group, commit wait time was approximately 5ms, and Paxos latency was about 9ms. As the number of replicas increased, latency remained relatively constant, but with a lower standard deviation.
- Terminating Paxos leaders resulted in a 10-second wait for a new leader to be elected. The Spanner paper emphasizes the importance of Paxos leader stability for higher availability. Shorter lease times could help in this scenario. In the event of leader failure, restarting the leader is generally preferred.
- For TrueTime, the error bound ε is roughly less than 4ms (p99) and 10ms (p999).

Google's F1 database, which was based on master-slave replicated MySQL, was replaced by Spanner. F1 maintains a history of all changes made to Spanner, which is itself written back to Spanner (change history).

The paper provides further internal details on how directories are also fragmented.

## 24.9. Spanner v/s CAP Theorem

Let's revisit the CAP theorem and examine how Spanner fits within its constraints. The information in this section is derived from Eric Brewer's paper [186].

Spanner, with its highly available Paxos groups for transaction management, might appear to violate the CAP theorem. However, it doesn't. Spanner prioritizes consistency. Its high availability is achieved at a significant maintenance cost borne by Google. Technically, it's still classified as a CP system. Spanner offers five 9s of availability.

### *24.9.1. Network Partitions*

Network partitions, which can lead to availability loss, account for a relatively small percentage (8%) of Spanner's outages. Common network issues include:

- Individual data centers becoming disconnected.
- Misconfigured bandwidth.

- Network hardware failures.
- One-way traffic failures.

Google's extensive experience in network engineering allows them to build a highly reliable network. Spanner's network leverages Google-owned infrastructure, including the software-defined B4 [187] network. Google's control over packet flow and routing enhances its resilience against network partitions.

*24.9.2. Handling Crashes*

- 2PC itself isn't crash-tolerant. If a participant fails, committing or aborting a transaction becomes challenging. This is an inherent limitation of 2PC. Spanner mitigates this somewhat by using Paxos groups for participants, ensuring high availability.
- Furthermore, during Paxos leader changes, a Spanner group essentially freezes until the lease expires. The paper suggests restarting the leader as a better solution than waiting for lease expiration.

*24.9.3. Handling Network Partitions*

2PC is not partition-tolerant; it cannot complete during a partition.

Even during partitions, Spanner transactions can proceed if the leader and a majority of a Paxos group remain on the same side of the partition.

Reads are generally more resilient. They can be performed at any replica that's caught up to the transaction's commit timestamp. Snapshot reads offer even greater robustness. However, if none of the replicas are sufficiently up-to-date (e.g., due to being on the non-majority side of a partition), reads will eventually fail.

### 24.10. TrueTime v/s CAP Theorem

Spanner's functionality heavily relies on TrueTime's accuracy and availability:

- **Transaction timestamps:** TrueTime provides the timestamps assigned to transactions.
- **Paxos leader leases:** Paxos group leader leases depend on TrueTime.
- **Snapshot isolation:** TrueTime's ability to generate consistent, ordered timestamps is fundamental to Spanner's versioning mechanism, which makes snapshot isolation possible. Without TrueTime, identifying data objects and ensuring a consistent snapshot (without partially applied updates) would be extremely difficult.

- **Schema changes:** TrueTime also facilitates schema changes.

Beyond Spanner, TrueTime has broader applications:

- **Cross-system snapshots:** TrueTime enables consistent snapshots across multiple independent systems without requiring 2PC. These systems can agree on a timestamp using TrueTime and then independently dump their state at that time.
- **Workflow timestamps:** TrueTime timestamps can be used as tokens passed through workflows (chains of events). Every system in the workflow can agree on the same timestamp.

TrueTime itself is not immune to the effects of network partitions. The underlying time sources (GPS receivers and atomic clocks) can drift apart, and as this drift increases, transactions may experience longer wait times before committing. Eventually, transactions may time out, impacting availability.

### 24.11. Paper Remarks

This paper stands as a landmark contribution to the field of distributed systems. Its density is a testament to the numerous groundbreaking concepts it introduced to the database community. Given the depth of its technical nuances, it requires careful study to fully appreciate; however, it remains essential reading for anyone deeply passionate about the architecture of distributed systems.

# 25. Eliminating Receive Livelock in an Interrupt-driven Kernel

(25) Mogul, J. C.; Ramakrishnan, K. K. Eliminating Receive Livelock in an Interrupt-Driven Kernel. ACM Transactions on Computer Systems (TOCS) 1997, 15 (3), 217–252.

We will now take a break from distributed systems and return to our discussion on operating systems. Earlier, we discussed microprocessors and memory; however, I/O devices play an equally important role in a computing system. For a machine in a data center, the most critical I/O component is the network. This chapter will focus on the inner workings of the network stack.

Jeff Mogul, a pioneering figure in computer science, authored this influential paper presented at USENIX Annual Technical Conference (ATC) '96. Since then, it has become a seminal work, sparking extensive discussion in academic circles. The paper presents techniques to eliminate livelock, a condition that can occur while processing incoming network packets. It explores the various ways operating systems historically handled I/O and offers improvements upon them. Additionally, because the network stack is exposed to the external world, it must manage flow control; therefore, this paper also explores the flow control techniques employed in modern internet systems.

In this insight, we will first establish how I/O devices operate - particularly how operating systems handle the signals they generate. Next, we will examine the Network Interface Card and the architecture of the Linux networking stack. We will then analyze how livelocks occur and, finally, explore the specific techniques introduced in the paper to eliminate them.

### 25.1. CPU, Memory, & I/O Communication

In a Von Neumann Architecture, the CPU and memory are core components, enabling Turing computation. Their communication relies on the system bus (e.g. PCIe), a set of electrical pathways connecting the CPU, memory, and I/O devices. The system bus comprises three primary logical components, as shown in Figure 108:

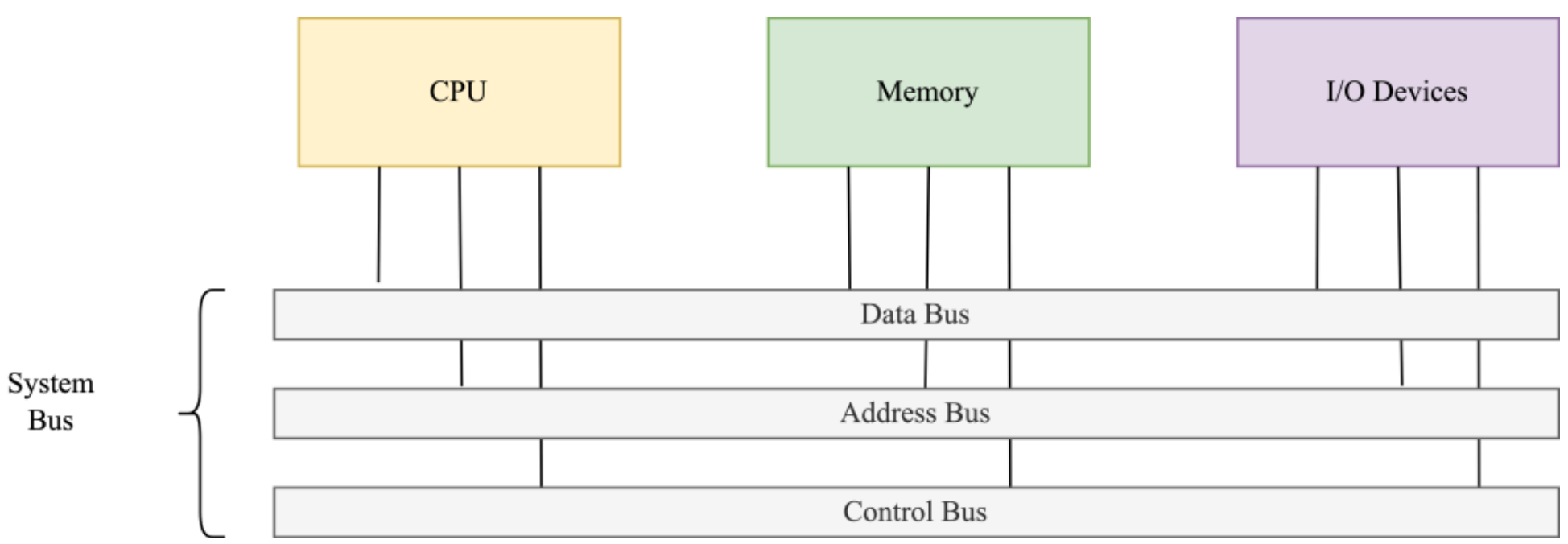

**Figure 108**: System bus.

- **Data Bus**: Bidirectional, carrying the actual data being transferred.
- **Address Bus**: Unidirectional, carrying memory addresses for data access.
- **Control Bus**: Carrying control signals for synchronization and coordination.

Modern systems employ hierarchical bus architectures to enhance communication speed, moving beyond the simple single-bus model.

## 25.2. Interrupt

An **interrupt** (or trap) is a signal that prompts the CPU to suspend its current instruction execution and instead execute a specific **Interrupt Service Routine (ISR)**.

### *25.2.1. Types of Interrupts*

Interrupts can be broadly categorized as follows:

- **Hardware Interrupts**: These are generated by hardware devices (via interrupt line as shown in Figure 109), compelling the microprocessor to jump to an ISR. A naive hardware implementation might directly modify the program counter to fetch the ISR's next instruction. However, this approach presents significant security vulnerabilities. A more common and secure method involves raising an **interrupt flag** on the control bus. The CPU's control unit detects this flag and, at the end of the current instruction cycle, handles the interrupt by branching to the ISR.

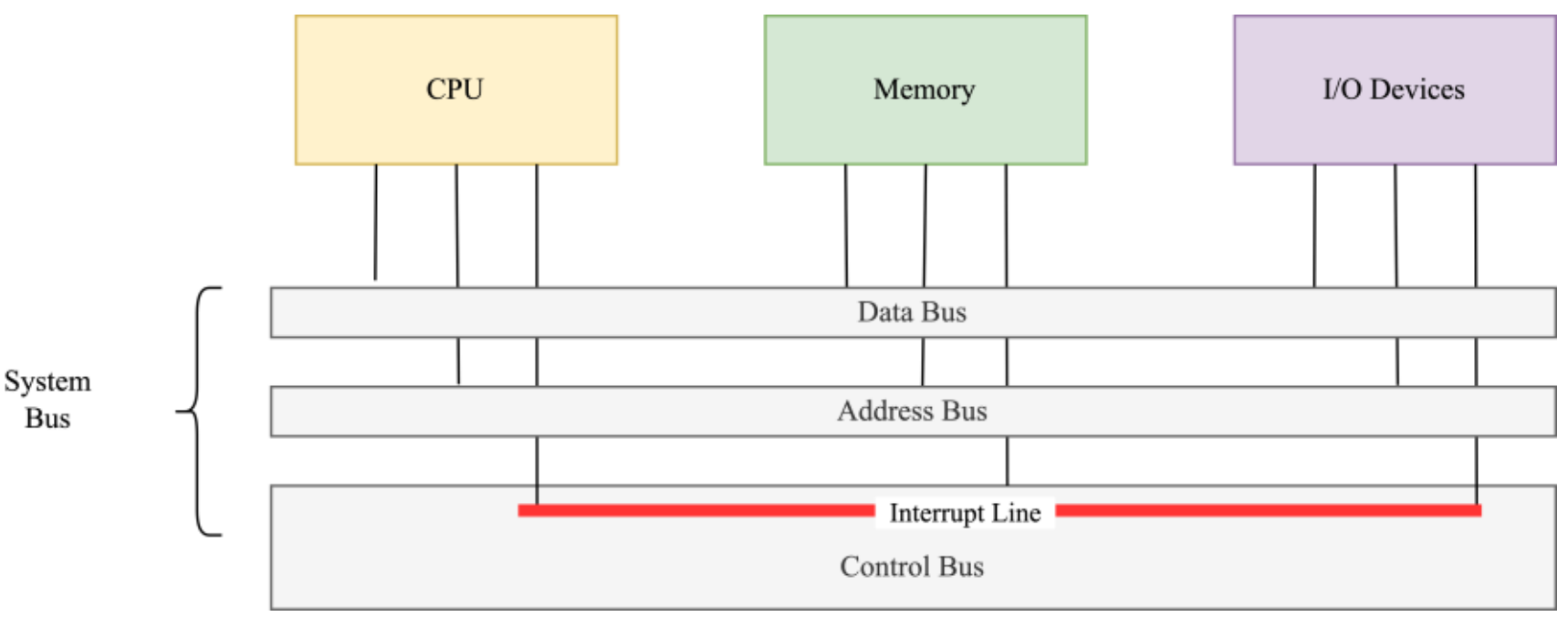

**Figure 109**: Interrupt line on control bus.

- **Software Interrupts**: These are signals that instruct the processor to jump to an ISR when specific conditions (evaluated by the processor) are met. For instance, a memory access resulting in a page fault triggers a software interrupt, forcing the CPU to transition from user mode (Ring 3) to kernel mode (Ring 0) to manage the fault. Similarly, a program causing a segmentation fault generates a software interrupt to the kernel.

*25.2.2. Interrupt Vector Table*

The Interrupt Vector Table (IVT) is a lookup table containing the memory addresses of ISR routines corresponding to hardware-generated interrupts. (The kernel is aware of the ISR addresses for software interrupts). The microprocessor uses the IVT to find the appropriate routine for a given hardware interrupt.

*25.2.3. Example: Keyboard Input*

When the keyboard receives an input, it generates an interrupt. The ISR reads the input and stores it in a keyboard buffer. The terminal program then reads from this buffer. This is shown in Figure 110.

*25.2.4. Interrupt-Driven Scheduling*

A scheduling system manages what executes on the microprocessor. Kernel CPU schedulers typically handle the scheduling of kernel and user threads.

However, as discussed, interrupts also dictate when an ISR is executed. Consequently, interrupt-driven scheduling is another form of CPU scheduling.

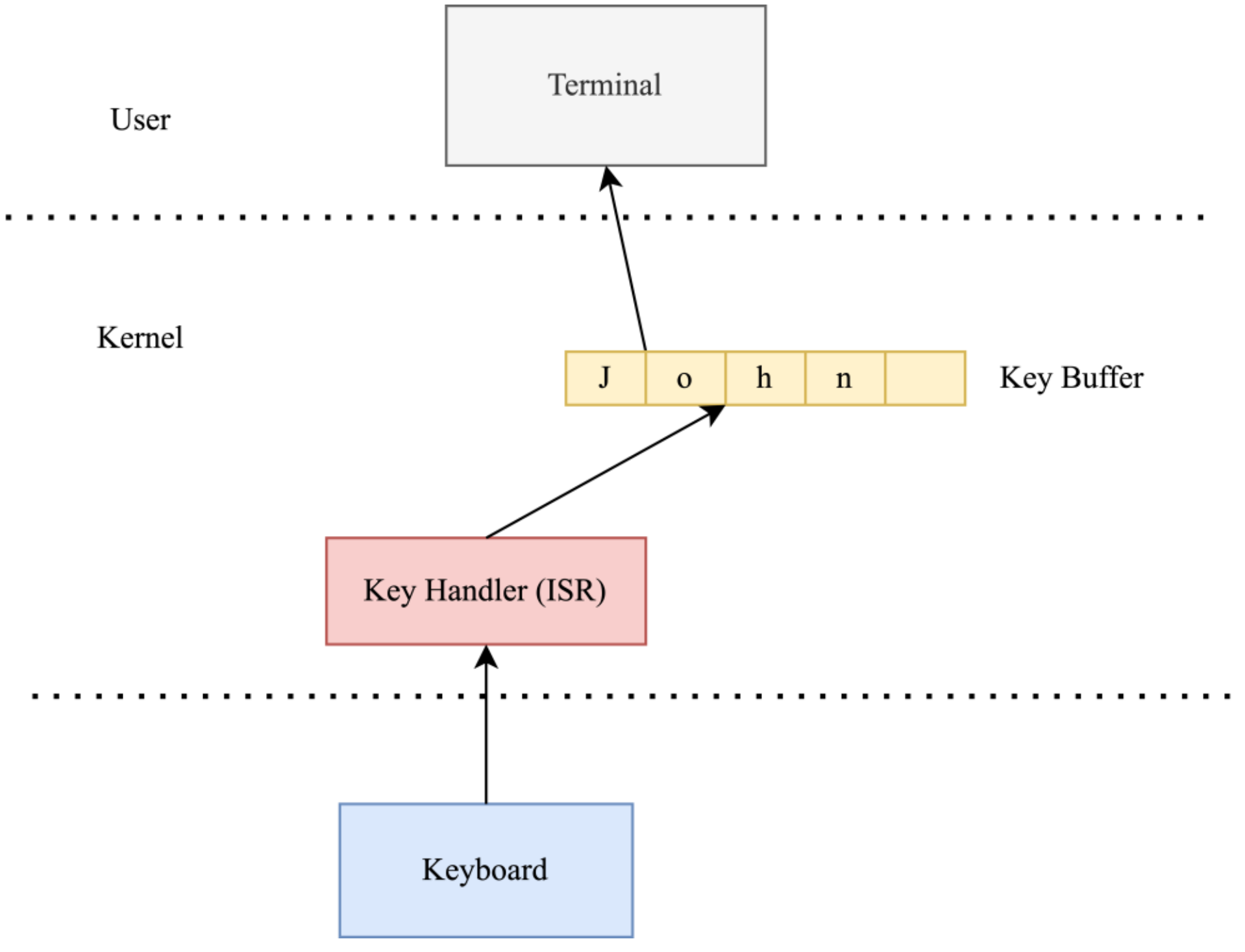

**Figure 110**: Keyboard input handling.

Therefore, we have different scheduling mechanisms: kernel CPU scheduling and interrupt-driven scheduling. Interrupt-driven scheduling, particularly those originating from hardware, generally takes higher precedence and executes at more privileged protection rings, such as level 0 or 1 called the **interrupt privilege level (IPL)**.

### 25.3. Network Interface Card

The Network Interface Card (NIC) is the I/O device that facilitates data reception and transmission over the network.

#### *25.3.1. NIC Interactions*

The NIC, like other I/O devices, interfaces directly with the system bus. Additionally, it incorporates an internal buffer memory. The interactions of NIC with CPU and memory is shown in Figure 111.

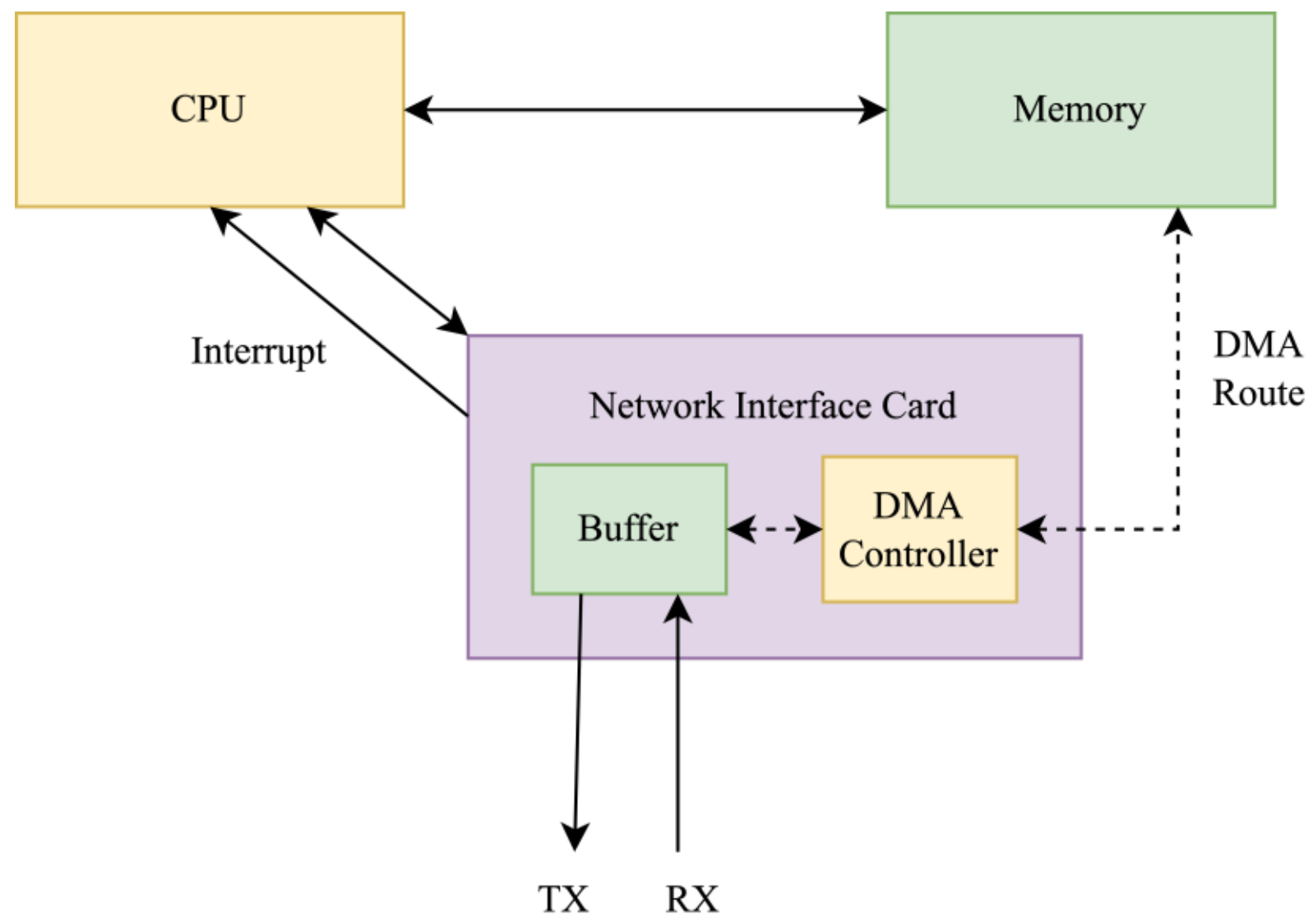

**Figure 111:** NIC interactions with CPU and memory.

**Reception (RX)**

- The NIC buffers incoming data into its internal memory.
- Upon buffer fullness, the NIC generates a maskable interrupt [188] to the CPU.
- The kernel, in response to the interrupt, copies the data from the NIC's buffer to RAM.
- Higher-layer protocols then process the data (e.g., TCP/IP copying data to socket buffers), and the application is notified.

**Transmission (TX)**

- The application writes data to a memory buffer.
- The kernel copies the data to the NIC's buffer.
- The NIC transmits the data over the network.

*25.3.2. Direct Memory Access*

*Not to be confused with RDMA.*

To improve efficiency, modern NICs, and other hardware like disk controllers and graphics cards, utilize Direct Memory Access (DMA).

This allows the NIC to directly read and write to system memory without CPU intervention. The NIC has an internal processor (DMA controller) that can manage these memory transfers. While DMA is in progress, the system bus is in use by the DMA controller, briefly making it inaccessible to the CPU. This offloads data transfer tasks from the CPU, allowing it to perform other operations concurrently.

### 25.4. Linux Network Stack

Networking stands out as one of the most intricate I/O operations within a computer system. This complexity stems from the layered nature of the network stack: the **MAC layer** where the NIC operates, the **IP layer** governing internet communication, the **Transport layer** where TCP/UDP functions, and the application-specific protocols used by various applications. Furthermore, unlike storage I/O, the RX/TX rates in networking are inherently unpredictable. Storage device controllers can regulate RX/TX speeds based on buffer availability. However, network streams involve interactions with numerous external machines beyond the local system's control.

Consider a DDoS attack scenario where a machine is flooded with requests. The NIC lacks the ability to control the incoming packet stream. Its influence over network capacity usage is also limited; the Ethernet connection might become saturated with incoming traffic. Consequently, the NIC's primary recourse in such situations is to begin dropping packets. Higher-level protocols then assume the responsibility for retransmissions and ensuring reliable delivery. This type of uncontrolled and potentially overwhelming input is a situation that is not encountered in I/O operations involving storage devices.

#### *25.4.1. Life of a RX Packet*

Let's dive deeper into the journey of a network packet as it arrives in a Linux system, as shown in Figure 112.

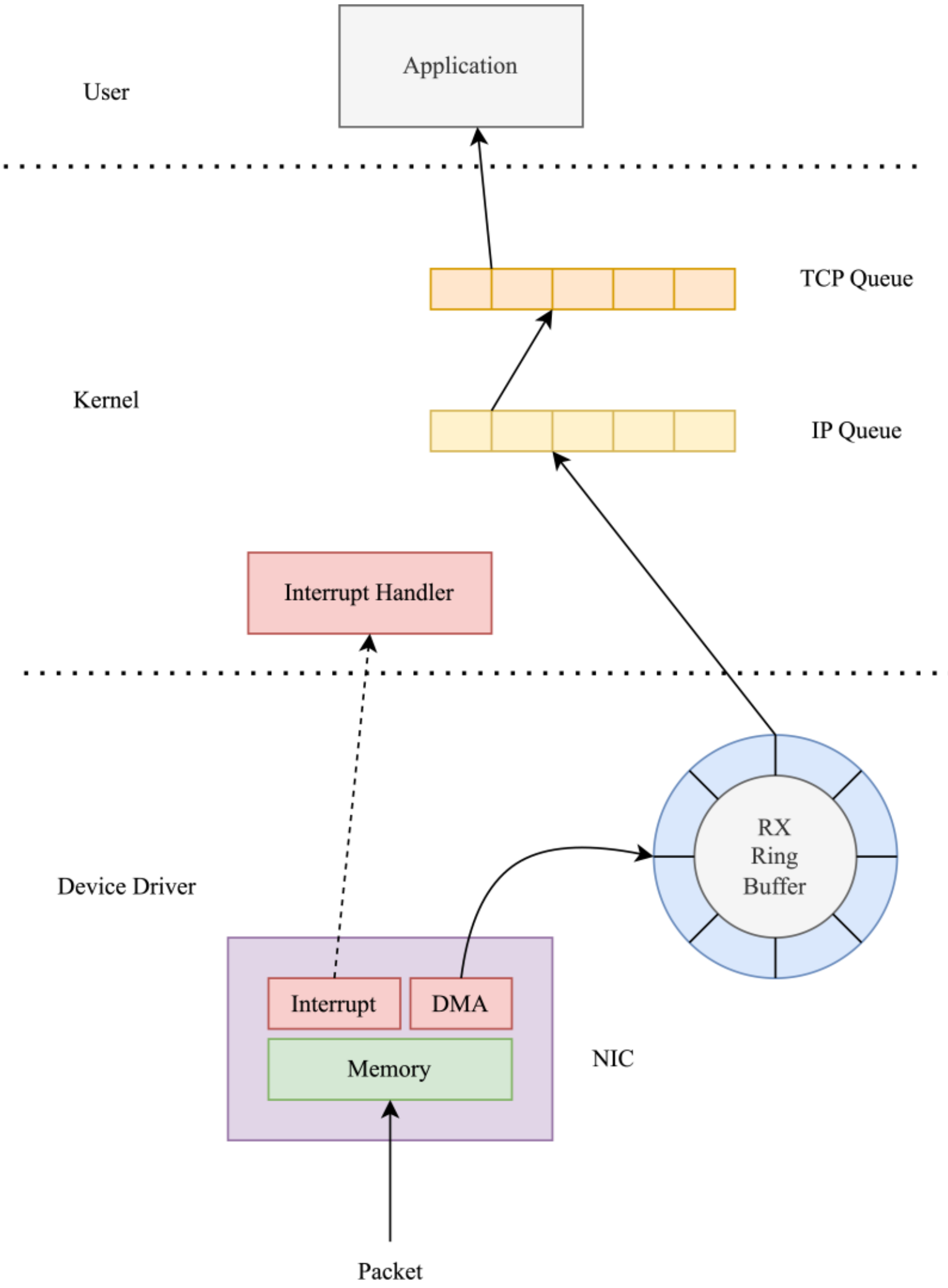

**Figure 112**: RX packet processing.

### Device Driver Layer

- The NIC temporarily holds the packet in a buffer memory. Alternatively, for efficiency, the NIC can employ **DMA** to directly deposit the packet into a pre-arranged circular buffer in system memory, known as a **RX ring buffer**.
- Once the packet is in its initial storage, the NIC signals the kernel's attention by generating a maskable interrupt. The interrupts are batched to amortize their cost.

#### Kernel Layer

- The ISR processes all packets in the NIC's memory or the RX ring buffer. It directs these packets towards their intended destinations within the kernel. For instance, data destined for a user application typically gets placed in the **TCP queue**. In contrast, packets intended for the system's firewall are routed to the **IP queue** only.

#### User Layer

- The user application thread reads the packets from the respective queue to the **socket buffer**.

### *25.4.2. Scheduling Overhead*

The reception process involves the following scheduling events:

- **Interrupt Scheduling**: This initial ISR is scheduled in response to hardware interrupt that reads the packets from NIC's memory or the RX ring buffer to the kernel queues.
- **User Thread Scheduling**: Finally, the user application thread is scheduled to read the packets from kernel queues to the socket buffer.

### *25.4.3. Data Copying*

During reception, the packet's data is typically copied the following number of times:

- **NIC to IP Queue**: The raw packet data is copied from the NIC's memory or the RX ring buffer into the IP queue within the kernel.
- **IP to TCP Queue**: If the packet is destined for a TCP socket, its data is copied from the IP queue to the corresponding TCP queue.
- **TCP Queue to User Space Socket Buffer**: Finally, when the user application reads the data, it's copied from the TCP queue in the kernel's memory space to the application's socket buffer in user space.

Modern zero-copy [189] techniques enhance network performance by bypassing the kernel for data transfer, allowing user-space threads direct access to the RX and TX ring buffers, thus eliminating kernel-to-user space data copies.

### 25.5. Receive Livelock

As previously mentioned:

- A NIC can uncontrollably receive a large volume of network packets, potentially generating numerous interrupts in quick succession.
- And interrupt-driven scheduling takes precedence over kernel scheduling.

Consequently, when network packets arrive, the processor is forced to spend its time processing the packets and placing them into appropriate kernel queues. However, if the kernel scheduler fails to schedule the corresponding kernel or user threads, the application will experience starvation.

This scenario can lead to a **receive livelock**, a state where the system cannot make meaningful progress because its entire processing capacity is consumed by handling receive interrupts.

Ideally, with unlimited CPU resources, an increase in input load would result in a linear increase in throughput. However, in the real world where CPU power is finite, the throughput will initially rise but then plateau and remain at a stable level. This sustained throughput is known as the **Maximum Loss-Free Receive Rate (MLFRR)**. At this rate, packets are not dropped, and the CPU's full potential is utilized.

Beyond the MLFRR, packet loss will occur. While some packet loss is acceptable as long as the system maintains MLFRR throughput, a livelock causes system thrashing, preventing the sustained processing of data at the MLFRR. These scenarios are shown in Figure 113.

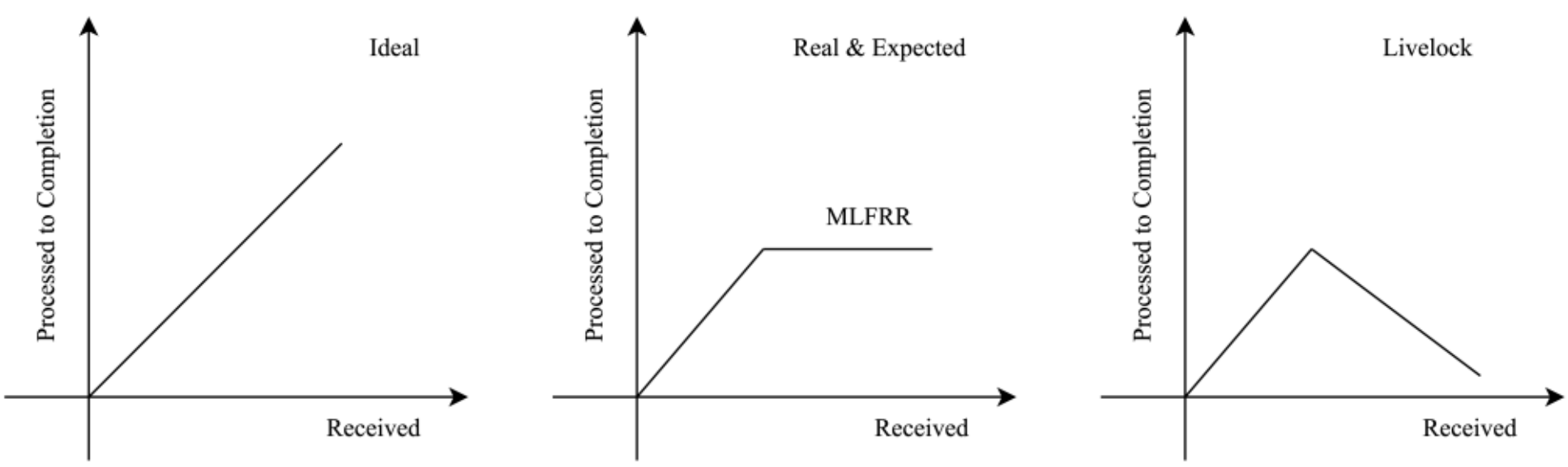

**Figure 113**: Packets received v/s processed in various scenarios.

Beyond the loss of throughput, a livelock manifests other problems, including:

- Increased end-to-end latency. While the delay in accepting a packet from the NIC might be reduced, the overall e2e latency, which includes packet processing, will increase.
- Ultimately, transmission throughput is also impacted. This occurs because even sending data requires the scheduling of kernel threads to move packets from kernel queues to the NIC.

### 25.6. Eliminating Livelock

The core system design and rate limiting principle we derive from this paper is **to discard new work at the earliest possible stage if the system lacks the capacity to handle it.**

In the context of receiving network packets, this translates to discarding packets at the NIC level itself when the system's processing capacity is saturated. This proactive dropping prevents unnecessary processing of packets that would ultimately fail to be handled completely, helping us achieve and sustain the MLFRR.

The authors implement this concept through two major techniques, as detailed below:

#### 1. Rate Limiting Interrupts

Interrupts are disabled under the following conditions:

- When the kernel queue is full. As a result, packets would be dropped directly at the NIC to avoid any processing overhead.
- When a certain amount of processing time has been spent processing packets. This allows kernel and user threads to make progress.

For the latter, the authors utilize a CPU timer to track the time spent processing packets within the ISR. If this duration becomes excessive, device interrupts are temporarily disabled and re-enabled only after the timer expires. Importantly, the timer interrupt has a higher priority than device interrupts, enabling it to interrupt ongoing interrupt handling. This mechanism ensures a more equitable distribution of CPU time.

**2. Polling Thread**

The authors introduce a spin polling thread. This thread is initiated upon receiving an interrupt, and once active, device interrupts are disabled. The polling is "spinning", meaning it involves continuous checks for new packets. This approach aims for the best of both worlds: low latency during periods of low RX rate (using interrupts) and sustained high throughput during high RX rates (using polling). The polling thread ceases operation and re-enables interrupts once no more packets are pending.

Crucially, the spin polling thread is subject to the regular scheduling decisions of the kernel scheduler, including round-robin scheduling, ensuring that other kernel and user threads continue to make progress.

Alternatively, interrupts could always be disabled and a continuously running polling thread could periodically process packets. However, this can introduce latency, potentially causing delays in packet delivery to applications.

### 25.7. Case of BSD Routers

The authors implemented these ideas within the BSD operating system for routers. Routers offer a simpler system to analyze compared to workstations, and they handle a substantial volume of packets for routing and forwarding. A router can be conceptualized as having multiple NICs, leading to multiple I/O queues at various levels.

Unlike general-purpose systems, routers primarily deal with the IP layer. Consequently, there are IP queues at the kernel level for IP packets.

- The receive interrupt handler copies the packet from input NIC's memory to **input IP queue**.
- The **IP forwarder** reads packets from the input IP queue and places them into the corresponding **output IP queue**.
- A transmit interrupt handler then reads from the output IP queue and forwards the packet to the designated output NIC.

This sequence is shown in Figure 6-2 in the paper.

As anticipated, the output packet rate decreases when the input packet rate exceeds a certain threshold, as shown in Figure 6-1 in the paper. This problem is exacerbated by the use of **screend**, a firewall program designed to block unwanted packets. While the IP forwarder operates in kernel space, **screend** runs in user space, resulting in a higher rate of packet drops.

*25.7.1. Eliminating Livelock*

The solution implemented is consistent with the principle discussed earlier: invest processing effort in a packet only if it were to be processed to completion.

- The authors entirely removed the input IP queue.
- The kernel employs a single polling thread, triggered by interrupts. This polling thread invokes handler procedures:
    - **Receive-packet callback procedure**: This procedure reads packets directly from the input NIC and places them into the appropriate output IP queues. The IP forwarder logic is integrated into this procedure.
    - **Transmit-packet callback procedure**: This procedure reads packets from the output IP queue and sends them out through the output NIC.
- Each of these callback procedures is assigned a fixed quota, defining the maximum number of packets it can process in one invocation.
- Each queue provides feedback on its occupancy level. When a queue reaches 75% capacity, the receive-packet callback is no longer executed. Instead, either the transmit-packet callback is executed or the system yields CPU time to user-space applications like **screend** to allow them to process the already queued packets.
- Finally, the polling thread as a whole periodically yields CPU time to allow other user-space processes to run.

*25.7.2. Evaluation*

The paper provides a relatively thorough evaluation of BSD routers with the fixes applied.

**Paper's Figure 6-3:** Polling without a limited quota (infinite quota) performs worse than the unmodified kernel. This is attributed to the receive-packet callback potentially processing a large number of input packets before encountering a full output IP queue and dropping them. The unmodified kernel performs better by dropping packets earlier at the input IP queue. However, polling with a defined quota successfully achieves the goal of sustaining the MLFRR.

**Paper's Figure 6-4:** Without feedback on output queue fullness, the performance is the same as an unmodified kernel. This is because packets are still likely to be dropped at the input queue of screend. With feedback, however, the polling thread proactively stops processing new input and yields to screend to clear its queue, allowing subsequent input processing.

**Paper's Figure 6-5:** This figure analyzes the impact of different polling quota values on performance (**sensitivity analysis**). Larger quotas are shown to be less effective and increase the perceived per-packet latency. A packet quota in the range of 10-20 is found to be optimal.

**Paper's Figure 6-6:** This figure presents similar analysis with feedback enabled. The results indicate that with effective feedback mechanisms, the specific polling quota becomes less critical, as the feedback itself limits the amount of work done in each procedure.

**Paper's Figure 7-1:** This figure examines the effect of yielding CPU time to user-space processes after a certain time interval. It aims to allow user-space processes to make progress. However, minor discrepancies are observed. These are potentially due to overhead from context switches, the lack of quota on the transmit procedure's processing (potentially monopolizing the CPU), and possible measurement errors.

### 25.8. Downside of Dropping Packets at NIC

Thus far, we've established that dropping packets at the NIC is generally optimal because it avoids investing resources in packets unlikely to be fully processed. However, this approach has a notable drawback: the drops are largely indiscriminate, and applications lack the ability to specify packet priority.

To address this, IP packets are often tagged with a **traffic class** (shown in Figure 114). Applications can also provide this traffic class information. The NIC can then read this traffic class and make intelligent dropping decisions based on the priority associated with each class.

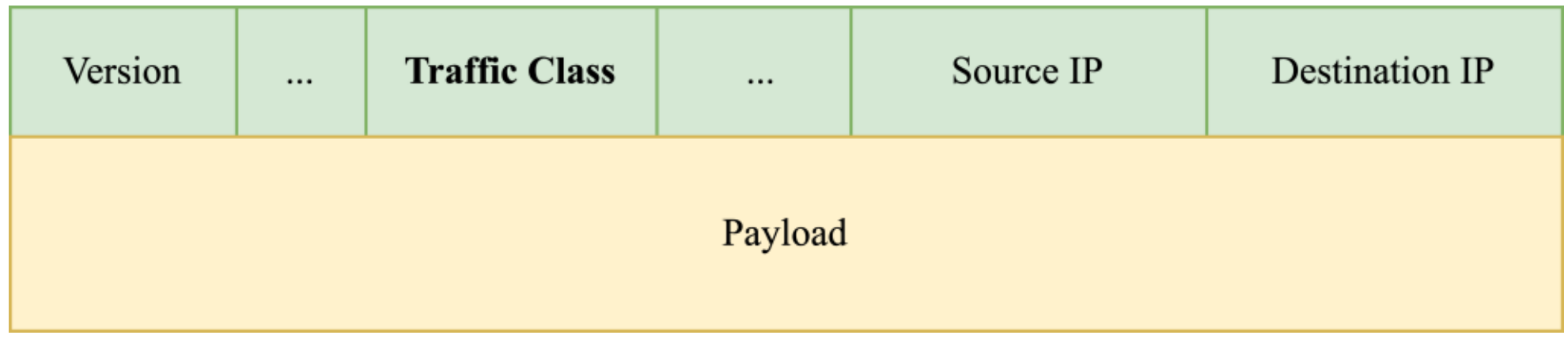

**Figure 114**: Traffic class in IP packet.

### 25.9. Paper Remarks

This paper is exceptionally well-written and accessible, making it easy for even beginners to grasp the problem statement and follow the elegant solution. It offers a clear and insightful walkthrough of the challenges and provides a compelling example of effective problem-solving. For anyone looking to deepen their understanding of the operating system I/O stack, the time invested in exploring its intricacies is highly rewarding.

# 26. CliqueMap: Productionizing an RMA-Based Distributed Caching System

(26) Singhvi, A.; Akella, A.; Anderson, M.; Cauble, R.; Deshmukh, H.; Gibson, D.; Martin, M. M.; Strominger, A.; Wenisch, T. F.; Vahdat, A. CliqueMap: Productionizing an RMA-Based Distributed Caching System. In Proceedings of the 2021 ACM SIGCOMM 2021 Conference; 2021; pp 93–105.

Almost all systems that query an underlying data store, such as a database, require a caching layer. The cache is responsible for storing a subset of the total data - specifically the most frequently accessed items. These systems are fast because the information is precomputed and stored in high-speed media, such as system memory. One of the most well-known examples of such a system is Memcached [133].

This paper, however, discusses CliqueMap, Google's caching solution. One major difference from Memcached is CliqueMap's use of Remote Direct Memory Access (RDMA) for reads. This allows it to scale significantly higher than traditional Memcached deployments.

This chapter has a dual intent: to explore how caching operates in distributed systems and to understand the mechanics of RDMA. We will take a closer look at RDMA that emerged in the 2010s as a cornerstone of high-performance networking.

In this insight, we will begin with RDMA, covering its history and various implementations. Then, we will explore core distributed caching concepts. Finally, we will examine the architecture of CliqueMap and see how it leverages RDMA to implement a high-throughput caching solution.

**Recommended Read**: **25. Eliminating Receive Livelock in an Interrupt-driven Kernel** where the networking stack was introduced.

### 26.1. Remote Direct Memory Access

Remote Direct Memory Access (RDMA) is a major cloud technology of today. Several systems make use of RDMA internally to speed up their processing.

### 26.1.1. *A Brief History*

The conceptual groundwork for RDMA was laid in the early 1990s. InfiniBand [190] emerged as a prominent technology that leveraged RDMA to achieve very high throughput and low latency.

Mellanox Technologies played a crucial role in driving the adoption of InfiniBand and RDMA, particularly in high-performance computing environments.

Efforts began to extend the benefits of RDMA to Ethernet networks, leading to the development of RDMA over Converged Ethernet [191] (RoCE). This technology made RDMA more accessible by enabling its use on standard Ethernet infrastructure.

RDMA is now essential in various applications, including High-Performance Computing, storage systems (e.g., NVMe over Fabrics), and machine learning and AI.

RDMA technologies continue to evolve, with advancements like GPUDirect RDMA, which enables direct data transfer between GPUs and RDMA-capable network adapters.

### 26.1.2. *Hardware RDMA*

Hardware RDMA enables direct access of memory regions without CPU intervention via NIC. For RDMA, the NIC should have the capability to support RDMA. DMA is supported by all NICs, however, RDMA is not supported by all. NICs from Mellanox are examples of the ones that support RDMA.

#### 26.1.2.1. Architecture

In the context of traditional networking stacks (shown in Figure 115), when an application seeks to retrieve data from another application running on a remote machine, it initiates a Remote Procedure Call (RPC). This RPC necessitates a traversal through the kernel and the NIC on the sender's side. Subsequently, the data request traverses the NIC and the kernel on the receiver's side. Following this, the application thread on the receiver's side is invoked, which then proceeds to read the requested data. Finally, this data is transmitted back through the same network stack, retracing the path.

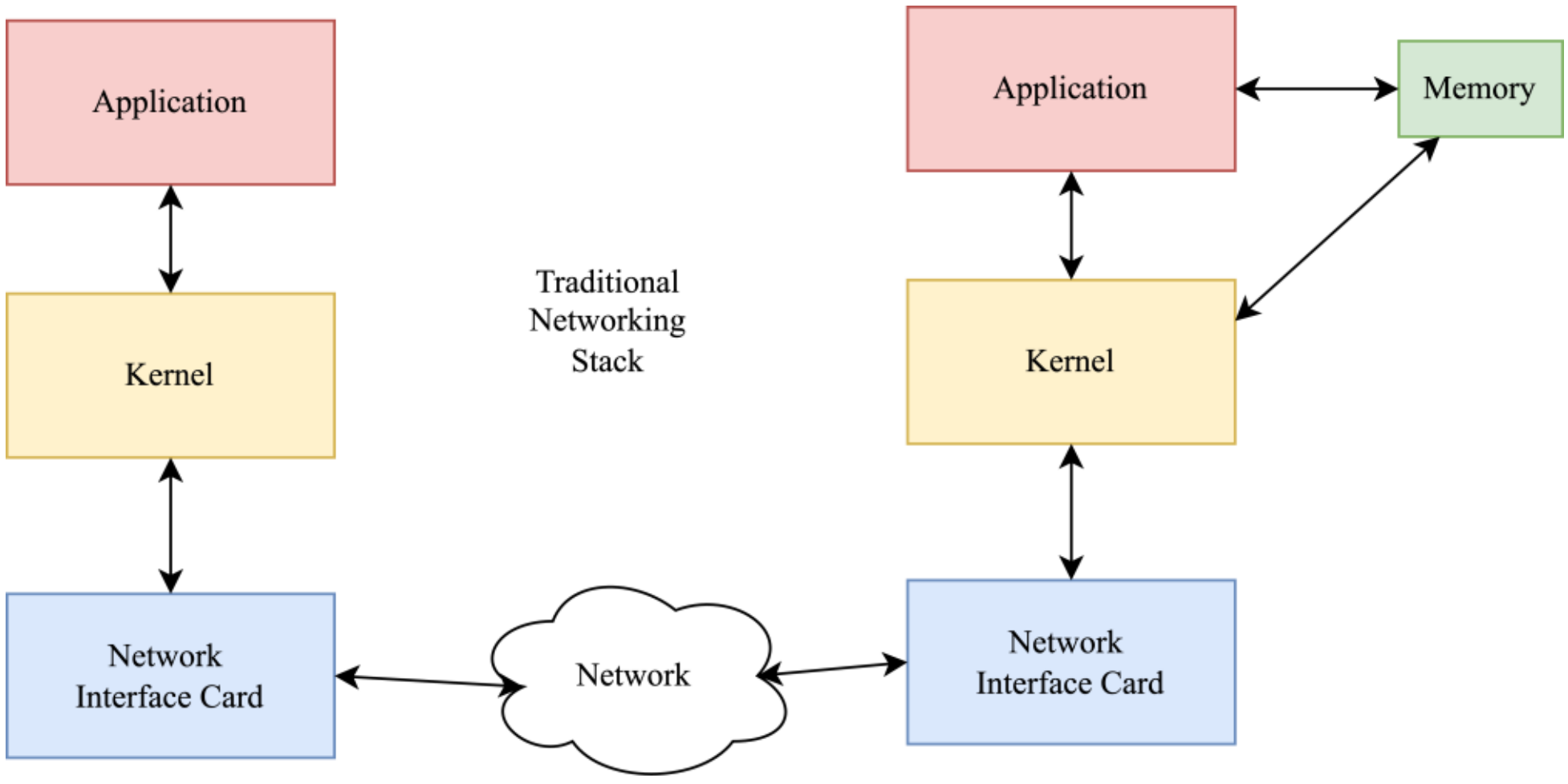

**Figure 115**: Traditional network stack.

Conversely, in the scenario utilizing hardware RDMA (shown in Figure 116), the NIC is granted direct access to designated memory regions. This remote direct access capability eliminates the need for CPU intervention, and consequently, bypasses the involvement of the kernel or application threads. The NIC itself possesses the capacity to directly read the necessary memory locations and transmit the response back.

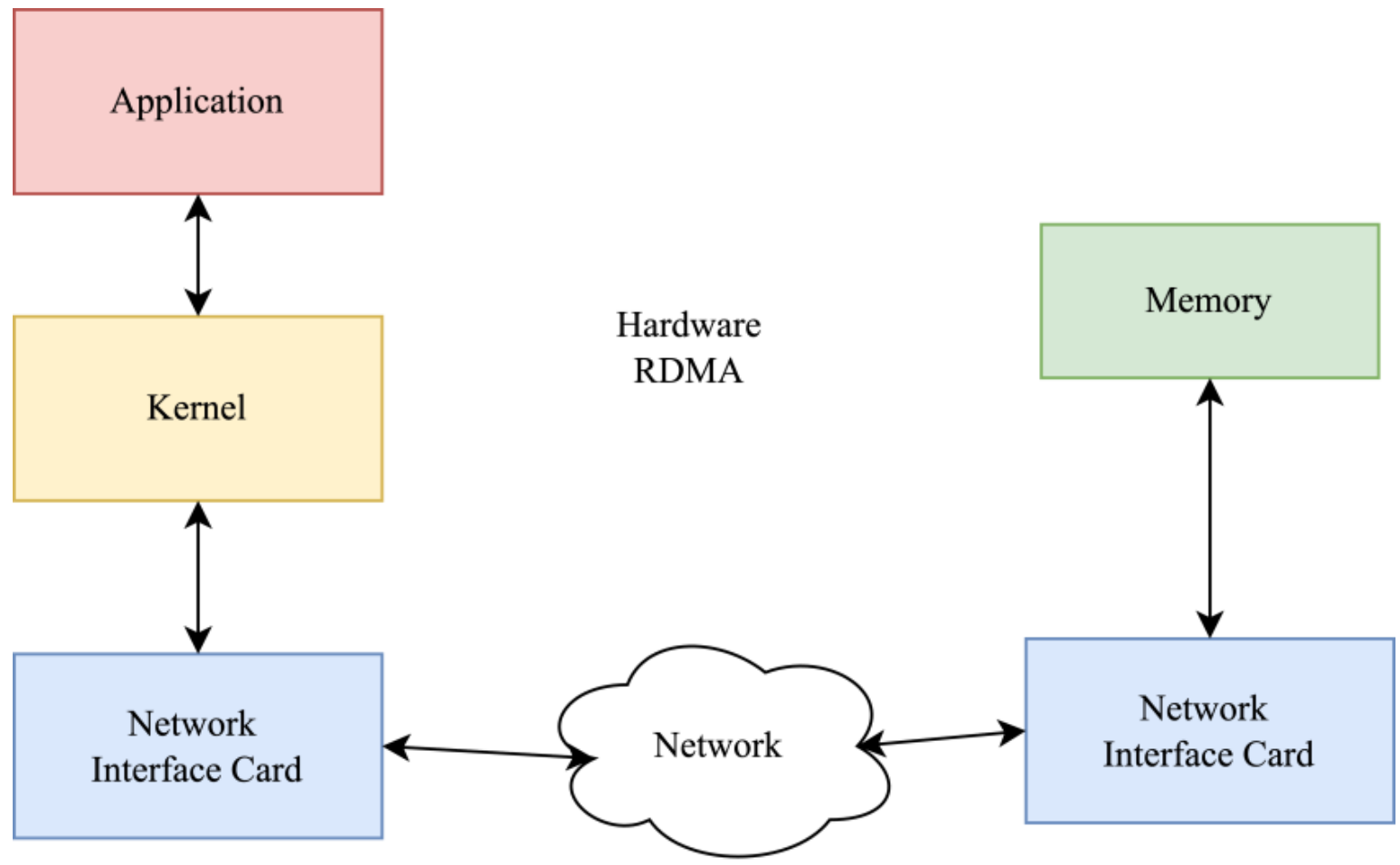

**Figure 116**: Hardware RDMA-based networking stack.

26.1.2.2. Implications

Hardware RDMA enables one-sided communication [192], one of the most valuable primitives in cloud networks.

While hardware RDMA delivers a substantial reduction in network latency, bypassing the operating system kernel and application layers on the receiver side introduces several significant implications:

**Increased Complexity in Client Protocol Implementation**

Although RDMA-enabled NICs gain direct access to memory regions, they lack the computational capabilities to execute logic. Specifically, they cannot parse and interpret the content of request messages. Consequently, the client application bears the responsibility of translating the request message into specific memory addresses that the NIC can directly read. This necessitates a more intricate client-side protocol, where the client must precisely format its requests to correspond to the target memory layout.

**Security Vulnerabilities**

Directly exposing memory regions introduces a significantly expanded attack surface. The absence of kernel intervention precludes the application of traditional security measures, such as encryption of request or response data over SSL-TLS. This lack of encryption, can severely restrict RDMA's applicability, confining it primarily to intra-cluster communication.

To mitigate these security concerns, many RDMA-capable NICs incorporate hardware-based encryption protocols, such as Google's PSP [193]. PSP, in particular, represents a crucial technological advancement beyond its role in RDMA, as it facilitates encryption without requiring CPU involvement. This significantly reduces CPU computational overhead.

*26.1.3. Software RDMA*

Software RDMA emerged as a response to security concerns associated with hardware RDMA. In contrast to hardware RDMA, software RDMA employs dedicated kernel threads, as shown in Figure 117. These threads mediate communication between the NIC and application-registered memory regions.

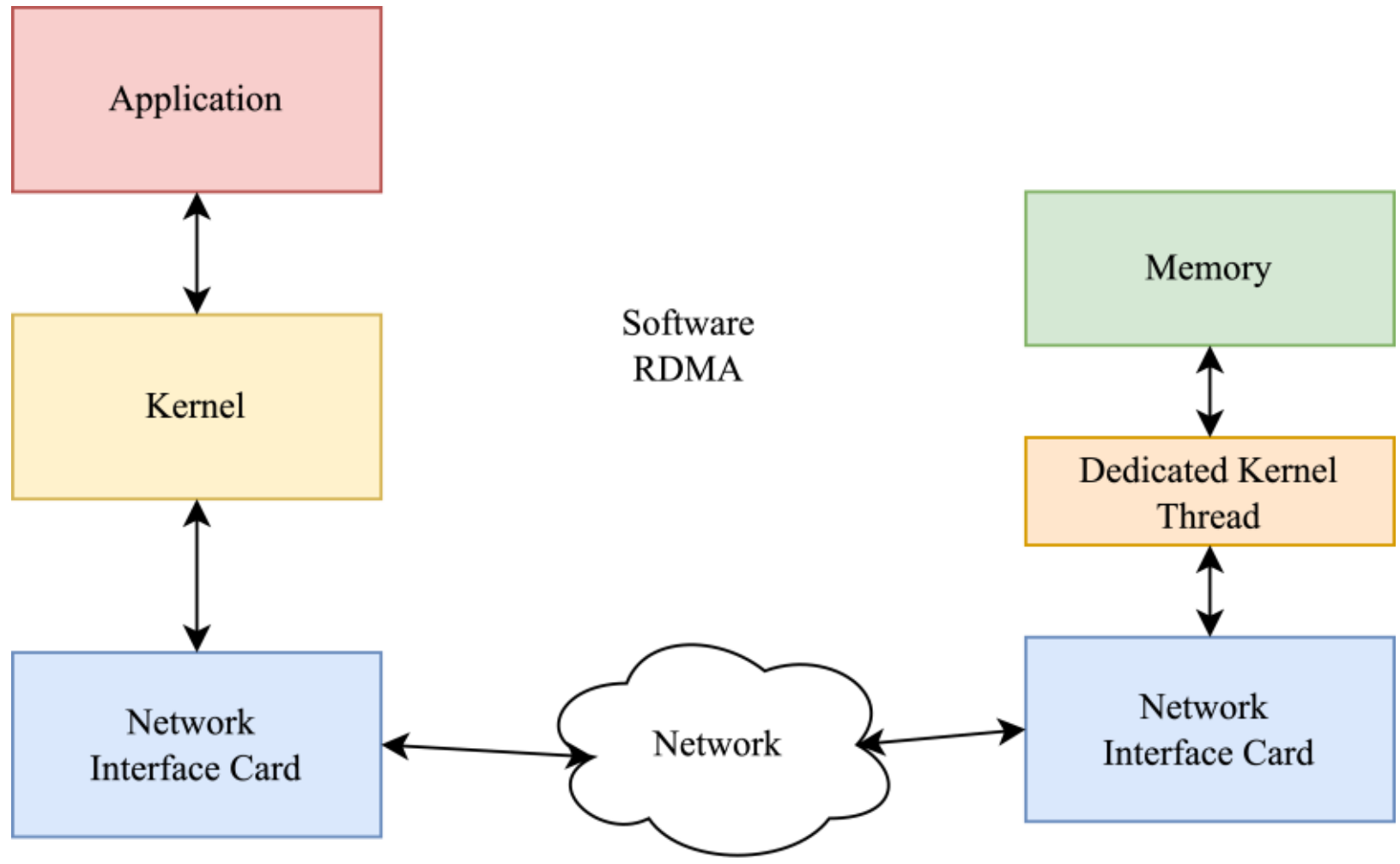

**Figure 117:** Software RDMA based networking stack.

This approach maintains the application thread bypass, eliminating user-space thread scheduling overhead. Additionally, the kernel threads possess the capability to execute logic and transmit data, enabling programmable network operations.

Software RDMA offers a sweet spot between hardware RDMA and the traditional network stack.

26.1.3.1. Pony Express

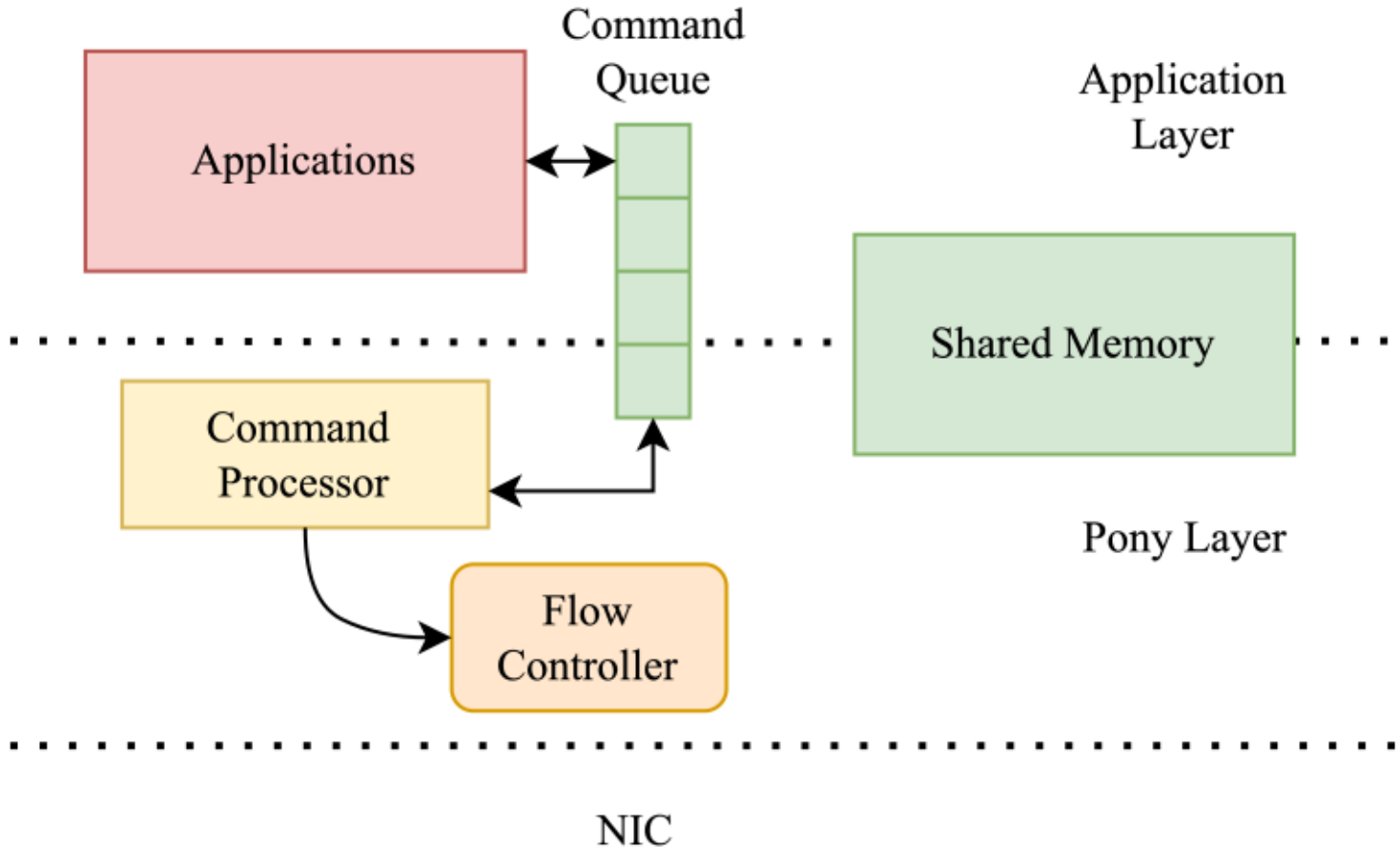

**Figure 118**: Pony Express.

Google's Snap [194] microkernel networking architecture, leverages a custom scheduler for CPU-NIC interactions, replacing interrupt-driven methods. This allows dedicated CPU cores to fully utilize NIC capacity, removing CPU bottlenecks. Snap's extensibility permits user-defined kernel code (**engines**) to handle network requests on these dedicated cores. Pony Express is one such engine.

Pony Express features a two-layered architecture, as shown in Figure 118: an application-facing layer that manages state and interacts with applications, including shared memory registration, and a lower layer that handles message processing and flow control.

Pony Express supports one-sided operations, such as RDMA reads, that extend beyond basic memory access. For example, it can perform hash table lookups if the application registers a hash map data structure.

Applications can utilize dedicated or shared Pony engine cores, communicating via shared command queues.

26.1.3.2. Performance

While traditional RPC achieve around 100,000 IOPS/core, Pony threads can reach 5M IOPS/core.

The throughput of Pony threads is influenced by the network's maximum transmission unit (MTU). Larger MTUs can improve NIC saturation but increase retransmission costs. Evaluations indicate that a single Pony thread can saturate 67 Gbps with a 5 KB MTU and 38 Gbps with a 1500B MTU. Considering modern 100 Gbps NICs, a few Pony threads can effectively saturate the NIC.

For conservative estimations, applications can assume 10 Gbps per Pony thread, translating to 250,000 IOPS per thread with a 5 KB MTU.

#### *26.1.4. Hardware vs. Software RDMA Comparison*

- Software RDMA can achieve higher user-defined IOPS due to the CPU's ability to execute custom logic, enhancing performance. For instance, software RDMA can implement custom flow control mechanisms to mitigate fabric pauses, a limitation of hardware RDMA. This demonstrates instances where software outperforms hardware, as seen in virtualization.

- Software RDMA offers significantly greater flexibility compared to hardware RDMA. Its reprogrammable nature allows for continuous innovation and adaptation even after deployment.
- Hardware RDMA, however, eliminates CPU overhead. While software RDMA incurs CPU costs, these costs are so minimal that it is almost zero.

#### *26.1.5. NIC - The Limiting Factor of Modern Networking*

In cloud environments, where machines commonly possess over 100 CPU cores, RDMA facilitates rapid NIC saturation. Consequently, the NIC has emerged as the primary bottleneck for network-intensive applications, especially in shared environments.

This limitation has spurred the development of Terabit Ethernet [195] to significantly increase NIC bandwidth. While 400 Gbps is achievable with current technology, reaching 1+ Tbps necessitates substantial architectural innovations, with Google and Meta heavily invested in this advancement.

### 26.2. Caching

Caching plays a crucial role in optimizing serving systems, which often rely on replication to handle high client request volumes. Caches are essentially volatile key-value stores where both keys and values are stored in memory only. There are no persistence guarantees, so one must not rely on them for data correctness. They should only be used for performance improvements.

#### *26.2.1. Cache Strategies*

Two primary caching strategies exist:

**1. In-Memory Cache**

This approach embeds the cache directly within each serving process, as shown in Figure 119.

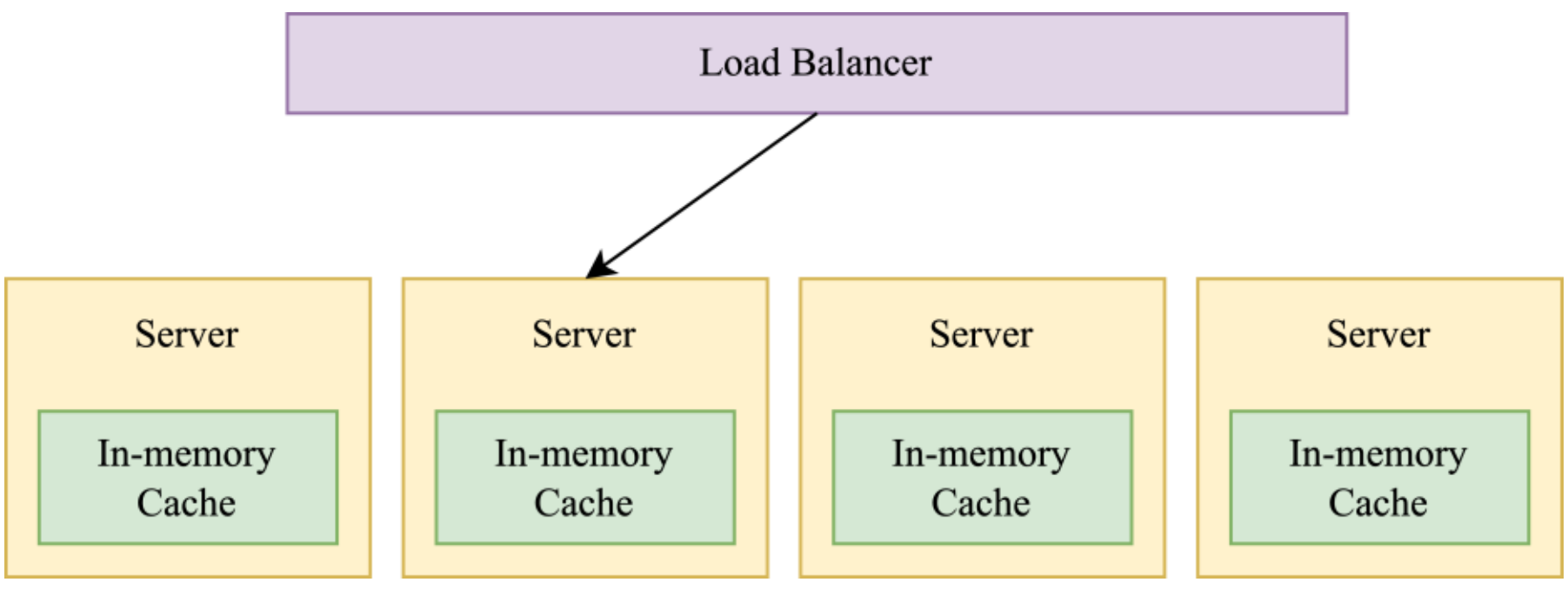

**Figure 119**: In-memory cache.

**Drawbacks:**

- Each replica incurs the full RAM cost of its local cache.
- Cached data is not shared across replicas, leading to redundant database fetches (e.g., if replica 1 caches key X, replica 2 must still retrieve it independently).

**2. Distributed Cache**

A separate, cluster-deployed caching layer is utilized, as shown in Figure 120.

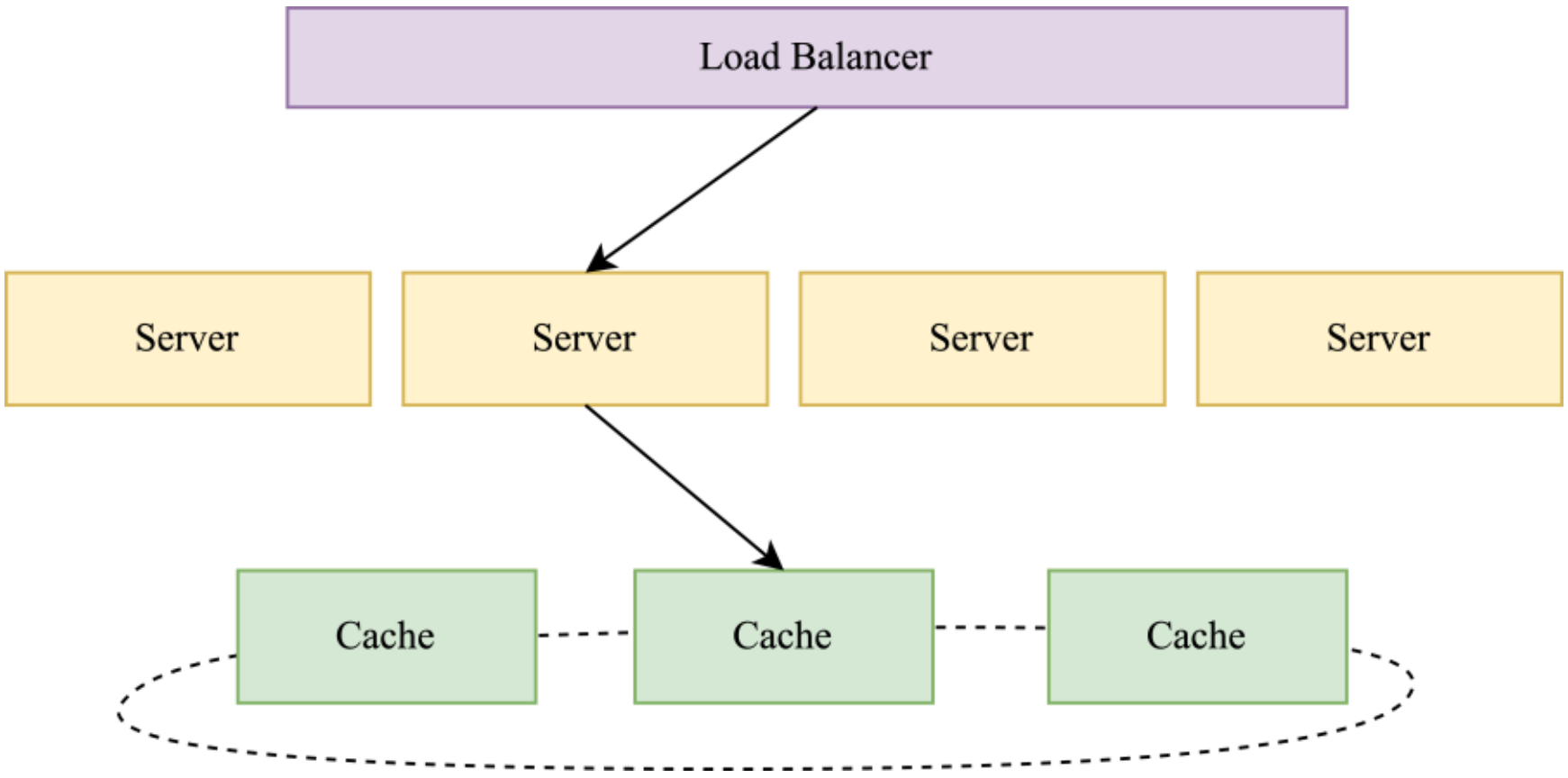

**Figure 120**: Distributed cache.

**Advantages:**

- Reduces memory overhead by storing each key-value pair only once.
- Enables data sharing across serving replicas.

**Disadvantage:**

- Introduces network latency as serving processes must communicate with the cache cluster. Numerous optimization techniques exist to mitigate this latency.

### *26.2.2. Cache Write Policies*

There are three different policies for writing to a cache.

**1. Write-Through**

In a write-through cache, every write operation updates both the cache and the backing data store (e.g., database) simultaneously. This ensures data consistency, as the cache always reflects the latest state of the data.

However, write operations can be slower due to the need to update both the cache and the database. This is very useful for read heavy workloads.

**2. Write-Back**

With a write-back cache, write operations are initially performed only on the cache. Updates to the backing data store are delayed and performed asynchronously, typically based on a timer or when a cache entry is evicted.

This significantly improves write performance, as writes are faster. However, it introduces the risk of data loss if the cache fails before the updates are written to the database. This is very useful for write heavy workloads.

**3. Write-Around**

Data is written directly to the backing store, bypassing the cache. This is used to prevent cache pollution, where infrequently written data fills the cache.

## 26.3. CliqueMap

CliqueMap, a widely used key-value store at Google, builds upon the foundational principles of Pilaf [196], an RMA-based storage system. While not developed internally, CliqueMap has become a critical component of Google's infrastructure.

Key architectural choices include:

- **Reads:** CliqueMap leverages RDMA for read operations, significantly reducing latency compared to traditional RPC. However, it falls back to RPC in certain scenarios.
- **Writes:** All write operations in CliqueMap are handled via RPC.

### *26.3.1. CliqueMap vs. Memcached*

CliqueMap offers several advantages over the industry-standard Memcached:

**1. RDMA-Powered Reads**

Unlike Memcached, which relies solely on RPC, CliqueMap utilizes RDMA for reads. At scale, RPC introduces considerable overhead (authentication, flow control, etc.), impacting performance. RDMA provides a more efficient approach, especially for read-heavy workloads, common in caching scenarios.

**2. Native Replication**

Memcached lacks built-in replication, requiring external solutions (e.g., Redis [132] wrappers) for redundancy. CliqueMap is designed with native replication capabilities, enhancing data availability and fault tolerance.

### *26.3.2. Data Organization*

A CliqueMap cluster employs multiple replicas. Each replica organizes data into distinct regions: a **data region** and an **index region**, as shown in Figure 1 in the paper.

#### 26.3.2.1. Data Region

The data region is segmented into slabs, and a slab-based allocator manages memory for value storage. Slab sizes are configurable to optimize for specific workloads. The memory allocated for slabs can be dynamically resized by the CliqueMap server threads (via RPC writes) to accommodate growing data.

Each data entry in the data region comprises:

- **Format:** A versioning identifier.
- **Key Length:** Specifies the size of the key.
- **Key:** The key itself.
- **Data Length:** Specifies the size of the value.

- **Data:** The value associated with the key.
- **Metadata:** Includes the version number, ensuring data consistency.
- **Checksum:** A checksum of the entire entry, crucial for data integrity.

26.3.2.2. Index Region

The index region organizes key-value pairs into buckets, forming a set-associative structure.

Each index entry contains:

- **KeyHash:** The hash of the key, used for bucket assignment.
- **VersionNumber:** A unique, monotonically increasing version number that's updated on every write.
- **Pointer:** The memory address of the corresponding value in the data region.

The **KeyHash** determines the bucket to which a key-value pair belongs, enabling efficient lookups.

### *26.3.3. Consistent Hashing*

The key-value pairs are divided among the replicas based on the hash of the key. Consistent Hashing comes into play although the paper doesn't mention which algorithm is used by CliqueMap internally.

### *26.3.4. Replication*

CliqueMap utilizes quorum replication, supporting:

- **R=1**: No replication (single replica).
- **R=2**: Replication across two replicas (for immutable key-value pairs).
- **R=3.2**: Replication across three replicas, requiring a majority of two for operations.

Given its complexity and reliability, the remainder of this chapter will focus mostly on the R = 3.2 mode.

### *26.3.5. Reading a Key-Value Pair (GET)*

Reading a key-value pair in CliqueMap involves the following steps:

- The key is hashed. The corresponding replica backends are identified based on the hash.
- Key-value is retrieved from the replicas using RDMA.
- Version numbers of the retrieved values are compared to ensure consistency.
    - In R=2 mode, versions must match.
    - In R=3.2 mode, a majority (2 out of 3) must match. The majority version number is the consistent version number.
- Finally, the client verifies the checksum of the fetched data. The retrieved key is compared to the original key to mitigate potential (though rare) hash collisions.

26.3.5.1. Read Strategies

CliqueMap employs several strategies for reads:

**1. Two Round Trips (2xR)**

Requires two RDMA round trips:

- Retrieving the bucket from the index region of all replicas using the key hash.
- Fetching the actual value from the preferred backend's data region.

The index entry is used to compare the version numbers and then the value is fetched from the data region of the preferred backend. This is optimal for large value sizes.

Pony Express is used for one-sided communication in both steps.

**2. Scan and Read (SCAR)**

Performs a single RDMA round trip. Pony Express reads the index region, retrieves the data region pointer, and returns the data entry. The Pony thread needs to execute some custom logic to make this possible. This consumes more server-side CPU but reduces latency. However, the value will be returned by all the replicas. This is preferable for small value sizes.

**3. RPC Fallback**

If RDMA reads cannot provide the value (as in case of overflow buckets, described later), a fallback to traditional RPC is used. This is significantly slower than 2xR and SCAR.

#### 26.3.5.2. Relaxed Reads

CliqueMap also offers relaxed reads, which prioritize performance over consistency. In this mode, the value is fetched from a single replica. This saves both server-side computation and client-side bandwidth consumption. However, the retrieved value may be outdated.

Essentially, this mode trades linearizability for reduced latency.

### *26.3.6. Writing to a Key-Value Pair (SET)*

#### 26.3.6.1. Total Order on Mutations

CliqueMap ensures total ordering of key-value pair updates, leveraging Google's TrueTime, a distributed time synchronization system. Each mutation is assigned a unique, monotonically increasing version number in the format **<TrueTime, Client ID, Sequence Number>**.

This versioning system means that mutation order is determined by the version number, not necessarily the arrival time of requests. While this might seem to violate external consistency, TrueTime guarantees that clients eventually generate higher version numbers for subsequent requests, maintaining external consistency.

#### 26.3.6.2. Erase Operations

The erase operation removes a key-value pair from the cache. To prevent conflicts with concurrent mutations, erase also uses version numbers. After an erase, the key-value pair is moved to a fully-associative **tombstone cache**. This ensures that reads for older versions of the key-value pair can still be served.

#### 26.3.6.3. Compare-And-Swap

Compare-And-Swap (CAS) is a crucial primitive for CliqueMap. It performs a mutation only if the current version number of the key-value pair matches a provided version number. This is essential for implementing transactional behavior on single key-value pairs (CliqueMap, like many key-value stores, doesn't support multi-key transactions).

The client reads the current value and its version number, and the subsequent mutation succeeds only if the version numbers match. This enables atomic updates and prevents race conditions.

*26.3.7. Cache Evictions*

As a volatile key-value store, CliqueMap can lose data without impacting system correctness. Eviction, the process of removing key-value pairs, is essential for accommodating new key-value pairs.

26.3.7.1. Eviction Types

CliqueMap employs two primary eviction types:

- **Capacity Evictions:** These occur when the cache reaches its memory capacity and needs to free space for new key-value pairs.
- **Associativity Evictions:** Due to the set-associative bucket structure, each bucket has a limited number of index entries. If numerous keys map to the same bucket, some keys must be evicted to make room for new entries.

26.3.7.1.1. Overflow Buckets

It is possible to address associativity evictions using overflow buckets. When a bucket overflows, new key-value pairs are stored in these overflow buckets. While this avoids immediate eviction, accessing data in overflow buckets incurs an RPC overhead.

26.3.7.2. Eviction Policies

CliqueMap supports standard eviction policies, such as Least Recently Used (LRU). However, the RDMA-based read operations present a challenge for tracking key access times.

To address this, clients communicate key access times to the server backends via separate batch RPCs. This allows the server to approximate LRU behavior despite the use of RDMA for reads.

### *26.3.8. Quorum Repairs*

In R=3.2 mode, each key is replicated across three replicas. The loss of one replica results in a **dirty quorum**, while the loss of two leads to an **inquorate state**. CliqueMap implements on-demand quorum repair to recover key-value pairs in these scenarios.

Quorum repair is also triggered when a replica restarts.

The repair process involves the following steps:

1. Replicas scan their cohorts (the other replicas holding the same key) to determine the current version of the key-value pairs.
2. If version numbers don't match, replicas fetch the latest version of the key.

This mechanism is also valuable for updating stale replicas, which might have missed a mutation (since R=3.2 only requires two successful writes out of three).

## 26.4. Evaluation

The authors conducted a comprehensive evaluation of the caching system, focusing on key performance metrics, particularly tail latency (p95, p99, p999). Minimizing tail latency is crucial for caching systems, as they underpin downstream applications, directly impacting overall system performance.

Key evaluation findings include:

- At 3 million operations per second (ops/sec), the p999 latency was ~10ms.
- R=3.2 mode demonstrated superior performance compared to R=1, particularly in tail latency. This is attributed to its ability to select the optimal server for read operations.
- 2xR outperformed SCAR for large value sizes.
- Data migration during quorum repair, performed via RPCs, introduced latency overhead.
- CliqueMap is optimized for read-heavy workloads. Higher read percentages resulted in lower observed latencies.
- **Hardware vs. Software RDMA:**
    - Pony Express demonstrated robust throughput and low latency even under high client load.

- Hardware RDMA significantly outperformed software RDMA, even when limited to the 2xR strategy (hardware RDMA only supported 2xR).

### 26.5. Paper Remarks

This paper presents a highly compelling caching architecture that stands out as one of the most promising solutions in recent academic literature. Beyond its specific caching innovations, the paper serves as an accessible and high-quality introduction to RDMA - a technology with profound implications for modern cloud infrastructure. The treatment of the subject is particularly insightful, offering a level of depth and clarity that distinguishes it from similar works by other industry leaders such as Facebook [197] and Twitter [198].

# 27. Don't Settle for Eventual: Scalable Causal Consistency for Wide-Area Storage with COPS

(27) Lloyd, W.; Freedman, M. J.; Kaminsky, M.; Andersen, D. G. Don't Settle for Eventual: Scalable Causal Consistency for Wide-Area Storage with COPS. In Proceedings of the Twenty-Third ACM Symposium on Operating Systems Principles; 2011; pp 401–416.

We now return to the world of databases one last time to discuss specialized systems in this and the following chapter. In this chapter, we focus on COPS, a NoSQL database designed to be causally consistent. As the CAP theorem dictates, it is impossible to achieve both strong consistency and availability simultaneously when network partitions are likely - a common reality for distributed databases. COPS, however, achieves causal consistency while remaining available. In fact, causal consistency is the highest consistency model achievable in an available system.

This paper was presented at Symposium on Operating Systems Principles (SOSP) 2011. The paper features contributions from prominent distributed systems researchers Wyatt Lloyd and Michael Freedman.

In this insight, we will begin with causal ordering and explain how it can be captured using logical clocks. In this process, we will explore elements from Leslie Lamport's classic paper, which describes how distributed systems can understand the partial ordering of events without the need for synchronized clocks. Next, we will examine caussal consistency in greater detail. Finally, we will drill down into the architecture of COPS, exploring how it ensures causal consistency through specific architectural choices and walkthrough examples.

### 27.1. Causal Ordering

In 1978, Leslie Lamport published "Time, Clocks, and the Ordering of Events in a Distributed System" [135], a seminal paper that significantly impacted distributed system design. This work, alongside Paxos and TLA+ [199], stands as one of Lamport's most influential contributions.

A fundamental challenge in distributed systems is clock synchronization. Perfect synchronization is unattainable, a fact rooted in both computer science and physics. However, the goal isn't perfect synchronization itself, but rather the ability to totally order events. With synchronized clocks, we could definitively determine the temporal order of events across different nodes. For example, if event **a** occurs on node 1 and event **b** on node 2, synchronized clocks would allow us to assign timestamps and directly compare them to establish which event happened first.

#### *27.1.1. The Partial Order*

Unfortunately, perfect global clock synchronization (outside of specialized solutions like Google's TrueTime) remains impractical. Consequently, we rely on the concept of **partial ordering**.

Let $\leq$ be a binary relation. Then $\leq$ is a partial order if:

- $\forall \mathbf{a}: \mathbf{a} \leq \mathbf{a}$ (**reflexivity**)
- $\forall \mathbf{a}, \mathbf{b}: (\mathbf{a} \leq \mathbf{b} \land \mathbf{b} \leq \mathbf{a}) \Rightarrow \mathbf{a} = \mathbf{b}$ (**antisymmetry**)
- $\forall \mathbf{a}, \mathbf{b}, \mathbf{c}: (\mathbf{a} \leq \mathbf{b} \land \mathbf{b} \leq \mathbf{c}) \Rightarrow \mathbf{a} \leq \mathbf{c}$ (**transitivity**)

A strict partial order (<) does not include the reflexivity condition.

Within a single system, ordering is achievable using a shared clock source and synchronization primitives like Linux futexes [200]. However, establishing order across distributed nodes requires different approaches.

Network messages offer a mechanism for partial ordering across distributed nodes. When node 1 sends a message to node 2, it includes its current timestamp (**T**). Node 2, upon receipt, knows that node 1's time was at least **T**. If node 2's local time is less than **T**, it adjusts its clock to **T**, ensuring subsequent events have timestamps greater than **T**, thereby establishing that all events after receipt of the message are ordered after the events before send of the message.

##### 27.1.1.1. Example

Consider the following event diagram where **a**, **b**, **c**, and **d** are 4 events happening across 2 nodes, as shown in Figure 121.

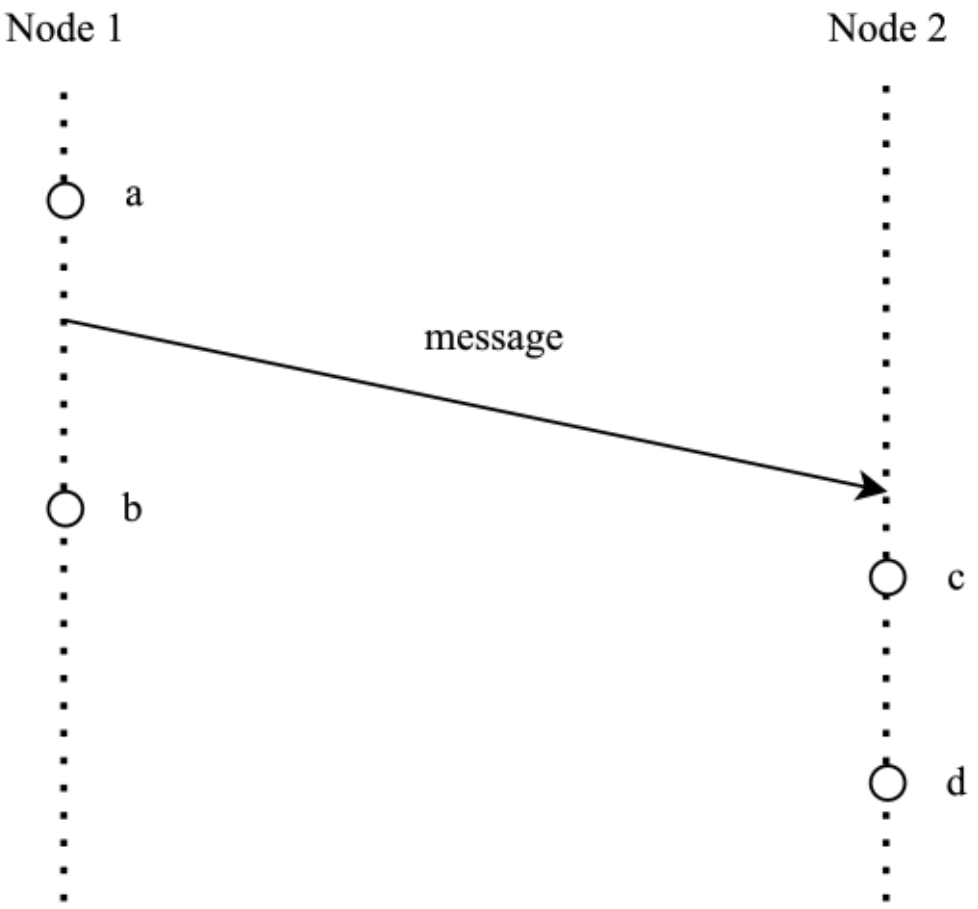

**Figure 121**: Partial order example.

It can be seen that:

- **a < b**, **c < d** (order within nodes)
- **a < c**, **a < d** (message causality)

However, the relative order of **b** to **c** or **d** remains undefined, as a result, we only know the partial order.

#### 27.1.1.2. Happened-Before Relation - The Causal Order

Lamport formalized this partial ordering with the **happened-before** relation (→) also known as the **causal order**:

1. If events **a** and **b** occur within the same node and **a** precedes **b**, then **a** → **b**.
2. If event **a** is the sending of a message and event **b** is its receipt, then **a** → **b.**
3. **Transitivity**: if **a** → **b** and **b** → **c**, then **a** → **c**.

This relation has the properties of a strict partial order**:**

- Not reflexive (an event doesn't happen before itself)
- Anti-symmetric
- Transitive (as defined above)

### *27.1.2. Logical Clocks*

Logical clocks aim to capture this partial ordering.

A logical clock $\mathbf{C}$ is a vector of clocks where each element $\mathbf{C}_i$ is a clock for a node $\mathbf{i}$. $\mathbf{C}$ assigns timestamps to events, such that if $\mathbf{a} \rightarrow \mathbf{b}$, then $\mathbf{C}(\mathbf{a}) < \mathbf{C}(\mathbf{b})$. **The converse isn't necessarily true.**

This condition is satisfied by:

- Within a node, if **a** → **b**, then $\mathbf{C}_i(\mathbf{a}) < \mathbf{C}_i(\mathbf{b})$.
- Across nodes, if **a (send)** → **b (receive)**, then $\mathbf{C}_i(\mathbf{a}) < \mathbf{C}_\mathrm{j}(\mathbf{b})$.

27.1.2.1. Counter-based Implementation

A simple implementation for logical clocks is based on counter variables:

- Each node maintains a counter.
- The counter increments for each local event.
- When a node **i** is sending a message, the current counter $\mathbf{C}_i$ is included.
- On receipt, the receiving node **j** updates its counter to $\mathbf{max}(\mathbf{C}_\mathrm{j}, \mathbf{C}_i + \mathbf{1})$.

Note that such counter-based logical clocks don't satisfy the strong clock condition (discussed below).

*27.1.3. Total Ordering from Partial Ordering*

Logical clocks provide partial ordering. Total ordering can be achieved by breaking ties using node IDs. For example, if **a** → **b** on node 1 and **c** → **d** on node 2, we can enforce all non-ordered events from node 1 before node 2 and arrive at **a** → **b** → **c** → **d**.

*27.1.4. Strong Clock Condition*

The strong clock condition requires:

- If **a** → **b**, then $\mathbf{C}(\mathbf{a}) < \mathbf{C}(\mathbf{b})$.
- If $\mathbf{C}(\mathbf{a}) < \mathbf{C}(\mathbf{b})$, then **a** → **b**.

Logical clocks cannot inherently satisfy this. For example, say an event **a** happened on node 1 and event **b** happened on node 2. The event **a** may have actually happened before event **b** but may be assigned a timestamp higher due to different counter positions on the two nodes.

However, it would be very hard for humans to reason about such cases where the machine believes that $\mathbf{C(b)} < \mathbf{C(a)}$, yet somehow **a** actually happened before **b**.

Lamport makes use of physical clocks to satisfy the strong clock condition and make it easy for humans to reason about the partial ordering of events. To achieve the strong clock condition using physical clocks, specific constraints are needed:

- Bounded individual clock drift ($\mathbf{k}$). For crystal clocks, $\mathbf{k} \leq \mathbf{10^{-6}}$.
- Bounded drift between clock pairs ($\boldsymbol{\varepsilon}$).
- Non-zero minimum network transmission time ($\boldsymbol{\mu}$).
- The receiver's clock upon receiving a message **m** must be greater than sender's clock before send, i.e., for all **i**, **j**, and **t**, where **t** is the real physical time:

$$\mathbf{C}_i(\mathbf{t} + \boldsymbol{\mu}) - \mathbf{C_j}(\mathbf{t}) > \mathbf{0}$$

This in turn requires the following condition to hold true (omitting derivation for brevity):

$$\frac{\boldsymbol{\varepsilon}}{\mathbf{1 - k}} \leq \boldsymbol{\mu}$$

When the above constraints are met (and it actually does in practice because clock drift is smaller than network delay), the strong clock condition is achieved by the following approach ($\mathbf{t}$ and $\mathbf{t'}$ are the real physical time of events):

- Say $\mathbf{T_m}$ is the time when a message was sent by node **i** and $\boldsymbol{\mu}$ is the time it took for the message to reach a node **j**.

$$\mathbf{T_m} = \mathbf{C}_i(\mathbf{t})$$

- Upon receiving **m** at time $\mathbf{t'}$, node **j** sets

$$\mathbf{C_j}(\mathbf{t'}) = \mathbf{max}(\mathbf{C_j}(\mathbf{t'}), \mathbf{T_m} + \boldsymbol{\mu})$$

**Recommended Read**: "Time, Clocks, and the Ordering of Events in a Distributed System" [135] to understand the derivation of the formulas in detail.

### 27.2. Causal Consistency

For a review of general consistency models, please refer to our previous discussion on ZooKeeper. In this chapter, we'll focus specifically on **causal consistency**, particularly within the context of single-value key-value stores.

Causal consistency mandates that all operations respect the causal ordering. All clients observe the writes happening in some causal order. In a causally consistent key-value store, this can lead to scenarios where multiple values exist for the same key, requiring eventual resolution.

Let's illustrate this with an example:

Imagine a key **X** initially holds the value 0. Node 1 performs two consecutive writes: **W(X, 1)** followed by **W(X, 2)**. Node 2 reads **X**, observes the value 1 (**R(X, 1)**), and then writes the value 3 (**W(X, 3)**).

The causal relationships are as follows, as shown in Figure 122:

- **W(X, 1) → R(X)** (read reflects the write)
- **W(X, 1) → W(X, 2)** (ordering within Node 1)
- **R(X) → W(X, 3)** (ordering within Node 2)

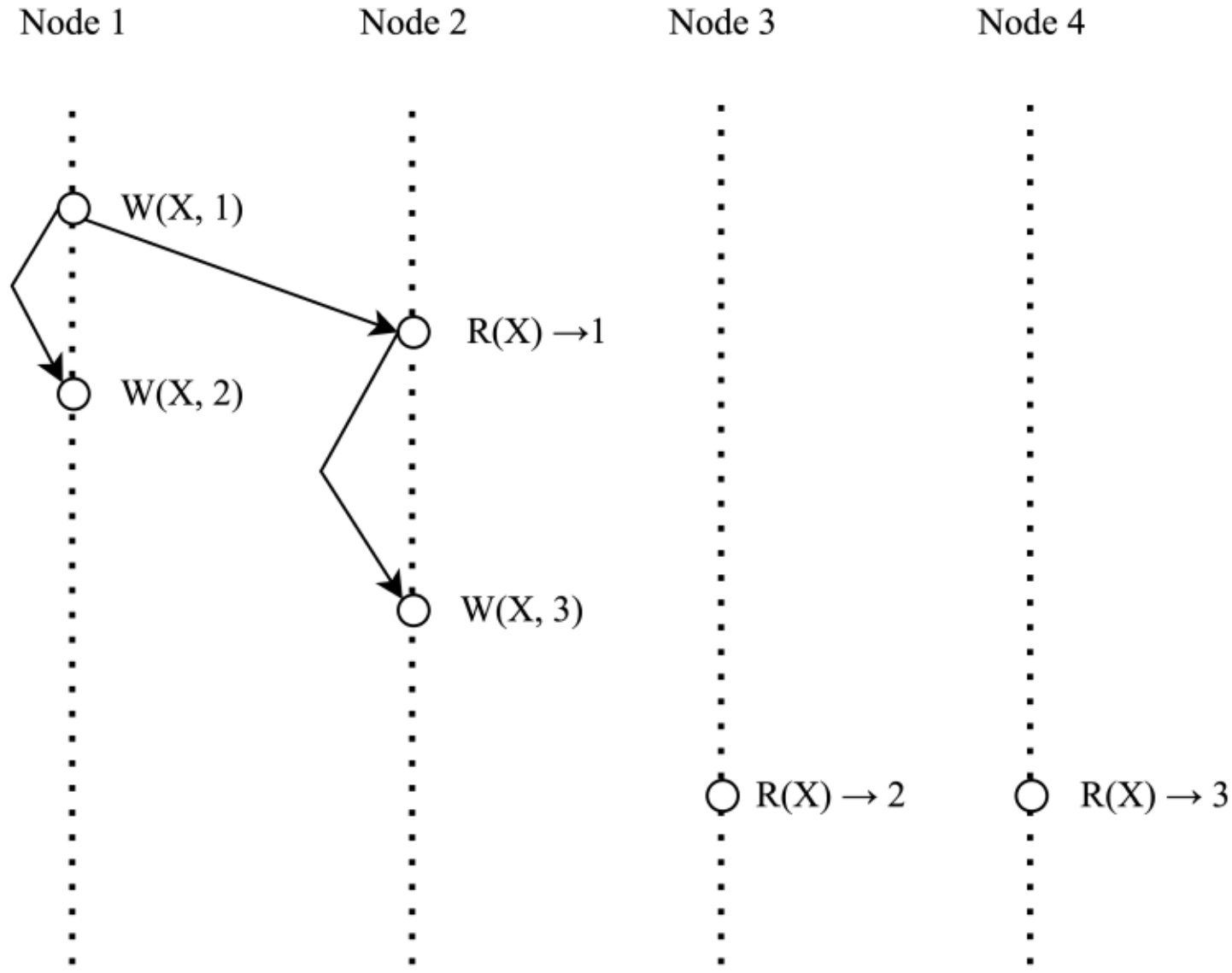

**Figure 122**: Causal relationships example.

Crucially, the order between **W(X, 2)** and **W(X, 3)** is undefined; they are not causally related. Therefore, both **X = 2** and **X = 3** are valid states within a causally consistent system. Thus, Node 3 may find the value to be 2 (**W(X, 3)** will be ordered before **W(X, 2)** from its perspective), and Node 4 may find the value to be 3 (**W(X, 2)** will be ordered before **W(X, 3)** from its perspective).

This contrasts with **sequential consistency**, which demands a single, definitive total order for every observer. In a sequentially consistent system, all reads must return either 2 or 3, depending on the system's chosen ordering (of course, given that the reads are ordered after all the writes).

**Eventual consistency** is even weaker than causal consistency. See **28. TAO: Facebook's Distributed Data Store for the Social Graph** to learn more.

## 27.3. Cluster of Order-Preserving Servers

Cluster of Order-Preserving Servers (COPS) was developed to address the **ALPS principles: Availability, Low-latency, Partition-tolerance, and Scalability**. It achieves this by sacrificing strong consistency in favor of causal consistency. This aligns with the CAP theorem, which demonstrates that strong consistency, availability, and partition tolerance cannot be simultaneously guaranteed. **Causal consistency (to be precise, it's real-time causal consistency), is the strongest consistency level achievable with availability and partition tolerance.**

While systems like Dynamo and TAO employ eventual consistency - the most relaxed model - due to its relative simplicity, COPS takes a more nuanced approach to offer causal consistency.

COPS has 2 different flavors - COPS and COPS-GT. To avoid ambiguity, COPS will henceforth refer to the enhanced COPS-GT version only.

### *27.3.1. API*

COPS provides a straightforward API:

- **put(key, val, context)**: Stores a value associated with a key.
- **get(key, context) → value**: Retrieves the value associated with a key.
- **get_trans(<keys>, context) → <values>**: Retrieves a consistent view of multiple key-value pairs.

Despite its simple API, reasoning about the returned values within a causally consistent system can be complex, contributing to the relative lack of widespread adoption for causal consistency. In fact, COPS never became more than an academic marvel. The computer science industry has been happy with eventual consistency.

### *27.3.2. Architecture*

COPS employs a network of globally distributed clusters (shown in Figure 123), each storing all key-value pairs.

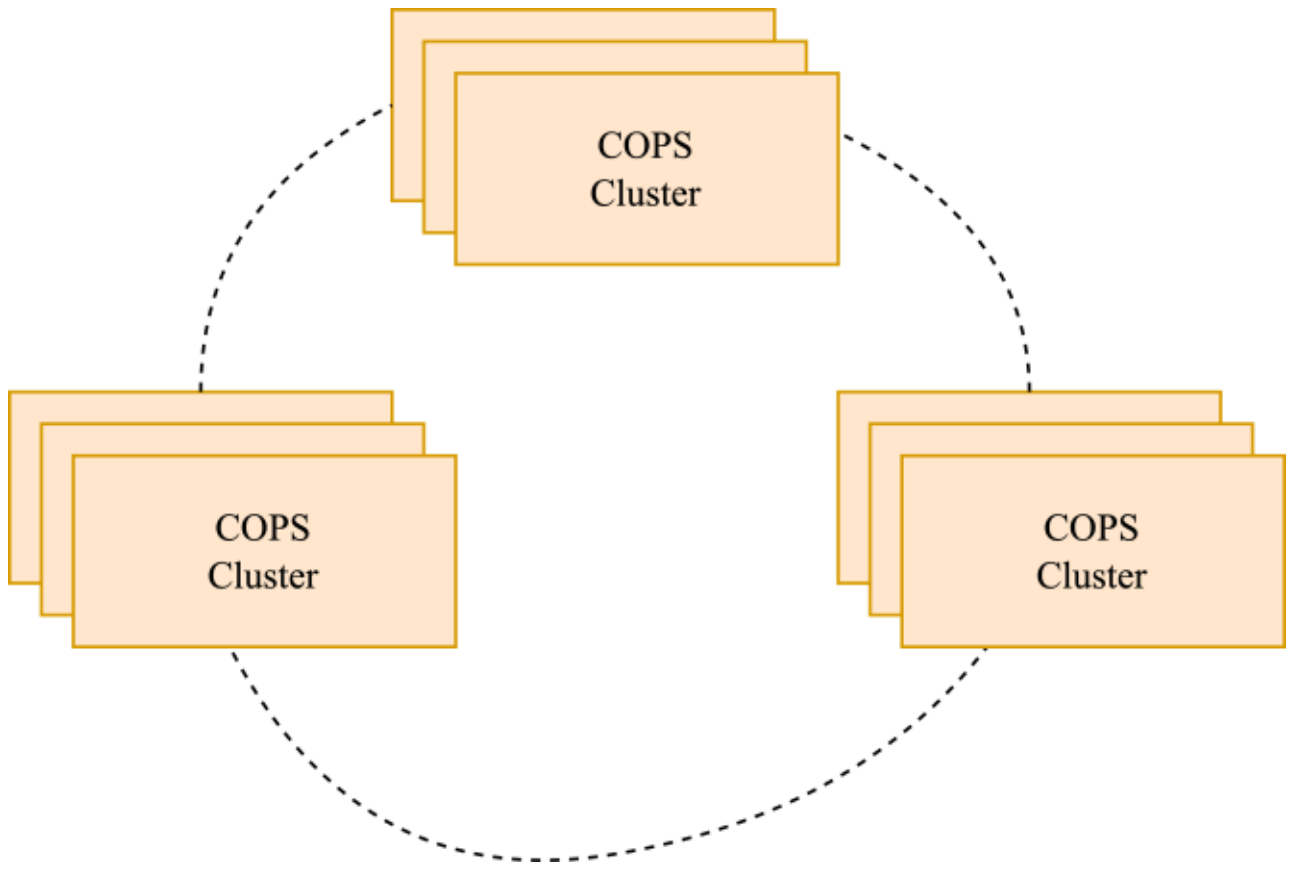

**Figure 123**: COPS cluster.

Each cluster functions as a linearizable key-value store. Within a cluster, key-value pairs are partitioned into **N** keyspaces and distributed across replicas using consistent hashing, as shown in Figure 124.

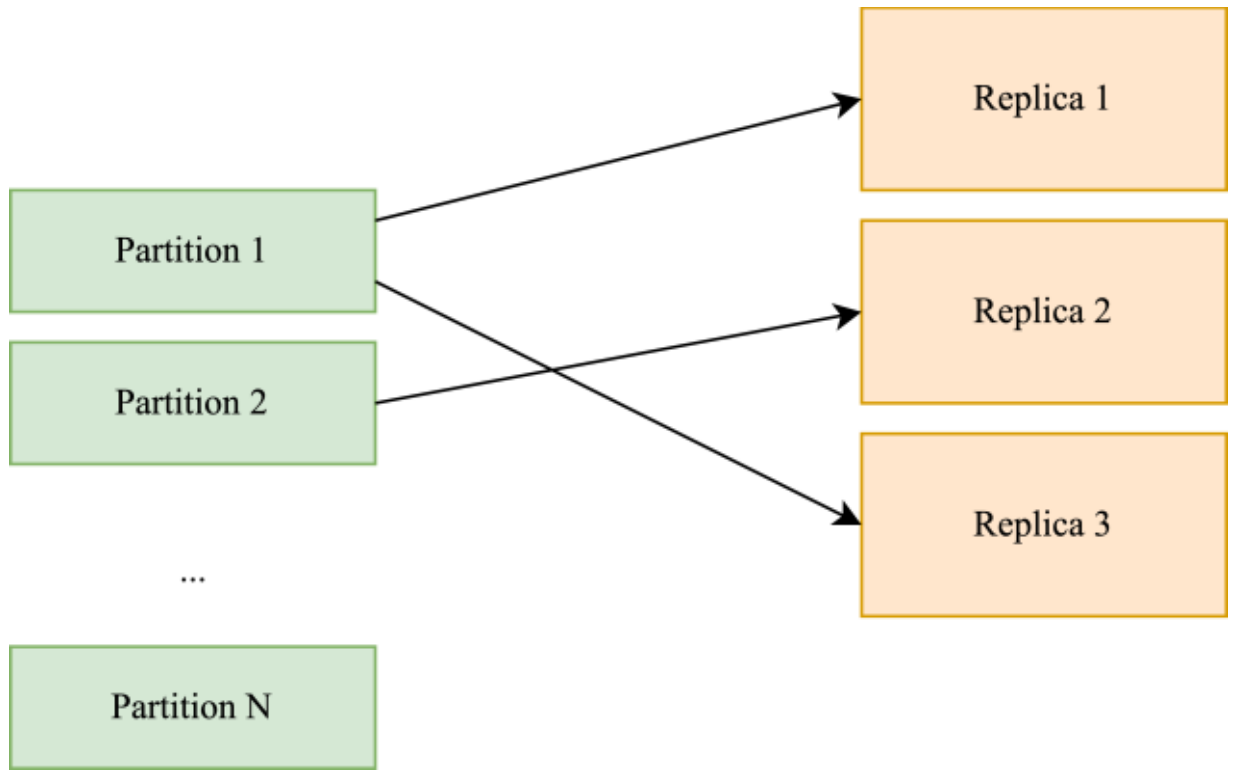

**Figure 124**: Key-space partition in COPS.

The cluster's linearizable storage utilizes the FAWN KV-store [201], with chain replication for each key-value pair. Client write requests are routed to the chain's head, while all read requests, including dependency checks (described below), are directed to the tail.

Each key-value pair is associated with metadata:

1. Version number
2. Dependencies (explained below) - A list of **<key, version>** pairs.

The system includes three internal APIs, not directly exposed to users, but utilized by the client library:

- **put_after(key, val, [deps]) → <bool, version>**
- **get_by_version(key, version) → <value, version, deps>**
- **dep_check(key, version) → bool**

Clients interact with the key-value store exclusively through a client library, which manages all communication.

### *27.3.3. Consistency Model*

The authors formally specify the **causal+ consistency** model in appendix A in the paper.

#### 27.3.2.1. Causal Order

COPS defines causal ordering (**~>**), similar to Lamport's happened-before relation, but tailored to its API:

- **Execution Thread:** Operations **a** and **b** within a single execution thread (node) are ordered as **a ~> b** if **a** precedes **b**.
- **Gets From:** If a **put** operation **a** is read by a **get** operation **b**, then **a ~> b**. This replaces Lamport's message-based causality with a read-write dependency.
- **Transitivity:** The **~>** relation maintains transitivity.

#### 27.3.2.2. Conflict Handling

Concurrent writes can lead to conflicting values across replicas, as they're permitted to diverge when writes lack causal ordering. In our previous example, **W(X, 2)** and **W(X, 3)** resulted in diverging values for X.

Conflicting values are undesirable. COPS resolves these divergences using a **handler function**. This function takes two conflicting values for a key and returns a single, converged value.

The handler function must be associative and commutative. A simple handler function can be **last-writer-wins** to impose an arbitrary order, forcing conflict resolution.

#### 27.3.2.3. Causal+ Consistency

This consistency model is slightly stronger than causal consistency and is called **causal+ consistency**.

Causal+ thus occupies a distinct position within the consistency spectrum, offering a stronger form of consistency than causal+ consistency, but weaker than strong consistency.

### *27.3.4. Versions and Dependencies*

Each value associated with a key is assigned a unique version. COPS uses Lamport timestamps (as described previously) to generate these versions. When combined with a last-writer-wins tie-breaking mechanism, this yields a global ordering for all writes to a key. A version number is thus represented as **<Lamport timestamp, node number>**.

Formally, for writes to keys **x** and **y** at versions **i** and **j** respectively, if $\mathbf{x}_i$ **~>** $\mathbf{y}_j$ ($\mathbf{x}_i$ causally precedes $\mathbf{y}_j$), then $\mathbf{i} < \mathbf{j}$.

#### 27.3.4.1. Progressing Property

Once a client reads a specific version for a key, subsequent reads will return that version or a later one. This is the progressing property guaranteed by causal+ consistency.

#### 27.3.4.2. Dependencies

Causal relationships introduce dependencies between key values.

If $\mathbf{x}_i$ **~>** $\mathbf{y}_j$, then $\mathbf{y}_j$ **depends on** $\mathbf{x}_i$. This means if a client reads $\mathbf{y}_j$, it must also be able to read **x** at version **i** or a later version. A single key can have dependencies on multiple versions of other keys.

COPS maintains these dependencies as a dependency graph stored internally.

An example of dependency graph is shown in Figure 125. Alphabets are keys and subscripts are version numbers.

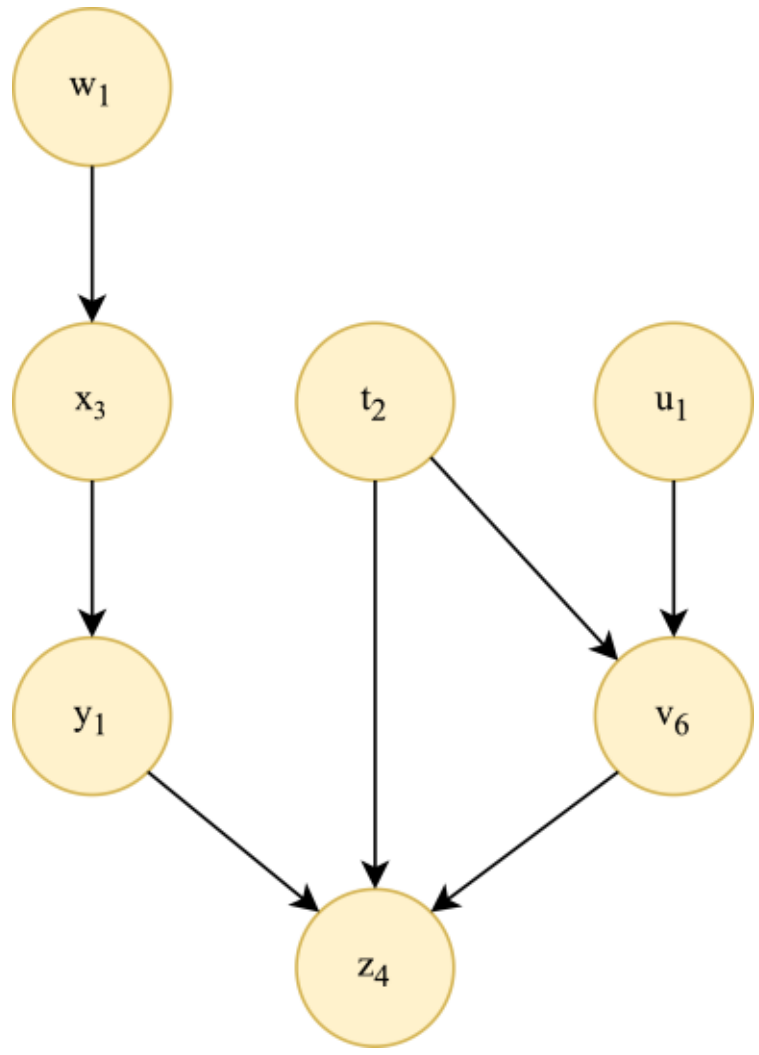

**Figure 125**: Dependency graph example in COPS.

27.3.4.3. Client Context

The client context tracks the client's causal knowledge, analogous to tracking received messages. It contains all previously read and written values within the client session.

- **Reads:** When a client performs a read, the read version for the key is added to the client context.
- **Writes:** When a client performs a write, the client provides the server with its context, ensuring the server is aware of the client's prior reads.

*27.3.5. Writing to a Key*

All write operations in COPS follow a two-phase process:

1. **Synchronous Local Write:** The write is immediately applied to the local cluster.
2. **Asynchronous Replication:** The write is subsequently propagated to all remote clusters.

Note that COPS clients perform only one write at a time; parallel **put** operations are not supported.

#### 27.3.5.1. Local Write

Internally, a **put(key, val, context)** call is translated into **put_after(key, val, [deps])**. This operation returns a boolean success indicator and assigns a version number to the key-value pair. The dependencies are derived from the client's context.

Upon receiving the **put_after** request, the server:

- Verifies that all dependent writes have been committed.
- Commits the current write.
- Assigns a version number.
- Returns **<success, version number>** to the client.

#### 27.3.5.2. Asynchronous Replication

Following the local write, the operation is streamed to remote clusters. The key owners in these clusters perform a similar dependency check using the **dep_check(key, version)** API to ensure that all dependent key-version pairs are committed before applying the write.

#### 27.3.5.3. Causal Ordering Guarantee

The write process ensures that all writes are applied in their causal order. This guarantees that if a client observes the effect of write **Y**, it will also observe the effect of any write **X** that causally precedes **Y** (**X** → **Y**).

### *27.3.6. Reading a Key*

To retrieve a single key, **get_by_version(key, version)** API is used. Setting **version** to **LATEST** fetches the most recent value. The API response includes the accessed version and a list of its dependencies.

Upon reception, the key-value pair and its associated dependencies are added to the context as dependencies.

### *27.3.7. Examples*

Let's run through some examples of how all these work.

**Example 1 (Figure 126)**

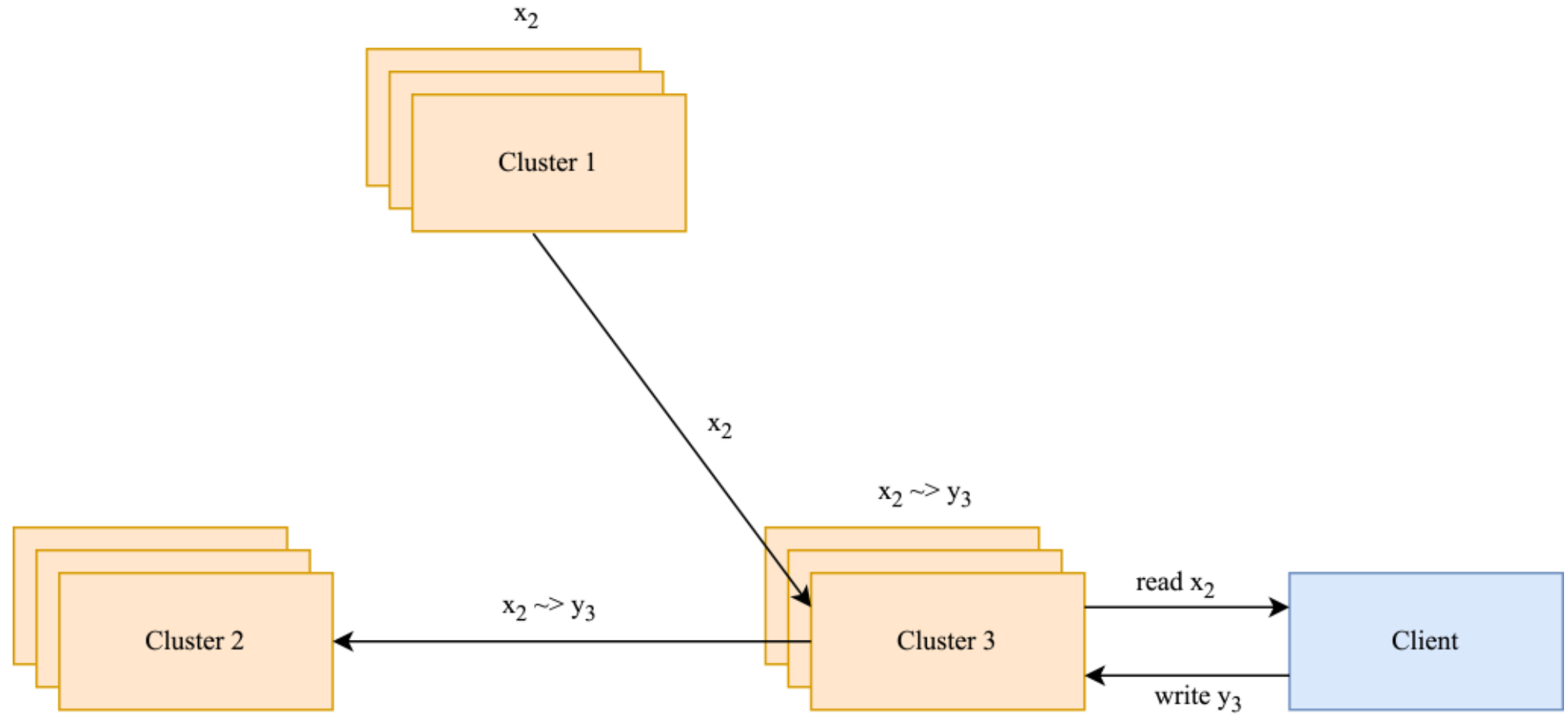

**Figure 126**: COPS example 1.

Suppose there are three clusters. A write, $\mathbf{x_2}$ (key **x**, version 2), occurs in cluster 1 and is propagated to cluster 3. A client in cluster 3 reads $\mathbf{x_2}$ and writes $\mathbf{y_3}$. The cluster forwards $\mathbf{y_3}$ to cluster 2. However, cluster 2 notices that $\mathbf{y_3}$ depends on $\mathbf{x_2}$ ($\mathbf{x_2}$ **~>** $\mathbf{y_3}$) and therefore must wait for cluster 1 to send $\mathbf{x_2}$ before $\mathbf{y_3}$ can be committed. This ensures that no clients in cluster 2 see $\mathbf{y_3}$ without seeing $\mathbf{x_2}$.

**Example 2 (Figure 127)**

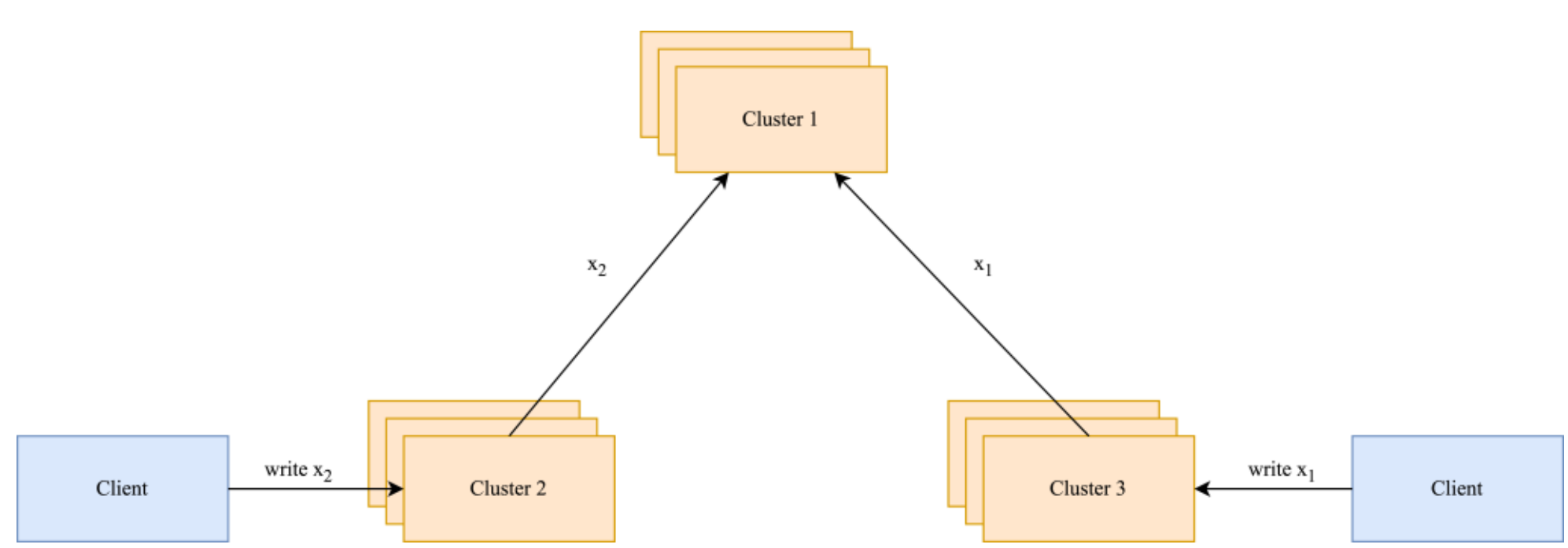

**Figure 127**: COPS example 2.

Let's consider a scenario where clients in clusters 2 and 3 simultaneously write $\mathbf{x_2}$ and $\mathbf{x_1}$, with no dependencies. Cluster 1 resolves this conflict using last-writer-wins, establishing a deterministic total order. Crucially, the outcome ($\mathbf{x_1}$ or $\mathbf{x_2}$) doesn't affect causal consistency as long as all clusters converge on the same value.

For example, say, $\mathbf{x_1}$ is the final resolved value. If $\mathbf{x_2}$ arrived first, cluster 1 clients will observe $\mathbf{x_2}$ followed by $\mathbf{x_1}$. Cluster 2 clients will experience a similar transition from $\mathbf{x_2}$ to $\mathbf{x_1}$. However, cluster 3 clients will never see $\mathbf{x_1}$, as $\mathbf{x_2}$'s value is superseded before they observe it. All clusters will converge to $\mathbf{x_1}$ in the final state.

*27.3.8. Reading Multiple Keys*

The **get_trans** addresses a critical issue: ensuring consistent multi-key reads in a system with causal dependencies. A naive approach of reading each key individually can lead to inconsistencies.

Consider keys **x** and **y**, initially at versions **i** and **j** respectively. Suppose a read is issued:

- The read for **x** retrieves $\mathbf{x_i}$.
- However before **y** is retrieved, there are two writes: write at version **k** to **x** causally followed by write at version **l** to **y** ($\mathbf{x_k \rightsquigarrow y_l}$).
- The read for **y** retrieves $\mathbf{y_l}$.

The result $\mathbf{<x_i, y_l>}$ is inconsistent. While each individual read adheres to causality, the combined snapshot violates it. A consistent read would return a snapshot like $\mathbf{<x_i, y_j>}$, $\mathbf{<x_k, y_j>}$, or $\mathbf{<x_k, y_l>}$.

The **get_trans(<keys>, context)** API resolves this through a two-phase process:

- **Phase 1**: Concurrent **get_version(key, LATEST)** calls retrieve initial versions. These may be inconsistent, but the returned versions and dependencies are used to calculate **causally corrected versions (CCVs)** for any inconsistent reads.
- **Phase 2**: **get_version(key, CCV)** is invoked for keys with incorrect versions, ensuring a consistent snapshot.

Re-examining the previous example, an initial read of $\mathbf{<x_i, y_l>}$ yields an inconsistency. However, $\mathbf{y_l}$ includes a dependency on $\mathbf{x_k}$. In Phase 2, $\mathbf{x_k}$ is retrieved, resulting in the consistent read $\mathbf{<x_k, y_l>}$.

Figure 7 in the paper details the **get_trans** algorithm, particularly the CCV generation process.

*27.3.9. Garbage Collection*

The system's garbage collection mechanism must address several key areas to maintain efficiency and consistency.

27.3.9.1. Old Versions

To support historical reads using the **get_by_version** API, old versions of data must be retained. To prevent unbounded growth, a transaction timeout (**trans_time**) of 5 seconds is implemented. Therefore, old versions need to be preserved for a duration of **trans_time** plus the maximum potential clock drift (**max_clock_drift**) within the distributed system. This ensures that reads within the timeout window can access the necessary historical data.

For example, in the dependency graph shown in Figure 128, $\mathbf{x_1}$ can be removed after **trans_time** before which there may be some **get_trans** operation trying to access it.

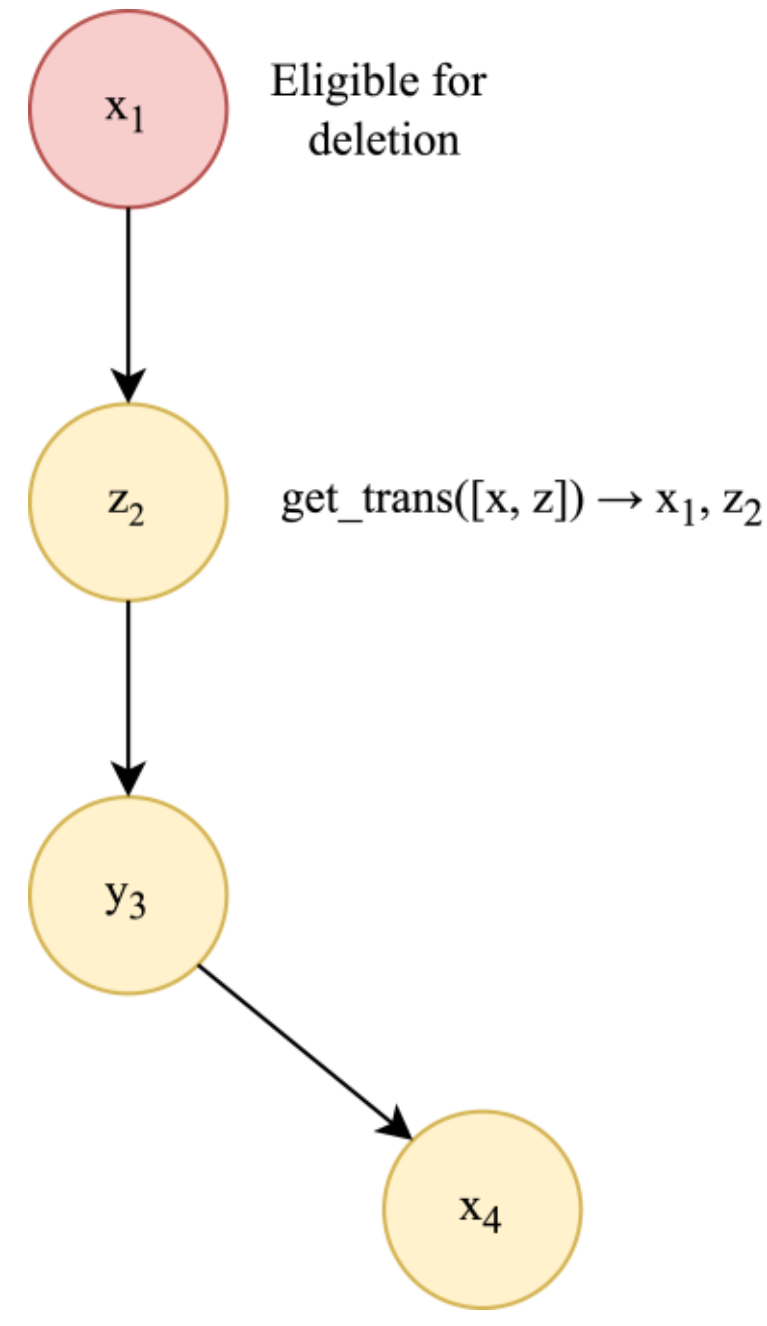

**Figure 128**: Old version garbage collection in COPS.

27.3.9.2. Dependencies

Beyond removing outdated versions, the system should also reclaim space occupied by fully replicated dependencies. A dependency, identified by a **<key, version>** pair, is considered fully replicated when it has been successfully copied to all replicas. After a designated **trans_time** period, a **never-depend** flag is applied to it, indicating it's safe for removal.

27.3.9.3. Client Metadata

Clients can safely remove fully committed dependencies from their local dependency graphs, mirroring the cluster replica cleanup process.

When a client detects a **never-depend** flag for a **<key, version>** pair (typically as a response to a **get_by_versions** request), the client can garbage collect that dependency and all its predecessors.

For example, in the dependency graph shown in Figure 129, when $\mathbf{y_1}$ is marked as never-depend, $\mathbf{w_1}$ and $\mathbf{x_3}$ are also garbage collected.

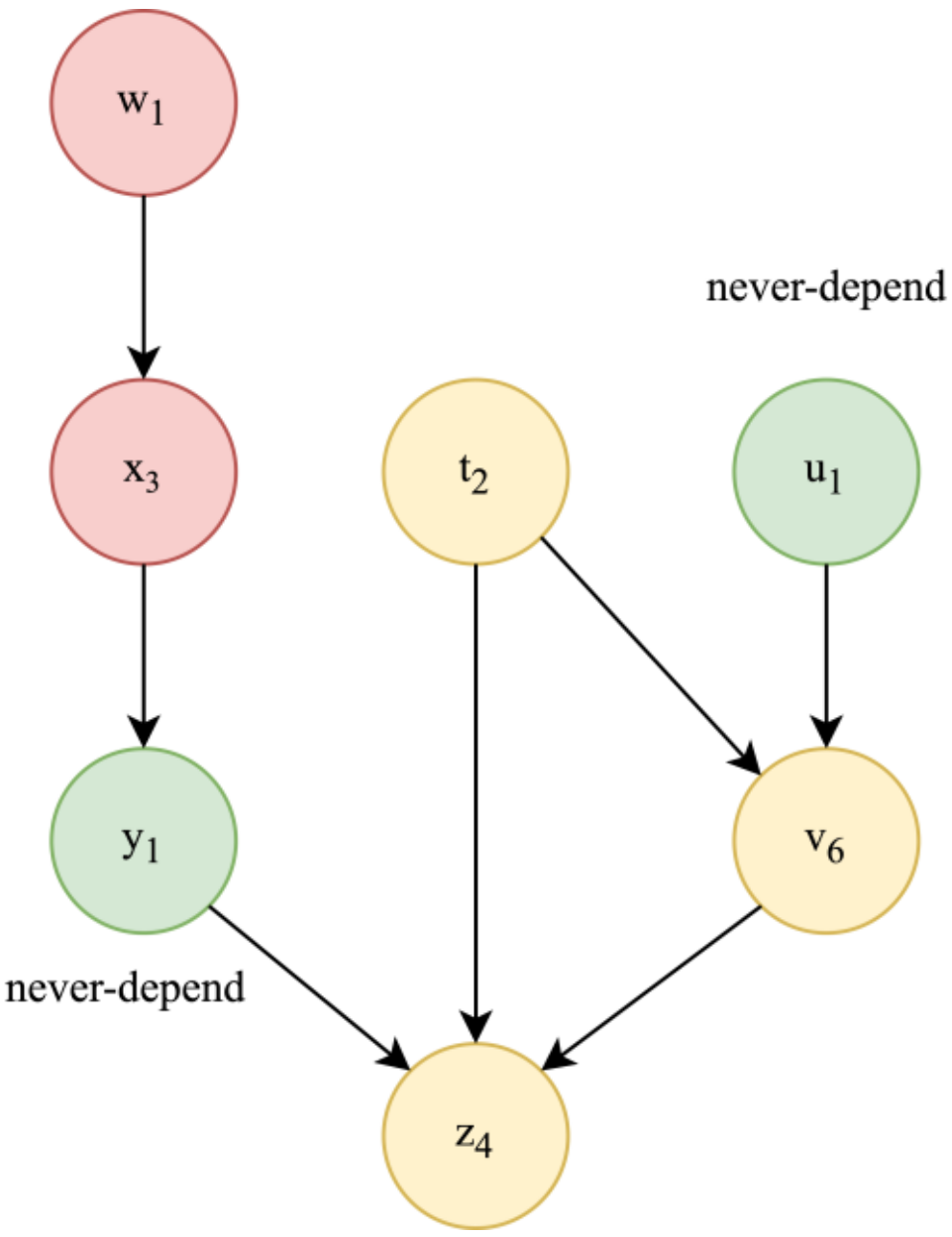

**Figure 129**: Client metadata cleanup in COPS.

Additionally we also have global checkpoint time. It represents the earliest timestamp for which at least one put_after operation is still pending. COPS nodes provide this checkpoint time to the client library in response to API calls. Clients can leverage this information to identify and clean up dependencies.

### 27.4. COPS - Conflict Detection

COPS - Conflict Detection extends the COPS family, beyond the vanilla and COPS-GT versions, by enabling client-defined conflict resolution through application-specific handlers, replacing the default last-write-wins approach.

### 27.5. Evaluation

Performance measurements at 52k gets/sec and 20k puts/sec yielded a 99.9th percentile tail latency of ~10ms for both **get** and **put** operations.

The authors observed the impact of varying the **put**-to-**get** ratio - increased **put** operations led to a decrease in throughput due to higher computational demands and a larger number of dependencies, which also negatively impacted **get** throughput.

Similarly, a higher **put** frequency resulted in reduced throughput, as the accumulation of dependencies up to the global checkpoint time increased. Conversely, a lower **put** frequency significantly reduced the number of dependencies.

### 27.6. Paper Remarks

This paper is a notoriously hard read. Its complexity is such that even understanding the final system's API requires a deep, foundational grasp of causality and distributed state. Due to this high barrier to entry and the overhead of its implementation, the paper’s direct impact on the industry has been limited. Nevertheless, for academics and distributed systems enthusiasts, it remains an essential and intellectually rewarding exercise.

# 28. TAO: Facebook's Distributed Data Store for the Social Graph

(28) Bronson, N.; Amsden, Z.; Cabrera, G.; Chakka, P.; Dimov, P.; Ding, H.; Ferris, J.; Giardullo, A.; Kulkarni, S.; Li, H.; others. TAO:Facebook's Distributed Data Store for the Social Graph. In 2013 USENIX Annual Technical Conference (USENIX ATC 13); 2013; pp 49–60.

Following our discussion of causal consistency in COPS, we now move to an eventually consistent NoSQL database for the final chapter of our database series. While SQL databases provide strong consistency and are excellent for multi-row transactions, eventually consistent data stores are preferred for their massive scalability.

This paper from Facebook was presented at the USENIX Annual Technical Conference (ATC) 2013. For understanding systems, it serves two purposes: first, it explains how eventually consistent databases are designed to scale effectively; and second, it demonstrates how a graph data structure can be layered on top of a distributed key-value store.

In this insight, we will review the various consistency models discussed throughout this book, using specific examples to understand each model in greater detail. We will then dive deep into TAO, exploring its graph API, how it handles reads and writes, and the specific consistency guarantees it provides.

### 28.1. Consistency Models Revisited

We've previously explored linearizability and sequential consistency in ZooKeeper, as well as causal consistency in COPS. In this discussion, we'll examine eventual consistency models and provide an illustrative example to summarize them, within the context of single-key distributed key-value stores.

The phrase "consistency model" in the context of a single data item read-write system means the following:

- Say clients are observing **all** the writes happening in the store. A consistency model determines the order in which the clients see the writes.
- After all the writes are applied, the consistency model determines what would be the effective state of the objects in the store.

The consistency model hierarchy is:

$$\textbf{Linearizable} > \textbf{Sequential} > \textbf{Causal} > \textbf{PRAM} > \textbf{Read} - \textbf{Your} - \textbf{Writes}\ (\textbf{RYW}) > \textbf{Eventual}$$

*28.1.1. Read-Your-Writes (RYW)*

This model guarantees that a client can observe the effect of its own writes.

*28.1.2. Example*

Consider the following example shown in Figure 130, with the initial state being, **{X: 0, Y: 0}**. The real-time ordering of events are:

1. **W(X, 1) or $\mathbf{W_1}$** - Client 1 sets key **X** to value 1.
2. **W(X, 2)** or $\mathbf{W_2}$ - Client 2 sets key **X** to value 2.
3. **W(Y, 3)** or $\mathbf{W_3}$ - Client 1 sets key **Y** to value 3.
4. **W(X, 4)** or $\mathbf{W_4}$ - Client 2 sets key **X** to value 4.

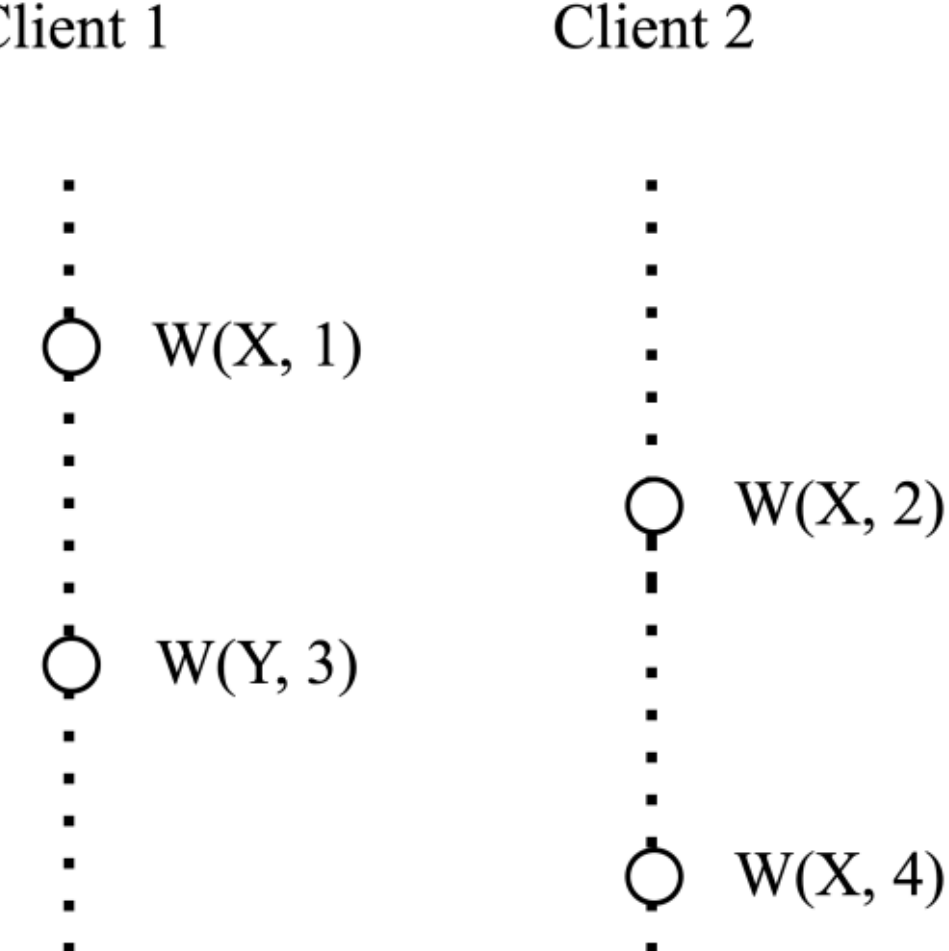

**Figure 130**: Consistency model example.

The causal relationships are:

1. $\mathbf{W_1} \rightarrow \mathbf{W_3}$ - events within the same system.
2. $\mathbf{W_2} \rightarrow \mathbf{W_4}$ - events within the same system.

Now we can define each consistency model.

28.1.2.1. Linearizability

- Order of writes: $\mathbf{W_1}$, $\mathbf{W_2}$, $\mathbf{W_3}$, $\mathbf{W_4}$. All clients will observe this and only this order.
- The final effective state will be **{X: 4, Y: 3}**.

28.1.2.2. Sequential Consistency

- Possible orders are all 4! permutations; **only one** of these orders will be chosen, and all clients will observe that same order:
  - $\mathbf{W_1, W_2, W_3, W_4}$
  - $\mathbf{W_1, W_2, W_4, W_3}$
  - $\mathbf{W_1, W_3, W_2, W_4}$
  - and so on.
- The final effective state will be either **{X: 1, Y: 3}** or **{X: 2, Y: 3}** or **{X: 4, Y: 3}** based on which total order was chosen.

**Note:** Up to the point of sequential consistency, the system maintains a deterministic order of writes. There will only be one order and hence only one final state. Beyond this, multiple values for the same key can exist without violating the consistency model. When such multiple values exist, the system can choose to present all the known values to the clients for the clients to decide.

28.1.2.3. Causal Consistency

- Possible orders are listed below (all orders in which $\mathbf{W_1}$ appears before $\mathbf{W_3}$ and $\mathbf{W_2}$ appears before $\mathbf{W_4}$); clients can observe any of these orders, and different clients may observe different orders:
  - $\mathbf{W_1, W_2, W_3, W_4}$
  - $\mathbf{W_1, W_2, W_4, W_3}$
  - $\mathbf{W_1, W_3, W_2, W_4}$
  - $\mathbf{W_2, W_1, W_3, W_4}$
  - $\mathbf{W_2, W_1, W_4, W_3}$
  - $\mathbf{W_2, W_4, W_1, W_3}$
- The final effective states will be **{X: 1, Y: 3}** and **{X: 4, Y: 3}**. Both are valid and the system can return different values for **X** without affecting consistency.

28.1.2.4. PRAM (or FIFO) Consistency

- Possible orders are listed below (all orders in which $\mathbf{W_1}$ appears before $\mathbf{W_3}$ and $\mathbf{W_2}$ appears before $\mathbf{W_4}$); clients can observe any of these orders, and different clients may observe different orders:
  - $\mathbf{W_1, W_2, W_3, W_4}$
  - $\mathbf{W_1, W_2, W_4, W_3}$
  - $\mathbf{W_1, W_3, W_2, W_4}$
  - $\mathbf{W_2, W_1, W_3, W_4}$
  - $\mathbf{W_2, W_1, W_4, W_3}$
  - $\mathbf{W_2, W_4, W_1, W_3}$
- The final effective states will be **{X: 1, Y: 3}** and **{X: 4, Y: 3}**. Both are valid and the system can return different values for **X** without affecting consistency.

For this example, PRAM consistency appears to be the same as causal consistency. To distinguish them, let's say there was a **R(X)** before $\mathbf{W_2}$ on Node 2, which fetched value 1, as shown in Figure 131. Then we have additional causal orders - $\mathbf{W_1}$ → **R(X)** and **R(X)** → $\mathbf{W_2}$. The final effective state will be **{X: 4, Y: 3}** for causal consistency.

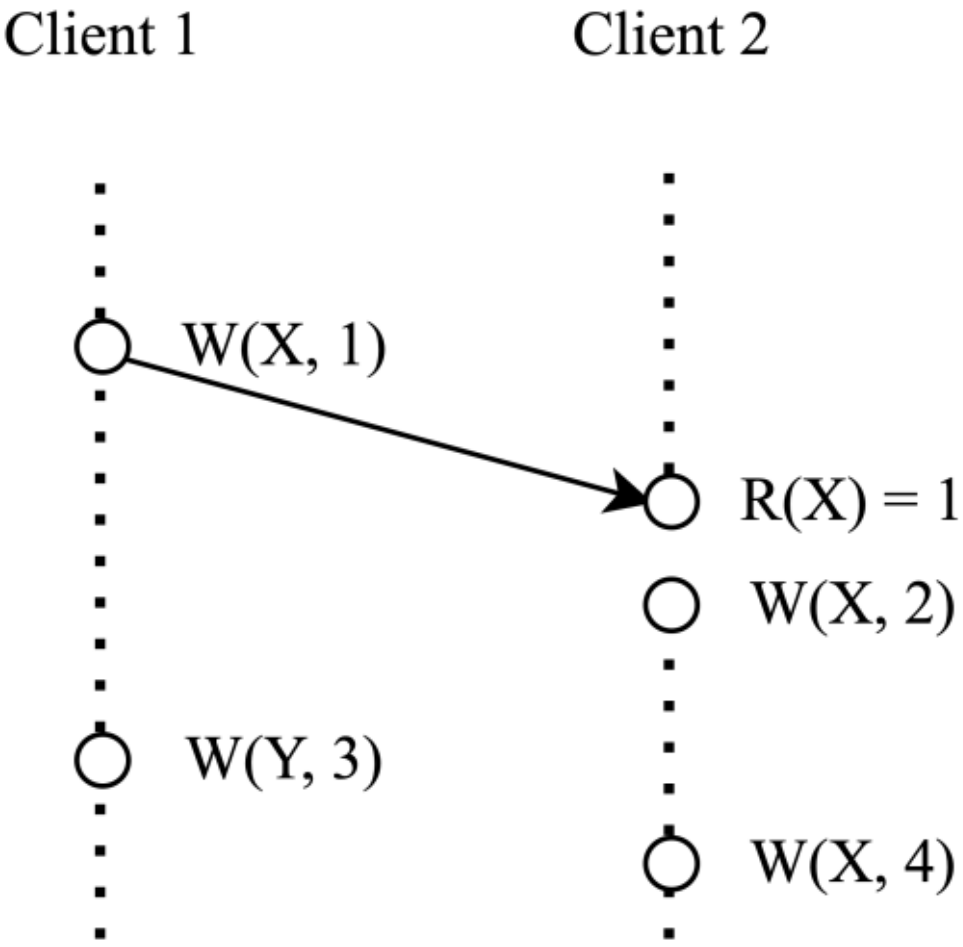

**Figure 131**: Causal v/s PRAM consistency example.

For PRAM consistency, however, all orders mentioned above would still be valid.

#### 28.1.2.5. RYW Consistency

This is where things start to differ slightly.

- Possible orders for all clients except client 1 and 2 are all possible 4! permutations; any ordering is fine.
- Possible orders for client 1 are those in which $\mathbf{W_1}$ is ordered before $\mathbf{W_3}$; client 1 will observe one of them.
- Similarly, possible orders for client 2 are those in which $\mathbf{W_2}$ is ordered before $\mathbf{W_4}$; client 2 will observe one of them.
- The final effective states for all other clients will be **{X: 1, Y: 3}**, **{X: 2, Y: 3}**, and **{X: 4, Y: 3}**. Different replicas in the data store may have different values, and that is completely fine.
- The final effective states for client 1 will be **{X: 1, Y: 3}**, **{X: 2, Y: 3}**, or **{X: 4, Y: 3}**; any one based on the order that it observed.
- The final effective states for client 2 will be **{X: 1, Y: 3}** or **{X: 4, Y: 3}**; any one based on the order that it observed.

#### 28.1.2.6. Eventual Consistency

- All orders are valid. They can be different for all the clients.
- The final effective states will be **{X: 1, Y: 3}**, **{X: 2, Y: 3}**, and **{X: 4, Y: 3}**.

### *28.1.3. Conflict Handling*

In all the examples above, we discussed scenarios involving causal, PRAM, RYW, and eventual consistency, where multiple values for the same key could exist. However, systems designed with these consistency models, such as Dynamo and COPS, typically implement a **conflict handler**. This conflict handler is associative and commutative, and helps converge multiple values to a single, consistent value.

In practical applications, this convergence process usually occurs very rapidly, resulting in only one effective and valid state being retained.

## 28.2. Read-After-Write Consistency

**Read-After-Write (RAW)** consistency primarily originated as an industry term used by companies like Amazon and Facebook.

It ensures that reads from any client after any write operation return the written value or a later version. This is essentially equivalent to linearizability, as it guarantees immediate visibility of a committed write across all clients.

However, in reality, **this is not linearizability**. It is just a descriptive way of stating a common technique used in industry to make writes visible to readers. To give a spoiler, the readers are redirected to the same replica where the writer performed the write. Thus, it is not exactly linearizability, in which all replicas would reflect the write. We will explore the implications of RAW consistency in the context of TAO shortly.

## 28.3. TAO

TAO is a graph database and widely deployed at Meta (formerly Facebook). It is distributed and offers eventual consistency. It is again a NoSQL database and is optimized for availability and performance over consistency.

## 28.4. Graph Model

In TAO, the graph structure is represented using **objects** (nodes) and **associations** (edges). Both objects and associations are fundamentally stored as key-value pairs, effectively building the graph model on a key-value abstraction.

### *28.4.1. Objects*

- The key is the object's unique identifier (**id**).
- The value is a structure containing the object type (**otype**) and a set of key-value pairs representing the object's attributes: **<otype, (key → value)>**. The **otype** determines how the internal key-value pairs within the value are interpreted.

### *28.4.2. Associations*

- The key is composed of the source object identifier (**id1**), the association type (**atype**), and the destination object identifier (**id2**): **<id1, atype, id2>**.
- The value includes a timestamp (**time**) and a set of key-value pairs representing the association's attributes: **<time, (key → value)>**. The **atype** allows for defining multiple relationship types between objects.

All objects and associations are stored as independent key-value pairs, with the interpretation of the value depending on the **otype** or **atype**. The system utilizes a limited set of object and association types.

The **time** included in the association value facilitates time-based sorting, a crucial feature for social media platforms like Facebook.

### 28.5. API

The TAO API, while presented as a graph interface, ultimately operates on the underlying key-value store.

#### *28.5.1. Write Operations*

- **association_add(id1, atype, id2, time, (k→v)*)**: Adds an association of type **atype** between objects **id1** and **id2**.
- **association_delete(id1, atype, id2)**: Deletes the specified association, which corresponds to deleting the key **<id1, atype, id2>**.
- **association_change_type(id1, atype, id2, newtype)**: Modifies the association type. This involves deleting the existing key **<id1, atype, id2>** and adding a new key **<id1, newtype, id2>** in a single operation.

#### *28.5.2. Read Operations*

- **assoc_get(id1, atype, id2set, high?, low?)**: Retrieves all associations of type **atype** between **id1** and objects within the **id2set**. This requires retrieving multiple key-value pairs from the database. Optionally, the client can specify the time range (**low** to **high**) for the associations.
- **assoc_count(id1, atype)**: Returns the number of associations of type **atype** originating from **id1**.
- **assoc_range(id1, atype, pos, limit)**: Returns a range of associations of type **atype** originating from **id1**, starting at position **pos** (when sorted temporally), and returning a maximum of limit associations.
- **assoc_time_range(id1, atype, high, low, limit)**: Returns associations of type **atype** originating from **id1**, within a specified time range (**low** to **high**), returning a maximum of **limit** associations.

## 28.6. Architecture

TAO's architecture is complex, reflecting its diverse range of use cases and the system's evolutionary development. This complexity is managed through a layered design and strategic sharding.

### *28.6.1. Sharding*

The entire dataset is divided into logical shards for distribution and scalability. The sharding is based on the object identifier. All associations from an object (**id1**) are part of the same shard.

### *28.6.2. Regional Deployment and Replication*

TAO is deployed across multiple regions, each acting as a leader for a subset of the logical shards, as shown in Figure 132.

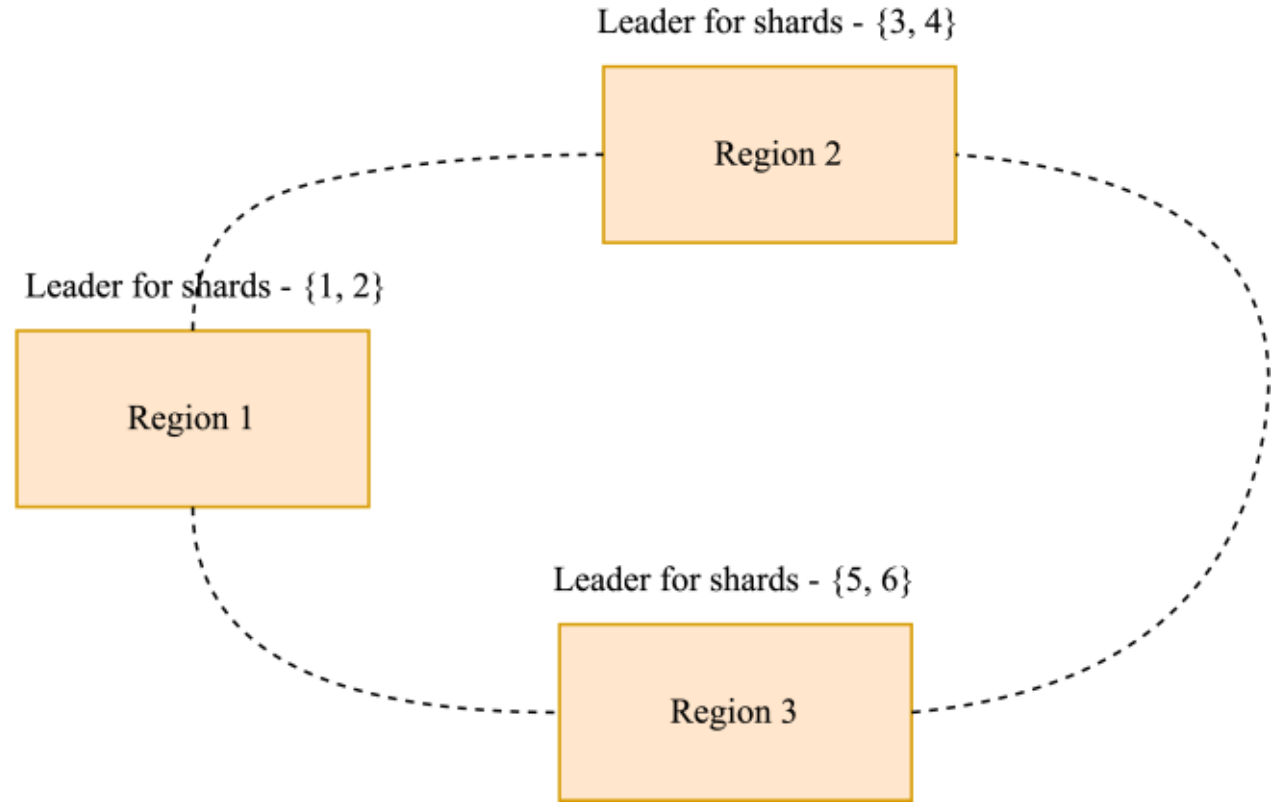

**Figure 132**: Multi-regional deployment in TAO.

Notably, each region stores all shards, regardless of its leadership role. This ensures data availability even if a region becomes temporarily unavailable.

Each region comprises the two layers - **storage layer** and **cache layer**.

### *28.6.3. Storage Layer (MySQL Database)*

All objects and associations are persisted in a sharded MySQL database. The API operations are translated into corresponding SQL queries. The storage layer within each region maintains all logical shards, with objects and associations stored in separate tables.

All objects are stored in one table, and all associations in another. The tables act as key-value stores where the value is serialized into a single column. The association table has an additional index on **<id1, atype, time>** to support range queries.

*28.6.4. Cache Layer*

This layer caches objects, association lists, and association counts, employing an LRU eviction policy. The cache is not simply a passive storage; it understands the semantics of the stored key-value pairs and can execute application logic.

The cache layer is organized hierarchically, as shown in Figure 133:

- **Follower Tier:** Client interactions occur at this tier, with multiple follower tiers per region.
- **Leader Tier:** This tier communicates directly with the storage layer.

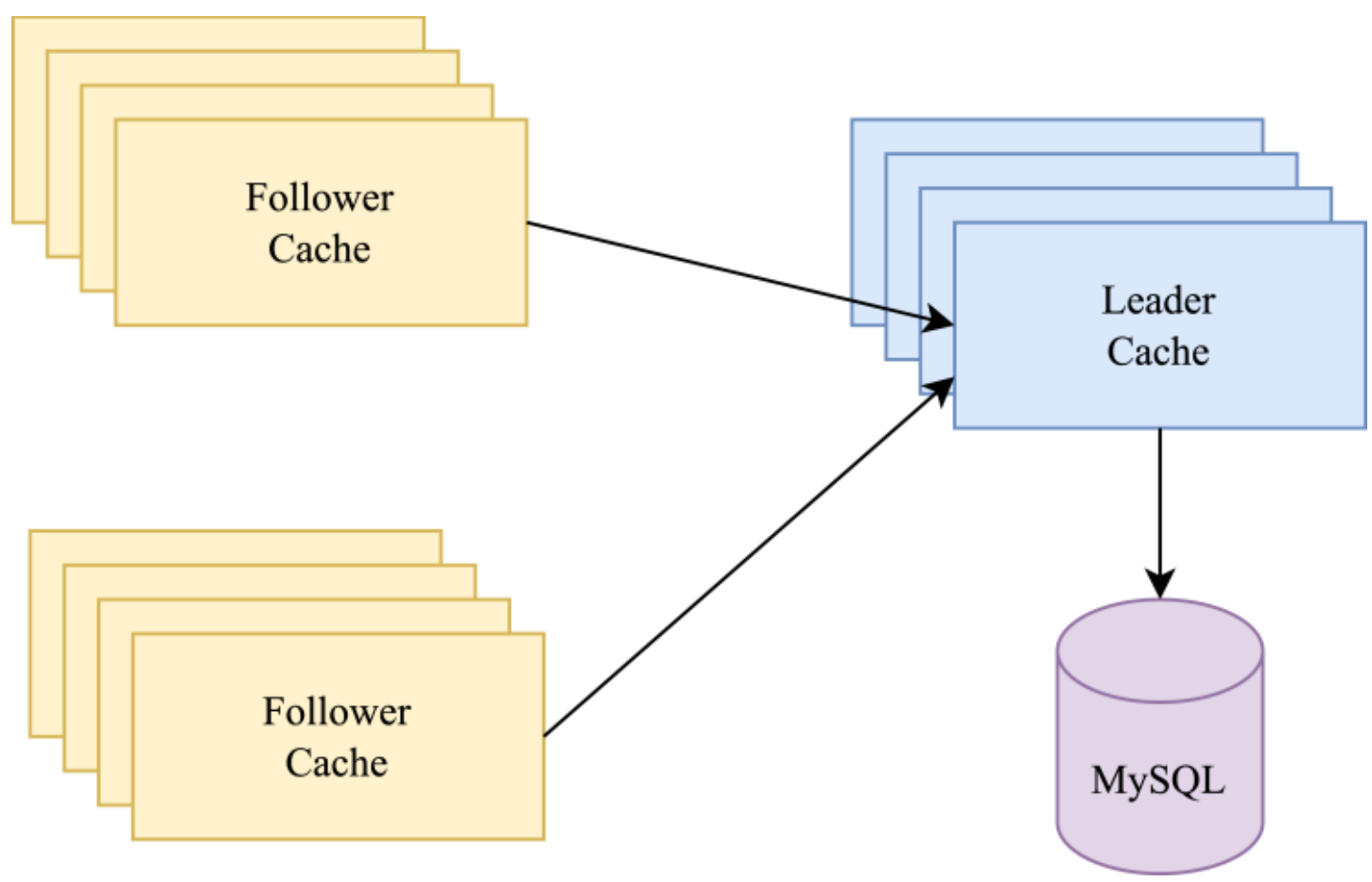

**Figure 133**: Multi-layer cache in TAO.

Each tier (follower and leader) maintains all shards. Consistent hashing distributes shards across servers within a tier. To manage high load on specific shards, those shards may be cloned and assigned to multiple servers.

Within each cache tier, TAO utilizes a slab allocator to manage memory. RAM is divided into arenas, each dedicated to a specific object or association type. This partitioning provides isolation for LRU eviction policies, ensuring that different data types do not interfere with each other. Further optimizations exist for small, fixed-size items like association counts.

The cache is responsible for storing the following data:

1. **Objects:** Objects, identified by their unique ID (<**id**>).
2. **Association Counts:** The number of associations for each combination of object identifier and association type (<**id, atype**>). This is crucial for efficiently executing **assoc_count** queries.
3. **Association Lists:** Lists of associations for each **<id, atype>** combination, ordered by time and typically limited to 6,000 entries. These lists are used to answer range queries. Queries beyond limit go to the database.

Note that these cached items directly reflect TAO's graph model. However, the underlying database operates as a simple key-value store.

#### *28.6.5. Scalability*

TAO's deployment model facilitates high scalability. Read operations are primarily handled by the local follower tier cache. Cache misses trigger queries to the region's storage layer (MySQL).

Write operations (described next) are propagated asynchronously to other regions.

### 28.7. Write Propagation

Write operations are propagated synchronously from the follower cache tier to the leader cache tier, which then updates the database in the storage layer.

Upon database write, a **changeset** is generated, encapsulating the modifications. For example, deleting an association results in a changeset that:

- Removes the association from the corresponding association list.
- Decrements the association count for the object and association type.

The changeset is applied synchronously along the path: **client → follower cache tier → leader cache tier → leader database**. However, updates to other cache tiers and databases are performed asynchronously, using an invalidation-based approach.

#### *28.7.1. Leader Cache Tier Invalidation*

After applying the update, the leader cache tier sends **invalidation messages** to all follower cache tiers within its region for the affected object. It also sends **refill messages** for association lists.

Importantly, a leader cache tier only sends invalidation messages within its own region.

### 28.7.2. Database Replication and Inter-Regional Invalidation

The database replicates the updates to non-leader regions. The database then sends invalidation messages to the leader cache tiers of those non-leader regions. These non-leader leader cache tiers then forward the invalidation messages to their respective non-leader follower cache tiers.

### 28.7.3. Example

Let's run through the paper's example to understand how writes work.

In paper’s Figure 2, solid lines represent data flow, while dotted lines indicate control messages, specifically invalidations and refills. Consider a write operation initiated by a client in a non-leader region.

The write request goes from the non-leader follower cache to the non-leader region's leader cache, and then to the leader cache. Along this path, the write is applied synchronously to each component. The leader cache performs the write to the leader database.

Following the database update, the leader cache sends invalidation messages to all follower caches within its own (leader) region.

Subsequently, the leader database replicates the write to the non-leader region's database. The leader database then sends an invalidation message to the leader cache within the non-leader region. This non-leader leader cache, in turn, propagates invalidation messages to all follower caches within its region.

**Note**: In the example above, the non-leader region's leader cache receives invalidation messages twice.

## 28.8. Consistency Model

While TAO is often described as eventually consistent, a deeper examination reveals a more nuanced consistency model.

### *28.8.1. Per-Key Sequential Consistency*

TAO leverages MySQL as its underlying storage. Each key-value pair is managed by a single MySQL database instance, which acts as the leader for that data. Although data is replicated across regions, a single leader is designated for each key-value pair at any given time. This design enables the serialization of writes for a specific key-value pair.

Because a single MySQL process acts as the leader for all operations on an object and its associations, atomic operations, such as *assoc_change_type*, can be implemented using SQL queries.

Additionally, every write is tagged with a version number. Follower replicas use these version numbers to maintain the same write order as the database, ensuring consistent observation of updates. Unlike Dynamo and COPS, TAO's automatic conflict resolution eliminates the need for explicit conflict handlers.

**Q. Is TAO Sequentially Consistent?**

NO! While writes for a *single* key-value pair are serialized, TAO does not guarantee sequential consistency across *all* key-value pairs. The single leader approach prevents conflicts within a specific key-value pair, but clients may observe updates to different keys in varying orders.

### *28.8.2. Read-Write Transactions*

TAO does not support traditional read-write transactions. This aligns with systems like Dynamo and COPS, which prioritize availability over strong consistency. All writes are blind-writes.

### *28.8.3. RAW Consistency*

Within a single cache tier (follower tier), TAO provides RAW consistency. When a write is successfully applied to the database, the local cache tier is updated. However, this RAW consistency is not guaranteed across different cache tiers. A client querying a different tier immediately after a write may not see the updated value.

## 28.9. Fault Tolerance

Given TAO's effective reliance on eventual consistency, its fault tolerance model is designed for simplicity and resilience.

### *28.9.1. Database Failures*

- **Leader Database Failure:** A non-leader database takes over as the new leader.
- **Non-Leader Database Failure:** Read requests are redirected to the leader database.

### *28.9.2. Cache Failures*

Cache invalidation messages are treated as best-effort deliveries. The system's consistency model is not compromised if these messages are lost; updates will eventually propagate. TAO makes best-effort delivery of invalidation messages to maintain cache consistency as much as possible.

- **Leader Cache Tier Failure:**
    - Follower cache misses are redirected to the database directly.
    - Write requests are redirected to another available leader cache tier replica, which then enqueues invalidation messages for the original leader.
- **Invalidation Failure:** The leader cache tier enqueues messages and delivers them when the follower becomes reachable.
- **Follower Cache Tier Failure:** Client requests are redirected to another follower cache tier, which results in the loss of RAW consistency for the client.

## 28.10. Evaluation

- TAO demonstrates remarkably high availability in production environments, reporting a 99.999% uptime.
- A single follower cache tier is capable of handling up to 600k QPS during peak hit rates. However, throughput decreases as the hit rate declines, due to expensive downstream data fetches for cache misses.
- The overall cache hit rate is 96.4%.

- Write latency within a single region is 12.1ms. For remote region writes, the latency increases to 74.4ms, with 58.1ms attributed to round-trip network latency.
- Despite its eventual consistency model, TAO exhibits low replication delays:
    - p85: 1s
    - p99: 3s
    - p999: 10s

### 28.11. Paper Remarks

This paper is a great read, offering a clear API specification and numerous practical lessons. A notable contrast exists between this industry paper and academic works like the COPS paper. This difference arises from their distinct motivations: academic papers prioritize theoretical excellence, while industry papers focus on delivering practical solutions. The TAO paper exemplifies this industry-driven approach, highlighting the real-world considerations that shape large-scale system design.

# 29. Mesos: A Platform for Fine-Grained Resource Sharing in the Data Center

(29) Hindman, B.; Konwinski, A.; Zaharia, M.; Ghodsi, A.; Joseph, A. D.; Katz, R.; Shenker, S.; Stoica, I. Mesos: A Platform for Fine-Grained Resource Sharing in the Data Center. In 8th USENIX Symposium on Networked Systems Design and Implementation (NSDI 11); 2011.

Data centers are expensive to set up and maintain. There are hundreds of machines organized as clusters. To optimize cost, data center resources must be shared across multiple workloads. Managing and sharing these resources effectively requires a data center-level operating system - also known as a cluster orchestrator - which is the central topic of this paper.

This paper, presented at Networked Systems Design and Implementation (NSDI) in 2011, comes from the UC Berkeley Systems Lab, with authorship by influential figures like Matei Zaharia (Spark's creator and CTO of Databricks [202]), Ali Ghodsi (CEO of Databricks), Scott Shenker, and Ion Stoica. UC Berkeley's Systems Lab is a powerhouse in computer systems research.

The concepts explored in these papers are closely intertwined with the development of several influential projects at UC Berkeley, including Spark [89], Delay Scheduling [203], Dominant Resource Fairness [204], and the technical report [205] detailing Mesos – all of which were being actively researched and built concurrently by the same authors.

In this insight, we will start by exploring cluster computing in detail, navigating through the various applications that typically run in a data center and discussing the vital role of a cluster orchestrator. This will be followed by a deep dive into resource allocation and scheduling strategies. Next, we will examine the differences between Hadoop (a MapReduce implementation) and Spark, , both of which are prominent frameworks for large-scale batch processing. We will also look into the Message Passing Interface, a framework used for large-scale scientific computations. Finally, we will analyze Mesos, discussing its architecture and how it manages a broad range of concurrent applications. We will walkthrough the detailed behavioral analysis of how Mesos handles various real-world scenarios, as presented by the authors.

## 29.1. Cluster Computing

A computing **cluster** comprises numerous interconnected machines, linked by a high-speed network, often referred to as a **fabric**. These clusters typically range from 100 to 10,000 machines. Users schedule their jobs on these machines.

### *29.1.1. Cluster Jobs*

Clusters are versatile platforms capable of executing a wide array of jobs such as **services**, **batch-processing jobs**, and **scientific computations**.

#### 29.1.1.1. Services

Users frequently deploy long-running services on clusters for high availability and scalability through replication across multiple machines. This is shown in Figure 134. Numerous essential systems, such as distributed file systems, lock services, and distributed databases, operate as a set of smaller, interconnected services known as **micro-services**. Each service is a set of replicated processes (called **tasks)** running on the cluster. Services represent the most common type of jobs scheduled by users in clusters hosted by cloud providers.

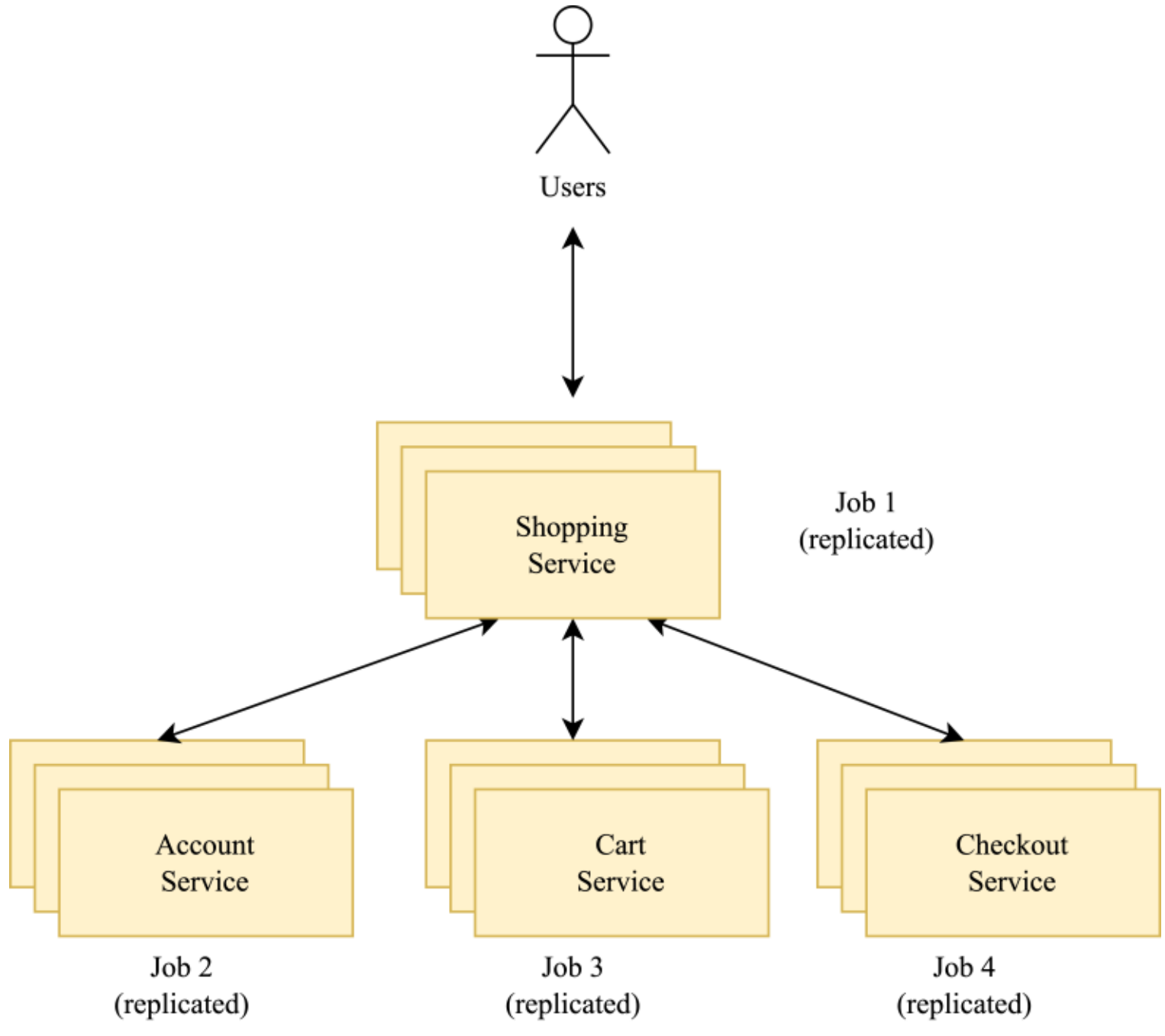

**Figure 134**: Microservices for shopping website.

#### 29.1.1.2. Batch Processing

The rise of Big Data has necessitated distributed processing as datasets outgrow the capacity of single machines.

Consequently, most modern data processing applications are executed across machines within a cluster. Typically, specialized frameworks (such as Hadoop [97]) run as services on these clusters, accepting data processing jobs as input and orchestrating their execution.

The paradigm of processing large datasets in parallel through phases was established with the advent of MapReduce, as shown in Figure 135. In this model, a user-submitted **data processing job** is decomposed into a series of smaller processing phases. Each of these phases is executed as a distinct **cluster job** (it's important to distinguish this "cluster job" from the higher-level "data processing job" submitted by the user). Furthermore, each job is broken down into individual **tasks**, with each task running as a separate process on the cluster. A task is responsible for processing a specific slice of the overall data.

**Note**: In Figure 135, **mappers** and **reducers** may run as a single job called **workers** consisting of a mix of map and reduce tasks.

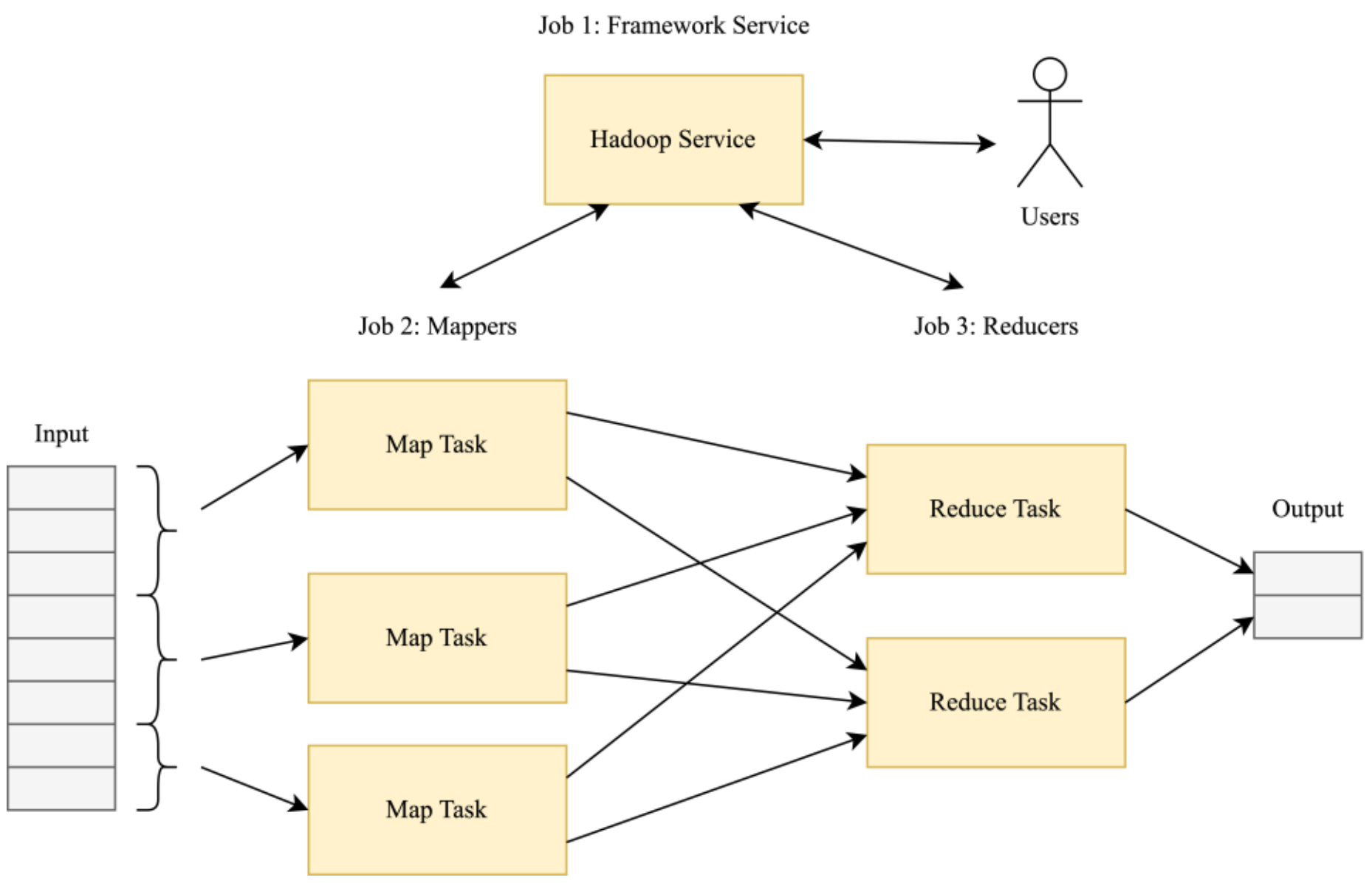

**Figure 135**: Map-Reduce batch processing jobs.

#### 29.1.1.3. Scientific Computations

Clusters are indispensable for tackling complex scientific problems, for instance, climate modeling, drug discovery, computational fluid dynamics, and astrophysics simulations. Clusters specifically designed for scientific computations are often High-Performance Computing (HPC) clusters, equipped with enhanced hardware like GPUs, and high-speed, low-latency networking technologies such as Remote Direct Memory Access (RDMA) and RDMA over Converged Ethernet (RoCE).

These HPC clusters are prevalent in academic research environments.

### *29.1.2. Containers*

The fundamental requirement of tasks running in a cluster is the ability to run on any machine. To ensure tasks can run on any machine within the cluster, it's crucial that they include all necessary dependencies. This is commonly achieved through containerization using technologies like Docker [48].

### *29.1.3. Cluster Orchestrator*

Clusters represent a powerful computing infrastructure, demanding significant investment for setup. Even tech giants like Google and Microsoft typically operate a limited number of clusters. Given the diverse range of jobs that need to be executed, efficient sharing of cluster resources among multiple users is important.

The finite machine resources within a cluster necessitate a mechanism for fair and effective allocation across different users. One initial approach, **static partitioning**, i.e., dividing machines among users, proves to be inflexible and frequently leads to underutilization of resources.

Consequently, a **cluster orchestrator** is essential for efficient cluster management. This orchestrator performs several key functions, as shown in Figure 136:

1. Task scheduling based on user requests.
2. Task status tracking and reporting to users.
3. Machine maintenance, including operating system upgrades and software management.
4. Resource usage monitoring and reporting.

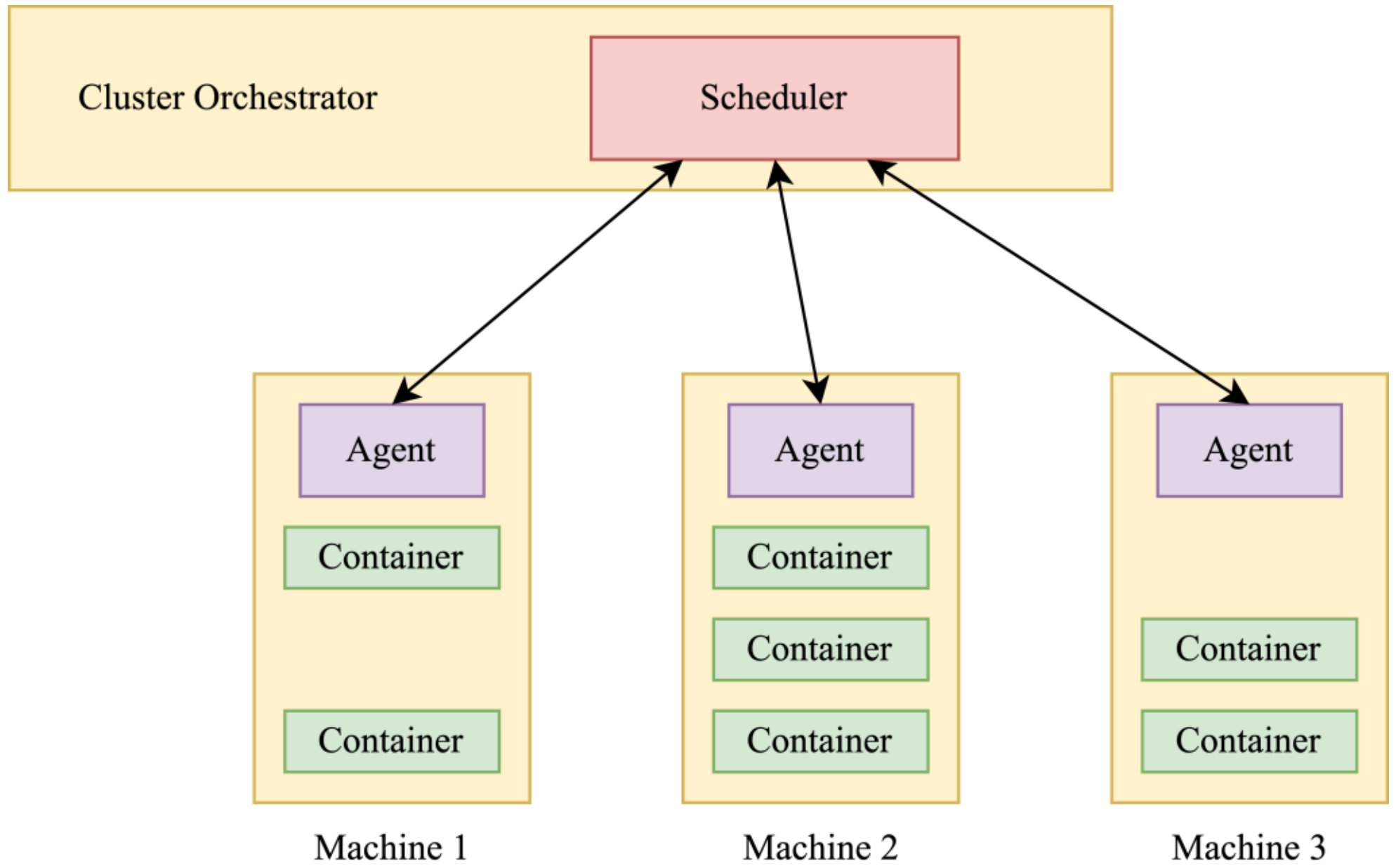

**Figure 136**: Cluster orchestrator.

29.1.3.1. Examples

Several sophisticated cluster orchestrators have been developed to manage and streamline resource allocation in large-scale computing environments. Examples include:

- **Borg** [206]**:** Google's internal cluster management system, designed to handle a vast array of workloads across its data centers. It's highly scalable and efficient, managing diverse applications, from long-running services to batch-processing jobs. Borg's design heavily influenced the development of Kubernetes.
- **Kubernetes (K8s)** [207]**:** An open-source container orchestration system, originally designed by Google, that automates the deployment, scaling, and management of containerized applications. It has become the industry standard for container orchestration. Kubernetes is especially popular for managing Docker containers, but it can also handle other container runtimes.

- **Docker Swarm:** Docker's native clustering and orchestration solution. It allows users to create and manage a cluster of Docker nodes, enabling the deployment and scaling of containerized applications.
- **Omega** [208]**:** A general-purpose cluster management system developed by Google, designed to address some of the limitations of Borg.
- **HashiCorp Nomad** [209]**:** A simple orchestrator that can deploy and manage both containerized and non-containerized applications.

29.1.3.2. Scheduler

The **scheduler** is the core component of an orchestrator, responsible for assigning tasks to available machines.

There are several challenges that cluster schedulers need to solve:

- **Colocation Interference:** Occurs when multiple tasks running on the same machine interfere with each other's performance, leading to unpredictable behavior.
- **Noisy Neighbors:** A specific type of colocation interference where one resource-intensive task negatively impacts the performance of other tasks sharing the same hardware. This is a widely-known problem faced by several programs running on clusters. See "The Tail at Scale" [210] by Jeff Dean.
- **Machine Heterogeneity:** Dealing with clusters composed of machines with varying hardware specifications and capabilities, making resource allocation and scheduling complex.

These issues have historically limited the adoption rate of cluster orchestrators in general. Notably, only Google's Borg has achieved exceptional machine utilization rates, ranging from 60% to 80%.

29.1.3.2.1. Types of Cluster Schedulers

Cluster schedulers come in various architectures:

**Monolithic Scheduler:** Example: Google's Borg. Monolithic schedulers have a single component handling all task placement decisions. Drawbacks include inflexible policies and rapid code growth, making maintenance challenging.

**<u>Two-Level Scheduler</u>:** Example: Apache Mesos. This model separates concerns into two distinct levels, as shown in Figure 137:

- The **schedulers**, responsible for making the actual task placement decisions.
- The **resource manager**, which manages cluster resources and offers them to the schedulers.

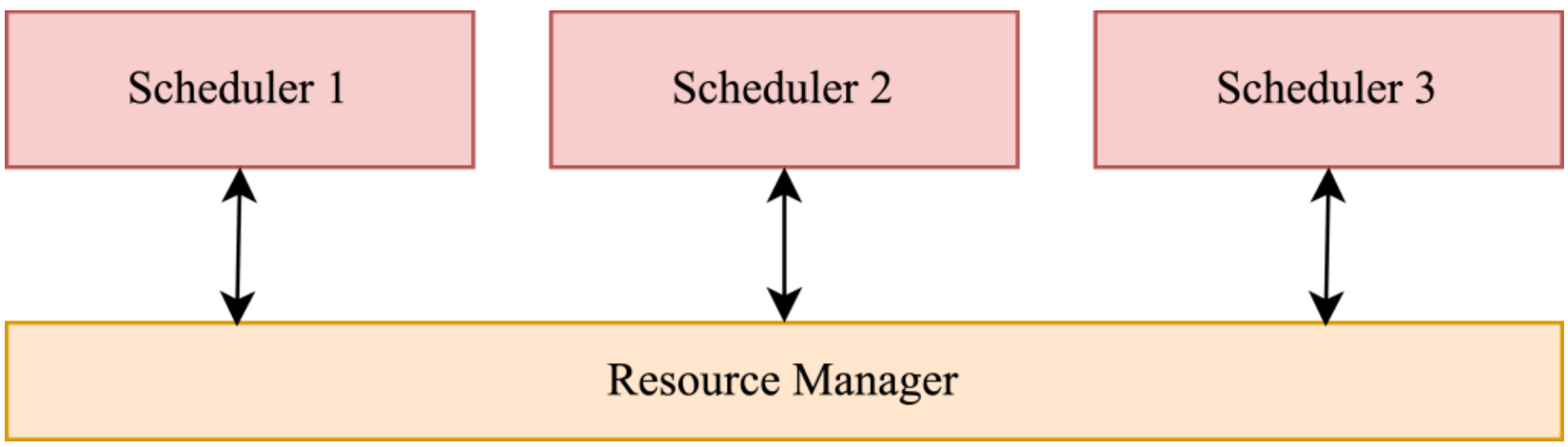

**Figure 137**: Two-level scheduler.

**<u>Shared-State Scheduler</u>:** Examples: Google Omega, HashiCorp Nomad. This model evolved from the two-level architecture to address its limitations. A key issue with two-level schedulers is that individual schedulers may have a limited view of the cluster, hindering optimal resource allocation. The shared-state model utilizes optimistic concurrency, where the cluster's state is modified through transactions. While this approach improves flexibility, it introduces the possibility of scheduling conflicts.

**Misc:**

- **Fully Distributed Scheduler:** Often employs simple random placement algorithms. Results in suboptimal placement and is primarily of academic interest.
- **Hybrid Placement:** Schedulers that combine aspects of the other scheduler types to try to achieve the best result.

#### 29.1.3.3. Agent

Apart from the scheduler, a cluster orchestrator has an agent running on each machine which receives tasks from the scheduler and executes them, reporting back their status and health.

## 29.2. Resource Allocation

Common examples of resources allocated to tasks include:

- **CPU cycles** - We measure the number of instructions that the task executes.
- **RAM** - We measure the amount of memory used by the task.
- **Network usage** - We measure the amount of bytes sent/received over the wire.

There are also other resources like GPU cycles and disk spindle usages. But we will focus our discussion on these three.

All these resources are scarce on cluster machines and it is important for tasks to share them. When a resource **X** must be shared among multiple tasks requesting it, a variety of algorithms can be employed to distribute it effectively.

### *29.2.1. Fair Allocation*

This algorithm allocates resources equally among all requests.

### *29.2.2. Weighted Fair Allocation*

This algorithm allocates resources proportionally to the requested amounts. Given allocation requests $\{\mathbf{x_1}, \mathbf{x_2}, ..., \mathbf{x_n}\}$ for a resource with total quantity $\mathbf{x}$, the resource allocated to job **i** is calculated as:

$$\mathbf{x} * \frac{\mathbf{x}_i}{\Sigma \mathbf{x_j}}$$

where $\Sigma \mathbf{x_j}$ represents the sum of all allocation requests.

### *29.2.3. Max-Min Fair Resource Allocation*

Max-min fair allocation prioritizes smaller demands, ensuring they receive their requested resources. Any remaining capacity is then distributed equally among larger demands.

Max-min allocation happens in rounds. Consider requests {2, 3, 4, 5} with a total capacity of 12.

- **Round 1**: Initial equal distribution - {3, 3, 3, 3}.
- **Round 2**: The smallest request is satisfied, so, the slack from the first request is allocated to others - {2, 3.33, 3.33, 3.33}.
- **Round 3**: The second smallest request is satisfied, so the slack from the second request is allocated to others - {2, 3, 3.5, 3.5}.
- **Round 4**: Final max-min fair allocation {2, 3, 3.5, 3.5}.

Max-min fair sharing typically leads to improved resource utilization.

### 29.3. Scheduling

**Resource allocation** deals with the distribution of available resources among competing tasks. Following allocation, **scheduling** determines the temporal order of distribution. Essentially, allocation answers **what** resources each task receives, while scheduling addresses **when** those resources are utilized.

Consider two tasks, each requiring 10,000 CPU cycles for completion and there is 1 physical CPU. Resource allocation would assign 10,000 cycles to each job. After that, scheduling dictates which job gets to use the resource first.

Common **non-preemptive scheduling algorithms**, such as **First-In, First-Out (FIFO)** and **Last-In, First-Out (LIFO)**, can be employed. However, non-preemptive approaches can lead to critical issues:

- **Starvation:** A long-running task can monopolize the CPU, preventing shorter tasks from executing for extended periods, or even indefinitely.
- **Inefficient Resource Utilization:** If a task encounters an I/O wait or other blocking event, the CPU remains idle until the task resumes, leading to wasted processing capacity.
- **Lack of Priority Handling:** Non-preemptive algorithms typically do not allow for prioritizing critical tasks, potentially causing delays in essential functions.

**Round-robin scheduling** is a **preemptive algorithm** where each task is allocated a fixed time slice, or quantum. Tasks execute for their assigned time slice and then yield the resources to the next task in the queue. This process repeats cyclically until all tasks complete execution.

### 29.4. The Time Dimension in Resource Allocation

By incorporating **time as a resource dimension**, we can more accurately model both resource allocation and scheduling simultaneously.

Instead of allocating a certain quantity of resources, we allocate **a resource usage rate over a specific time period**. For instance, CPU resources can be represented as **<CPU rate, time>** instead of raw CPU cycles, and network usage as **<bandwidth, time>** instead of total bytes transferred/received.

CPU rate is essentially clock rate (each clock executes one instruction). It is fixed for all the CPUs on a machine. So, essentially, <CPU rate, time> boils down to **<Num CPUs, time>**. The resource allocation of **<2 CPUs, 5s>** means that a task can use the 2 CPUs for 5 seconds.

If we fix the time period to one second, the allocated CPU count directly corresponds to the CPU usage within that second. It can be fractional as well. For example, say the resource allocation is **<0.2 CPU>**, then the task can execute 0.2 CPU worth of instructions in a second. Assuming the clock rate is 1 Ghz, then the task can execute 200M instructions in a second.

Similarly, the allocated bandwidth within a second reflects the network usage in that second.

Modern schedulers allocate CPU resources as a rate instead of number cycles. Cloud providers, such as Google Cloud, offer **virtual CPUs (vCPUs)** which are the same as CPU, however, in heterogeneous clusters with varying CPU clock rates, vCPUs provide a standardized abstraction for CPU capacity.

An analogy can be drawn with **Joule** (energy consumed) and **Watt** (energy consumption rate per second):

- **Joule: CPU cycles :: Watt: CPU/vCPU**
- **Joule: Network usage :: Watt: Bandwidth**

When resource allocation is defined in terms of rate, resource allocation algorithms naturally become scheduling mechanisms, such as:

- **Fair Scheduling:** Distributes available CPU or bandwidth equally among tasks. For CPU, this means each task receives an equal portion of the total available CPU. For bandwidth, it means an equal share of the network's capacity. This is essentially **round-robin**.
- **Weighted Fair Scheduling:** Allocates resources proportionally based on assigned weights. Tasks with higher weights receive a larger share of the CPU or bandwidth. This is essentially **weighted round-robin**.
- **Max-Min Fair Scheduling:** Prioritizes tasks with lower allocations, bringing them up to a fair share before distributing excess resources.

**Note**: RAM is allocated as a static quantity and not as a rate. This is because RAM usage exists regardless of a task's execution state - running or preempted (blocked).

### 29.5. Delay Scheduling

In distributed data-processing, there is one more resource that needs to be considered for scheduling - the data that is used for processing. Simply allocating resources without considering **data locality** can lead to tasks being scheduled on nodes where their required data is not readily available. This necessitates data transfer across the network, significantly impacting performance due to increased latency and network bandwidth consumption.

To mitigate this, delay scheduling [203] is employed (initially introduced for Hadoop). This technique introduces a deliberate delay before scheduling a task. Instead of immediately assigning the task to the first available resource, the scheduler waits for a short period, hoping that a node with the task's preferred data (i.e., local data) becomes available. If a local resource becomes free within the delay window, the task is scheduled there. Otherwise, it is scheduled on a remote node.

Delay scheduling statistically improves performance substantially by reducing data transfer and lowering latency.

### 29.6. Bin Packing Problem

The bin packing problem is a classic NP-hard optimization challenge: given a set of items with varying sizes and a finite number of bins (or containers), each with a fixed given capacity, the goal is to pack all items into the fewest possible bins. This problem arises frequently in computer science and other fields.

#### 29.6.1. Examples

- **Memory Allocation:** Allocating fixed-size pages of RAM to processes, minimizing the number of physical memory pages used.
- **Scheduling in Cloud Computing:** Packing tasks with varying resource demands (CPU, memory, storage) into machine instances to optimize resource utilization and reduce costs.
- **File Storage:** Storing files of different sizes on storage devices with fixed capacities, minimizing the number of storage devices required.
- **Network Packet Packing:** Packing network packets of variable sizes into fixed size frames for transmission.

#### 29.6.2. Approximation Algorithms

Even though the problem is NP-hard, there are several approximation algorithms to solve the problem:

- **First-Fit:** Places each item into the first bin where it fits.
- **Best-Fit:** Places each item into the bin with the smallest remaining capacity that can accommodate it.
- **Worst-Fit:** Places each item into the bin with the largest remaining capacity.
- **Almost-Worst-Fit:** Similar to worst fit, but avoids the absolute worst fit bin.
- **Next-Fit:** Places each item into the last bin it checked. If the item doesn't fit, a new bin is opened.
- **Next-k-Fit:** A variation of Next-Fit, that checks the last **k** bins.

These algorithms provide approximate solutions with certain **error bounds**, indicating how far the solution might be from the optimal:

- **Next-Fit:** Has an error ratio of 2, meaning it may use up to twice the optimal number of bins.
- **First-Fit, Best-Fit, Almost-Worst-Fit:** Has an error ratio of approximately 1.7.
- **Worst-Fit:** Has an error ratio of 2.

These error bounds represent worst-case scenarios, and in practice, these algorithms often perform significantly better. The choice of algorithm depends on the specific application and the trade-off between solution quality and computational cost.

### 29.7. Hadoop v/s Spark

Hadoop, launched in 2006, emerged as the open-source counterpart to Google's MapReduce, mirroring its core functionality of breaking down big data processing jobs into map and reduce tasks. Notably, the Hadoop project also developed the Hadoop Distributed File System [83], drawing inspiration from Google File System.

In contrast, Spark was specifically conceived with machine learning algorithms, such as logistic regression [211], in mind. The original whitepaper [89] elaborates on this motivation, which can be summarized as follows:

Consider the iterative nature of logistic regression. A fundamental step involves computing the following linear combination in the map stage:

$$\boldsymbol{\beta} + \mathbf{W_1} * \mathbf{x_1} + \mathbf{W_2} * \mathbf{x_2} + \ldots + \mathbf{W_n} * \mathbf{x_n}$$

Here, $\boldsymbol{\beta}$, $\mathbf{W_1}$, $\mathbf{W_2}$, ... $\mathbf{W_n}$ are model parameters and $\mathbf{x_1}$, $\mathbf{x_2}$, ... $\mathbf{x_n}$ are the input features from each record. This calculation feeds into a sigmoid function to produce a probability, which will be close to either 0 or 1. The deviation of this predicted probability from the true label is then calculated as the error. In the reduce stage, these errors are aggregated across the entire dataset to determine the overall error of the current model.

Logistic regression is inherently iterative. The algorithm repeatedly refines the model parameters in each round until the error converges to a minimum. The final set of weights constitutes the trained logistic regression model. This entire process represents a single perceptron (or logistic regression unit) within a deeper neural network architecture, where multiple such units are arranged in layers.

Spark's key innovation lies in its use of **Resilient Distributed Datasets** (RDDs). RDDs are fault-tolerant, immutable, distributed collections of data that can be stored in memory for the duration of a job. RDDs can originate directly from the input dataset or be the result of applying map or reduce operations to existing RDDs. By keeping RDDs in memory (with mechanisms to spill over to disk in case of memory pressure), Spark avoids the overhead of repeatedly reading data from storage. Due to the deterministic nature of map and reduce functions, RDDs can always be recomputed on demand if necessary.

This in-memory processing of RDDs is the primary differentiator between Spark and Hadoop. In Hadoop, each iteration of a machine learning training process necessitates reading the input data from storage and processing it anew. Conversely, Spark allows the input data to reside in memory as RDDs, readily available for subsequent processing rounds, leading to significant performance gains for iterative algorithms like logistic regression.

### 29.8. Message Passing Interface

**Message passing** is a programming model where independent processes coordinate by exchanging data through messages.

Each process operates within its own memory space, and communication happens via explicit sending and receiving of data.

The Message Passing Interface (MPI) is a standardized library that implements this model efficiently. Key benefits of MPI include:

- MPI libraries are highly optimized to deliver excellent performance.
- MPI leverages the fastest available communication method:
    - Using shared memory for communication between processes on the same machine.
    - Using TCP/IP for communication between processes on different nodes.
    - Taking advantage of RDMA, if available, for highly efficient data transfer between nodes.
- MPI typically guarantees reliable delivery of messages and preserves the order in which they are sent.

The primary application of MPI lies in building parallel programs deployed on HPC clusters. An MPI parallel program launches multiple processes that communicate with each other through message passing. Each of these processes represents a task that needs to be scheduled on the cluster's machines. The MPI scheduler is responsible for managing task placement and execution.

### 29.8.1. Example

A popular example of an MPI program is Strassen's Matrix Multiplication [212] (divide and conquer). Each task runs on a smaller slice of matrix and the results are merged together, as shown in Figure 138.

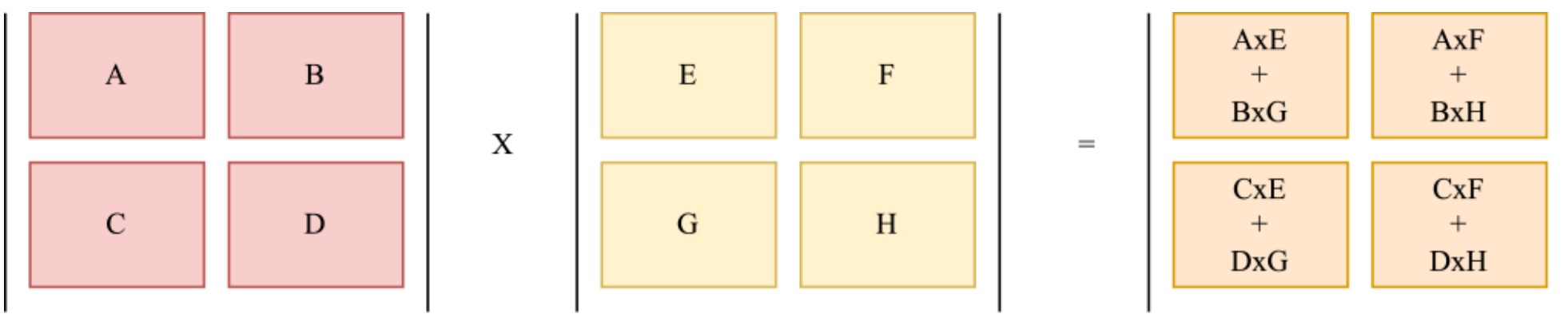

**Figure 138**: Strassen's matrix multiplication.

Indeed, many parallel computing algorithms can be written in MPI and executed in a distributed environment such as HPC that are optimized for communications.

### 29.8.2. MPI v/s MapReduce

Many computations that MPI can perform can also be executed in a distributed MapReduce environment. However, MapReduce has significant communication overhead. MPI, on the other hand, is optimized for communication and hence should be the preferred choice for computationally heavy programs that require substantial communication. Many programs that are well-suited for parallel computing are also better suited for MPI than MapReduce.

## 29.9. Mesos

Mesos is a two-level scheduler designed for various workloads such as Hadoop and MPI, referred to as **frameworks**, running within a cluster. Each framework runs certain user-submitted jobs on the cluster. Each job spawns several **tasks**. For instance, consider a Hadoop framework executing a word count job submitted by user. This would necessitate scheduling numerous map-reduce tasks. Similarly, consider an MPI framework running a ray tracing job, which requires several sub-tasks to operate concurrently for processing.

### 29.9.1. Use Cases

Mesos is optimized for data processing and scientific computations. It is not intended for long-running services. Downstream frameworks are incentivized to utilize:

- **Short tasks:** That is, tasks that run to completion in a short-time.
- **Elastic scaling:** That is, maintaining operation regardless of available resource fluctuations.

### *29.9.2. Architecture*

Mesos' architecture is shown in Figure 2 in the paper. Mesos features a replicated **master** for high availability, with mastership coordinated through ZooKeeper. There are multiple **schedulers** each for a particular framework. Each scheduler registers its resource requirements with the master.

In addition to the master and schedulers, an **executor** (akin to an agent) runs on each machine in the cluster. There are different executors for each framework. These executors are responsible for receiving and executing the tasks assigned to them.

Mesos supports multiple frameworks concurrently. For instance, Hadoop and MPI frameworks can run simultaneously on the same cluster, with each framework having its own dedicated scheduler.

#### 29.9.2.1. The Resource Broker

Essentially, the Mesos master acts as a resource broker, facilitating the allocation of resources between the schedulers and the executors. Executors report the available resources on their respective machines back to the Mesos master in the form of **resource offers**. The master then employs a resource allocator to distribute these resources to the registered schedulers. The entire flow is shown in Figure 3 in the paper.

The schedulers then allocate tasks for these resources which are forwarded to the executors.

### *29.9.3. Resource Allocation*

Each scheduler specifies a **filter** defining resources it will accept. Filter can be of two types:

- **Minimum Resources** - Only offer nodes with at least **R** resource free. Any allocation below this threshold will be rejected by the scheduler.

- **Preference List** - Only offer nodes from a list **L**. Any allocation on nodes outside of the list will be rejected. This is used for specifying preferences (e.g. for data locality).

The resource allocation algorithm allocates resources to the schedulers. The algorithm is pluggable and so can be replaced with a custom one. By default, a variant of max-min called the Dominant Resource Fairness [204] algorithm is used for allocating resources as per filters. Only the allocations passing the filters are sent to the scheduler.

Mesos can also claim back resources by revoking the tasks. In these terms, Mesos also has a concept of **guaranteed allocations** to a framework. Tasks won't be revoked for frameworks with utilization less than guaranteed allocations.

#### *29.9.4. Isolation*

Multiple frameworks can run their tasks on the same machine and they are isolated by leveraging Linux containers which can isolate CPU, memory, and network bandwidth.

### 29.10. Behavior Analysis

This paper stands out by including a dedicated section on detailed behavior analysis, which is usually not found in other works. Note that the paper only provides the result. The detailed derivation of the results is presented in their technical report [205].

For analysis, the authors differentiates between two framework types:

1. **Elastic:** These frameworks can dynamically adjust their resource allocation. Increased resources lead to faster processing, but the framework remains functional even with fewer resources. Hadoop exemplifies this type.
2. **Rigid:** These frameworks require a fixed amount of resources to execute. MPI, which needs all resources to run parallel programs, falls into this category.

For behavior analysis purposes, the cluster is conceptually divided into **N** equivalent slots. For instance, if each machine has 5 CPUs and 10 GB of RAM and is divided into 5 slots, each slot represents resources equivalent to <1 CPU, 2 GB RAM>. The authors assume each task occupies a single slot, implying uniform resource requirements across tasks.

The analysis focuses on three key metrics:

1. **Framework ramp-up time:** The duration required to acquire all allocated resources for a job.
2. **Job completion time:** The total time a framework takes to complete all its jobs.
3. **System utilization:** The overall resource utilization of the cluster.

The authors also consider various types of task distribution, as shown in Figure 139:

1. **Constant:** All tasks within a job have the same completion time.
2. **Exponential:** Task completion times vary according to an exponential distribution.
3. **Bi-modal**: Tasks' duration have two values - short and long.

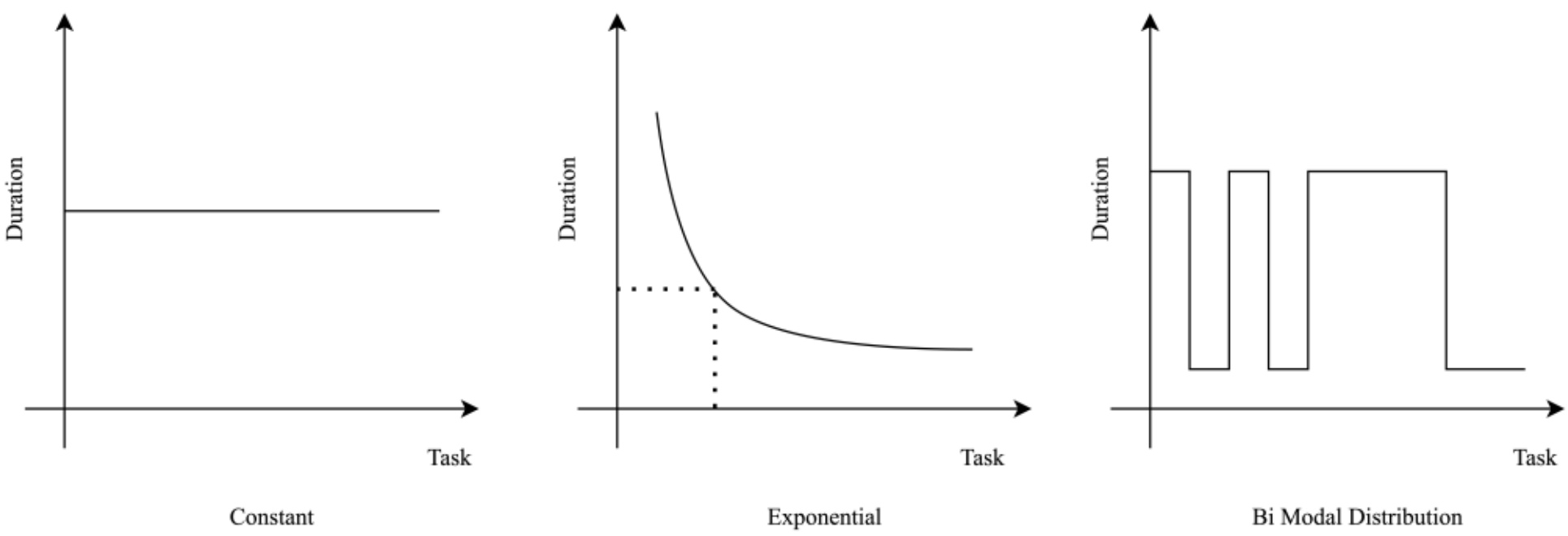

**Figure 139**: Different types of task duration distribution.

*29.10.1. Homogeneous Case*

In the homogeneous scenario, all frameworks exhibit the same task distribution (either constant or exponential).

Consider a framework allocated **k** slots. If the total computation required for framework is $\boldsymbol{\beta} * \mathbf{k} * \mathbf{T}$, where $\boldsymbol{\beta}$ is a constant multiplier, $\mathbf{T}$ is the mean task duration, and **k** slots are available, the computation will complete in $\boldsymbol{\beta} * \mathbf{T}$ time.

The detailed derivations are available in the technical report and summarized in Table 2. Intuitively:

- For a constant distribution, the ramp-up time is **T** (the time for each task). This is because after time **T**, all existing slots will be freed, including the **k** slots (all allocated resources). For an exponential distribution, the ramp-up time is longer (approximately $\mathbf{T * \ln k}$) as longer tasks would take more time to release their slots.
- For an elastic framework, the completion time is $\mathbf{T + \beta * T}$ . The initial **T** accounts for the time to acquire the necessary slots after the first task is scheduled (see the technical report for a detailed derivation). For rigid frameworks, the completion time is higher as no task can begin till ramp-up is complete.
- Rigid frameworks exhibit lower utilization because their tasks cannot commence until the entire ramp-up phase is complete.

In essence, elastic frameworks and constant task distributions lead to better performance. Non-rigid frameworks with exponential task distributions perform the worst.

*29.10.2. Heterogeneous Tasks*

When task distributions are heterogeneous, such as a bimodal distribution with both short and long tasks, short tasks can experience starvation behind long-running tasks. The authors suggest the following scheduler strategies to mitigate this:

1. **Random task assignment:** Probabilistically increases the chance of scheduling short tasks. The probability depends on the fraction $\boldsymbol{\varphi}$ of the long tasks to total tasks. The probability that at least 1 short task will be scheduled on a machine with **S** slots is $\mathbf{(1 - \varphi^S)}$.
2. **Reserve slots for short tasks:** Similar to creating separate queues in HPC.

*29.10.3. Placement Preferences*

Each scheduler can have a preferred list of slots (essentially, preferred nodes). However, not all scheduler frameworks can be guaranteed their preferred slots. In such instances, Mesos can employ **lottery scheduling** to distribute preferred slots randomly.

Lottery scheduling is an instantaneous form of weighted fair allocation. In this context, the weight is the total intended allocation of a framework ($\mathbf{s}_i$).

## 29.11. Implementation

- Hadoop's fine-grained map and reduce tasks align well with Mesos's resource management model. Notably, Hadoop already provides a pluggable API for developing schedulers. To enhance data locality, delay scheduling was implemented within these schedulers themselves.
- For MPI integration, the Torque scheduler was adapted for Mesos. When MPI jobs are submitted to the queue, the Torque scheduler begins accepting resource offers from Mesos. Upon accepting an offer on a specific node, tasks from the queued jobs are executed there. Given that Torque tasks are resource-intensive, Torque employs guaranteed allocation and avoids accepting resources beyond its reservation to prevent unnecessary task revocations.
- Finally, Spark was also integrated with Mesos. Spark executors store the results of map-reduce-like operations (i.e., RDDs) in memory. These cached RDDs can be reused in subsequent iterations, significantly accelerating processing.

## 29.12. Evaluation

The paper presents a remarkably thorough evaluation, a level of detail rarely encountered in research publications.

- In the macrobenchmark section, Mesos is compared against static partitioning. The results demonstrate that Mesos achieves superior cluster utilization of CPU resources when Hadoop, Spark, and Torque are run concurrently.
- Torque's execution time takes a hit, due to its non-elastic nature.
- Hadoop, particularly with a large mix of jobs, exhibits the most significant performance gains due to its elasticity. However, smaller Hadoop jobs experience a performance degradation. This is attributed to the resource broker layer within Mesos, which introduces additional overhead.
- Mesos enables frameworks to scale up dynamically by leveraging resources when other frameworks have lower demands.

- While Mesos introduces a slight overhead, the impact is minimal. For instance, an MPI job took 50.9s without Mesos and 51.8s with Mesos, while Hadoop execution times were 160s versus 166s, respectively. In both cases, the overhead is less than 4%. Nevertheless, this overhead becomes substantial for small Hadoop jobs.
- Mesos performs effectively with Hadoop jobs that benefit from data locality. Employing delay scheduling reduces the running time by half. With a 5-second delay in scheduling, 100% data locality is achieved.
- The Spark framework, a key integration within Mesos, demonstrates significantly better performance than Hadoop. While the first ML training iteration yields comparable results to Hadoop, subsequent iterations show substantial improvements, with 10x reduction in running time.
- Mesos exhibits low overhead even at a scale of 50,000 nodes, indicating high scalability.
- The Mean Time To Recovery (MTTR) for the master node is approximately 4-8 seconds.

### 29.13. Paper Remarks

The authors' deep expertise in large-scale data processing systems is evident throughout this paper, offering valuable insights into the challenges and solutions encountered in this domain. Consequently the system developed by the authors represents a truly significant achievement in computing. This paper is highly recommended as a foundational text for understanding the specific classes of problems - and the corresponding constraints - that define modern cluster computing environments.

# 30. Autopilot: Workload Autoscaling at Google Scale

(30) Rzadca, K.; Findeisen, P.; Swiderski, J.; Zych, P.; Broniek, P.; Kusmierek, J.; Nowak, P.; Strack, B.; Witusowski, P.; Hand, S.; others. Autopilot: Workload Autoscaling at Google. In Proceedings of the Fifteenth European Conference on Computer Systems; 2020; pp 1–16.

Almost all applications running in data center clusters and the cloud are scalable - a technique through which an application can handle more work without impacting performance by adding resources. All cloud providers offer autoscaling, which automatically scales a user's application based on certain observed metrics corresponding to the needs of the application.

This paper from Google was presented at EuroSys 2020. It introduces Autopilot, Google's internal autoscaling solution for jobs running on Borg, Google's in-house cluster orchestrator.

In this insight, we will first start with Borg. Then, we will go through scaling concepts, exploring the different types. Next, we will dive into the details of Autopilot. We will understand the various statistical concepts and models that power vertical scaling. Finally, we will explore the different strategies for horizontal scaling.

**Recommended Read**: **29. Mesos: A Platform for Fine-Grained Resource Sharing in the Data Center** where cluster computing and resource allocation was introduced.

### 30.1. Borg

Borg [206] is a cluster orchestrator developed by Google for managing its clusters. The design of Kubernetes was significantly influenced by Borg.

#### *30.1.1. Jobs and Tasks*

A Borg cluster consists of roughly 100 to 10,000 physical machines connected through a high-speed network fabric. Users submit jobs for execution on these machines, and these jobs are categorized into several types:

1. **Services:** These are long-duration jobs that frequently constitute components of larger systems, such as those employing a microservices architecture. A service is generally replicated, and an individual instance is called a **task**.
2. **Batch:** These jobs are typically composed of short-lived data-processing tasks.

A task executes as a containerized process on the cluster machines.

Given Google's extensive infrastructure for internet-scale services, most jobs running within Borg are services. Services occupy the majority of machine capacity, with batch jobs utilizing the remaining available capacity.

### 30.2. Scaling

Every job running in a cluster/cloud requires resources. The most critical of these are **CPU** and **RAM.**

Note that with the rise of data-intensive and compute-intensive applications (such as ML training), memory bandwidth [213] is also becoming a crucial resource. Similarly, **network bandwidth** is critical for many network-intensive applications. However, this paper focuses solely on CPU and RAM, assuming an abundance of network and memory bandwidth. These assumptions will likely evolve with changes in the computing landscape.

When users submit jobs to cluster/cloud, they need to specify resource requirements. However, these specifications can often be inaccurate, either overestimating or underestimating the actual needs of the job. For instance, if a service experiences a 5x increase in user requests compared to the expected volume, its resources will need to scale up. Conversely, if the service receives 10x fewer requests than anticipated, resources should be scaled down.

#### *30.2.1. Types*

Resource scaling for a job can be categorized into two primary types:

- **Vertical Scaling:** Increasing or decreasing the resources allocated to each task, as shown in Figure 140.
- **Horizontal Scaling:** Adding or removing tasks, as shown in Figure 141.

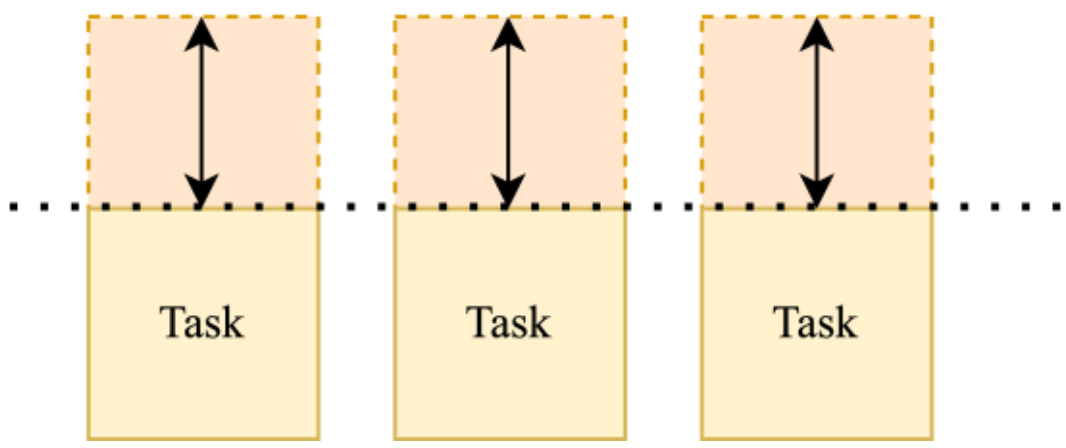

**Figure 140**: Vertical scaling.

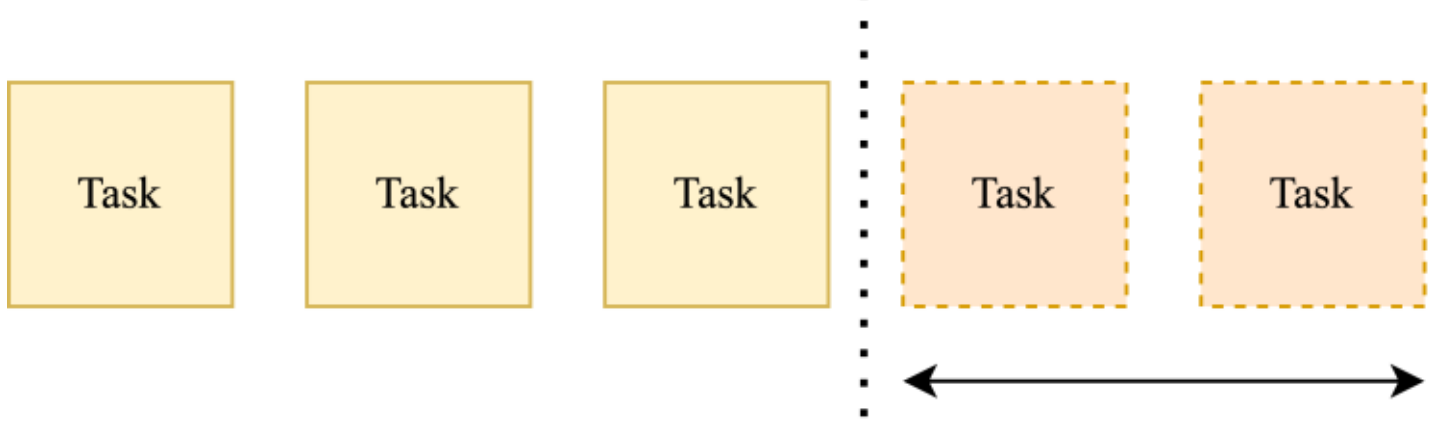

**Figure 141**: Horizontal scaling.

**Q. Which to prefer?**

The decision is context-dependent. It hinges on numerous factors, and there is no universally ideal solution.

Cluster orchestrators often favor vertical scaling as it tends to keep the number of tasks lower, which can expedite the computation of optimal resource assignments. However, there are applications that have to scale out horizontally because the NICs on individual machines can't handle the required network capacity.

*30.2.2. Limits & Oversubscriptions*

Even with resource scaling, establishing limits is crucial. For example, users should define upper bounds for the RAM and CPU that each task can consume. A user might set a task limit to <2 CPU, 4 GB RAM>.

These limits are often based on the worst-case observed scenario, which may not accurately reflect a task's actual, typical resource needs. The average resource utilization can be significantly lower than the defined limits. Many tasks exhibit diurnal usage patterns, with high usage during specific hours of a day and low usage at other times.

To address this discrepancy, cluster/cloud providers employ **oversubscription** on machines. Consider a machine with a capacity of 256 GB RAM and 64 CPU, hosting tasks with the following limits:

- 20 CPU, 100 GB RAM
- 40 CPU, 50 GB RAM
- 40 CPU, 100 GB RAM

All these tasks can be scheduled on the same machine, with CPU resources being oversubscribed under the assumption that not all tasks will simultaneously utilize their maximum allocated CPU. Figure 142 shows example usage of tasks over time.

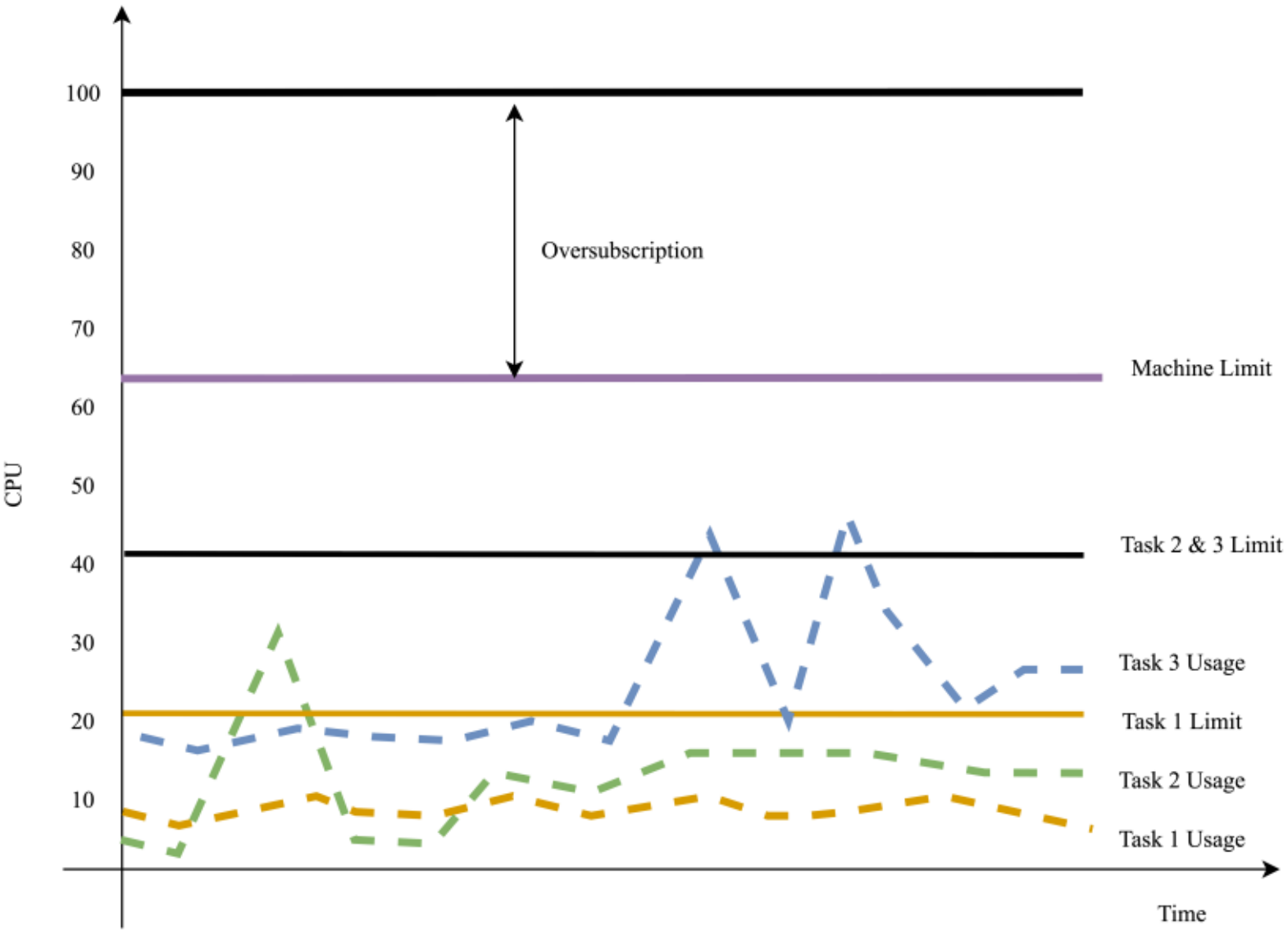

**Figure 142**: Example task usages under oversubscription.

Oversubscription is vital; without it, machines often remain underutilized.

Furthermore, these limits are often **soft limits**. For example, if the first task uses 50 CPU while other tasks are idle, this is permissible (soft limiting). However, soft limiting transitions to **hard limiting** when there's resource contention. If other tasks begin to use their allocated 40 CPU, the first task will experience preemption to prevent it from exceeding its 20 CPU limit. The presence of priorities further complicates this dynamic.

#### 30.2.3. Enforceable vs. Non-enforceable Resources

CPU is an enforceable resource. If a task exceeds its hard CPU limit, it will be descheduled to enforce adherence to the limit. In contrast, RAM limits are non-enforceable. Cluster machines typically have limited swap space (due to diskless [214] architecture choices), so if a task surpasses a certain hard RAM limit, it must be Out-Of-Memory (OOM) terminated.

#### 30.2.4. Chargeback Model

Users consuming resources on a cluster/cloud are typically charged for their usage, as computing operation isn't free. There are real costs associated with acquiring machines, setting up networks, and maintaining real estate, among other expenses.

A straightforward chargeback model could involve charging users based on the resource limits they set. However, this approach might not lead to competitive pricing. Charging users based on their actual resource usage might seem more equitable. Yet, even usage-based charging presents challenges, as a user's average usage can be significantly lower than their peak demand. For example, a user might average 1 GB of RAM usage but occasionally spike to 100 GB. Ultimately, cluster/cloud providers must provision for these peak demands. Therefore, charging solely on average usage would be unfavorable for service providers.

A more effective approach could be to charge users based on their usage plus a portion of the oversubscription cost, distributed proportionally among the tasks.

In the example above, if the actual average CPU usages are 10, 20, and 25, then:

- Base charge for user 1: 10
- Base charge for user 2: 20
- Base charge for user 3: 25

Slack CPU: 64 - (10 + 20 + 25) = 9

Oversubscription Tax for user 1: (20 / (20 + 40 + 40)) * 9 = 1.8
Oversubscription Tax for user 2: (40 / (20 + 40 + 40)) * 9 = 3.6
Oversubscription Tax for user 3: (40 / (20 + 40 + 40)) * 9 = 3.6

### 30.3. Autopilot

Autopilot is designed to automate the scaling of resources for jobs running on Borg.

The primary goal of Autopilot is to automate two key aspects of resources for a job:

- Setting resource limits for individual tasks (**vertical autoscaling**) and
- Adjusting the number of running tasks (**horizontal autoscaling**).

Automating the setting of resources is especially important because these resources directly determine the cost charged back to users. Therefore, optimizing this process can lead to substantial cost reductions for users.

#### *30.3.1. Architecture*

Autopilot's architecture is based on a triple closed-loop control system. One loop manages horizontal scaling, while the other two independently control per-task CPU and memory resources.

Each job is treated as an isolated entity, meaning there is no learning or information sharing across different jobs.

The data flow within this closed-loop system is as follows (shown in Figure 1 in the paper):

1. Resource usage data of a task is continuously logged and exported in the form of time series. For instance, the CPU utilization of a specific task over time.
2. The resource usage time series for all tasks of a job are fed into **recommender systems**. These recommenders analyze the data and generate suggested job resource limits, which are then passed to the Autopilot service.
3. The **Autopilot service** takes these recommended job limits and communicates them to an **actuator** component. The actuator then translates these job limits into task limits and task count, and sends them to the Borgmaster.
4. Finally, the Borgmaster processes these updates and takes appropriate action, such as adjusting the resource allocation of the affected tasks or adjusting the number of tasks.

### 30.4. Per-task Vertical Autoscaling

As previously mentioned, vertical autoscaling automates the setting of task resource limits, freeing users from manual configuration. The goal is to determine limits that prevent potential issues - such as performance degradation due to preemptions and OOM errors caused by overly restrictive RAM limits - while also optimizing resource utilization to reduce costs for users. Fundamentally, this is an optimization problem.

Autopilot addresses this challenge using two primary approaches:

- **Moving Window Recommenders**
- **Machine Learning**

Irrespective of the approach used, the goal is to find the resource limits for the job.

Let's delve into the details of their operation. The paper provides a comprehensive explanation of the mathematical details.

To understand the underlying mechanisms, we need to introduce some notation. The core of the system relies on time-series analysis.

Let:

- $\mathbf{r}_i[\boldsymbol{\tau}]$ represent the resource usage value exported by task **i** at time $\boldsymbol{\tau}$.
- $\mathbf{s}_i[\mathbf{t}]$ represent a distribution of all data points in $\mathbf{r}_i$ within the time window $[\mathbf{t} - \mathbf{5\ minutes}, \mathbf{t}]$. This distribution is essentially a histogram composed of discrete buckets.

If we denote the range of the $\mathbf{k^{th}}$ bucket as $[\mathbf{b}[\mathbf{k}-\mathbf{1}], \mathbf{b}[\mathbf{k}])$, then $\mathbf{s}_i[\mathbf{t}][\mathbf{k}]$ represents the frequency (count) of data points from task **i**'s resource usage within $[\mathbf{b}[\mathbf{k}-\mathbf{1}], \mathbf{b}[\mathbf{k}])$ between time $[\mathbf{t} - \mathbf{5\ minutes}, \mathbf{t}]$.

For CPU usage, the monitoring system utilizes a fixed set of 400 buckets, and a task's CPU usage distribution ( $\mathbf{s}_i[\mathbf{t}]$ ) can have non-zero frequencies across multiple buckets.

In contrast, for memory usage, the distribution $\mathbf{s}_i[\mathbf{t}]$ typically has a significant non-zero frequency in only one bucket - the one corresponding to the highest observed memory usage. This is because memory usage tends to be relatively stable over short timeframes. Even when objects are deallocated, the underlying memory pages might remain assigned to the process due to memory allocators like TCMalloc, which cache pages for potential reuse.

We can derive the resource usage distribution for an entire job, $\mathbf{s}[\mathbf{t}]$, by summing the individual task distributions:

$$\mathbf{s}[\mathbf{t}][\mathbf{k}] = \sum_i \mathbf{s}_i[\mathbf{t}][\mathbf{k}]$$

Essentially, this involves summing the frequencies of the corresponding $\mathbf{k}^{\mathbf{th}}$ bucket across all tasks belonging to the job. The data collection period ($\boldsymbol{\tau}$) is approximately 1 second. Consequently, each individual task distribution $\mathbf{s}_i[\mathbf{t}]$ is derived from roughly 300 data points within the 5-minute window, and the job distribution $\mathbf{s}[\mathbf{t}]$ incorporates approximately 300 × (number of tasks) data points.

Let's illustrate the mathematical formulation with an example. Consider a job comprising two tasks.

The first task exhibits the following CPU usage over a 5-minute window:

- 10 CPU for 2.5 minutes (150 seconds)
- 20 CPU for 2.5 minutes (150 seconds)

The second task shows the following CPU usage over the same 5-minute window:

- 20 CPU for 2.5 minutes (150 seconds)
- 30 CPU for 2.5 minutes (150 seconds)

Assume the CPU usage bucket boundaries are defined as {0, 10, 20, 30, 40}, resulting in the following buckets in a distribution: [0, 10), [10, 20), [20, 30), [30, 40). Then,

$\mathrm{s}_0[\mathrm{t}][0] = 0$
$\mathrm{s}_0[\mathrm{t}][1] = 150$ # number of seconds in 2.5 minutes.
$\mathrm{s}_0[\mathrm{t}][2] = 150$
$\mathrm{s}_0[\mathrm{t}][3] = 0$

$\mathrm{s}_1[\mathrm{t}][0] = 0$
$\mathrm{s}_1[\mathrm{t}][1] = 0$
$\mathrm{s}_1[\mathrm{t}][2] = 150$
$\mathrm{s}_1[\mathrm{t}][3] = 150$

Finally s[t] is [0, 150, 300, 150].

*30.4.1. Moving Window Recommenders*

The core idea behind moving window recommenders is to analyze samples ($\mathbf{s}[\boldsymbol{\tau}]$) to make a recommendation. Importantly, not all these samples are treated equally; more recent samples are given higher significance. This is achieved using a weighting function:

$$\mathbf{W}[\boldsymbol{\tau}] = \mathbf{2}^{\frac{-\boldsymbol{\tau}}{\mathbf{t}_{1/2}}}$$

where $\boldsymbol{\tau}$ is the time of a sample. $\mathbf{t}_{1/2}$ (the half-life of the weight) is 12 hours for the CPU.

With 12 hours, the weights would follow a pattern where the most recent 12 hours samples have a weight of 1, the next 12 hours samples (12 to 24 hours old) have a weight of 0.5, the subsequent 12 hours samples (24 to 36 hours old) have a weight of approximately 0.25, and so on, effectively halving the weight every 12 hours.

The paper defines three distinct types of windowing recommendations each applied based on the application's use case.

30.4.1.1. Peak

This is simply the maximum value observed across all buckets in the last **N** samples. This is used for memory for jobs with minimal OOM tolerance.

30.4.1.2. Weighted Average

First, we calculate the mean of the sample $\mathbf{s}[\boldsymbol{\tau}]$ at a given time $\boldsymbol{\tau}$. This is done using the standard formula for the mean of a distribution:

$$\mathbf{s}[\boldsymbol{\tau}]_{\mathbf{mean}} = \sum_{\mathbf{j}} \frac{(\mathbf{b}[\mathbf{j}] * \mathbf{s}[\boldsymbol{\tau}][\mathbf{j}])}{\sum_{\mathbf{j}} \mathbf{s}[\boldsymbol{\tau}][\mathbf{j}]}$$

where $\mathbf{b}[\mathbf{j}]$ is a representative value for the $\mathbf{j}^{\mathbf{th}}$ bucket and $\mathbf{s}[\mathbf{t}][\mathbf{j}]$ is the frequency of that bucket.

To incorporate the historical trend captured by the moving window, a weighted average of the sample mean values is computed:

$$\textbf{Weighted Average} = \frac{\sum_{\textbf{all samples}}(\textbf{weight} \; * \; \textbf{sample mean})}{\sum_{\textbf{all samples}} (\textbf{weight})}$$

The above is a simplified formulation of equation (2) in section 3.2.2.

Weighted average is used for CPU limits of batch jobs, which typically tolerate CPU throttling during peak usage.

30.4.1.3. j[th] Percentile of Adjusted Usage

To calculate this, we first adjust the frequency of each bucket in the historical data by multiplying it with the corresponding weight and the bucket's value. This creates a **load-adjusted distribution**. We compute the $\mathbf{j^{th}}$ percentile directly from this load-adjusted distribution.

$$\textbf{Adjusted frequency of bucket k} = \mathbf{b}[\mathbf{k}] * \sum_{\textbf{all samples}} (\textbf{weight} * \textbf{sample}[\mathbf{k}])$$

The above is a simplified formulation of equation (3) in section 3.2.2. Let's understand what multiplying frequency by $\mathbf{b}[\mathbf{k}]$ means. Going by our previous example, say the buckets are [0, 10), [10, 20), [20, 30), [30, 40). Let the corresponding **b** be [0, 10, 20, 30], i.e., taking the minimum in each bucket. Say there is only one sample - the most recent one - and its distribution is [0, 150, 300, 150]. Then, after load adjustment of the sample, we will get a new distribution, [0, 1500, 6000, 4500]. The earlier numbers {0, 150, 300, 150} represented pure frequencies. Now, the numbers {0, 1500, 6000, 4500} carry hints of actual usages. The p95 of the former distribution is 38, whereas the p95 of the latter distribution is 38.67. The latter is better suited because it represents the 95th percentile of the usage that would be observed.

The $\mathbf{j^{th}}$ percentile is used to set CPU limits for serving jobs (typically p95 or p90), RAM limits for jobs with low OOM tolerance (p98), and RAM limits for jobs with intermediate OOM tolerance (p60).

*30.4.2. ML Recommenders*

The ML recommender for vertical autoscaling involves several steps, but the underlying calculations are relatively straightforward. The system utilizes **N** different models (ensemble of models), each characterized by a decay rate ($\mathbf{d_m}$) and a safety margin ($\mathbf{M_m}$).

Step 1 to 4 determine the most optimal limit for the job found by each model. Step 5 and 6 then find the best model with the most optimal limit for the job.

**Step 1: Calculate Overrun Cost**

The overrun cost, $\mathbf{o}(\mathbf{L})[\mathbf{t}]$, represents the cost incurred if the resource limit is set to $\mathbf{L}$ at time $\mathbf{t}$, and the actual usage exceeds this limit. It's calculated as a decaying sum of past overruns:

$$\mathbf{o}(\mathbf{L})[\mathbf{t}] = (\mathbf{1} - \mathbf{d_m}) * \mathbf{o}(\mathbf{L})[\mathbf{t} - \mathbf{1}] + \mathbf{d_m} * (\textstyle\sum_\mathbf{j} \mathbf{s}[\mathbf{t}][\mathbf{j}] \textbf{ such that } \mathbf{b}[\mathbf{j}] > \mathbf{L})$$

The first term, $(\mathbf{1} - \mathbf{d_m}) * \mathbf{o}(\mathbf{L})[\mathbf{t} - \mathbf{1}]$, represents the decayed overrun cost from the previous time step. The decay factor $(\mathbf{1} - \mathbf{d_m})$ ensures that older overrun costs have a diminishing impact.

The second term, $\mathbf{d_m} * (\sum_\mathbf{j} \mathbf{s}[\mathbf{t}][\mathbf{j}] \textbf{ such that } \mathbf{b}[\mathbf{j}] > \mathbf{L})$, accounts for the current overrun. It sums the frequencies $\mathbf{s}[\mathbf{t}][\mathbf{j}]$ of all buckets $\mathbf{j}$ where the bucket's value $\mathbf{b}[\mathbf{j}]$ is greater than the proposed limit $\mathbf{L}$. This sum is then weighted by the model's decay rate $\mathbf{d_m}$.

**Step 2: Calculate Underrun Cost**

The underrun cost, $\mathbf{u}(\mathbf{L})[\mathbf{t}]$, represents the cost of underutilization if the limit is set to $\mathbf{L}$ at time $\mathbf{t}$, meaning the allocated resources are more than the actual usage. The formula is analogous to the overrun cost:

$$\mathbf{u}(\mathbf{L})[\mathbf{t}] = (\mathbf{1} - \mathbf{d_m}) * \mathbf{u}(\mathbf{L})[\mathbf{t} - \mathbf{1}] + \mathbf{d_m} * (\textstyle\sum_\mathbf{j} \mathbf{s}[\mathbf{t}][\mathbf{j}] \textbf{ such that } \mathbf{b}[\mathbf{j}] < \mathbf{L})$$

**Step 3: Determine the Best Limit for Each Model**

Each model $\mathbf{m}$ attempts to find a candidate limit $\mathbf{L_m}'[\mathbf{t}]$ that minimizes a weighted sum of the current overrun cost, underrun cost, and the cost of changing the limit from the previous time step ($\mathbf{L_m}'[\mathbf{t} - \mathbf{1}]$):

$$\mathbf{L_m}'[\mathbf{t}] = \mathbf{argmin_L}(\mathbf{w_o} * \mathbf{o}(\mathbf{L})[\mathbf{t}] + \mathbf{w_u} * \mathbf{u}(\mathbf{L})[\mathbf{t}] + \mathbf{w_{\Delta L}} * \boldsymbol{\Delta}(\mathbf{L}, \mathbf{L_m}'[\mathbf{t} - \mathbf{1}]))$$

where:

$\mathbf{w_o}$ is the weight for the overrun cost.
$\mathbf{w_u}$ is the weight for the underrun cost.
$\mathbf{w_{\Delta L}}$ is the weight for the cost of changing the limit.

$\mathbf{\Delta(L, L_m{}'[t-1])}$) is an indicator function that equals 1 if the proposed limit $\mathbf{L}$ is different from the previous best limit for model $\mathbf{L_m{}'[t-1]}$, and 0 otherwise. This term penalizes frequent limit changes.

**Step 4: Apply Safety Margin**

After finding the optimal limit $\mathbf{L_m{}'[t]}$ based on cost minimization, each model adds its specific safety margin $\mathbf{M_m}$ to arrive at its final recommended limit $\mathbf{L_m[t]}$:

$$\mathbf{L_m[t] = L_m{}'[t] + M_m}$$

**Step 5: Calculate the Cost of Each Model's Recommendation**

To evaluate the performance of each model over time, a decaying cost $\mathbf{c_m[t]}$ is calculated:

$$\begin{aligned}\mathbf{c_m[t]} &= \mathbf{d * (w_o * \text{cost of overrun by model m}} \\ &\quad \mathbf{+ w_u * \text{cost of underrun by model m}} \\ &\quad \mathbf{+ w_{\Delta L} * \text{cost of change})} \\ &\quad \mathbf{+ (1-d) * c_m[t-1]}\end{aligned}$$

The cost of change would be 1 if the chosen limit $\mathbf{L_m[t]}$ is different from the previously applied limit, and 0 otherwise. The decay parameter $\mathbf{d}$ gives more weight to recent costs.

**Step 6: Choose the Best Model and Set the Limit**

Finally, the system selects the model $\mathbf{m}$ that minimizes a combined cost, taking into account the model's historical cost, the cost of switching models, and the cost of changing the limit:

$$\mathbf{L[t] = argmin_m(c_m[t] + w_{\Delta m} * \Delta(m[t-1], m[t]) + w_{\Delta L} * \Delta(L[t-1], L_m[t]))}$$

where:

$\mathbf{w_{\Delta m}}$ is the weight for the cost of changing the model.

$\mathbf{\Delta(m[t-1], m[t])}$ is 1 if the chosen model at time $\mathbf{t}$ is different from the model chosen at the previous time step, and 0 otherwise. $\mathbf{\Delta(L[t-1], L_m[t])}$ is 1 if the limit $\mathbf{L_m[t]}$ proposed by the chosen model is different from the limit set at the previous time step ($\mathbf{L[t-1]}$), and 0 otherwise.

The final resource limit $\mathbf{L[t]}$ is then set to the limit recommended by the chosen model ($\mathbf{L_m[t]}$).

30.4.2.1. Model and Hyperparameters

**Model Parameters** ($\mathbf{d_m}$, $\mathbf{M_m}$): Each of the $\mathbf{N}$ models has its own fixed decay rate and safety margin. By creating a diverse set of models with different parameter values, the system can choose the one that best adapts to the workload's behavior.

**Hyperparameters** ($\mathbf{d}$, $\mathbf{w_{\Delta L}}$, $\mathbf{w_{\Delta m}}$, $\mathbf{w_o}$, $\mathbf{w_u}$): These are higher-level parameters that control the learning process and the decision-making of the recommender system. They need to be tuned or learned through experimentation to optimize the overall performance of the autoscaling mechanism.

### 30.5. Horizontal Autoscaling

Horizontal autoscaling in Autopilot is triggered by the following metrics:

- **CPU Utilization:** Similar to vertical autoscaling, users can define a lookback window (**T**). Within this window, a chosen percentile or the maximum CPU usage is considered the required usage. Users also specify their desired average utilization. Based on the ratio of the required usage to the average utilization, Autopilot determines the necessary number of tasks for the job.
- **Target Size (User-Defined Function):** Alternatively, users can provide their own function to calculate the desired number of tasks. This function can take various input metrics, such as QPS, into account.

*30.5.1. Strategies*

Horizontal scaling carries inherent costs. Allocating new machines for additional tasks takes time. Furthermore, during the scaling process, fluctuations in the triggering metrics (like CPU utilization or QPS) can occur, potentially leading to instability and further unnecessary scaling actions. To mitigate these issues, Autopilot employs several strategies:

1. **Deferred Downscaling:** Upscaling is performed immediately when needed. However, downscaling is delayed for a specific period. Once new tasks are started, they remain operational for a stabilization period. This is beneficial because a temporary load spike might recur shortly after it subsides, in which case the scaled-up capacity would already be in place. The trade-off is potentially higher resource consumption during the deferral period.
2. **Slow Decay:** During downscaling, the number of tasks is reduced gradually rather than abruptly. This helps to avoid sudden drops in capacity and potential disruptions.
3. **Defer Small Changes:** Horizontal scaling can have disruptive side effects on certain types of applications. For example, a distributed cache using consistent hashing to distribute data might experience significant re-sharding and temporary availability impacts even with the addition or removal of a single task. Therefore, small scaling changes are often deferred to avoid these disruptions.
4. **Limited Growth:** Even during upscaling events, the maximum rate of task growth is limited. This prevents the rapid creation of a large number of new tasks, which could themselves become unavailable during their startup phase and exacerbate instability.

Figure 143 is an example of how these strategies play out.

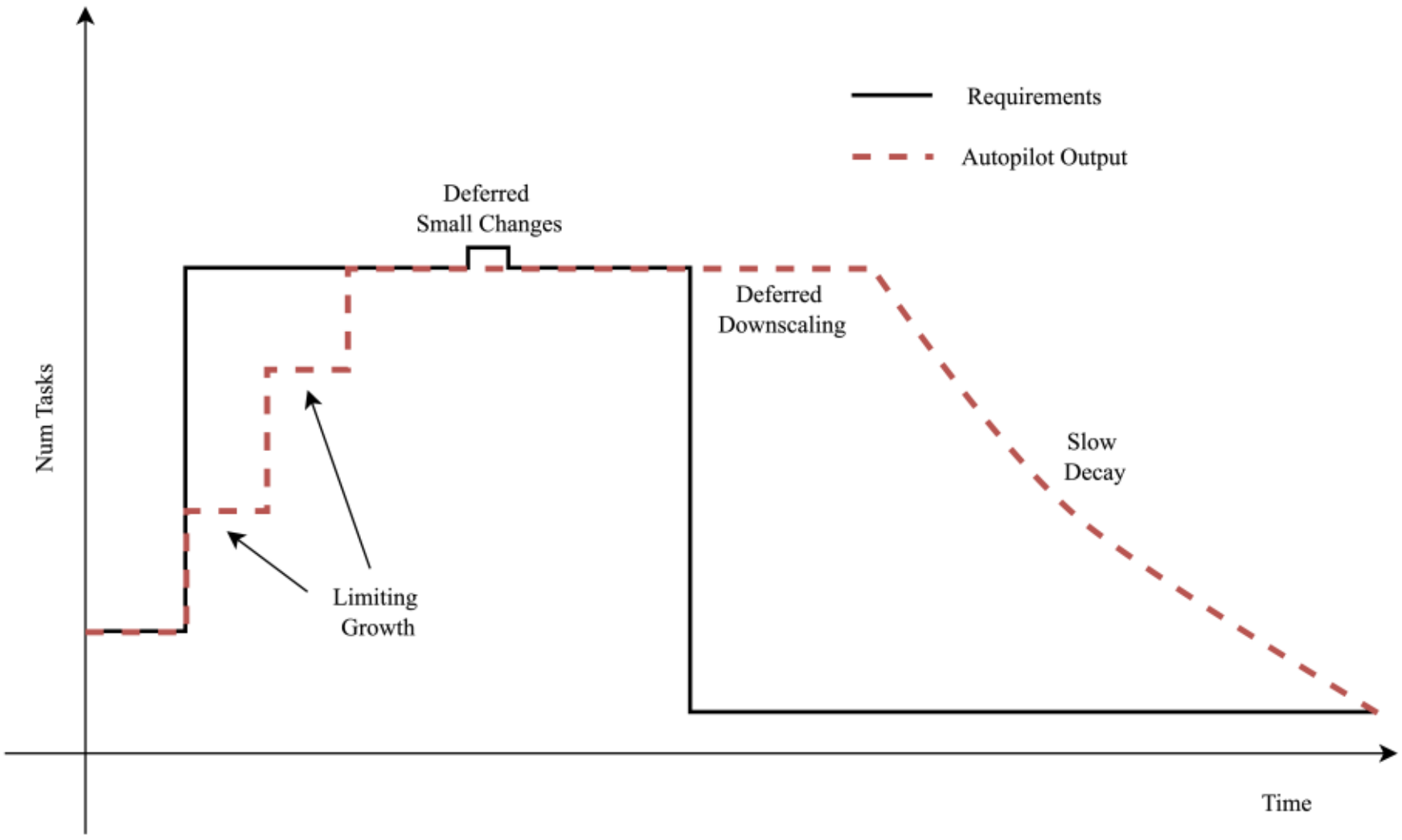

**Figure 143**: Horizontal scaling strategies.

### 30.6. Evaluation

The study evaluates the system's performance directly using real-world workloads. To track this, the authors employ the following metrics:

- **Footprint:** Calculated as the normalized sum of average task limits for a job, expressed in the number of machines (where $\textbf{number of machines} = \textbf{RAM bytes} / \textbf{memory per machine}$).
- **Relative slack:** Defined as $(\textbf{limit} - \textbf{usage}) / \textbf{limit}$ .
- **Absolute slack:** The difference between limit and usage, normalized to the number of machines.
- **Relative OOM rate:** The ratio of the number of OOM events to the total number of tasks.
- **Number of job-days without OOMs.**

The results show:

- Non-autopiloted jobs exhibit a relative slack ranging from 46% to 60%, while autopiloted jobs show a tighter range of 23% to 31%, with ML recommenders demonstrating superior performance.
- Autopilot significantly reduces absolute slack to approximately 500 machines compared to 12,000 machines for non-autopiloted jobs, resulting in estimated savings of tens of millions of USD for Google. Figure 3(c) in the paper shows that Autopilot is indeed responsible for the smaller footprints observed in autopiloted jobs (rather than sampling errors).
- Migrating 500 previously non-autopiloted jobs led to substantial performance improvements, saving 1,708 machines over a four-month period.
- OOM events are infrequent in autopiloted jobs, with 99.5% experiencing no OOMs. Figure 5 in the paper is somewhat difficult to interpret. The y-axis is the percentage of days when no OOMs were observed. It indicates that all algorithms achieved zero OOMs on at least 97% of days, with the moving window approach exhibiting the fewest days with OOMs.
- Figure 6 in the paper reveals an inverse relationship between memory slack and relative OOM rate. Moving window recommenders excel at managing OOMs while maintaining low slack, whereas ML recommenders tend to reduce limits too aggressively, leading to more OOMs.

- Recommendations demonstrate stability, with no changes occurring on 70% of job-days and only 6 to 7 changes observed at the 99th percentile (Figure 7 in the paper).
- Long-running jobs exhibit lower slack compared to new jobs, suggesting that Autopilot cautiously adjusts limits for new jobs to mitigate the risk of OOMs.

### 30.7. Paper Remarks

The mathematical details driving the vertical autoscaling in this paper are central to its contribution. The formulation might seem daunting, however, it becomes easy when broken down into its constituent parts. The fact that this approach operates successfully across Google's incredibly diverse and large-scale infrastructure highlights its robustness.

Sections 5 and 6 offer valuable insights into the strategies employed to ensure widespread adoption of the system. These lessons are often invaluable when designing infrastructure, highlighting the importance of demonstrating the system's value to client teams to drive impact.

# 31. F1 Query: Declarative Querying at Scale

(31) Samwel, B.; Cieslewicz, J.; Handy, B.; Govig, J.; Venetis, P.; Yang, C.; Peters, K.; Shute, J.; Tenedorio, D.; Apte, H.; others. F1 Query: Declarative Querying at Scale. Proceedings of the VLDB Endowment 2018, 11 (12), 1835–1848.

In previous chapters, we focused heavily on the core database engine, learning how it handles reads, writes, and transactions. In this chapter, we turn our attention to another vital component: query processing. Query processing is crucial for several data applications, such as real-time analytics, business intelligence, and report generation. Query processing has been extensively studied. It is significantly more straightforward on standalone databases. In distributed settings, however, it becomes highly complex; a comprehensive study of distributed query processing could easily fill an entire book.

This paper from Google was presented at Very Large Data Bases (VLDB), the premier global database conference, in 2018. It presents F1 Query, Google's most advanced query processing engine. The paper features a large number of authors and is incredibly dense. Because of its broad scope, the paper primarily provides high-level overviews of its various components rather than exhaustive technical minutiae.

In this insight, we will begin by defining a query engine and its high-level components. Since most queries involve more than a single data source or table, we will then examine the various types of joins that query engines must execute. Following that, we will explore how materialized views generated from query results are useful. We will then dive into the details of a critical query compiler stage - the query optimizer - which is responsible for reducing execution costs. Finally, we will go through the architecture of the F1 query engine, exploring its different execution modes and how it performs distributed joins at scale.

**Recommended Read**: **10. MapReduce: Simplified Data Processing on Large Clusters**

## 31.1. Query Engine

**SQL** (Structured Query Language) stands as the standard formal language for data extraction and processing. A **query engine**, then, is the system that executes these SQL queries.

### *31.1.1. Query Engines vs. SQL Databases*

SQL databases manage the storage of data in a relational format and handle transactions on that data. They inherently provide atomicity, concurrency control, durability, and transaction isolation. In contrast, a query engine's role is to extract a view of the data from a SQL database based on the user's formal query.

In traditional, standalone databases, the query engine is integrated directly within the SQL database itself. For instance, in PostgreSQL [107], the query parser, planner, and executor are all components of the core database system.

However, distributed SQL databases often separate the query engine into a distinct layer. In such architecture, query engines operate as independent units that can interface with various underlying SQL databases. These databases can be OLTP (like Spanner or Aurora) or OLAP (such as Napa/Mesa or Snowflake). Query engines can also be used with NoSQL databases.

### *31.1.2. Examples of Query Engines*

Several query engines exists, many offered through cloud:

- **Dremel** [85]**:** A pioneering query engine developed by Google, which later formed the foundation for Google Cloud's BigQuery [84].
- **F1 Query:** The successor to Dremel, currently in use within Google.
- **Apache Hive** [99]**:** Originally developed by Facebook for querying and managing large datasets residing in distributed storage.
- **Amazon Athena** [120]**:** Interactive query service provided by Amazon Web Services that enables querying data in Amazon S3 [58] using standard SQL.
- **Apache Drill** [215]**:** An open-source, schema-free SQL query engine for Big Data exploration.

### *31.1.3. Core Components of Query Engines*

While the specific architectures of query engines vary based on their intended use, they generally share the following high-level components:

- **Executor (Driver)**: It parses and analyzes the incoming query, optimizes it, breaks it down into executable tasks (often for a parallel processing framework like MapReduce), and assigns these tasks to available workers.
- **Workers**: These are the processes responsible for carrying out the tasks required by a query.
- **Metadata Store**: This component holds crucial information such as the location of data, access methods, and user-defined functions (UDFs).

Let's illustrate the components with an example query:

```
SELECT Department, COUNT(*)
FROM Professors
WHERE STARTS_WITH(Name, "s")
GROUP BY Department;
```

This query aims to count the number of professors whose names begin with "s", grouped by their department. Here **STARTS_WITH** is a UDF.

Consider a scenario where the underlying **Professors** table is sharded across multiple database instances, with each instance potentially holding data for different departments. The metadata store provides information about the shard locations of the **Professors** table. The metadata store also provides the executor with the definition of the UDF.

Upon receiving this query, the executor optimizes and translates the SQL query into a MapReduce pipeline. The executor would then generate map tasks that operate on the shards. It's even possible to have multiple map tasks for the same department if its corresponding data slice is large.

The map function for this query might look conceptually like this:

```
map(Professor p):
  if STARTS_WITH(p.Name, "s"):
    emit (p.Department, 1)
```

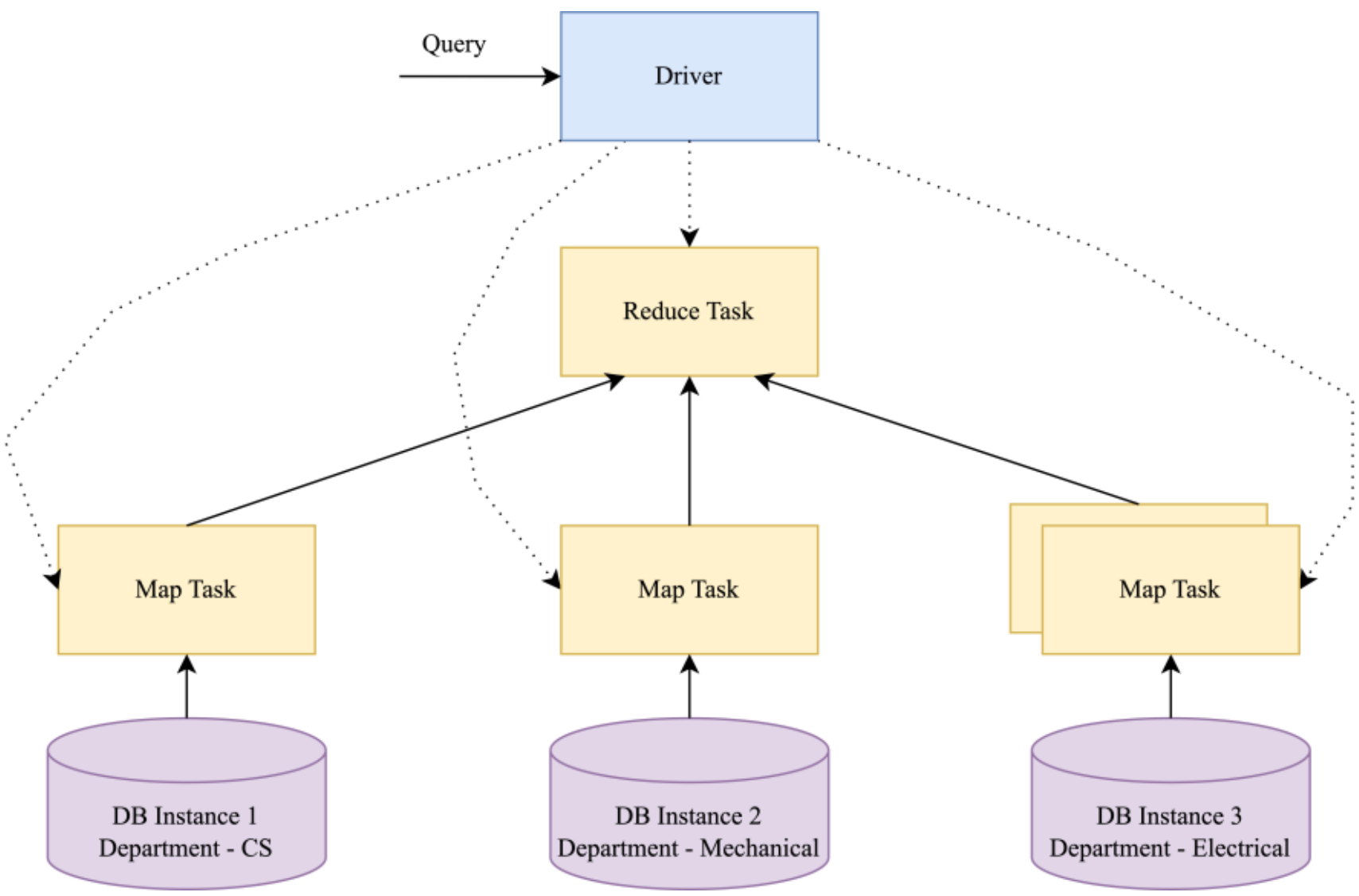

**Figure 144**: Example of MapReduce pipeline for query execution.

Finally, a reducer would aggregate the results from all the map tasks, performing a count for each department. Figure 144 shows the working of the MapReduce pipeline.

### 31.2. Read Consistency

Regardless of whether you're using a SQL or NoSQL database, query engines themselves cannot guarantee a consistent read without support from the underlying data store.

Consider our previous example to illustrate this potential issue. When a query is executed across multiple workers, each worker might begin processing at different times. This can lead to inconsistencies. For instance, one mapper might read its data instance before certain transactions are committed, while another mapper reads its instance after those same transactions.

Let's say the first mapper reads the first instance for the *Computer Science* department. Subsequently, a transaction adds a new row to the first instance for the *Computer Science* department with a professor whose name begins with 's'. Simultaneously, another transaction adds a row to the second instance for the *Mechanical* department, also with a professor whose name begins with 's'. Finally, the mapper reading from the second instance performs its read.

The problem here is that the first mapper's read doesn't reflect either of the new transactions, while the second mapper's read includes both the transactions. This results in an inconsistent view of the data.

To achieve consistent reads, the underlying data store needs to be versioned and multi-row transactional. Google Spanner is an example of such a database. It globally orders all transactions in real-time. Consequently, when a query is executed in Spanner, all reads occur at a specific point in time, providing a consistent snapshot of the data (a.k.a snapshot isolation). Even with concurrent modifications, the data returned by the query will represent the state of the database at the chosen query time.

### 31.3. SQL Joins

A frequent operation in data processing involves combining two relations based on a related attribute, followed by further processing of the resulting dataset. **Joins** are therefore a fundamental operation in query engines.

Query engines employ several strategies to perform join operations, each optimized for different data characteristics.

#### *31.3.1. Lookup Joins*

These are particularly effective when joining a small table with a significantly larger one. The query engine reads the entirety of the smaller table and then performs targeted lookups in the larger table for matching rows based on the join condition, as shown in Figure 145.

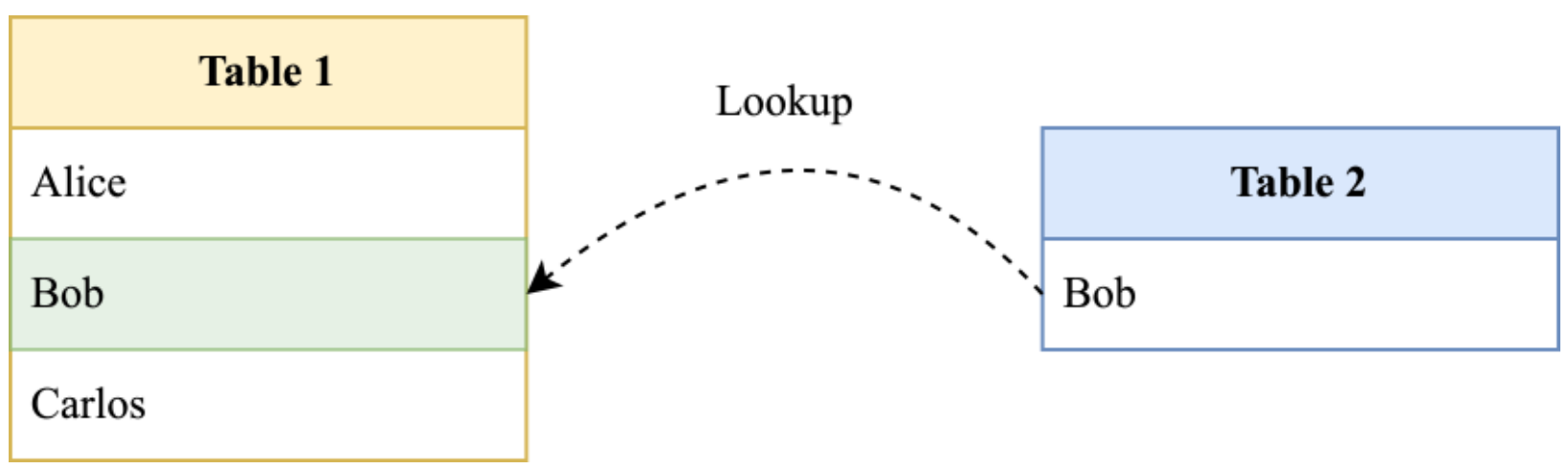

**Figure 145**: Lookup join.

**Note:** For efficient lookup joins, the larger table must be indexed on the columns specified in the join condition. This allows for rapid retrieval of matching rows.

### *31.3.2. Hash Joins*

Hash joins are a common and versatile join strategy that proceeds in the following steps (shown in Figure 146):

- The query engine reads both tables and identifies the smaller one.
- An in-memory hash table is constructed using the join key(s) from the smaller table. If the hash table exceeds available memory, portions of it may be spilled to disk.
- The larger table is then scanned row by row. For each row in the larger table, the join key(s) are used to probe the hash table of the smaller table to find matching rows.

### *31.3.3. Merge Joins*

Merge joins are efficient when both tables involved in the join are already sorted (or have indexes that enforce an ordering) on the columns participating in the equality join condition. The algorithm operates as follows (analogous to merge step in merge sort), shown in Figure 147:

- The query engine verifies the existence of suitable indexes on the equality columns of both tables.
- The rows of both tables are then merged based on the sorted order of the join columns.
- For each match found in the join columns, the corresponding rows are combined in the result. If there are multiple matching rows in either table for a given join key, a cartesian product of those matching rows is generated.
- If the join column values do not match, the row with the smaller value is discarded, and the next row from that table is considered. This process continues until all rows from both tables have been processed.

**Note**: The examples provided above illustrate the concept of an **inner join**. The underlying principles remain consistent across other join types like **outer**, **left outer**, and **right outer joins**. The key difference lies in how non-matching rows are handled. In contrast to inner joins, these other join types preserve non-matching rows from one or both of the tables involved in the join.

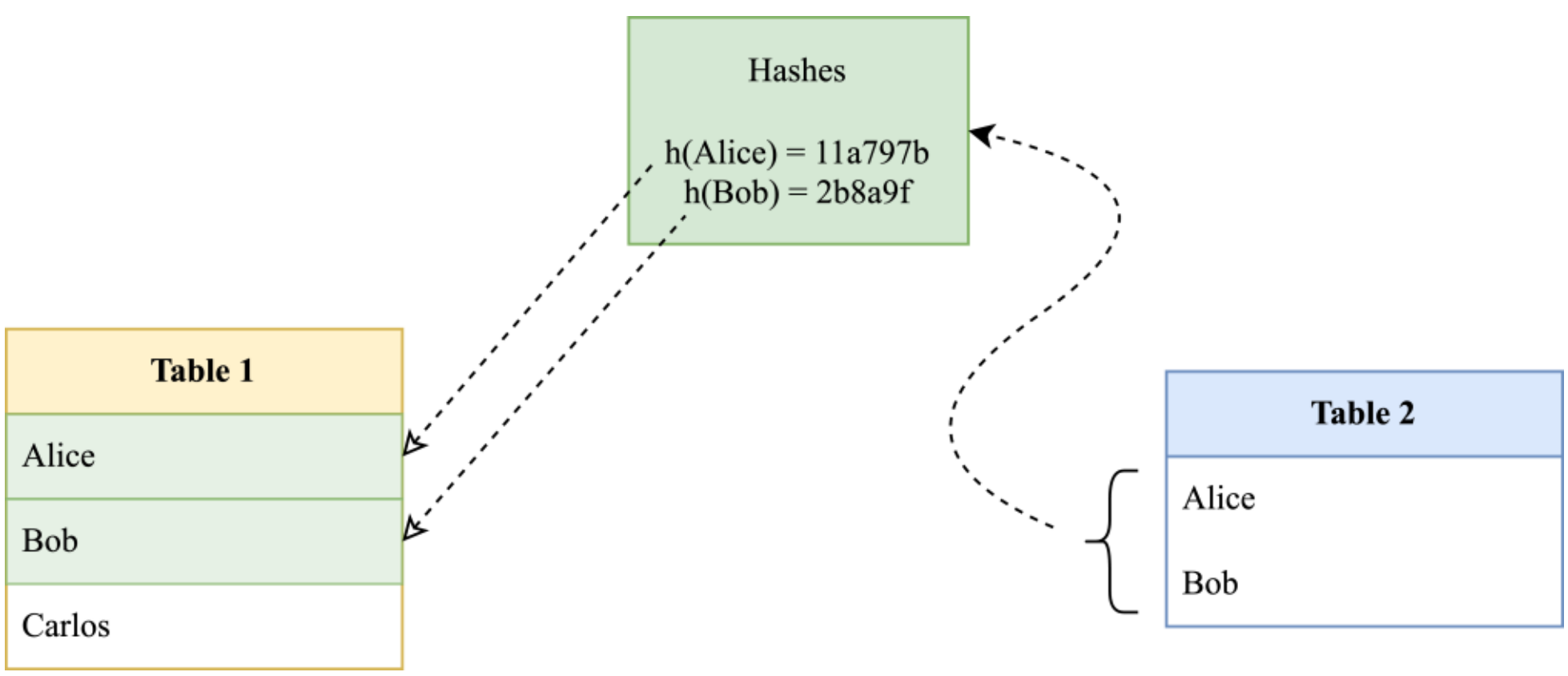

**Figure 146**: Hash join.

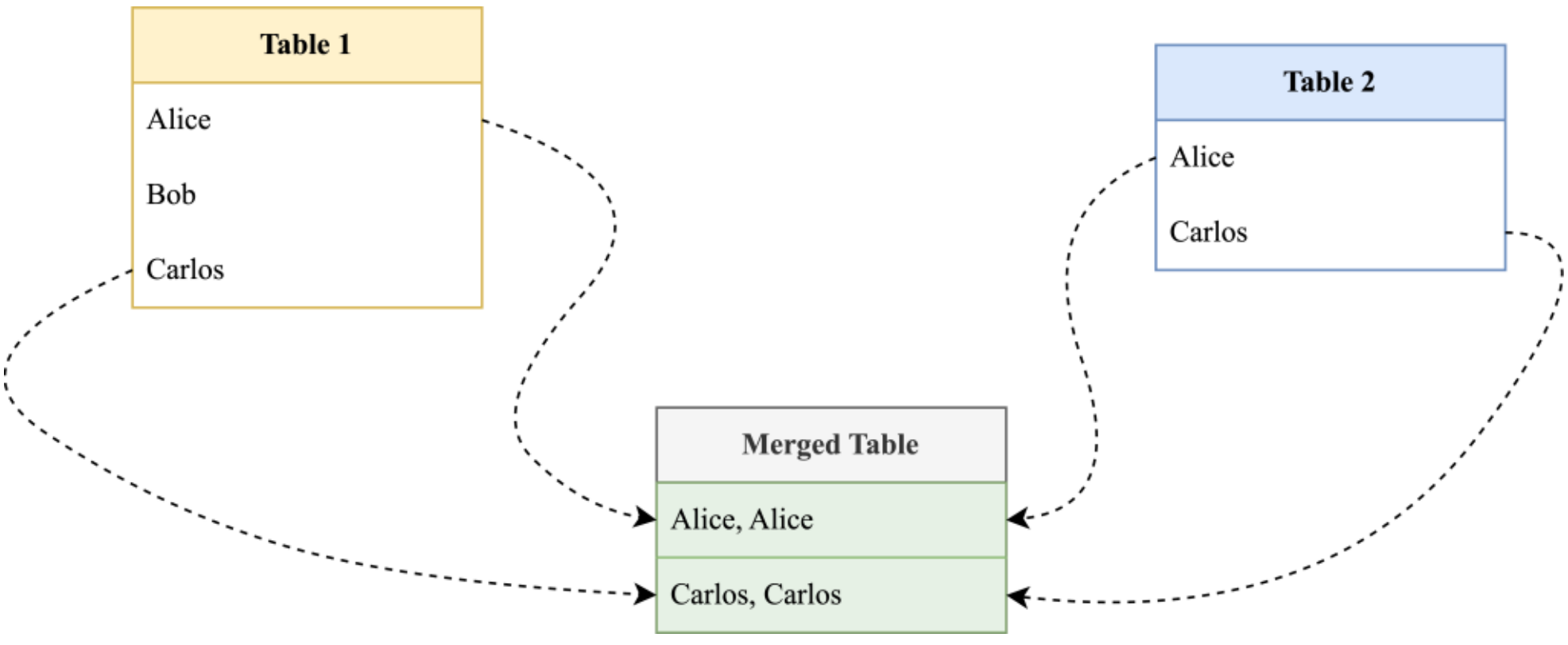

**Figure 147**: Merge join.

### 31.4. View Materialization

Once a query is executed, the resulting data is typically sent back to the client. However, query engines also often have the capability to **materialize** these results into a persistent table stored within the same database as the original data.

These materialized tables, commonly known as **materialized views**, offer significant performance benefits for subsequent queries that can leverage the pre-computed data.

A prime use case is the creation of roll-up tables in analytical databases, as shown in Figure 148. By materializing aggregated data into these roll-up tables, future analytical queries against this summarized information execute much faster.

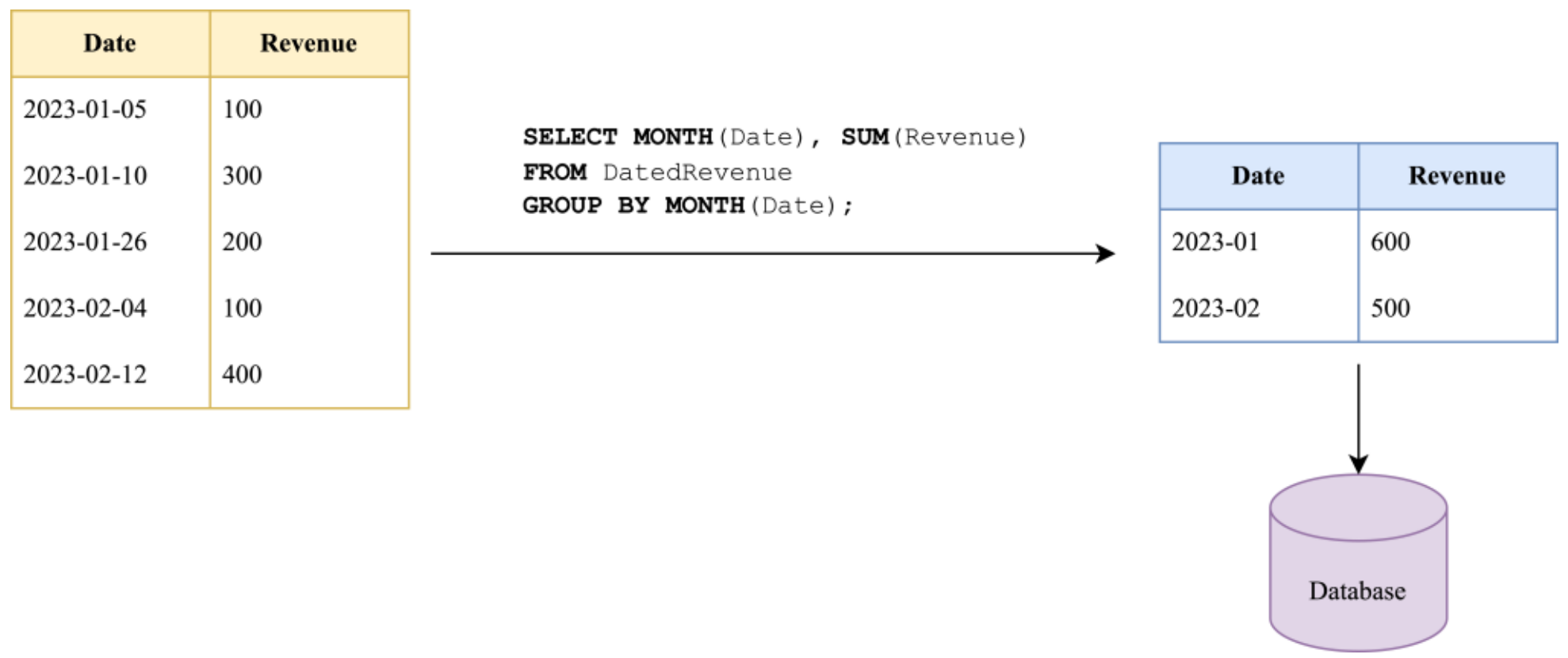

**Figure 148**: View materialization.

Another application arises when dealing with complex queries involving joins across multiple tables, which can be computationally expensive to run repeatedly. In such scenarios, materializing the result of the join into a physical table can reduce the execution time for subsequent identical or similar queries.

A key drawback is the need to refresh the materialized table whenever the underlying source data is updated. This requires re-executing the defining query to bring the materialized view in sync with the latest changes. Consequently, materialization is most beneficial for data that is relatively static, like those in data warehouses.

### 31.5. Query Optimization

Query optimizers employ several techniques to enhance query execution efficiency. Let's explore some key strategies:

#### *31.5.1. Filter Pushdown*

This optimization involves relocating filter conditions earlier in the query execution plan, ideally to the data source itself.

**Example**: Consider the query:

```
SELECT * FROM Customers WHERE City = "New York";
```

Without filter pushdown, the query engine would retrieve all customer records and then filter them at the client to identify those in *New York*. This approach can be inefficient, especially in distributed environments. Filter pushdown optimizes this by sending the `WHERE City = "New York"` condition directly to the data source. Consequently, when a worker reads from a database instance, it only receives the rows matching the filter.

*31.5.2. Constant Folding*

This technique involves evaluating constant expressions during the compilation phase rather than at runtime.

**Example**: The query:

```
SELECT 2 * 3 + 6;
```

is optimized to:

```
SELECT 12;
```

*31.5.3. Outer Join Narrowing*

This optimization aims to reduce the number of rows processed during an outer join operation by applying filters to the outer join condition.

**Example**: Consider the query:

```
SELECT * FROM (Customers c
LEFT JOIN Orders o
ON c.CustomerId = o.CustomerId)
WHERE o.OrderDate > '2023-01-01';
```

Although a left outer join is used to retrieve all **Customers**, the `WHERE o.OrderDate > '2023-01-01'` condition can be pushed down to the **Orders** table before the join operation. This limits the number of order records that need to be considered in the join.

*31.5.4. Sort Elimination*

Sorting is often necessary for SQL operators like **ORDER BY**, **GROUP BY**, **UNION**, and **INTERSECT**. Sort elimination strategies avoid unnecessary sorting by leveraging existing indexes or restructuring the query execution plan. For instance, if data is already sorted based on an index, an explicit **ORDER BY** clause on the same column might be redundant.

*31.5.5. Common Plan Deduplication*

A subplan represents a smaller, reusable component of a query that can be executed independently. By identifying and eliminating redundant subplans, the query optimizer can save processing time.

**Example**: Consider the following query:

```
SELECT *
FROM Products p
JOIN OrderDetails od ON p.ProductId = od.ProductId
JOIN Orders o ON od.OrderId = o.OrderId
WHERE o.OrderDate BETWEEN '2023-01-01' AND '2023-12-31' AND
p.Price > 100;
```

This can be optimized by extracting the common filtering condition on Products:

```
WITH FilteredProducts AS (
  SELECT *
  FROM Products
  WHERE price > 100
)
SELECT *
FROM FilteredProducts fp
JOIN OrderDetails od ON fp.ProductId = od.ProductId
JOIN Orders o ON od.OrderId = o.OrderId
WHERE o.OrderDate BETWEEN '2023-01-01' AND '2023-12-31';
```

In this optimized version, the filtering of products with a price greater than 100 is performed only once in the **FilteredProducts** common table expression, which is then used in the subsequent joins.

*31.5.6. Materialized View Rewrites*

If a materialized view contains the data needed to answer a query, the optimizer can rewrite the original query to directly use the materialized view, significantly reducing execution time.

*31.5.7. Constraint Propagation*

Query optimizers utilize declarative constraints defined on the database schema (such as primary keys, foreign keys, unique constraints, and non-null constraints) to infer properties about the data and reduce the search space for a query. For example, if a **JOIN** is performed on a foreign key column that is also defined as **NOT NULL**, the optimizer might be able to make assumptions about the presence of matching rows.

*31.5.8. Attribute Pruning*

This optimization eliminates the retrieval of unnecessary columns from the query execution plan.

**Example**: For the query:

```
SELECT p.ProductId, p.ProductName, o.OrderDate
FROM Products p
JOIN Orders o ON p.ProductId = o.ProductId;
```

The query optimizer will ensure that only the *ProductId*, *ProductName* from the **Products** table, and *OrderDate* from the **Orders** table are fetched during the execution, avoiding the overhead of retrieving other columns that are not required in the final result.

### 31.6. F1 Query

Google developed F1 Query as a powerful SQL query engine, initially for its internal F1 [184] database. Its origins lie in the demanding data landscape of Google Ads. The sheer scale of Ads data necessitated highly scalable infrastructure, and F1 Query emerged as a solution. F1 Query is the successor to Dremel, another pioneering data processing technology developed at Google.

This versatile engine can query data across a diverse range of Google's storage systems. It provides a unified SQL interface over OLTP databases like Spanner and F1, OLAP systems such as Napa/Mesa, NoSQL databases like Bigtable, and even simple tabular data in formats like Record I/O, Capacitor, CSV files, and Google Sheets. This cross-functional capability even allows joins between data residing in different systems like Spanner and Bigtable, although this might come with consistency trade-offs. Importantly, F1 Query presents all data as relational tables with standard data types, abstracting away the underlying storage format, even for non-relational data (e.g., representing graph data as tables). F1 Query's architecture is also extensible. Clients have the flexibility to define their own data source and sink adapters, tailoring it to specific needs.

Query results are typically returned directly to the client. However, F1 Query also supports materialization, allowing clients to persist query outputs as Record I/O or Capacitor files on Colossus. Furthermore, results can be temporarily sessioned for efficient reuse in subsequent operations.

### *31.6.1. Working with Structured Data*

Beyond standard SQL, F1 Query offers a rich set of functions for handling structured data - columns containing multiple primitive data types, conceptually similar to C++ structs. These primitive types include common ones like **int32**, **int64**, **string**, **bool**, and **float**. Structured data can then be composed of these, for instance:

```
struct OrderData {
  string name;
  string address;
  float total_micros;
};
```

Accessing specific fields within structured data is straightforward using the familiar dot (.) operator. For example:

```
SELECT OrderId, OrderData.name FROM Orders;
```

F1 Query also supports arrays of primitive types (e.g., **Array<int32>**) and complex nesting, such as arrays of structs or structs containing arrays. Handling arrays requires a bit more finesse. F1 Query provides the **UNNEST** operator for this purpose, which we will explore in more detail later.

Finally, F1 Query can also seamlessly work with popular serializable formats like protocol buffers as structured data types. XML and JSON are also supported but protocol buffers are far more efficient than them [216].

### *31.6.2. Extensible*

F1 Query SQL offers a rich feature set, but its flexibility extends further, allowing for custom enhancements through **user-defined functions (UDFs)**, **user-defined aggregate functions (UDAs)**, and **table-valued functions (TVFs)**.

These extensions can be implemented directly in SQL or leveraging other languages like the scripting language Lua.

31.6.2.1. User-Defined Functions (UDFs)

UDFs are scalar functions that operate on individual projected values. The term "scalar" implies that these functions are stateless, meaning their output depends solely on the input value for each row. Consider this example:

```
SELECT DATE(timestamp) FROM Orders;
```

Here, **DATE** is a UDF that transforms the timestamp into a more readable date format.

31.6.2.2. User-Defined Aggregate Functions (UDAs)

In contrast to stateless UDFs, UDAs are stateful functions, typically implemented with **Initialize**, **Accumulate**, and **Finalize** steps. Notably, UDAs can be constructed using UDFs that share a persistent state across rows.

```
SELECT COUNT_DISTINCT_DATE(*) FROM Orders;
```

Here, **COUNT_DISTINCT_DATE** represents a UDA designed to count the number of distinct dates within the **Orders** table.

31.6.2.3. Table-Valued Functions (TVFs)

TVFs differ significantly by accepting an entire relation (a table or the result of a subquery) as input and producing another relation as output. Consequently, TVFs are typically used within the **FROM** clause of a query.

```
SELECT * FROM EXTRACT_LAST_DAY(Orders);
```

In this case, **EXTRACT_LAST_DAY** is a TVF that returns a table containing orders placed on the last day. Note that even TVFs ultimately expand into standard SQL queries. This underlying SQL representation allows the entire query, including the expanded TVF logic, to be subject to F1 Query's powerful optimization engine.

```
SELECT * FROM (SELECT * FROM Orders WHERE DATE(timestamp) =
CURRENT_DATE());
```

### *31.6.3. Architecture*

The overall architecture is shown in Figure 1 in the paper. Users interact with F1 Query via a **client library**, which sends requests to a pool of dedicated servers known as **F1 servers**. A central **F1 master** orchestrates all F1 servers.

Each F1 server takes on the responsibility of processing incoming queries, which includes optimization. The optimized query is then broken down into a series of tasks that are distributed to **F1 workers** for execution. Notably, both F1 servers and F1 workers are designed to be stateless.

The F1 workers are strategically distributed across multiple data centers. To optimize data access, if a worker initially receives a query for a dataset that isn't locally available, the client is transparently redirected to another worker that is geographically closer to the data's source and destination.

The F1 workers make use of **UDF servers** which store all the custom functions. UDF servers process bulk inputs to amortize the cost.

### *31.6.4. Query Execution Phases*

The execution of a query in F1 Query proceeds through distinct phases, shown in Figure 2 in the paper:

- **Optimization**: The raw query from the user is optimized using various techniques discussed previously.
- **Physical Plan**: The optimized query is converted to a physical execution plan, for example, map-reduce tasks.
- **Execution**: The tasks are executed according to the selected mode - **interactive** or **batch**.

### *31.6.5. Operators*

The F1 workers can be conceptually viewed as MapReduce counterparts, with their functionalities aligning with standard SQL operators:

- **<u>SCAN</u>:** This operator is responsible for reading data directly from the underlying data source. It functions similarly to a **Map** in the MapReduce paradigm.

- **AGGREGATION:** This operator performs data aggregation, potentially based on specific keys. This mirrors the role of a **Reduce** in MapReduce.
- **SORT:** This operator handles the sorting of data streams originating from upstream operators. In the MapReduce analogy, this corresponds to a **Shuffle**.

The complete execution pipeline for a given query is constructed as a tree of these F1 operators, with the corresponding F1 workers distributed as needed to perform each stage of the processing.

**Example:** Figure 3 in the paper provides an example of a query and how it translates to the different operators.

*31.6.6. Execution Mode*

Two execution modes exist: **interactive** and **batch**. Interactive execution is well-suited for applications that issue individual, short-lived queries. Conversely, batch execution is intended for applications that execute longer, recurring queries.

31.6.6.1. Interactive Execution

This mode prioritizes speed by keeping all data in memory and transmitting results through RPC streams. If a failure occurs, the client library automatically retries the entire query. Interactive execution can be implemented in two ways:

- **Single-threaded:** A single process handles the complete query execution, with all operators running within a single thread.
- **Distributed:** The query execution plan is broken down into **fragments** and executed by a cluster of F1 worker processes. Multiple F1 workers can be involved at each stage of the query execution.

When a query is executed across multiple F1 workers, for each operator, the input data must be divided among them. This partitioning is handled differently depending on the operator:

- **SCAN Operator Partitioning:** The input for the SCAN operator is divided into N partitions based on the table's primary key. For example, a range-based query might have each partition correspond to a specific segment of the row space. These partitions are then distributed to the available F1 workers.

- **AGGREGATION Operator Partitioning:** The input for the AGGREGATION operator is partitioned based on the key(s) being aggregated, which may or may not be the same as the table's primary key. The SCAN operator is aware of the aggregation key and routes its output to the correct AGGREGATION worker instances. Furthermore, the SCAN operator may perform some initial aggregation steps before sending the data.

31.6.6.2. Batch Execution

Many ETL pipelines leverage F1 Query in batch execution mode for continuous data extraction. In this mode, queries are executed reliably using the MapReduce framework. Given that these queries are expected to be long-running, potentially taking several hours to complete, it is essential to checkpoint the progress to prevent data loss in the event of worker failures. This feature is provided by the MapReduce framework.

Another key difference between interactive and batch modes lies in worker activation. In interactive mode, all F1 workers are immediately active to process results from downstream operators. Conversely, in batch mode, workers become active only upon the completion of the downstream operators.

Long-running queries need to be stored in a registry, known as the **query registry**, which is built on top of Spanner. A **query scheduler** then schedules these queries on a **query executor** that manages the MapReduce worker pool. This flow is shown in Figure 7 in the paper.

### 31.7. Join Processing

Joins are fundamental to any query engine. F1 Query implements several types of joins: **lookup**, **hash**, and **merge joins**, and also supports **array joins** for structured data.

### *31.7.1. Array Join*

Array joins are common in query engines that support structured data. When a structured column contains an array of data, it often needs to be unfolded. An example is shown in Figure 149.

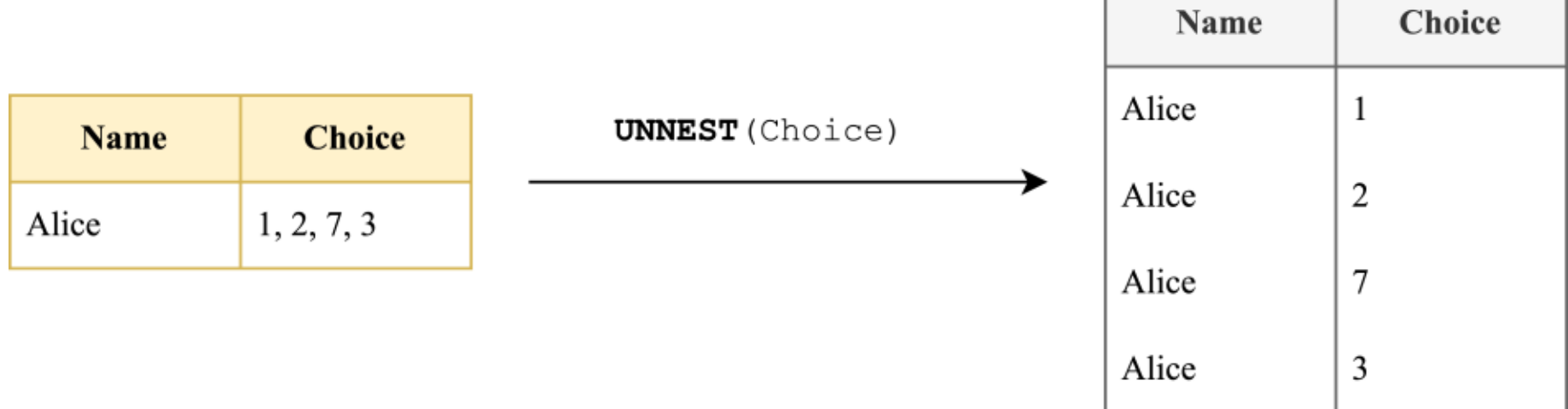

**Figure 149**: Array join.

### *31.7.2. Join Type Selection*

F1 Query infers the optimal join strategy based on the tables involved. The original F1 database paper [184] illustrates the F1 Query's join type selection process.

Again, note that, F1 was built to handle Google Ads' immense data demands - driven by the planet-scale traffic it serves (virtually every web page visit triggers multiple queries to Ads servers).

Consider the following schema:

- **Creative** (<*CustomerId*, *CreativeId*>): Stores all creative assets (text, images, etc.) associated with a customer.
- **AdgroupCreative** (<*CustomerId*, *CampaignId*, *AdgroupId*, *CreativeId*>): Represents an ad, linking a creative to a specific ad group within a campaign for a customer. An ad is uniquely identified by <*AdgroupId*, *CreativeId*> within a <*CustomerId*, *CampaignId*> shard.
- **AdClick** (<*AdgroupId*, *CreativeId*, *ClickId*>): Records ad clicks, with each ad potentially having numerous clicks.

Let's examine the join operation: <**AdClick**> ⊕ <**AdgroupCreative**> ⊕ <**Creative**>

**Step 1**: <**AdClick**> ⊕ <**AdgroupCreative**>: F1 Query employs a lookup join here as the **AdClick** table will only have a subset of ads that were clicked. The system looks up corresponding <*AdgroupId, CreativeId*> entries in the **AdgroupCreative** table.

**Step 2**: <Step 1> ⊕ <**Creative**>: A hash join is utilized in this step. The **Creative** table is assumed to be readily accessible in memory, making a hash join an efficient choice for this final stage.

*31.7.3. Example Execution*

The paper offers an example of how a join is processed by F1 Query, specifically joining **Ads** and **Clicks** tables, as shown in Figure 5 in the paper.

The **Ads** table is stored in an OLTP F1 database, while **Clicks** are stored in an OLAP Mesa database. Separate F1 workers are designated to extract data from these sources. A dedicated F1 worker, resembling a reducer, receives the streamed results. These workers are sharded to aggregate values based on a specific key.

In this case, F1 Query decides to use a hash join. The hash of the **Ads** table is stored in-memory in a hash table. When the clicks are streamed back from F1 workers on Mesa, they are joined in-memory with the Ads data and then aggregated. Finally, the aggregated results are streamed back to the F1 server.

*31.7.4. Performance Considerations*

In hash joins, partitioning the hash space across different workers can lead to skew. This occurs because the distribution of data might cause some hash slices to receive significantly more results than others, making some workers run hotter. To mitigate this, F1 Query supports **broadcast hash join**, which copies the entire hash to all workers, eliminating partitioning.

In lookup joins, the lookup happens directly within the same fragment. For instance, a SCAN fragment will perform the lookup on the other (larger) table. Since each worker will look up the larger table, it's efficient to perform a bulk lookup of all keys derived from the smaller table. However, skew can still occur if some partitions from the smaller table join with too many values in the larger table. The query optimizer can dynamically repartition the input on the fly. **Dynamic range repartitioning** is beneficial and often outperforms static partitioning strategies.

### 31.8. F1 Query Optimizer

The F1 query optimizer is a notably intricate part of the F1 query system with several stages as shown in Figure 150.

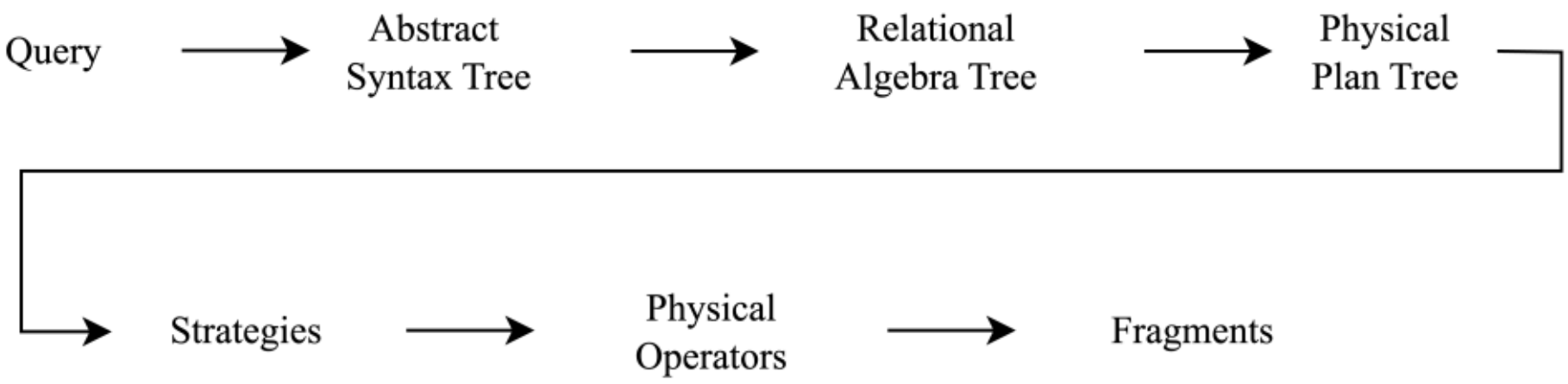

**Figure 150**: Query optimization stages.

In essence, the query optimizer takes the initial query and builds an **abstract syntax tree** (AST). It likely then uses context-free grammar [217] rules to convert this AST into a **relational algebra plan**. This plan is then optimized through a series of rewriting steps. As mentioned earlier, F1 query utilizes several rewrite rules, such as filter pushdown, constant folding, attribute pruning, constraint propagation, outer join narrowing, sort elimination, common subplan deduplication, and materialized view rewrites.

The resulting optimized relational algebra plan tree is converted to a **physical plan tree** that is organized into modules called **strategies**. Each strategy is responsible for converting the logical operators from the relational algebra into concrete **physical operators**, where each physical operator is represented by a class that manages its data.

Ultimately, the physical operators are assigned to a set of fragments ready for execution.

### 31.9. Production Numbers

Section 8 offers a glimpse into the real-world application of F1 Query within Google:

- Interactive queries executed in centralized mode demonstrate impressive performance, with 99% completing in under one second. The system handles a substantial load of 500k interactive queries per second, totaling 40 billion queries daily. Interactive queries running in distributed mode typically experience longer processing times.

- Batch processing handles approximately 55,000 queries each day. However, the sheer volume of data processed by these batch jobs is immense, scaling to tens of petabytes daily. While most batch queries finish within minutes, some can take several hours to complete due to the data intensity.
- Despite a rapid doubling of the query volume every quarter, the execution latency for interactive queries has remained stable. This consistent performance underscores the remarkable scalability of the F1 Query system.

### 31.10. Paper Remarks

This paper represents a significant challenge, requiring multiple readings to fully grasp its intricate details. The density of the text, combined with the specialized nature of query engine literature, creates a steep learning curve for those new to the database domain. However, the depth of knowledge gained from a thorough study of this material is immense. This paper is highly recommended for anyone interested in the engineering behind Google Ads - one of the world's most massive software systems - and the methods used to tackle Big Data challenges at a planetary scale.

# 32. Napa: Powering Scalable Data Warehousing with Robust Query Performance at Google

The previous chapters provided an extensive study of databases as essential stores for structured data. This chapter examines data warehouses and their role in managing analytical data. Data warehouses are fundamental components of the modern data landscape; with the rise of data analytics, they serve as the primary repositories for historical, aggregated, and multi-dimensional data used for complex trend analysis and strategic decision-making.

Napa represents the next generation of planet-scale data warehousing at Google, succeeding Mesa [218]. As a critical infrastructure for analytics workloads, Napa stores enormous datasets for various internal tenants. The extensive list of contributors to the research underscores the massive collaborative effort required to build such a system. This paper was presented at the Very Large Data Bases (VLDB) 2021 conference.

In this insight, we will go through the foundational data warehouse concepts, including underlying data models and the physical formats used for storage. Then we will explore the Napa architecture, detailing the various performance-cost-consistency tradeoffs the system supports and the specific optimizations it employs to maintain efficiency at scale.

### 32.1. Data Warehouse

Data warehouses are SQL-compatible databases designed primarily for analytical data storage. Data warehouses, like SQL databases, offer data consistency. However, they typically aren't updated in real-time. In fact, most data warehouses rely on ETL pipelines to feed data into them.

Consider the following example of a data warehouse that contains two relational tables.

**HourlyRevenue:** This table stores the revenue generated from orders received each hour. ETL pipelines periodically update this table by pulling data from various sources like logs and OLTP databases, as shown in Figure 151. These pipelines tail the sources such that it ensures that duplicate updates are avoided. Importantly, because updates are periodic, the data warehouse might not reflect the very latest hour's revenue in real-time.

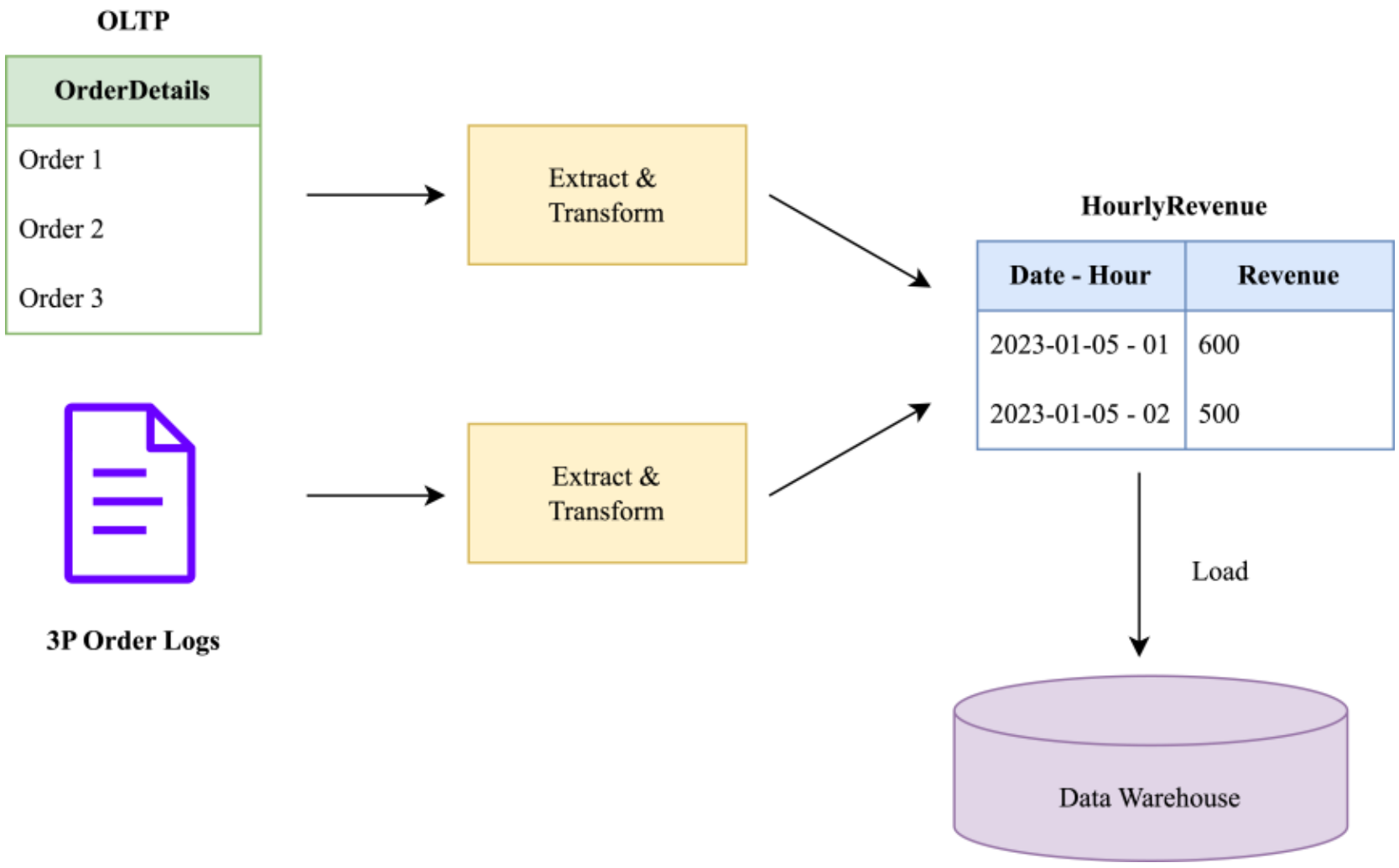

**Figure 151**: ETL pipeline for HourlyRevenue.

**DailyRevenue:** Similar to **HourlyRevenue**, this table stores daily revenue figures. To maintain consistency, these daily numbers are essentially materialized views derived from the **HourlyRevenue** table. There are two main approaches to materializing these views:

- **Real-time Materialization:** Any update to the **HourlyRevenue** table immediately updates the **DailyRevenue** table, ensuring strong consistency between the two. This technique is rare as it can slow down the ingestion pipeline.
- **ETL-based Materialization:** A separate ETL pipeline aggregates the hourly revenue data for each day and updates the **DailyRevenue** table, as shown in Figure 152.

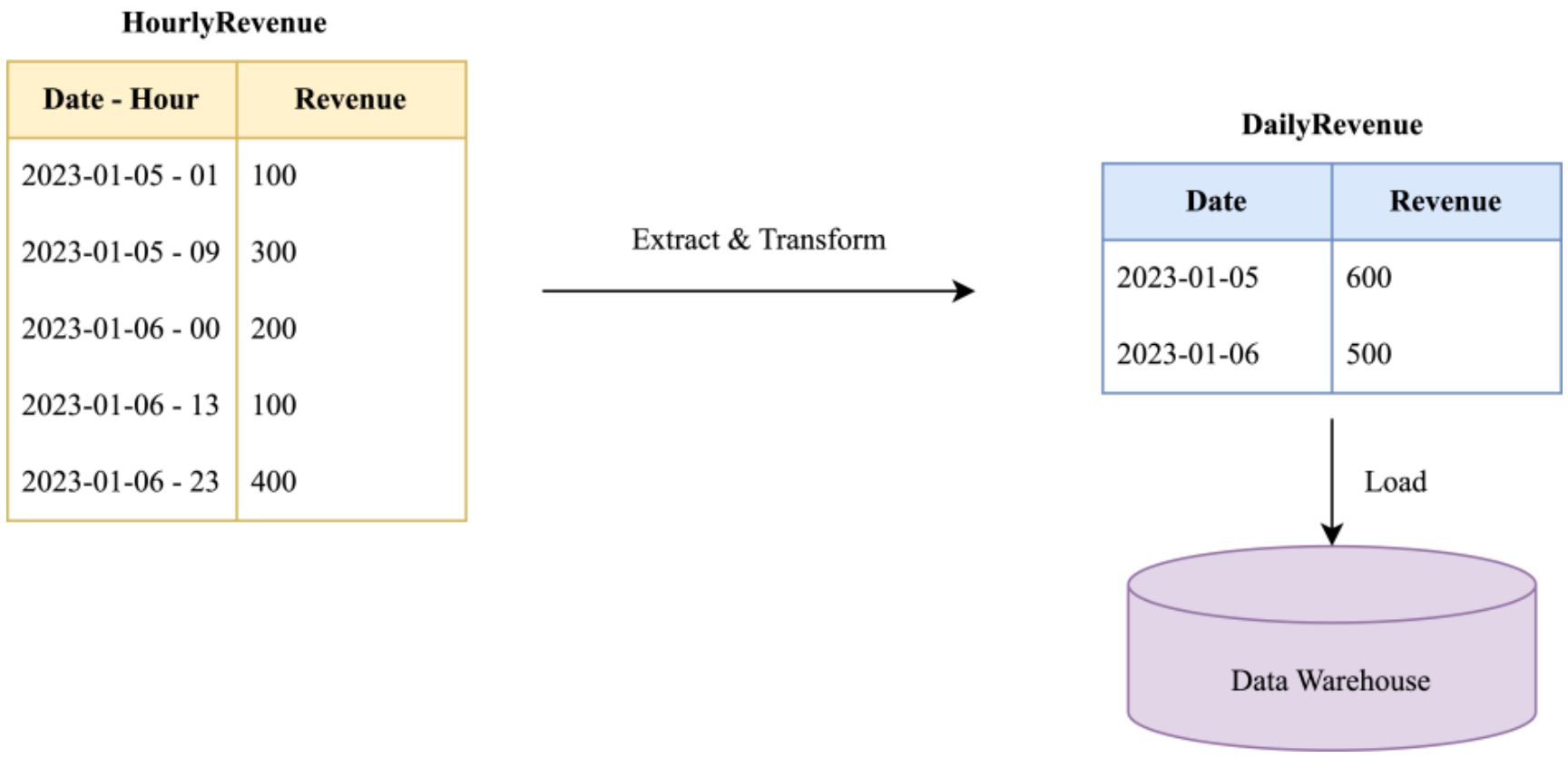

**Figure 152**: ETL pipeline for DailyRevenue.

Note that data warehouses are designed to typically store numeric data suitable for aggregation, rather than arbitrary data types like structs. For storing diverse data types, other database solutions like SQL or NoSQL databases are more appropriate.

### *32.1.1. Data Model*

Let's explore how data is represented and stored within a data warehouse. The section summarizes a few crucial points; one may find more details in the Mesa paper [218] by Google.

Data is organized into tables. The tables are not strictly relational, instead, the underlying model leans towards a key-value structure. Both key and value can encompass multiple columns. But crucially, value columns must be of an **aggregatable** type, typically numeric (such as integers or floats).

Data warehouses almost invariably have aggregated views built on top of some base tables. Consider an hourly table with columns **<Date-Hour, Revenue>**. A common pattern is to have a corresponding **<Date, Revenue>** table. This daily revenue table effectively drops the "Hour" granularity and aggregates the revenue across all hours within each day. Such an aggregated table is called a **rollup table**, while the original, more granular table is referred to as the **drilldown table**.

To facilitate the creation of these rollup tables, the data model must define an **aggregator function** for each value column. This function specifies how values should be combined when aggregating across different key values.

Formally, an aggregator function $\mathbf{F}: \mathbf{V} \times \mathbf{V} \rightarrow \mathbf{V}$ typically exhibits the following properties:

- Associative: $\mathbf{F}(\mathbf{v_0}, \mathbf{F}(\mathbf{v_1}, \mathbf{v_2})) = \mathbf{F}(\mathbf{F}(\mathbf{v_0}, \mathbf{v_1}), \mathbf{v_2})$
- Commutative: $\mathbf{F}(\mathbf{v_0}, \mathbf{v_1}) = \mathbf{F}(\mathbf{v_1}, \mathbf{v_0})$

This concept extends to tables with multiple value columns. Consequently, for a table with **N** value columns, we would have **N** corresponding aggregate functions.

*32.1.2. Differential Storage Scheme*

The core principle behind analytical data storage is often to record data as a sequence of changes, or deltas.

For instance, imagine an ETL pipeline runs and captures the revenue for hour **X**. Let's say the first run records **X → $100**. When the ETL pipeline runs again, it detects new orders (not present in the previous run) that contribute to the revenue for hour **X**. This second run now reports **X → $150**.

Instead of simply overwriting the previous value, the data can be stored as a series of deltas:

**X → +$100, +$50**

This method of storage is known as a **differential storage scheme**.

*32.1.3. Deltas*

A delta is essentially a file that contains keys along with the corresponding change in value for each column.

Each time an ETL pipeline runs to bring data into the data warehouse, it generates a new delta. This newly created delta is then applied to both the base table and any associated materialized views. Figure 153 shows an example of deltas for **HourlyRevenue**.

**Delta 1**

| Date - Hour | Revenue |
|---|---|
| 2023-01-05 - 08 | +50 |
| 2023-01-05 - 09 | +30 |

**Delta 2**

| Date - Hour | Revenue |
|---|---|
| 2023-01-05 - 07 | +10 |
| 2023-01-05 - 08 | +100 |
| 2023-01-05 - 09 | +70 |

**Delta 3**

| Date - Hour | Revenue |
|---|---|
| 2023-01-05 - 09 | +100 |

**Figure 153**: Deltas for HourlyRevenue.

The entire analytical database is therefore built upon a sequence of these deltas. These deltas are often organized in a structure similar to a Log-Structured Merge (LSM) tree. This involves a hierarchy of delta files, where higher levels contain progressively larger deltas formed by merging smaller deltas from lower levels, as shown in Figure 154. The merging process for each key utilizes the aggregator functions defined earlier. The largest, merged delta residing at the bottommost level is referred to as **base delta/table**.

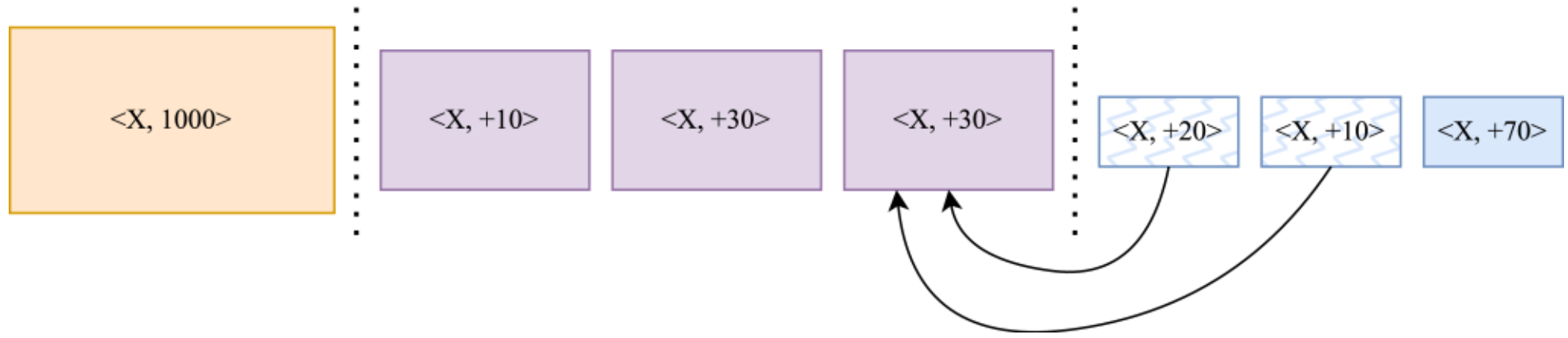

**Figure 154**: LSM tree from deltas.

A significant advantage of this delta-based approach is the ability to execute a query against any prefix of the delta sequence. This will yield a consistent snapshot of the data as it existed at that specific point in time.

Each table independently maintains its data deltas. Consequently, within the database, every table has its own LSM tree structure. Collectively, these individual LSM trees constitute what can be described as a **log-structured merge forest**.

### *32.1.4. Physical Format*

Delta files are typically organized in a columnar format. Within a delta file, the keys are segmented into blocks. Each of these blocks is then stored column by column, often employing compression techniques to reduce storage space.

Alongside each delta data file, an index file is maintained. This index provides pointers to the different blocks within the corresponding data file. Furthermore, to accelerate query performance, each data file block also stores the actual keys present within it. The physical format of deltas is shown in Figure 155.

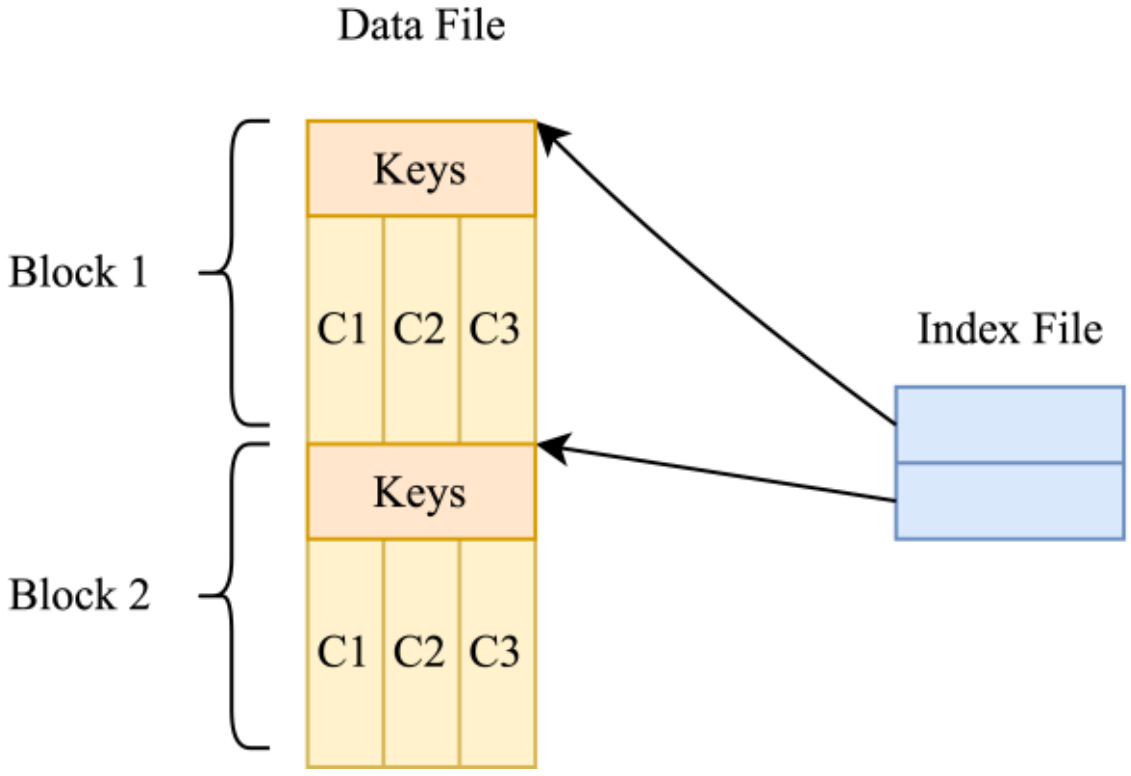

**Figure 155**: Physical format of deltas.

The lookup process then involves:

1. Performing a binary search on the index file to locate the relevant block.
2. Conducting a binary search within the key list of the identified data file block.

*32.1.5. Consistency*

Applying incremental updates (deltas) to base tables and their associated materialized views can occur asynchronously. Tables might be updated before their corresponding roll-up views. Additionally, in sharded data warehouses replicated across clusters, updates can arrive at different shards independently and out of logical order.

Despite this potential for temporary inconsistencies, it's crucial that queries return consistent results. For instance, the total daily revenue calculated by aggregation from an hourly table must always align with the revenue recorded in the daily summary table, even though the daily table is updated offline periodically.

A straightforward approach to ensure this consistency is to restrict queries to the point where all relevant deltas have been applied across all shards, tables, and views, as shown in Figure 156. This simplification is viable because data warehouses do not need real-time updates, which significantly eases their design compared to distributed SQL databases.

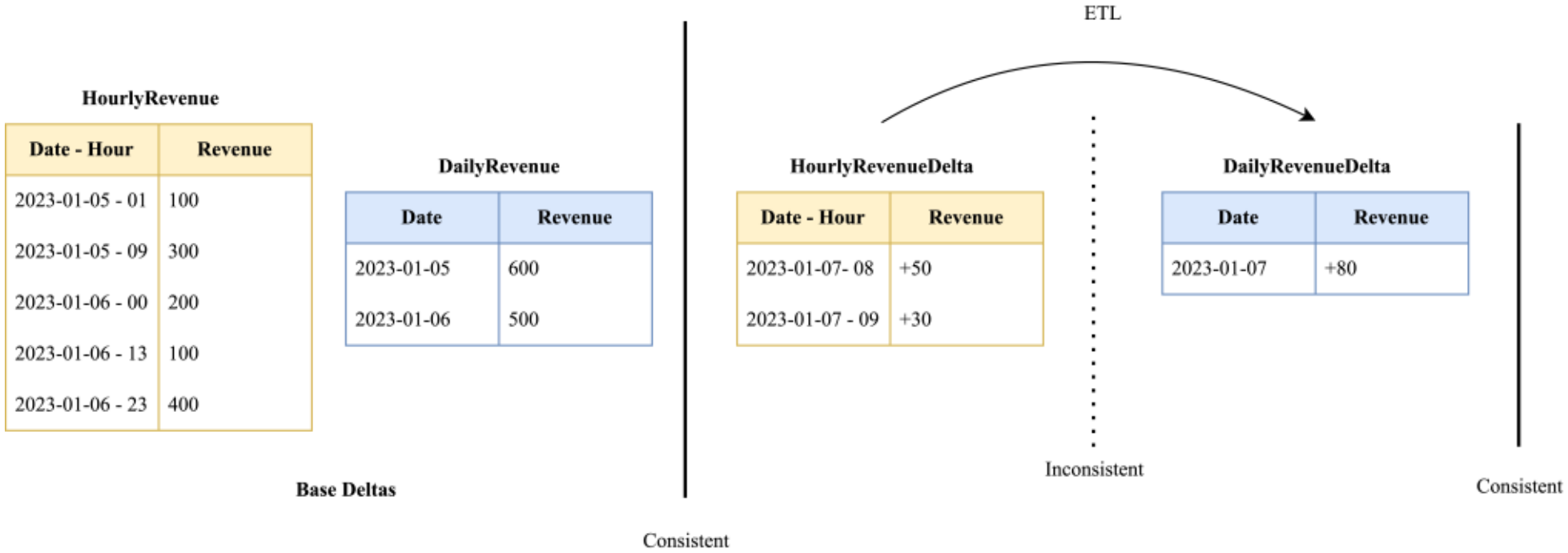

**Figure 156**: Points of consistency in data warehouses.

In Figure 156, the consistent data points are queryable. Conversely, the inconsistent point, lacking the necessary updates for the corresponding daily table view, cannot be queried.

### 32.2. Napa

Napa, Google's next-generation, large-scale data warehouse, builds upon the foundation of Mesa. Its architecture divides data processing into distinct stages, as shown in Figure 157.

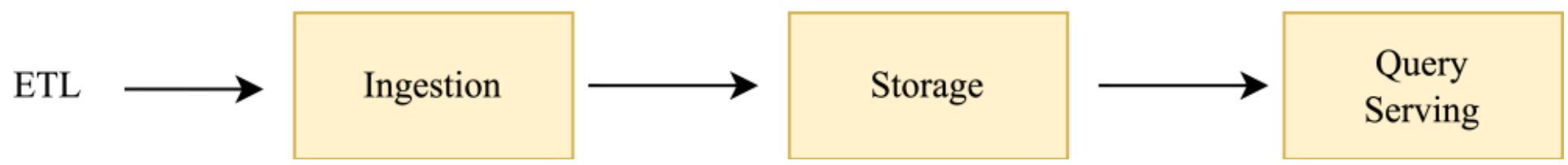

**Figure 157**: Processing stages in Napa.

This decoupling is strategically designed to optimize for a primary objective: achieving low-latency and consistent query performance.

The ingestion phase in Napa prioritizes high throughput, generating data deltas rapidly. However, unmerged deltas can become inefficient and can significantly increase query costs.

Ultimately, Napa aims to strike a balance between three crucial factors:

1. **Data Freshness:** The speed at which updates become available for querying.
2. **Query Latency:** The speed at which queries are executed and results are returned.
3. **Cost:** The resources expended to ensure timely data updates.

Achieving all three simultaneously - **high freshness**, **low query latency**, and **low cost** - is inherently challenging. The trade-offs are as follows:

- **High Freshness + Low Cost = High Query Latency:** Prioritizing immediate data availability with limited resources leads to slower query execution.
- **Low Cost + Low Query Latency = Low Freshness:** Focusing on efficient querying with minimal expenditure necessitates delaying data updates.
- **High Freshness + Low Query Latency = High Cost:** Delivering real-time data with fast query responses requires significant resource investment.

Sacrificing data freshness allows for low-latency and cost-effective querying. Traditional data warehouses, including Mesa, exemplify this by batching updates and applying them periodically (e.g., hourly). This consolidation optimizes data for querying; for instance, merging all deltas for a table reduces the need to access multiple delta files during lookups. Consequently, updates are intentionally delayed to enhance query performance and minimize costs.

Conversely, prioritizing high ingestion rates can compromise query performance. The ETL pipeline operates at a high frequency, resulting in queries needing to process and combine data from numerous deltas, thus increasing latency.

The paper also suggests that data warehouses might opt to sacrifice consistency. However, this assertion is questionable, particularly considering that many applications query multiple tables concurrently, making data consistency a critical requirement.

#### *32.2.1. Queryable Timestamp*

The **queryable timestamp (QT)** marks the point in time up to which all data deltas for a table have been consistently applied and are ready for querying. In essence, QT signifies the data freshness available for querying on a given table, even if it's distributed across multiple storage replicas.

The freshness of a table can be quantified as the duration between the current time and the queryable timestamp: $\mathbf{Now}() - \mathbf{QT}$. This is shown in Figure 5 in the paper.

The QT value advances from a previous point ($\mathbf{X}$) to a new point ($\mathbf{Y}$) when the data ingested within the time interval ($\mathbf{Y} - \mathbf{X}$) has undergone optimization to meet the desired query performance standards.

Many client queries come with specific latency requirements, which, in turn, restrict the number of data deltas that can be accessed and merged to generate a response within the acceptable timeframe. Considering these constraints, the QT represents the most recent delta such that querying all deltas from the oldest up to this point will adhere to the stipulated latency limits and complete within the expected duration.

Note that a QT is maintained individually for each table. The QT of an entire database (when consistency across all its tables is a prerequisite) is determined by the minimum QT value among all the tables within that database.

*32.2.2. Architecture*

Napa is a distributed data warehouse; indeed, the immense scale required to serve Google necessitates a distributed architecture.

Napa's architecture is logically separated into two distinct planes: the **data plane**, which handles the storage and retrieval of the actual data, and the **control plane**, responsible for managing the associated metadata, essentially QT.

**Ingestion servers**, residing within the data plane, receive data from ETL pipelines. This incoming data is then transformed into deltas and persisted on Colossus (formerly the Google File System).

Once data ingestion is complete, the QT for the affected table is advanced in the control plane. This QT metadata is durably stored in Spanner, Google's globally consistent distributed SQL database.

When the F1 query engine receives a query, it first consults the control plane to determine the optimal QT. This QT represents the most recent point in time up to which data can be consistently read from the database while adhering to the query's latency requirements. Subsequently, the query engine sends the request to **delta servers** that utilize the data deltas up to this determined QT to execute the query.

The entire flow is shown in Figure 158.

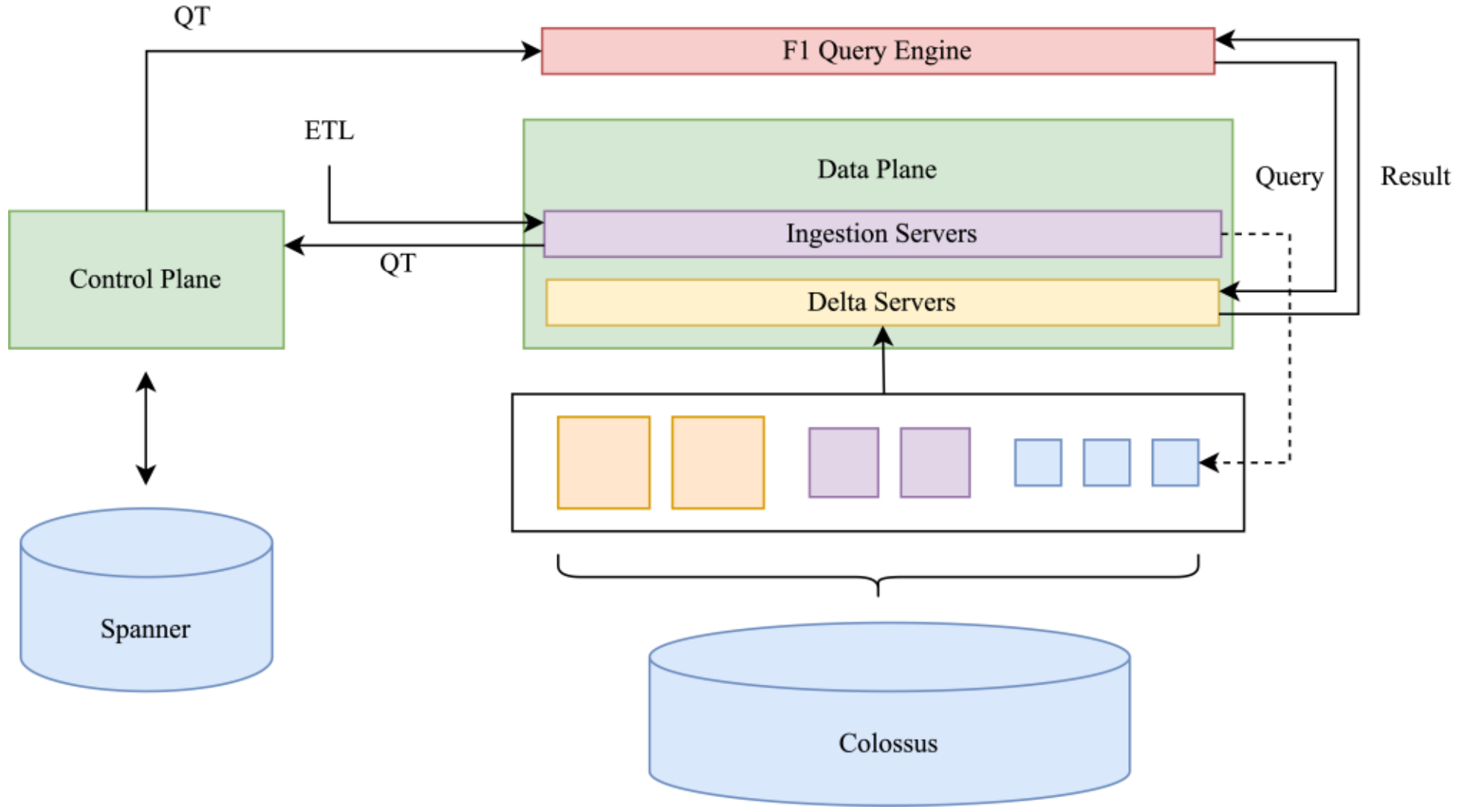

**Figure 158**: Napa's architecture.

### 32.2.3. Storage Structure

The underlying storage structure mirrors that of Mesa. Data for each table is organized as a collection of deltas, and these deltas across all tables in a database collectively form an LSM forest.

The primary storage format for these deltas is columnar, where retrieving each column necessitates a separate I/O operation. For smaller deltas, an optimized format called PAX [219] (Portable Archives) is utilized, enabling all records to be read with a single I/O operation.

### 32.2.4. View Maintenance

From this point forward, the paper delves into significant technical detail, particularly concerning Napa's multi-layered optimizations for achieving high performance. Understanding some of these intricate strategies likely requires expertise in database internals.

Upon a table update, downstream views are asynchronously refreshed using the F1 query engine. Note that Napa views can be formed by joining multiple upstream tables. Because a database's QT only advances when the QTs of all its derived tables and views also advance, an efficient pipeline is crucial for maintaining data consistency.

#### 32.2.4.1. Data Skew

Parent tables are typically sharded based on their key or a key prefix. However, views often operate on a reduced keyspace due to the exclusion of certain key columns and the aggregation of value columns. This reduction can lead to data skew in the views.

Consider a parent table with the key columns **<A, B, C, D>** and a view with the key **<B>**. If the parent table's sharding is based on the prefix **<A>**, and the majority of entries have **B=1** or **B=2**, then the view will likely experience hotspotting on the shards corresponding to these frequent B values.

#### 32.2.4.2. View Sort Order

The materialization of a view requires different sorting strategies depending on the relationship between the view's key and the base table's key:

- **Prefix Key:** If the view's key is a prefix of the base table's key (e.g., base key **<A, B, C, D>**, view key **<A, B>**), no sorting is necessary.
- **Partial Prefix Key:** If the view's key is a partial prefix (e.g., base key **<A, B, C, D>**, view key **<A, B, D>**), the non-prefix components of the view's key (i.e., **D**) need to be sorted before reduction. However, this scenario might not introduce significant sorting requirements.
- **Non-Prefix Key:** If the view's key shares no prefix with the base table's key (e.g., base key **<A, B, C, D>**, view key **<D, C, A>**), substantial sorting is required.

To understand the necessity of sorting in the non-prefix case, consider the transformation **<A, B, C, D>** → **<D, C, A>** where aggregation occurs on **B**. Consider the case of following base table keys and their transformation:

<1, 1, 1, 1> → **<1, 1, 1>**
<1, 1, 1, 2> → <2, 1, 1>

<1, 1, 1, 3> → <3, 1, 1>

...

<1, 2, 1, 1> → **<1, 1, 1>**

...

The key **<1, 1, 1>** appears multiple times after transformation, potentially, in different shard locations. Consequently, the view must be sorted based on its key **<D, C, A>** before it can be correctly constructed and aggregated.

### *32.2.5. Compaction*

The compaction process merges **N** input deltas through an N-way merge. The merging operation stops as soon as any one of the **N** deltas is fully processed (i.e., all its keys have been merged).

### *32.2.6. Query Serving*

Napa employs several key optimizations to accelerate query performance. Let's examine them individually.

#### 32.2.6.1. View-Based Querying

Napa preferentially uses pre-computed views over base tables to serve query results whenever feasible. This can significantly reduce the computational overhead of query execution.

#### 32.2.6.2. Filter Pushdown

The F1 query engine pushes query filters down to the Delta servers. This minimizes the volume of data transferred across the network, improving efficiency.

#### 32.2.6.3. B-tree Indexing for Delta Pruning

Napa maintains a B-tree index on the data stored in deltas. This index allows the system to efficiently identify and skip deltas that are unlikely to contain the requested keys, reducing the amount of data that needs to be scanned.

#### 32.2.6.4. Optimized QT Lookup

Every query requires looking up the current QT in the control plane. Usually, this involves a snapshot read on Spanner for the latest value. Napa optimizes this by periodically fetching the QT and storing it in a distributed metadata cache. This approach introduces a degree of staleness, as the cached QT might not always reflect the absolute latest state but speeds up query processing.

#### 32.2.6.5. Distributed Read-Through Cache

All data reads pass through a distributed read-through cache. This cache helps share read operations for the same data blocks across different parts of a query. For instance, in self-joins or queries with subqueries, multiple components might access the same index files in deltas, benefiting from the cached data.

#### 32.2.6.6. Prefetching Cache Layer

Napa utilizes another distributed caching layer focused on prefetching data blocks. This prefetching can be either offline or online.

- **Offline Prefetching**: For frequently queried tables, blocks are proactively fetched into the cache as soon as the QT advances.
- **Online Prefetching**: If a shadow query execution has already fetched relevant data blocks, the main query executor can leverage this cached data. The details for shadow query execution are missing in the paper.

#### 32.2.6.7. Lazy Merging Across Deltas

When a query is executed, a Delta server reads data from multiple relevant deltas and merges the results before returning the response to the client. This process is described as optimized; however, the mechanism of this optimization is unclear, and the explanation is somewhat vague.

From what can be understood, the authors are likely generating subqueries from the main query such that each involved delta server reads only one delta.

#### 32.2.6.8. Hybrid File Format

Napa's file format is primarily columnar for efficient analytical processing. However, it also incorporates PAX for faster lookups in small files.

#### 32.2.6.9. Read Hedging

To mitigate the impact of slow replicas, Napa employs **read hedging**. If a response from an initial replica takes longer than expected, a parallel request is sent to another replica containing the same data. The fastest response is then used.

### 32.3. Production Metrics

The paper dedicates a significant portion to design details, leaving insufficient space for a thorough evaluation of the system. The evaluation section primarily presents a limited set of production metrics:

- Query latency decreases with an increasing number of materialized views. This is anticipated, as materialized views enable queries to avoid costly operations like joins or sorts on frequently accessed data paths.
- Query latency increases with a greater number of deltas involved in a query. This is attributed to the increased I/O and merging operations required by the delta servers.
- When the view maintenance pipeline experienced issues, query latency remained unaffected, although data freshness was compromised. This stability was due to the use of QT for query execution.
- Napa finds application across various domains, including:
    - Internal experimentation and analysis clients, which prioritize query performance and low cost over strict data freshness.
    - Another application type that prioritizes lower resource costs, accepting higher query latency as a trade-off.
    - External, user-facing applications that prioritize both high data freshness and high query performance, incurring higher costs to achieve this.

### 32.4. Paper Remarks

The paper provides a wealth of information, though certain concepts are likely more accessible to database domain experts. For those less familiar with the specific intricacies of the field, some sections may lack immediate clarity. Despite this, this paper and its predecessor on Mesa offer a comprehensive introduction to the internal mechanics of large-scale data warehouses.

**Further Reading**: "The Snowflake Elastic Data Warehouse" [123]. Once the foundational concepts from the Google papers are grasped, the Snowflake architecture serves as a logical and relatively straightforward progression in understanding cloud-native data warehousing.

# 33. Photon: Fault-tolerant and Scalable Joining of Continuous Data Streams

(33) Ananthanarayanan, R.; Basker, V.; Das, S.; Gupta, A.; Jiang, H.; Qiu, T.; Reznichenko, A.; Ryabkov, D.; Singh, M.; Venkataraman, S. Photon: Fault-Tolerant and Scalable Joining of Continuous Data Streams. In Proceedings of the 2013 ACM SIGMOD international conference on management of data; 2013; pp 577–588.

In the previous chapter, we learned how various types of data joins are implemented in a distributed query engine. However, data is not always static - it is sometimes streamed and must be joined with other data sources in real time. This process is known as a streaming join. In this chapter, we will build upon our previous knowledge to explore how streaming joins work.

Presented at ACM Special Interest Group on Management of Data (SIGMOD) 2013, this paper from Google details streaming join in Google Ads, a platform known for its planet-scale data processing. The year 2013 marked a significant period for stream processing, as Google was concurrently developing MillWheel [91] and Dataflow, foundational technologies that influenced the creation of Apache Flink and Apache Beam [92].

In this insight, we will begin by defining streaming joins and the various methods used to implement them. Next, we will briefly explore how online advertising works - a primary and elegant application of streaming joins. We will then walk through the internals of Photon, examining its specific requirements and how those needs are met to ensure low-latency, near-accurate realtime event processing. As a bonus, we will also cover Continuous Query Language, a popular framework in the streaming ecosystem.

**Must Read**:

- **11. Apache Flink™: Stream and Batch Processing in a Single Engine**
- **12. The Dataflow Model: A Practical Approach to Balancing Correctness, Latency, and Cost in Massive-Scale, Unbounded, Out-of-Order Data Processing**

### 33.1. Streaming Joins

Query engines enable users to query data from a variety of sources. If the retrieved data is in a relational format, it can be joined based on common keys. This process, where joins are performed on static data, is termed a **static join**.

An alternative approach is the **streaming join**, where at least one of the data sources is a continuous event stream rather than static data. Streaming joins can be categorized into two types:

- **Stream-Stream Join**: This involves joining two independent streams of data.
- **Stream-Table Join**: Here, a data stream is joined with a static dataset (table).

#### *33.1.1. Ordered Stream*

These are data streams where all events are ordered according to the join key. Joining such streams is straightforward, resembling a merge join operation on two tables with sorted keys, as shown in Figure 159.

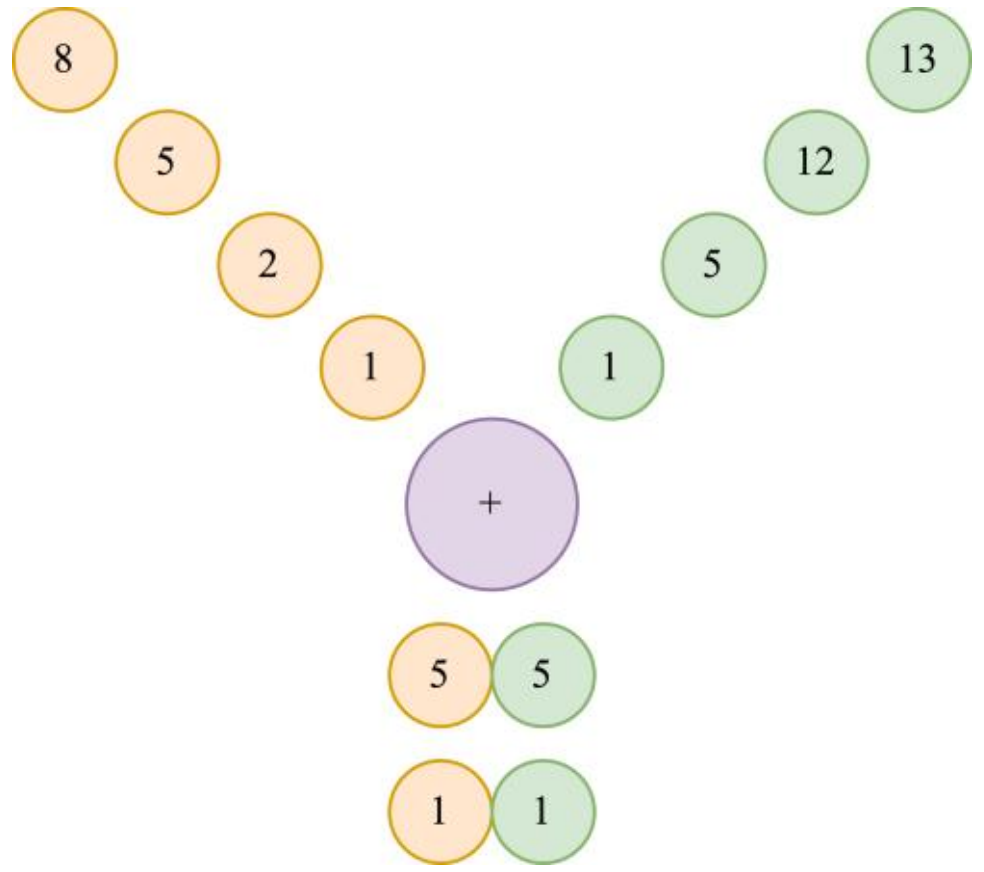

**Figure 159**: Join on ordered streams.

When a join key doesn't find a match, the outcome depends on the join type: an inner join will discard the unmatched event, while an outer join will pair it with NULL values.

#### *33.1.2. Unordered Stream*

These are streams where at least one of the event streams has events arriving out-of-order with respect to the join key when compared to the other stream.

#### 33.1.2.1. Case: Only One Stream Unordered and No Delayed Events

It is slightly easier to handle cases in which only one of the streams is unordered. A watermark is required on the unordered stream to mark the end of events up to a specific key. This watermark signals to the processing engine the potential arrival of corresponding events from the other stream.

The join operation remains conceptually similar to a merge join. However, watermarks play a vital role in determining when to finalize and trigger the join windows, as shown in Figure 160.

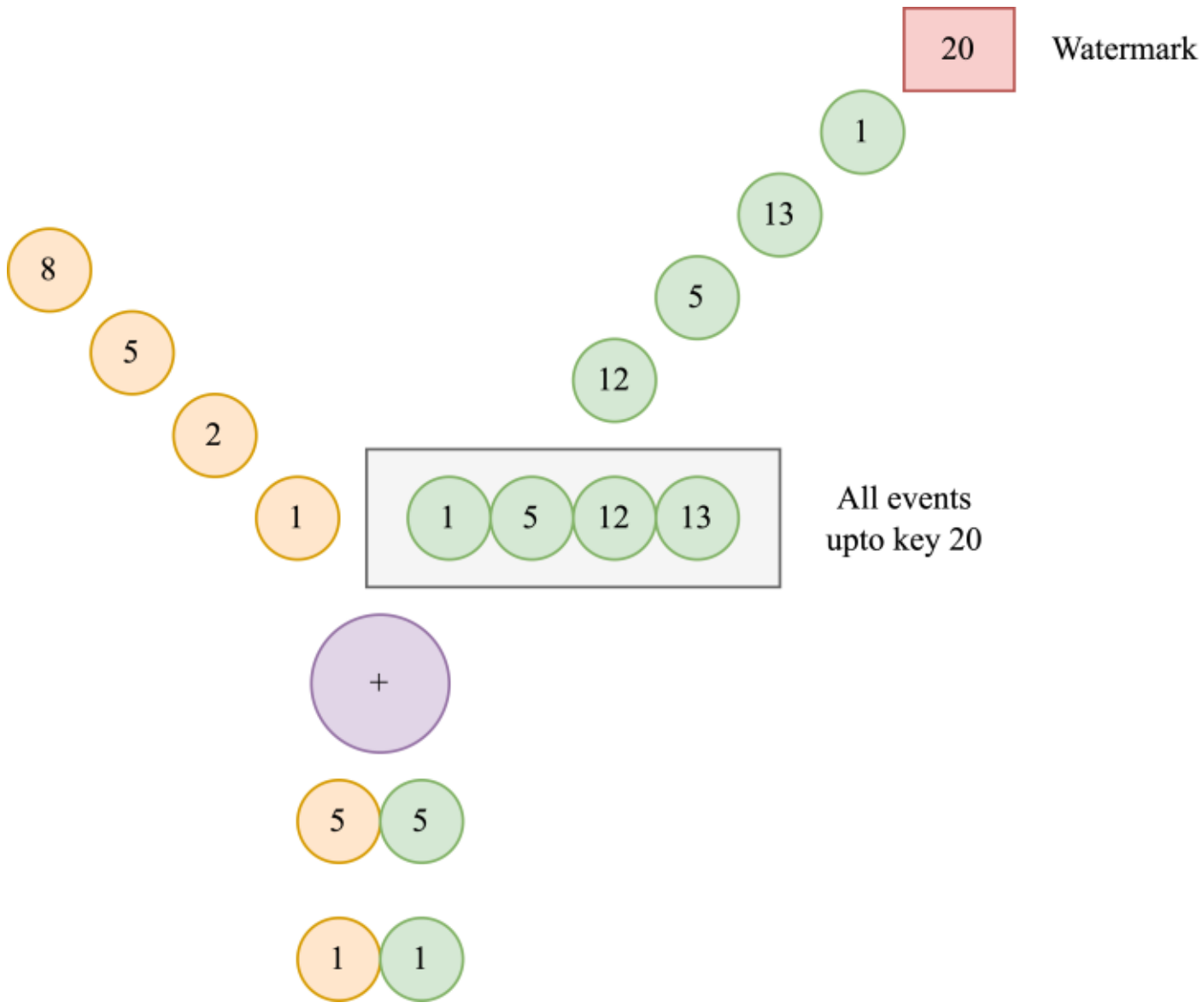

**Figure 160**: Join on unordered streams with watermarks.

Watermarks can be as simple as timeouts.

#### 33.1.2.2. Case: Delayed Events or Both Streams Unordered

Streams can experience delayed event arrival. In the earlier example, an event with a timestamp preceding the watermark of 20 could arrive after the watermark has already been processed. Furthermore, scenarios can involve both input streams being unordered. In both of these situations, stream materialization becomes necessary. A common destination for this materialization is a relational table, which stores all events from the streams. As events are added to one table, a lookup is performed on the other to fire all joined events, as shown in Figure 161.

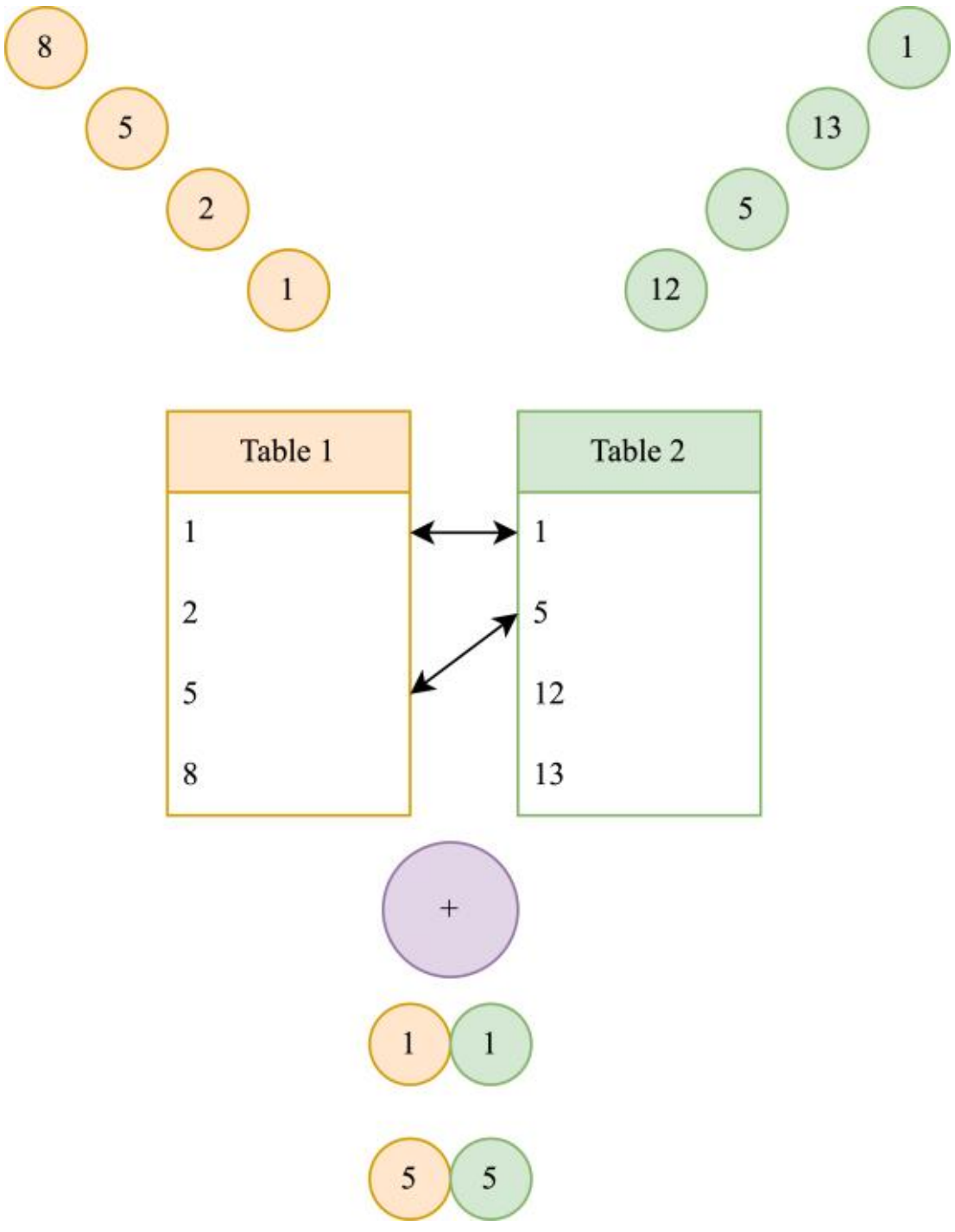

**Figure 161**: Join on unordered streams.

However, this approach can lead to unbounded table growth if the number of distinct keys is high and the join operation is sparse. Several optimizations can be employed based on specific user needs:

- **Watermark-Based Pruning**: Watermarks can be used to selectively remove rows from the materialized table. By maintaining the latest watermarks from both streams, the table can be pruned to retain only events newer than these watermarks. For instance, in the previous example, if a watermark of 20 is received on the right stream, all events in the left table with a key less than 20 can be discarded.
- **Single-Join Optimization**: If events in one stream are guaranteed to join with exactly one event in the other stream, only the unjoined events from the former stream need to be persisted. Continuing the example, if each event in the right stream joins with exactly one event in the left stream, only the right-stream events that haven't found a match in real-time need to be stored in the right table. Later, when a join occurs, those matched events can be removed immediately.

### 33.2. Online Advertising

Online advertising stands out as a prominent application of streaming joins. Before delving into Photon, it's helpful to understand the basics of how Google's advertising works.

Google's frontend servers deliver ads to users viewing webpages. When a user clicks on an ad, this click event is recorded by a separate server, distinct from the ad-serving server. The paper further elaborates on the architectural reasons behind this separation of concerns, primarily driven by performance considerations.

Given Google Ads' massive scale, it handles millions of queries per minute. Advertisers are typically charged based on the clicks their ads receive.

Fundamentally, this involves two continuous event streams:

- **Query Event Stream**: Comprising all queries served by the frontend servers.
- **Click Event Stream**: Encompassing all ad clicks collected by click servers.

Both streams share a common key, **query_id**, which serves as the basis for the necessary join operation. The key is composed of **<timestamp, server ip, process id>** and so it is possible to define a total order on the keys.

### 33.3. Photon

Photon is a streaming join system deployed within Google Ads. Its primary function is to join web search queries with corresponding user click data.

Let's analyze the characteristics of the input streams:

- The query stream is usually ordered but might contain delayed events.
- The click stream is unordered (based on **query_id**) and can include late-arriving events.

Effectively, both streams exhibit characteristics of unordered streams. However, a key advantage is that **each click event is guaranteed to join with exactly one query event**. Once a click event has been successfully joined, it can be discarded.

The design of Photon aims to achieve the following:

- **At-most-once semantics:** Ensuring that each click event is processed no more than once to prevent overcharging.
- **Near real-time accuracy:** Maximizing the number of click events joined with their corresponding query events with minimal delay.
- **Eventual exactly-once semantics**: Guaranteeing that all late-arriving click events are eventually processed.

The joined output from Photon is critical for various low-latency applications, including billing, click statistics and analysis, spam detection, and budget control. Consequently, Photon must process joins with latencies in the order of seconds.

### *33.3.1. Strawman*

A straightforward approach to joining these events would involve a single server ingesting all data. All incoming query events would be stored in a table. Upon the arrival of a click event, the system would perform a lookup in the query event table to find a match and output the joined result, as shown in Figure 162.

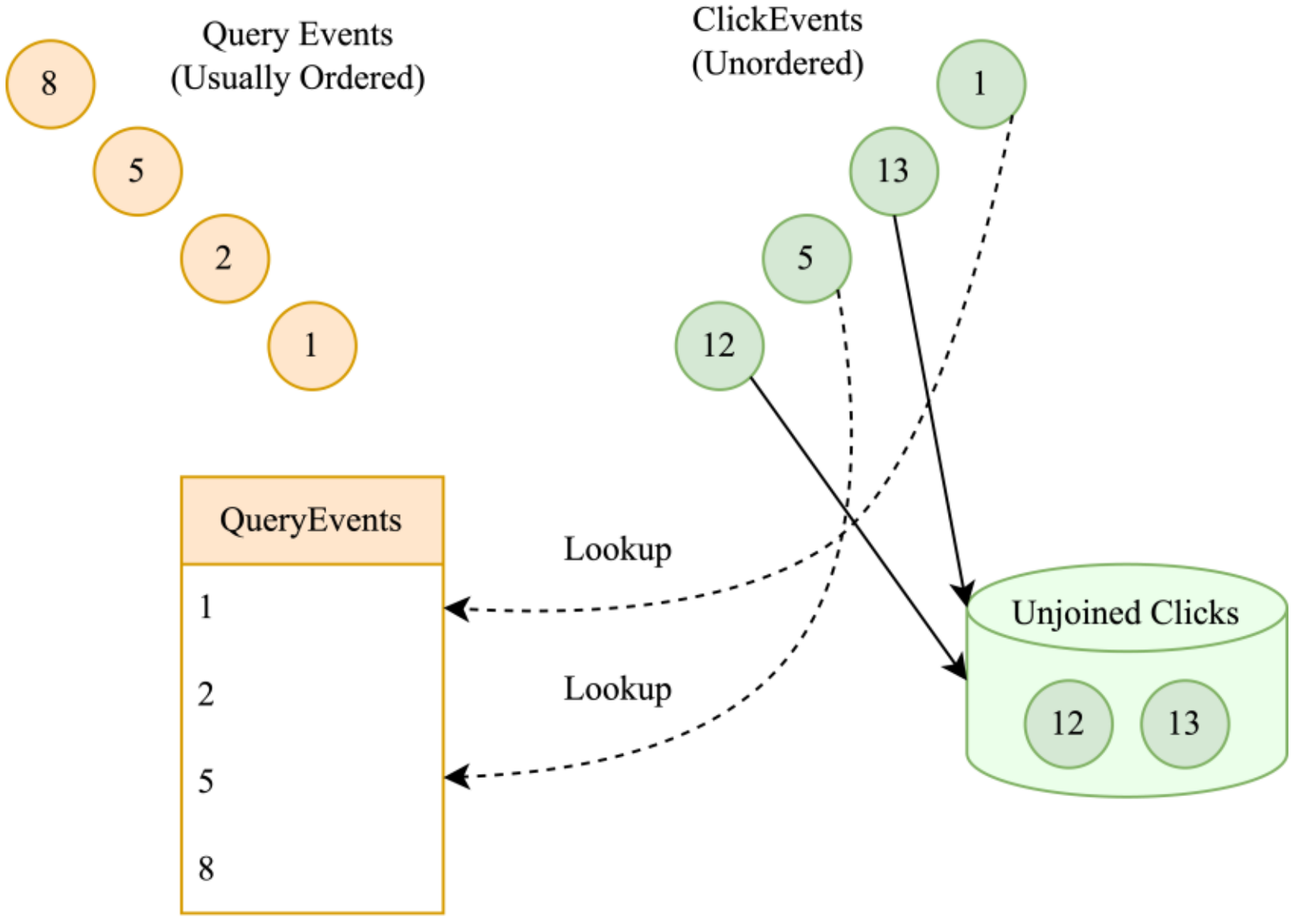

**Figure 162**: Strawman for queries (unordered) and clicks (unordered) join.

If a click event doesn't immediately find a corresponding query event (potentially due to the query event arriving late), the click event would be temporarily held and the lookup retried when the delayed query event arrives.

While this basic design is conceptually sound and could function for the Google Ads use case, it falls short in providing the necessary fault tolerance and scalability required to handle Google Ads' immense volume of millions of clicks per minute.

*33.3.2. Architecture*

On a high level, Photon is designed for multi-homing, providing resilience against datacenter-level failures.

All query events are stored in a dedicated **query event store**. The Photon pipeline operates across multiple data centers, coordinating through **IdRegistry**, as shown in Figure 2 in the paper. This component ensures, in a fault-tolerant manner, that each event is processed exclusively within a single datacenter.

Within each datacenter (shown in Figure 5 in the paper), a **Dispatcher** processes incoming click events. The Dispatcher continuously attempts to process each click event until successful, guaranteeing at-least-once processing. It then forwards the click event to the **Joiner**. The Joiner's role is to look up the corresponding query in the query event store and perform a crucial step: inserting the click's ID into the IdRegistry. This insertion ensures the at-most-once processing of each click. Finally, the joined event is outputted to the downstream sink.

*33.3.3. IdRegistry*

IdRegistry is a key-value store that maintains consistent replication across multiple datacenters. This helps with at-most-once guarantee for click event processing.

Internally, IdRegistry is built upon **PaxosDB**, a fault-tolerant, distributed key-value store. PaxosDB utilizes the Paxos algorithm to ensure a single master and linearizability for all updates, which are routed through the Paxos leader. This resembles ZooKeeper in its function of providing consistent coordination, albeit with a raw key-value data model instead of a hierarchical file system. A key difference is that while ZooKeeper is typically confined to a datacenter or zone, IdRegistry is globally replicated. Operating Paxos at this global scale presents significant challenges, which Photon addresses through:

**1. Server-Side Batching**

Committing larger groups of click events together, leveraging the system's tolerance for slight delays to improve efficiency.

**2. Sharding**

Partitioning the IdRegistry into multiple independent PaxosDB instances, each responsible for a specific key range. However, sharding can introduce consistency challenges during changes. To mitigate this, IdRegistry employs TrueTime.

TrueTime provides guarantees on clock skew across machines, ensuring that the difference is bounded by a small delta ($\epsilon$). An event intended for time **T** will certainly have occurred by $\mathbf{T} + \epsilon$. Consequently, when a shard change is scheduled at time **T**, it will be completed by $\mathbf{T} + \epsilon$, and all shards will use the new sharding information. This allows for consistency even during sharding updates, using the old modulo before $\mathbf{T} + \epsilon$ and the new modulo at or after $\mathbf{T} + \epsilon$. This approach mirrors the strategy in Spanner for managing Paxos leader transitions.

*33.3.4. Fault Tolerance*

Let's consider potential failure scenarios to understand the design considerations for each component.

**Q. What if the Dispatcher fails before processing a click event and sending it to the Joiner?**

This scenario poses no harm. Click events are not directly streamed for joining but are stored as logs. These logs are further divided into smaller files, and the Dispatcher maintains its processing state by periodically recording the file offset up to which events have been processed.

Regardless of when a Dispatcher crashes, the worst-case outcome is a restart that reprocesses click events from the beginning of the last recorded offset. However, the Dispatcher's responsibility is to ensure at-least-once processing, so this reprocessing is acceptable.

**Q. What if the Joiner fails after receiving a click from the Dispatcher?**

- **Before recording in IdRegistry:** This is effectively a no-op. The Dispatcher will retry sending the click event to another Joiner instance.
- **After recording in IdRegistry but before outputting the joined event**: This is where the value for a key stored in IdRegistry becomes crucial. The IdRegistry stores a token against a click which uniquely identifies the joiner which added the event. When a Joiner restarts and receives the same click event (due to the Dispatcher's at-least-once delivery), it will attempt to record the click in IdRegistry again with the same token. IdRegistry will return success as the token would match.

Despite this mechanism, it's still possible for joined events to be missed in the output. This can occur if a subsequent retry from the Dispatcher is handled by a different Joiner, whose attempt to add the click to IdRegistry will be rejected (as it would be already present and the token won't match). To guarantee eventual exactly-once semantics, a separate offline process analyzes both the IdRegistry and the output sink. This analysis identifies any missing joined events, which are then added to the output.

Another potential approach would have been to ensure atomicity between writing to IdRegistry and writing the joined event to the output sink, possibly through a two-phase commit protocol. However, the paper argues that the join process involves adapters and several user-defined annotations. These computations are likely expensive, making it costly to re-execute them if a two-phase commit transaction were to fail.

**Q. What happens if a query event arrives late?**

In this case, the Dispatcher stores the corresponding click events in the local Google File System. This ensures that these "orphaned" click events will eventually be processed at least once when the late query events finally arrive and are processed by a subsequent Joiner.

*33.3.5. The Exactly-Once Semantics*

The Dispatcher's at-least-once delivery combined with IdRegistry's at-most-once guarantee combined with offline processing collectively provides Photon's exactly-once processing semantics.

*33.3.6. Misc Optimizations*

Photon incorporates several key optimizations to enhance its performance.

33.3.6.1. Pruning Old Keys

As previously mentioned, a click event joins with at most one query event. Therefore, once a click event has been successfully joined, its entry in IdRegistry can be removed. However, before removal, it's crucial to ensure these click events won't reappear from the Click Logs.

To manage the size of IdRegistry, a garbage collection policy is in place, set to retain entries for N days.

33.3.6.2. Dispatcher IdRegistry Lookup

To avoid unnecessary processing, the Dispatcher performs an initial lookup in IdRegistry before forwarding a click event to a Joiner. If the click is already present, it is assumed the event has been processed.

33.3.6.3. EventStore

The EventStore implementation features a two-tiered architecture: **CacheEventStore** and **LogsEventStore**.

CacheEventStore leverages the temporal locality of events, where recent query events are likely to be associated with recent clicks. This in-memory cache, sharded by the hash of **query_id**, stores several minutes' worth of query events, significantly reducing disk I/O.

LogsEventStore provides persistent storage for all events in log files. It also maintains a log file map to enable rapid lookup of an event's offset within the log files based on its **query_id**.

### 33.4. Production Numbers

- Photon achieves an end-to-end latency (p90) of 7-8 seconds from event log to join completion.
- Server-side batching effectively reduces the number of transactions on PaxosDB.
- The Dispatcher's lookup on IdRegistry saves resources. Joiners have only about 5% overlap in the events they process.
- Less than 0.0001% of joined events are missed by Joiners after registering in IdRegistry.

### 33.5. Bonus: Continuous Query Language

Continuous Query Language [220] (CQL), a SQL-based language tailored for streaming data, originated at Stanford University. In CQL, a stream is defined as an infinite collection of elements, each represented by a tuple **<T, S>**, where **S** denotes the data tuple and **T** is its associated timestamp. A relation, also known as an **instantaneous** relation, represents a mapping of tuples at a specific moment in time.

CQL employs three primary types of operators to process these data structures, as shown in Figure 163:

- **Stream-to-Relation (SoR):** The sole operator in this category is the window operator. It applies various windowing techniques to transform a continuous stream into a temporary relation.
- **Relation-to-Relation (RoR):** The standard SQL operators such as selection, projection, and aggregation, which operate on relations to produce new relations.
- **Relation-to-Stream (RoS):** This type of operator computes the difference between the current relation and the preceding one, effectively converting changes in a relation back into a stream.

By combining these operators, CQL establishes a relational algebra framework capable of seamlessly operating across both streams and relations.

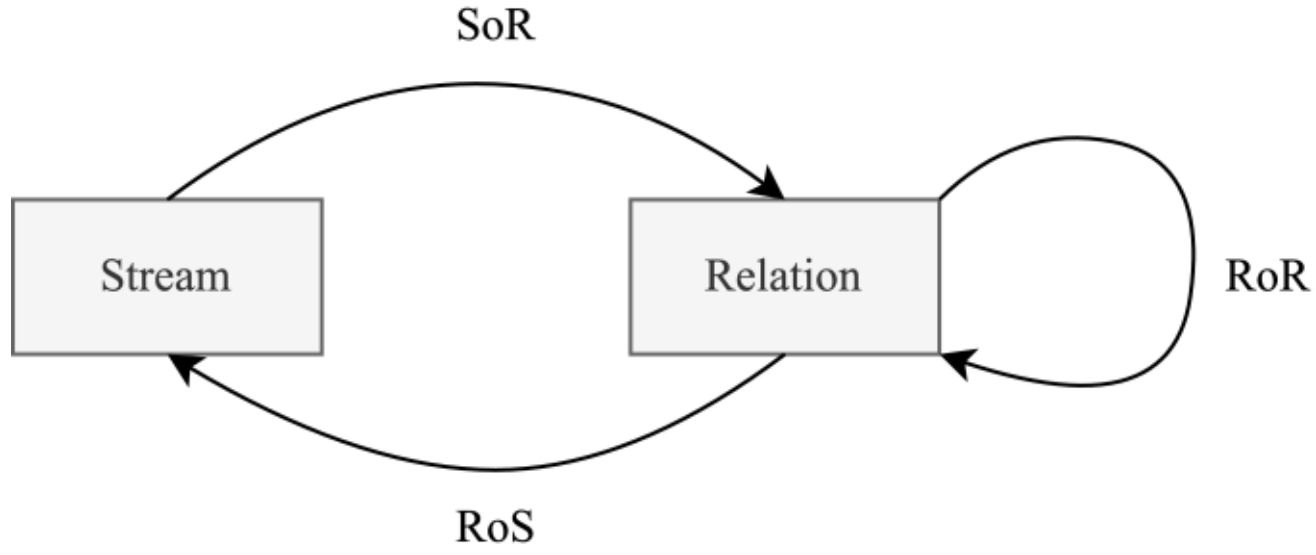

**Figure 163**: CQL operators.

Consider the following CQL example:

```
SELECT page, COUNT(*) FROM VISITS (RANGE 1 HOUR) GROUP BY page;
```

In this query, the CQL engine continuously monitors the incoming stream of page visits. For every event within the last hour, it groups the visits by the page attribute and calculates the count for each page. This computation results in a relation that reflects the aggregated visit counts for each page within the specified time window. The frequency at which this resulting relation is output depends on the specific application requirements. For instance, an output period of one minute would indicate that the aggregated results are generated and presented every minute.

### 33.6. Paper Remarks

From a theoretical perspective, Photon's core architecture is straightforward, adhering to the fundamental principles of a basic stream-joining system. However, the true significance of this paper lies in its documentation of the rigorous engineering and sophisticated optimizations necessary to operate at Google's unprecedented scale. For readers already well-versed in distributed systems and stream-processing mechanics, the paper provides a clear, accessible case study.

# 34. CRDTs: Consistency without Concurrency Control

(34) Letia, M.; Preguica, N.; Shapiro, M. CRDTs: Consistency without Concurrency Control. RR-6956, INRIA 2009.

In the previous chapters, we explored transactions in data stores. A transaction serves as an atomic action across multiple data items. Maintaining consistency requires concurrency control, ensuring the final state is as if each transaction were executed in isolation. There are several concurrency control mechanisms, such as locking and multi-versioning.

However, what if the operations performed on a data store possess specific mathematical properties? For instance, if an operation only involves adding a value to an existing number, it is considered commutative. In such cases, is traditional concurrency control still necessary? This question forms the cornerstone of Conflict-free Replicated Data Types (CRDTs).

Authored in 2009, this paper from National Institute for Research in Computer Science and Automation was presented at the IEEE International Conference on Distributed Computing Systems (ICDCS). It introduces compelling design concepts for distributed systems that operate without global synchronization.

In this deep dive, we will first define CRDTs formally and explore their core principles. Next, we will examine several well-known CRDTs before focusing on the Ordered Set - a popular non-trivial CRDT - and analyzing how TreeDoc can be utilized for its implementation.

### 34.1. Conflict-free Replicated Data Types

**Conflict-free Replicated Data Types (CRDTs)** ensure data consistency across multiple replicas without the need for explicit conflict resolution:

- **Conflict-free**: Indicating that concurrent updates from different replicas will always converge to the same final state.
- **Replicated**: Highlighting that the data is distributed and maintained across several independent nodes.

Imagine a shared data structure that exists on multiple nodes. To manage updates to this data structure consistently, CRDTs typically provide the following core interface:

```
interface CRDT {
   Apply(T update);
   Merge(CRDT other);
};
```

Here, we have two essential operations:

- **Apply(update)**: This operation incorporates a local update to the data. The update is combined with the current local state using a specific operator, and the result becomes the new local state.
- **Merge(other)**: This operation integrates the state of another CRDT instance (other) into the current local state. This merging also utilizes a specific operator, resulting in a new local state that reflects both previous states.

The fundamental principles behind CRDTs are:

- All updates originating from clients are initially applied locally.
- Subsequently, these updates are propagated asynchronously to other replicas through various communication protocols. The crucial guarantee is that once all updates have been delivered and applied across all nodes, the final state of the data structure will be identical everywhere.

To achieve this eventual consistency regardless of the order or timing of updates, the operator used in both **Apply** and **Merge** must possess specific mathematical properties:

- **Commutativity**: For updates $\mathbf{a}$ and $\mathbf{b}$, $\mathbf{f}(\mathbf{a},\mathbf{b}) = \mathbf{f}(\mathbf{b},\mathbf{a})$. This allows nodes to process updates in any sequence.
- **Associativity**: For updates $\mathbf{a}$, $\mathbf{b}$, and $\mathbf{c}$, $\mathbf{f}(\mathbf{a},\mathbf{f}(\mathbf{b},\mathbf{c})) = \mathbf{f}(\mathbf{f}(\mathbf{a},\mathbf{b}),\mathbf{c})$. This ensures that merging states from different nodes is consistent.
- **Idempotency**: Applying the same update multiple times has the same effect as applying it once. This prevents duplicate updates from causing unintended changes in the data's state.

A diverse range of CRDT implementations exists, each tailored for different data structures and update semantics.

### *34.1.1. Grow-Only Counter*

The Grow-Only Counter (G-Counter), is a straightforward CRDT that allows only increasing integer values. Its simplicity makes it a foundational example.

**Implementation**

- Each replica maintains a map (**s**) where keys are node identifiers and values are the current counter observed from that node.
- The **Apply()** operation increments the counter value associated with the local node:

$$\mathbf{s}[\mathbf{self}] = \mathbf{s}[\mathbf{self}] + \mathbf{1}$$

- The **Merge()** operation updates each node's counter by taking the maximum of the locally known value and the value from another replica:

$$\mathbf{s}[\mathbf{i}] = \mathbf{max}(\mathbf{s}[\mathbf{i}], \mathbf{other}.\,\mathbf{s}[\mathbf{i}])\ \forall \text{ nodes } \mathbf{i}$$

The nature of the merge operation ensures that G-Counters are commutative, associative, and idempotent.

### *34.1.2. Positive-Negative Counter*

To accommodate both increments and decrements, we can extend the G-Counter concept with the Positive-Negative Counter (PN-Counter). A naive extension of the **Apply()** operation to include decrements would violate idempotency.

**Implementation**

To maintain idempotency, a PN-Counter employs two G-Counters per node: a positive counter for increments and a negative counter for decrements.

The **Merge()** operation takes the maximum value for each node's positive counter and the maximum value for each node's negative counter. The overall value of the PN-Counter is the sum of all positive counter values minus the sum of all negative counter values.

*34.1.3. Grow-Only Unordered Set*

The Grow-Only Unordered Set mirrors the behavior of the G-Counter but for sets. It only allows adding elements.

**Implementation**

- Each replica maintains a single set.
- The Apply() operation adds new elements to the local set using a union operation.
- The Merge() operation combines sets from different replicas using set union.

Due to the properties of set union, Grow-Only Unordered Sets are commutative, associative, and idempotent.

*34.1.4. 2-Phase Unordered Set*

Analogous to the PN-Counter, the 2-Phase Unordered Set (2P-Set) introduces the ability to remove elements. It maintains two sets: an **added** set and a **removed** set.

**Implementation**

- Adding an element places it in the **added** set.
- Removing an element adds it to the **removed** set. Once an element is in the **removed** set, it cannot be added back.
- The **Merge()** operation takes the union of the **added** sets and the union of the **removed** sets from the merging replicas. The effective set contains elements present in the merged **added** set but not in the merged **removed** set.

A significant drawback of the 2P-Set is that the **removed** set can grow indefinitely.

*34.1.5. Last Writer Wins*

The Last Writer Wins (LWW) strategy, often applied to sets or key-value stores, uses timestamps to resolve conflicts. Each operation is associated with a timestamp.

LWW can be used to implement a 2P-Set without the unbounded growth of the **removed** set. When an element is added or removed, the operation is timestamped. During a merge, if two conflicting operations (add and remove) for the same element are encountered, the operation with the later timestamp wins.

The LWW strategy is particularly crucial in eventually consistent key-value stores like Dynamo. When concurrent writes occur for the same key, replicas eventually converge to the value associated with the latest timestamp. This ensures that a single, consistent version of each key's value is maintained across the distributed system.

### 34.2. Sequence CRDTs

While the CRDTs discussed so far (counters and basic sets) illustrate fundamental principles, they often fall short of the requirements of more complex applications. A particularly interesting and non-trivial CRDT is the ordered set, which is the central topic of discussion in this paper.

An **ordered set** is a data structure where the elements maintain a specific sequence that isn't inherent to the elements themselves. Unlike a sorted set where order is determined by element properties, the order in an ordered set is explicitly defined by client operations. For instance, an ordered set could contain the English alphabet in a non-alphabetical order: **{d, a, c, b}**. **The elements themselves are treated as opaque values.**

The key challenge with ordered sets lies in allowing clients to specify the precise location where a new element should be inserted within the existing order. Consider an ordered set **{a, b}**. A client might request the insertion of element **c** between **a** and **b**, resulting in the new order **{a, c, b}**.

To achieve this fine-grained control over ordering, ordered sets typically require clients to tag each element with a unique identifier. This identifier space must be unbounded, meaning that for any two identifiers **A** and **B**, there exist infinitely many identifiers **Z** such that **A < Z < B**. These identifiers serve as the basis for determining the position of elements within the ordered set.

Consequently, each element in the ordered set effectively becomes a pair: **<ID, element>**, where the **ID** dictates its position in the sequence.

Let's revisit the example of inserting **c** between **a** and **b**. If **a** has an identifier of **1** and **b** has an identifier of **2**, then when inserting **c**, the client can assign it any identifier between **1** and **2**, such as **1.5**. This approach inherently ensures the CRDT properties:

- **Commutativity**: The order of concurrent insertions with distinct identifiers doesn't affect the final ordered set.
- **Associativity**: Merging sets in different orders yields the same final ordered set because each element's position is determined by its unique identifier.
- **Idempotency**: Adding the same **<ID, element>** pair multiple times has the same effect as adding it once, due to the set-like nature of the underlying data structure (we don't have duplicates with the same ID).

*34.2.1. Resolving Identifier Conflicts*

Despite the use of unbounded identifiers, conflicts can still arise. For example, if two different nodes concurrently attempt to insert elements **c** and **d** between **a** and **b** and both assign the same identifier, say **1.5**, to their respective elements. Several strategies can be employed to resolve such conflicts:

- **Timestamping (Last Writer Wins)**: Similar to the LWW approach discussed earlier, each insertion operation can be timestamped. In case of an identifier collision, the element associated with the later timestamp takes precedence in the ordering.
- **Disambiguator ID**: Each node is assigned a unique identifier. This identifier can be used to break ties.

*34.2.2. TreeDoc: Leveraging Trees for Ordered Sets*

In our previous examples, we conceptually used real numbers as identifiers to represent the ordering of elements. However, relying on floating-point numbers in computer systems presents challenges due to potential precision limitations. This could lead to a finite identifier space, violating the crucial unbounded property required for seamlessly inserting elements between any two existing ones.

A more robust approach involves using binary strings as identifiers. Binary strings offer a natural way to represent order and can easily guarantee an unbounded identifier space.

Consider two elements with binary identifiers **10** and **11**. We can always generate new identifiers that fall lexicographically between them, such as **100**, **101**, **110**, and **111**, effectively creating space for four new elements in between. This process can be repeated indefinitely, ensuring an unbounded identifier space.

#### 34.2.2.1. Tree Representation in TreeDoc

These binary string identifiers can be elegantly organized into a tree structure, which is the core idea behind **TreeDoc**. Each binary string identifier corresponds to a path from the root of the tree. The order of the identifiers (and thus the elements) is determined by performing an infix traversal of this tree.

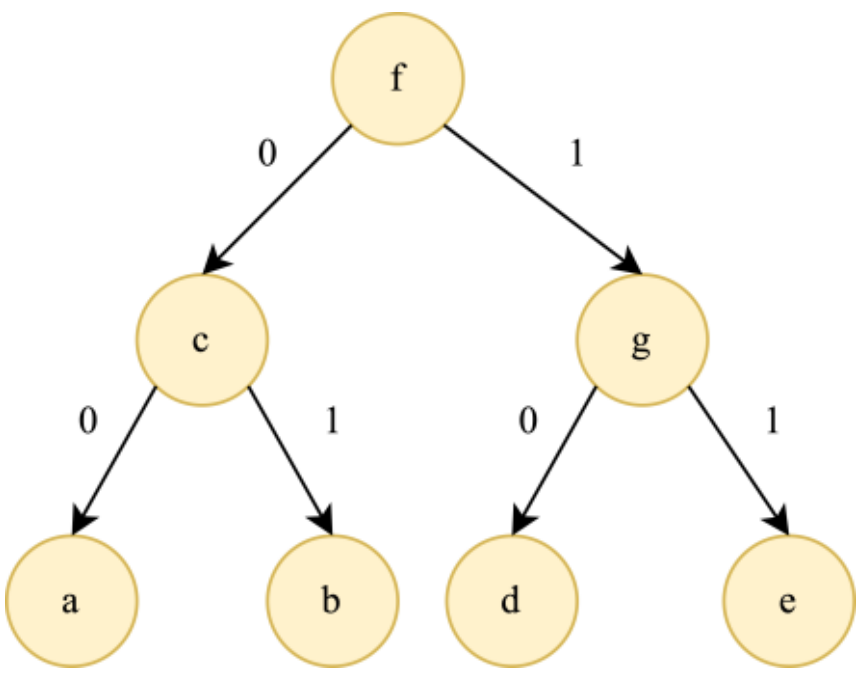

**Figure 164**: A TreeDoc.

In Figure 164, the order of elements is - **{a, c, b, f, d, g, e}** with IDs **`00`, `0`, `01`, ``, `10`, `1`, `11`** respectively**.**

#### 34.2.2.2. Insertions

When inserting a new element between two elements with existing identifiers **X** and **Y**, the TreeDoc algorithm traverses the tree to find the leftmost available position between the nodes corresponding to **X** and **Y**. The new element is inserted at that position.

#### 34.2.2.3. Handling Concurrent Insertions with the Same Identifier

A potential conflict arises when multiple writers concurrently attempt to insert elements and inadvertently assign them the same identifier (resulting in the same position in the tree). To address this, a node in the TreeDoc tree isn't a single element but rather a multi-node. This multi-node can contain multiple elements that share the same tree position (same ID), as shown in Figure 165. The order among these co-located elements within the multi-node is then determined using a disambiguator ID, as discussed earlier (e.g., using node identifiers to establish a consistent secondary ordering).

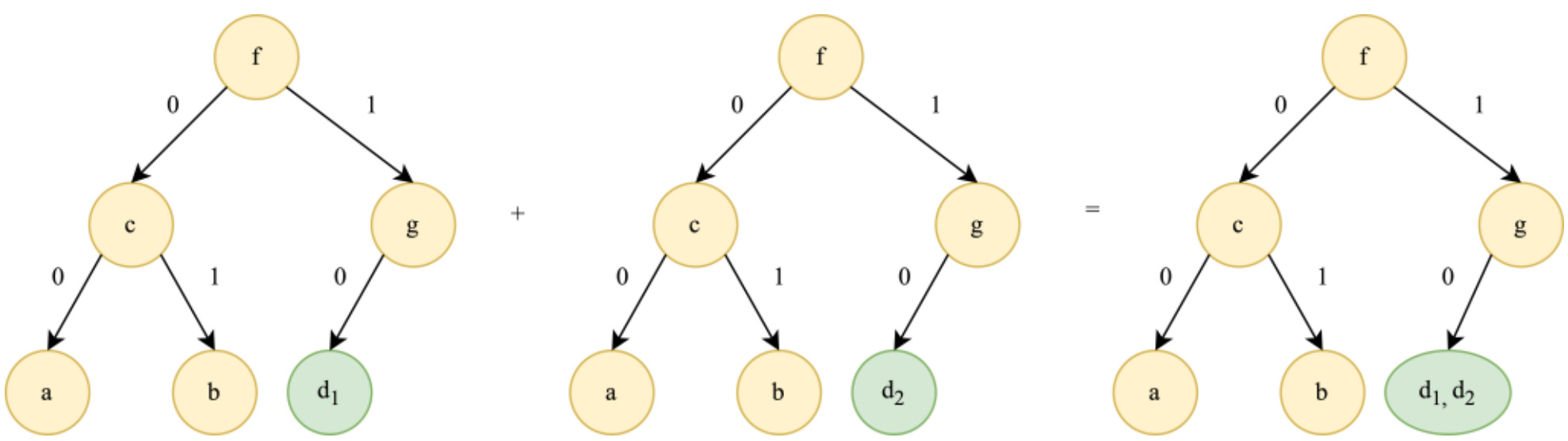

**Figure 165**: Concurrent insertion in TreeDoc.

In Figure 34.2, both $\mathbf{d_1}$ and $\mathbf{d_2}$ share the same node position. The order of elements is {**a, c, b, f,** $\mathbf{d_1}$**,** $\mathbf{d_2}$**, g**}.

34.2.2.4. Deletions in TreeDoc

Instead of physically removing the corresponding node from the tree, the node associated with the element's identifier is simply marked as a tombstone. During an infix traversal, tombstoned nodes are effectively skipped, thus representing the deletion of the element while preserving the tree structure and the identifiers of other elements.

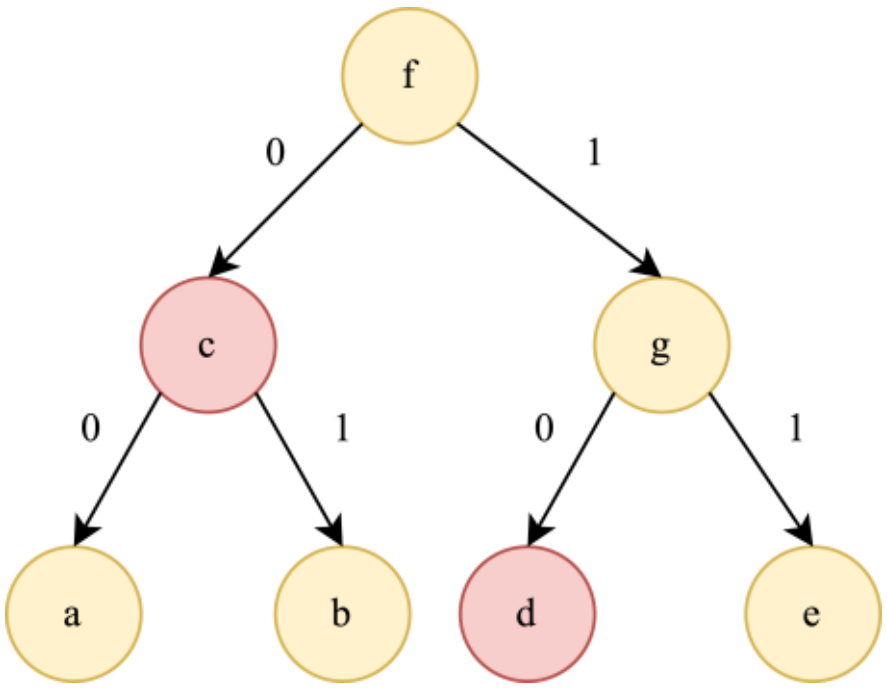

**Figure 166**: Deletion in TreeDoc.

In Figure 166, the order of elements is {**a, b, f, g, e**}.

34.2.2.5. Rebalancing

Over time, the TreeDoc structure may become skewed or accumulate an excessive number of tombstoned elements. To address this, a rebalancing operation called **flattening** is performed. This process restructures the tree and, importantly, assigns new identifiers to all elements.

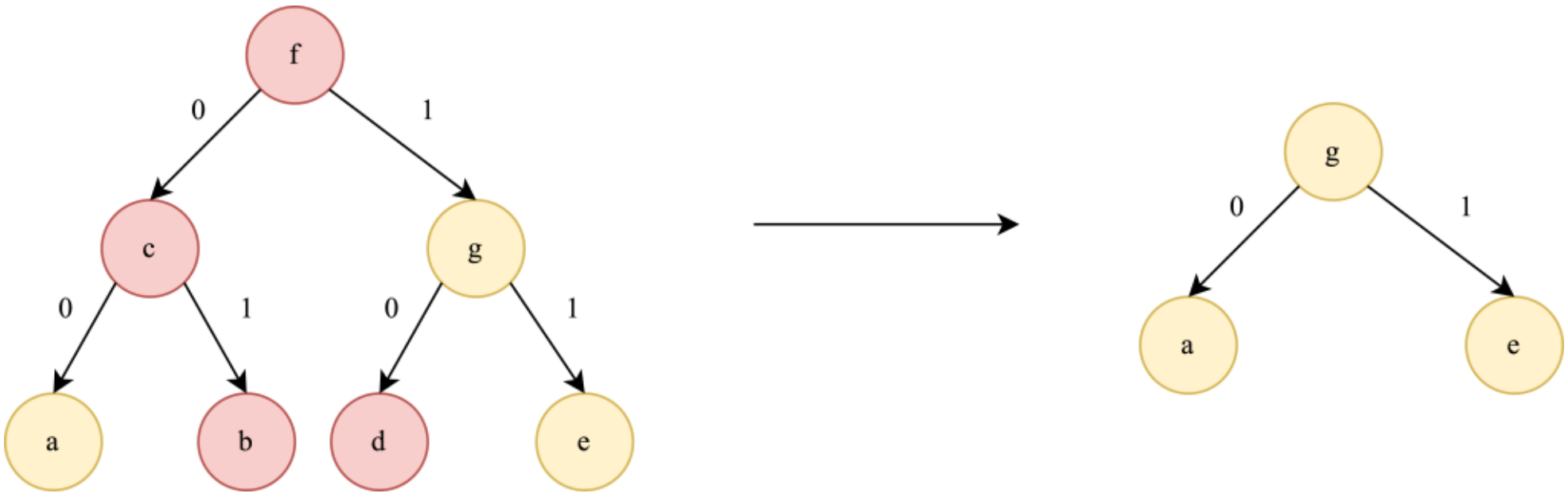

**Figure 167**: Flattening a TreeDoc.

In Figure 167, the identifiers for **a**, **g**, and **e** change from `**00**` to `**0**`, `**1**` to ``, and `**11**` to `**1**` respectively.

Because flattening necessitates changes across all elements and their identifiers, it requires a consistent state across all replicas. To achieve this, the data structures must be locked on all replicas while flattening is in progress. The authors utilize Paxos-Commit [221], a commit protocol built upon a consensus algorithm, to ensure this consistency, effectively managing the transaction required for the flattening operation. Note that Paxos-Commit itself is a commit protocol, not a consensus mechanism.

*34.2.3. Simultaneous Text-Edit*

TreeDoc represents an ordered set of elements, making it particularly well-suited for collaborative text editing by multiple users concurrently.

Consider a scenario where the current shared text is **abc**.

- Replica 1 performs an insertion of the character **e** between **a** and **b**.
- Simultaneously, Replica 2 performs an insertion of the character **f** between **a** and **b**.

Upon asynchronous broadcast of these operations, the final state of the shared text across all replicas will be updated to **aefbc**, as shown in Figure 168.

The insertions of **e** and **f** represent a concurrent conflict. This conflict is resolved based on disambiguators associated with each replica. Since the disambiguator for Replica 1 is less than the disambiguator for Replica 2, the insertion from Replica 1 (**e**) is ordered before the insertion from Replica 2 (**f**).

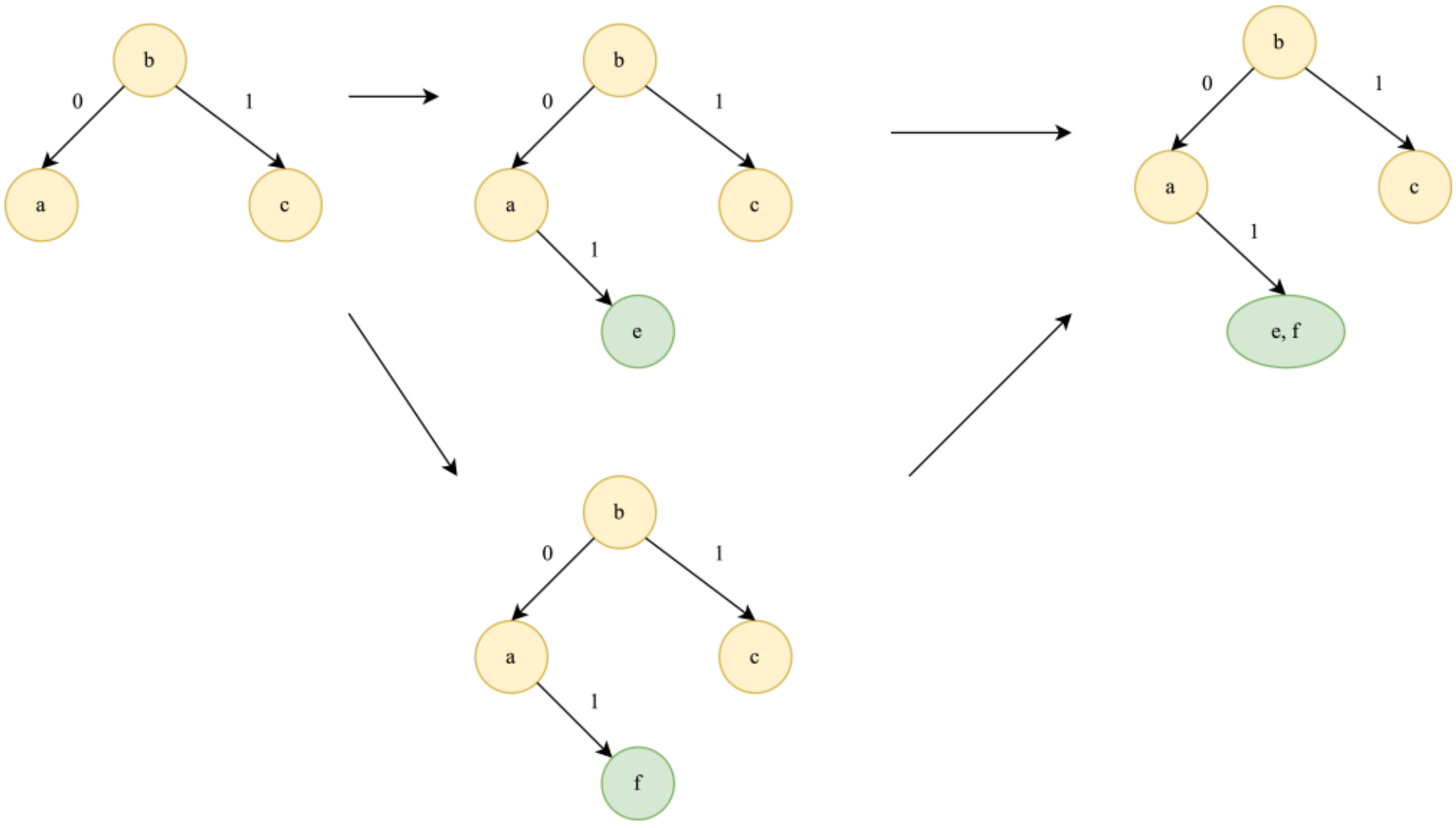

**Figure 168**: Simultaneous text-edit.

*34.2.4. Can Sequence CRDTs help in Distributed Logs?*

No, Sequence CRDTs are not suitable for distributed logs. Distributed logs require that all replicas observe every operation in precisely the same sequential order (a total order). Sequence CRDTs, such as ordered sets, only guarantee eventual convergence to the same final state across replicas; they do not enforce a consistent order of operations.

CRDTs operate within a mathematical structure called a **semi-lattice**, characterized by a partial order of operations. Each replica can observe any valid linearization of this partial order, leading to eventual consistency. However, different replicas can (and often do) observe these operations in different sequences.

Consider the example in Figure 169. Assume three operations, **a**, **b**, and **c**, are performed on a distributed data structure.

- Replica 1 observes the updates in the order: **{a}**, then **{a, b}**, then **{a, b, c}**.
- Replica 2 observes the updates in the order: **{b}**, then **{b, c}**, then **{a, b, c}**.
- Replica 3 observes the updates in the order: **{c}**, then **{a, c}**, then **{a, b, c}**.

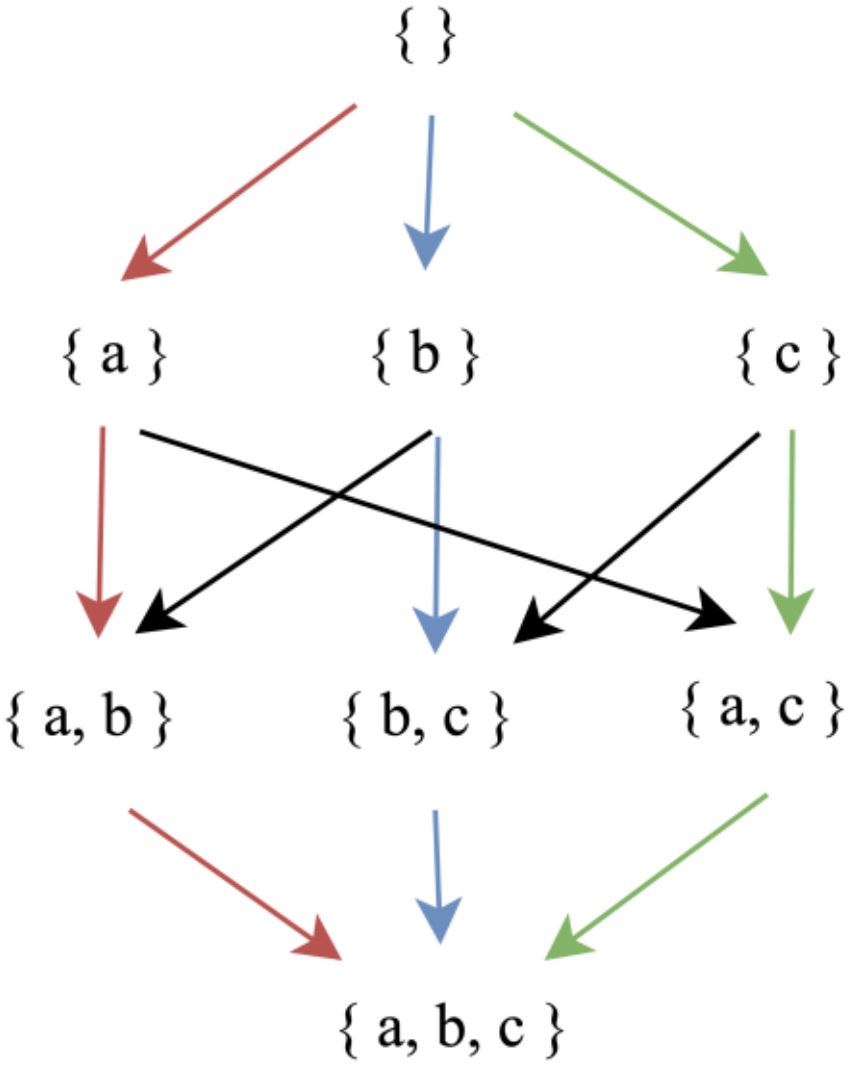

**Figure 169**: Ordered set semi-lattice.

As this example illustrates, although all replicas eventually reach the same final state **{a, b, c}**, they process the individual operations (**a**, **b**, and **c**) in different orders. This lack of a globally consistent order of operations makes Sequence CRDTs unsuitable for distributed logs, where the order of events is critical for consensus and maintaining a consistent history.

### 34.3. Large-Scale Implementation of Sequence CRDT

The authors have deployed TreeDoc at a significant scale, accommodating clients globally. A key performance consideration for sequence CRDTs is the flattening operation, which necessitates a commit. To enable efficient commits across geographically dispersed nodes, the authors strategically segmented their implementation into two distinct site types:

- **Nebula Sites**: These geographically distributed sites primarily handle the real-time streaming of local updates both back to the centralized core site and to other nebula sites.
- **Core Site**: This centralized site aggregates updates from all nebula sites and also disseminates updates back to them. Critically, only nodes within the core site participate in the flattening operation, which requires a commit protocol.

The system operates in discrete time periods called **epochs**. Each site operates within a specific epoch, and the epoch number is incremented after every flattening operation performed by the core site. This epoch-based system ensures that the ID space is unique for each epoch. Importantly, while the core site is undergoing the flattening process, nebula sites can continue to accept and process local updates.

Communication and update propagation are streamlined based on the current epoch. Any two sites residing in the same epoch can directly exchange updates. This is feasible because within a given epoch, the ID space remains consistent, guaranteeing that all insertions and deletions are correctly applied and remain commutative.

Following a successful flattening operation by the core site, it broadcasts an update signaling the new epoch. This can lead to following scenarios:

- **A nebula site sends an update from a previous epoch**: The core site will reject such updates, as it only accepts updates corresponding to the current epoch. The nebula site will need to undergo its own flattening process and receive new IDs for its pending updates before they can be successfully transmitted to the core site in the new epoch.
- **The core site sends updates from the new epoch to a nebula site**: The nebula site must acknowledge the epoch change and prepare to accept updates from the new epoch. This typically involves the nebula site also performing a flattening operation to align its internal state with the new ID space before applying the incoming updates from the core site's new epoch.

Let's walk through an example to understand the concepts (detailed algorithm can be found in section 4 in the paper).

Say, the core site had the following CRDT - **a (`00`)**, **b (`0`)**, and **c (``)** which is flattened to a new ID space - **a (`0`)**, **b (``)**, and **c (`1`)**, as shown in Figure 170(a).

Say the nebula sites also had the same CRDT, however, there were 2 updates:

- **b (`0`)** was deleted.
- **d (`1`)** was added.

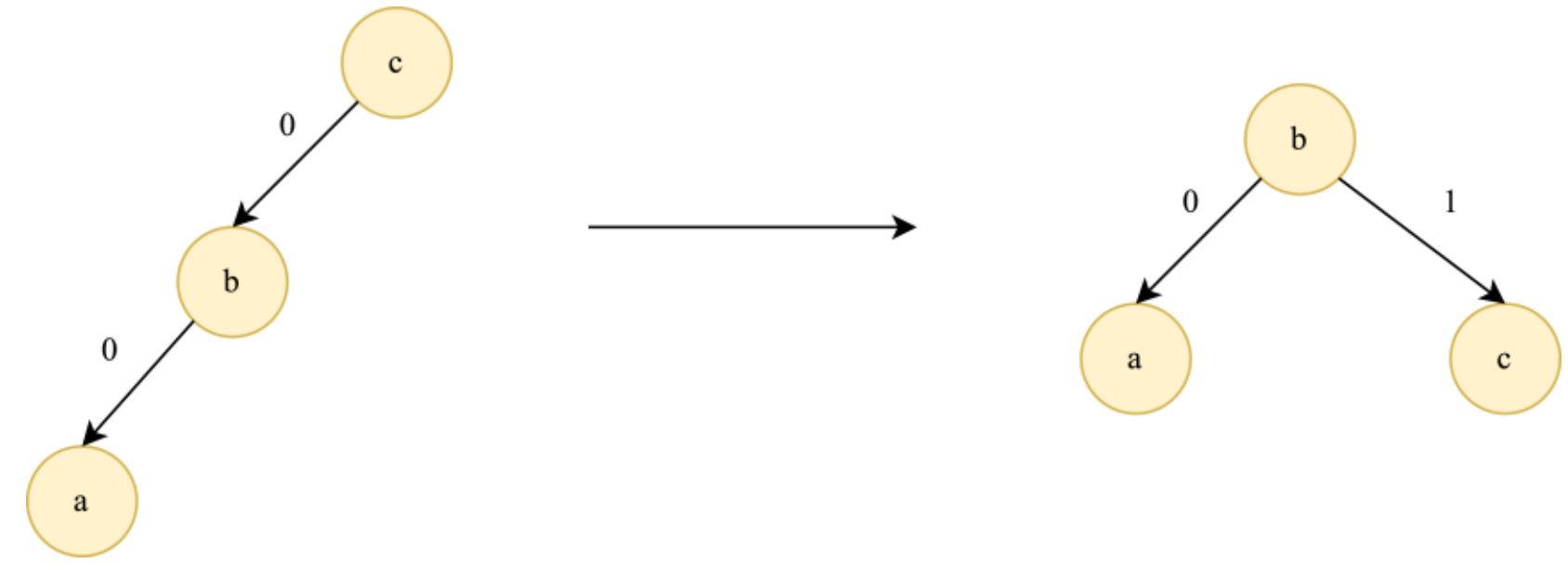

**Figure 170(a)**: Example core site flattening.

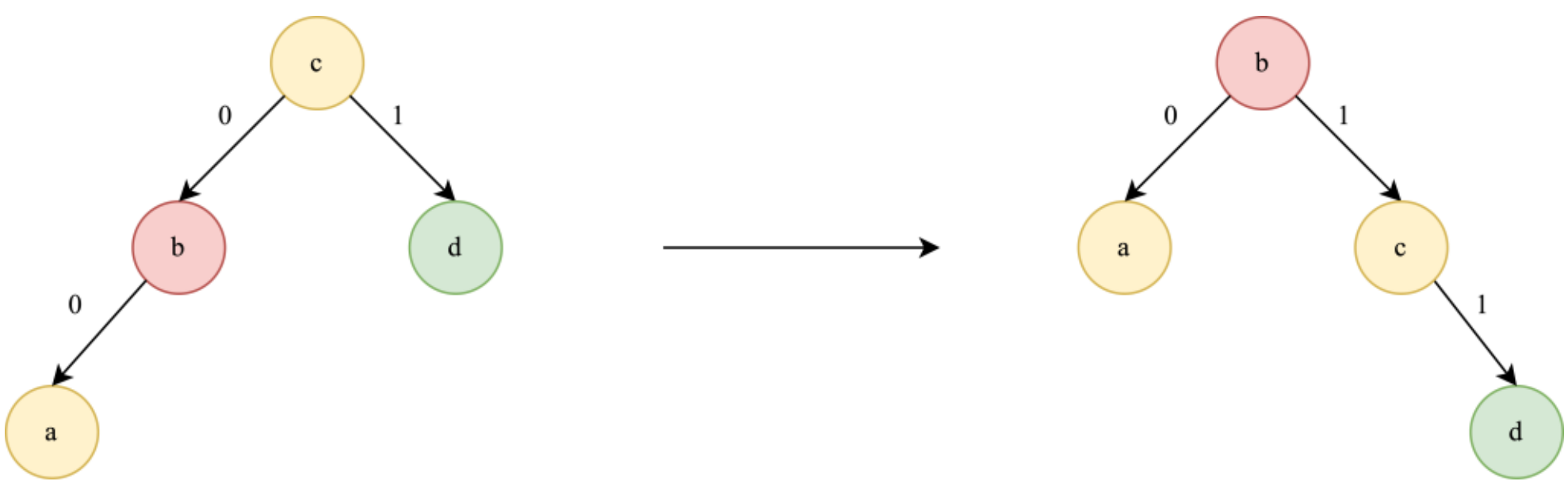

**Figure 170(b)**: Nebula updates after flattening.

If the nebula sites send **delete `0`** and **add <`1`, d>**, then those updates will be rejected by the core site (otherwise it will end up deleting **a** and failing to add duplicate ID **`1`**). Upon detection of this rejection failure, the nebula sites would run flattening, thereby computing the new ID space. Post flattening, the insertion of **d** will also be assigned a new ID in the new space. Then the updates corresponding to the new ID space will be sent. In the example, the updates will be **delete ``** and **add <`11`, d>**, as shown in Figure 170(b).

### 34.4. Paper Remarks

This paper is pretty straightforward and stands as one of the most concise entries in this collection. It provides a foundational look at non-trivial CRDT implementation. Given the industry-wide adoption of CRDTs, it is an essential resource for understanding how to achieve consistency without global synchronization.

# 35. Microsecond Consensus for Microsecond Applications

(35) Aguilera, M. K.; Ben-David, N.; Guerraoui, R.; Marathe, V. J.; Xygkis, A.; Zablotchi, I. Microsecond Consensus for Microsecond Applications. In 14th USENIX Symposium on Operating Systems Design and Implementation (OSDI 20); 2020; pp 599–616.

In the CliqueMap chapter, we explored how Remote Direct Memory Access (RDMA) serves as a backbone for an in-memory key-value store. In this chapter, we will explore how RDMA can be leveraged to implement an algorithm for a more sophisticated problem: consensus. Consensus is a non-trivial problem, and achieving it in distributed systems is impossible and can only be done under certain assumptions. The algorithms that achieve consensus are often hard to understand - for example, Paxos. This paper takes the concepts of Paxos and implements them using RDMA to achieve microsecond-level consensus. This is very useful in applications demanding microsecond latency.

Presented at USENIX Operating System Design and Implementation (OSDI) '20, this influential paper from VMware Research and EPFL (Swiss) has since received significant attention within the distributed systems community. It was authored by Marcos Aguirela, a leading researcher at VMware.

In this insight, we will start with the latency requirements of modern microservices systems. We will also examine latency vs. throughput in detail and understand the trade-offs between them. Next, we will revisit RDMA - this time understanding hardware-based RDMA in greater detail. Finally, we will review what consensus is and see how Mu solves the consensus problem using a Paxos-like algorithm implemented using RDMA to achieve microsecond latency.

### 35.1. Microservices Latency

Almost all backend systems are composed of a multi-level microservices architecture, as shown in Figure 171.

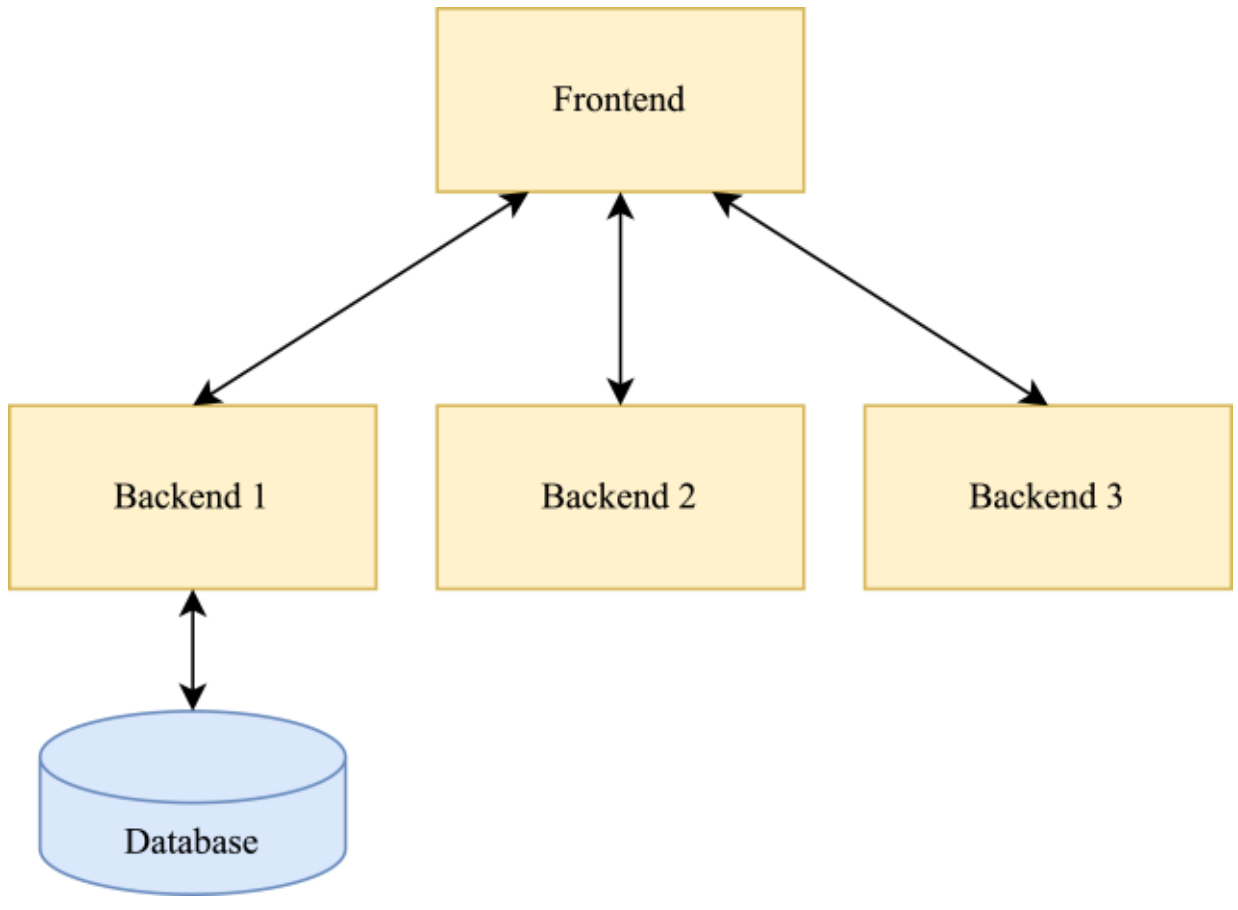

**Figure 171**: A microservices based system.

If 99% of requests to the backends are fast, the overall speed of the application will still be slow, as any single backend can introduce latency. For example, consider a frontend service that depends on five backend services, each with a good 99% latency. The probability that a random request will be fast is calculated as $0.99^5$, which equals approximately 0.95 or 95%. This calculation still doesn't consider that the frontend service itself may run into issues. For example:

- **CPU Scheduling**: The thread processing responses from the backend services may not be scheduled on time by the CPU scheduler. Additionally, the usual Linux context switch time is 5ms; even if the thread is scheduled on time, it may take a latency hit.
- **Page Fault**: The thread processing the response may run into a page fault and get descheduled by the operating system.
- **Network Congestion**: The network path used by the thread to communicate with backend services or clients might experience congestion.

To optimize the latency of the system, the application needs to optimize each backend to achieve four or five nines of availability, meaning that only when the p9999 or p99999 latency is low will the overall latency be good.

Jeff Dean has an excellent article [210] on what it takes to optimize for tail latency in a cloud-like environment where thousands of jobs compete for resources. Researchers have explored various approaches to optimize the tail latency of applications, including:

- Steering interrupts to different CPU cores to improve processing efficiency.
- Implementing custom kernel scheduling policies to prioritize certain tasks.
- Using custom paging mechanisms, like huge pages [222], to reduce TLB misses and improve memory access times.

### 35.2. Latency v/s Throughput

**Latency** and **throughput** are two important measures of system performance. Latency defines the time it takes for a unit of work to be completed, and throughput defines the amount of work that can be done by the system per unit of time.

To understand this, consider a well-known analogy: a pipe through which water flows, as shown in Figure 172.

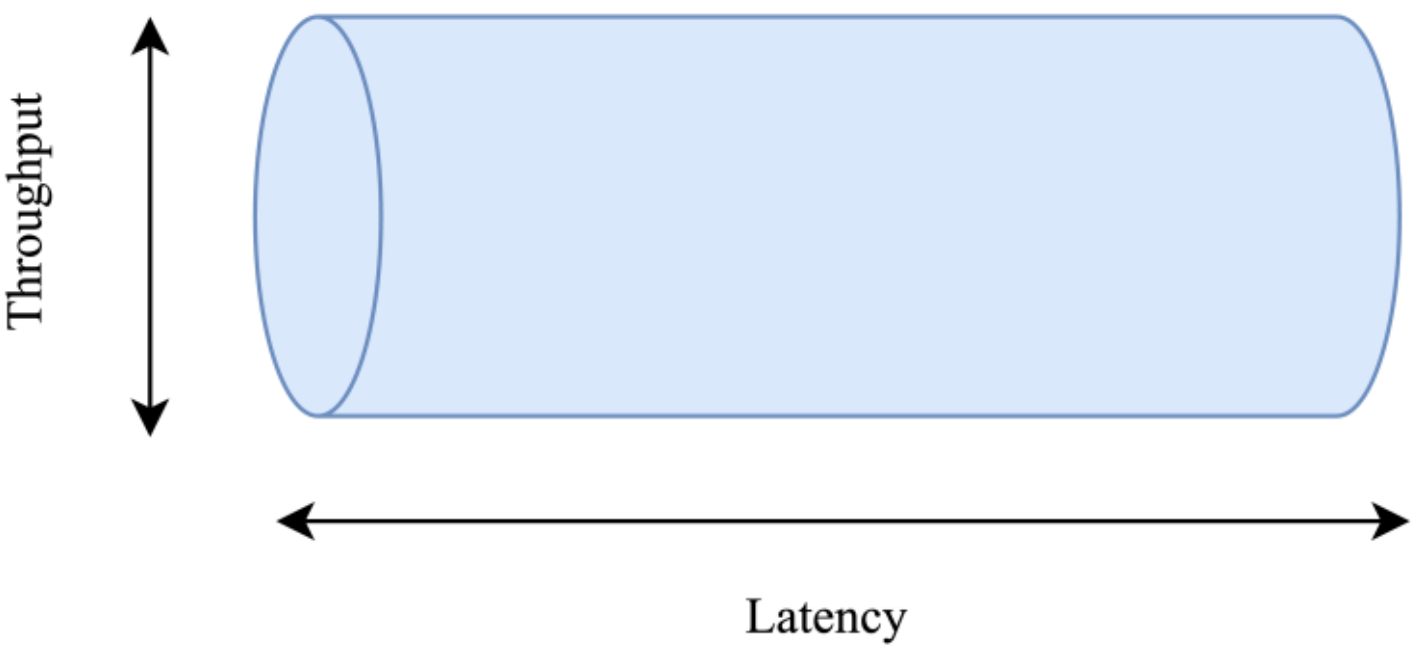

**Figure 172**: Water pipe analogy for latency v/s throughput.

Ignoring friction and viscosity, and imagining water flowing through the pipe at a constant speed, the latency of each water molecule would be the length of the pipe divided by the speed. The throughput of the pipe would be determined by its diameter. The larger the diameter, the more water can flow. Throughput is expressed as volume per unit of time; in terms of water, it could be gallons per second.

- Latency can impact throughput because if water flows faster, more water can pass through per unit of time.
- Conversely, throughput can also impact latency. If the diameter is small, less water can flow through, potentially causing a backlog or queue of water molecules waiting. This waiting time in the queue is part of the total latency.

Coming back to computer science, let's consider work items each requiring a fixed amount of CPU processing to complete. The latency of a work item would be the time it takes for its processing to finish after submission. If a work item requires 100ms of processing and there is no queuing, all work items will take 100ms to complete.

However, when a CPU is processing a work item, other submitted work items may be stalled. With one CPU, we achieve a throughput of 10 items/second. But not all work items will have the same latency; the worst-case latency (for the last work item in a sequence) can be 1 second. If, however, work items are submitted only after the previous one completes, there would be no queuing, and the latency would be uniformly 100ms for all.

Adding more CPUs (assuming no communication overhead between them) can linearly increase throughput. With six CPUs, we could achieve a throughput of 60 items/second.

In both cases (the pipe example and the number of CPUs), there is a hard limit on the physical, tangible resources. The number of CPUs cannot be unbounded. This limit on physical resources imposes a maximum throughput beyond which work items will experience queuing no matter how fast or slow they are submitted, thereby impacting latency. In the example above, where work items take 100ms of CPU time and there are 6 CPUs, the maximum supported throughput without impacting the 100ms latency is 60 items per second (assuming work items are submitted sequentially). Beyond this throughput, queuing will occur, and latency will increase.

In an ideal system, latency remains constant as throughput increases up to a certain point. Beyond this point, latency begins to increase as work items experience queuing. In a real-world scenario, well-designed systems strive to approximate this ideal behavior, although some deviation is inevitable due to factors such as CPU scheduling delays and context switching. In contrast, poorly designed systems struggle to maintain throughput, and latency degrades rapidly as throughput increases. The throughput v/s latency for different systems is shown in Figure 173.

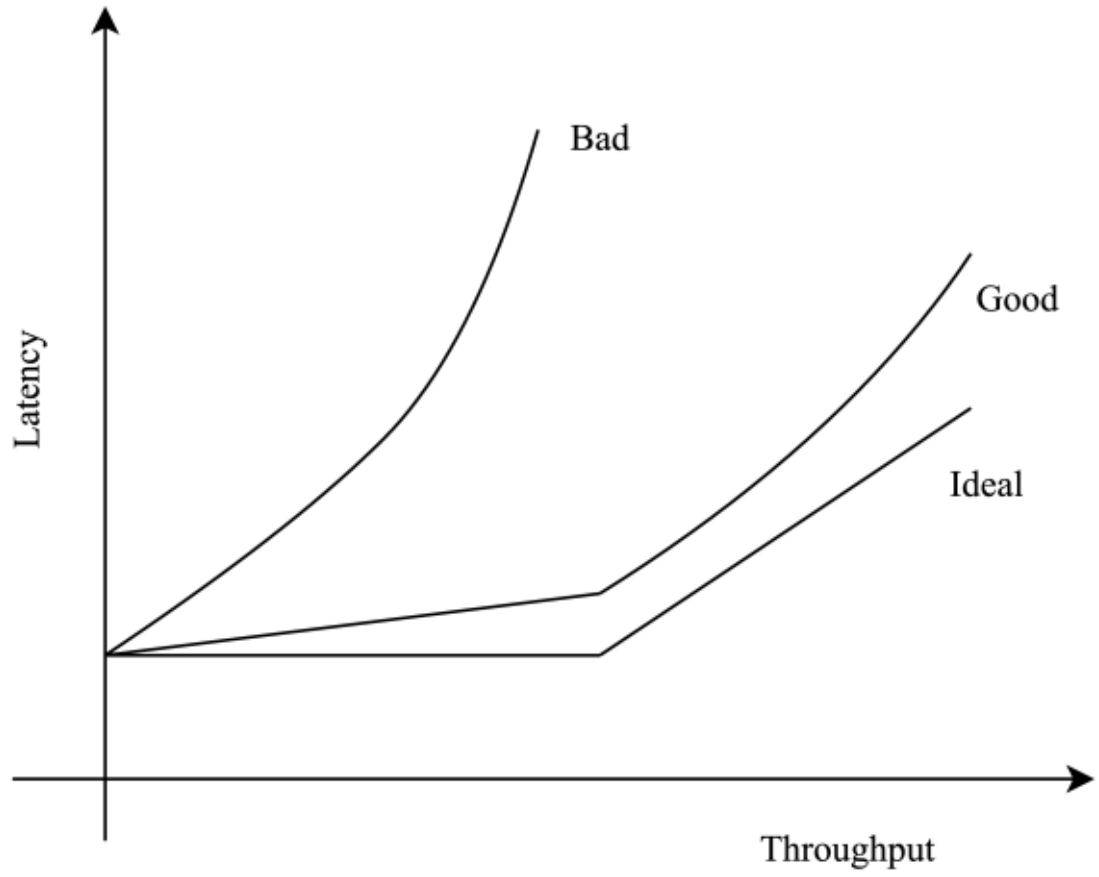

**Figure 173**: Variation of latency with throughput for different systems.

### 35.3. Non-Uniform Memory Access

On a Non-Uniform Memory Access [223] (NUMA) machine socket (shown in Figure 174), multiple CPUs each possess local L1 and L2 caches, while sharing a common L3 cache. The caches are accessed for reads and writes before the memory.

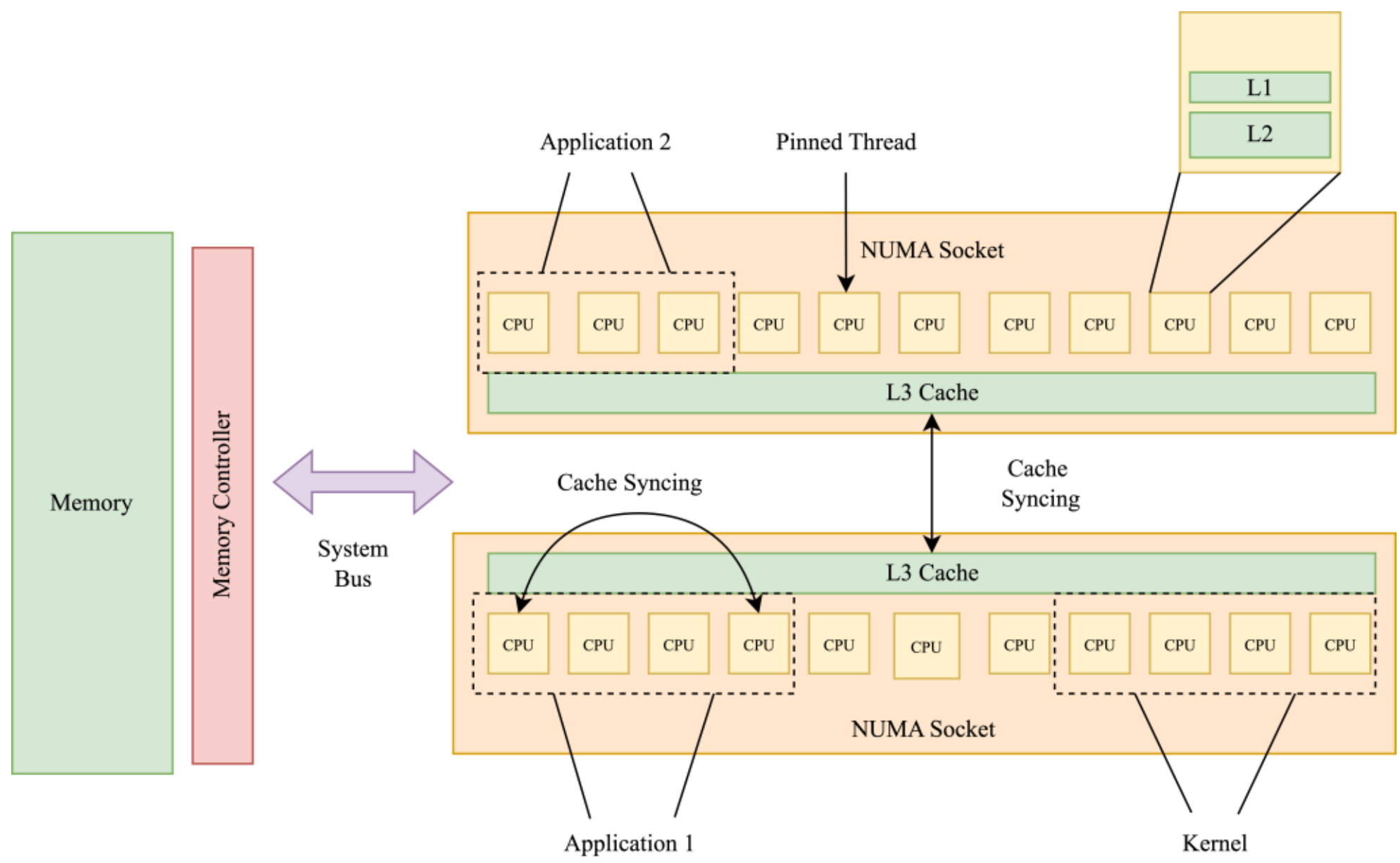

**Figure 174**: NUMA architecture.

Threads are executed on these CPUs. Typically, threads belonging to the same application are scheduled to run on a shared set of CPU cores to leverage data sharing within the caches, establishing thread affinity to a CPU. It is also possible to "pin" a thread to a specific CPU core, dedicating that core and eliminating scheduling latency for that thread. However, when multiple threads of an application execute on different CPUs that do not share L1/L2 caches, these lower-level caches can become inconsistent for the same memory location. Hardware is responsible for maintaining cache coherence. This synchronization is often performed lazily, introducing a latency of approximately 400 ns on the access path, depending on the inter-CPU distance.

For a given memory location, read operations preferentially access the cache before resorting to RAM. Similarly, write operations are initially performed only on the cache (in write-back mode). Subsequently, these cache modifications are synchronized with RAM through two primary mechanisms:

- **Memory Barriers:** A specific instruction that enforces the synchronization of cache pages to RAM. Memory barriers are crucial for ordering memory operations and are therefore essential in the implementation of locking primitives.
- **Cache Eviction:** When a cache page is evicted to make space for new data, its contents, including any writes, are written back to RAM.

Almost all computers (servers and workstations) are NUMA machines.

### 35.4. Remote Direct Memory Access

**Recommended Read**: **26. CliqueMap: Productionizing an RMA-Based Distributed Caching System** where RDMA, particularly, Software RDMA was introduced in detail. In this chapter, we will focus on Hardware RDMA only.

As shown in Figure 175, in traditional networking stacks, when an application intends to retrieve data from a counterpart application operating on a remote machine, it initiates a Remote Procedure Call (RPC). This RPC execution requires a pathway through the operating system kernel and the NIC on the sending host. Subsequently, the data request traverses the NIC and the kernel of the receiving host. Following this kernel-level processing, the appropriate application thread on the receiving machine is invoked to access and read the requested data from its memory.

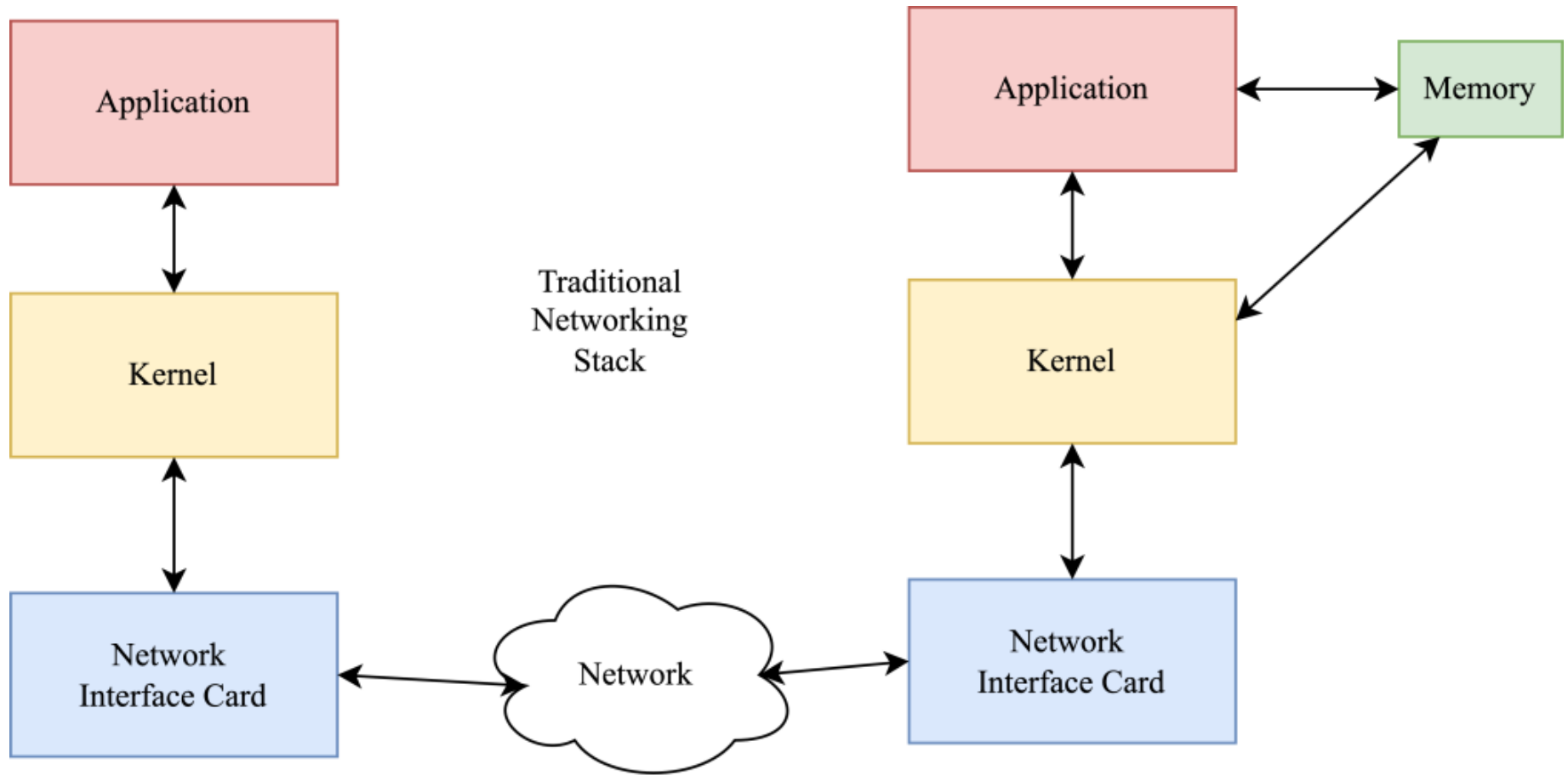

**Figure 175**: Traditional networking stack.

In contrast, as shown in Figure 176, when employing RDMA, the NIC is granted direct access to specifically designated memory regions on the remote machine. This capability for remote direct access fundamentally eliminates the necessity for CPU intervention and, as a consequence, completely bypasses the involvement of both the operating system kernel and application threads in the data transfer process. The NIC itself possesses the inherent ability to directly read the required memory locations on the remote system and transmit the corresponding response back to the requesting application.

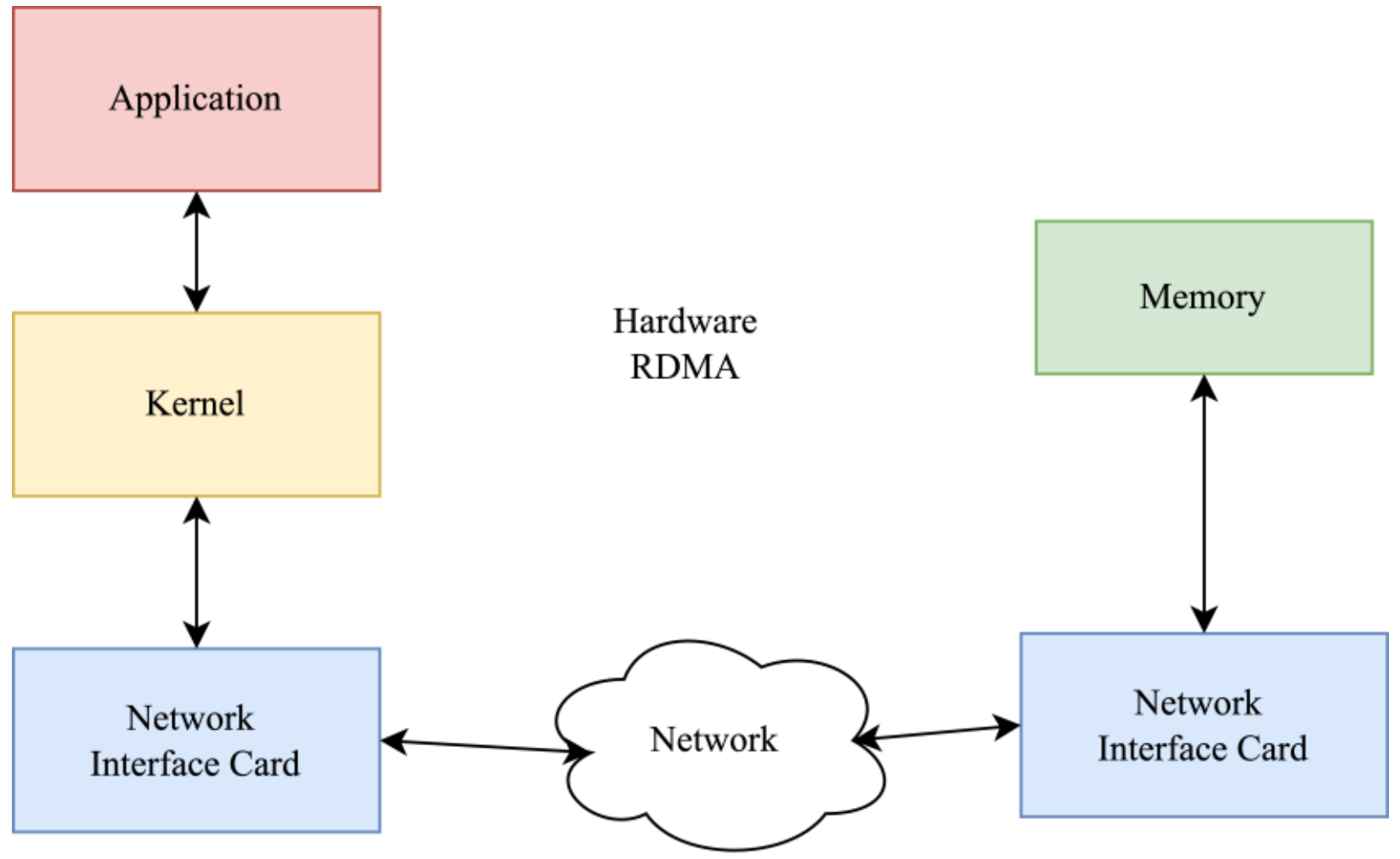

**Figure 176**: Hardware RDMA-based networking stack.

Let's take a detailed look behind RDMA mechanics.

*35.4.1. Memory Registration*

Not all of the host's memory is available for remote access via RDMA. Only specific portions of the host's memory that is explicitly registered with the NIC can be accessed remotely, as shown in Figure 177. This registration process makes these designated memory regions accessible to the NIC's RDMA controller, which can then perform read and write operations directly on those memory locations without CPU intervention.

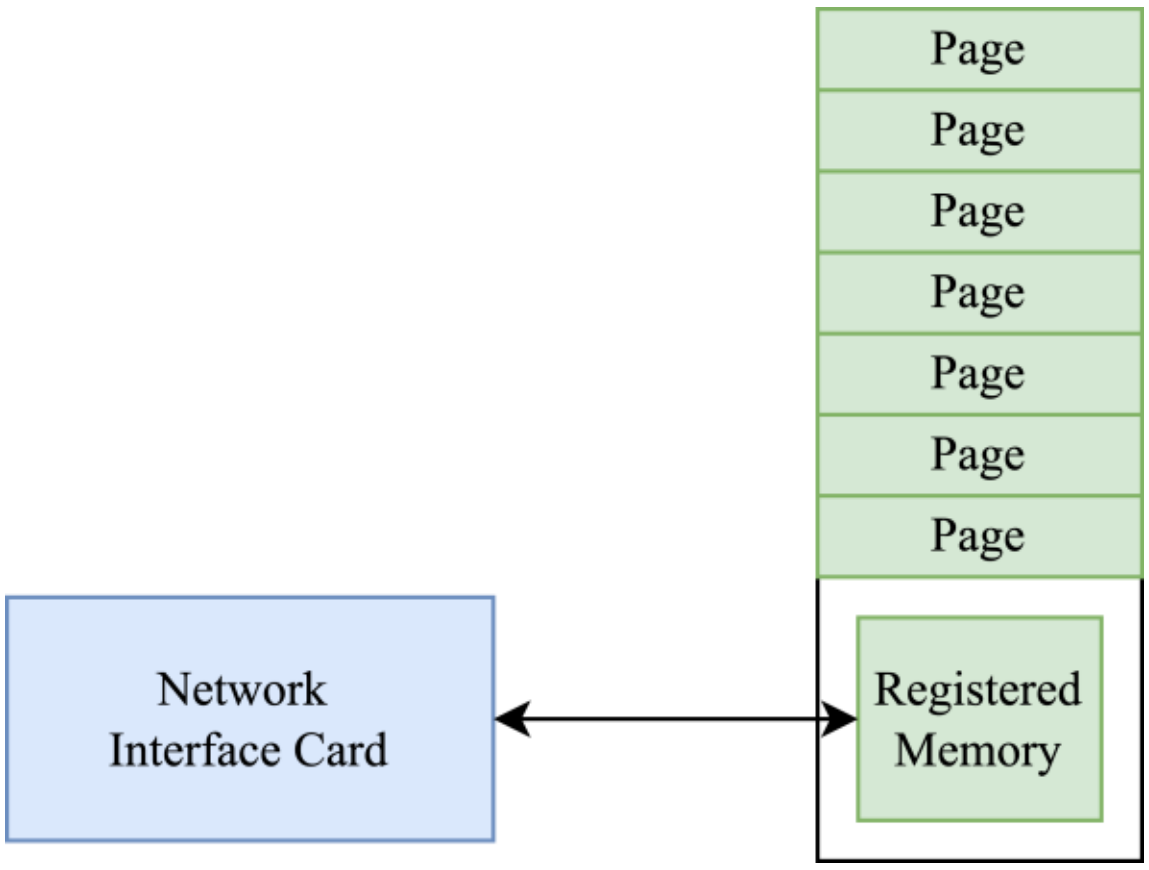

**Figure 177**: Registered memory for RDMA.

The registered memory locations are usually "pinned", i.e., they are always resident in the memory and are never paged out.

*35.4.2. Queue Pairs*

A Queue Pair (QP) represents the fundamental communication endpoint of an RDMA connection, as shown in Figure 178. Any work request submitted to one end of a QP can be processed and received at the other end of the paired QP on the remote machine. Conceptually, a QP comprises two distinct queues: a **Send Queue** for initiating outgoing messages and a **Receive Queue** for handling incoming messages.

Messages placed onto the Send Queue are essentially commands directed towards the NIC on the local host. The commands may also have pointers to the memory location where the payload can be read from.

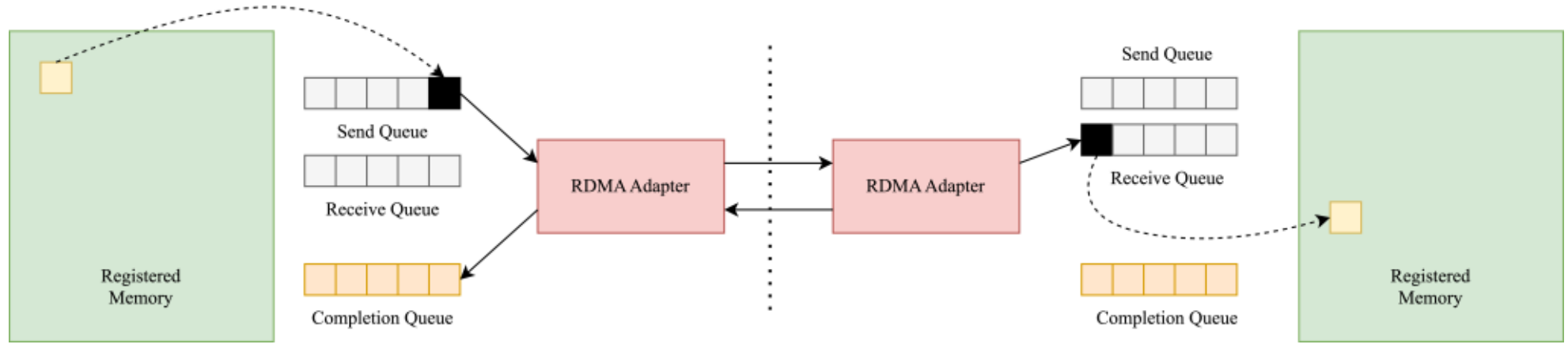

**Figure 178**: Queue pairs in Hardware RDMA.

The client's CPU encodes these commands and places them into the Send Queue of the QP. Subsequently, the host's NIC hardware takes over and executes these commands. The NIC communicates with the remote NIC to create a similar command on the Receive Queue. Finally, the payload is transferred.

*35.4.3. RDMA Commands*

NIC's RDMA controller supports certain primitive commands such as:

- **Read**: Retrieve data from a specified memory address on the remote host.
- **Write**: Transfer and store data to a designated memory address on the remote host.
- **Atomics**: Perform atomic memory operations like **Compare-and-Swap (CAS)** on a given memory address or **Fetch-and-Increment** a numerical value at a specific location.

Higher level commands can be built on top of these primitives.

*35.4.4. RDMA Connection Types*

RDMA supports several variations of connection semantics, offering different trade-offs in terms of reliability and connection establishment:

**1. Reliable Connection (RC)**

Prioritizes reliable data delivery. It leverages acknowledgements generated by the NIC hardware to ensure that commands are successfully received by the remote peer.

Furthermore, RC guarantees the ordering of message delivery. To achieve this reliability, the underlying hardware implements mechanisms for retransmissions (retries), acknowledgements, and sequence number tracking, all managed directly by the NIC.

RC in RDMA is the functional equivalent of TCP in traditional networking.

**2. Unreliable Connection (UC)**

In contrast to RC, UC does not provide guarantees for reliable delivery or ordering of messages.

**3. Unreliable Datagram (UD)**

A connectionless mode of communication, similar to UDP in traditional networking. No dedicated channel is established before communication. UD offers lower overhead but lacks the reliability and ordering guarantees of RC.

### *35.4.5. Access Checks*

To ensure memory protection and controlled access in RDMA environments, each registered memory region is associated with an **access flag**. This flag explicitly defines the permissible operations on that memory region, specifying whether it can be read from or written to remotely.

Similarly, each QP also carries an access flag that defines the allowed access privileges to the remote memory regions when operations are initiated through that specific QP.

Collectively, the access flags associated with both the memory regions and the Queue Pairs provide a robust mechanism for implementing various access control policies, regulating how clients can interact with the host's registered memory.

### *35.4.6. Memory Regions and NUMA Access*

When a CPU writes a payload to a registered memory location and initiates an RDMA write request to a remote computer, the memory write is immediately synchronized with RAM using memory barriers [224].

However, consider a scenario where a payload is received via RDMA, and the NIC writes it to memory. It is possible that a subsequent CPU read might only access the L1/L2 cache, potentially missing the newly written payload if the cache hasn't been updated.

There are two primary solutions to address this potential inconsistency:

- Designate the memory regions accessed by RDMA as uncacheable. This forces all CPU reads to fetch data directly from RAM.
- Leverage PCIe coherence mechanisms, such as Compute Express Link [225], which ensures that memory writes by I/O devices are synchronized with the CPU's cache lines.

### 35.5. Consensus

**Recommended Read**: **18. Paxos Made Simple** for a more detailed introduction to consensus and state machines.

Consensus algorithms address the fundamental challenge of achieving agreement on a single value among distributed nodes. These algorithms must satisfy the following properties:

- **Termination (Liveness):** Every non-faulty node eventually decides on a value.
- **Integrity (Safety):** If all non-faulty nodes initially propose the same value, that value must be the decided value.
- **Agreement (Safety):** All non-faulty nodes agree on the same decided value.

There are no consensus protocols that can guarantee all these properties (the FLP impossibility). Paxos is one of the consensus algorithms that can guarantee safety, fault tolerance, and liveness to a high degree of probability.

One of the most critical applications of consensus is in generation of distributed logs for a **replicated state machine** (RSM). A **distributed log** is a sequence of ordered instructions that an RSM executes. Each log entry represents an instruction for the machines, ensuring that all nodes in the RSM execute these instructions in the same order.

For example, Figure 179 shows an RSM with a single variable **X** stored in it. The instructions for the state machines can be **blind-writes (X := Y)** or **read-writes (IF v(X) = Y THEN X := Z)**.

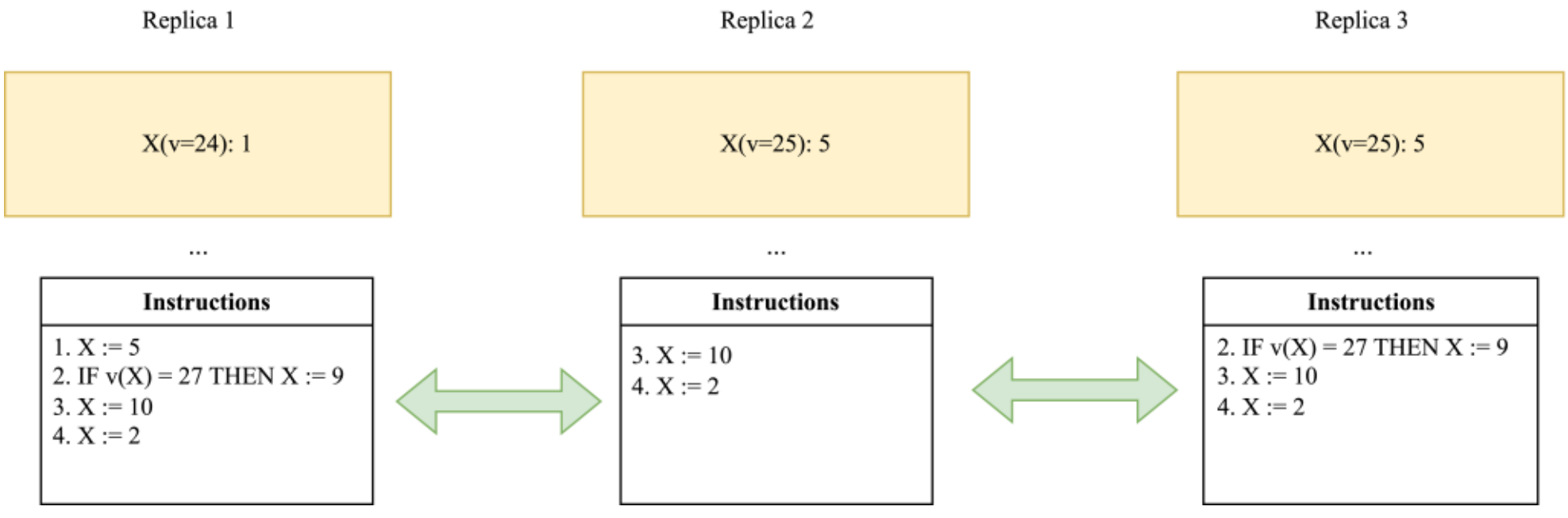

**Figure 179**: Distributed logs in RSM.

Note that different nodes of an RSM may be at different positions in the execution log but will finally converge to the same final state.

### 35.6. Microsecond Applications

Within the scope of this discussion, a pertinent question arises: is microsecond-level latency truly necessary? Human perception is generally limited to delays exceeding 10ms. However, the focus on microsecond-level latency is driven by several factors:

- As previously explained, latency exhibits an additive characteristic in layered microservices systems. Higher-level systems depend heavily on the performance of underlying components. Even small latencies at each stage accumulate, eventually resulting in an overall latency that becomes noticeable to humans.
- Furthermore, this paper highlights additional examples of systems where low latency is critical. These include certain key-value stores and high-frequency trading platforms (although achieving success in HFT often necessitates even finer-grained latency at the nanosecond level).

### 35.7. Mu

Mu, a microsecond consensus system, leverages RDMA as its fundamental communication mechanism to achieve high-performance agreement.

Its architecture, as shown in Figure 1 in the paper, comprises two distinct planes:

- **Replication Plane**: Dedicated to the writing of distributed logs.
- **Background Plane**: Handles management tasks, including leader election.

All inter-node communication within Mu relies on RDMA. The primary communication flow is as follows:

- Clients utilize RDMA to directly write their messages into server memory. The RC RDMA connection guarantees message delivery.
- Servers periodically poll their memory to detect incoming messages and subsequently process them.

This RDMA-centric communication model applies uniformly to both the replication and background planes.

#### *35.7.1. Roles*

Each node in the Mu system operates in one of two roles: **leader** or **follower**.

#### *35.7.2. RDMA Memory Regions*

The RDMA memory space is partitioned into two dedicated memory regions: one for the replication plane and another for the background plane.

##### 35.7.2.1. Permissions

- All nodes possess write permissions to each other's background plane memory region.
- However, write access to the replication plane memory region is exclusively granted to the current leader only.

##### 35.7.2.2. Replication Plane Memory Region

This region contains the following data:

- **Proposal Number**: The next proposal number to be used, maintained only on the leader node.
- **Logs**: The ordered entries of the distributed log. To manage storage, logs are written in a circular buffer, with older entries being overwritten once all nodes have acknowledged reaching a specific log position.

35.7.2.3. Background Plane Memory Region

This region stores:

- **Leader Details**: Information about the current leader. Additionally, the leader maintains a heartbeat sequence for monitoring.
- **Permission Details**: Information regarding access rights within the system.

*35.7.3. Replication Plane*

Let's delve into the workings of Mu's replication plane, with the core logic detailed in Listing 2.

At its heart, the authors have implemented the Paxos algorithm utilizing RDMA. For those familiar with Paxos, the implementation follows a straightforward pattern:

1. A **proposer** (termed "leader" in the paper) aims to propose a log entry for a specific log position (slot) it perceives as empty. **Acceptors** (referred to as "followers") respond with either an acknowledgment (ACK) or any value currently occupying that slot.
2. Subsequently, the proposer proposes a value for the slot. If all acceptors indicated an empty slot, the proposer's original value is proposed. Otherwise, the proposer must propose the value returned by the acceptors.

**Important**: If the proposer's initial value isn't accepted for a given slot, it retries with the next available slot position.

A key characteristic of Mu is that the acceptors' (i.e., followers') CPUs are not directly involved in the proposal and acceptance phases. Both these steps are executed entirely by the proposer (leader). Follower CPUs periodically fetch the committed logs and apply them to their state machines.

We will discuss the leader election in the next section. For now, let's assume a leader has obtained write permission to the replication plane memory region from a majority of followers, termed **confirmed followers**. It's possible to have multiple nodes believing they are the leader, for instance, if one leader believes it has a majority while some followers have granted write permission to a new leader without the original leader's awareness.

#### 35.7.3.1. Data Structures

Each node maintains a **Log** struct with the following fields:

- **minProposal**: The minimum Paxos proposal number seen by this node. A proposer can only issue new proposals with a number exceeding all known **minProposal** values.
- **FUO (First Unoccupied)**: The index of the first log slot that does not yet contain a value. All log entries up to (but not including) FUO are considered stable and can be directly applied.
- **slots[]**: An array containing log entries in the format **<proposal number, value>**. The proposal number indicates the proposal under which the value (the actual log entry) was accepted. The proposal number may change but the value, once committed, can never change throughout the working of the algorithm.

#### 35.7.3.2. Core Algorithm

Let's examine Listing 2 in the paper, the core algorithm:

- **Lines 13-15**: As previously mentioned, a proposer continuously executes the two phases of Paxos - PREPARE and ACCEPT - until its proposed value is accepted into a log slot. The next log entry is targeted for the FUO position.
- **Lines 18-24**: The leader determines the highest proposal number it must use for the current FUO slot (based on its local FUO and knowledge of other nodes' **minProposal**). It then writes this proposal number to all confirmed followers. Simultaneously, it reads any existing value at that FUO slot on any of the followers. This mirrors the Paxos PREPARE phase, where acceptors acknowledge or return the current value of the target slot.
- **Lines 25-28**: These are critical. Following the PREPARE phase, if any follower returned a value for the FUO slot, the leader must use that same value in the subsequent ACCEPT phase. The leader then increments its FUO and retries the process for the next slot.
- **Lines 31-33**: If the FUO slot is found to be empty on all confirmed followers, the leader writes its proposed log entry to that position. If a race condition occurs (e.g., another leader has already written to that slot), the write operation will abort. The proposer then needs to retry for the next available slot.

- **Lines 34-35**: Importantly, the proposer only relinquishes its attempt when it has successfully written its own proposed value to some log slot.

#### 35.7.3.3. Impact of Multiple Leaders on Consensus

Multiple leaders do not compromise the safety of the consensus in Mu. Paxos, by its design, prioritizes safety over guaranteed termination (as dictated by the FLP impossibility result).

Consider a scenario with two nodes believing they are the leader. This can arise when a majority of followers grant write permission to a new leader while the previous leader still considers itself the leader. In this situation, the original leader will lose its write permission on a majority of nodes, **which will share at least one node with the new majority**. Consequently, the original leader will be unable to perform any writes – neither proposal numbers nor log entries. It can still read information and update its local state.

Indeed, the write permission to the replication plane memory region acts as a crucial safeguard, even without direct CPU involvement in the data transfer. This write permission is managed through mechanisms that do involve the host's CPU.

#### 35.7.3.4. Catch-up Mechanism

New leaders and followers that have fallen behind in the log can catch up by reading the FUO value from all other nodes and then blindly applying all log entries up to the highest observed FUO to their local logs (as shown in Listings 3 and 4). Subsequently, they set their local FUO to this highest known value. In fact, any aspiring leader must perform this catch-up process before assuming the leadership role.

This mechanism works because if the FUO is incremented on even a single replica, the corresponding committed command will eventually be reflected on all other replicas. This is guaranteed by the monotonic nature of proposal numbers and the fundamental properties of Paxos. All log slots up to the FUO on every replica will always contain the exact same committed command.

#### 35.7.3.5. Omitting the Prepare Phase

An optimization employed in Mu is the potential omission of the PREPARE phase immediately after a leader is elected. At this point, it's known that the leader has been granted write permission by a confirmed majority of followers. Therefore, the leader can directly attempt to write its value to the next available slot. The **minProposal** value is also already established. If the leader subsequently loses its leadership, this will be detected when its write attempt during the ACCEPT phase fails on the majority of nodes.

#### 35.7.3.6. Follower Commit Read

Followers continuously poll the log memory region to read new entries. However, they need a mechanism to determine how far they can safely read. Since the leader doesn't directly update the FUO on follower nodes, followers rely on a **canary byte** embedded within each log entry. This canary byte serves two critical purposes:

- It acts as a marker, allowing followers to identify the beginning of a complete log entry.
- It ensures that a log entry is read only when the entire entry has been written. This is important because RDMA writes across the entire log entry memory region might not be a single atomic operation. Therefore, the leader writes the log entry in several steps and then atomically sets the canary byte to signal completion.

### *35.7.4. Background Plane*

#### 35.7.4.1. Leader Election

Let's ask a fundamental question: Does the replication algorithm described earlier guarantee safety, liveness, and fault tolerance? No. While it ensures safety and fault tolerance (inherent properties of Paxos), it does not guarantee liveness. This limitation is the same inherent challenge present in the Paxos protocol itself.

Following Leslie Lamport's approach with Paxos, Mu employs a leader to facilitate faster progress. Interestingly, even leader election in Mu is achieved through Paxos, by appending an entry to the distributed log. Consequently, leader election itself is also susceptible to the same liveness issues.

Despite this, the presence of a leader significantly optimizes the algorithm's speed. To enhance the probability of achieving liveness, the authors introduce randomness in the form of timeouts.

35.7.4.1.1. Timeout Implementation

Timeouts are implemented using a heartbeat mechanism. Each node maintains a local counter that increments periodically. This counter's value is written to a specific memory address within the background plane. Other nodes monitor these counters via RDMA reads. If a counter continuously increases, the corresponding node is considered alive.

The authors define two threshold values - **failure** and **recovery** - to determine if a node is considered dead or has recovered.

35.7.4.1.2. Impact on Correctness

Does this heartbeat mechanism affect the correctness of the consensus? No. Similar to standard Paxos, Mu remains inherently safe regardless of the number of active leaders. Timeouts are solely employed to introduce an element of randomness, aiming to ensure liveness with a high probability.

35.7.4.1.3. Miscellaneous Optimizations

The authors have implemented several optimizations to minimize latency. One notable effort involves the leader election thread relinquishing its leadership bid if it detects that the replication thread is stalled and not making progress.

35.7.4.2. Permission Management

Mu employs various strategies for managing permissions. Re-registering memory regions becomes increasingly expensive (e.g., 100ms for a 4 GB region registration) as the memory region size grows, making it undesirable for maintaining low tail latency.

Instead, Mu primarily utilizes:

- **Changing the access flags on the QP (fast path)**: This provides a quick way to modify access permissions.

- **Re-initiating the QP (slow path)**: This operation resets the access to read-only and is used as a fallback mechanism when the fast path is insufficient.

## 35.8. Evaluation

The paper presents a compelling evaluation section that warrants closer examination.

### *35.8.1. Setup*

Mu's performance is assessed in two primary deployment modes: **standalone** and **attached**.

In the standalone configuration, a single thread continuously calls the propose() function within a tight loop, isolating the core consensus mechanism. Conversely, the attached mode integrates Mu as a component within a complete end-user application.

The evaluation reveals that the standalone mode exhibits slightly superior performance compared to the attached mode. This advantage stems from the standalone mode's ability for the CPU to directly submit work items to the NIC with minimal overhead. In contrast, the attached mode introduces an intermediary layer, potentially hindering this direct interaction.

Within the attached mode, two sub-configurations are explored: **direct** and **handover**. In the direct mode, the application's primary thread is responsible for executing the replication logic. This allows for relatively direct communication with the NIC, bringing its performance closer to the standalone mode, though not identically. The slight performance difference arises because multi-threaded applications can introduce resource contention and the application thread might migrate across CPU cores, leading to cache misses.

The handover mode, on the other hand, offloads the actual replication tasks to a dedicated background thread, separate from the application's threads (potentially a shared replication thread for multiple application threads). This background thread runs on a different CPU than the applications' thread. Consequently, when this background thread needs to access common memory, cache coherence protocols are invoked to maintain data consistency, introducing a latency penalty of around 400ns.

#### 35.8.1.1. Systems Under Comparison

The evaluation benchmarks Mu against several other RDMA-based replication protocols, including APUS [226], Hermes [227], and DARE [228]. Additionally, it compares Mu's impact when integrated with applications requiring microsecond-level consensus, such as LiQ, Memcached, Redis, and HERD.

### *35.8.2. Results*

#### 35.8.2.1. Replication Latency Comparison

- Figure 3 in the paper demonstrates that the standalone mode consistently achieves the lowest replication latency across various payload sizes, outperforming all other evaluated modes within Mu.
- When compared to other replication systems (as shown in Figure 4 in the paper), Mu integrated with Redis or Memcached exhibits superior performance, achieving p99 latencies of less than 2μs.

#### 35.8.2.2. End-to-End Latency Comparison

- Figure 5a in the paper illustrates that when Mu is integrated with LiQ, the replicated performance remains remarkably close to the unreplicated performance, introducing a minimal overhead of approximately 2μs.
- Similarly, integration with HERD shows that the replicated performance closely mirrors the unreplicated performance. Notably, Mu-replicated demonstrates better performance and more stable tail latency compared to DARE-replicated.
- For key-value stores like Memcached and Redis, Mu integration leads to better overall performance, particularly in terms of tail latency.

#### 35.8.2.3. Failover Time

The measured failover time is impressively low, at less than 1ms. The majority of this time is attributed to failure detection. This suggests an aggressive polling mechanism for detecting node failures.

35.8.2.4. Throughput

To maximize throughput, Mu employs request batching. With a batch size of 128 requests, 8 concurrent outstanding requests (a queue depth of 8), and a 64-byte payload per request, the median observed latency per operation is 17μs.

This translates to a calculated throughput of approximately 48 Gbps using the formula: (128 batch size * 8 outstanding requests * 64 bytes/request * 8 bits/byte * 2 replicas * $10^6$ μs/seconds) / 17μs.

As is typical with such systems, the latency remains relatively stable up to a certain throughput threshold. For Mu, this inflection point is around 45 operations per microsecond. The authors attribute this limitation to memory operations becoming the bottleneck for the replication thread beyond this point. Increasing the number of queued (outstanding) requests can potentially improve throughput but at the cost of increased latency due to longer queueing delays.

## 35.9. Paper Remarks

The paper is well-written, effectively establishing the necessary background before detailing the design. The evaluation is thorough, and the problem addressed carries broad applicability across numerous systems. For readers with a solid understanding of RDMA, the methodology is straightforward to follow.

# 36. ORCA: A Distributed Serving Systems for Transformer-Based Generative Models

(36) Yu, G.-I.; Jeong, J. S.; Kim, G.-W.; Kim, S.; Chun, B.-G. Orca: A Distributed Serving System for Transformer-Based Generative Models. In 16th USENIX Symposium on Operating Systems Design and Implementation (OSDI 22); 2022; pp 521–538.

It is time to turn our attention to modern-day system design, which would be incomplete without a distributed Machine Learning (ML) serving system. ML models are used in several modern-day systems. For serving inferences from these models, there is usually a dedicated system that exposes a microservice upon which all other services rely. In this chapter, we will explore the design of one such serving system for the Transformer model - a cornerstone ML model in Generative Artificial Intelligence.

This paper was presented at USENIX Operating System Design and Implementation (OSDI) '22 by Seoul National University and offers an excellent introduction to modern ML serving systems.

In this insight, we will start by discussing what hardware accelerators are and how they are becoming crucial in modern computing. Then, we will explore GPUs - the most popular hardware accelerators today - in great detail, discussing their specific architectures and specifications. Next, we will perform a deep dive into Transformer models. We will also touch upon pipelined parallelism, a concept found in distributed parallel data processing and compiler design. Finally, we will examine Orca to see its specific strategies, design, architecture, and pipelining used for optimizing Transformer serving.

### 36.1. Hardware Accelerators

Modern computing systems rely on hardware accelerators to speed up computationally intensive tasks. These specialized components execute complex instruction sets far more efficiently than a general-purpose CPU using RISC instructions.

Accelerators typically connect to the system bus via a standard PCIe interface, where they receive data and instructions from the CPU.

Common examples of hardware accelerators include:

- **Graphics Processing Units (GPUs)**: Originally designed for Single Instruction, Multiple Data (SIMD) processing, making them ideal for image manipulation. GPUs are crucial for training and inference in AI and machine learning (ML) applications.
- **Field-Programmable Gate Arrays (FPGAs)**: These integrated circuits offer post-manufacturing reconfigurability. They consist of a grid of reconfigurable logic blocks and programmable interconnects, enabling users to implement custom digital circuits tailored to specific applications.
- **Cryptographic Accelerators**: Co-processors specifically engineered to handle computationally demanding cryptographic operations.

## 36.2. Graphics Processing Units

GPUs were initially designed for rendering graphics. However, they evolved into powerful parallel processors. CPUs typically use a Single Instruction, Single Data (SISD) approach, whereas GPUs excel at SIMD computations. This means they can apply the same operation to many data points simultaneously, which is ideal for tasks like image processing, where identical computations are performed on numerous pixels.

### *36.2.1. The Rise of General-Purpose GPUs*

A significant shift occurred in 2007 with the introduction of NVIDIA's TESLA [229] architecture. This marked the advent of the first **General-Purpose GPUs (GPGPUs)**, specifically designed for non-graphics computations. Unlike consumer-oriented GeForce cards, TESLA GPUs were built for high-performance SIMD operations. This innovation paved the way for subsequent generations of powerful GPU architectures like the V100 (2017), A100 (2020), and H100 (2022), which are now widely used in data centers and for scientific computing. One can program TESLA and later GPUs using CUDA.

### *36.2.2. Architecture*

A GPU's power comes from its highly parallel architecture, as shown in Figure 180. GPUs are composed of many **Streaming Multiprocessors (SMs)**, often referred to as **cores**. For example, a V100 GPU has 80 SMs. Each SM, in turn, contains multiple **subcores**. A V100, for instance, has four subcores per core, as shown in Figure 181.

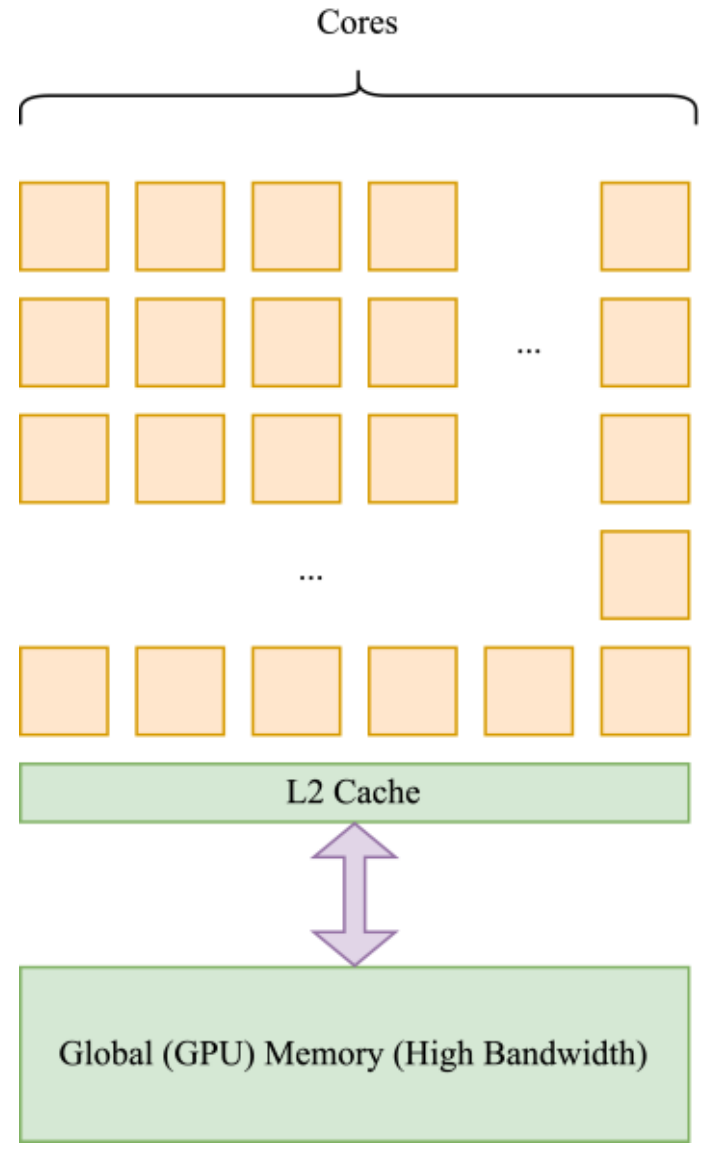

**Figure 180**: GPU architecture.

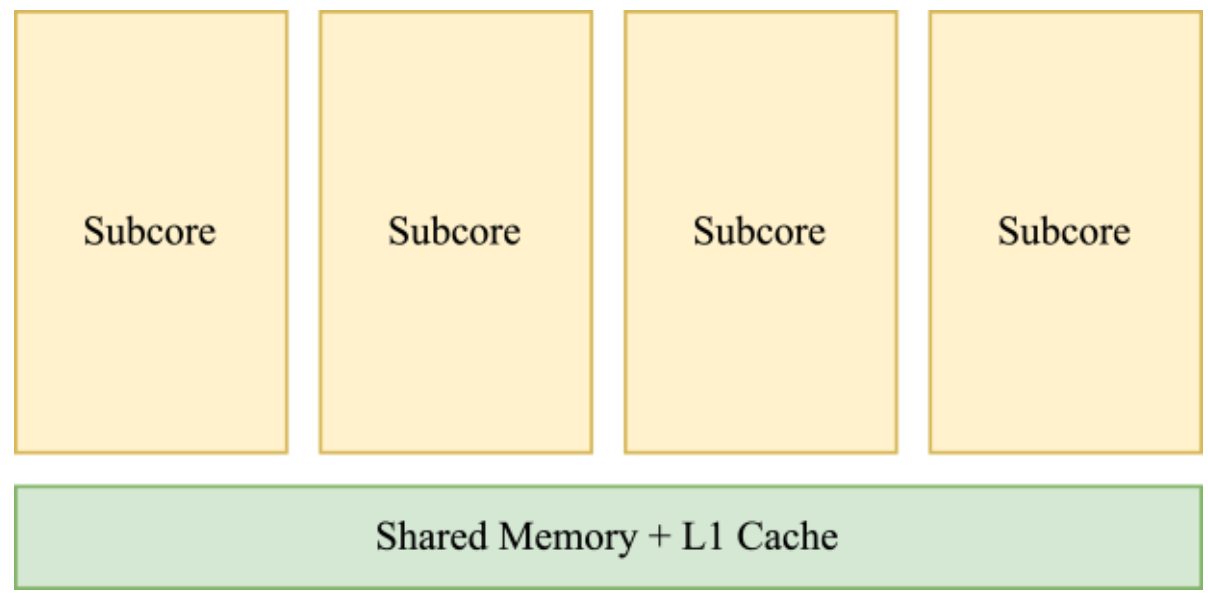

**Figure 181**: A GPU core.

Each subcore is equipped with specialized units to handle various computations, as shown in Figure 182:

- Fetch/decode units for instructions.
- Arithmetic units for integer and floating-point operations.
- Tensor units for matrix computations, crucial for deep learning.
- Load/store units for memory access.
- A set of warps, where a **warp** is a collection of registers.

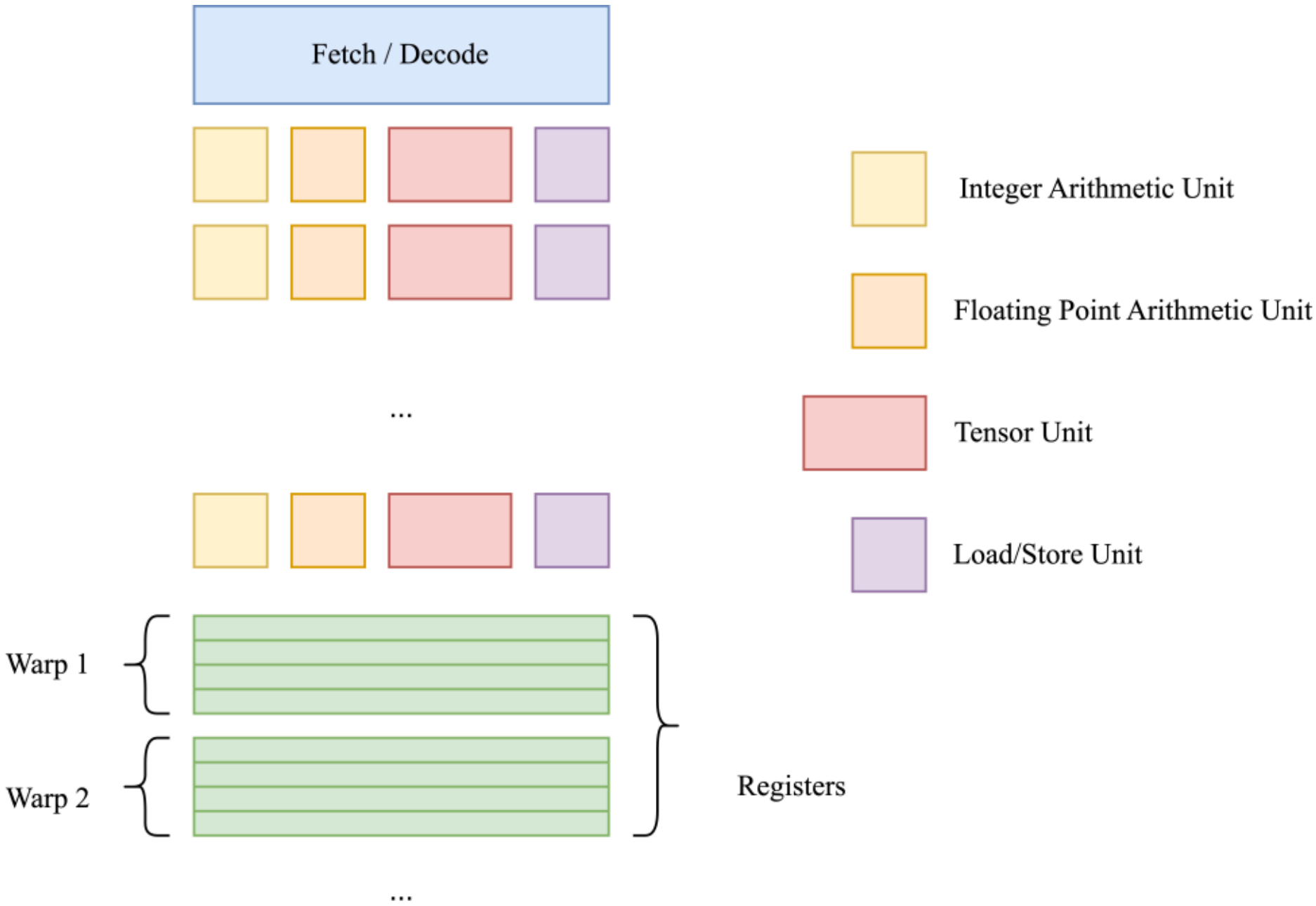

**Figure 182**: A GPU sub-core.

A CUDA thread's state is stored in registers organized into warps. These threads execute a single, common instruction simultaneously on different data, embodying the SIMD principle.

For example, In V100, each warp can store the state of 32 CUDA threads, as shown in Figure 183. There are 16 such warps per subcore. This allows a V100 subcore to execute 512 threads in parallel; an entire core can execute 2,048 threads concurrently, and the entire GPU can execute 163,840 threads in parallel.

| | | | | | |
|---|---|---|---|---|---|
| R0 | 0 | 1 | 2 | ... | 31 |
| R1 | 0 | 1 | 2 | ... | 31 |
| R2 | 0 | 1 | 2 | ... | 31 |
| ... | | | | | |

**Figure 183**: A register warp in GPU.

### 36.2.3. Memory Model

GPUs feature an on-chip memory hierarchy similar to a Non-Uniform Memory Access [223] architecture, with three levels:

- **Global memory**: Accessible by all subcores.
- **Shared memory**: Accessible only within a specific SM (core).
- **Local memory**: Accessible only by a single thread.

### 36.2.4. CUDA

CUDA (Compute Unified Device Architecture) is a programming language used to develop GPU kernels. A kernel is a routine specifically compiled to run on accelerators like GPUs.

Consider a simple loop like

```
for (int i = 0; i < n; i++)
    a[i] = 2 * a[i];
```

The body of the loop is a prime example of a kernel, as it involves the same operation applied to multiple data points.

When a program runs, it typically starts on the CPU. To utilize the GPU, the CPU first copies the necessary data to the GPU's global memory. Then, it initiates the creation of threads on the GPU.

For very large datasets (e.g., billions of data items), the data is divided into blocks. Each block is mapped to an SM core in a round-robin fashion, with the context stored in warp registers. This allows for efficient distribution of work across the numerous parallel processing units of the GPU. While the entire dataset resides in global memory, individual threads can copy portions to their respective shared or local memory for faster access during computation.

### 36.2.5. Synchronization Primitives

To manage parallel execution and data consistency, CUDA provides synchronization primitives:

- **Barriers**: Used to synchronize threads within a thread block. For instance, one thread might copy data from global to SM memory, and the other threads in that block will wait until the copy is complete.
- **Atomics**: Used to synchronize access to shared and global variables, ensuring that concurrent writes or reads don't lead to data corruption.

### *36.2.6. Specifications*

GPUs have grown increasingly powerful over time as shown in Table 4.

**Table 4**: Specifications of popular GPUs.

| Feature | V100 (2017) | H100 (2022) |
|---|---|---|
| **CUDA Cores** | 5,120 | 7,296 |
| **Transistors** | 21.1 billion | 80 billion |
| **Process Node** | 12nm | 4nm |
| **Power** | 250W | 350W |
| **FP32 Performance** | 14.1 TFLOPS | 25 TFLOPS |
| **Memory Capacity** | 16GB | 80GB |
| **Memory Bandwidth** | 900 GB/s | 1,280 GB/s |

**Note**: The listed memory bandwidth refers to on-die bandwidth within the GPU itself. Bandwidth between the system and GPU memory is dependent on the PCIe interface speed (e.g., PCIe Gen 5 offers around 200 GB/s).

### *36.2.7. GPU ↔ GPU Communication*

NVIDIA Collective Communications Library [230] (NCCL), pronounced "nickel", is a software library that facilitates communication between GPUs. NCCL is the GPU counterpart to the Message Passing Interface library, which is widely used for communication in traditional High-Performance Computing systems over regular networks.

However, NCCL isn't designed for communication over regular networks (such as ethernet). Instead, it's optimized for a dedicated hardware network, specifically NVLink, which directly connects GPUs within a machine in a mesh network for fast data exchange. The CPU ↔ GPU communication happens over PCIe whereas the GPU ↔ GPU communication happens over NVLink, as shown in Figure 184.

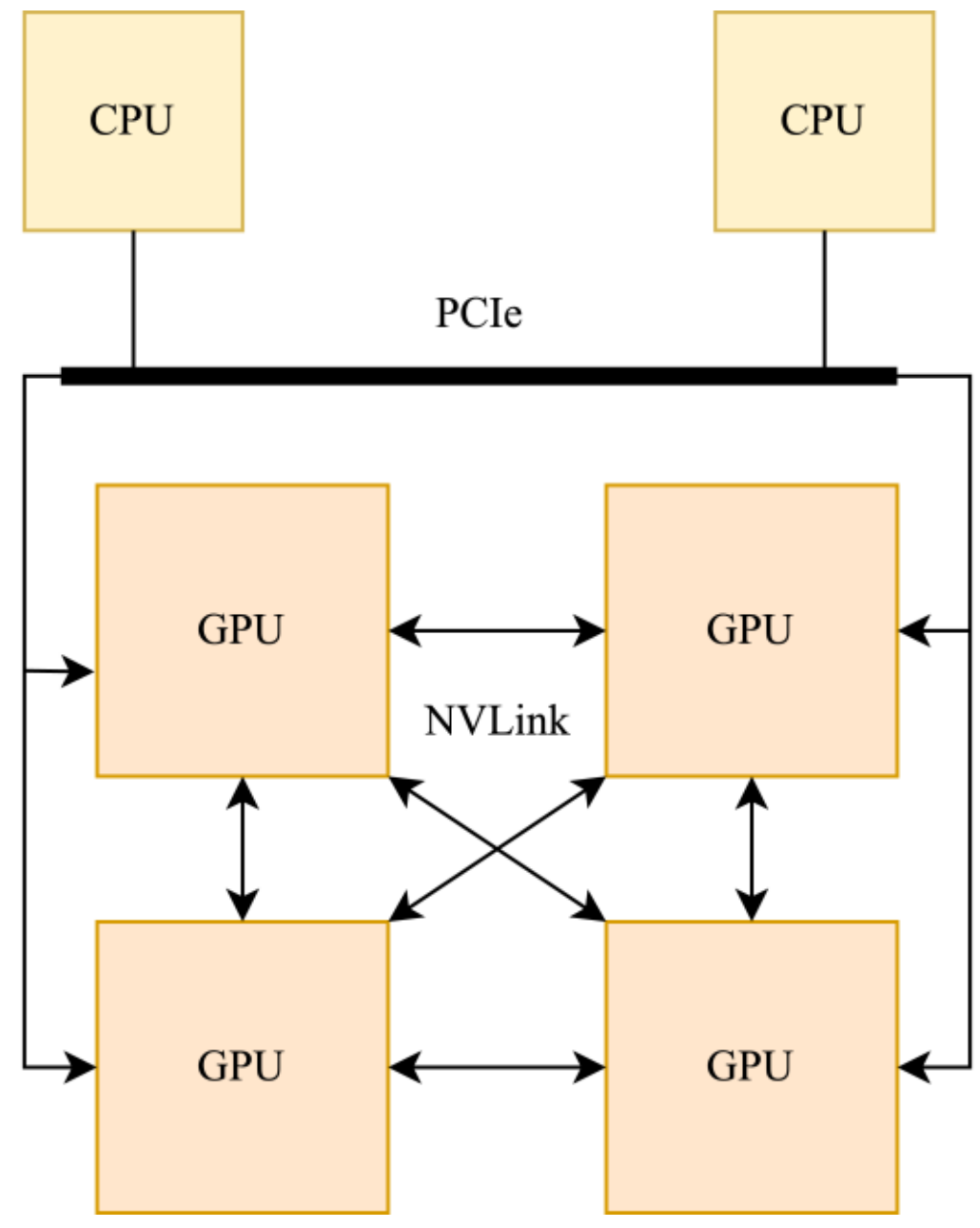

**Figure 184**: CPU ↔ GPU and GPU ↔ GPU communication.

### 36.3. ML Serving

Once a ML model is trained, the real work of ML serving (or model serving) begins. ML serving is an infrastructure challenge focused on taking that trained model, packaging it up, and deploying it as an API. The goal is to make ML models readily available to generate predictions at scale.

Every ML serving system operates with two distinct layers:

- **The Serving Layer (Frontend)**: Responsible for receiving incoming prediction requests from other systems and then handling the responses back to them.
- **The Execution Layer (Backend)**: Stores the actual ML model and runs the inputs through it to generate the output, which are then passed back to the serving layer.

For ML models like Transformers (a widely used modern architecture), specific systems have emerged to handle each layer. For the serving layer, popular choices include Triton [231] and TensorFlow Serving [232]. For the execution layer, FasterTransformer [233] and LightSeq [234] are frequently used.

### 36.4. Transformers

**Disclaimers**:

- This section offers a concise overview of only the most crucial aspects.
- There are many variations of Transformers used for a variety of applications like language processing, image processing, etc. Here we are going to explore a simple transformer for language processing only.
- The section assumes that readers have a basic understanding of neural networks.

A transformer takes one or more input tokens and produces a single output token. Mathematically, a token is represented as an **embedding** which is as a **H**-dimensional vector, where **H** (often called the embedding size or **hidden model state** size) is the total number of features.

At its core, a Transformer processes information through a series of interconnected layers, as shown in Figure 185.

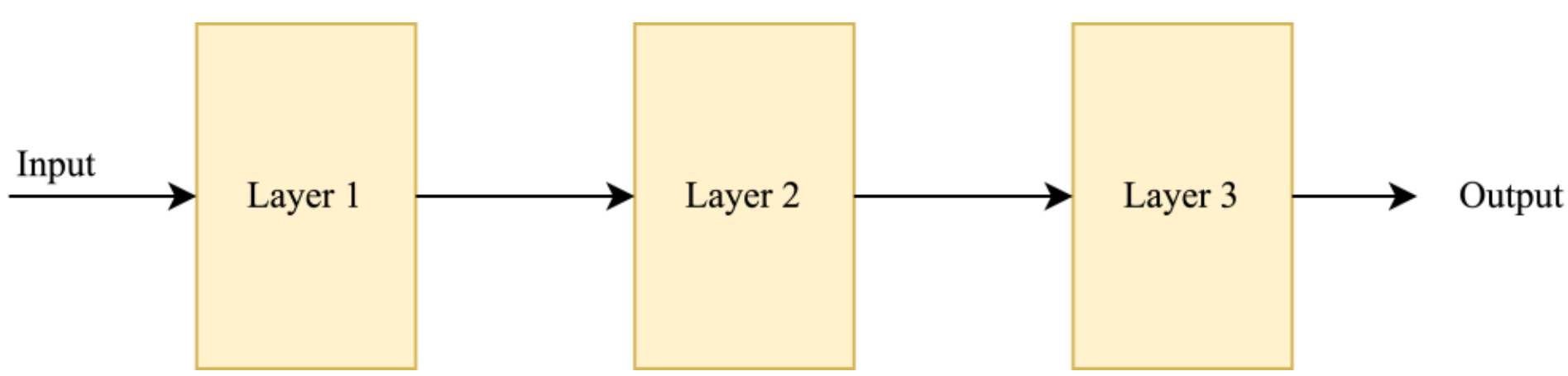

**Figure 185**: Transformer layers.

Inside each layer, there are several interconnected neural networks. All these layers transform one or more **H**-dimensional input vectors into another **H**-dimensional vector, primarily by applying the learned weights to them and then performing some normalization. An example transformer layer is shown in Figure 1(b) in the paper.

#### *36.4.1. The Attention Layer*

The **Attention** layer is a critical component within Transformer-based neural networks. For text (token-based) data, a **single-headed attention layer** operates with three key elements:

- **<u>Query (Q)</u>**: What is being looked at? In the context of text processing, this represents the current token that the attention mechanism is focusing on.
- **<u>Keys (K)</u>**: What are the elements to pay attention to? These are all the tokens in the sequence that the query might attend to.
- **<u>Values (V)</u>**: What are values associated with the keys?

Let's use the sentence: "a quick brown fox jumps over a lazy dog" to illustrate. When the attention layer processes the word "jump", it's essentially asking: "How relevant is 'jump' to every other word in this sentence?" It then evaluates this relationship with respect to all the tokens: ["a", "quick", "brown", "fox", "jump", "over", "a", "lazy", "dog"].

There are also **multi-headed attention layers** that consider multiple tokens as queries simultaneously, however, we'll focus on the single-headed version here for simplicity.

#### 36.4.1.1. The Mathematics Behind Attention

At the heart of the attention mechanism are three core matrices:

- **Query (Q)**: A **1 x H** dimensional vector (or matrix). This is essentially the input to the attention layer, representing the token whose relationships with others we're trying to understand.
- **Key (K)**: An **L x H** dimensional matrix, where **L** is the total number of tokens.
- **Value (V)**: Another **L x H** dimensional matrix.

For each of the **L** tokens being evaluated against the query, its individual key and value are **H**-dimensional vectors. These are typically generated by multiplying the token's embedding with weight parameters ($\mathbf{W_k}$ and $\mathbf{W_v}$). These parameters are learned during the model's training phase. Because there are **L** tokens to consider, the keys and values form **L x H** dimensional matrices.

While both keys and values are derived from tokens, they serve distinct purposes. The key is used to calculate how relevant a token is to the query (its "attention weight"). The value, on the other hand, represents the features or content associated with that token, which will be combined based on those attention weights.

The primary goal is to compute

$$\mathbf{softmax}(\mathbf{QK^T})\mathbf{V}$$

Let's break down this calculation:

- $\mathbf{QK^T}$: This matrix multiplication results in a **1 x L** dimensional matrix. This matrix quantifies how much "attention" the query should pay to each of the **L** tokens.
- $\mathbf{softmax}(\dots)$: Applying the softmax function to this **1 x L** matrix normalizes these attention scores, ensuring they sum up to 1.
- $\mathbf{softmax}(\mathbf{QK^T})\mathbf{V}$**:** Finally, multiplying these normalized attention scores by the value matrix (**V**) yields a **1 x H** dimensional matrix (or a vector). This resulting vector is the output of the attention layer for the given input query.

Essentially, the attention layer takes an **H**-dimensional input vector and produces an **H**-dimensional output vector, effectively transforming the query based on its relevance to all tokens.

*36.4.2. Attention Layer for Autoregressive Models*

In autoregressive models, the model used in **Generative Pre-trained Transformers (GPT)**, the goal is to generate text one token at a time, where each new token is predicted based on the preceding context. When Transformers are used for this task, a technique called **causal masking** is applied. This means that during the attention calculation, a token can only "attend" to (or learn from) tokens that came before it in the sequence, not tokens that appear later.

For an attention layer in GPT, the components are defined slightly differently:

- **Query (Q)**: This is the embedding of the last token generated, serving as the context for predicting the next one.
- **Keys (K)**: These are the keys for all tokens that precede the token currently being predicted.
- **Values (V)**: These are the values for all tokens that precede the token currently being predicted.

The output of this attention layer will be an **H**-dimensional vector. This vector essentially represents the embedding of the next word in the sequence, ready to be processed further to determine the actual token.

#### 36.4.2.1. Example Walkthrough

Let's solidify the concepts we've discussed by walking through an example - the one from the paper.

In paper's Figure 1(a), each node represents a Transformer layer. During each iteration of this process, a context and a query are supplied as input, and a single token is produced as output.

Consider the following progression:

- **Iteration 1**:
  - Query: "this"
  - Context: "I think this"
  - Output: "is"
- **Iteration 2**:
  - Query: "is"
  - Context: "I think this is"
  - Output: "great"
- **Iteration 3**:
  - Query: "great"
  - Context: "I think this is great"
  - Output: <EOS> (End of Sequence token)

The context provided in each step is subsequently transformed into the keys and values utilized by the attention layer. Notably, with each successive iteration, the dimensionality (**L**) of both the keys and values progressively increases.

**Note**: The paper differentiates between an initiation phase and an increment phase. In the initiation phase, a multi-head attention layer processes all the initial context tokens. For simplicity, however, we'll focus solely on single-head attention in all phases.

### *36.4.3. Batching*

Processing only one **<input, context>** pair at a time within each iteration significantly underutilizes the capacity of GPUs, which are highly efficient at parallel matrix operations. To boost throughput, processing typically involves batching multiple requests together.

Here's how the matrix dimensions change with batching:

- **Query (Q)**: A **B × H** dimensional matrix.
- **Key (K)**: A **B × L × H** dimensional matrix.
- **Value (V)**: A **B × L × H** dimensional matrix.

Here, **B** represents the batch size (the number of <input, context> pairs processed simultaneously). Each "kernel" operation, which constitutes a single iteration, processes an entire batch. For effective batching, a critical requirement is that the input shapes of all matrices within the same batch must be consistent. Since **H** (the embedding dimension) is typically constant, the crucial constraint is that **L** (the number of tokens or context length) must be identical across all inputs in the batch. This means all input sequences within a batch must have the same length for batching to work efficiently.

### 36.5. Pipelined Parallelism

Pipelined parallelism is a widely used technique, particularly in compiler design for parallel programs, and has also found its way into ML model compilers for efficient model implementation.

The core idea is simple: a large task is broken down into sequential stages, as shown in Figure 186. Think of it like an assembly line, similar to how cars are manufactured. As one work item (e.g., a car chassis) finishes stage $\mathbf{S}_i$ (e.g., welding) and moves on to stage $\mathbf{S}_{i+1}$ (e.g., painting), a succeeding work item (the next car chassis) can immediately enter stage $\mathbf{S}_i$ . This keeps all stages busy, increasing overall throughput.

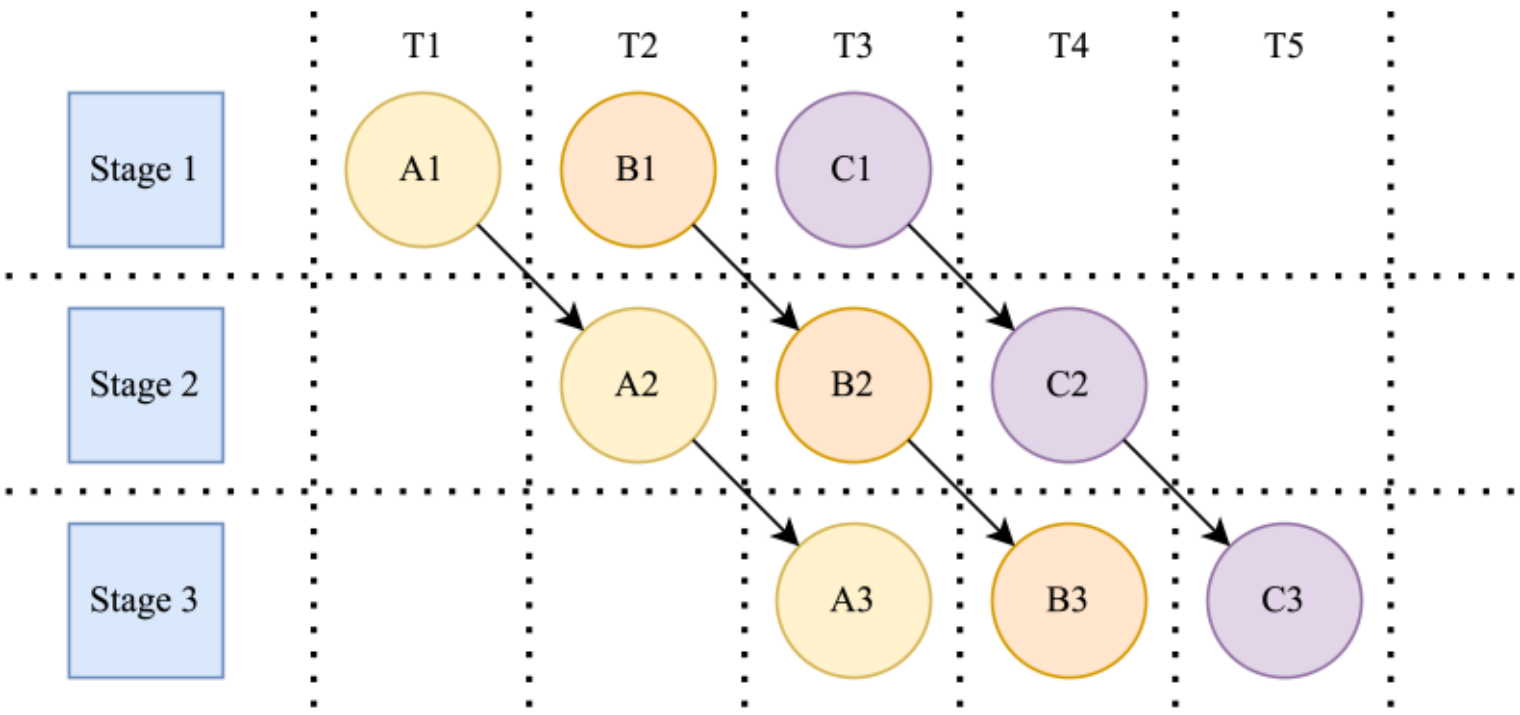

**Figure 186**: Pipelined parallelism.

We'll explore later how this technique is specifically implemented within different ML execution engines.

## 36.6. Orca

Orca is a serving system designed specifically for Transformer-based ML models. What sets Orca apart is its tight integration of the serving and execution engines into a single, unified system. This unique design delivers significant speedups for individual requests and achieves high overall throughput by maximizing the utilization of GPUs.

We'll start with a conceptual overview of Orca's goals before diving deeper into its architecture.

### *36.6.1. Design*

Conceptually, Orca operates as a sophisticated scheduler for inference requests, where each request involves generating a sequence of tokens. The processing for a given request concludes once all its tokens have been generated through successive iterations of the transformer model.

#### 36.6.1.1. Limitations of Existing Transformer Serving Systems

Existing transformer serving systems (like FasterTransformer) commonly employ **request-level scheduling**. This means the execution engine processes requests in fixed batches, where every request within a batch starts with the same number of context tokens. Multiple iterations are then run until all requests in the batch conclude with an <EOS> token.

This approach presents two significant disadvantages:

- **Batch Bottleneck**: The performance of an entire batch is bottlenecked by the slowest or longest-running request within it. If one request requires generating significantly more tokens than others (e.g., 100 times more), all other requests in that batch will be delayed until the longest one completes.
- **Forced Padding**: To create uniform batches, all requests are typically forced to have an equal number of tokens. Existing systems achieve this by padding shorter contexts to match the length of the longest context in the batch, leading to wasted computation and memory.

### 36.6.1.2. Solution 1: Iteration-Level Scheduling

To mitigate the latency issues, Orca introduces **iteration-level scheduling**. Instead of processing a full request end-to-end, Orca picks requests from a pool, processes them for only a single transformer iteration (generating one new token), and then returns them to the request pool, as shown in Figure 4 in the paper. This allows requests that generate an <EOS> token to be immediately returned to the user, significantly improving latency for shorter outputs.

With iteration-level scheduling, it becomes crucial to manage the **Key-Value (KV) cache** for each request. For every attention layer and for each token in a request's context, the corresponding KV cache data must be stored on the GPU. This state is retained in the **Attention Manager** until the request fully completes. This retention is essential for incrementally computing KV pairs across different iterations, avoiding redundant computations.

#### 36.6.1.2.1. Request Selection for Scheduling

Orca's execution engine maintains a pool of requests eligible for the next iteration run (i.e., those awaiting their next token). Requests are selected based on two primary factors:

- **Maximum Batch Size**: This is primarily constrained by the GPU's computational capacity, which depends on its core and subcore count.
- **Available GPU Memory**: Model parameters, code, and other static elements consume a constant amount of memory. The most variable element of memory consumption is the KV cache stored in the Attention Manager. Each request has an upper bound on the maximum number of tokens it can generate, which in turn sets an upper bound on its memory footprint within the Attention Manager.

As described in Algorithm 1, the scheduler:

1. Randomly picks a request from the pool until the batch reaches its maximum size (Lines 21-23).
2. Selects the requests for processing if their maximum token slots in the Attention Manager does not exceed the total available Attention Manager slots (Lines 24-26).

3. Once selected, these requests are processed through one iteration of the transformer. After the iteration completes, the newly generated token for each request is appended to it. Requests marked as finished (Line 13-14) are removed from the batch, and their reserved slots in the Attention Manager are freed, allowing new requests to be added.

#### 36.6.1.3. Solution 2: Selective Batching

Orca implements selective batching across different neural network layers. All layers except for the attention layer can process inputs in batches, as they typically operate on matrices of a consistent Batch Size (**B**) x **H**.

However, the attention layer is different; it works with matrices of dimensions **B** x Context Length (**L**) x **H**, where **L** can vary between requests within a batch. To accommodate this, Orca **Split**-s the input for each individual request before it enters the attention layer. After the attention layer processes these split inputs, their outputs are then **Merge**-d back together before being passed to the subsequent neural network layers. This is shown in Figure 5 in the paper.

### *36.6.2. Architecture*

Orca, like all modern ML serving and training systems, employs a distributed architecture to leverage the power of multiple GPUs for efficient inference.

Traditionally, ML serving systems could be scaled by simply creating **N** copies of the entire model across **N** GPUs and distributing incoming requests among these copies. However, with the increasing size and complexity of contemporary ML models, they often exceed the memory capacity of a single GPU. This necessitates a more sophisticated approach where both the model's trained weight parameters and the inference requests themselves must be split.

#### 36.6.2.1. Partitioning

Orca utilizes two primary methods for model splitting:

- **Inter-layer split**: This approach distributes different layers of the model across various GPUs. The output of one layer on one GPU is then communicated as input to the next layer on another GPU.

- **Intra-layer split**: This method involves splitting within a single layer. ML models are fundamentally composed of weight matrices and bias vectors. These often share a common dimension, for instance $\mathbf{H}$, representing the hidden state size or embedding vector length. Intra-layer splitting divides these weights and biases along this $\mathbf{H}$ dimension. For example, a key matrix $\mathbf{W_k}$ with dimensions $\mathbf{(1, H)}$ could be split into $\mathbf{(1, H_1)}$, $\mathbf{(1, H_2)}$, and $\mathbf{(1, H_3)}$, where $\mathbf{H_1} + \mathbf{H_2} + \mathbf{H_3} = \mathbf{H}$. Similarly, the input query vector is also split. Essentially, this approach distributes the computation of embedding features.

For instance, consider a scenario (shown in Figure 6 in the paper) where an inter-layer split is combined with intra-layer parallelism. A given request might first enter GPU 1, GPU 2, and GPU 3. Each of these GPUs would be responsible for processing a specific portion (e.g., 1/3) of the embeddings for the first and second layers of the model. The results from these three GPUs are then sent to GPU 4, GPU 5, and GPU 6, respectively, which then process the third and fourth layers of the model. This coordinated effort allows all six GPUs to collectively process the entire model for the request.

36.6.2.2. The Request Execution Flow

The complete execution flow is shown in Figure 7 in the paper. The **scheduler** creates a batch of requests and sends the batch over to the **engine master**. The master then sends these requests, along with crucial control messages, to the first **worker**'s controller. The control message for each request specifies the index of the next token to be generated, which is vital for identifying the correct input for the current iteration.

The first worker's controller schedules this work across its GPUs, taking into account any configured intra-layer parallelism. It passes the request ID and current token index to the GPUs, allowing them to correctly identify inputs and manage key-value pairs within the attention layers.

After scheduling, the first controller forwards the requests and control messages to the next controller in the processing pipeline, which then similarly schedules the work on its GPUs. GPUs in subsequent layers wait for the completion of the preceding layers, receiving their inputs. Once all GPUs in a given layer have completed their processing, they send the newly generated tokens back to their respective controllers, which then forward them to the engine master.

This distributed approach allows Orca to handle the massive computational demands of modern ML models by efficiently distributing the model across multiple high-performance GPUs.

#### 36.6.2.3. Communication

GPUs communicate their outputs to subsequent GPUs in the processing chain using NCCL. For communication between nodes (e.g., engine master to controller messages), Orca relies on gRPC.

### *36.6.3. Pipelined Parallelism*

Let's explore how pipelining works in both Orca and FasterTransformer. Both systems partition the transformer model into distinct stages, where each stage typically represents a transformer layer.

#### 36.6.3.1. Pipelining in Orca

Orca achieves pipeline parallelism by feeding the next batch through the layers as the previous batch exits those layers. This is made possible by Orca's iteration-level scheduling. With this approach, an entire batch is designed to pass through all layers only once before being sent back to the scheduler for potential rescheduling.

As depicted in paper's Figure 8, for instance, batch **CD** can enter Stage 1 in Orca as soon as batch **AB** leaves stage 1. This continuous flow helps keep the pipeline busy.

#### 36.6.3.2. Pipelining in FasterTransformer

In contrast, FasterTransformer's execution engine processes an entire batch to completion before that batch leaves the engine. This means all requests within a batch must be fully processed. Consequently, only a single batch moves through the different stages at any given time. This can lead to underutilization of GPU capacity, as stage $\mathbf{S}_{i-1}$ remains idle while stage $\mathbf{S}_i$ is in use.

To mitigate this, FasterTransformer employs **microbatching**, where larger batches are split into smaller microbatches. These microbatches then move through the stages in a pipelined fashion, offering a key advantage - keeps all parts of the GPU busy and occupied, improving overall hardware utilization. However, this approach comes with certain drawbacks:

- It doesn't fundamentally solve the problem of latency. The execution engine's output still waits for all requests across all microbatches to complete, due to its request-level scheduling.
- It can negatively impact latency. Each batch effectively has to pass through more stages. If there are $\mathbf{N}$ microbatches, this can add up to $\mathbf{N-1}$ additional stages of sequential processing for the entire batch to complete.

## 36.7. Evaluation

### *36.7.1. Microbenchmarks*

In this scenario, iteration-level scheduling was disabled in Orca for a direct comparison with FasterTransformer.

- Orca performed slightly worse compared to FasterTransformer. This was mostly because Orca doesn't use batching for Attention layers, which increases the overhead.
- However, as the model size increases, which requires its deployment across multiple GPUs and hence also necessitates cross-CPU communication, Orca outperforms FasterTransformer by 47% (175B model, 16 GPUs). This is primarily because of Orca's use of NCCL for cross-GPU communication, reserving it for data only. Control messages are on a separate (gRPC) channel.

### *36.7.2. Synthetic Requests*

- The per-request latency increased as the model size increased due to increased communication across GPUs.
- Due to miscellaneous overheads, the per-request latency also increased as the throughput increased up to a point, after which the latency shot up too high.
- Orca offered much higher throughput support than FasterTransformer. In both the 175B and 341B models, Orca supported at least 2 req/s without latency significantly shooting up.

- With the 101B model and very low throughput, FasterTransformer performed better than Orca in terms of latency. This was because FasterTransformer used batching in Attention layers.

### *36.7.3. Effect of Batch Size*

As batch size increased, Orca offered higher throughput. The per-request latency remained constant across different batch sizes. The tipping point for each batch size was at different throughput levels. With a batch size of 32, 6 req/s throughput was achievable.

As batch size increased, FasterTransformer also provided better throughput. Microbatching was not very useful; in fact, maximum throughput was achieved by setting microbatch sizes equal to the batch size. Increasing batch size can help, but this also increases the likelihood of including a request that has significantly more tokens, thereby slowing down the entire pipeline.

## 36.8. Paper Remarks

This paper is an excellent starting point for those intending to build ML serving systems. It provides a detailed examination regarding the construction of a system for the Transformer, a cutting-edge ML model. Through this analysis, it provides valuable insights into how such systems can be optimized. While simplification often involves the decomposition of monolithic systems, the paper demonstrates that achieving scale through optimization can sometimes benefit from tight integration. This paper is highly recommended to all ML and systems enthusiasts. Readers with a solid background in neural networks and computer systems will find it easy to follow.

# 37. The Honey Badger of BFT Protocols

(37) Miller, A.; Xia, Y.; Croman, K.; Shi, E.; Song, D. The Honey Badger of BFT Protocols. In Proceedings of the 2016 ACM SIGSAC conference on computer and communications security; 2016; pp 31–42.

In the previous chapters, we explored several consensus algorithms; however, they were all designed for non-byzantine faults. In this chapter, we will focus on protocols designed for public or permissioned environments where Byzantine faults can occur. Thousands of researchers have explored this complex field, motivated by its critical impact on blockchain and cryptocurrency economics. Developing solutions here is particularly challenging, as the literature demands a deep understanding of logical constructs and advanced cryptography.

This paper was presented at ACM Special Interest Group on Security, Audit and Control (SIGSAC) 2016, a premier conference in computer security. Authored by Andrew Miller et al. and affiliated with leading global universities, the paper is inherently technical and academically focused.

In this insight, we will first revisit the different failure types and network models in distributed systems. We will also revisit consensus and briefly summarize the different protocols available. Then, we will perform a deep dive into consensus under Byzantine faults. We will discuss the different protocols that have been developed over the years in both non-permissioned environments, such as Bitcoin, and permissioned environments, such as Dolev-Strong, PBFT, and Streamlet. Next, we will perform a deep dive into several foundational concepts that form the basis for the HoneyBadger protocol, including reliable broadcast and its implementation using erasure encoding, common coin and its implementation using threshold cryptography, binary agreement and its implementation using common coin, and common subset and its implementation using reliable broadcast and binary agreement. We will also explore Tor, a foundational technology for privacy. We will then put all the building blocks together to see how HoneyBadger functions.

Readers should be advised that the content is complex. Given the considerable depth of the research, the insights will address only the most salient aspects to facilitate an understanding of BFT protocols generally and the specific innovations introduced by HoneyBadger. The chapter won't go through the proofs of the different algorithms.

**Recommended Read**: **20. Paxos Made Simple** where consensus and related concepts were introduced.

### 37.1. Failure Types

There are two primary failure modes commonly observed in distributed systems:

- **Crash Failures**: These occur when a node terminates unexpectedly, potentially due to hardware malfunctions, software errors, or power outages.
- **Byzantine Failures**: These involve nodes exhibiting arbitrary or malicious behavior, such as sending incorrect data or deviating from protocol specifications.

In private networks, which are controlled and confined environments, only crash failures typically need to be considered. However, public networks, like those used by internet-scale systems and decentralized protocols such as blockchains, must account for Byzantine failures. Byzantine failures are a superset of crash failures.

This chapter will focus on Byzantine failures.

### 37.2. Network Models

When discussing distributed systems, we typically encounter four primary network models, each with different assumptions about message delivery times:

- **Synchronous**: In this model, there's a fixed upper bound, **Δ**, on how long any message can take to be delivered. This means a message sent at time **t** is guaranteed to arrive by **t + Δ**.
- **Weak Synchronous**: Here, the upper bound on message delay, **Δ**, isn't constant; it can increase over time. So, the delay bound itself is a function of time.
- **Partial Synchronous**: This model is asynchronous for an initial period but eventually becomes synchronous. There's an upper bound, **Δ**, on message delay that holds true only after a specific event called the **Global Stabilization Time (GST)**.
- **Asynchronous**: This is the most permissive model, with no upper bound on message delay. Messages are guaranteed to eventually reach their destination, but there's no fixed time limit for delivery.

These models represent a spectrum of assumptions, from the most restrictive (synchronous) to the most permissive (asynchronous):

$$\textbf{Synchronous} > \textbf{Weak Synchronous} > \textbf{Partial Synchronous} > \textbf{Asynchronous}$$

As we move from right to left, we are increasing our assumptions about the network's behavior. Real-world networks most closely resemble the asynchronous model.

### *37.2.1. Partial Synchrony*

The Partial Synchronous model is particularly interesting because it bridges the gap between purely synchronous and asynchronous systems. It posits that there exists a time bound, $\boldsymbol{\Delta}$, and a special event, the **Global Stabilization Time (GST)**, with these characteristics:

- The network's adversary must eventually cause the $\mathbf{GST}$ event to occur.
- Any message sent at time $\mathbf{t}$ will be delivered by $\boldsymbol{\Delta} + \mathbf{max}(\mathbf{t}, \mathbf{GST})$. This means:
  - If the message is sent after $\mathbf{GST}$ ($\mathbf{t} > \mathbf{GST}$), it will arrive by $\mathbf{t} + \boldsymbol{\Delta}$.
  - If it's sent before or at $\mathbf{GST}$, it will arrive by $\mathbf{GST} + \boldsymbol{\Delta}$.

In essence, a partially synchronous system behaves asynchronously until GST and then switches to behaving synchronously after $\mathbf{GST}$. This model might seem impractical, but it's actually a good representation of how real-world networks behave. Most networks experience outages for a limited time before stabilizing and resuming normal operation.

## 37.3. Consensus

Consensus algorithms address the fundamental challenge of achieving agreement on a single value among distributed nodes. These algorithms must satisfy the following properties:

- **Termination (Liveness)**: Every non-faulty node eventually decides on a value.
- **Integrity (Safety)**: If all non-faulty nodes initially propose the same value, that value must be the decided value.

- **[Alternative] Weak Validity (Safety)**: If a non-faulty node decides a value, then that value must be proposed by some non-faulty node. Weak Validity is weaker than Integrity.

- **Agreement (Safety)**: All non-faulty nodes agree on the same decided value.

Consensus algorithms can generate a totally ordered list of values, which are essential for:

- **Distributed logs**: Crucial for systems like replicated state machines that need to execute instructions in a specific order.
- **Ordered transactions**: Necessary for systems such as blockchain (as we'll discuss later).

### *37.3.1. No Perfect Algorithm Exists! - The FLP Impossibility*

FLP impossibility result is presented in "Impossibility of Distributed Consensus with One Faulty Process" [163]. This theorem establishes that:

**In an asynchronous network, no consensus protocol can simultaneously guarantee safety, liveness, and fault tolerance.**

### *37.3.2. Known Algorithms*

Although, no perfect algorithms exists, there are several algorithms for consensus under different network and failure model assumptions shown in Table 5.

**Table 5**: Summary of various consensus protocols.

| | Network | Safety | Liveness | Crash Faults | Byzantine Faults |
|---|---|---|---|---|---|
| **Paxos** | Asynchronous | Yes | No | Yes | No |
| **Raft** | Asynchronous | Yes | No | Yes | No |
| **PoW/PoS** | Asynchronous | No* | No* | Yes | Yes |
| **Dolev-Strong** | Synchronous | Yes | Yes | Yes | Yes |
| **PBFT** | Synchronous/ Partial-Synchronous | Yes | Yes | Yes | Yes |
| **Streamlet** | Partial-Synchronous | Yes | No* | Yes | Yes |
| **HoneyBadger** | Asynchronous | Yes | No* | Yes | Yes |

* - They actually offer those properties, but only with high probability.

In this chapter, our focus will be on protocols that can handle Byzantine faults. See **20. Paxos Made Simple** for a crash-failure-only consensus protocol.

### 37.4. Consensus under Byzantine Faults

Reaching consensus is significantly more challenging when Byzantine faults are possible, compared to dealing only with crash faults. Since participants on public networks can't always be trusted to behave correctly, Byzantine faults are a critical consideration.

A prime example of a system requiring Byzantine fault-tolerant consensus is blockchain. A **blockchain** is essentially a chain of blocks, each containing transactions, as shown in Figure 187. These transactions represent values that have been agreed upon through a consensus mechanism, and thus must be totally ordered within the chain.

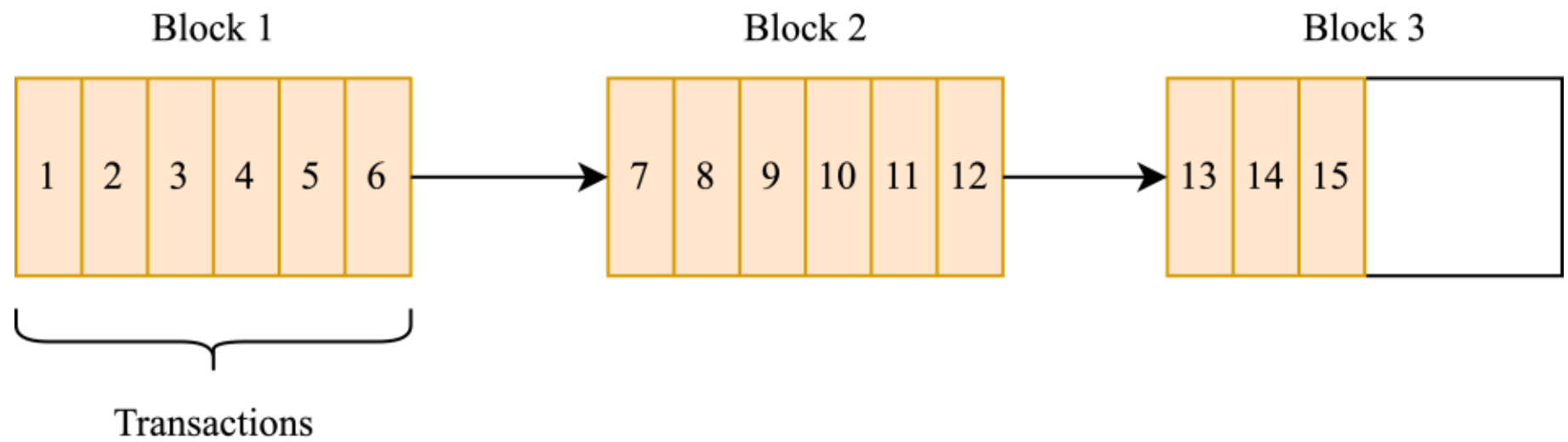

**Figure 187**: A blockchain.

All modern Byzantine networks under assume makes use of public key cryptography with a trusted party responsible for issuing the

We can broadly categorize Byzantine based consensus protocols into two main types: **non-permissioned** and **permissioned**.

#### *37.4.1. Non-Permissioned Protocols*

In a non-permissioned network, any node can join and participate without needing explicit authorization. These protocols are designed to operate in an open, decentralized environment, typically across an asynchronous internet setting and in the presence of Byzantine faults.

Examples include:

- **Proof-of-Work (PoW)**: Bitcoin is the most well-known example. Nodes mine new blocks by solving complex cryptographic puzzles, securing the network.
- **Proof-of-Stake (PoS)**: Ethereum is a prominent example. Participants stake their cryptocurrency as collateral to validate transactions and create new blocks.

Both PoW and PoS protocols are built to achieve consensus and maintain network integrity even when some participants act maliciously (Byzantine faults) within an asynchronous internet environment. However, these protocols can only maintain their consensus properties (safety and liveness) under specific assumptions. For instance, the primary security vulnerability of PoW is the **51% attack** [235]. If malicious actors gain control of 51% or more of the network's total computing power (hashrate), they can compromise the network's safety. This majority control allows them to create a longer, alternative blockchain, enabling them to reverse confirmed transactions (like double-spending) or censor new ones.

Additional challenges:

- **Vulnerability to Internet-Scale Attacks**: They can be susceptible to widespread attacks like Sybil attacks, where a single entity creates numerous fake identities to gain disproportionate influence.
- **Reliance on Cryptocurrencies for Incentives**: These networks require native cryptocurrencies to incentivize participants (miners or validators) to secure the network.
- **Slow Consensus**: Reaching final consensus can be very slow, sometimes taking several hours for values to be irreversibly confirmed.

*37.4.2. Permissioned BFT Protocols*

Permissioned BFT protocols are designed for environments where participation is restricted and known. These protocols aim to solve consensus within a defined set of participants. They offer varying guarantees regarding safety, liveness, and fault tolerance depending on the underlying network model.

All permissioned BFT protocols assume that there can be **f** faulty nodes out of a total of **N** nodes in the network. This assumption is vital, as consensus is impossible if all nodes are Byzantine. Even crash consensus protocols (like Paxos) assume that the majority of nodes are non-faulty. Typical values of **f** are $\mathbf{N-1}$, $\mathbf{N/2}$, $\mathbf{N/3}$.

Additionally, in a permissioned BFT, nodes communicate over **authenticated channels**. This means if Node **A** gets a message seemingly from Node **B**, it can be sure Node **B** actually sent it; adversaries can't forge/replay these messages. However, these communication channels don't necessarily need to be encrypted. The content of messages can be sent unencrypted, even if adversaries can read them.

BFT protocols are typically evaluated based on their latency, throughput, and scalability. Their complexity, in turn, is measured by the number of messages exchanged and the overall communication overhead.

Here are some notable permissioned BFT protocols:

- Dolev-Strong (1983)
- Practical Byzantine Fault Tolerance (1999)
- Streamlet (2020)
- HoneyBadger (2016)

Let's explore some of the protocols before diving into HoneyBadger.

37.4.2.1. Dolev-Strong (Synchronous)

The Dolev-Strong [236] protocol is a foundational BFT protocol designed for **N** nodes to agree on a value, tolerating up to **f** malicious nodes ($\mathbf{f} < \mathbf{N}$). It guarantees both safety and liveness under a synchronous network model.

The protocol unfolds in $\mathbf{f+1}$ rounds:

- **Round 0**: A designated leader broadcasts its chosen value.
- **Rounds 1 to f**: Non-faulty nodes verify received messages. If valid, they add their signature and re-broadcast the message.
- **Round $\mathbf{f+1}$**: Each non-faulty node uses the final set of authenticated messages to deterministically decide on the value.

Dolev-Strong has a message complexity of $O(N^2)$ and takes $\mathbf{f+1}$ time steps to complete. Messages also grow in size as they accumulate signatures. While theoretically significant, its synchronous network assumption severely limits its practical applicability in the real-world.

#### 37.4.2.2. Practical Byzantine Fault Tolerance (Synchronous & Weak-Synchronous)

Practical Byzantine Fault Tolerance [237] (PBFT) is a seminal consensus algorithm designed to operate in synchronous and weak-synchronous network models. PBFT is used in Hyperledger Fabric [238].

PBFT assumes a total of $\mathbf{N}$ nodes, where at most $\mathbf{f}$ of these can be Byzantine nodes. To guarantee safety and liveness, at least $\mathbf{2} * \mathbf{f} + \mathbf{1}$ nodes must be non-faulty. This requires $\mathbf{N} \geq \mathbf{3} * \mathbf{f} + \mathbf{1}$.

PBFT uses a leader-based approach where one replica acts as the primary (leader) and others are backups. It employs a three-phase commit protocol for client requests:

- **Pre-Prepare**: The primary proposes a sequence number to all backups in a PRE-PREPARE. This phase is crucial for ensuring the primary is honest about the order of requests.
- **Prepare**: Backups verify the PRE-PREPARE message and broadcast their agreement to all other replicas in PREPARE messages.
- **Commit**: Once a supermajority of PREPARE messages are received, replicas broadcast COMMIT messages, signaling their final agreement.

It relies heavily on broadcast messages (O($\mathrm{N}^2$) message complexity in its normal operation), which limits its scalability to a relatively small number of replicas (typically dozens to a few hundreds).

##### 37.4.2.2.1. PBFT under Asynchrony

Under an asynchronous network, PBFT is still safe (just like all consensus protocols, safety over liveness holds). However, liveness is affected. The paper provides a network scheduler (see section A in Appendix) which makes PBFT violate liveness.

#### 37.4.2.3. Streamlet (Partial Synchronous)

In 2020, Streamlet [239] was released which is a super simple consensus protocol that works under partial synchrony. Note that the proof is complex to understand (try the paper as a challenge!).

Streamlet assumes that $\mathbf{f < N/3}$. Additionally, it assumes synchronized clocks (required for liveness).

Streamlet proceeds in blocks instead of individual transactions (i.e. individual values). For consensus, one can think of the "blocks" as values themselves. In Streamlet, once a proposed block accumulates votes from at least $\mathbf{2 * N/3}$ distinct players, it becomes **notarized**.

The algorithm proceeds as follows:

- In each epoch, the designated leader for that epoch proposes a new block. This proposed block always extends from the longest notarized chain the leader has seen. If there are multiple longest notarized chains, the leader can break ties arbitrarily.
- Every other node in the network (voter) observes the leader's proposal. If the proposed block extends from one of the longest notarized chains that the voter has seen, the voter casts a vote (a digital signature on the proposed block). Importantly, non-faulty nodes typically vote for the first valid proposal they see from the epoch's leader.

If three blocks are notarized in consecutive epochs along a notarized chain, then all blocks up to the second-to-last of these three blocks are confirmed (finalized), as shown in Figure 188.

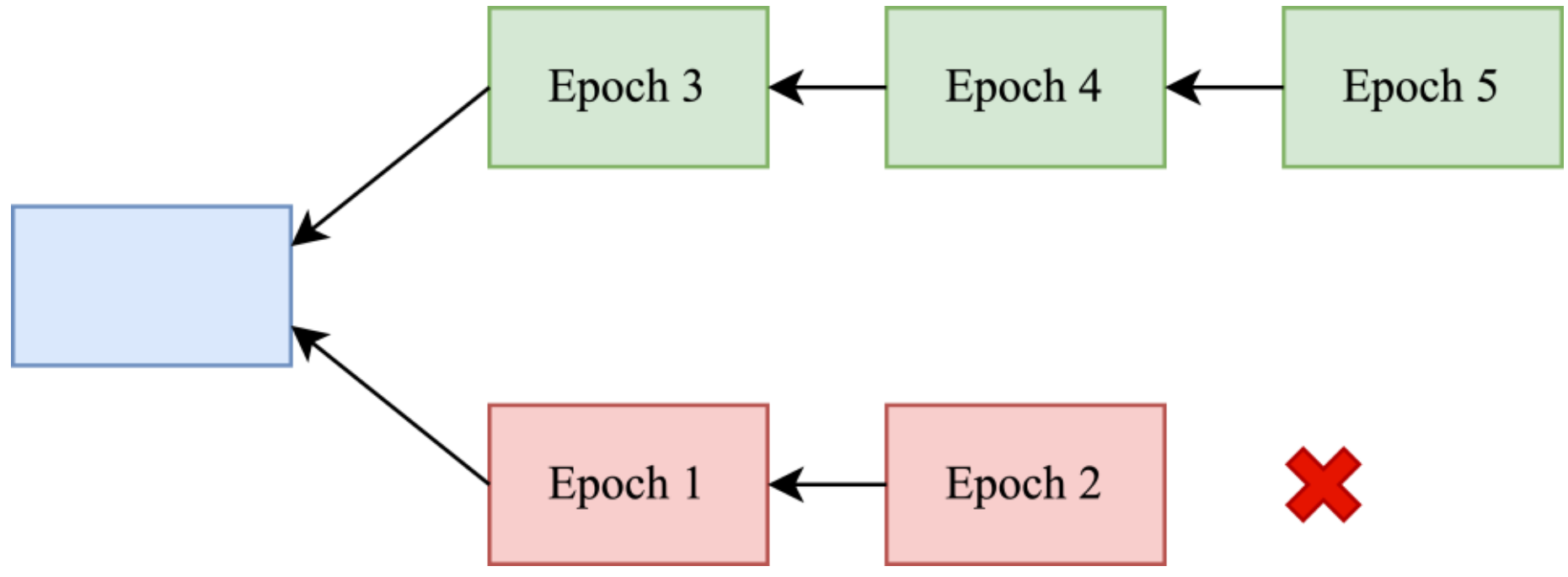

**Figure 188**: Block finalization in Streamlet.

37.4.2.3.1. Liveness in Streamlet

Streamlet is always safe under all conditions. However, Streamlet depends on synchronized clocks for liveness which may not always hold.

### 37.5. Broadcast

In distributed systems, broadcast is how one node sends a message to all other nodes.

#### *37.5.1. Ordering Guarantees*

Broadcast mechanisms offer different guarantees about the order in which messages are received:

- **Total Order**: All messages from all senders are received in the exact same global order by every recipient.
- **Causal Order**: Messages are delivered in an order that respects their causal relationships (if message A causes message B, A is delivered before B).
- **Single-Source FIFO**: Messages from the same sender are received in the order they were sent by all recipients.

#### *37.5.2. Reliability Guarantees*

Beyond ordering, broadcasts also provide guarantees about reliability:

- **Integrity**: Only receive messages that were actually sent.
- **No Duplicates**: Won't receive the same message more than once.
- **Non-Faulty Liveness**: Messages sent by non-faulty nodes will be received by all other non-faulty nodes.
- **Faulty Liveness**: If a faulty node sends a message, either all non-faulty nodes receive it, or none do.

#### *37.5.3. Types of Broadcast*

The guarantees discussed above combine to define different types of broadcast:

- **Reliable Broadcast**: Provides only the reliability guarantees, with no specific ordering.
- **Atomic Broadcast**: Guarantees both total order and reliability.
- **FIFO Broadcast**: Combines single-source FIFO order with reliability.

#### *37.5.4. Atomic Broadcast*

An atomic broadcast (ABC) guarantees the following properties:

- **Validity (Liveness):** If a non-faulty node broadcasts a message, then it eventually delivers it.
- **Integrity (Safety):** Each node delivers a message at most once, and only if the message was actually broadcast by some node.
- **Agreement (Safety):** If a message is delivered by some non-faulty node, then the message is eventually delivered by every other non-faulty node.
- **Total Order (Safety)**: All non-faulty nodes agree on the exact same delivery order of messages.

#### 37.5.4.1. ABC and Consensus

ABC is equivalent to solving consensus. This means that the FLP impossibility result applies to ABC: it's not possible to achieve ABC in a completely asynchronous network if even one node can fail.

### *37.5.5. Reliable Broadcast*

A reliable broadcast (RBC) guarantees the following properties (dropping the total order property of ABC):

- **Validity (Liveness):** If a non-faulty node broadcasts a message, then it eventually delivers it.
- **Integrity (Safety):** Each node delivers a message at most once, and only if the message was actually broadcast by some node.
- **Agreement (Safety):** If a message is delivered by some non-faulty node, then the message is eventually delivered by every other non-faulty node.

RBC may seem the same as consensus, however, there are subtle differences. RBC focuses on getting a single, specific message from one sender to all non-faulty nodes, ensuring everyone consistently receives that same message. In contrast, consensus is about a group of nodes, each starting with its own proposed value, collectively agreeing on one final value from those proposals.

RBC is achievable in asynchronous networks and in presence of Byzantine faults. Bracha's Protocol [240] is an example that demonstrates how to achieve Byzantine fault-tolerant RBC. Here's a simplified Figure of its phases (assuming $\mathbf{f < N/3}$):

- A node sends an **(SEND, m)** message to all other nodes.
- Upon receiving an **(SEND, m)** message, nodes respond by sending an **(ECHO, m)** message to everyone.
- Upon receiving enough **(ECHO, m)** messages, nodes send a **(READY, m)** message to all others. This indicates a sufficient number of peers have echoed the message.
- A node delivers **m** once it has received $\mathbf{2 * f + 1}$ **(READY, m)** messages.

Perhaps, the agreement property is the one that makes this protocol more sophisticated than a simple message sent to all the nodes. Due to this, all nodes need to acknowledge to each other that they are ready to send a message. Say the size of the message is |**v**|, then the complexity will be $O(N^2|v|)$.

There is another variant of RBC called **terminating RBC** which adds a termination property to RBC. Even that is not equivalent to consensus.

### 37.6. Erasure Encoding

Erasure encoding, also known as **Forward Error Correction (FEC)**, is a powerful technique for protecting data during storage or transmission. It adds redundancy to a message, allowing the original data to be perfectly reconstructed even if some parts are lost or corrupted.

Here's how it works (shown in Figure 189): An original file or message is divided into **k** data blocks. From these **k** blocks, **m** additional parity blocks are computed. The total number of blocks then becomes $\mathbf{n = k + m}$.

- **k**: Number of original data blocks.
- **m**: Number of generated parity blocks.
- **n**: Total number of blocks ($\mathbf{n = k + m}$).

To retrieve the original data, you only need any **k** of the **n** total blocks.

**Reed-Solomon** codes are the most widely used erasure codes, providing the maximum possible fault tolerance for a given amount of redundancy.

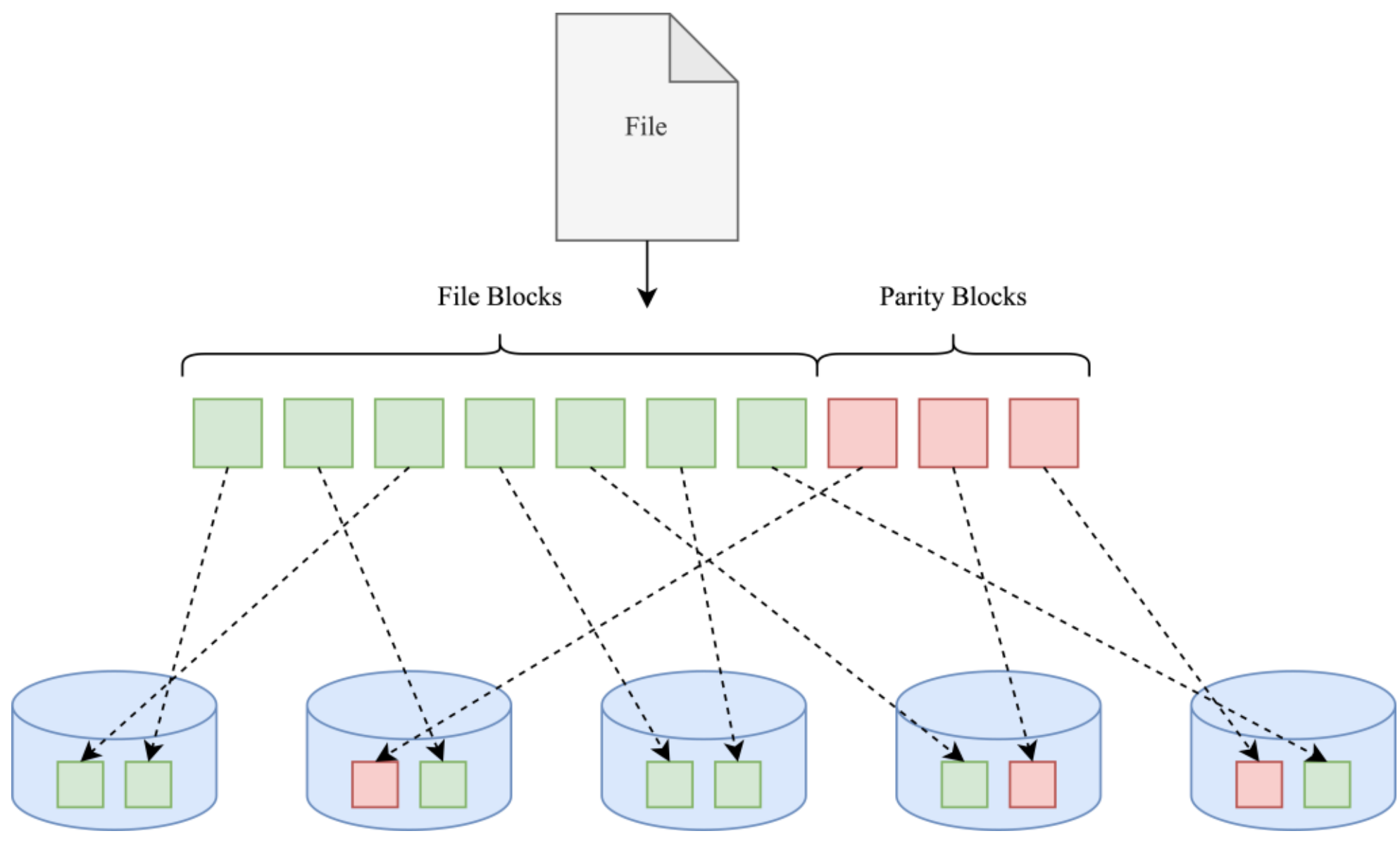

**Figure 189**: Erasure encoded file.

### *37.6.1. RBC with Erasure Encoding*

Erasure encoding can be cleverly applied to RBC to enhance efficiency and robustness.

#### 37.6.1.1. Algorithm

A message is broken into $\mathbf{N-2*f}$ data blocks, and $\mathbf{2*f}$ parity blocks are computed, resulting in $\mathbf{N}$ total blocks.

The node roughly follows these steps (Figure 2 in the paper):

- The sender distributes each of the $\mathbf{N}$ encoded blocks to a different node.
- Nodes that receive a block then echo it by sending their received block to all other nodes.
- Once a node receives $\mathbf{N-f}$ distinct echoed blocks, it's guaranteed that at least $\mathbf{N-2*f}$ of them are valid (since at most $\mathbf{f}$ Byzantine nodes could send incorrect blocks). With $\mathbf{N-2*f}$ valid blocks, the original message can be reconstructed.

37.6.1.2. Ensuring Data Integrity with Merkle Trees

A critical challenge arises: how can a node be sure that the **N − f** blocks it received are actually correct and not corrupted by Byzantine nodes, which would lead to bad data upon reconstruction? This is where a **Merkle tree** becomes essential.

The solution involves:

- The sender first computes a Merkle tree of all **N** encoded blocks (data and parity). A **jth** node receives only the **jth** branch of the merkle tree from root to leaf as part of the initial broadcast, as shown in Figure 190.

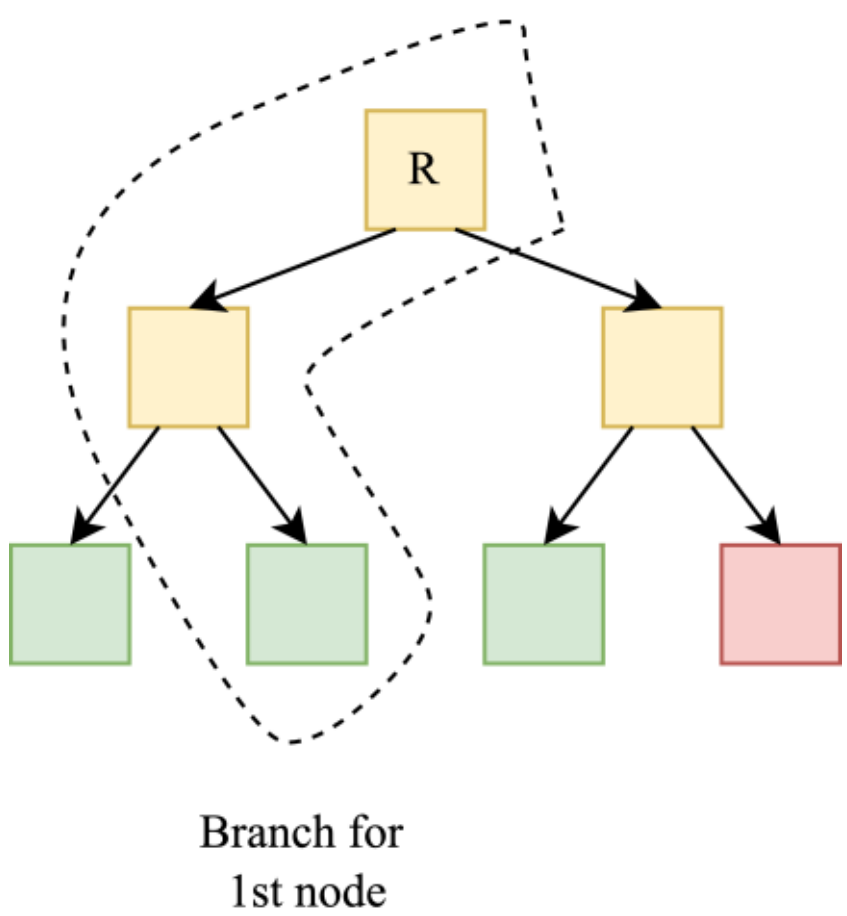

**Figure 190**: Merkle tree of encoded blocks.

- When a node receives **N − f** blocks and corresponding branches during the ECHO phase, it can now use any of the **N − 2 ∗ f** blocks to interpolate all other blocks. Then, it can verify the integrity of the received blocks by computing the Merkle root hash. If it doesn't match the Merkle root hash, RBC is aborted.

37.6.1.3. Complexity Analysis

Let **|v|** be the size of the message. The complexity of this erasure-encoded RBC scheme can be estimated as:

- SEND: Roughly O(|v|) as the message is split and parts are sent.

- ECHO: This is the most dominant. It involves **N** nodes each sending parts of the message to **N** other nodes, leading to $O(N|v|)$ for the message parts themselves. Additionally, the Merkle tree verification adds a factor of **log N**, leading to an overall $O(\lambda N^2 \log N)$ for **NxN** communication. The additional $\boldsymbol{\lambda}$ factor is the cryptographic security parameter.

The overall complexity is thus approximately $O(N|v| + \lambda N^2 \log N)$. This provides an improvement when the message size **|v|** >> **Nlog N**.

### 37.7. Threshold Encryption

Threshold Public-Key Encryption (**TPKE**) lets anyone encrypt a message using a single public key (**PK**). Decrypting it requires the cooperation of a threshold of participants, each holding a share of the secret key. Specifically, if there are **N** total secret key shares, at least $\mathbf{f + 1}$ of them are required to decrypt the message, protecting against up to **f** Byzantine faults.

#### *37.7.1. API*

- **TPKE.Setup($\mathbf{1^\lambda}$) → PK, {$\mathbf{S_i}$}**: Generates the public key and distributes **N** unique secret key shares ($\mathbf{S_i}$) among the participants.
- **TPKE.Enc(PK, m) → C**: Encrypts a message **m** into ciphertext **C** using the public key.
- **TPKE.DecShare($\mathbf{S_i}$, C) → $\boldsymbol{\sigma_i}$**: A participant uses their secret share $\mathbf{S_i}$ to create a partial decryption $\boldsymbol{\sigma_i}$. This $\boldsymbol{\sigma_i}$ doesn't reveal the message on its own.
- **TPKE.Dec(PK, C, {i, $\boldsymbol{\sigma_i}$}) → m**: Combines at least $\mathbf{f + 1}$ partial decryption shares ($\boldsymbol{\sigma_i}$) to fully reconstruct the original plaintext message **m**.

#### *37.7.2. The Mathematics Behind It*

At its core, TPKE leverages Shamir's Secret Sharing Scheme [241]. This mathematical technique allows a secret (like the private key) to be divided into multiple pieces, such that only a predefined minimum number of pieces can reconstruct the original secret.

In a Byzantine system, non-faulty nodes will only release their decryption shares when the protocol demands it. Byzantine nodes, however, can share their shares among themselves. That's why, with up to **f** Byzantine nodes out of **N** total, it's essential to require at least $\mathbf{f + 1}$ decryption shares for successful decryption.

## 37.8. Common Coin

A common coin is a distributed protocol that enables a set of nodes to agree on a shared random bit (0 or 1). All participants will ultimately concur on the single value of this random bit.

The common coin has two essential properties:

- **Agreement**: All non-faulty nodes will output the same random bit.
- **Unpredictability**: The value of the random bit cannot be predicted by an adversary until a sufficient number of non-faulty nodes have contributed to its generation.

### *37.8.1. Realizing a Common Coin with Threshold Cryptography*

A common coin can be efficiently implemented using TPKE, specifically leveraging its underlying threshold signature scheme.

Instead of encrypting data, we use the threshold signature capabilities of a TPKE setup. In an **(N, f)** threshold signature scheme:

- There are **N** unique signature shares distributed among **N** parties.
- Any $\mathbf{f+1}$ of these **N** signature shares are required to reconstruct a complete, valid signature on a message.

Here's how this works to generate a common coin (due Cachin [242], in paper's Figure 12):

- Each non-faulty node multicasts its threshold signature share (**ThresholdSign(** $\mathbf{PK}$**,** $\mathbf{SK}_i$**,** **sid)**), where **PK** is the public key, $\mathbf{SK}_i$ is its individual secret key share, and **sid** is a unique, agreed-upon nonce that serves as the name for this specific common coin instance.
- Upon receiving $\mathbf{f + 1}$ valid signature shares for the **sid**, a node can combine them to form a complete signature. It then verifies this complete signature using **ThresholdVerify(PK, ...)**.

The bits of this final, verified signature (on **sid**) serve as the source of the common random bits for the coin. Since the signature is deterministically derived from inputs that are hidden until the threshold is met, its value is unpredictable to an adversary beforehand but universally agreed upon once reconstructed. This method ensures that all non-faulty nodes will eventually agree on the same random bits, as they will all reconstruct the identical signature from the same set of $\mathbf{f + 1}$ valid shares.

### *37.8.2. Complexity*

Common coin requires only one round of asynchronous communication. The communication cost is $O(\lambda N)$ where $\boldsymbol{\lambda}$ is the cryptographic security parameter (length of the verified signature).

## 37.9. Binary Agreement

Binary agreement is a specialized form of consensus where distributed nodes agree on a single 0 or 1 bit. It shares the core properties of any robust consensus protocol:

- **Termination (Liveness)**: Every non-faulty node eventually decides on a value.
- **Validity (Safety)**: The decided value must be proposed by at least one non-faulty node.
- **Agreement (Safety)**: All non-faulty nodes agree on the same decided bit.

### *37.9.1. Consensus and Binary Agreement*

Consensus is more general than binary agreement. Yet, binary agreement faces the same fundamental challenge: the FLP Impossibility. This means a deterministic binary agreement algorithm cannot guarantee both safety and liveness in an asynchronous network if even one node fails.

### *37.9.2. Asynchronous Binary Agreement with a Common Coin*

In an asynchronous network, the critical challenge for binary agreement (and general consensus) is ensuring liveness. As seen in protocols like Paxos, randomization is key to achieving this. A common coin provides this necessary randomization, allowing nodes to agree on a shared random bit to break deadlocks and ensure progress.

Here's a high-level overview of an Asynchronous Binary Agreement (ABA) protocol using a common coin (shown in paper's Figure 11, due Moustefaoui [243]):

**Initialization**: Each node $\mathbf{P}_i$ starts with an initial proposal $\mathbf{v}_i \in \{0, 1\}$.
**Repeated Rounds (k = 1, 2, ...)**:

- Each node $\mathbf{P}_i$ broadcasts its current estimate $\mathbf{e}_i$ for the bit. Upon receiving a sufficient number of valid estimates, nodes ECHO them.
- $\mathbf{P}_i$ collects votes. If it sees a clear majority (e.g., $\mathbf{2} * \mathbf{f} + \mathbf{1}$ for 0 or 1 in a Byzantine setting), it updates its estimate.
- If $\mathbf{P}_i$ has unambiguous support for a value (e.g., $\mathbf{N} - \mathbf{f}$ votes for 0), it tries to decide that value and broadcasts a DECIDE message.
- If $\mathbf{P}_i$ cannot make a deterministic decision (e.g., it sees support for both 0 and 1, or insufficient support for either), it invokes a common coin sub-protocol. This sub-protocol outputs a shared random bit $\mathbf{c} \in \{\mathbf{0}, \mathbf{1}\}$ that all non-faulty nodes will agree on. $\mathbf{P}_i$ uses the common coin $\mathbf{c}$ to update its estimate for the next round.
- If a node receives a DECIDE message for a value from enough peers (e.g., $\mathbf{2} * \mathbf{f} + \mathbf{1}$), it terminates and outputs that value.

This iterative process, driven by the random input from the common coin, guarantees probabilistic liveness: the protocol will eventually terminate for all non-faulty nodes, even if the exact number of rounds isn't fixed.

The communication complexity is $O(\lambda N^2)$ (the $\boldsymbol{\lambda}$ factor comes from the use of common coins) with a high probability of completing within O(1) rounds. Additionally, it can complete within O(k) rounds with a probability of $\mathbf{1} - \mathbf{2}^{-\mathbf{k}}$.

### 37.10. Common Subset

This is again a specialized form of consensus. Each node starts with its own private set of values. The goal is for all non-faulty nodes to eventually agree on a single, final set of values such that:

- **Termination (Liveness)**: All non-faulty nodes eventually output a set of values.
- **Agreement (Safety)**: All non-faulty nodes output the exact same set of values.
- **Validity (Safety)**: The output set must only contain values that were initially proposed by some non-faulty node. It doesn't contain any values from Byzantine nodes.

*37.10.1. Asynchronous Common Subset*

Asynchronous Common Subset (ACS) is a fundamental building block that can be implemented using two key primitives: RBC and ABA. RBC ensures that all non-faulty nodes receive the same set of messages from a particular sender. ABA is then used to decide which of these sets to collectively finalize. Note that ABA can begin before all RBCs are complete due to the asynchronous nature of communication (no known **Δ** for message delivery).

37.10.1.1. Ben-Or Algorithm

The Ben-Or protocol (as shown in Figure 4 in the paper) is a classic example of this implementation.

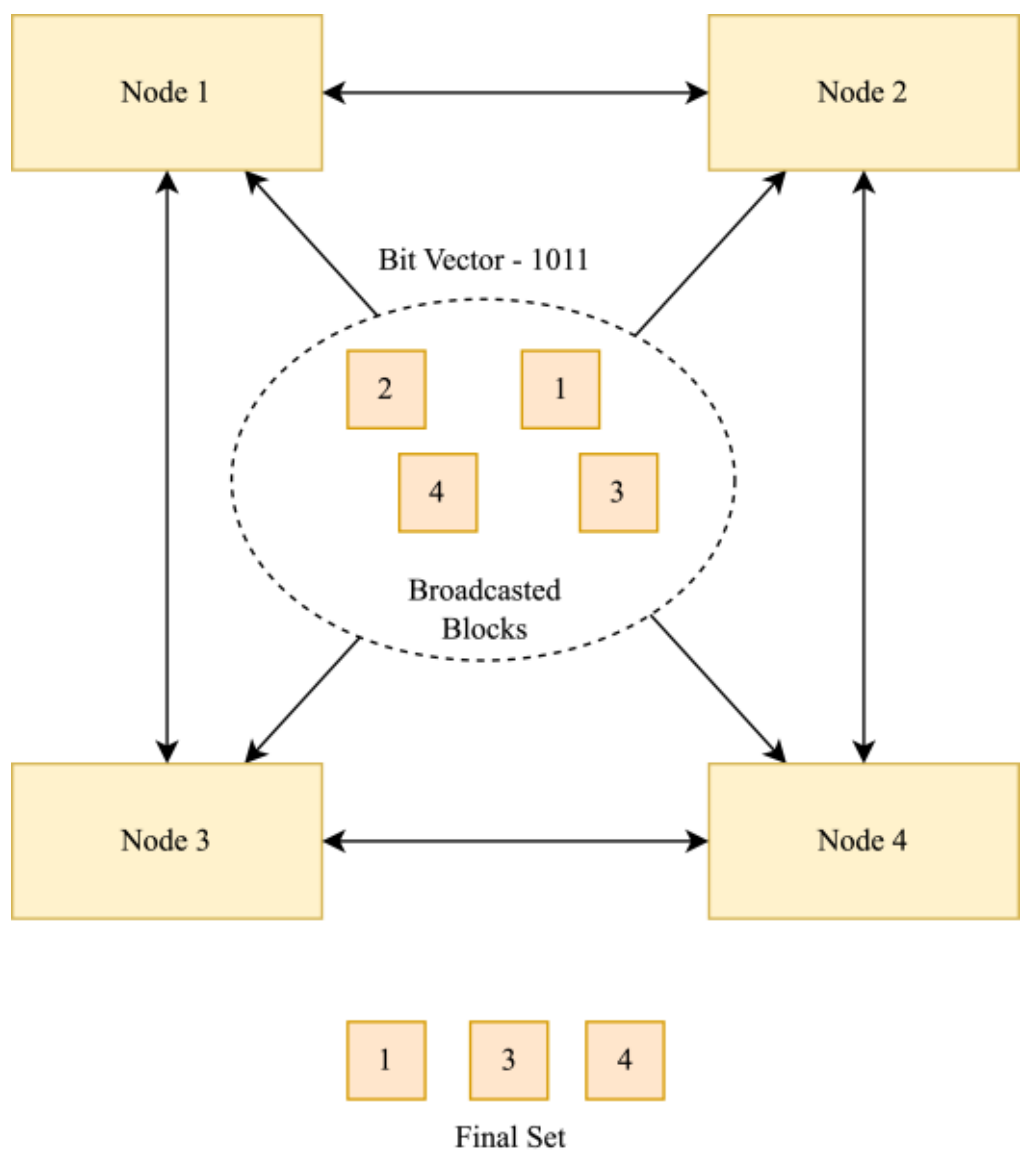

**Figure 191**: Ben-Or algorithm.

As shown in Figure 191, every node **i** simultaneously initiates an RBC ($\mathbf{RBC}_i$) to broadcast its message. Concurrently, ABA ($\mathbf{BA}_i$) instances are run to agree on **N** bits, where each bit corresponds to whether a specific node's broadcasted message should be included in the common subset. If $\mathbf{BA}_i$ for node **i** is finalized as 1, that message will be included.

A key aspect of Ben-Or is that RBC and ABA can run in parallel. If a node observes at least $\mathbf{N-f}$ bits in the bit vector set to 1, it can assume that any unreceived broadcasts from other nodes will have their corresponding bits set to 0. This mechanism accounts for Byzantine failures, where up to **f** nodes might not send any messages at all. While it might seem counterintuitive for ABA to finish before RBC, this is an expected behavior of the protocol and doesn't hinder its correctness.

37.10.1.2. A Critical Downside: Censorship

Despite its utility, the discussed ACS protocol has a significant drawback: censorship vulnerability. Since the protocol can conclude when only $\mathbf{N - f}$ bits in the bit vector are 1, an adversary can continuously block messages (and therefore transactions) from a single non-faulty node. This prevents that node's message from ever being included in the common agreed-upon set, effectively censoring its contribution.

37.10.1.3. Complexity Analysis

Total cost of RBC = $\mathbf{N} * \mathbf{Cost\ of\ single\ RBC}$ = $O(N^2|v| + \lambda N^3 \log N)$
Total cost of ABA = $\mathbf{N} * \mathbf{Cost\ of\ single\ ABA}$ = $O(\lambda N^2)$
Total cost = $O(N^2|v| + \lambda N^3 \log N)$ given $|\mathbf{v}| >> \boldsymbol{\lambda}$.

**37.11. Tor**

Tor is a public service on the internet that enables anonymous message exchange, ensuring that a message's recipient can't identify the sender's address. This makes it an excellent tool for bypassing censorship by centralized services.

At its core, Tor functions as a distributed multi-hop proxy system. It uses a routing protocol called **onion routing**. When a client sends a message, it typically travels through multiple hops - for instance, three - before reaching its destination. The client first obtains the public keys of these hops from the Tor public directory. Then, the message undergoes a layered encryption process, as shown in Figure 192:

- The message is initially encrypted with the server's public key.
- This encrypted message, along with the server's IP, is then encrypted with the public key of the last hop.
- This process continues backward, with each layer of encryption corresponding to a successive hop.

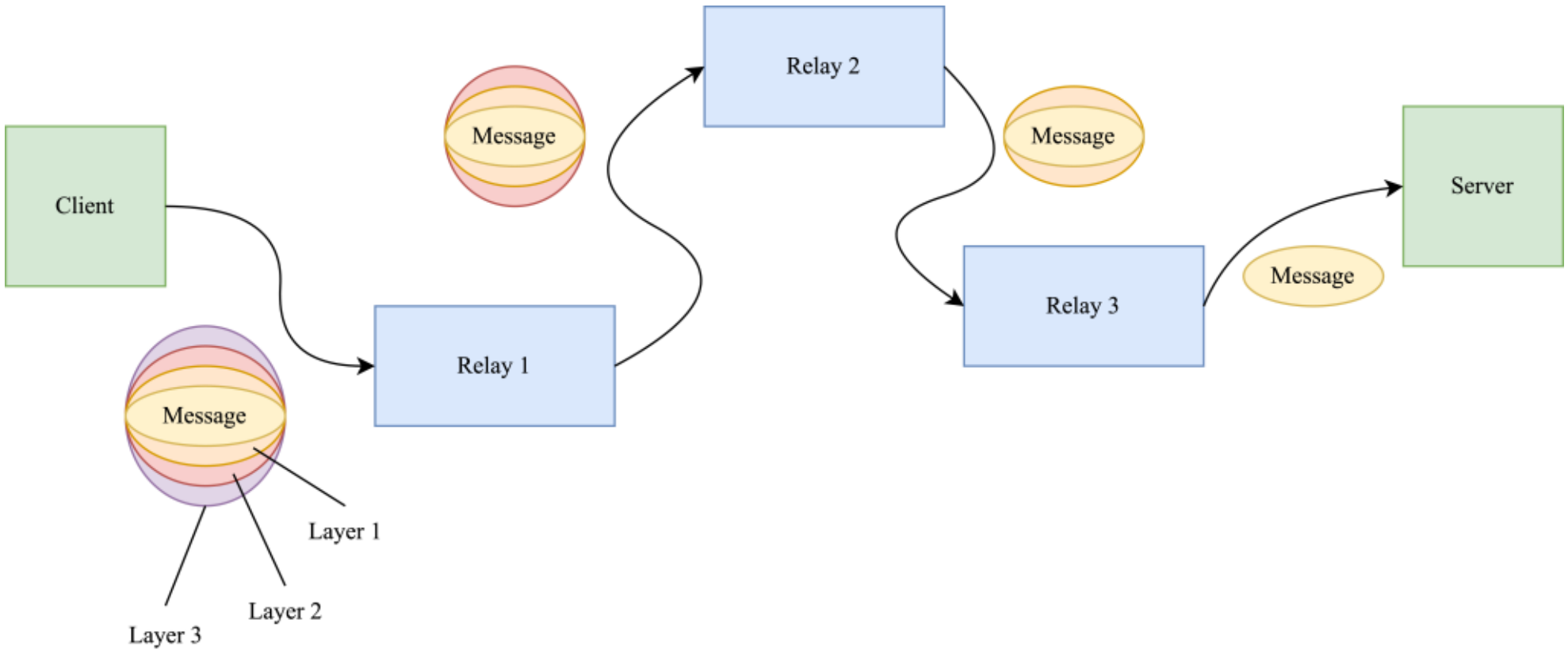

**Figure 192**: Onion routing.

The result of this onion routing is that each hop in the relay only knows the immediately preceding and succeeding hops, not the entire route. Similarly, the final server only sees the last hop that delivered the message, effectively concealing the original sender's identity.

### 37.12. HoneyBadger

HoneyBadger is a consensus protocol designed for:

- Asynchronous networks.
- Up to $\mathbf{f < N/3}$ Byzantine faults.
- Trusted setup (i.e. permissioned setup).

HoneyBadger ensures safety under all conditions. However, unlike non-randomized protocols, which cannot guarantee liveness under asynchronous conditions (FLP impossibility), HoneyBadger leverages randomized algorithms to achieve probabilistic liveness. This means it ensures liveness with a high probability, as its underlying components, such as ACS, rely on common coins that also provide probabilistic liveness.

#### *37.12.1. Goals*

Beyond simply achieving consensus in an asynchronous environment, HoneyBadger aims for two key objectives:

- **High Transaction Throughput**: The protocol prioritizes the volume of transactions processed over individual transaction latency. It achieves this by batching and processing large sets of transactions simultaneously, leading to communication efficiency.
- **Censorship Resilience**: As discussed earlier, standard ACS is susceptible to censorship. HoneyBadger counters this vulnerability by integrating TPKE, which requires a minimum number of parties to decrypt messages, thus preventing individual malicious actors from censoring transactions.

The problem of atomic broadcast in BFT systems can be reduced to ACS if transactions are batched. ACS provides an unordered set of sub-batches from each node, which are then ordered using a predefined property (e.g., lexicographically).

*37.12.2. The Algorithm*

Figure 1 in the paper illustrates the core HoneyBadger algorithm. Here's a breakdown of its steps:

- **Step 1: Batching and Encryption**: Each node randomly selects a batch of transactions of size **B** from its local pool. This selection can be based on priority or FIFO. The selected batch forms a message, which is then encrypted using the public key of the threshold encryption scheme, i.e., **TPKE.Enc(PK, m)**. This encryption ensures that the message cannot be decrypted unless at least $\mathbf{f + 1}$ parties cooperate.
- **Step 2: ACS Input**: The encrypted messages from all nodes are then input into the ACS protocol. The final vector is agreed-upon (encrypted) messages containing the transactions.
- **Step 3: Decryption**: For every selected encrypted message, each node obtains a decryption share. These shares are then broadcast using RBC.
- **Step 4: Block Generation**: Once all selected messages are decrypted, the transactions within them are sorted according to a common order (e.g., by a generated ID or hash). This final sorted list of transactions forms a new block in the blockchain.

The crucial sorting step in Step 4 is what transforms ACS (a common set determination algorithm) into an atomic broadcast or total order broadcast consensus algorithm, enabling the final block to be consistently ordered across all non-faulty nodes.

*37.12.3. Complexity*

Cost of ACS = $O(N^2|v| + \lambda N^3 \log N)$
Cost of decryption = $O(\lambda N^2)$

Note, the batch size from each node, $|\mathbf{v}|$, is set to $\Omega$ ($\lambda N \log N$) and hence the total message size across all nodes is $\Omega$ ($\lambda N^2 \log N$).

Cost of ACS = $O(\lambda N^3 \log N)$
Total cost = $O(\lambda N^3 \log N)$

Section 4.5 has more analysis and appendix B has some detailed proofs for them.

**37.13. Evaluation**

HoneyBadger's performance characteristics reveal that it's not optimized for small batch sizes. The primary communication overhead stems from RBC and the decryption process in threshold cryptography.

Figure 5 in the paper illustrates that the communication cost remains relatively constant even as batch sizes increase up to 128 transactions. Beyond this point (for **N = 8, f = 2**), the batch size becomes the dominant factor in the total communication cost. Interestingly, with a larger number of nodes, the batch size becomes less of a dominant cost factor. For instance, with **N = 128** and **f = 32**, the batch size doesn't dominate the cost until it reaches 16,000 transactions.

The authors conducted experiments on Amazon EC2, though note that the test environment lacked authentication setup (which would incur additional overhead in a production environment with TLS). As shown in paper's Figure 6, throughput consistently increased with larger batch sizes, even when varying the number of nodes (with **f = N/4**). Specifically, the system achieved 20,000 transactions/second with 40 nodes and 1,500 transactions/second with 104 nodes.

While throughput increases, so does latency, as depicted in Figure 7 in the paper. Larger batch sizes necessitate longer processing latencies, with transaction latencies observed in the order of minutes. The positive slope of this plot indicates that throughput can still be increased, avoiding full saturation.

Compared to PBFT, HoneyBadger achieves higher throughput, even though their asymptotic communication costs are similar. This is because HoneyBadger distributes communication across all nodes, whereas PBFT tends to create hotspots around leader nodes.

To assess HoneyBadger's performance over Tor, the authors set up HoneyBadger nodes on single machines and configured communication paths through Tor, using five random relay hops. HoneyBadger achieved a transaction volume of 800 transactions/second. However, latency proved to be highly variable; a single message could be delayed by up to 316 seconds, with an average delay of 12 seconds (variance = 2208). Crucially, HoneyBadger's asynchronous design allowed it to tolerate this significant network variance.

### 37.14. Paper Remarks

This paper is incredibly challenging, often requiring multiple readings to grasp its inherent complexities. The material underscores the vastness and density of the blockchain space, especially regarding proof construction, which is a significant mental exercise. Readers are advised to make use of write-ups on Decentralized Thoughts [244] blog to aid comprehension.

# 38. Availability in Globally Distributed Storage Systems

(38) Ford, D.; Labelle, F.; Popovici, F.; Stokely, M.; Truong, V.-A.; Barroso, L.; Grimes, C.; Quinlan, S. Availability in Globally Distributed Storage Systems. In Proceedings of the 9th USENIX Symposium on Operating Systems Design and Implementation; 2010.

In data centers, failures are inevitable. The jobs are relatively easy to restore upon failures - as binaries can simply be executed on different machines. However, protecting data is more complex. How do we ensure data availability during outages, such as a single machine becoming unavailable or, in worse cases, an entire region becoming unreachable due to network faults? This is extremely crucial for organizations like cloud providers, as maintaining user trust depends heavily on the high availability of data. This chapter will address these challenges and the solutions used to ensure data resilience.

This paper, published by Google in 2010 and presented at Operating System Design and Implementation (OSDI) 2010, is quite challenging due to its highly mathematical nature. The paper covers two independent main topics:

- The first section examines the **availability** and **Mean Time To Failure** of individual hardware components like **disks** and **nodes**. It also delves into **failure bursts** and **domain-related failures** (such as rack failures), which are significant sources of correlated failures.
- The second, entirely independent section discusses how to handle failures and thrive. It builds:
    - Probabilities of state transitions for stripes accounting for all possible failure burst sizes. It uses cell simulation to verify these probabilities.
    - A **Markov model** that incorporates the calculated probabilities, where each failure is independent but can impact multiple nodes and chunks, thus reflecting correlated failures.
    - A **non-Markov model** with constant failure and recovery rates, treating chunk failures as independent for less accurate calculations.

### 38.1. Part 1: Availability

In a typical data center cluster, thousands to tens of thousands of machines host millions of **nodes**, which are essentially processes running on these machines. Often, a group of these nodes is dedicated to a specific distributed system, like storage. For example, some nodes might be part of a distributed file system, storing large volumes of data.

This paper specifically examines the availability of distributed storage systems. Storage nodes in a distributed system can become unavailable for a variety of reasons, each with different implications for recovery time and data integrity:

- **Node Restarts**: These are the quickest to recover from, typically taking only tens of seconds. It's often just a matter of restarting a process.
- **Planned Reboots**: For maintenance, repairs, or upgrades, machines (running the node) may undergo planned reboots. These can take anywhere from several minutes to a few hours.
- **Unplanned Reboots**: Unexpected events like kernel crashes also lead to reboots. Similar to planned reboots, these can cause downtime ranging from several minutes to hours.
- **Hardware Corruption**: Data can become corrupted even on seemingly healthy hardware. Studies have shown a small but present rate of checksum mismatches, indicating corruption in 1 in $10^6$ to $10^7$ data blocks in Google's data centers.
- **Hardware Failures**: Physical hardware failures are a significant concern. The annual replacement rate for disks ranges from 2% to 4%. When considering an entire fleet of storage nodes, the overall failure rate is between 3.9% and 8.3% annually. These failures necessitate more extensive recovery procedures, potentially involving data reconstruction from replicated or erasure-coded chunks.

From paper's Figure 3, Node restarts and planned reboots are quite common occurrences. Unplanned reboots also happen, albeit less frequently.

To quantify availability, two key metrics are used:

- **Average Availability**: This is calculated as the total uptime divided by the sum of total uptime and total downtime.

$$\textbf{Average Availability} = \frac{\sum_i \textbf{Uptime}}{\sum_i \textbf{Uptime} + \sum_i \textbf{Downtime}}$$

- **Mean Time To Failure (MTTF)**: This metric represents the average time a system operates correctly before a failure occurs.

$$\textbf{MTTF} = \frac{\textbf{Number of Failures}}{\textbf{Total Uptime}}$$

A closer look at the causes of storage node unavailability, as shown in Figure 4 in the paper, reveals that planned reboots are the most frequent, followed by node restarts.

Table 2 in the paper provides insights into the MTTF for various components:

- A disk has an MTTF of 10-50 years.
- A node has a significantly shorter MTTF of 4.3 months.
- A rack has an MTTF of 10.2 years.

*38.1.1. Data Replication*

In a distributed storage system, files are broken down into **stripes**, and these stripes are then replicated into **chunks** to ensure data availability and fault tolerance. There are two popular methods for replicating file chunks: **replicated encoding** and **erasure encoding**.

38.1.1.1. Replicated Encoding

Replicated encoding is where file chunks are simply duplicated, with each copy stored on a separate storage node, as shown in Figure 193. A single node can hold chunks from various stripes.

Common replication factors include **r = 2** or **r = 3**. The storage space required directly increases with the replication factor.

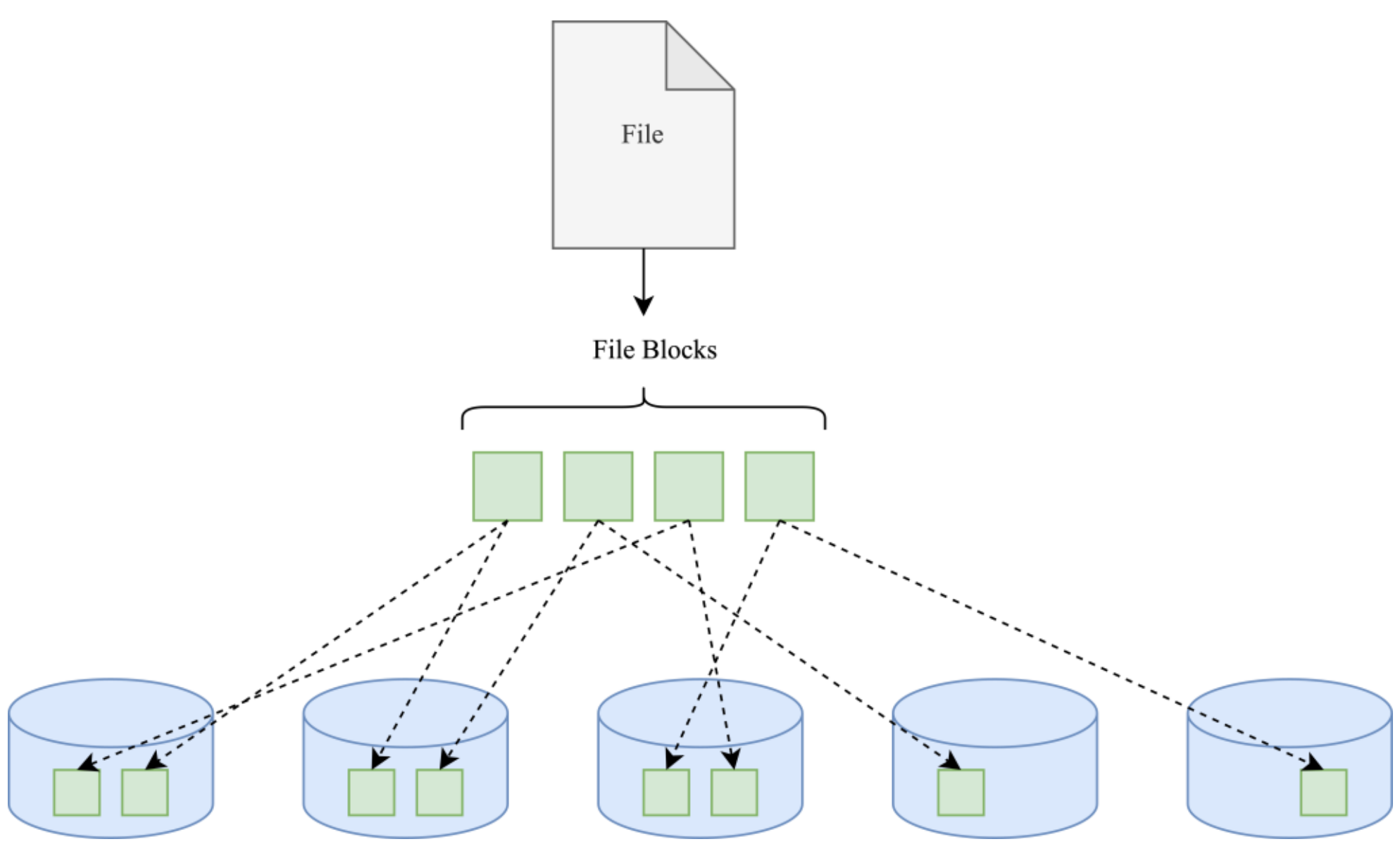

**Figure 193**: Replicated encoding.

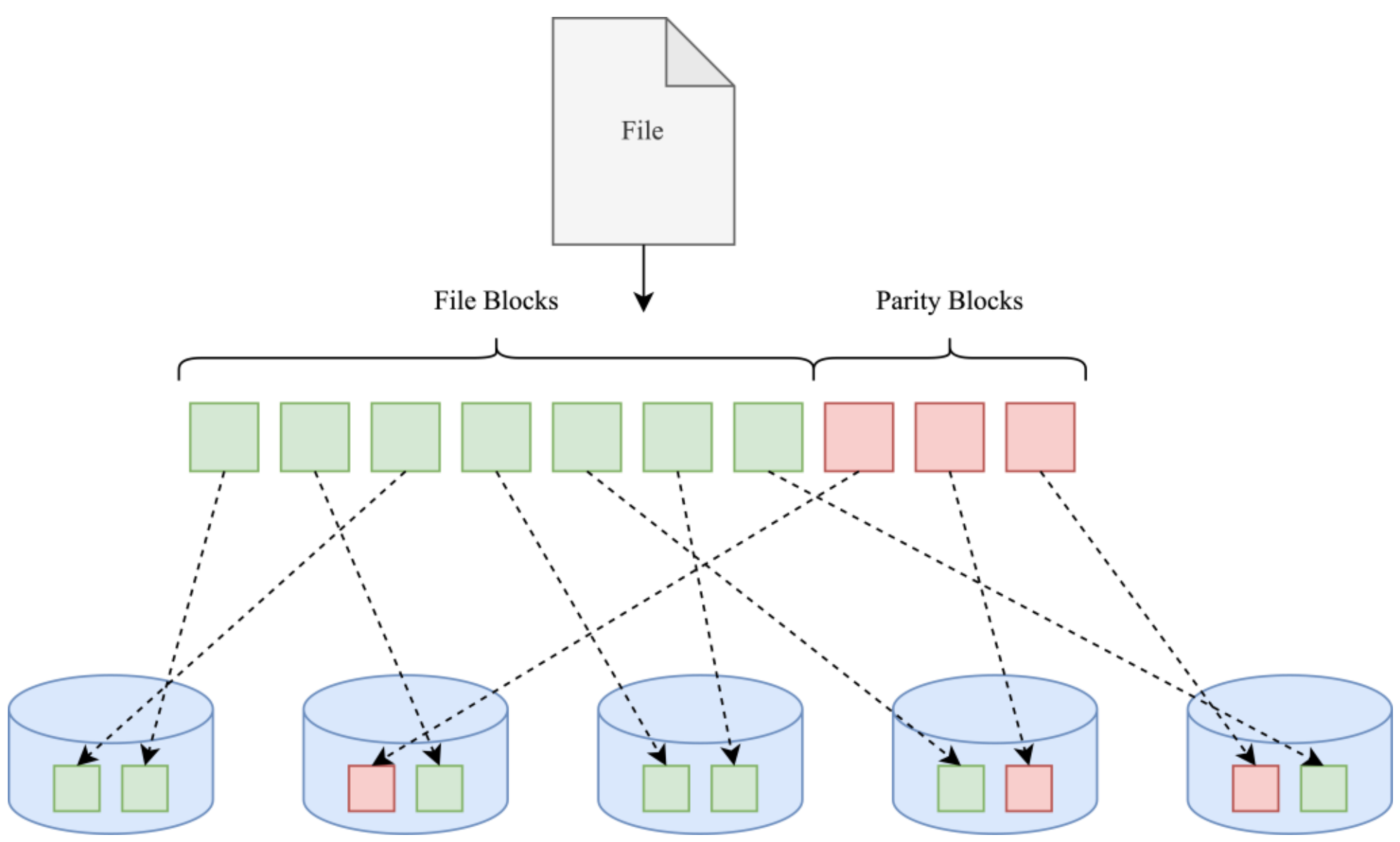

**Figure 194**: Erasure encoding.

#### 38.1.1.2. Erasure Encoding

A file is divided into **k** data blocks. From these **k** blocks, **m** additional parity blocks are computed. The total number of blocks then becomes **n** = **k** + **m**, as shown in Figure 194.

- **k**: Number of original data blocks.
- **m**: Number of generated parity blocks.
- **n**: Total number of blocks (**n** = **k** + **m**).

The key advantage is that to retrieve the original data, you only need any **k** of the **n** total blocks. This means you can lose up to **m** blocks and still recover your data.

**Reed-Solomon codes** are the most widely used erasure codes, offering the maximum possible fault tolerance for a given amount of redundancy.

### *38.1.2. Correlated Failures*

Correlated failures occur when a single event or root cause leads to a large number of nodes failing simultaneously or in quick succession. These are particularly impactful in distributed storage environments because they can wipe out entire sections of data.

For example, when a top-of-rack (ToR) switch goes down, every storage node connected to that switch within the rack becomes unavailable, leading to the loss of all data stored on them. Similarly, a power outage affecting a specific area or data center can cause widespread, correlated failures.

#### 38.2.2.1. Failure Burst

A failure burst refers to multiple component failures grouped together within a short timeframe. The authors defined a failure burst as a cluster of failures where the time gap between any two successive failures is less than 120 seconds, as shown in Figure 5 in the paper. While individual failures within a burst must be close (within 120 seconds), the total duration of a burst itself can exceed 120 seconds.

As illustrated in Figure 6 in the paper, when analyzing the percentage of node failures for different clustering window sizes (with a minimum failure size of 10 nodes), the curve tends to flatten out after 120 seconds. This suggests that 120 seconds is an optimal burst size for effectively capturing most clustered failure events.

With this 120-second burst size, the likelihood of a purely random (non-correlated) failure being included in a burst is relatively low, at about 8%. The probability of a random failure being part of a burst involving at least 10 nodes is a minuscule 0.068%. These low probabilities indicate that failure bursts are effective at pinpointing large, correlated failures rather than just isolated random events.

Figure 7 in the paper visualizes different types of failure bursts. The steep curve represents scenarios like power outages, where many nodes fail rapidly in quick succession. The slower, more even-paced curve often points to controlled activities such as rolling reboots or upgrade processes.

#### 38.1.2.2. Rack Affinity

A data cluster consists of machines organized into racks. Within each rack, all machines are interconnected in a mesh topology. For communication across racks, machines connect via ToR switches, as shown in Figure 195. These ToR switches are often single points of failure and a significant source of outages, highlighting the importance of quantifying their failure rate.

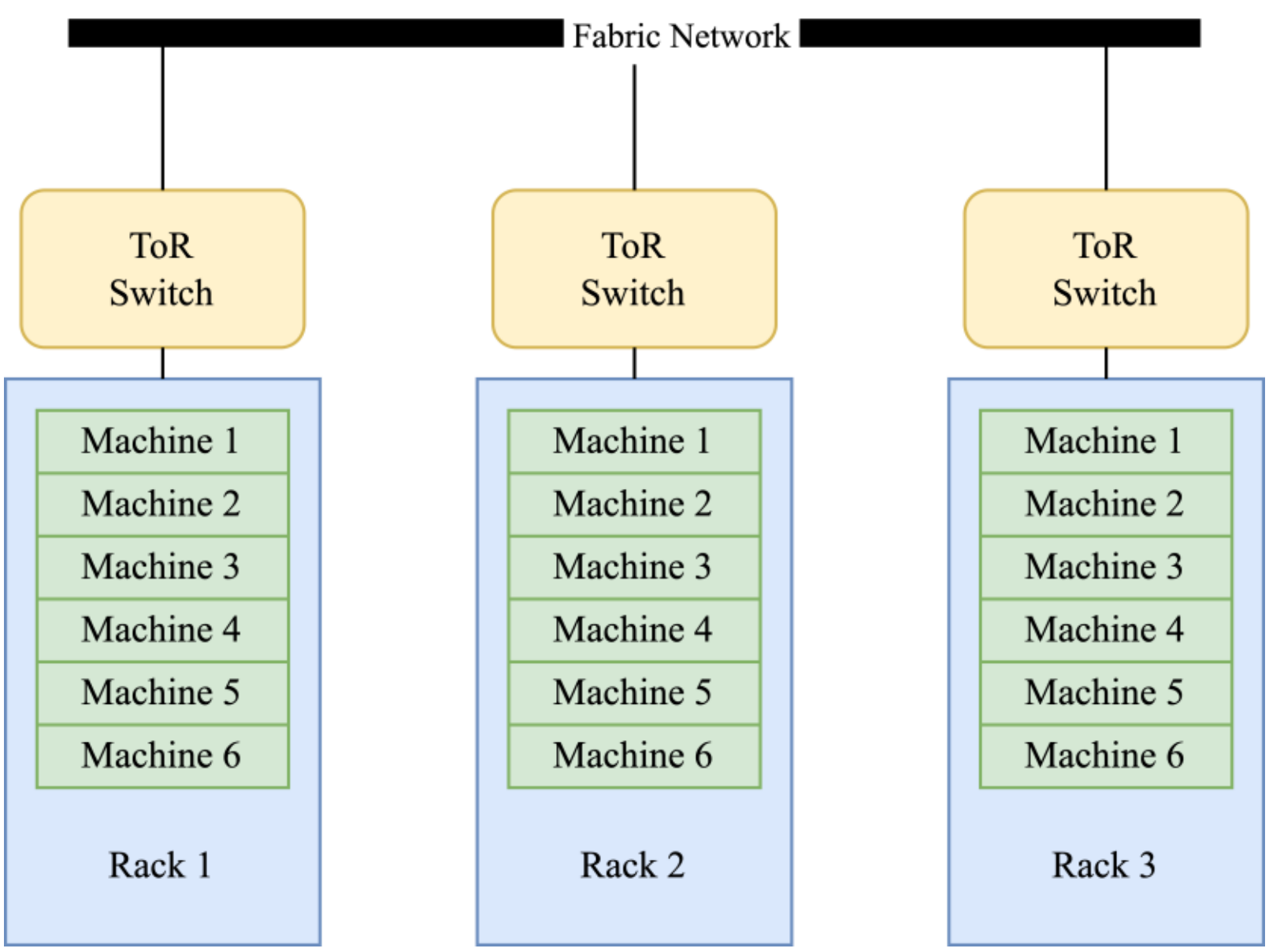

**Figure 195**: ToR switches.

The **Rack Score (R)** quantifies the concentration of failures of **N** nodes within a data center rack. It's calculated as:

$$\mathbf{R} = \sum_i \frac{\mathbf{k}_i * (\mathbf{k}_i - \mathbf{1})}{\mathbf{2}}$$

where $\mathbf{k}_i$ is the number of failures on rack **i**. **R** will always be an integer. A larger **R** value indicates a higher concentration of failures within a single rack.

**Rack Affinity** for a given rack score **R** is a probability metric defined as:

$$\textbf{Rack Affinity} = \mathbf{P}(\textbf{rack score of a randomly chosen N nodes} < \mathbf{R}) + \mathbf{0.5} * \mathbf{P}(\textbf{rack score of a randomly chosen N nodes} = \mathbf{R})$$

In simpler terms, it measures how likely it is for a random failure event of the same size (**N**) as the observed failure to result in a rack score less than or equal to the observed **R**.

Higher Rack Affinity (closer to 1):

- Implies a greater probability of observing the given rack score or a lower one.
- This means the observed **R** value is large for a given number of failed nodes (**N**).
- Hence, it indicates that the failure burst is highly concentrated within one or a few racks.

Lower Rack Affinity (closer to 0.5):

- Suggests that the observed **R** value is relatively small for a given **N**.
- Hence, it indicates that the failure burst is more spread out across multiple racks.

This probability (Rack Affinity) can be efficiently calculated using dynamic programming, considering the total number of failure nodes (**N**) and the observed Rack Score (**R**).

### 38.2. Part 2: Handling Failures

To safeguard against data unavailability caused by failures, data replication across different nodes is crucial. When a node fails, a recovery process begins to restore the data chunks stored on it. These chunks are recovered from their existing copies on other nodes. The recovery queues prioritize data stripes that have lost the most chunks. The speed of this recovery is constrained by the bandwidth of individual disks, nodes, and even entire racks.

### *38.2.1. Rack-Aware Placement Policy*

As previously mentioned, rack failures are a common cause of storage node loss. To mitigate this, a **rack-aware placement policy** ensures that each chunk of a data stripe is replicated on nodes located in different racks.

This policy allows us to define the probability that **k** chunks out of **n** are affected by failures as:

$$\mathbf{P} = \frac{\text{Total number of ways to place a stripe of size } \mathbf{n} \text{ in a cell}}{\text{Number of ways to place a stripe of size } \mathbf{n} \text{ with } \mathbf{k} \text{ failures}}$$

This probability can be calculated using dynamic programming.

Figure 10 in the paper illustrates the expected MTTF of a data stripe under different failure burst sizes:

- **Small**: Up to 0.1% of nodes impacted.
- **Medium**: 0.1% to 1% of nodes impacted.
- **Large**: 1% to 10% of nodes impacted.

The data shows that as the failure burst size increases, the stripe's MTTF significantly decreases by orders of magnitude. A higher replication factor, such as RS = (20, 10), provides greater resilience to failures. Conversely, with a replication factor of R = 1, the MTTF is considerably smaller.

#### 38.2.1.1. Cell Simulation

The authors also conducted a simulation of node failures to assess their impact on stripe availability (defined as the number of stripes unavailable for 15 minutes). The results, displayed in Figure 11 in the paper, demonstrate that the combinatorial calculations (in the previous section) closely match the actual observed unavailability.

*38.2.2. Markov Chains*

A **stochastic process** is a mathematical concept that describes a sequence of random variables within a probability space, often interpreted as evolving over time.

A Markov chain is a specific type of stochastic process. Its key characteristic is that the probability of any future event depends only on its most recent state, not on the sequence of events that led to it.

Consider a two-state (**A** and **B**) Markov process as an example, as shown in Figure 196.

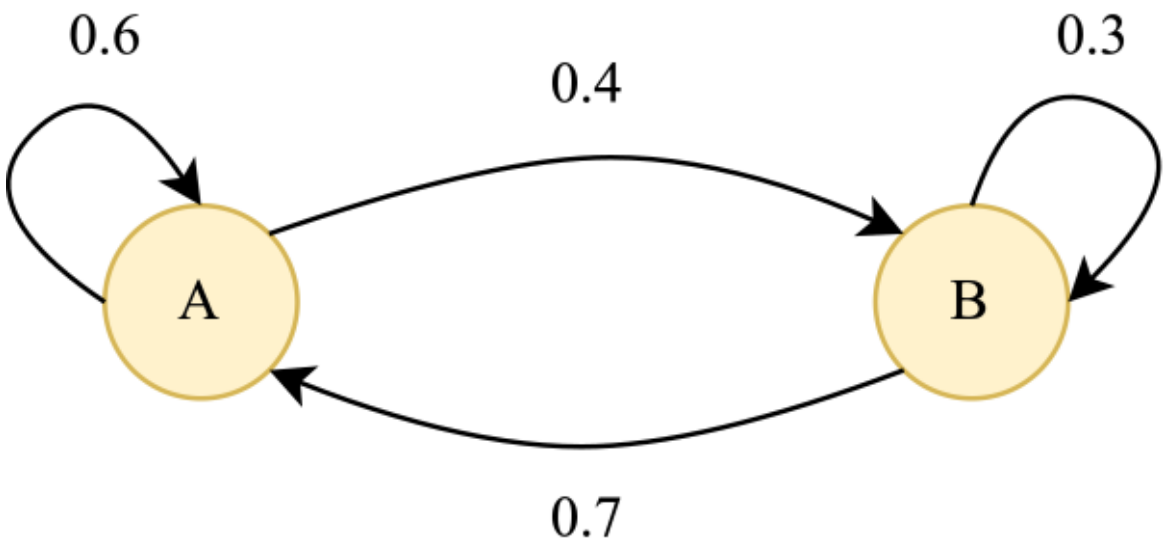

**Figure 196**: A two-state Markov process.

**States**: We can represent the current state as a vector. For instance, **[ 1 0 ]** means there's a 100% probability of being in state **A**. An initial state could also be probabilistic, like **[ 0.25 0.75 ]**.

**Transitions**: The movement between states is governed by a transition matrix (**T**). For example:

$$\mathbf{T} = \begin{bmatrix} \mathbf{0.6} & \mathbf{0.7} \\ \mathbf{0.4} & \mathbf{0.3} \end{bmatrix}$$

In this matrix, each column represents the probabilities of transitioning from a specific state, and these probabilities must sum to 1. For instance, from state **A** (first column), there's a 60% chance of staying in **A** and a 40% chance of moving to **B**.

**Predicting Future States**: With the current state vector ($\mathbf{S}_i$), one can predict any future state ($\mathbf{S}_{i+1}$) using the formula:

$$\mathbf{S}_{i+1} = \mathbf{T} \times \mathbf{S}_i$$

To find the state after **n** steps ($\mathbf{S_n}$) from an initial state ($\mathbf{S_0}$):

$$\mathbf{S_n} = \mathbf{T^n} \times \mathbf{S_0}$$

Interestingly, the state vector as **n** approaches infinity ($\mathbf{S_\infty}$) converges to an eigenvector [245] of the transition matrix **T**. This eigenvector represents the steady-state probabilities of being in each state after a very long time, regardless of the initial starting point.

38.2.2.1. Markov Model for Stripe Availability

Previous failure models, which focused on calculating probabilities, considered failure bursts of various sizes but didn't account for any correlations between them. In contrast, the Markov model introduces correlated failures, like those affecting multiple disks or machines.

The Markov model makes a key assumption: individual correlated failure events are independent. For example, a single disk failure might impact multiple data chunks (thereby making it a correlated event), however, the model assumes this failure is independent of any other disk failures. This assumption places the model's robustness in a specific range:

- It's weaker than a "perfect" model where failure events are interdependent. Consider a scenario where one disk fails, and others from the same manufacturing batch follow shortly after their warranty expires - the Markov model doesn't fully capture this dependency.
- It's stronger than a model where every single chunk failure is considered independent. We'll explore that specific case in more detail later.

In this Markov model, a stripe transitions into an unavailable state once fewer than **r** chunks remain available, as shown in Figure 197.

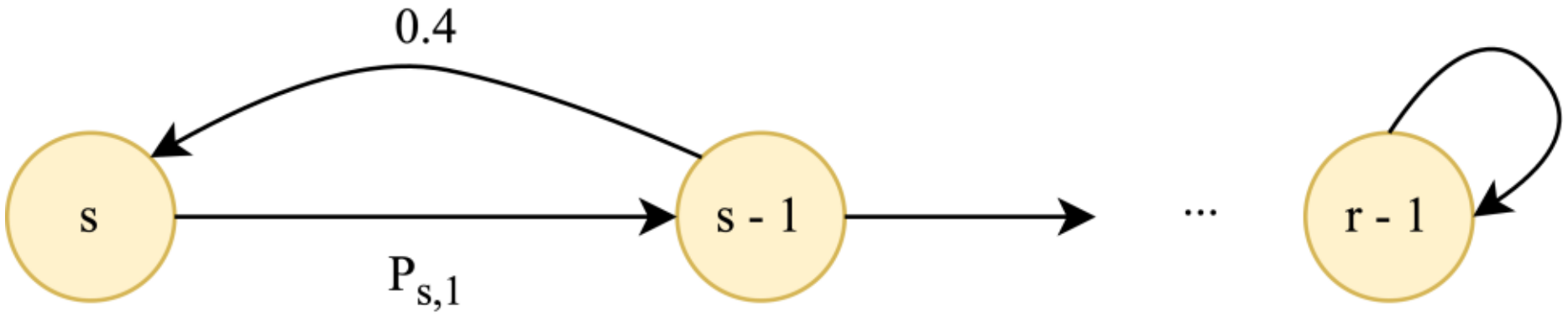

**Figure 197**: Markov model for stripe availability.

Let's define the probabilities and rates of these transitions:

$\mathbf{P}_{i,\mathbf{j}}$ represents the probability of transitioning from state $\mathbf{i}$ to state $\mathbf{j}$ due to a random failure. This failure results in the loss of $\mathbf{i} - \mathbf{j}$ chunks.

$\boldsymbol{\lambda}$ is the overall failure rate across all possible independent failures (like node and disk failures) and $\boldsymbol{\lambda}_{i,\mathbf{j}}$ is the rate of transition for stripes from state $\mathbf{i}$ to state $\mathbf{j}$. Since each failure has a $\mathbf{P}_{i,\mathbf{j}}$ probability and all failures are considered independent, we can express this as:

$$\boldsymbol{\lambda}_{\mathbf{i},\mathbf{j}} = \boldsymbol{\lambda}\mathbf{P}_{\mathbf{i},\mathbf{j}}$$

The rate of transitioning from any available state $\mathbf{i}$ to the unavailable state $\mathbf{r} - \mathbf{1}$ (meaning the stripe becomes unavailable) is the sum of probabilities of all possible transitions from $\mathbf{i}$ that result in $\mathbf{r} - \mathbf{1}$ or fewer chunks. This is expressed as:

$$\boldsymbol{\lambda}_{i,\mathbf{r}-\mathbf{1}} = \boldsymbol{\lambda} \sum_{\mathbf{j}=\mathbf{0}}^{\mathbf{r}-\mathbf{1}} \mathbf{P}_{i,\mathbf{j}}$$

This means we sum the probabilities of going from state $\mathbf{i} \rightarrow \mathbf{i} - \mathbf{1}$, $\mathbf{i} \rightarrow \mathbf{i} - \mathbf{2}$, $\mathbf{i} \rightarrow \mathbf{i} - \mathbf{3}$, and so on, all the way down to $\mathbf{r} - \mathbf{1}$ (the threshold for unavailability).

The authors also model recovery as a serial recovery process operating at a constant per-chunk rate, $\boldsymbol{\rho}$. Using both the failure rate ($\boldsymbol{\lambda}$) and this recovery rate ($\boldsymbol{\rho}$), we can calculate the mean time to failure for the system. This calculation essentially involves determining the expected time for a random walk [246] to reach the $\mathbf{r} - \mathbf{1}$ state, signifying a failure.

38.2.2.2. Findings

The authors employed a Markov model to calculate the MTTF of data stripes under various independent correlated failures, yielding following observations:

**Model Accuracy in Failure Bursts**

The model effectively captures the impact of failure bursts. For instance, in a scenario where dozens of nodes failed, the actual MTTF was 1.76e+6 days, while the model predicted 5e+6 days. This demonstrates that the model's predictions are accurate in terms of magnitude.

### Rack Failures

The model successfully differentiates between failure bursts that threaten availability (by spanning multiple racks) and those that do not. In an event involving multiple rack failures, the model predicted an MTTF of 5.77e+8 days, compared to an actual MTTF of 29.52e+8 days.

### Recovery Rate Impact

The authors also theoretically examined a scenario where chunks fail independently. In this case, the transition rate from state $\mathbf{i}$ to $\mathbf{i - 1}$ is $\mathbf{i\lambda}$, where $\boldsymbol{\lambda}$ is the failure rate of a single chunk. This is because any of the $\mathbf{i}$ available chunks can fail independently. The transition rate from state $\mathbf{i}$ to $\mathbf{i + 1}$ is $\boldsymbol{\rho}$.

Interestingly, the stripe MTTF derived from this independent failure model (though the derivation is not included in the paper) reveals that reducing the recovery rate can significantly improve the MTTF. Specifically, if the recovery rate is reduced by a factor of $\boldsymbol{\mu}$, the stripe MTTF increases by a factor of $\boldsymbol{\mu}^{\mathbf{s-r}}$.

### Impact of Correlated Failures on Unavailability and Replication

The presence or absence of correlated failures profoundly impacts unavailability and the effectiveness of replication:

- **Without correlated failures**: A 10% reduction in recovery time leads to a 19% decrease in unavailability. The effect of replication also greatly enhances availability.
- **With correlated failures**: A 90% reduction in recovery time only results in a 6% decrease in unavailability. Furthermore, replication does not increase the MTTF as significantly when correlated failures are present.

## 38.3. Paper Remarks

This paper, while conceptually straightforward, is mathematically dense. The main challenge lies in the omission of numerous mathematical derivations, which, due to space constraints, makes it difficult to fully grasp the intricacies of the models. Nevertheless, it's an excellent resource for understanding how cloud storage systems achieve data availability.

# Epilogue

Our journey began with the bedrock of systems: operating systems. We traced the path from microprocessors and memory to the virtualization of those resources, and followed the evolution of storage and network from traditional to their modern variants.

We also scaled upward into data processing, examining the shift from batch to streaming models. We navigated the landscape of data stores - learning SQL, NoSQL, and specialized databases - and also saw how the data processing concepts converge in the mechanics of distributed query processing and data joins.

On the theoretical side, we dissected the nuances of consensus through various protocols and analyzed the role of the two-phase commit in database integrity. We also grappled with the "hardness" of clock synchronization, uncovering why time remains one of the greatest hurdles in distributed systems.

In the realm of cloud computing, we explored the serverless era, including object stores, serverless computation, and the orchestration systems that handle various workloads.

Finally, we looked at the cutting edge: how machine learning systems utilize hardware accelerators and how decentralized protocols like IPFS and blockchain are implemented.

I recognize that this journey was often daunting. Many of the papers we studied are notoriously difficult, but by wrestling with them, you have significantly leveled up your technical intuition.

The journey does not end here; in many ways, this is just the beginning. The world of computing systems is expanding at an unprecedented pace. Every shift in AI, quantum, or blockchain is an invitation for systems engineers to rethink established designs and build what’s next. You have made it through the fire - you are now ready to take this knowledge and make your mark on the world.

I hope you enjoyed viewing the world of systems through the lens of these foundational papers. I wish you the very best in your next adventure.